%% file: thesis.tex
\documentclass[phd,final,oneside]{qmul_thesis}
\usepackage{graphics}
\usepackage{graphicx} 
\usepackage{float}
\usepackage{myfigcap}
\usepackage{amsmath}
\usepackage{amssymb}
\usepackage{amsthm}
\usepackage{slashed}
\usepackage[usenames,dvipsnames,pdf]{pstricks}
\usepackage{epsfig}
\usepackage{pst-grad} % For gradients
\usepackage{pst-plot} % For axes
\usepackage{pstricks-add}
\usepackage[off]{auto-pst-pdf} 

\renewcommand{\baselinestretch}{1.5}

\input{mydefs}
\input{./vb.tex}

\input{./feynman}
\psfull

\title{Second Order QED Processes in an Intense Electromagnetic Field}
\author{Anthony Francis Hartin} 
\date{2006}

\begin{document}
\maketitle \clearpage

%\tableofcontents
%
\title{Second Order QED Processes in an Intense Electromagnetic Field}
\author{Anthony Francis Hartin}
\bibliographystyle{alpha}

\begin{abstract}
\input{abstract}

\end{abstract} 

\begin{acknowledgements}
\input{acknowledgements}
\end{acknowledgements} 

\setcounter{tocdepth}{2}
\tableofcontents
\listoftables 
%\listoffigures

\chapter{Introduction}
\input{chap1} %introduction
\chapter{General Theory}
\input{chap2}
\chapter{Cross section Calculations}
\input{chap3}

\chapter{SCS in a circularly polarised electromagnetic field - Results and Analysis}
\input{chap4}
\chapter{STPPP in a circularly polarised electromagnetic field - Results and Analysis}
\input{chap5}
\chapter{External Field Electron Propagator Radiative Corrections}
\input{chap6}
\chapter{SCS and STPPP Resonance Cross sections}
\input{chap7}

\chapter{STPPP in the Beam Field of an $e^+e^-$ Collider}
\input{chap8}

\chapter{Conclusion}
\input{chap9}

\addcontentsline{toc}{chapter}{Appendices}
\appendix
\input{appendicies}

\addcontentsline{toc}{chapter}{Bibliography}

\bibliography{my_bib}

\end{document}

%% file: mydefs.tex
\newcommand{\nwcd}{\newcommand}
\nwcd{\dint}{{\displaystyle \int}}
\nwcd{\dsum}{\sum}
\nwcd{\limfunc}{\mbox}
\nwcd{\NEG}{\slash}
\nwcd{\nts}{\negthinspace}
\nwcd{\nms}{\negmedspace}
\nwcd{\nths}{\negthickspace\negthickspace}
\nwcd{\vep}{\varepsilon}
\nwcd{\undl}{\underline}
\nwcd{\dd}{\ddagger}
\nwcd{\pr}{\prime}
\nwcd{\witl}{\widetilde}
\nwcd{\bm}{\pmb}
\nwcd{\mvb}{\mathversion{bold}}
\nwcd{\lami}{(\lambda_{i})}
\nwcd{\lamf}{(\lambda_{f})}
\renewcommand{\bar}[1]{\setbox0=\hbox{$#1$}
                       \setlength{\unitlength}{\wd0}
                       \smash{\buildrel \line(1,0){0.7} \over #1}}
\def\dbr#1{\setbox0=\hbox{$#1$} 
           \setlength{\unitlength}{\wd0}
           \smash{\buildrel \overline{\line(1,0){0.7}} \over #1}}
\nwcd{\balph}[1]{\setbox0=\hbox{$#1$}
                       \setlength{\unitlength}{\wd0}
                       \smash{\mathrel{\mathop{\line(1,0){0.7}} \atop #1}}}
\def\dalph#1{\setbox0=\hbox{$#1$} 
           \setlength{\unitlength}{\wd0}
           \smash{\mathrel{\mathop{\overline{\line(1,0){0.7}}} \atop #1}}}
\nwcd{\half}{{\textstyle \frac{1}{2}}}
\nwcd{\sql}[1]{#1\hspace{-1.5ex}\raisebox{-1.0ex}{$\sim$}}
\nwcd{\squ}[1]{#1\hspace{-1.5ex}\raisebox{-1.4ex}{$\sim$}}
\nwcd{\eapbf}{\frac{e^{2}a^{2}}{4(k\bar{p})(kp_{f})}}
\nwcd{\eapbi}{\frac{e^{2}a^{2}}{4(k\bar{p})(kp_{i})}}
\nwcd{\eapdbf}{\frac{e^{2}a^{2}}{4(k\dbr{p})(kp_{f})}}
\nwcd{\eapdbi}{\frac{e^{2}a^{2}}{4(k\dbr{p})(kp_{i})}}
\nwcd{\pipbrp}{[(p_{i}\bar{p}_{r\pr})+m^{2}]}
\nwcd{\pipbr}{[(p_{i}\bar{p}_{r})+m^{2}]}
\nwcd{\bpr}{\bar{p}_{r}}
\nwcd{\dpr}{\dbr{p}_{r\pr}}
\nwcd{\pmsqb}{\bar{p}^{2}_{r}-m^{2}}
\nwcd{\pmsqbp}{\bar{p}^{2}_{r\prime}-m^{2}}
\nwcd{\pmsqdb}{\dbr{p}^{2}_{r}-m^{2}}
\nwcd{\pmsqdbp}{\dbr{p}^{2}_{r\prime}-m^{2}}
\nwcd{\mat}[1]{$\; #1 \;$}
\nwcd{\gmu}{\gamma_{\mu}}
\nwcd{\gnu}{\gamma_{\nu}}
\def\Tr{\mathop{\rm Tr}\nolimits}       % Tr for functional trace
\nwcd{\gmaok}{\gamma_{\mu}\slashed{a}_{1}\slashed{k}}
\nwcd{\gmatk}{\gamma_{\mu}\slashed{a}_{2}\slashed{k}}
\nwcd{\gnaok}{\gamma_{\nu}\slashed{a}_{1}\slashed{k}}
\nwcd{\gnatk}{\gamma_{\nu}\slashed{a}_{2}\slashed{k}}
\nwcd{\aokgm}{\slashed{a}_{1}\slashed{k}\gamma_{\mu}}
\nwcd{\atkgm}{\slashed{a}_{2}\slashed{k}\gamma_{\mu}}
\nwcd{\aokgn}{\slashed{a}_{1}\slashed{k}\gamma_{\nu}}
\nwcd{\atkgn}{\slashed{a}_{2}\slashed{k}\gamma_{\nu}}
\nwcd{\pmsq}{\displaystyle \left[\frac{\slashed{p} + m}{p^{2} - m^{2}}
\right]}
\nwcd{\pbr}{\left[\frac{\slashed{\bar{p}}_{r} + m}{\bar{p}_{r}^{2} - m^{2}}
\right]}
\nwcd{\pbrp}{\left[\frac{\slashed{\bar{p}}_{r\prime} + m}{\bar{p}_{r\prime}
^{2} - m^{2}}\right]}
\nwcd{\pdbr}{\left[\frac{\slashed{\dbr{p}}_{r} + m}{\dbr{p}_{r}^{2} - m^{2}}
\right]}
\nwcd{\pdbrp}{\left[\frac{\slashed{\dbr{p}}_{r\prime} + m}{\dbr{p}_{r\prime}
^{2} - m^{2}}\right]}

%% file: vb.tex
\font\twelvevb=cmmib10 scaled \magstep1
\font\eightvb=cmmib7 scaled \magstep1
\font\sixvb=cmmib5 scaled \magstep1
\skewchar\twelvevb='177 \skewchar\eightvb='177 \skewchar\sixvb='177
\newfam\vbfam \textfont\vbfam=\twelvevb \scriptfont\vbfam=\eightvb
\scriptscriptfont\vbfam=\sixvb

%% file: feynman.tex
\gdef\Feynmanlength{\setlength{\unitlength}{0.01pt}}  % Say \Feynmanlength

\global\newcount\LINETYPE                     
\global\newcount\LINEDIRECTION
\global\newcount\LINECONFIGURATION
\newcommand{\LTYPE}{\LINETYPE}
\newcommand{\LDIR}{\LINEDIRECTION}

\global\LINETYPE=1  \global\LINEDIRECTION=0  \global\LINECONFIGURATION=0
\global\newcount\fermion    \fermion=1
\global\newcount\scalar     \scalar=2
\global\newcount\photon     \photon=3
\global\newcount\gluon      \gluon=4
\global\newcount\especial   \especial=5

\global\newcount\REG            \global\REG=0
\global\newcount\FLIPPED        \global\FLIPPED=1
\global\newcount\CURLY          \global\CURLY=2
\global\newcount\FLIPPEDCURLY   \global\FLIPPEDCURLY=3
\global\newcount\FLAT           \global\FLAT=4
\global\newcount\FLIPPEDFLAT    \global\FLIPPEDFLAT=5
\global\newcount\CENTRAL        \global\CENTRAL=6
\global\newcount\FLIPPEDCENTRAL \global\FLIPPEDCENTRAL=7
             
\global\newcount\SQUASHEDGLUON  \global\SQUASHEDGLUON=8

\newcount\adjx \adjx=0
\newcount\adjy \adjy=0
\global\newdimen\BIGPHOTONS     \BIGPHOTONS=0pt  %  DEFAULT:  10 & 11-PT PHOTONS
\global\newdimen\THICKPHOTONS     \THICKPHOTONS=0pt  %  FOR E-W PHOTONS 
\global\newdimen\THICKPHOTONSWITCH    \THICKPHOTONSWITCH=0pt
\gdef\THICKPHOTONTEST{
\THICKPHOTONSWITCH=0pt
\ifdim\THICKPHOTONS=0pt \relax
  \else \ifnum\LTYPE=3
           \ifnum\LDIR=2 \THICKPHOTONSWITCH=1pt \fi % THICK \E PHOTON
           \ifnum\LDIR=6 \THICKPHOTONSWITCH=1pt \fi % THICK \W PHOTON
        \fi
\fi
}  % end of THICKPHOTONTEST

\global\newcount\phantomswitch   \global\phantomswitch=0
\global\newcount\stemlength   \global\stemlength=275   % Default STEM length.
\global\newcount\absstemlength        % A copy of STEM length.
\global\newcount\stemlengthx          % FOR STEMS on particle lines
\global\newcount\stemlengthy          % FOR STEMS on particle lines
\newdimen\FRONTSTEM  \FRONTSTEM=0pt   % FOR STEMS
\newdimen\BACKSTEM   \BACKSTEM=0pt    % FOR STEMS
\newdimen\EITHERSTEM \EITHERSTEM=0pt  % FOR STEMS
\global\newcount\arrowlength                % FOR ARROWS
\global\newdimen\ATTIP   \global\ATTIP=0pt  % FOR ARROWS
\global\newdimen\ATBASE  \global\ATBASE=1pt % FOR ARROWS
\global\newcount\unitboxnumber  % SHOWS THE NUMBER OF `UNIT BOXES' IN LINE
\global\newcount\unitboxnumberpo  % One more than \unitboxnumber (in GLUONSETUP)
\global\newcount\particlelengthx  % THE X-LENGTH OF THE PARTICLE LINE
\gdef\plengthx{\particlelengthx}
\global\newcount\particlelengthy  % THE Y-LENGTH OF THE PARTICLE LINE
\gdef\plengthy{\particlelengthy}  
\global\newcount\boxlengthx  % THE X-LENGTH OF THE BOX:  abs(plengthx) usually
\global\newcount\boxlengthy  % THE y-LENGTH OF THE box:  abs(plengthy) usually
\global\newcount\particleadjustx  % Replaces \gluonadjustx, \scalaradjustx etc.
\global\newcount\particleadjusty  % Replaces \gluonadjusty, \scalaradjusty etc.
\global\newcount\particlelength   % The LENGTH of a particle line BOX (x)
\global\newcount\particlefrontx
\gdef\pfrontx{\particlefrontx}
\global\newcount\PFRONTx
\global\newcount\particlefronty
\gdef\pfronty{\particlefronty}
\global\newcount\PFRONTy
\global\newcount\particlebackx
\gdef\pbackx{\particlebackx}
\global\newcount\particlebacky
\gdef\pbacky{\particlebacky}
\global\newcount\particlemidx
\gdef\pmidx{\particlemidx}
\global\newcount\particlemidy
\gdef\pmidy{\particlemidy}
\global\newcount\seglength  \global\newcount\gaplength
\global\gaplength=850  %default
\global\seglength=1416  % Length of each seg not including `ends' for attachment
\global\newcount\Xone    \global\newcount\Yone    % user co-ords (\Xone,\Yone)
\global\newcount\Xtwo    \global\newcount\Ytwo    % user co-ords (\Xtwo,\Ytwo)
\global\newcount\Xthree  \global\newcount\Ythree  % user's (\Xthree,\Ythree)
\global\newcount\Xfour   \global\newcount\Yfour   % user co-ords (\Xfour,\Yfour)
\global\newcount\Xfive   \global\newcount\Yfive   % user co-ords (\Xfive,\Yfive)
\global\newcount\Xsix    \global\newcount\Ysix    % user co-ords (\Xsix,\Ysix)
\global\newcount\Xseven  \global\newcount\Yseven  % user's (\Xseven,\Yseven)
\global\newcount\Xeight  \global\newcount\Yeight  % user's (\Xeight,\Yeight)
\newsavebox{\lastline}  %  Default name for an unnamed particle line.
\global\newcount\numlineparts   % Num of pieces each unitbox of the line needs
\global\newcount\upperlineadjx  \upperlineadjx=0  %Default
\global\newcount\upperlineadjy  \upperlineadjy=0  %Default
\global\newcount\lowerlineadjx  \lowerlineadjx=0  %Default
\global\newcount\lowerlineadjy  \lowerlineadjy=0  %Default
\global\newcount\thirdlineadjx  \thirdlineadjx=0  %Default
\global\newcount\thirdlineadjy  \thirdlineadjy=0  %Default
\global\newcount\fourthlineadjx \fourthlineadjx=0  %Default
\global\newcount\fourthlineadjy \fourthlineadjy=0  %Default
\global\newcount\unitboxwidth   \unitboxwidth=1000%Default
\global\newcount\unitboxheight  \unitboxheight=0  %Default
\global\newcount\numupperunits  \numupperunits=8  %Default
\global\newcount\numlowerunits  \numlowerunits=8  %Default
\global\newcount\numthirdunits  \numthirdunits=8  %Default
\global\newcount\numfourthunits \numfourthunits=8  %Default
\global\newcount\fermioncount   \global\fermioncount=0    
\global\newcount\scalarcount    \global\scalarcount=0    
\global\newcount\photoncount    \global\photoncount=0    
\global\newcount\gluoncount     \global\gluoncount=0    
\global\newcount\especialcount  \global\especialcount=0    
\global\newcount\vertexcount    \global\vertexcount=-1
\global\newcount\XDIR
\global\newcount\YDIR
\gdef\SETDIR{  % SETS THE DIRECTIONS
\ifcase\LDIR 
     \global\XDIR=0  \global\YDIR=1   %\N  case.
\or  \global\XDIR=1  \global\YDIR=1   %\NE case.
\or  \global\XDIR=1  \global\YDIR=0   %\E  case.
\or  \global\XDIR=1  \global\YDIR=-1  %\SE case.
\or  \global\XDIR=0  \global\YDIR=-1  %\S  case.
\or  \global\XDIR=-1 \global\YDIR=-1  %\SW case.
\or  \global\XDIR=-1 \global\YDIR=0   %\W  case.
\or  \global\XDIR=-1 \global\YDIR=1   %\NW case.
\else\DIRECTERROR 
\fi}  % END OF \SETDIR
\gdef\moduloeight#1{
\ifnum#1>7 \global\advance #1 by -8 
\relax
\moduloeight#1 
\relax
\else \relax  
\fi}
\gdef\multroothalf#1{\global\multiply #1 by 7071 \global\divide #1 by 10000}
\gdef\negate#1{\global\multiply #1 by -1}

\gdef\slanttest(#1,#2){ 
\ifodd\LDIR
\multiply #1 by 7071  \divide #1 by 10000
\multiply #2 by 7071  \divide #2 by 10000
\fi
}
\gdef\gslanttest(#1,#2){
\ifodd\LDIR
\multroothalf#1
\multroothalf#2
\fi
}
\gdef\setplength{ % calcs length of particle line
\global\particlelengthx=\unitboxwidth
\global\particlelengthy=\unitboxheight
\global\multiply \particlelengthx by \unitboxnumber
\global\multiply \particlelengthy by \unitboxnumber
\global\advance \particlelengthx by \particleadjustx
\global\advance \particlelengthy by \particleadjusty
}
\gdef\boxlengthdefault{  % DEFAULT FOR BOX SIZES IN \drawas
\global\boxlengthx=\plengthx
\global\boxlengthy=\plengthy
\ifnum\plengthx<0 \global\multiply\boxlengthx by -1 \fi
\ifnum\plengthy<0 \global\multiply\boxlengthy by -1 \fi
}
\gdef\rearcoords{  %  CALCULATES THE CO-ORDINATES OF THE BACK OF PARTICLE LINE
\global\particlebacky=\particlefronty 
\global\particlebackx=\particlefrontx 
\global\advance \particlebackx by \particlelengthx
\global\advance \particlebacky by \particlelengthy
}
\gdef\midcoords{  %  CALCULATES THE CO-ORDINATES OF THE MID OF PARTICLE LINE
\global\particlemidy=\particlefronty
\global\particlemidx=\particlefrontx
\global\stemlengthx=\particlelengthx  % Convenient variables not being used
\global\stemlengthy=\particlelengthy  
\global\divide\stemlengthx by 2
\global\divide\stemlengthy by 2
\global\advance \particlemidx by \stemlengthx
\global\advance \particlemidy by \stemlengthy
}
\gdef\setcoords(#1,#2,#3)(#4,#5,#6)[#7,#8]{  
\global\upperlineadjx=#1
\global\lowerlineadjx=#2
\global\thirdlineadjx=#3
\global\upperlineadjy=#4
\global\lowerlineadjy=#5
\global\thirdlineadjy=#6
\global\unitboxwidth=#7
\global\unitboxheight=#8
}
\gdef\drawoldpic#1(#2,#3){  % DRAWS PRE-SAVED PICTURE
\global\particlefrontx=#2
\global\particlefronty=#3
\rearcoords  
\midcoords
\put(#2,#3){\usebox{#1}}
}
\gdef\drawsavedline`#1' as #2[#3#4](#5,#6)[#7]{
\global\LINETYPE=#2
\global\LINEDIRECTION=#3
\global\LINECONFIGURATION=#4
\global\particlefrontx=#5
\global\particlefronty=#6
\global\unitboxnumber=#7  
\selectcase
\rearcoords% moved from before selectcase.
\midcoords
\ifnum\phantomswitch=0 \drawas{#1}\fi
}

\gdef\drawas#1{
\global\savebox{#1}(\boxlengthx,\boxlengthy){
\setlength{\unitlength}{0.01pt}
\begin{picture}(\boxlengthx,\boxlengthy)
\multiput(\upperlineadjx,\upperlineadjy)(\unitboxwidth,\unitboxheight)
{\numupperunits}{\upperunitbox}
\ifnum\numlineparts > 1  %  If the line needs 2 parts per unit or more
\multiput(\lowerlineadjx,\lowerlineadjy)(\unitboxwidth,\unitboxheight)
{\numlowerunits}{\lowerunitbox}  
\fi
\ifnum\numlineparts > 2  %  If the line needs 3 parts per unit or more
\multiput(\thirdlineadjx,\thirdlineadjy)(\unitboxwidth,\unitboxheight)
{\numthirdunits}{\thirdunitbox}  
\fi
\ifnum\numlineparts > 3  %  If the line needs 4 parts per unit or more
\multiput(\fourthlineadjx,\fourthlineadjy)(\unitboxwidth,\unitboxheight)
{\numfourthunits}{\lowerunitbox}  
\fi
\end{picture} }
\global\PFRONTx=\pfrontx  \global\PFRONTy=\pfronty   %save this value
\SETFRONTSTEM
\THICKPHOTONTEST
\ifdim\THICKPHOTONSWITCH=1pt\global\advance\PFRONTy by 20  \fi
\put(\PFRONTx,\PFRONTy) {\usebox{#1}}   %\pfrontX,Y=\particlefrontx,y
\ifdim\THICKPHOTONSWITCH=1pt
\global\advance\PFRONTy by -40
\put(\PFRONTx,\PFRONTy) {\usebox{#1}}   % The second \E or \W photon ->thicker
\global\advance \PFRONTy by 20  %re-adjust:  advanced by -20 in total above.
\fi  %End of \ifdim\THICKPHOTONSWITCH=1
\SETBACKSTEM
\seglength=1416   \gaplength=850   % Re-set \SCALR defaults.
}
\gdef\drawandsaveline`#1' as #2[#3#4](#5,#6)[#7]{
\global\newsavebox{#1}
\drawsavedline`#1' as #2[#3#4](#5,#6)[#7]
}

\gdef\drawline#1[#2#3](#4,#5)[#6]{   % Draw line but don't name it.
\drawsavedline`\lastline' as #1[#2#3](#4,#5)[#6]}

\gdef\TYPEERROR{\message{*** ERROR IN PARTICLE TYPE SELECTION ***}
\message{+++ Try with line type \fermion,\scalar,\photon,\gluon 
(see manual) +++}\SETERR}
\gdef\DIRECTERROR{\SETERR\message{*** ERROR IN PARTICLE DIRECTION SELECTION ***}
\message{+++ Try again with direction N, NE, E, SE  etc. or see manual +++}}
\gdef\UNIMPERROR{\message{*** ERROR IN PARTICLE OPTIONS SELECTION ***}
\message{
+++ The requested options combination has not yet been implemented +++}\SETERR}
\gdef\SETERR{\gdef\upperunitbox{{\tiny Error}}  % PRINTS `error' in diagram.
\gdef\lowerunitbox{\relax}
\gdef\thirdunitbox{\relax}
}
\gdef\neglengthcheck{\ifnum\unitboxnumber < 1 
\message{   *** ERROR:  PARTICLE OF NEGATIVE OR ZERO LENGTH REQUESTED. ***   }
\message{   ***         TAKING ABSOLUTE VALUE. ***   }\negate\unitboxnumber \fi}
\gdef\selectcase{  
\neglengthcheck   %  check for particles of negative length.
\SETDIR  
\ifcase\LINETYPE
\TYPEERROR  % \LINETYPE=0 case.
\or \selectfermion  % \LINETYPE=1 case.
\or \selectscalar   % \LINETYPE=2 case.
\or \selectphoton   % \LINETYPE=3 case.
\or \selectgluon    % \LINETYPE=4 case.
\or \selectespecial % \LINETYPE=5 case.
\else \TYPEERROR \fi  }
\gdef\selectfermion{
\ifnum\fermioncount=0 \input feynman/fermionsetup \fi   
\global\advance\fermioncount by 1  % Counts number of fermions drawn. 
\ALLfermion   
}
\gdef\selectscalar{
\ifnum\scalarcount=0 \input feynman/scalarsetup \fi   
\global\advance\scalarcount by 1  % Counts number of scalars drawn. 
\ALLscalar
}
\gdef\selectphoton{   % RECURSIVELY RE-DEFINED IN PHOTONSETUP(23+).TEX.
\ifnum\photoncount=0 \input feynman/photonsetup  \fi   
\selectphoton
}
\gdef\selectgluon{   % RECURSIVELY RE-DEFINED IN GLUONSETUP(25+).TEX.
\ifnum\gluoncount=0 \input gluonsetup  \fi
\selectgluon
}
\gdef\selectespecial{\UNIMPERROR}
\gdef\checkvertex{ %immediately re-defines \drawvertex,\vertexcap,\linkvertex...
\ifnum\vertexcount=-1   \input vertex  \fi}
\gdef\drawvertex#1[#2#3](#4,#5)[#6]{\checkvertex\drawvertex#1[#2#3](#4,#5)[#6]}
\gdef\vertexcap#1{\checkvertex\vertexcap#1}
\gdef\vertexcaps{\checkvertex\vertexcaps}
\gdef\vertexlink#1{\checkvertex\vertexlink#1}
\gdef\vertexlinks{\checkvertex\vertexlinks}
\gdef\stemvertex#1{\checkvertex\stemvertex#1}
\gdef\stemvertices{\checkvertex\stemvertices}
\gdef\flipvertex{\checkvertex\flipvertex}
\global\arrowlength=349  % Length of arrow
\gdef\drawarrow[#1#2](#3,#4){
\global\LDIR=#1
\SETDIR
\global\boxlengthx=#3  % Just a convenient variable name.  No significance.
\global\boxlengthy=#4  % The Arrow co-ordinates.
\ifdim#2=1pt  % CASE \ATBASE WHERE THE CO-ORDS ARE AT THE ARROWS BASE.
\adjx=\arrowlength      \adjy=\arrowlength
\multiply\adjx by \XDIR \multiply\adjy by \YDIR  % Set in \SETDIR
\slanttest(\adjx,\adjy)
\global\advance\boxlengthx by \adjx    \global\advance\boxlengthy by \adjy
\fi
\ifnum\phantomswitch=0\put(\boxlengthx,\boxlengthy){\vector(\XDIR,\YDIR){0}}\fi
}  % END OF \drawarrow.
\gdef\SETFRONTSTEM{
\EITHERSTEM=\FRONTSTEM   \advance\EITHERSTEM by \BACKSTEM
\ifdim\EITHERSTEM>0pt
\global\stemlengthx=\stemlength   \global\stemlengthy=\stemlength   
\global\absstemlength=\stemlength   
\SETDIR
\gslanttest(\stemlengthx,\stemlengthy)
\gslanttest(\absstemlength,\REG)  % the \REG is to use up the parameter space.
\ifnum\XDIR=0 \stemlengthx=0 \fi
\ifnum\YDIR=0 \stemlengthy=0 \fi
\global\multiply\stemlengthx by \XDIR
\global\multiply\stemlengthy by \YDIR
\ifdim\FRONTSTEM=1pt 
\ifnum\phantomswitch=0
          \put(\pfrontx,\pfronty){\line(\XDIR,\YDIR){\absstemlength}}\fi
\global\advance\plengthx by \stemlengthx
\global\advance\plengthy by \stemlengthy
\global\advance\PFRONTx by \stemlengthx   
\global\advance\PFRONTy by \stemlengthy
\global\advance\pmidx by \stemlengthx
\global\advance\pmidy by \stemlengthy
\global\advance\pbackx by \stemlengthx
\global\advance\pbacky by \stemlengthy
\ifnum\LTYPE=3
\global\photonfrontx=\PFRONTx  \global\photonfronty=\PFRONTy
\global\photonbackx=\pbackx    \global\photonbacky=\pbacky
\fi  % END LTYPE
\ifnum\LTYPE=4
\global\gluonfrontx=\PFRONTx  \global\gluonfronty=\PFRONTy
\global\gluonbackx=\pbackx    \global\gluonbacky=\pbacky
\fi  % END LTYPE
\fi  % END FRONTSTEM
\fi  % END EITHERSTEM
}    % end of \SETFRONTSTEM
\gdef\SETBACKSTEM{
\ifdim\BACKSTEM=1pt 
\ifnum\phantomswitch=0
       \put(\pbackx,\pbacky){\line(\XDIR,\YDIR){\absstemlength}}\fi
\global\advance\plengthx by \stemlengthx
\global\advance\plengthy by \stemlengthy
\global\advance\pbackx by \stemlengthx
\global\advance\pbacky by \stemlengthy
\fi  % END BACKSTEM
\global\stemlength=275  \FRONTSTEM=0pt  \BACKSTEM=0pt % Reset default switches.
}    % END OF \SETBACKSTEM 
\gdef\drawloop#1[#2#3](#4,#5){  %RECURSIVE.  
\input loops  % contains loops definitions
\drawloop#1[#2#3](#4,#5)}
\Feynmanlength  % Set length scale to centipoints.

%% file: abstract.tex
%\LaTeXparent{thesis.tex}

Some non linear, second order QED processes in the presence of intense plane 
electromagnetic waves are investigated. Analytic expressions with 
general kinematics are derived for Compton scattering and $e^+e^-$ pair 
production in a circularly polarised external electromagnetic field. Special 
kinematics, including collinear photons and vanishing external field intensity, 
are employed to show that the general expressions reduce to expressions 
obtained in previous work. The differential cross sections were 
investigated numerically for photon energies up to 50 MeV, external field 
intensity parameter $\nu^2$ to value 2, and all scattering angles. The variation of full 
cross sections with respect to external field intensity was also established.

The presence of the external field led to resonances in the 
Compton scattering and pair production differential cross sections. 
These resonances were investigated by calculating the electron self energy 
in the presence of the external field. Numerical analysis of the external field 
electron self energy showed agreement with previous work in appropriate limits. 
However the more general expressions were utilised to calculate resonance widths. 
At resonance the differential cross sections were enhanced by several orders 
of magnitude. The resonances occurred for values of external field intensity 
parameter $\nu^2 <1$, lowering the limit of $\nu^2 \sim 1$ at which 
point non linear effects in first order external field QED processes become important.
Generally, full cross sections increased with increasing external field
intensity, though peaking sharply for Compton scattering and levelling off for pair production.

An application was made to non linear background studies at $e^+e^-$ 
linear colliders. The pair production process and electron self energy 
were studied for the case of a constant crossed electromagnetic field. It 
was found that previous analytic expressions required the external field to be 
azimuthally symmetric. New analytic expressions for the more 
general non azimuthally symmetric case were developed and a numerical 
parameter range equivalent to that proposed for future linear 
collider designs was considered. The resonant pair production cross section 
exceeded the non resonant one by 5 to 6 orders of magnitude. Extra background pair 
particles are expected at future linear collider bunch collisions, raising previous estimations.

%% file: acknowledgements.tex
%\LaTeXparent{thesis.tex}

The Physics Department at Monash University provided me with many years of employment and the 
inspiration to pursue a career in physics. Drs Harry S Perlman and Gordon Troup provided 
much encouragement to pursue work in the field of QED and were coauthors on more than one 
occasion. Harry's cigar smoke always provided notice of his presence in the department. 
Thanks also to Dr. Peter Derlet who was a group member worthy of emulation and whose latex 
outlines I made use of. 

My thanks to John Dawkins, former Australian Minister for Higher Education in the 
Australian Labour government. His attacks on the Higher Education sector in 1987 and 1988 
introduced me to political activism and my development as a human being. Tony Blair would 
have been proud of him. John convinced me that the pursuit of knowledge is always more 
important than the pursuit of profit.

The Physics Department at Queen Mary provided employment and the opportunity to revisit my 
thesis. Particular thanks must go to my supervisor Prof Phil Burrows and the FONT research 
group. Phil always took me seriously and provided years of financial, moral and physics 
support. The ILC was the practical application to which some modest theoretical work in 
this thesis could be directed.

Above all, thanks must go to my family and my partner. My parents, brother and sisters never 
stop believing in me and encouraging me to complete my work. My partner, with constant love 
and support, endured the writing up period with never a word of complaint.

%% file: chap1.tex
Quantum electrodynamics (QED) is one of the most successful physics 
theories of the last century. A measure of that success is in terms of the range of phenomena
described and the accordance of its numerical calculations with experimental results. The
presence of an external field has the effect of introducing a new range of
QED phenomena. Analytic study of these external field
phenomena is of interest to provide further QED predictions that can be tested 
experimentally. This then is the motivation for this thesis which attempts a
detailed theoretical evaluation of some second order QED processes in the presence of an 
intense electromagnetic field.

To achieve this aim, in the first instance, Chapter 1 is devoted to a review
of the literature which deals with the subject of interest. Section 1.1
provides a broad sweep of QED since its inception, dealing in some detail
with the development of the theory in regards to the external field, and in
particular the external electromagnetic field. Sections 1.2 and 1.3 deal,
respectively and in greater detail, with the first and second order QED
processes in the presence of an external electromagnetic field. The space
devoted to the first order processes is larger than may otherwise have been
expected for two reasons. The techniques developed for the first order
external field processes serve as the basis for the more complicated second
order external field processes. Secondly, limiting cases of second order
processes bear a direct relationship to first order processes and provide an
important test for the correctness of the second order calculations which
are more complex and therefore more open to error.

Section 1.4 provides a review of the experimental attempts at measurement of
Intense Field Quantum Electrodynamics (IFQED) phenomena. So far these 
experimental efforts have been confined to the first order processes. 
A brief review of laser systems is also provided,
which are a source of intense, polarised electromagnetic fields, 
and which have been used in IFQED experimental studies.
Finally in section 1.5 we determine the particular problems to be studied in this
thesis and the means by which numerical predictions of experimentally
observable parameters will be arrived at over the course of the remaining
chapters.

\bigskip\

\section{QED and the external field}

QED is the theory of interactions involving electrons, positrons and photons.
Containing at its heart a wave particle duality, QED had twin origins in the
electromagnetic field equations of \cite{Maxwell92} and the discovery, due to
\cite{Planck01} and \cite{Einstein05}, that the electromagnetic field is
quantised. The development of the theory was spurred on by electron beam
experiments which revealed the wave nature of the electron \cite{Davisson27,Thomson27} 
in apparent contradiction to its initial discovery as a particle \cite{Thomson97}.

The differing strands were first drawn together into a relativistic quantum
field theory by \cite{Dirac28a,Dirac28b,HeiPau29,HeiPau30,Fermi32}. 
However the proposed, quantised interacting field equations proved
extremely difficult to exactly solve. A way forward was provided for the
case of QED by the weak coupling of the electron and photon fields and the
expansion of the field equations in powers of the coupling constant\footnote{
The coupling constant for QED is the fine structure constant $\alpha$ 
which is $\sim\frac{1}{137}$.} which allowed perturbation theory to
be employed.

With the theory of QED in place, theoretical calculations of the basic QED
interactions were performed. The theoretical description of the scattering
of an electron and photon, which was experimentally discovered by \cite{Compton23}, 
were first written down by \cite{KleNis28}.
\cite{BreWhe34} developed the equations connected with the production
of an electron-positron pair from the interaction of two photons. \cite{Moller32} and 
\cite{Bhabha35} described, respectively, electron-electron
scattering and electron-positron scattering. For a general historical review
of the developments of this early period see \cite{Pais86}.

Perturbation theory however was limited by the fact that only the first term
in the perturbation series gave results that were in agreement
with experiment. All further terms led to meaningless divergences.
Various methods of removing the divergences were developed. 
The main methods included renormalisation of the
electron mass and charge to take into account the effect of the
Dirac-Maxwell field interaction on the fundamental parameters of the theory,
the introduction of cut-off parameters which presume the incorrectness of
the theory at very high energies, and various regularisation procedures such
as that due to \cite{PauVil49}. A thorough review of issues involved with divergences is
contained in chapters 9 and 10 of \cite{JauRoh76}.

The experimental discovery of the electron anomalous moment \cite{KusFol47,KusFol48} 
and the Lamb shift \cite{LamRut47} spurred on further theoretical 
developments of QED in the late 1940's. Two main developmental strands emerged. 
A reformulation of the fundamental field equations which aided the program of renormalisation 
was developed \cite{Tomonaga46,Tomonaga47,Tomonaga48,Schwinger48a,Schwinger48b}. In this view 
wave functions develop from one space-like surface to 
another resulting in equations which are covariant at each stage of calculation. This is known as 
the proper time method.

The second reformulation of QED, based on earlier work by \cite{Stuckleberg43}, was due to Feynman. This
reformulation pictured portions of a mapped out space-time in which QED interactions take place. 
Expressions containing the Feynman matrix element solutions could be written down directly with the aid of
diagrams \cite{Feynman48a,Feynman48b,Feynman49a,Feynman49b}. The equivalence of the Schwinger and Feynman
reformulations was proved \cite{Dyson49}. It is Feynman's reformulation of QED that will serve as the
basis for the theoretical work in this thesis.

The problem of the interaction of an external field with an electron was
first attempted by \cite{Thomson33} who calculated the solution for the orbit
of a non relativistic electron moving in the field of a monochromatic plane
electromagnetic wave. However the advent of the quantised relativistic theory presented difficulties for a
rigorous treatment of the external field. The interactions with each particle
of a quantised external field lead to impossibly complex calculations. A semi classical approximation 
which, for example, treated the external field as classical and neglected the photon-external field
interaction, proved necessary. Such calculations proceeded with the solution of the Dirac
equation for an electron embedded in the external classical field. These solutions were
found for a constant crossed electric and magnetic field 
\cite{Schwinger51} and for a plane wave electromagnetic field \cite{Volkov35}. 
For external fields in which the Dirac equation could not be solved exactly, the Born
approximation was required. The Born approximation consists of a further expansion of the QED matrix
elements in powers of the coupling to the external field. For a
review of the basic theory associated with QED and the external field see
chapters 14 and 15 of \cite{JauRoh76}.

One of the first consequences of the external field in QED was the possible
polarisation of the vacuum into electron and positron pairs. \cite{Uehling35}
investigated Dirac positron theory for the case of an external electrostatic
field. The existence of a formula for the charge induced by a charge
distribution implied polarisation of the vacuum. Deviations from
Coulomb's law were investigated for the scattering of heavy particles and
shifts in energy levels for atomic electrons. \cite{Schwinger48b} applied their
proper time method to the problem of vacuum polarisation by a prescribed
electromagnetic field, and \cite{Valatin51} reinvestigated the method Dirac
and Heisenberg introduced to deal with the appearance of divergent integrals
connected with vacuum polarisation.

\cite{KhaRoh63} showed that the Feynman-Dyson formulation of QED
leads to vacuum polarisation terms that violate gauge invariance. They found
the inconsistency to lie in the unjustified interchange of integrations and
limiting processes. With a more careful integration procedure, divergent
integrals were avoided along with the need for cut-off procedures or appeals
to invariance for undefined integrals. \cite{Ferrell73} extended the work of
\cite{KhaRoh63} by showing that calculations of the vacuum polarisation
tensor which allow for gauge invariance at each step, result in the divergent
counter-term which was introduced in the original calculation to keep the photon mass
zero.

One of the other fundamental external field problems considered was that of
electron motion in an external electromagnetic field. 
Various authors considered the original Volkov solution as an 
infinite sum of contributions related to the number of
external field photons that interact with the electron. \cite{Zeldovich67}
interpreted this as the electron obtaining a quasi-level structure in the
external field. Other authors extended the work of Volkov, with \cite{Sengupta67}
solving the Dirac equation for an electron in an external field
consisting of two polarised plane electromagnetic waves, \cite{Bagrov74,Bagrov75}
discussing the exact solution of a relativistic electron interacting with a
quantised and a classical plane wave travelling in the same direction, and
\cite{Fedorov75} proposing a method of constructing a complete orthonormal
system for the electron wave function for an electron embedded in a
quantised monochromatic electromagnetic wave.

The electron propagator in an external electromagnetic field was the subject
of study by \cite{Schwinger51} and \cite{Valatin51}, who both obtained a
proper-time representation. Schwinger considered the case of a constant
crossed electromagnetic field whereas \cite{ReiEbe66} found an
expression for the exact Greens function of the electron in the presence of
an intense, circularly polarised electromagnetic plane wave. \cite{ReiEbe66}'s result, written as a 
single integration over the electron
4-momentum and as an infinite sum of products of Bessel functions, revealed
that the electron under the influence of the external field gained a mass
increment above the field free mass (see also \cite{BroKib64}). 
\cite{Ritus72} obtained another representation for the electron propagator in the presence of a 
plane wave electromagnetic field in terms of Volkov functions, the properties of which
were investigated by \cite{Mitter75}. The Ritus representation will be used in this thesis. 
Other work on the external field electron
propagator included \cite{Fedorov75} who obtained the electron Green's function
for an electron in a quantised monochromatic electromagnetic wave.

Consideration of IFQED in a reference frame moving almost at the speed of
light was found useful for avoiding gauge problems. This method, which was 
developed independently by \cite{KogSop70}, and \cite{NevRoh71},
involved the construction of a new time variable which forms light-like
surfaces rather than the usual space-like surfaces. \cite{Bjorken71}
formed a new physical picture which considered the electron as composed of bare constituents which interact with 
each other and the external field to form a final state. They applied
this formulation to elastic electron and photon scattering,
bremsstrahlung and pair production.

An alternative to the semi classical method of incorporating the external
field into QED, is the method of coherent states. The semi classical method
implied that a discrete number of external field photons contribute to the scattering
process and a quantum treatment of external field photons was suggested. Such a quantised 
treatment was rendered less formidable by coherent states which allowed the 
replacement of quantum field operators by their classical
counterparts \cite{Glauber63}. Work on fundamental external field QED processes using
coherent states was performed by \cite{Derlet95}, and a general review of
coherent states is presented by \cite{Zhang90}.

An approximation to the semi classical method is possible for external field
QED processes in which electron energies are ultra relativistic. For these
electron energies the solutions to the Dirac equation in the presence of the
external field could be replaced by their classical counterparts. With this
approximation the first order external field QED processes were evaluated
for the cases of an external magnetic field \cite{Baier68}, an external
Coulomb field \cite{Baier69} and an external electromagnetic field \cite
{Baier72}. The method was also applied to the some of the second order
processes, in particular vacuum polarisation \cite{Baier75b} and electron
and photon elastic scattering \cite{Baier75,Baier76}.

\bigskip\

\section{First order external field QED processes}

The development of the laser in the early 1960's stimulated further research
into interactions involving electrons, photons and an external
electromagnetic field. Studies of the first order 
IFQED processes are examined in this section. First order processes are represented by Feynman diagrams 
with one vertex and an odd number of fermion and boson lines in initial and final states. 
A 1969 review of the work done on finite order IFQED processes was provided by \cite{Eberly69}.

Several papers in the first half of the 1960's considered the first order
external field process in which an electron embedded in a plane wave laser
field scatters a single photon. This process is dependent on the intensity
of the external field and is referred to as High Intensity Compton
scattering (HICS).

The first attempts at obtaining IFQED cross sections dealt with the 
interaction of external field and electron as given by terms in a
perturbation series expansion. The $n\text{th}$ term in such a perturbation 
series corresponds to the external field contributing $n$ photons to the 
process. \cite{Fried61} considered non linear processes in which
two or three external field photons interact with the electron in the
initial state. Comparison of the matrix elements of these processes yield
the expansion parameter of the perturbation series, which approaches unity
for high intensity of the external field \cite{Stehle63}. 
\cite{Vachaspati62,Vachaspati63} obtained similar results 
by considering the scattering process using classical theory. 
Much later, the HICS process in which $n+N$
photons contribute in the initial state and $n$ in the final state, was also
considered in the context of perturbation theory \cite{Kormendi84}. Perturbation terms in the transition
amplitudes were found to diverge for vanishing external field intensity. This difficulty was avoided by
use of special kinematics or appropriate 
approximations. For instance, \cite{Fried63} considered
the HICS process using approximate solutions to the Dirac Hamiltonian
obtained by \cite{BloNor37} in which negative energy states are neglected.

A common semi classical approximation used exact solutions, of the
Dirac equation for fermions embedded in a classical plane electromagnetic
wave \cite{Volkov35}. With use of Volkov solutions, the transition amplitude of the HICS process decomposed 
into an infinite sum of incoherent amplitudes corresponding to the number of laser
field photons that contribute to the process. Each contribution to the
transition amplitude produces a final state which can be thought of as an harmonic of the external 
field photons. The $n$th harmonic contribution is
proportional to the $n$th power of the external field intensity parameter 
$\nu ^2$. For laser intensities that were foreseeable in the 1960s, the
intensity parameter was much less than one, and only the first few harmonic
contributions to the transition amplitude were considered. In the limit of
vanishing intensity of the external field, the first harmonic contribution to
the cross section reduces to the Klein-Nishina formula for single external
field photon scattering from the asymptotically free electron 
\cite{BroKib64,NikRit64a,Goldman64}.

The transition amplitude of the HICS process is dependent on the state of
polarisation of the external field. \cite{NikRit64a} considered the case of a linearly
polarised external electromagnetic field. The transition probability contained an 
infinite summation of complicated functions which were evaluated in limiting cases
only. In contrast a circularly polarised external field produces Bessel functions, the
properties of which are well known \cite{NarNikRit65,BroKib64}. A circular polarised external field 
introduces an azimuthal symmetry into the HICS process which results in analytically 
less complicated HICS cross sections \cite{Mitter75}. 
The HICS process for the case of an elliptically polarised external electromagnetic
field was considered by \cite{Lyulka75}.

One important debate that took place concerned the dependency of a frequency shift in the 
scattered photon on the intensity of the external laser field. In the
semi classical approximation, various authors found an expression for the
energy-momentum of the scattered photon dependent both on the number of
laser photons that contributed to the process, and the intensity of the
laser field \cite{BroKib64,Goldman64}.

This semi classical expression for a frequency shift was dismissed on the grounds that 
external field boundary conditions were omitted. The HICS cross section was
reevaluated making use of the adiabatic switching hypothesis in which the external  field 
was represented as a linear combination of monochromatic occupation number states with boundary 
conditions consisting of asymptotically free electrons and photons (at $t=\pm \infty $).
A distinct analytic expression for an intensity dependent frequency shift was obtained 
\cite{FriEbe64}.

In turn, the \cite{FriEbe64} result was brought into question by appealing to the correspondence 
principle in that the semi classical and fully quantum mechanical treatments of the HICS 
process should yield the same result in the classical limit.
Neglecting radiative corrections, it was indeed found that a quantum
mechanical treatment of the external field as coherent states gave the same 
intensity dependent frequency shift as obtained
in the semi classical treatment \cite{Frantz65}.
To complete the rebuttal, the results of \cite{FriEbe64} were explained as
arising from the physically problematic, adiabatic switching of spatially infinite external 
field states. Correct boundary conditions for the switching on and off
of the external field were obtained by considering finite wave trains. In such a
case the intensity dependent frequency shift re-emerged \cite{Kibble65}. 
An intensity dependent frequency shift could be interpreted in
terms of a Doppler shift produced by the electron in the external field
acquiring an average velocity in the direction of propagation of the external field
photons. In such an interpretation the way the laser is switched on or off
is irrelevant.

The existence of the frequency shift mechanism allows the HICS process to be
used as a generator of high energy photons \cite{Milburn63,Bemporad65,Sandorfi83}. 
The polarisation properties of these high energy photons
are of fundamental importance in nuclear physics applications \cite{GriRek83}. 
\cite{Grinchishin82,Ginzburg83a} considered polarisation effects of the
interaction of an electron with an external laser field in the lowest order
of perturbation theory. With the polarisation states of the electron
expressed in Stokes parameters, and those of the photon in helicity states,
the differential cross section of the process was obtained for general kinematics 
\cite{Ginzburg83a} and in the rest frame of the initial electron \cite{Mcmaster61}.
Later papers expressed the polarisation state of the scattered HICS
particle as a function of polarisation states of the initial particles for a circularly polarised 
external field \cite{GriRek83,Tsai93}.

One point of interest concerning the HICS process was the impact that a
second external field would have on the scattering. The problem of photon
emission by an electron in a bi-chromatic electromagnetic field was
considered in the lowest order of perturbation theory. There was a significant enhancement of the 
cross section of the scattering
process, proportional to the ratio of frequencies of the two external fields
\cite{PraVac68}. In later work, the second external field was
considered exactly by writing the external field four-potential in the
Volkov solution as a sum of two co-directional plane wave fields with different frequencies. 
In the limit of weak intensity of one
external field, calculations indicated a ten percent enhancement of the
scattering cross section for a range of intensities of the second external field
\cite{GucGus75}. The same process was re-examined non relativistically with allowance for initial 
electron momentum in light of contemporary experimental work \cite{Ehlotzky87}.

Since intense external fields are usually supplied experimentally by intense laser beams,
photon depletion of the laser beam by a single HICS multi-photon
process can be neglected \cite{Mitter79}. However most experimental work also use 
electron beams (see for example \cite{EngRin83}) and photon depletion
may become significant. If such is the case the external field cannot be treated 
as a classical field with constant photon number density and the semi classical method breaks down. 
An approach in which the external laser field is considered quantised from the outset, becomes 
necessary \cite{BerVar81a,Becker88,Becker89}. 

Using the method of coherent states developed by \cite{Glauber63}, the HICS process was considered 
using an electron wave function solution of the Dirac equation for an external field consisting of one 
quantised, circularly polarised electromagnetic mode. Expressions obtained for the
frequency of the scattered photon and the transition probability for the
HICS process reduced to those obtained in the semi classical approximation
when the expectation value of the external field photon number was very
large \cite{BerVar81b}. The transition probability of the HICS process
depends strongly on the state of polarisation of the external quantised
field. For linearly polarised modes, the external field photons form
squeezed states and for circular polarisation they form coherent states \cite{GazShe89}.

In other work the HICS process for a non relativistic electron retarded by a Coulomb field was 
considered \cite{Lebedev70}. Phonon scattering of
conduction band electrons in the presence of an intense electromagnetic
field was studied by \cite{BuiOle67}. \cite{Becker81b} examined the
radiation emitted by a relativistic electron moving in a dispersive non
absorptive medium under the influence of an electromagnetic wave. \cite{VarEhl84}
considered the HICS in an external field consisting of a strong homogeneous
magnetic field and an intense microwave field.

The other first order IFQED process to be reviewed is the production of an electron and 
positron from an initial state consisting of one photon and an external field.
This process is referred to as One Photon pair production (OPPP).

\cite{Reiss62} was the first to consider the OPPP process in the semi classical approximation 
with a plane wave electromagnetic field. \cite{Reiss62} calculated transition probabilities by 
considering perturbations to the solution of the Dirac equation in the experimental field.
In contemporary work several authors considered the same process using Volkov
solutions in the Feynman-Dyson formulation of QED. Expressions obtained for
the OPPP process contained many of the same features present in the HICS
process. The transition amplitudes obtained were an infinite sum of incoherent
amplitudes corresponding to the number of external field photons that
combine with the initial photon to produce the pair. Indeed, the transition
probability of the OPPP process was obtained directly from the transition
probability of the HICS process by an exchange of particle momenta via the
substitution rule \cite{NikRit64a}. In the limit of vanishing external
field intensity parameter $\nu ^2$, the transition probability of the OPPP
process reduces to that of the Breit-Wheeler process for two photon pair
production. For vanishing frequency of the external field the Toll-Wheeler result for the absorption 
of a photon by a constant electromagnetic field was obtained \cite{Reiss62,NikRit64a}. 
\cite{NikRit67} investigated the OPPP process in the limit of vanishing ratio of external
field frequency parameter $\frac{\omega}{m}\equiv \frac{\hbar\omega}{mc^2}$ to intensity parameter $\nu^2$.

The state of polarisation of the external electromagnetic field has a
significant effect on the OPPP process, as it does for the HICS process.
Transition probabilities for an initial photon polarised parallel and
perpendicular to a linearly polarised external electromagnetic field were
obtained by \cite{Reiss62} and \cite{NikRit64a}. A circularly polarised
electromagnetic field was considered by \cite{NarNikRit65}, and the more general
case of elliptical polarisation by \cite{Lyulka75}. Spin effects were dealt
with by \cite{Tegnov68}. \cite{Baier76} also
considered the OPPP process for an elliptically polarised external
electromagnetic field. They obtained a new representation for the transition
probability in terms of Hankel functions, by considering the imaginary part
of the polarisation operator in the external field.

The transition probability of the OPPP process passes through a series of
maxima and minima as the intensity of the external field increases \cite{NarNikRit65}. 
The increase in external field intensity has a limiting effect
on the process, increasing the likelihood that more external field photons
will contribute to the pair production, but increasing the lepton mass and
thus the energy required to produce the pair \cite{Becker91}.

The dependence of the OPPP process on the spectral composition of the
external electromagnetic field was of interest to several authors. An
external electromagnetic field consisting of two co-directional linearly polarised waves of
different frequencies and orthogonal planes of polarisation, yielded a transition 
probability of similar structure to that obtained for a monochromatic electromagnetic field 
\cite{Lyulka75}. \cite{Borisov77} considered an external field of similar form with two
circularly polarised wave components. \cite{Borisov77} obtained transition probabilities
which, in the limit of vanishing frequency of one electromagnetic wave component,
reduced to those for an external field consisting of a circularly polarised field and a 
constant crossed field \cite{ZhuHer72}.

The advent of new high powered laser beams in the 1990s led to a renewed flurry of analytic work. 
The HICS process was considered numerically for laser intensities up to $10^{21}\;\text{Wcm}^{-2}$ and for 
elliptical polarisation of the external field \cite{Panek02,Panek03}. A series of papers 
considered the first order IFQED processes for one or two external fields of elliptical polarisation 
\cite{Roschupkin00}. Complete polarisation effects of all contributing particles were 
studied for a circularly polarised external field and arbitrary polarisation of all 
other particles \cite{Ivanov03}.

The OPPP process was also considered in external fields other than electromagnetic waves. For 
example, OPPP in a magnetic field was considered by \cite{Schwinger54,DauHar83,Beskin84} and an 
external field consisting of one electromagnetic wave and one magnetic wave was
considered by \cite{Oleinik72}.

\section{Second order external field QED processes}

The second order IFQED processes provide an abundance of phenomena for study. 
These are represented by Feynman diagrams with two vertices and include Compton scattering,
Two Photon pair production, Two Photon Pair Annihilation, M\"oller scattering and 
electron self energy. The second order IFQED processes require the external field photon propagator 
\cite{Schwinger51,Oleinik67} or the external field electron propagator which is available either in 
a proper time representation \cite{Oleinik68} or a Volkov representation 
\cite{NikRit64a,NikRit64b,Mitter75}. A 1975 review of work done on second order 
IFQED processes was provided by \cite{Mitter75}.

External field Compton scattering or stimulated Compton scattering (SCS) was first 
considered with an external field consisting of a linearly polarised 
electromagnetic wave. The cross section was calculated in the non relativistic limit of small photon energy and 
external field intensity for a reference frame in which the initial electron is at
rest. Resonant singularities in the
cross section due to the poles of the electron propagator in the external
field being reached for physical values of the energies involved. This
contrasted with Compton scattering in the absence of the external field which
does not contain these singularities. The singularities were interpreted in terms 
of a quasi-level energy structure \cite{Zeldovich67} for the electron in the 
external field. Resonance takes place when the energy of the incident or scattered photon is
approximately the difference between two of the electron quasi-levels. The cross section resonances 
were avoided by inserting the external field electron self energy into the electron propagator. 
The resonant cross section exceeded the non resonant cross section by several orders of
magnitude \cite{Oleinik67}.

\cite{AkhMer85} considered the SCS cross section in a linearly polarised external electromagnetic 
field and wrote down the SCS matrix element for a circularly polarised external field. This
calculation was performed for the special case where the momentum of the
incoming photon is parallel to the photon momentum associated with the
external field. \cite{AkhMer85} avoided the resonant infinities found by
\cite{Oleinik67} by considering a range of photon energies for which resonance did
not occur.

The two photon, electron-positron pair production process in the presence of
an external electromagnetic field or stimulated two photon pair production
(STPPP) has remained uncalculated. However \cite{KozMit87} dealt with the process 
in a strong magnetic field for the case in which the energy of each of the photons is alone 
insufficient to produce the pair. The cross section obtained also contains resonances. 

The calculation of the SCS cross section revived discussion on the existence of an intensity 
dependent frequency shift in the scattered photon. Drawing on the earlier debate, 
the intensity dependent frequency shift was tied to the choice of boundary
conditions for the electron wave function. An intensity dependent frequency shift emerged 
analytically as long as the electron and the external field always remained coupled 
\cite{Vonroos66}.

\cite{OleSin75} used a modified form of the Volkov
solution which allows for the adiabatic switching on and off of the external
field at times in the remote past and remote future. This procedure turned out to be 
equivalent to equating the quasi electron momentum with the
free field electron momentum and introducing an electron mass shift.
Allowance for adiabatic switching lead to a modified energy-momentum
conservation law for the first order HICS process. As a result the scattered photon frequency 
differed by several orders of magnitude to previous calculations \cite{BroKib64,NikRit64a}. 
This result was significant since previously it was thought that the intensity dependent frequency 
shift would be a small effect \cite{Vonroos66}.

\cite{Belousov77} applied the adiabatically modified Volkov solution to the second order 
SCS process and obtained an expression for the frequency shift of the scattered photon. 
The resultant, adiabatically modified, SCS cross section expressions contained frequency 
shifted resonances. [Bel84] considered the SCS process with adiabatic boundary conditions in two 
inertial reference frames to show that the scattered photon frequency shift is a relativistic effect.

Several papers have dealt with the second order M\"oller process (the scattering of two electrons) 
in the presence of a strong electromagnetic field. \cite{Oleinik67} 
considered the process in an external field consisting of a
linearly polarised electromagnetic wave. As in other IFQED processes, the external field modifies 
the energy-momentum conservation of the scattering process and contributes external field quanta.
\cite{Oleinik67} found it necessary to calculate the transition probability in a centre of mass-like 
reference frame in which the external field quanta are absorbed in the initial 
electron momenta. Under certain conditions the M\"oller scattering transition probability
acquired terms corresponding to electron attraction, giving rise to the
possibility of electron pairing in an electromagnetic field \cite{Oleinik67}.

\cite{Bos79a} performed a calculation of the M\"oller scattering cross section in a circularly 
polarised electromagnetic plane wave with a greater emphasis on numerical results. These 
calculations were performed in a non relativistic energy regime with a reference frame in which 
incoming electrons have opposite and equal quasi-momentum. Numerical investigations calculated 
the SCS differential cross section with variation of the intensity of the external
field, the geometry of the scattering process, and the number of external
field quanta that participate in the process. The SCS differential cross section differed 
considerably from basic Compton scattering in experimentally accessible regions 
\cite{Bos79a,Bos79b}.

Further investigations in the 1980's considered the external field M\"oller process in a relativistic 
regime and with a low intensity, elliptically polarised, external electromagnetic field. 
Mathematically simpler differential cross section expressions were obtained. The analytic 
cross sections indicated an intensity dependent electron mass shift as expected, however the 
contribution of external field quanta was neglected in the low intensity limit \cite{Roschupkin84}.
In contrast with M\"oller scattering in a Coulomb field, the M\"oller scattering in
an electromagnetic field contained terms which suppressed the cross
channel of the scattering cross section \cite{FedRos84}.

The possibility that IFQED differential cross sections could contain
resonant infinities was recognised soon after the initial first order
calculations were performed. Increasing intensities of available lasers led
to the consideration of cross section terms involving contributions from two
or more external field quanta. It was recognised that employment of
perturbation theory for these contributions would lead to infinite electron
propagation functions. The solution proposed was the inclusion of the electron self energy
\cite{NikRit65}.

\cite{Oleinik67} was the first to encounter these resonant infinities in a
calculation of the external field M\"oller scattering differential cross section. 
The obtained propagator poles coincided with transitions between the electron energy quasi-levels
of the electron embedded in the electromagnetic wave, in direct analogy to the
resonance scattering of light by atoms \cite{Mitter75}. The infinities were removed by recalculating 
the differential cross section using a photon propagator corrected for the photon self energy. 
The differential cross section infinities were rendered finite and the resultant resonant 
peaks exceeded the non resonant differential cross section by several orders of magnitude.
The calculation of resonant cross sections was extended to the SCS
process by inclusion of the electron self energy into the electron propagator 
\cite{Oleinik68}. 

\cite{Fedorov75} considered, as an approximation to the second order SCS process,
the resonant scattering of an electron in the field of two external
electromagnetic waves. The induced resonance width was calculated by summing
an infinite series of resonance terms and was greater than that obtained by \cite{Oleinik68}. 
Resonant scattering in two external fields allowed for multiple transitions
between two distinct sets of electron energy states. Whereas \cite{Oleinik68} made an analogy 
with spontaneous resonant emission, \cite{Fedorov75} made an analogy with stimulated resonant 
emission.

\cite{Bos79a} provided a more extensive evaluation of the resonant M\"oller
scattering in an external field by calculating the width, spacing and heights of the
differential cross section resonance peaks. The resonant differential
cross section exceeded the non resonant differential cross section by several
orders of magnitude as stated by \cite{Oleinik67}, but only under certain conditions. 
Numerical calculations showed that, at the time, conditions that produced resonant 
cross sections would be difficult to reach experimentally. More recently, the resonant M\"oller 
process was revisited analytically and numerically for high radiation powers in a external field of 
elliptical polarisation \cite{Panek04,Roschupkin96}.

The calculation of the resonant cross sections of second order IFQED processes required the 
calculation of the electron and photon self energies in the presence of an external field. These
external field self-energies are also second order IFQED processes.

\cite{Ritus70} was one of the first to consider the electron and photon self energy 
processes in an external electromagnetic field. Both the electron and
photon mass shifts in a constant crossed electromagnetic wave were calculated.\footnote{A constant 
crossed electromagnetic wave is one in which the electric and magnetic field vectors are 
constant, equal in magnitude and orthogonal.}
These calculated mass shifts became appreciable as the external field intensity reached that 
of Schwinger's characteristic field \cite{Schwinger54}. The method used to calculate the 
Mass Operator began by calculating the total probability for radiation in the 
external field. The imaginary part of the mass operator was then obtained via the
optical theorem and was proportional to the intensity of the external field. The electron 
mass operator was also dependent on spin and an anomalous magnetic moment was calculated.

The expression obtained for the electron mass operator in a constant
crossed electromagnetic field was confirmed by calculating the
mass operator directly using Schwinger's equation. This approach
allowed a better assessment of the analytic properties of the external field
mass operator and vacuum polarisation operator. The analytic properties revealed that
these operators, in contrast to the non external field case, are transcendental 
functions of the momenta squared and depend non trivially on the dynamic variable. 
The Green's functions obtained from these operators contained an infinite number of poles 
which depend on external field variables \cite{Ritus72}. In a later paper, the
mass correction to the elastic scattering amplitude of the electron in a constant crossed 
electromagnetic field was calculated and the probability for two
photon emission by the electron in the external field, the mass correction
to the probability for one photon emission, and the mass correction to the
anomalous magnetic moment of the electron were all found \cite{MorRit75}.

Other work on the mass operator for an electron embedded in a constant
crossed electromagnetic wave was performed by \cite{Narozhnyi79}. Asymptotic expressions 
were obtained to calculate the corrected electron propagator to third order in the fine
structure constant which yielded an estimation of the
lower bound for an IFQED expansion parameter.

The electron mass operator in a circularly
polarised electromagnetic field was calculated and used to obtain a corrected external field 
electron propagator \cite{BecMit76}. This corrected propagator was written 
explicitly to first order in the fine structure constant in both numerator and
denominator. The symmetry of the circularly polarised external field led to a matrix 
structure for the mass operator and electron propagator that was almost diagonal. 
\cite{BecMit76} obtained numerical results for the real and imaginary parts of the 
electron mass shift with the aid of approximation formulae.

An external field consisting of a constant crossed electromagnetic field and
an elliptically polarised electromagnetic plane wave was considered in a
calculation of the mass operator for both an unpolarised \cite{Klimenko90a}
and a polarised \cite{Klimenko90b} electron. The analytic form of this mass operator 
reduced to that of previous work by allowing each of the electromagnetic waves making up the 
external field to go to zero in turn.

The presence of an external electromagnetic field alters the photon self energy, the
photon propagator and the probability of polarisation of the vacuum.
A plane wave electromagnetic field alone does not polarise the vacuum, but the
presence of another agent such as a second field or a photon leads to
real effects \cite{Toll52}.

Most of the early work on external field vacuum polarisation was performed
with a constant crossed electromagnetic field. \cite{Sannikov67,Sannikov95} 
considered photon elastic scattering. The effect on the photon was a change of 
polarisation with the direction of propagation remaining unchanged. \cite{Narozhnyi69} 
calculated the vacuum polarisation operator in a constant crossed field and made radiative 
corrections to the photon propagator. A more general treatment was provided by 
\cite{BatSha71} who used relativistic, gauge and charge invariance to write an eigenvector
representation for the vacuum polarisation operator. \cite{Bialynicka70} used a slightly 
varying external electromagnetic field of otherwise arbitrary form.

The presence of an intense electromagnetic field system renders the vacuum a 
birefringent medium. This view emerges from the non linear properties of Maxwell's equations when
allowance is made for virtual electron-positron pair creation \cite{KleNig64}. 
\cite{Narozhnyi69} showed that two waves with different dispersion laws can propagate in an
external field and that the two refractive indices can be determined. The
direction of propagation of the two waves coincide and their polarisation is
orthogonal. \cite{Bialynicka70} calculated the group and phase velocities of both propagation 
modes.

A mass correction to the elastic scattering amplitude
of a photon in a constant crossed electromagnetic field
was calculated to order $\alpha^2$. The mass correction was used to 
obtain the probability of pair production \cite{MorNar77}.
Renormalisation group methods were used to
study external field vacuum polarisation for asymptotic forms of a static
electromagnetic wave \cite{ColWei73,Kryuchkov80}.

\cite{BecMit75} obtained an expression for the vacuum polarisation tensor
with a circularly polarised electromagnetic field. The calculation was performed with 
light-like coordinates which are used in null-plane formulations of quantum electrodynamics
\cite{NevRoh71}. \cite{BecMit75} used a proper time representation for the external field 
electron Green's function. This lead to expressions for the vacuum polarisation tensor
which have proved cumbersome to use \cite{AffKru87}.
This form for the vacuum polarisation tensor contained integrations with 
infinitely bounded integrations of Hankel functions. These integrations were performed 
by using asymptotic expressions for the Hankel functions and imposing
limits on the intensity of the external field. The effects of vacuum
polarisation on a photon in a circularly polarised external field were approximately 
described by a vacuum with two complex indices of refraction. Numerical 
calculations showed that vacuum birefringence was a small effect unless the photon which 
probes the vacuum has very high energy.

Vacuum polarisation studies were performed in other types of external fields. 
The scattering of circularly polarised waves in a Coulomb field was considered for low 
intensity and frequency of the electromagnetic wave \cite{Yakovlev67}.
Analytical properties of the photon polarisation tensor in an external
magnetic field using the Furry picture were investigated \cite{Shabad75}. The vacuum 
polarisation tensors in a constant electromagnetic field was calculated using a technique 
borrowed from string theory \cite{Schubert01}.

\section{Experimental Work}

There have been few attempts at experimental detection of IFQED processes until comparatively 
recently. This was mainly due to the availability of ultra intense electromagnetic 
fields which are required to produce detectable phenomena particular to IFQED.
The main laboratory source of intense electromagnetic fields are
lasers. Indeed it has been the ongoing development of the laser that has 
provided the impetus for the theoretical calculations reviewed in this chapter. It is 
therefore worth devoting some space to discussing ultra intense lasers and their relevant
experimental parameters

The most common laser in use for present day experiments is the so called $T^3$ or 
Table Top Terawatt laser system \cite{Becker91,Bamber99}. These laser
systems produce ultra intense, ultra short pulses via the technique of
Chirped Pulse Amplification (CPA). CPA begins with an ultra short low energy pulse from a 
standard mode-locked optical laser such as a Nd:YAG or a dye laser. The normal operation for 
an optical laser involves many lasing modes, the number of which is determined by the gain 
bandwidth $\triangle\nu _g$ which in turn is determined by the uncertainty relations 
(\cite{MilEbe88} pg.355). The modes of oscillation are locked together via 
suitable electronic circuitry to produce single peaks of higher
amplitude and shorter duration.
The mode-locked oscillations are passed through a
non linear refractive medium (such as an optical fibre) to introduce a small time dependence 
to the carrier frequency of the mode locked oscillations. This "chirping" allows the laser 
pulse to be temporally stretched. The pulse is then amplified to modest energies and passed 
through a dispersive medium which compresses and produces an ultra intense pulse. 
The resultant laser intensities can exceed $10^{18}\,\text{Wcm}^{-2}$ \cite{EsaSpr92}.
CPA has the advantage of avoiding undesirable high field
effects like self-focusing which can result in a beam of much smaller
intensity or a smaller spot size.

Using the CPA technique a 20 TW peak power, 1.2 ps pulse was generated
from a mode-locked Nd:YAG oscillator. Initially, 120 ps pulses at a 76 MHz were coupled into 
a single mode optical fibre which provided the dispersive medium of which stretch and 
linearly chirped the pulses. The chirped pulse was amplified by passes through a 
Nd:silicate medium resulting in pulse energies of as much as 70 J \cite{Sauteret91}.

An experimental proposal at the Stanford Linear Accelerator Center (SLAC) 
included a CPA high intensity laser beam. A seed pulse from a Nd:YLF laser could be 
amplified to in excess of 1 J and intensities in excess of $10^{19}\,\text{Wcm}^{-2}$. Since 
frequency tripling of nanosecond pulses could be achieved with $\sim 80\%$
efficiency, an ultra intense UV beam ($\gamma =350\,nm$) with peak energy flux $\sim 
4\times10^{17}\,\text{Wcm}^{-2}$ would also be available \cite{McDonald91}.

An alternative to the CPA technique is the production of ultra intense laser beams
via direct amplification of a series of lasing media. For instance, a 248 nm seed pulse was 
generated via a Nd:YAG-pumped, mode-locked dye laser, pre-amplified to a gain of $10^6$ and 
frequency doubled via passage through a KDP crystal. The seed pulse passed through
an electron beam-pumped KrF medium to produce a 248 nm, 4 TW (390 fs,
1.5 J) peak power laser pulse with intensity $10^{19}\,\text{Wcm}^{-2}$ \cite{Watanabe89}.
A similar system producing 248 nm pulses with intensity 
$\sim 2\times\,10^{19}\,\text{Wcm}^{-2}$ 
could be focused to a spot size of diameter $1.7 \,\mu\text{m}$ obtaining intensities in excess of 
$10^{20}\,\text{Wcm}^{-2}$ \cite{Luk89}.

The arrival of ultra intense lasers for which IFQED phenomena could feasibly be detected 
led to a number of experimental studies. A Neodym-glass laser, focused to an intensity of $1.7\times 
10^{14}\;\text{Wcm}^{-2}$ was brought into collision 
with a low energy (500-1600 eV) electron beam to generate various harmonics. The
greatest yield was obtained for second harmonic photons at $0.032\pm 0.003$
photons per laser pulse for an electron energy of 1600 eV. 
These results provided only an order of magnitude confirmation of approximate theoretical
cross sections \cite{EngRin83}. A modification to this experimental configuration was proposed in order 
to take advantage of a distinct forward-backward asymmetry in observed cross sections \cite{PunLeu89}.

\cite{Federici80} produced 5-78 MeV $\gamma$-rays with intensity $10^4$-$10^5$ 
photons s$^{-1}$ and
an energy resolution of 1-10\% from the interaction of an argon ion
laser and electrons produced from the ADONE storage ring at Frascati.
\cite{Ginzburg83a} proposed the production of a similar high energy,
high luminosity $\gamma $-beam using parameters appropriate for the VLEPP
and SLAC linear colliders. 

Back scattered radiation produced from collisions of intense laser beams with electron beams
was studied using a cold fluid model. The radiation occurred in odd harmonics of the incident laser 
frequency and the strength of the harmonics was strongly dependent on the incident laser
frequency \cite{EsaSpr91}.

Interaction of an incident laser source of wavelength 1$\mu$m, pulse
length 2 ps and energy 20 J per pulse, with an electron beam from
an RF linear accelerator, produced hard x-rays (photon energy = 50-1200 keV) with 1 ps
pulse length and 6$\times 10^9$ photons per pulse. Electron beams from a
betatron resulted in hard x-rays with a more moderate photon flux and spectral
brightness \cite{Sprangle92}.

Other studies concentrated on the photon-plasma interactions generated by collisions of laser beams with 
solid targets. Intense laser pulses of power
density $10^{16}\,\text{Wcm}^{-2}$ were directed onto a silicon target to produce 12 nm,
2 ps x-ray pulses with a conversion efficiency of approximately 0.3\%. 
It was estimated that 1.5 nm,10 fs pulses could be
produced at a conversion efficiency in excess of 10\% \cite{Murnane91}. 
More recently the Rutherford-Appleton Laboratory's VULCAN laser system was focused to
$10^{19}\,\text{Wcm}^-2$ onto 
a lead target. Up to 4 harmonics were observed in the emission spectrum \cite{Watts02}.

\cite{EsaSpr92} discussed the possibility of experimentally investigating new IFQED phenomena
such as the optical guiding of laser pulses, wake
fields, laser frequency amplification and relativistic harmonic generation.
\cite{Hora88} proposed a new type of accelerator driven by the frequency and phase
modulation of the two interacting ultra intense lasers. A 100 TW laser pulse power 
could produce a particle acceleration of 600 GeV cm$^{-1}$.
Beamsstrahlung radiation produced by the proximity of intense electron and
positron beams and the resultant intense electromagnetic fields was modelled
in order to determine transverse beam sizes from observed beamsstrahlung
fluxes at SLAC \cite{Zeimann91}. 

Experimental study of IFQED processes in intense Coulomb fields can be achieved via
heavy-ion collisions. \cite{Greiner85} provides a review of IFQED studies involving heavy ion collisions. 
For instance, one experimental study scattered electrons from Argon in the presence of a CO$^2$ laser pulse. 
The kinetic energy of scattered electrons indicated the absorption of up to 
11 photons were observed at theoretically predicted rates \cite{KroWat73,Weingartshofer83}.

\cite{Mikaelian82} discussed the possibility of
observing the scattering of light by light at SLAC using a 19.5 GeV $\gamma$-beam 
and 4.66 eV laser light. Using the form of the electric and magnetic field produced by an ideal lens 
and a given focal length, it was calculated that a single $e^+e^-$ pair could be produced from a focused ruby laser 
pulse of duration $10^{-11}\,\sec $ and total power $10^{19}$ W \cite{BoiWol64,BunTug70}. 
Order of magnitude cross section estimates of external field pair prod near a 
Coulomb centre and by a single photon indicated that experimental observation of
these processes would be practicable \cite{Becker91}.

The experimental program proposed by \cite{McDonald91} was completed and reported 
on by the end of the 1990's. A Nd:glass laser with peak intensities of 
$\sim 0.5\times10^{18}\,\text{Wcm}^-2$ and a 46 GeV electron beam were used to observe the first order 
non linear OPPP and HICS processes. 
Up to four external field photons were found to contribute to the process in excellent agreement with 
theoretical predictions. It was hoped that the mass spectrum of $e^{+}e^{-}$ pairs 
produced from the OPPP process may shed some
light on cross section peaks observed in previous heavy ion collision
experiments \cite{Bamber99}.

Another program of experimental work used a  1.053 $\mu$m laser focused to a 
5 $\mu$m spot size to generate a peak laser intensity of 
$\sim 10^{18}\,\text{Wcm}^-2$. A longitudinal shift in scattered electron momenta due to 
multi-photon contributions from the laser field was observed. Also observed was the predicted electron 
mass shift due to the presence of the external field \cite{Meyerhofer95,Meyerhofer96}.

\section{The Present Work}

The second order IFQED processes have generally received less attention than
the first order IFQED processes. Without taking resonances into account, the cross sections of 
the second order IFQED processes are diminished by a factor of the fine structure
constant compared to the cross sections of the first order processes and, at first glance appear 
experimentally less viable. However, the potential resonances in the second order IFQED cross sections 
provide much motivation for study. These resonances are interesting phenomena in their 
own right and may be experimentally more viable than the first order IFQED processes.
The subject matter of the this thesis involves consideration of some of
the second order IFQED processes and their resonances.

Some of the standard theory of QED required as a basis for further calculations is presented in 
Chapter 2. The Feynman formulation of S-matrix theory was used to perform the 
cross section calculations. A derivation of the Volkov wave function for fermions in an external, 
plane wave electromagnetic field is provided, and the Volkov solutions are used in the Ritus representation of the 
external field electron propagator. The electron self energy, optical theorem and 
regularisation procedures will all be used to calculate radiative corrections.

As part of the original work presented in this thesis, stimulated Compton
scattering (SCS) with a circularly polarised external electromagnetic field,
arbitrary kinematics and without recourse to a non relativistic regime, is
considered. A circularly polarised external electromagnetic field results in well known Bessel functions 
appearing in the cross section. Circularly polarised electromagnetic field are easily achieved 
experimentally using intense laser beams.
The stimulated two photon pair production (STPPP) process in the
same circularly polarised external field is also considered with the aid of expressions obtained
for the SCS calculations and by use of crossing symmetry which links the
two processes. The analytic expressions for both SCS and STPPP processes are contained in
Chapter 3.

Numerical results and analysis of the SCS and STPPP differential cross sections are presented
in Chapters 4 and 5 respectively. Analysis is presented in terms of the discrete 
contribution of external field quanta to the process. A subset of the complete parameter space in which 
the differential cross sections vary, is examined. Since experimental validation is important, parameter 
sets which maximise the STPPP and SCS differential cross sections are a priority.

One crucial feature that emerged from the literature is the
possibility of resonances in the second order IFQED processes.
The calculation of resonant SCS and STPPP cross sections require the
electron self energy in the presence of the external field. Though this self
energy exists in the literature for the case of a circularly polarised
electromagnetic field, the given proper time representation proves
technically difficult to include in the calculations in this thesis. In chapter 6 an alternative calculation 
of the external field electron self energy using dispersion relations is provided, and a 
representation in terms of infinite summations of Bessel functions is obtained. The correctness of the 
result will be established with use of the optical theorem and the well known HICS 
differential cross section.

The results of Chapter 6 enable an analytic and numerical calculation of the SCS and 
STPPP resonant cross sections to be performed in Chapter 7.
Attention is given to the requirements of any future attempt at
experimental measurement. Radiative corrections to first order in the fine
structure constant are included in the external field electron propagator
and an external field regularisation and renormalisation procedure is
outlined.

Research and development of $e^+e^-$ colliders involve background studies
of pair production processes. These processes occur in the midst of focused
electron and positron bunches which produce intense bunch fields. These fields are almost
plane wave and contain constant electric and magnetic components that are mutually orthogonal to the 
direction of propagation of the bunches. 
It is of interest therefore to perform the STPPP calculation in the presence of a constant crossed 
electromagnetic field. The differential cross section is calculated using real beam parameters proposed 
for future linear collider designs, with the aim of estimating whether there will be a significant 
increase in expected background pairs. This is the subject of Chapter 8 which draws on the preceeding 
chapters.

In chapter 9 the work of all preceeding chapters is drawn together in conclusion.

%% file: chap2.tex
\section{Introduction}
\label{c2intro}

Presented in this chapter is some of the basic QED theory required to perform
the cross section calculations of the IFQED processes to be considered in this thesis.

Basic matters such as the metric to be used, the units employed and 
certain normalisations are stated in section \ref{c2units}. Section 
\ref{c2bip} presents the Lagrangian of the interacting Maxwell and Dirac fields. IFQED 
calculations are most conveniently performed in the Bound Interaction Picture. The standard 
S-matrix theory describing the time evolution of the State vector is outlined in 
section \ref{c2smat}.
Section \ref{c2feyn} discusses the Feynman diagrams for which the task 
of extracting the desired transition of states from the iteration solution of
the S-matrix is considerably simplified. The crossing symmetry of the 
S-matrix will be used to simplify the calculation of related IFQED processes. This 
is described in section \ref{c2sub}. Section \ref{c2cross} outlines how the differential 
cross section is obtained from squaring the matrix element and introducing the phase integral. 
The differential cross section requires summation over all fermion spins and photon 
polarisations which introduces a trace calculation of products of Dirac $\gamma$ matrices. This 
is explained in section \ref{c2sum}.

Section \ref{c2ext} defines the 4-potential of the external fields required for the IFQED 
calculations. Those described are a circularly polarised and a constant crossed plane 
electromagnetic wave. The derivation of the Volkov solution of the Dirac equation in an external plane wave 
field of general form is outlined in section \ref{c2volk}. The resulting Volkov $E_p$ functions 
are used to construct the Ritus form of the external field (bound) electron propagator.

Section \ref{c2rad} discusses the radiative 
corrections to the external field electron propagator. These corrections are necessary 
since the presence of the external field leads to propagator poles. The radiative corrections 
produce a shift in intermediate electron energy to complex values. An expression for the 
electron energy shift in an external field is obtained in section \ref{c2efees}. The radiative 
corrections require regularisation and renormalisation in order to remove divergences. The 
usual, non external field version of these is presented in section \ref{c2regn}. The optical 
theorem described in section \ref{c2opt} relates elastic scattering amplitude to 
transition probability will be useful for validation of the self energy calculations in Chapter 
6.

The content of this chapter draws on several texts including 
\cite{Schweber62,AkhBer65,JauRoh76,Nachtmann90,ItzZub80,ManSha84,GreRei02,BerLifPit82,Muirhead65}.

\medskip\
\section{Units, normalisation constants, notation  and metric}
\label{c2units}

In this thesis Planck (natural) units are used in which the speed of 
light in a vacuum $c$, reduced Planck constant $\hbar$ and Coulomb force constant 
$\dfrac{1}{4\pi\epsilon_0}$ are all equal to 1. The metric to be used, $g_{\mu\nu}$, has signature
(1,-1,-1,-1) so that for any contravariant 4-vector $x^\nu =(x^0,\sql{x})$, a covariant 4-vector is
formed via $x_\mu=g_{\mu\nu}x^\nu=(x_0,-\sql{x})$.

Dirac bispinors are $u(p)$ for electrons and $v(p)$ for positrons. Projection Operators 
$\Lambda^{\pm}(p)$ pick out electron or positron bispinors from linear combinations. The symbol 
$e(k)$ represent photon polarisation 4-vectors.

Normalisation constants $\sqrt{\frac{m}{V\epsilon_p}}$ for Dirac bispinors, and 
$\sqrt{\frac{1}{2V\omega}}$ for polarisation 4-vectors are chosen so that the 
probability of finding either a fermion of mass $m$ and energy $\varepsilon_p$, or a photon of 
energy $\omega$ in a box of Volume V, is unity (see for example \cite{Muirhead65}).

A particular representation for Dirac $\gamma$ matrices is unimportant since only their 
anti-commutation and hermicity properties are required. The "slash" notation will be used to 
denote products of 4-vectors and Dirac $\gamma$ matrices so that $\slashed{a}=\gamma^\mu a_\mu$. 

The script letters $\Im$ and $\Re$ will be used to refer to the imaginary and real parts of some 
quantity. The convention of implied summation is assumed so that 

\begin{equation}
x^\mu x_\mu \equiv \dsum\limits_0^3 x^\mu x_\mu
\end{equation}

\bigskip\
\section{The Bound Interaction Picture}
\label{c2bip}

The theory of quantum electrodynamics describes the interactions between the
quantised Maxwell and Dirac fields. The description requires the free
Maxwell and free Dirac Lagrangian densities $L_M$ and $L_D$.

\begin{equation}
\label{c2.eq1} 
\begin{array}{cll}
& L_M = -\frac 14\,F^{\mu\nu }F_{\mu\nu } \quad & \text{where} \quad 
F_{\mu\nu}=\partial_{\mu}A_{\nu}-\partial_{\nu}A_{\mu} \\[10pt] 
& L_D = \bar \psi (i\slashed{\partial}-m)\psi \quad & \text{where} \quad 
\slashed{\partial}=\gamma^{\mu}\dfrac{\partial}{\partial x_\mu}
\end{array}
\end{equation}

\medskip\ 

The Lagrangian densities entail the free Maxwell and free Dirac field
equations specifying the space-time evolution of the Maxwell field operator $%
A_\mu (x)$ and the Dirac field operator $\psi (x)$.

\begin{equation}
\label{c2.eq2} 
\begin{array}{ccc}
\partial^{2}_{\mu}A_\mu (x) & = & 0 \\[10pt] 
(i\slashed{\partial}-m)\psi (x) & = & 0 
\end{array}
\end{equation}

\medskip\ 

The solutions to equations \ref{c2.eq2} are written in terms of operators
that create (denoted by the superscript $+$) or destroy (denoted by the
superscript $-$) an electron, positron or photon at space-time point $x_\mu $%
. The electron and positron field operators are written in terms of Dirac
bispinors $u_s(p)$ and $v_s(p)$, and the photon field operators are written
in terms of polarisation 4-vectors $e_r^\mu (k)$.

\begin{equation}
\label{c2.eq3} 
\begin{array}{ccccccc}
\psi _p^{\mu +}(x,p) & = & \sqrt{\dfrac m{V\epsilon _p}}\,u_s^\mu (p)\,e^{-ipx}
&  & \psi _p^{\mu -}(x,p) & = & \sqrt{\dfrac m{V\epsilon _p}}\,v_s^\mu
(p)\,e^{ipx} \\[10pt]

\bar \psi _p^{\mu +}(x,p) & = & \sqrt{\dfrac m{V\epsilon _p}}\,
\bar u_s^\mu (p)\,e^{ipx} &  & \bar \psi _p^{\mu -}(x,p) & = & \sqrt{\dfrac
m{V\epsilon _p}}\,\bar v_s^\mu (p)\,e^{-ipx} \\[10pt] 

A^{\mu +}_{k}(x) & = & \sqrt{\dfrac 1{2V\omega }}\,e_r^\mu (k)\,e^{-ikx} &  & A^{\mu-}_{k}(x) & 
= & \sqrt{\dfrac 1{2V\omega }}\,e_r^\mu (k)\,e^{ikx} 
\end{array}
\end{equation}

\medskip\ 

Interaction between the Maxwell and Dirac fields in the presence of an external 
electromagnetic field specified\footnote{the explicit form of $A^{e}_{\mu}$ will
be considered in section \ref{c2ext}.} by the 4-potential $A^{e}_{\mu}$, is usually viewed in the
bound interaction picture \cite{Furry51}. The bound interaction picture provides a way of 
describing the interacting system of free Maxwell field, free Dirac field and the external field such
that the solutions of the combined field equations are as simple as possible.

In the bound interaction picture the total Lagrangian density for the
interacting Maxwell and Dirac fields is a sum of the free Maxwell Lagrangian
density $L_M$, the bound Dirac Lagrangian density $L_{BD}$ and the
interaction Lagrangian density $L_I$

\begin{equation}
\label{c2.eq4} 
\begin{array}{rcl}
L & = & L_M+L_{BD}+L_I \\  
&  &  \\ 
\text{where\quad \quad }L_{BD} & = & \bar \psi (x)[(i\slashed{\partial}-e 
\slashed{A}^e)-m]\psi (x) \\ L_I & = & -e\bar \psi (x)\slashed{A} \psi
(x) 
\end{array}
\end{equation}

\medskip\ 

In the bound interaction picture the interaction between the Dirac field and external field
is implicit in a bound Dirac Lagrangian $L_{BD}$. Then the interaction Lagrangian $L_I$ describes 
interaction between bound Dirac and Maxwell fields. A solution for the bound Dirac field 
operator is required and the Maxwell field operator is left unchanged.

\newpage\
\section{S-matrix Theory}
\label{c2smat}

The system of interacting fields viewed in the Bound Interaction Picture can
be specified by a state vector $\left| \Phi (t)\right\rangle _F$. The time
evolution of such a state vector from a time $t_0$ in the past, can be
written with the aid of a unitary operator $U(t,t_0)$ which itself evolves
in time

\begin{equation}
\label{c2.eq5} 
\begin{array}{rcl}
\left| \Phi (t)\right\rangle _F & = & U(t,t_0)\,\left| \Phi
(t_0)\right\rangle _F \\[20pt]  
\text{where\quad \quad }i\dfrac d{dt}U(t,t_0) & = & H_I(t)\,U(t,t_0) \\ 
H_I(t) & = & -\int L_I(x_\mu )\; d^3x 
\end{array}
\end{equation}

\medskip\ 

There is no exact solution for these equations. However an approximation can
be made by an expansion in powers of the interaction Hamiltonian $H_I$. The
approximation is valid because the coupling between the Maxwell and Dirac fields is proportional 
to the fine structure constant, and is therefore weak. The result is written with the aid of the Dyson 
chronological product $P$

\begin{equation}
\label{c2.eq6} 
\begin{array}{rl}
U(t,t_0) = & \dsum\limits_{n=0}^\infty \dfrac{(-i)^n}{n!}\dint_{\nths t_0}^tdt_1%
\dint_{\nths t_0}^tdt_2\ldots \dint_{\nths t_0}^tdt_n \;P[H_I(t_1)H_I(t_2)\ldots H_I(t_n)] \\[15pt]

\text{where} \quad\quad & P[H_I(t_1)H_I(t_2)\ldots H_I(t_n)]=H_I(t_{i_1})H_I(t_{i_2})\ldots
H_I(t_{i_n}) \\
& \quad\quad t_{i_n}\leq t_{i_{n-1}}\leq \ldots \leq t_{i_1} 

\end{array}
\end{equation}

\medskip

The S-operator and S-matrix are defined respectively in equation \ref{c2.eq7}. These describe the time evolution of the interacting system from
an initial state\ $\left| i(t_0)\right\rangle $ in the remote past to
a final state $\left| f(t)\right\rangle$ in the remote future.

\begin{equation}
\label{c2.eq7} 
\begin{array}{rl}
S & = \lim \limits_{t\rightarrow \infty }\;\lim \limits_{t_0\rightarrow-\infty }U(t,t_0) 
\\[10pt]  
S_{fi} & = \left\langle f(\infty )\right| S\left| i(-\infty )\right\rangle 
\end{array}
\end{equation}

\medskip

Substituting in the expressions for the $U$-operator in equation \ref{c2.eq6}
and extracting integrations over spatial variables for the Dyson
chronological product, the iteration solution for the S-operator can be
written in terms of the interaction Lagrangian

\begin{equation}
\begin{array}{rl}
\label{c2.eq8}

S &= \dsum\limits_{n=0}^\infty S^{(n)} \quad\quad \text{where} \quad\quad S^{(0)}=1 \\[10pt] 

S^{(n)} &=\dfrac{(-i)^n}{n!}\dint_{\nths -\infty }^\infty d^4x_1\dint_{\nths -\infty }^\infty
d^4x_2\ldots \dint_{\nths -\infty }^\infty d^4x_nP[L_I(x_1)L_I(x_2)\ldots L_I(x_n)] 

\end{array}
\end{equation}

\medskip

Since pairs of Dirac operators in the Dyson chronological product above
occur at identical space-time points, $S^{(n)}$ can be written in terms of
the time ordering product $T$

\begin{equation}
\label{c2.eq9}
\begin{array}{rl}

S^{(n)} & =\dfrac{(-i)^n}{n!}\dint_{\nths -\infty }^\infty
d^4x_1\dint_{\nths -\infty }^\infty d^4x_2\ldots \dint_{\nths -\infty }^\infty
d^4x_n \, T[L_I(x_1)L_I(x_2)\ldots L_I(x_n)] \\[10pt]

\text{where} & \\
& \begin{array}{ccll}
T[M(t_1)N(t_2)] & = & \left\{ 
\begin{array}{c}
M(t_1)\;N(t_2)\quad 
\text{if}\quad t_1>t_2 \\ -N(t_1)\;M(t_2)\quad \text{if}\quad t_2>t_1 
\end{array}
\right\} & \text{fermions} \\[20pt]
 T[M(t_1)N(t_2)] & = & P[M(t_1)N(t_2)] & \text{bosons} 

\end{array}
\end{array}
\end{equation}

\medskip

The second order terms of the $S$-operator iteration solution constitute the second order 
processes. A subset of these represent the specific interactions that will be studied in 
this thesis.

\bigskip\
\section{Wicks theorem and Feynman diagrams}
\label{c2feyn}

The expression for the S-operator obtained in section \ref{c2smat} provides a
perturbative expansion in which terms of the desired order can be extracted.
The second order terms are written

\begin{equation}
\label{c2.eq31}S^{(2)}=-\frac 12\int_{-\infty }^\infty d^4x_1\int_{-\infty
}^\infty d^4x_2 \; T[L_I(x_1)L_I(x_2)] 
\end{equation}

\medskip\ 

Each field operator present in equation \ref{c2.eq31} consists of a sum of 
creation and destruction operators. For a specific transition 
$\left|i\right\rangle \rightarrow \left| f\right\rangle $ all that is 
required are the correct number of destruction 
operators for states $\left| i\right\rangle$ to vanish and the correct number of
creation operators for states $\left| f\right\rangle$ to exist. Selection
of the desired set of operators is simplified by use of the normal product $N$
which places creation operators to the left of destruction operators.
The normal product can be introduced into equation \ref{c2.eq31} by Wick's
Theorem which allows the time ordering product $T$ to be written as a sum of
the normal products $N$ and vacuum expectation values. For instance the time ordering of two field 
operators A,B is

\begin{equation}
\label{c2.eq31b}
T[A(x_1),B(x_2)]=N[A(x),B(x)]+ 
\left\langle 0\right| T[A(x_1)B(x_2)]\left|0\right\rangle
\end{equation}

\medskip\

The complete $S^{(2)}$ expansion which involves the time ordering of six field operators yields 
eight normal product terms, some of which are identical under permutation 
operations (\cite{ManSha84} p.109-110). There are two identical normal product
terms for both Compton scattering and pair production so the S-matrix for these 
processes gains a factor of two. Additionally, either of the two Maxwell field operators can 
absorb initial photons leading to two terms in, for instance, the Compton Scattering S-matrix

\begin{equation}
\label{c2.eq32} 
\begin{array}{ccr}
S_{fi} & = & - e^2 \dint_{\nths -\infty}^\infty d^4x_1 \; d^4x_2\; 
\left\{ \overline{\psi}_f(x_2) A_{f}^{\mu+}(x_2)
\underbrace{\psi (x_2) \overline{\psi }(x_1)} A_{i}^{\nu-} (x_1)\psi_i (x_1)\right. \\
&  &  \\  &  & \left. + \overline{\psi}_{f}(x_2) A_{i}^{\nu-}(x_2)
\underbrace{\psi(x_2)\overline{\psi}(x_1)}
A_{f}^{\mu+}(x_1)\psi_{i}(x_1) \right\} \\  &  &  
\end{array}
\end{equation}

\medskip 

The fermion contraction $\underbrace{\psi (x_2)\overline{\psi }(x_1)}$ is a vacuum 
expectation value 
$\left\langle 0\right| T[\psi (x_2)\overline{\psi }(x_1)]\left|0\right\rangle$ 
from which the fermion propagator is derived. 

Equation \ref{c2.eq32} can be written down directly with the aid of 
Feynman diagrams which graphically represent the transition of states desired.
Straight lines represent fermion operators, wavy lines represent photon
operators and straight lines running between two vertices represent fermion
propagators. The Dirac equation can also be solved for fermions in the presence of an 
external field (see section \ref{c2volk}) and these solutions are represented within 
Feynman diagrams by double straight lines. The complete rules pertaining to Feynman diagrams 
can be found in the standard texts (see for example \cite{Muirhead65} pg.326 or \cite{ManSha84}
section 7.3)

\bigskip\
\section{Crossing symmetry}
\label{c2sub}

\begin{figure}[!b]
\centerline{\includegraphics[height=6cm,width=9cm]{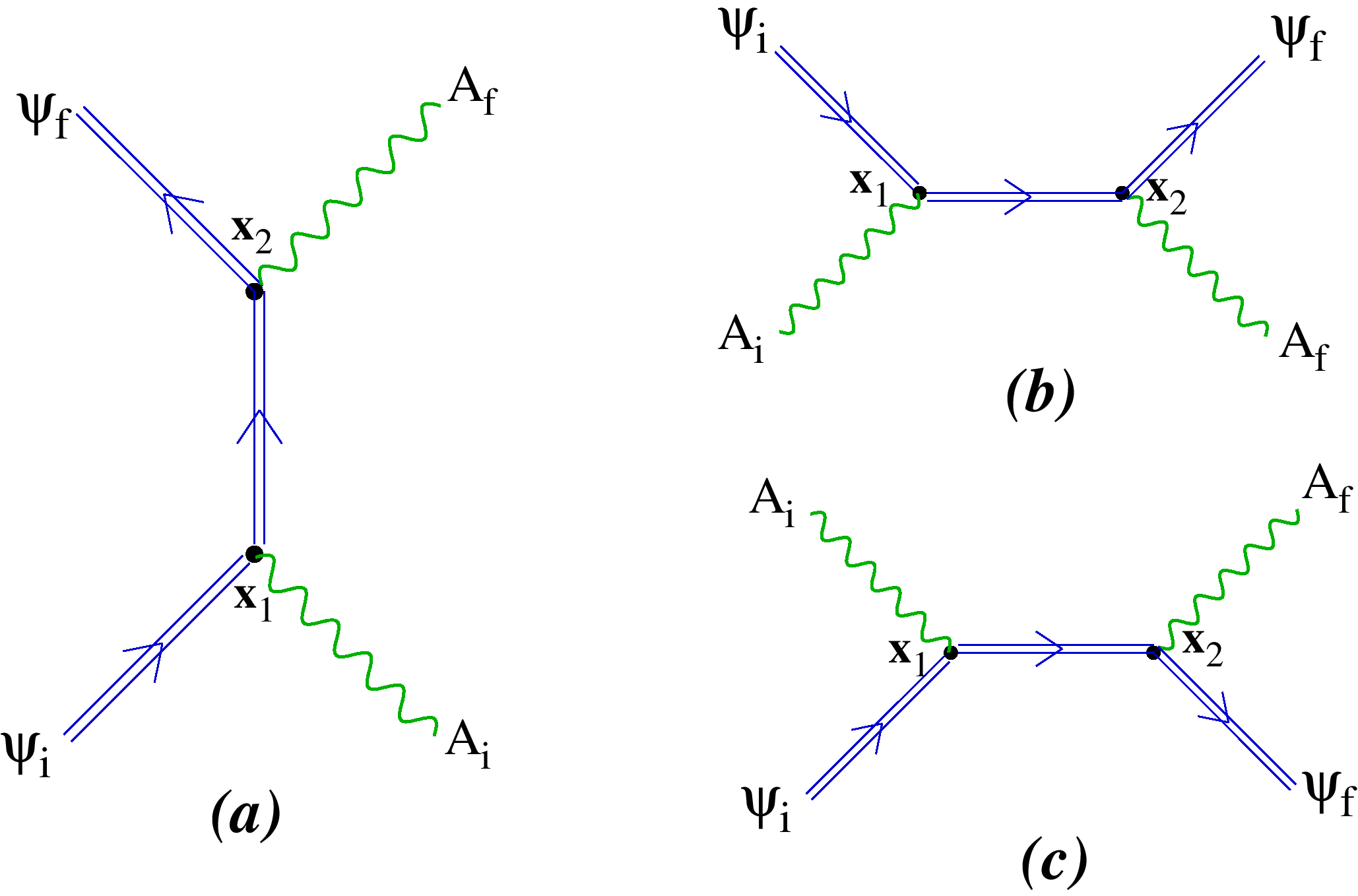}}
\caption{\bf Feynman diagrams for (a) Compton scattering (b) $e^{+}e^{-}$ 
pair production, and (c) $e^{+}e^{-}$ Pair Annihilation.}
\label{c2.fig2}
\end{figure}

The structure of the iteration solution in equation \ref{c2.eq9} is such that the $S$-matrix 
is invariant under certain symmetry operations. Particular sets of these symmetry operations permit a 
detailed evaluation of the $S$-matrix for one scattering process to be used, after suitable
substitutions of 4-momenta, to obtain results for other related scattering processes. This crossing
symmetry entails a set of rules that govern this exchange of 4-momenta.

A crossing symmetry rests on the premise that the annihilation of a
particle is equivalent to the creation of its anti-particle and is most clearly
expressed using the Feynman diagrams.
Consider two Feynman diagrams $M,M'$ which involve electrons with
momenta $p,p'$, positrons with momenta $q,q'$ and photons with momenta $k,k'$. If $M,M'$
differ in only one external line such that an outgoing particle in $M'$ is an ingoing 
particle in $M$ (or vice versa) then a set of correspondences can be found (table 2.1).

\medskip\

\begin{table}[!h]
\label{c2subt1}
\center{
\begin{tabular}{|c|c|c|} \hline
$\text{diagram }M'$ & $\text{diagram }M$  & correspondence \\ \hline\hline
$k'\text{ out}$ & $k\text{ in}$ & $k'\leftrightarrow -k \quad 
e^{\prime }\leftrightarrow e$ \\ \hline
$p'\text{ out}$ & $q\text{ in}$ & $p'\leftrightarrow -q \quad 
u(p')\leftrightarrow v(p) \quad \Lambda_{-}(p')\leftrightarrow \Lambda_{+}(q)$ \\ \hline
$q'\text{ out}$ & $p\text{ in}$ & $q'\leftrightarrow -p \quad 
v(q')\leftrightarrow u(p) \quad \Lambda_{+}(q')\leftrightarrow \Lambda_{-}(p)$ \\ \hline
\end{tabular} }
\caption{\bf\bm Crossing symmetry correspondences.}
\end{table}

The symbols $u(p),v(p),e(k),\Lambda _{-}(p),\Lambda _{+}(p)$  were introduced in section 
\ref{c2units} and respectively are the electron and positron bispinors, the photon
4-polarisation and the projection operators \cite{JauRoh76}.

The Feynman diagrams depicted in figure \ref{c2.fig2} represent Compton scattering, pair 
production and pair annihilation. These three processes are all related by crossing symmetries. 

\bigskip\

\section{Summation over spin and polarisation states}
\label{c2sum}

A calculation of the cross section of a transition from states $\left|
i\right\rangle $ to states $\left| f\right\rangle $ requires a summation
over fermion spin states and photon polarisation states. Initial states have to
be averaged.

The fermion spin states are expressed by Dirac bispinors and for Compton scattering 
and pair production appear in the S-matrix in the form

\begin{equation}
\label{c2.eq33} 
\begin{array}{rcl}
S_{fi}=\overline{u}_r^\alpha (p_f)\;Q_{\alpha \beta }\;u_s^\beta (p_i) &  & 
\text{Compton scattering} \\[10pt]
S_{fi}=\overline{u}_r^\alpha (p_{-})\;P_{\alpha \beta}\;v_s^\beta (p_{+}) &  & 
\text{pair production} 
\end{array}
\end{equation}

\medskip\ 

$Q_{\alpha \beta }$ and $P_{\alpha \beta }$ are functions of Dirac 
$\gamma $-matrices, and $u_s(p)$ and $v_s(p)$ are Dirac bispinors of the electron
and positron respectively. Leaving aside averaging for now, the sum over initial and final spin states in the square 
of the matrix element is written

\begin{equation}
\label{c2.eq34} 
\begin{array}{rcl}
\dsum\limits_{r,s}\;\overline{u}_r^\alpha (p_f)\;Q_{\alpha \beta
}\;u_s^\beta (p_i)\;\overline{u}_s^\gamma (p_i)\;\overline{Q}_{\gamma \delta
}\;u_r^\delta (p_f) &  & \text{Compton scattering} \\[10pt]
 \dsum\limits_{r,s}\;
\overline{u}_r^\alpha (p_{-})\;P_{\alpha \beta }\;v_s^\beta (p_{+})\; 
\overline{v}_s^\gamma (p_{+})\;\overline{P}_{\gamma \delta }\;u_r^\delta
(p_{-}) &  & \text{pair production} 
\end{array}
\end{equation}

\medskip\ 

Equation \ref{c2.eq34} can be rearranged and written in terms of products of
Dirac bispinors which yield the projection operators $\Lambda _{\alpha \beta
}^{+}(p)$ and $\Lambda _{\alpha \beta }^{-}(p)$.

\begin{equation}
\label{c2.eq35} 
\begin{array}{ccccc}
\dsum\limits_ru_r^\alpha (p)\,\overline{u}_r^\beta (p) & = & \dfrac 1{2m}(%
\slashed{p}+m)_{\alpha \beta } & \equiv & \Lambda _{\alpha \beta }^{+}(p) \\ 
\dsum\limits_rv_r^\alpha (p)\,\overline{v}_r^\beta (p) & = & \dfrac 1{2m}(%
\slashed{p}-m)_{\alpha \beta } & \equiv & \Lambda _{\alpha \beta }^{-}(p) 
\end{array}
\end{equation}

\medskip\ 

The combination of projection operators and $\gamma$-matrix functions, $\Lambda_{\alpha \beta}Q_{\alpha \beta}$ 
will result in a summation over diagonal terms (a trace) which in turn yields
a function of scalar products of the 4-vectors taking part in the process.

Photon polarisation 4-vectors $e_r^\alpha (k)$ appear in the S-matrix 
as slash vectors. Both the Compton and pair production processes involve two photons and the
trace contains four polarisation vectors whose polarisation states can be
denoted by subscripts $i$ and $j$.

\begin{equation}
\label{c2.eq37}\dsum_i\dsum_j\limfunc{Tr}\left[ \ldots \slashed{e}_i(k_1)
\ldots \slashed{e}_j(k_2)\ldots \overline{\slashed{e}}_j(k_2)\ldots 
\overline{\slashed{e}}_i(k_1)\ldots \right] 
\end{equation}

\medskip\ 

Polarisation 4-vectors are extracted from the trace leaving behind Dirac $\gamma $-matrices.%

\begin{equation}
\label{c2.eq38}\dsum_ie_i^\alpha (k_1)\overline{e}_i^\beta
(k_1)\dsum_je_j^\mu (k_2)\overline{e}_j^\nu (k_2)\; \limfunc{Tr}\left[ \ldots
\gamma _a\ldots \gamma _\mu \ldots \gamma _\nu \ldots \gamma _\beta \ldots
\right] 
\end{equation}

\medskip\ 

Polarisation 4-vectors can be viewed in a reference frame such that
longitudinal, transverse and scalar components can be separated. For free photons the 
summation over polarisation states reduces to a summation over
transverse components. With $g^{\alpha\beta}$ representing the metric tensor and $n$ a time like 
unit vector

\begin{equation}
\label{c2.eq39}\dsum_{i=1}^2e_i^\alpha (k_1)\overline{e}_i^\beta
(k_1)=-g^{\alpha \beta }-\frac 1{(kn)^2}\left[ k^\alpha k^\beta
-(kn)(k^\alpha n^\beta +k^\beta n^\alpha )\right] 
\end{equation}

\medskip\ 

\enlargethispage*{2cm}
For a real photon ($k^2=0$) gauge invariance requires that $-g^{\alpha\beta}$ be the only
non vanishing term on the right hand side of equation \ref{c2.eq39} and the summation 
over polarisation states (equation \ref{c2.eq38}) reduces to

\begin{equation}
\label{c2.eq40} \limfunc{Tr}\left[ \ldots \gamma _\mu \ldots \gamma _\nu
\ldots \gamma _\nu \ldots \gamma _\mu \ldots \right] 
\end{equation}

\medskip\

Both the fermion spins and photon polarisations consist of two states. 
Since both Compton scattering and pair production contain two each of these particles 
in their initial states, then initial state averaging results in an additional factor of $\frac{1}{4}$. 

The final result after all summing and averaging is 

\begin{equation}
\label{c2.eq36} 
\begin{array}{rcl}
\dfrac{1}{16m^2}\limfunc{Tr}\left[ (\slashed{p}_f+m)Q(\slashed{p}_i+m)\overline{Q}%
\right] &  & \text{Compton scattering} \\[10pt]
-\dfrac{1}{16m^2}\limfunc{Tr}\left[ (\slashed{p}_{-}+m)
P(\slashed{p}_{+}-m)\overline{P}\right] &  & \text{pair production} 
\end{array}
\end{equation}

\medskip\

\section{The transition probability and the scattering cross section}
\label{c2cross}

Numerical values quantifying the particular interaction of interest are
obtained via the transition probability $W$ or the scattering cross section $%
\sigma $. The transition probability and scattering cross section are both derived from 
the $T$-matrix specifying a transition from an initial state $i$ of total 
4-momentum $P_i$ to a final state $f$ of total 4-momentum $P_f$. The $T$-matrix is related 
to the $S$-matrix by

\begin{equation}
\label{c2.eq10}S_{fi}=\delta _{fi}+i(2\pi )^4\,\delta (P_i-P_f)\,T_{fi} 
\end{equation}

\medskip\ 

In any scattering process a specific initial state can result
in any number $N_f$ of final states. If the scattering process takes place 
in a spatial volume $V$ and consists of asymptotically free initial particles, an 
interaction region and asymptotically free final states,
the small number of final states of 3-momentum lying between $p$ and $p+dp$
available to a single particle is given by

\begin{equation}
\label{c2.eq11}dN_f=\frac V{(2\pi )^3}\,dp 
\end{equation}

\medskip\ 

The cross section of the scattering is defined by a product of the
transition probability per unit of space-time $W$, the flux density of
initial particles $J_i$ and the number of final states

\begin{equation}
\label{c2.eq12} 
\begin{array}{rl}
d\sigma_{fi} & = \dfrac W{J_i}\,dN_f \\ 
\text{where} \quad \quad W & = (2\pi )^4\,\delta (P_i-P_f)\,\overline{\dsum\limits_i}
\dsum\limits_f\left| T_{fi}\right| ^2 
\end{array}
\end{equation}

\medskip\ 

For scattering processes involving two initial particles with relative velocity $v$, 
the flux density is $J_i=\dfrac{v}{V^2}$. The two IFQED processes to be studied in 
Chapter 3 both have two particles in the initial state. The SCS process has a 
photon $(\omega _i,k_i)$ and electron $(\epsilon _i,q_i)$ in the initial state and a photon 
$(\omega _f,k_f)$ and electron $(\epsilon _f,q_f)$ in the final state. 
The STPPP process has two photons $(\omega _1,k_1)$, $(\omega _2,k_2)$ in the 
initial state and an electron $(\epsilon_{-},q_{-})$ and positron 
$(\epsilon_{+},q_{+})$ in the final state. After extracting the
normalisation constants of field operators from the $T$-matrix the
differential cross sections for the SCS and STPPP process can be written

\begin{equation}
\label{c2.eq13} 
\begin{array}{rl}
d\sigma _{fi} &=\dfrac 1{v(2\pi )^2}\dfrac{m^2}{4\epsilon _i\omega _i}\, 
\delta(P_i - P_f)\overline{\dsum\limits_i}\dsum\limits_f\left| T_{fi}\right| 
^2\,\dfrac{dp_f\,dk_f}{\epsilon _f\,\omega _f} \quad\quad \text{SCS process} \\[10pt]  

d\sigma _{fi} &=\dfrac 1{v(2\pi )^2}\dfrac{m^2}{4\omega _1\omega _2}\, 
\delta(P_i - P_f)\overline{\dsum\limits_i}\dsum\limits_f\left| T_{fi}\right| 
^2\,\dfrac{dp_{-}\,dp_{+}}{\epsilon _{-}\,\epsilon _{+}} \quad\quad \text{STPPP process} 

\end{array}
\end{equation}

\bigskip
\section{The external field}
\label{c2ext}

There will be two external fields considered in this thesis; a circularly polarised 
electromagnetic field and a constant crossed electromagnetic field. The experimental source of 
circularly polarised fields are intense laser beams. The idealisation of the laser beam as a 
plane wave of constant energy and intensity is reasonable given the monochromaticity and
coherence properties of the laser and since photon depletion of the beam can be neglected. 

The circularly polarised external field can be described by a 4-potential $A_\mu ^e(x)$ 
which is a sum of sinusoidal components containing the 4-potential components 
$a_{1\mu}, a_{2\mu}$ and the 4-momentum of the external field $k_{\mu}$

\begin{equation}
\label{c2.eq17}A_\mu ^e(x)=a_{1\mu }\cos (kx)+a_{2\mu }\sin (kx) 
\end{equation}

\medskip\ 

The 4-potential $A_\mu^e(x)$ can be viewed with a gauge in which the time (zeroth) 
component vanishes. The Lorentz gauge condition $(kA^e)=0$ then requires the 3-vectors 
$\sql{a}_1$, $\sql{a}_2$ and $\sql{k}$ to be mutually orthogonal. With $a^2$ being 
the square of the magnitude of the 3-vectors $a_{1\mu}$ and 
$a_{2\mu}$, and with a minus sign resulting from the choice of metric (see section 
\ref{c2units}), scalar products of 4-vectors $a_{1\mu }$, $a_{2\mu }$ and $k_\mu$ satisfy 

\begin{equation}
\label{c2.eq18} 
\begin{array}{ccc}
(a_1a_1)=(a_2,a_2) & = & -a^2 \\ 
(a_1k)=(a_2k)=(kk) & = & 0 
\end{array}
\end{equation}

\medskip\ 

A subsequent identity which any 4-vectors $Q_\mu $ and $P_\mu$ satisfy can be found

\begin{equation}
\label{c2.eq19}
\begin{array}{rl}
\text{If} \quad\quad & (kQ),(kP) = 0 \\
\text{then} \quad\quad & (a_1Q)\,(a_1P)\;+\;(a_2Q)\,(a_2P) = a^2(QP)
\end{array}
\end{equation}

\medskip\ 

The source of the constant crossed electromagnetic fields considered in this thesis are 
dense, ultra relativistic fermion bunches. A plane wave representation for these bunches is valid as 
long as the bunch profile is not significantly disrupted. The constant crossed electromagnetic field 
4-potential contains only one component and satisfies

\begin{equation}
\label{c2ext.last}
\begin{array}{ccc}
A^{e}_{\mu}(x) &=& a_{1\mu} (kx) \\
(a_1 a_1) &=& -a^2 \\
(a_1 k) &=& 0
\end{array}
\end{equation}

\medskip\
\section{The Volkov solution and the Bound Electron Propagator}
\label{c2volk}

In the bound interaction picture the external field is taken into
account by the electron - external field interaction. A solution of
the Dirac equation modified by the inclusion of the external field operator 
$A_\mu ^e(x)$ is required. The necessary modification is achieved by the 
replacement of the momentum operator $\partial $$_\mu $ with 
$\partial $$_\mu +ieA_\mu ^e$. The quadratic Dirac equation with an external field is obtained by
applying the operator $i\slashed{\partial}-e\slashed{A}^e +m$ to the linear Dirac equation with
external field (equation \ref{c2.eq2}). It is written in terms of the antisymmetric spin tensor
$\sigma^{\mu\nu}$ and the field tensor of the external field $F^{e}_{\mu\nu}$

\begin{equation}
\label{c2.eq21}
\begin{array}{rl}

&\left[ (p-eA^e)^2-m^2-\frac{ie}2F^{e}_{\mu \nu }\sigma^{\mu \nu}\right] \psi _V(x,p)=0 \\[20pt]
\text{and} \quad\quad & F^{e}_{\mu \nu }=k_\mu \dfrac{\partial A_\nu }{\partial \phi }-k_\nu 
\dfrac{\partial A_\mu }{\partial \phi } \\[10pt]
&\text{where}\quad\quad \phi =k^\mu x_\mu

\end{array}
\end{equation}

\medskip\ 

Equation \ref{c2.eq21} can be solved exactly when the external field is a classical 
electromagnetic wave \cite{Volkov35}. The general solution is a product of the bispinor 
solution $u_{s}(p)$ to the free Dirac equation $(p-m)u_s(p)=0$ and a function $F(\phi)$ 
whose form is to be determined

\begin{equation}
\label{c2.eq22}
\psi _V(x,p)=u_s(p)\,F(\phi ) 
\end{equation}

\medskip\

Expansion of operator products in equation \ref{c2.eq21} and substitution of the general solution
$\psi_{V}(x,p)$ yields a first order differential equation in $F(\phi)$. With use of the Lorentz
gauge condition, the solution is 

\begin{equation}
\label{c2.eq23}F(\phi )=\exp \left[ \frac e{2(kp)}\slashed{k}\slashed{A}^e\right] \exp 
\left( -ipx -i\int_0^{kx}\left[ \frac{(A^ep)}{(kp)}-\frac{e^2A^{e2} }{2(kp)}\right] d\phi \right) 
\end{equation}

\medskip\ 

By expanding $\exp [e/2(kp)]$ in a power series and using $(kA^e)=0$ , the
Volkov solution for electrons is written in equation \ref{c2.eq24}. 
The positron solution is obtained by substituting the anti-particle bispinor $v(p)$ and 
setting the momentum negative $p\rightarrow -p$.

\begin{equation}
\label{c2.eq24} 
\begin{array}{rl}
\psi _V(x,p) & = E_p(x)\;u_s(p) \\[10pt] 
E_p(x) & = \left[ 1+\dfrac e{2(kp)}\slashed{k}\slashed{A}^e\right] \exp \left(
-ipx -i\dint_{\nths 0}^{kx}\left[ \dfrac{e(A^ep)}{(kp)}-\dfrac{e^2A^{e2}}{2(kp)}%
\right] d\phi \right) 
\end{array}
\end{equation}

\medskip\ 

The final element required in a calculation of second order process in the
presence of the external field are the particle propagators which link the
destruction of initial states at space-time point $x_\mu $ with the creation
of final states at $x_\mu ^{\prime }$. In this work only the bound electron propagator 
$G^e(x,x')$, which can be constructed from products of the Volkov
electron propagator in direct analogy to the non external field 
case\footnote{See for example \cite{ManSha84} pgs 73-74.}, is required. $G^e(x,x')$ can be 
written as the Green's function solution of the equation

\begin{equation}
\label{c2.eq25}(\slashed{p}-e\slashed{A}^e-m)G^e(x,x^{\prime })=\delta
(x-x^{\prime })
\end{equation}

\medskip\ 

An integral representation for $G^e(x,x^{\prime })$ has been obtained for a
linearly polarised external electromagnetic field using a proper time method
\cite{Schwinger51}. However for this thesis a representation obtained by
\cite{Ritus72}, which consists of the non external field fermion propagator
sandwiched between Volkov $E_p$ functions \cite{Mitter75}, was chosen.

\begin{equation}
\label{c2.eq26}
G^e(x,x^{\prime })=\frac 1{(2\pi )^4}\int d^4p \,E_p(x)
\frac{\slashed{p}+m}{p^2-m^2}\bar E_p(x^{\prime }) 
\end{equation}

\medskip

In the absence of the external field, the Volkov $E_p$ functions reduce to
simple exponential functions, and the free fermion propagator is recovered.
That equation \ref{c2.eq26} is a solution of equation \ref{c2.eq25} is due
to the orthogonality properties of the $E_p$ functions \cite{Mitter75}

\begin{equation}
\label{c2.eq27} 
\begin{array}{rl}
\dfrac 1{(2\pi )^4}\dint E_p(x)\,\overline{E}_p(x^{\prime })\,d^4p & = 
\delta ^4(x-x^{\prime }) \\[10pt]
\dfrac 1{(2\pi )^4}\dint E_p(x)\,\overline{E}_{p^{\prime }}(x)\,d^4x & = 
\delta ^4(p-p^{\prime }) 
\end{array}
\end{equation}

\medskip\ 

Explicit forms for the Volkov $E_p$ functions are obtained by substituting the explicit 
forms for the external field 4-potentials to be considered.

\begin{equation}
\label{c2.eq28} 
\begin{array}{rclr}

E_p(x) &=& \left[ 1+\dfrac e{2(kp)}\left( 
\slashed{k}\slashed{a}_1\cos (kx)+\slashed{k}\slashed{a}_2\sin (kx)\right) \right] 
\quad\quad\quad &\text{circularly polarised} \\[10pt] 
&\times & \exp \left[ -iqx-ie\dfrac{(a_1p)}{(kp)}\sin (kx)+
ie\dfrac{(a_2p)}{(kp)}\cos (kx)\right]& \\[20pt]

E_p (x) &=& \left[ 1 + \dfrac{e}{2 (kp)} \slashed{k} \slashed{a} (kx)  \right] 
 & \text{constant crossed} \\[10pt]
& \times & \exp \left[ - iqx + i \dfrac{e^2 a^2}{2 (kp)}(kx) -
i \dfrac{e (ap)}{2 (kp)} (kx)^2 - i \dfrac{e^2 a^2}{6 (kp)} (kx)^3 \right]&
\end{array}
\end{equation}

\medskip\ 

A modification of the electron momentum is suggested by gathering together terms such that

\begin{equation}
q_\mu = p_\mu +\frac{e^2a^2}{2(kp)}\,k_\mu 
\end{equation}

\medskip\

Indeed the Volkov function as a whole can be written in terms of $q_\mu$ due to the 
properties of the external field 4-potential set out in equations \ref{c2.eq18} and 
\ref{c2ext.last}. The fermion momentum modified by the external field can be interpreted as a
shift in the fermion mass given by

\begin{equation}
\label{c2.eq30}m^2+e^2a^2\equiv m_{*}^2 
\end{equation}

\medskip
\section{Radiative corrections to the bound electron propagator}
\label{c2rad}

The S-matrix expansion allows for radiative corrections proportional to
powers of the fine structure constant. For instance Compton
scattering can include in its intermediate state the creation of a photon 
immediately followed by its destruction. This radiative correction is the electron self 
energy. Inclusion of the electron self energy in the IFQED processes results in a modified 
bound electron propagator which can be labelled $G_{RC}^e(x,x')$.
The electron self energy in the presence of the external field $\Sigma ^e(x,x')$
is defined by a product of the bound electron propagator without radiative corrections
$G^e(x,x')$ and the free photon propagator $D(x,x')$

\begin{equation}
\label{c2.eq41}
\begin{array}{rl}
\Sigma ^e(x,x^{\prime }) & = ie^2\gamma^\mu G^e(x,x^{\prime })\gamma _\mu D(x,x^{\prime }) 
\\[20pt]  
\text{where\quad \quad }D(x,x^{\prime }) & = -\dfrac 1{(2\pi )^4}\dint 
d^4k\;\dfrac 1{k^2+i\epsilon }\,e^{-ik(x-x')}
\end{array}
\end{equation}

\medskip\ 

The radiatively corrected bound electron propagator $G_{RC}^e(x,x^{\prime })$ is written as the 
Green's function solution of the equation

\begin{equation}
\label{c2.eq42}
(\slashed{p}-e\slashed{A}^e-m-\Sigma^e(x,x'))G_{RC}^e(x,x') =  \delta (x-x') 
\end{equation}

\medskip

The solution to this equation proves cumbersome and a simpler method of including the
radiative corrections is obtained by calculating the effect on the
intermediate electron energy levels \cite{BecMit76}.

\bigskip
\section{The external field electron energy shift}
\label{c2efees}

Taking radiative corrections into account, electron energy states are quasi-stationary due to a 
constant interaction between the electron and the cloud of virtual photons that surround it. 
The energy levels are broadened, gaining a small finite shift $\Delta \epsilon _{p,s}$. 
For the purposes of this thesis an expression for this energy level shift
when the electron also interacts with an external electromagnetic field is
required.

The wave function $\psi _V^{\Delta \epsilon _{p,s}}(x_2)$ of an electron of
momentum $p$ and spin $s$ at the space-time point $x_2$ , interacting with a
quantised Maxwell field and classical plane wave $A^e(x_2)$ satisfies the
equation (\cite{AkhBer65} pg.758)
 
\begin{equation}
\label{c2.eq43}\left( \slashed{\partial}-ie \slashed{A}^e(x_2)+m\right) \psi _V^{\Delta \epsilon
_{p,s}}(x_2)=-\int d^4x_1 \;\Sigma^e(x_2,x_1)\psi _V^{\Delta \epsilon _{p,s}}(x_2)
\end{equation}

\medskip\ 

The electron energy level shift can be separated from the electron wave
function by writing the solution of equation \ref{c2.eq43} as a product of
an exponential function and the Volkov wave function

\begin{equation}
\label{c2.eq44}\psi _V^{\Delta \epsilon _{p,s}}(x_2)\equiv e^{-i\,\Delta
\epsilon _{p,s}\,t_2}\;\psi _V(x_2,p)
\end{equation}

\medskip\ 

Inserting the general form of the solution expressed by equation \ref
{c2.eq44} into equation \ref{c2.eq43}, and operating on the right by the
adjoint of the Volkov wave function, $\psi _V^{\dagger }(x_2,p)$

\begin{equation}
\label{c2.eq45}-i\,\Delta \epsilon _{p,s}\int d^4x_2\,\psi _V^{\dagger
}(x_2,p)\,\psi _V(x_2,p)=-\int d^4x_1d^4x_2\,\psi _V^{\dagger
}(x_2,p)\,\Sigma^e(x_2,x_1)\,\psi _V(x_1,p)e^{-i\,\Delta \epsilon _{p,s}\,t_1}
\end{equation}

\medskip\ 

Using the orthogonality properties of the Volkov functions $E_p(x)$, the integration over
space-time variable $x_2$ on the left hand side of equation \ref{c2.eq45}
yields a function of the electron energy, electron mass and the space-time volume
$\frac{\epsilon _p}mVT$. Assuming that the width of the electron energy
levels is small, the substitution $\Delta \epsilon _{p,s}\rightarrow 0$ can
be made on the right hand side of equation \ref{c2.eq45} and the electron
energy level shift can be written

\begin{equation}
\label{c2.eq46}\Delta \epsilon _{p,s}=\frac m{\epsilon _p}\frac 1{VT(2\pi
)^4}\int d^4x_1d^4x_2\,\bar \psi _V(x_2,p)\,\Sigma ^e(x_2,x_1)\,\psi _V(x_1)
\end{equation}

\bigskip\
\section{Non external field regularisation and renormalisation}
\label{c2regn}

Calculation of the external field electron energy shift $\Delta \epsilon _{p,s}$ 
encounters a difficulty due to the presence of divergences
in the external field electron self energy. In this section the
removal of these divergences in the absence of the external field is considered. 
A discussion of the impact of the external field is left to chapters 6 and 7.

The removal of divergences is a three step process. Firstly, regularisation
involves a modification of QED so that the electron self energy remains
finite to all orders of perturbation theory. The momentum dependent 
electron self energy is written as an 
integration over photon 4-momentum $k_\mu$ \cite{ManSha84}.

\begin{equation}
\label{c2.eq47}\Sigma (p)=\frac{e^2}{(2\pi )^4}\dint d^4k\;\frac 1{k^2}\,
\frac{2\slashed{p}-2\slashed{k}-4m}{(p-k)^2-m^2}
\end{equation}

\medskip\ 

The integration contains an ultraviolet divergence at its upper bound and an infrared
divergence as the momentum reaches the mass shell. Regularisation is the introduction of
appropriate limits which allow the integration to proceed. A small ficticious photon mass
$i\varepsilon$ is added to the denominators and either a large cut-off is used for the upper
bound of the integration or the photon propagator is replaced \cite{PauVil49,Wein95}

\begin{equation}
\dfrac{1}{k^2+i\varepsilon}\rightarrow \dfrac 1{k^2-\Lambda ^2+ i\varepsilon}
\end{equation}

\medskip\ 

QED is a renormalisable field theory and renormalisation proceeds from the recognition that
the interaction of the Dirac and Maxwell fields modifies the properties of the free fields.
The bare field operators (labelled by subscript 'B') are modified such that

\begin{equation}
\begin{array}{ccl}
\psi=Z^{-1/2}_{2}\psi_B \quad&\quad A^{\mu}=Z^{-1/2}_{3}A_{B}^{\mu} \\[10pt] 
e=Z^{1/2}_3 e_B \quad&\quad m=m_B + \delta m
\end{array}
\end{equation}

\medskip\ 

The cofactors $Z_2$, $Z_3$ and $\delta m$ are determined from the condition that the
propagators of the renormalised fields have identical poles and residues to that of the
propagators of the free fields with the absence of interactions. The electron self energy is
included in the electron propagator via a sum of loop diagrams, the result being

\begin{equation}
\label{c2.eq49}
G(p)=(\slashed{p}-m-\Sigma(p)+i\varepsilon)^{-1}
\end{equation}

\medskip

The renormalised fields lead to a renormalised electron self energy $\Sigma_{RC}(p)$ and
renormalised electron propagator $G_{RC}(p)$ 

\begin{equation}
\label{c2.eq48}
\begin{array}{ccl}
\Sigma(p) & \rightarrow  & \Sigma_{RC}(p)=\Sigma(p)-(Z_2 -1)(\slashed{p}-m)+Z_2 \delta m \\ 
G(p) & \rightarrow  & G_{RC}(p)=[Z_2(\slashed{p}-m)-Z_2 \delta m -\Sigma(p)]^{-1} 
\end{array}
\end{equation}

\medskip\ 

The pole in $G_{RC}(p)$ should occur at $\slashed{p}=m$ with residue 1. With the aid of
L'Hopital's residue rule \cite{Boas83}, the renormalisation cofactors are

\begin{equation}
\begin{array}{rcl}
Z_2\delta m &=& -\Sigma(\slashed{p}\rightarrow m) \\
Z_2 &=& 1 + \Sigma '(\slashed{p}\rightarrow m)
\end{array}
\end{equation}

\medskip\

At first inspection $Z_2$ turns out to contain an infrared divergence. This is dealt with by
taking into account emission of soft photons in external lines of the electron self energy 
process. The ultimate result is that $Z_2 = 1$. The final regularised electron propagator is

\begin{equation}
G_{RC}(p)=(\slashed{p}-m-\Sigma(p)+\Sigma(\slashed{p}\rightarrow m)+i\varepsilon)^{-1}
\end{equation}

\bigskip\
\section{The optical theorem}
\label{c2opt}

The unitary properties of the $S$-operator allow a connection to be found
between the elastic scattering amplitude%
\footnote{i.e. scattering from an initial state $\left| i\right\rangle$ through 
intermediate states (labelled $\left| f\right\rangle$) to final states identical to initial 
states $\left| i\right\rangle$.} of an arbitrary scattering process and the
transition probability from initial states $\left| i\right\rangle $ to final
states $\left| f\right\rangle $. This so called optical theorem was developed for quantum 
scattering from the original expression relating refractive index to forward scattering. In this 
thesis it will be used to relate electron self energy to Bremsstrahlung in an external field.

Since the $S$-operator is Hermitian, the following relation holds

\begin{equation}
\label{c2.eq14}\left\langle i^{\prime }\right| S^{\dagger }S\left|
i\right\rangle =\dsum\limits_f\,\left\langle f\right| S^{\dagger }S\left|
i^{\prime }\right\rangle ^{*}\,\left\langle f\right| S^{\dagger }S\left|
i\right\rangle 
\end{equation}

\medskip\ 

The $T$-matrix can be substituted for for the S-matrix using equation \ref{c2.eq10}. If states 
$\left| i\right\rangle $ and $\left| i^{\prime }\right\rangle $
are such that their total momentum is equal, then equation \ref{c2.eq14} becomes

\begin{equation}
\label{c2.eq15}\frac 1{2i}\left[ \left\langle i^{\prime }\right| T\left|
i\right\rangle -\left\langle i\right| T\left| i^{\prime }\right\rangle
^{*}\right] =\frac 12\dsum\limits_f\,(2\pi )^4\,\delta
(P_f-P_i)\,\left\langle f\right| T\left| i^{\prime }\right\rangle
^{*}\left\langle f\right| T\left| i\right\rangle 
\end{equation}

\medskip\ 

The final result relating the imaginary part of the elastic scattering amplitude $T_{elastic}$
with the transition probability $W$, is obtained by identifying $\left|
i\right\rangle $ with $\left| i^{\prime }\right\rangle $

\begin{equation}
\label{c2.eq16}\Im T_{elastic}=\frac 12W 
\end{equation}

\medskip\

%% file: chap3.tex
\section{Introduction}

In this chapter analytic expressions are developed for the differential
cross sections of two IFQED processes, stimulated Compton
scattering (SCS) and stimulated two photon pair production (STPPP) in an external circularly 
polarised electromagnetic field.

In sections \ref{c3smat} - \ref{c3evscs} analytic expressions are found for the differential
cross section of the SCS process without recourse to specific kinematics or
restricted energy regimes. The SCS differential cross section is considered 
in a reference frame in which the initial electron is at rest $(\sql{p}_i=0)$, and the Lorentz gauge
condition in which the 4th component 
of the external field 4-potential vanishes. The analytic expressions for both the SCS and the 
STPPP differential cross sections are lengthy and issues involved with extracting numerical 
calculations from them are considered in section \ref{c3numev}.

In section \ref{c3limit} the analytic expressions for the SCS differential cross section in two 
limiting cases are considered. These limiting cases are important for comparison with previous 
results by other authors. In the limit of vanishing external field the SCS 
expressions reduce to the Klein-Nishina differential cross section for non external field 
Compton scattering. The second case limits kinematics such that the initial photon is parallel 
to the direction of propagation of the external field. This second limiting case was 
considered by \cite{AkhMer85} and small differences in the analytic expressions, which will 
prove to be significant, will be found.

In section \ref{c3pprod} the STPPP process is considered using a centre of mass-like reference 
frame. A crossing symmetry (see section \ref{c2sub}) diagrammatically links the SCS 
and STPPP processes and allows much of the analytic work of sections \ref{c3smat} - 
\ref{c3evscs} to be reused. The STPPP differential cross section so obtained is considered in 
the limit of vanishing external field and the analytic expressions reduce to the Breit-Wheeler 
differential equation for non external field pair production.

The presence of the external field has the effect of shifting the fermion rest mass resulting in a 
corrected fermion momentum which is a function of external field parameters. In this chapter 
the corrected momentum of any free fermion $p_x$ is denoted by the symbol $q_x$. 

\bigskip
\section{The stimulated Compton scattering (SCS) matrix element}
\label{c3smat}

\begin{figure}[!b]
\label{c3.comp.fig1}
\centerline{\includegraphics[height=4.5cm,width=9cm]{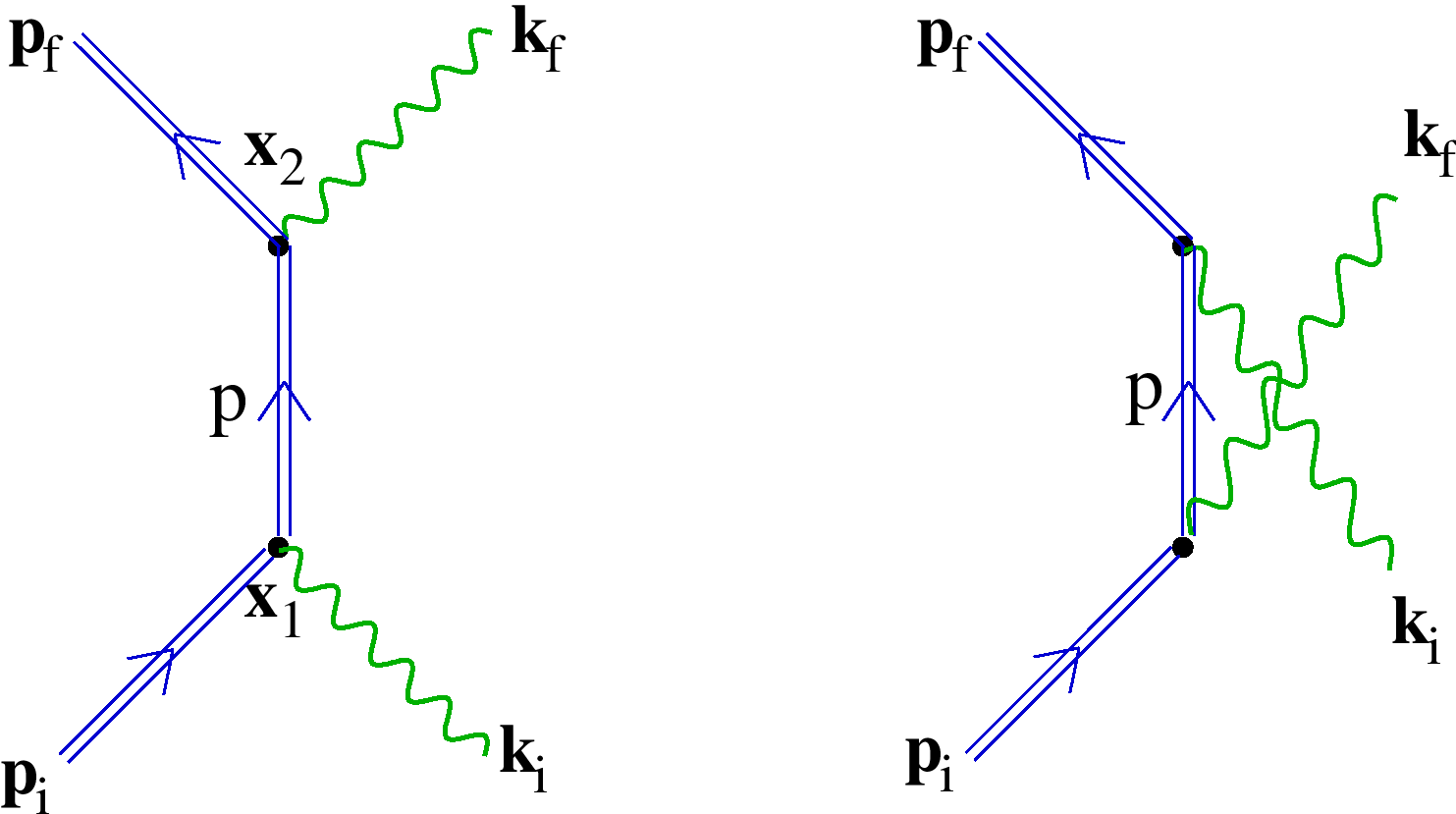}}
\caption{\bf Feynman diagrams for stimulated Compton scattering.}
\end{figure}

Stimulated Compton scattering is the QED process in which a photon $k_i$ together with a
discrete number of photons from the external field $lk$ combine with an
electron $p_i$ to produce a final state consisting of a photon $k_f$ and an
electron $p_f$. The SCS process can be represented graphically by the two topologically
different Feynman diagrams shown in figure \ref{c3.comp.fig1}. The double
straight lines indicate the electron embedded in the external
electromagnetic field.

%----------------other attempts at the feynman diagrams--------------
%\begin{figure}[t]
%\hbox{ 
%\psfig{figure=./tex/feyndiag/stim_comp1.eps,height=6.0cm,width=6.5cm}
%       \hspace{1cm}
%       
%\psfig{figure=./tex/feyndiag/stim_comp2.eps,height=6.0cm,width=6.5cm}}
%\caption{\bf Feynman Diagrams for stimulated Compton scattering.}
%\label{c3.comp.fig1}
%\end{figure}

%\begin{figure}[t]
%\input{./tex/figures/comp.tex}
%\centerline{\psfig{figure=./tex/feyndiag/stim_comp1.eps,height=6.5cm,width=7cm}}
%\caption{\vb Feynman Diagrams for stimulated Compton scattering.}
%\label{c3.comp.fig1}
%\end{figure}
%-------------------------------------------------------------------------

The matrix element of the scattering process can be written from Wick's
theorem with the aid of figure \ref{c3.comp.fig1} and the usual rules for
Feynman diagrams.

\begin{equation}
\label{c3.smat.eq1}
\begin{array}{rl}
S_{fi}^e & = -e^2 \dint d^4x_1d^4x_2 \\[10pt]
& \times \;\;
\biggl\{\overline{\psi} _{V}(x_2,p_f) \overline{\slashed{A}} _{k_f}(x_2)G^e(x_2,x_1)
\slashed{A}_{k_i}(x_1)\psi _{V}(x_1,p_i) \\[10pt]

& + \;\; \overline{\psi}_{V}(x_2,p_f) \overline{\slashed{A}}_{-k_i}(x_2)G^e(x_2,x_1)
\slashed{A}_{-k_f}(x_1)\psi _{V}(x_1,p_i)\biggr\}
\end{array}
\end{equation}

\medskip\ 

The Volkov wave functions $\psi_{V}$ and bound electron propagator $G^e$ 
(section \ref{c2volk}) and the Maxwell wave functions $A_k(x)$ (equation \ref{c2.eq3}), are 
substituted into equation \ref{c3.smat.eq1}. An extra integration over intermediate 4-momentum 
$p$ appears and the integrand is a function of Volkov $E_p$ functions, 4-momenta and Dirac 
$\gamma$ matrices sandwiched between free Dirac bispinors $\bar{u}(p_f)$ and $u(p_i)$.

\begin{equation}
\label{c3.smat.eq2} 
\begin{array}{rl}

S_{fi}^e & = -e^2 \dint d^4x_1d^4x_2d^4p \\[10pt]  
& \times \;\; \bar u(p_f) 
\biggl\{\bar E_{p_f}(x_2) \slashed{e}(k_f) e^{ik_fx_2}E_p(x_2)\pmsq\bar
E_p(x_1)\slashed{e}(k_i) e^{-ik_ix_1} E_{p_i}(x_1) \\[10pt]  
& + \;\; \bar E_{p_f}(x_2)\slashed{e}(k_i) e^{-ik_ix_2} E_p(x_2)\pmsq\bar 
E_p(x_1) \slashed{e}(k_f) e^{ik_fx_1}E_{p_i}(x_1)\biggr\} u(p_i) 
\end{array}
\end{equation}

The integrations over $x_1$ and $x_2$ are performed by writing the 
dependence on $x_1$ and $x_2$ in the integrand as an infinite summation of Bessel functions 
\cite{NarNikRit65}. In that case, $x_1$ and $x_2$ appear entirely as an exponential dependence, 
and the integrations over $x_1$ and $x_2$ yield a product of delta functions which
contain particle 4-momenta within their arguments.
The part of the integrand containing $x_1$ in equation \ref{c3.smat.eq2} is

\begin{equation}
\begin{array}{rl}

\bar E_p(x_1)\slashed{e}(k_i) E_{p_i}(x_1)e^{-ik_ix_1} &= \left[ 1+\dfrac
e{2(kp)}\slashed{a}_1\slashed{k}\cos \phi _1+\dfrac e{2(kp)}
\slashed{a}_2\slashed{k}\sin \phi _1\right] \\[10pt]

& \times \;\; \slashed{e}(k_i) \left[ 1-\dfrac e{2(kp_i)}\slashed{a}_1\slashed{k}\cos
\phi _1-\dfrac e{2(kp_i)}\slashed{a}_2\slashed{k}\sin \phi _1\right] e^{-iF(x_1)} \\[20pt]

\text{where} \quad\quad F(x_1) & = \alpha _{1i}\sin \phi _1-\alpha _{2i}\cos \phi
_1-qx_1+q_ix_1+k_ix_1 \\[10pt] 
\phi _1 & = (kx_1) \\[10pt] 
\alpha _{1i} & = e\left[ \dfrac{(a_1 p_i )}{(kp_i )}-\dfrac{(a_1 p)}{(kp)}\right] 
\end{array}
\end{equation}

\medskip\

Equivalent expressions are found for the part of the integrand containing $%
x_2$. Using trigonometric double angle formula to write the function $F(x)$
in the form $z\sin (\phi -\phi _0)$ and using well known properties of
Bessel functions \cite{Watson22}, the dependence on $x_1,x_2$ is
expanded in a Fourier series \cite{NarNikRit65}.

\begin{equation}
\label{c3.smat.eq3} 
\begin{array}{rl}
(1,\cos \phi _1,\sin \phi _1)\,e^{-iz_i\sin (\phi _1-\phi _{0i})} & = 
\sum\limits_{r=-\infty }^\infty (M_1,M_2,M_3)e^{-ir\phi _1} \\[10pt]
(1,\cos \phi _2,\sin \phi _2)\,e^{iz_f\sin (\phi _2-\phi _{0f})} & = 
\sum\limits_{s=-\infty }^\infty (N_1,N_2,N_3)e^{is\phi _2} \\[15pt]

\text{where} \quad\quad M_1 & = J_r(z_i)e^{ir\phi _{0i}} \\
M_2 & = \frac 12\left[ J_{r-1}(z_i)e^{-i\phi _{0i}}+J_{r+1}(z_i)e^{i\phi
_{0i}}\right] e^{ir\phi _{0i}} \\
M_3 & = \frac i2\left[ J_{r-1}(z_i)e^{-i\phi _{0i}}-J_{r+1}(z_f)e^{i\phi
_{0i}}\right] e^{ir\phi _{0i}} \\
N_1 & = J_s(z_f)e^{-is\phi _{0f}} \\
N_2 & = \frac 12\left[ J_{s-1}(z_f)e^{i\phi _{0f}}+J_{s+1}(z_f)e^{-i\phi
_{0f}}\right] e^{-is\phi _{0f}} \\
N_3 & = -\frac i2\left[ J_{s-1}(z_f)e^{i\phi _{0f}}-J_{s+1}(z_f)e^{-i\phi
_{0f}}\right] e^{-is\phi _{0f}} \\[15pt]

\;z_i = \sqrt{\alpha _{1i}^2+\alpha _{2i}^2} & z_f =  \sqrt{\alpha
_{1f}^2+\alpha _{2f}^2} \\[10pt] 
\cos \phi _{0i} = \dfrac{\alpha _{1i}}{z_i} &
\cos \phi _{0f} = \dfrac{\alpha _{1f}}{z_f} \\[10pt] 
\sin \phi _{0i} = \dfrac{\alpha _{2i}}{z_i} &  
\sin \phi _{0f} = \dfrac{\alpha _{2f}}{z_f}

\end{array}
\end{equation}

\medskip\ 

Substituting these Fourier expansions into equation \ref{c3.smat.eq2},
grouping together the exponential dependence, and expanding products of
4-vectors, the expression for the matrix element becomes

\begin{equation}
\label{c3.smat.eq5} 
\begin{array}{rl}
S_{fi}^e & = -e^2 \dsum\limits_{rs}\dint d^4x_1 \,d^4x_2 \,d^4q \\[10pt]
& \times \;\; \bar u(p_f) \biggl\{B_{rs}(\slashed{e}(k_f),N,p,p_f)\left[ 
\dfrac{\slashed{p}+m}{p^2-m^2}\right] B_{rs}(\slashed{e}(k_i),M,p_i,p)e^{-iG} \\[10pt]  
& + \;\; B_{rs}(\slashed{e}(k_i),N,p,p_f)\left[ \dfrac{\slashed{p}+m}{p^2-m^2}\right] 
B_{rs}(\slashed{e}(k_f),M,p_i,p)e^{-iH}\biggr\}u(p_i) \\[20pt]

\text{where} \quad G & = (q_i+k_i+rk-q)x_1-(q_f+k_f+sk-q)x_2 \\[10pt]
H & = (q_i-k_f+rk-q)x_1-(q_f-k_i+sk-q)x_2 \\[10pt]
B(\slashed{e}(k_f),N,p,p_f) & = \slashed{e}(k_f) N_1-\dfrac{e}{2(kp)}\left( 
\slashed{e}(k_f) \slashed{a}_1\slashed{k}N_2+\slashed{e}(k_f) 
\slashed{a}_2\slashed{k} N_3 \right) \\[10pt]

+ & \dfrac e{2(kp_f)}\left( \slashed{a}_1\slashed{k}\slashed{e}(k_f) N_2+
\slashed{a}_2\slashed{k}\slashed{e}(k_f) N_3\right) -
\dfrac{e^2a^2}{4(kp)(kp_f)}\slashed{k}\slashed{e}(k_f)\slashed{k} N_1

\end{array}
\end{equation}

\medskip 

$q,q_i$ and $q_f$ represent the shifted electron momentum ($q=p+\frac{e^2a^2}{2(kp)}k$). It is 
convenient to make a transformation of integration variable $d^4p\rightarrow d^4q$. 
This shift has been used before in 2nd order IFQED calculations \cite{Oleinik68} 
and the Jacobian of the transformation is unity (see Appendix \ref{jacob}).

The transformation $(p,p_i,p_f) \rightarrow (q,q_i,q_f)$ is made in the integrand of equation 
\ref{c3.smat.eq5} with the aid of the orthogonality of 4-vectors $a_{1\mu},a_{2\mu}$ and $k_{\mu}$. The 
following equivalences hold

\begin{equation}
\begin{array}{rcl}
B(\slashed{e}(k_f),N,p,p_f) & \equiv & B(\slashed{e}(k_f),N,q,q_f) \\[10pt] 
e\left[ \dfrac{(a_n p_m )}{(kp_m )}-\dfrac{(a_n p)}{(kp)}\right] & \equiv &
e\left[ \dfrac{(a_n q_m )}{(kq_m )}-\dfrac{(a_n q)}{(kq)}\right] 
\;\; \text{where} \;\; n=1,2 \;\; m=i,f\\[10pt]
 p^2-m^2 & \equiv & q^2-m_{*}^2 \;\; \text{where} \;\; m_{*}^2=m^2 + e^2 a^2 \\[10pt] 
\slashed{p}+m & \equiv & \slashed{q}-\dfrac{e^2a^2}{2(kq)}\slashed{k}+m 
\end{array}
\end{equation}

\medskip\ 

Integration over $x_1,x_2,q$ (in that order) yields a single Dirac
delta function, the argument of which expresses the conservation of
4-momentum for the SCS process. 
Mathematically, $r$ external field photons contribute in the initial state and $s$ 
contribute in the final state. Physically, however, it is the total
number $l=r-s$ of external field photons contributing to the process that holds interest. 
Finally, the matrix element can be written

\begin{equation}
\label{c3.smat.eq4}
\begin{array}{rl}
S_{fi}^e & = -e^2 (2\pi)^8 \dsum\limits_{rs}\delta ^4(q_i+k_i+(r-s)k-q_f-k_f)\;\bar u(p_f)\ Q\ 
u(p_i) 
\\[20pt]
\text{where} \quad Q & =B(\slashed{e}(k_f),\bar N,\bar q,q_f)\left[ 
\frac{\slashed{\bar{q}}_r-\frac{e^2a^2}{2(k\bar{q})}\slashed{k}+m}{\bar q_r^2-m_{*}^2}\right] 
B(\slashed{e}(k_i),\bar M,q_i,\bar q) \\[10pt]
  & + \;\; B(\slashed{e}(k_i),\dbr{N},\dbr{q},q_f)\left[ 
\frac{\slashed{\dbr{q}}_r-\frac{e^2a^2}{2(k\dbr{p})}\slashed{k}+m}
{\dbr{q}_r^2-m_{*}^2}\right] B(\slashed{e}(k_f),\dbr{M},q_i,\dbr{q}) \\[20pt]

\text{and} \quad\quad & \bar q_r = q_i+k_i+rk \\[10pt]
& \dbr{q}_r = q_i-k_f+rk

\end{array}
\end{equation}

\medskip\

Notationally, the symbols $\bar M,\bar N$ and $\dbr{M},\dbr{N}$ refer to the
functions $M,N$ defined in equation \ref{c3.smat.eq3}, with 4-momentum $q$
replaced by $\bar q_r$ and $\dbr{q}_r$ respectively.

The sum over initial and final states of the square of the S-matrix element $%
{\displaystyle \sum_{if}} |S_{fi}^e|^2$ is written as a
trace of products of Dirac $\gamma $-matrices with the aid of the summation rules for Dirac 
bispinors and photon polarisations. A factor of $\frac{1}{8m^2}$
is introduced by an average over two electron spin states and the sum over bispinors. An 
extra factor of $\frac{1}{2}$ is obtained from an average of possible initial photon polarisations.
Four infinite summations over indicies $r,s,r',s'$ are introduced. The product of Dirac delta functions leads
to $s=s'$. Three infinite summations remain and after making a 
shift in summation variable $\dsum\limits_s\rightarrow \dsum\limits_l$, the square of the matrix element is

\enlargethispage*{1cm}
\begin{equation}
\label{c3smat.eqlast}
\begin{array}{rl}
\dsum\limits_{if} |S_{fi}^e| ^2 & =\dfrac{e^4}{16m^2} (2\pi)^8 \dsum\limits_l \,\delta^4(q_i+k_i+lk-q_f-k_f)\,
\dsum\limits_{rr'}\,\Tr\,Q \\[10pt]
\text{or} \quad \dsum\limits_{if} |T_{fi}^e| ^2 & = \dfrac{e^4}{16m^2} \dsum\limits_{lrr'} 
\,\limfunc{Tr}\,Q \\[20pt]
\text{where} \quad\quad \Tr \,Q = & \Tr \,Q_1(\bar q_r,\bar q_{r'})+
\Tr\,Q_1(\dbr{q}_r,\dbr{q}_{r' })+\Tr\,Q_2(\bar q_r,\dbr{q}_{r\prime })+
\Tr\,Q_2^{*}(\bar q_r,\dbr{q}_{r'}) \\[20pt]

\text{and} \quad\quad 
Q_1(\bar q_r,\bar q_{r'}) & = (\slashed{p}_f+m)B(\gamma^\mu ,\bar N_r,\bar q,q_f)\left[ 
\dfrac{\slashed{\bar{q}}_r-\frac{e^2a^2}{2(k\bar{q})}\slashed{k}+m}{\bar q_r^2-m_{*}^2}\right] 
B(\gamma^\nu ,\bar M_r,q_i,\bar q) \\[12pt]

& \times \;\; (\slashed{p}_i+m)\tilde B(\gamma _\nu ,\bar M_{r'},q_i,\bar q)\left[
\dfrac{\slashed{\bar{q}}_{r'}-\frac{e^2a^2}{2(k\bar{q})}\slashed{k}+m}{\bar{q}_{r'}^2-m_{*}^2}\right]^{*}
\tilde B(\gamma _\mu ,\bar N_{r'},\bar q,q_f) \\[15pt]

Q_2(\bar q_r,\dbr{q}_r') & = (\slashed{p}_f+m)B(\gamma^\mu ,\bar N_r,\bar q,q_f)\left[ 
\dfrac{\slashed{\bar{q}}_r-\frac{e^2a^2}{2(k\bar{q})}\slashed{k}+m}{\bar q_r^2-m_{*}^2}\right] 
B(\gamma^\nu ,\bar M_r,q_i,\bar q) \\[12pt]
  
& \times \;\; (\slashed{p}_i+m)\widetilde{B}(\gamma _\nu ,\dbr{M}_{r'},q_i,\dbr{q})\left[ 
\dfrac{\slashed{\dbr{q}}_{r'}-\frac{e^2a^2}{2(k\dbr{q})}\slashed{k}+m}{\dbr{q}_{r'}^2-m_{*}^2}\right]^{*}
\widetilde{B}(\gamma _\mu ,\dbr{N}_{r'},\dbr{q},q_f)

\end{array}
\end{equation}
\enlargethispage*{1cm}
\medskip\

The full trace expressions for $\limfunc{Tr}\;Q_1$ and $\limfunc{Tr}\;Q_2$
are written down in Appendix \ref{trace}.

\section{The SCS phase space integral}

The calculation of the SCS cross section requires an integration over final
states of particle momentum $(\epsilon _f,\sql{q_f})$ and $%
(\omega _f,\sql{k}_f)$ . This integration is referred to as the phase space integral

\begin{equation}
\label{c3.phase.eq1}
\int \frac{d^3\sql{k}_f\;d^3\squ{q}_f}{\omega _f\;\epsilon _f} \, 
\delta ^4(q_i+k_i+l\,k-q_f-k_f) 
\end{equation}

\medskip\ 

Work done on the first order IFQED processes in circularly polarised field dealt with the phase
space integral by making a suitable change of integration variables. Azimuthal symmetry of the
scattering process combined with a centre of mass reference 
frame, resulted in a single integration over a simple function of scalar 
products \cite{NarNikRit65}. \cite{AkhMer85}, in their work on 
second order external field Compton scattering, also made use of an azimuthal symmetry by 
restricting their results to a subset of possible kinematics in which the initial photon momentum
is collinear with that of the external field.
With arbitrary kinematics, however, the second order SCS process in circularly
polarised external field does not contain a symmetry with respect to the azimuthal angle 
$\varphi _f$. Additionally, the centre of mass reference frame results in relatively complicated
expressions for particle energy-momenta and scattering angles. 

\begin{figure}[h]

\centerline{\includegraphics[height=6cm,width=6cm]{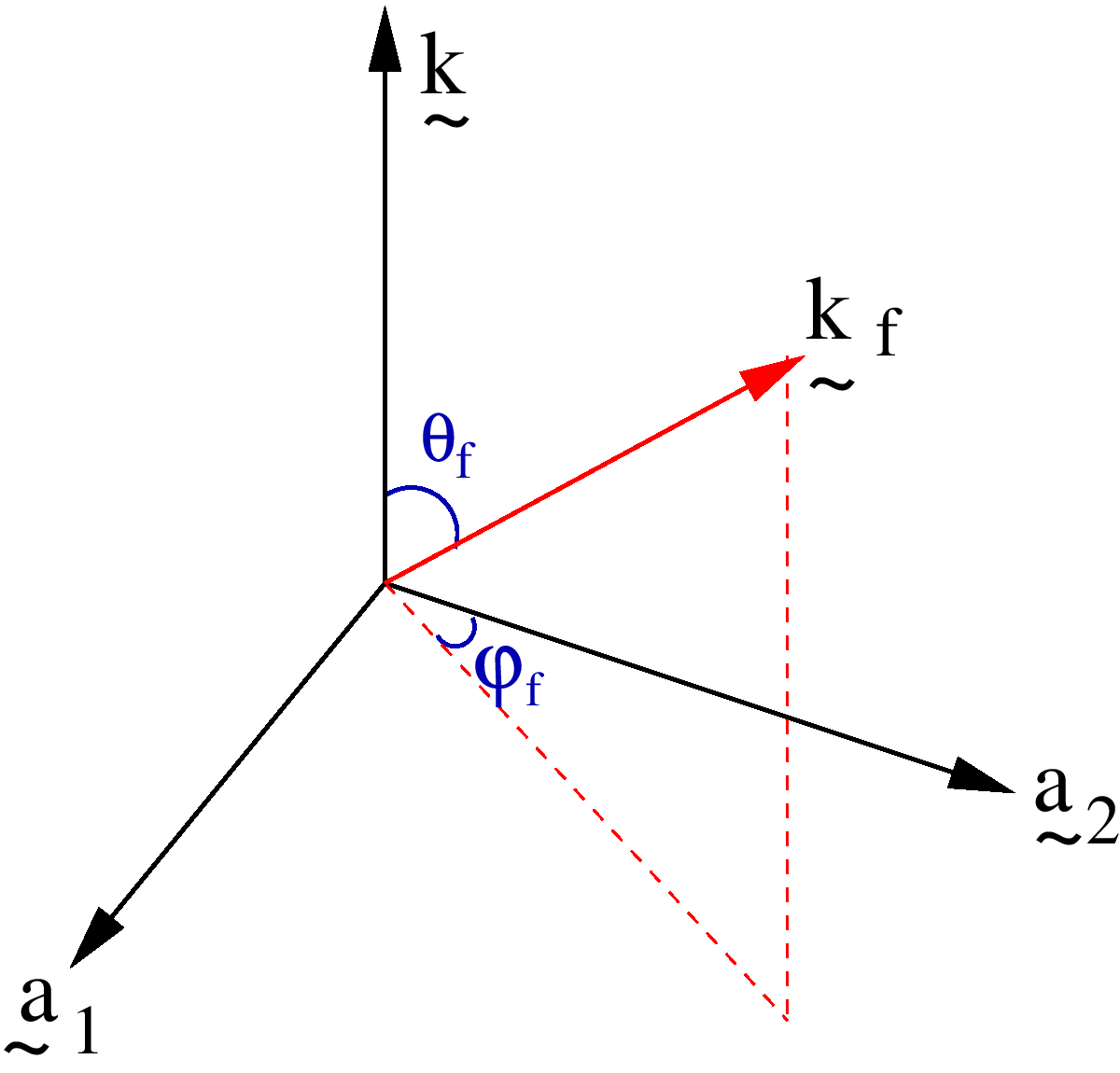}}
\caption{\bf scattering Geometry for stimulated Compton scattering}
\label{c3phase.fig1}
\end{figure}

A simpler transformation is obtained by considering the laboratory reference frame and defining an 
element of solid angle $d\Omega_{k_f}$(=$\,d\cos\theta_f \,d\varphi_f$) into which the final SCS photon is scattered. The 
direction of the scattering is defined by angles $\theta_f,\varphi_f$ depicted in figure 
\ref{c3phase.fig1}.

An integration variable transformation from free electron 3-momentum to bound electron 3-momentum 
is required first, using the result that the Jacobian of the 4-vector transformation  
$d^4p_f\rightarrow d^4q_f$ is unity (see Appendix \ref{jacob})

\begin{equation}
\label{c3.phase.eq2} 
\begin{array}{c}
\dfrac{d^3\squ{p_f}}{\epsilon _{p_f}} = 2\dint
\,d^4p_f\;\Theta (\epsilon _{p_f})\,\delta (p_f^2-m^2) = 2\dint
\,d^4q_f\;\Theta (\epsilon _{q_f})\,\delta (q_f^2-m_{*}^2) = \dfrac{d^3 
\squ{q_f}}{\epsilon _{q_f}} 
\end{array}
\end{equation}

\medskip\ 

and finally the phase space integral transforms as

\begin{equation}
\label{c3.phase.eq3}\int \frac{d^3\sql{k_f}\;d^3\sql{%
q_f}}{\omega _f\;\epsilon _f}\delta ^4(q_i+k_i+l\,k-q_f-k_f)\equiv\dint \;\,\frac{%
d\Omega _{k_f}\;\omega _f^2}{m\omega _f+(k_ik_f)+\left[ l+\frac{e^2a^2}{%
2m\omega }\right] (kk_f)} 
\end{equation}

\bigskip\
\section{Symbolic evaluation of the SCS cross section}
\label{c3evscs}

With the results of the last two sections and the knowledge that the relative velocity
of the two initial SCS particles $p_i$ and $k_i$ is unity in the laboratory frame,
the SCS differential cross section is written with the aid of the general expression (equation 
\ref{c2.eq13}) as 

\begin{equation}
\label{c3.symev.eq1}
\frac{d\sigma }{d\Omega _{k_f}}=\left( \frac {e^2}{8\pi m}\right)^2\left( \frac m{\omega _i}\right) 
\sum\limits_{lrr'} \frac{\omega _f^2}{m\omega_f+(k_ik_f)+[l+\frac{e^2a^2}{2(kp_i)}] (kk_f)}\;\Tr\,Q 
\end{equation}

\medskip\ 

Before the calculation can proceed further, $\Tr\,Q$, which was defined in equation 
\ref{c3smat.eqlast} must be evaluated. This evaluation, though not theoretically difficult, is 
cumbersome due to its length. A simple algebraic
expansion of the 4 traces yields approximately 105,000 terms. Analytic means of simplifying the 
trace expressions are useful.

The trace of products of large numbers of Dirac $\gamma $-matrices and terms
like $(\slashed{p}+m)$ can be simplified by successive application of the
Kahane algorithm and the Becker-Schott rule \cite{Kahane68}. The
calculation of the traces in this thesis was automated with the aid of the 
computer software package Mathematica, and an add-on package FeynCalc \cite{Mertig91}.

Performing the entire trace in one pass and using inbuilt functions in Mathematica to 
simplify results, proved tediously time consuming. The computational time required was
shortened by algebraically expanding the trace as a series of trace
coefficients of products of the functions $M$ and $N$ denoted as

\begin{equation}
X_{ij}^r\equiv \frac{M_{ir}N_{jr}}{(p_r^2-m^2)}\quad \quad \text{where}\quad
\quad i\text{ and }j=1,2,3 
\end{equation}

\medskip\ 

Using the invariance of a trace of products of Dirac $\gamma $-matrices
under reversal and cyclic permutation of its elements, many of these trace
coefficients are identical under exchange of some of the four-vectors
involved in the calculation.

In total, there are 17 independent trace coefficients associated with $\Tr\,Q_1$ 
and 15 independent trace coefficients associated with $
\Tr\,Q_2$ . Each independent trace coefficient has, as a
co-product, a function of $M_{ir}$ and $N_{jr}$ which can be simplified using well known 
addition formulae for Bessel functions \cite{NarNikRit65,Watson22} and the relations

\begin{equation}
\label{c3.symev.eq2} 
\begin{array}{rcl}

\alpha _{1i}M_2+\alpha _{2i}M_3 & = & rM_1 \\ 
\alpha _{1f}N_2+\alpha _{2f}N_3 & = &(r-l)N_1 \\
J_{n-1}(z)+J_{n+1}(z) &=& \dfrac{2n}zJ_n(z) 

\end{array}
\end{equation}

\medskip

It was necessary to develop many relations of the form set out in equation \ref{c3.symev.eq2} 
in order to complete the full trace calculation. These relations
are presented in Appendix \ref{appMM}. The trace coefficients themselves,
along with the results of their calculation are presented in Appendix \ref{trace}.

In order to obtain numerical values for the SCS differential cross section,
it was necessary to write all scalar products in terms of initial state quantities $\omega ,\omega
_i,l,\theta _i,\nu ^2$ and the final state scattering angles $\theta _f$ and 
$\phi _f$. From the conservation of momentum for the SCS process,
the final electron 4-momentum $p_f$ can be written in terms of other
particle 4-momenta, and the energy of the final photon can be written

\begin{equation}
\label{c3.symev.eq4}
\omega _f=\frac{(q_ik_i)+l(k\bar p)}{m+\left( \frac{e^2a^2}{2m}+l\omega
\right) \left( 1-\cos \theta _f\right) +\omega _i\left( 1-\cos \theta
_i\,\cos \theta _f\;-\sin \theta _i\,\sin \theta _f\,\cos \phi _f\right) } 
\end{equation}

\medskip\ 

The external field 4-potentials $a_{1\mu }$ and $a_{2\mu }$ appear in the
differential cross section expressions in linear combinations of scalar products with other 
4-vectors such that the identities of section \ref{c2ext} can be applied.

As a consequence, the arguments of the Bessel functions appearing in the
differential cross section, which depend on the external field 4-potentials,
can be written

\begin{equation}
\label{c3.symev.eq3}
\begin{array}{ccc}
\bar z_i^2 & = & \dfrac{e^2a^2}{(k\bar p)^2}\dfrac{(kk_i)}{(kp_i)}\left[
2(q_ik_i)-m_{*}^2\dfrac{(kk_i)}{(kp_i)}\right] \\[10pt] 
\bar z_f^2 & = & \dfrac{e^2a^2}{(k\bar p)^2}\dfrac{(kk_f)}{(kp_f)}\left[
2(q_fk_f)-m_{*}^2\dfrac{(kk_f)}{(kp_f)}\right] \\[10pt] 
\dbr{z}_i^2 & = & \dfrac{e^2a^2}{(k\dbr{p})^2}\dfrac{(kk_f)}{(kp_i)}\left[
2(q_ik_f)-m_{*}^2\dfrac{(kk_f)}{(kp_i)}\right] \\[10pt] 
\dbr{z}_f^2 & = & \dfrac{e^2a^2}{(k\dbr{p})^2}\dfrac{(kk_i)}{(kp_f)}\left[
2(q_fk_i)-m_{*}^2\dfrac{(kk_i)}{(kp_f)}\right] 
\end{array}
\end{equation}

\bigskip\
\section{Numerical evaluation of the SCS cross section}
\label{c3numev}

The amount of computing time required to complete a numerical calculation of the SCS
differential cross section depends, critically, on the convergence of the infinite summations
over integers $r$ and $r^{\prime }$. Consequently it is of interest to
examine the behaviour of the functions ${\displaystyle \sum_rX^r}$ and 
${\displaystyle \sum_{r^{\prime }}X^{r^{\prime }}}$ which mainly account for
the appearance of $r$ and $r'$ in the differential cross section
expressions.

The $r,r'$ dependence of the functions $X^r,X^{r'}$
obtained by substitution of the expressions for $M,N,\bar p$ and $\dbr{p}$
(equation \ref{c3.smat.eq3}, \ref{c3.smat.eq4} and Appendix \ref{appMM}) appears in
the order of a product of squares of Bessel functions, an exponential
dependence in the numerator, and a linear dependence in the denominator

\begin{equation}
\label{c3.numev.eq1}\sum_{r=-\infty }^\infty J_r(z_i)J_{r-l}(z_f)\frac{%
e^{-ir\phi }}{r+a} 
\end{equation}

\medskip\ 

These summations were discussed by \cite{Bos79a} in connection with a study
of M\"oller scattering in a circularly polarised electromagnetic field,
however no analytic solution is known. It is of interest,
then, to discuss the asymptotic forms of the summation. Both the functions $%
e^{-ir\phi }$ and $\frac 1{r+a}$ can be expanded in a Taylor series, but
neither of these expansions provide any particular benefit. The asymptotic
forms of the Bessel functions are determined by the numerical regimes of the
order and argument \cite{AbrSte65}.

\begin{equation}
\begin{array}{llll}
J_r(z) & \sim & \dfrac 1{\Gamma (r+1)}\left( \dfrac zr\right) ^r & \quad 
\text{where}\quad r\geq 0\quad \text{and}\quad 0\leq z\ll r \\[10pt] 
J_r(z) & \sim & \sqrt{\dfrac 2{\pi z}}\;\cos \left( z-\dfrac{r\pi }2-\dfrac \pi 4\right) & 
\quad \text{where}\quad z\gg r 
\end{array}
\end{equation}

\medskip\ 

The four Bessel function arguments in equation \ref{c3.symev.eq3} are
directly proportional to either $\nu ^2$ and $\cos \,\theta _i$, 
or $\nu ^2$ and $\cos \,\theta _f$ . Numerical evaluations will include 
parameter regimes where $\nu ^2$ is not small and the
angles $\theta _i$ and $\theta _f$ range fully from $0$ to $2\pi $.
Therefore the asymptotic forms for the Bessel functions cannot be used.
However, the series in equation \ref{c3.numev.eq1} converges rapidly for $|r|>z_i,z_f$
\cite{Bos79a} and this convergence was used to limit the numerical evaluation of the 
summation over all $r$.

Any future work in external field QED involving fermion
solutions in a circularly polarised electromagnetic plane wave will involve
infinite summations of the type discussed here. In QED processes of higher
order than those considered in this work the computational time required to
complete the summations numerically may become a severely limiting factor.
For this reason, in Appendix \ref{sumsol}, some original work on
analytic solutions of infinite summations involving squares of Bessel
functions is presented, in the hope that it may provide some impetus to a future analytic
solution of the summation expressed by equation \ref{c3.numev.eq1}.

\bigskip
\section{The SCS cross section in various limiting cases}
\label{c3limit}

Previous work done by other authors on the SCS differential cross section has been 
performed with special kinematics, non relativistic energy regimes and/or
asymptotically small external field intensity. For the purposes of
comparison and validation it is of interest to examine the general SCS differential
cross section developed in the previous sections for various limiting cases.

\cite{AkhMer85} perform most of their work on the SCS process with a
linearly polarised external electromagnetic field. They also write
down an expression for the square of the matrix element of the SCS process
with a circularly polarised external electromagnetic field, though they
provide no numerical results. \cite{AkhMer85} consider kinematics in
which the direction of propagation of the incoming photon is parallel to the
direction of propagation of the external field photons. These kinematics imply that 

\begin{equation}
\label{c3.limit.eq1}
\begin{array}{rcl}
\sql{\hat k_i}-\sql{\hat k} &=& 0\\ 
(kk_i),(a_1k_i),(a_2k_i)&=&0 \\
\bar z_i= \dbr{z}_f &=& 0 \\ 
\dbr{z}_i^2=\bar z_f^2\;(\equiv z^2) &=& \dfrac{e^2a^2}{(kp_i)^2} 
\dfrac{(kk_f)}{(kp_f)}\left[ 2(q_fk_f)-\dfrac{(kk_f)}{(kp_f)}\right] 
\end{array}
\end{equation}

\medskip\ 

With application of these relations to the general results obtained in
section \ref{c3smat} for the square of the SCS S-matrix element, the infinite summations over $r$ and $r^{\prime }$ can be performed immediately and the resulting 
expressions are greatly reduced in length. 

\begin{figure}[!b]
\label{c3lim.fig1}
\centerline{\includegraphics[height=8cm,width=10cm]{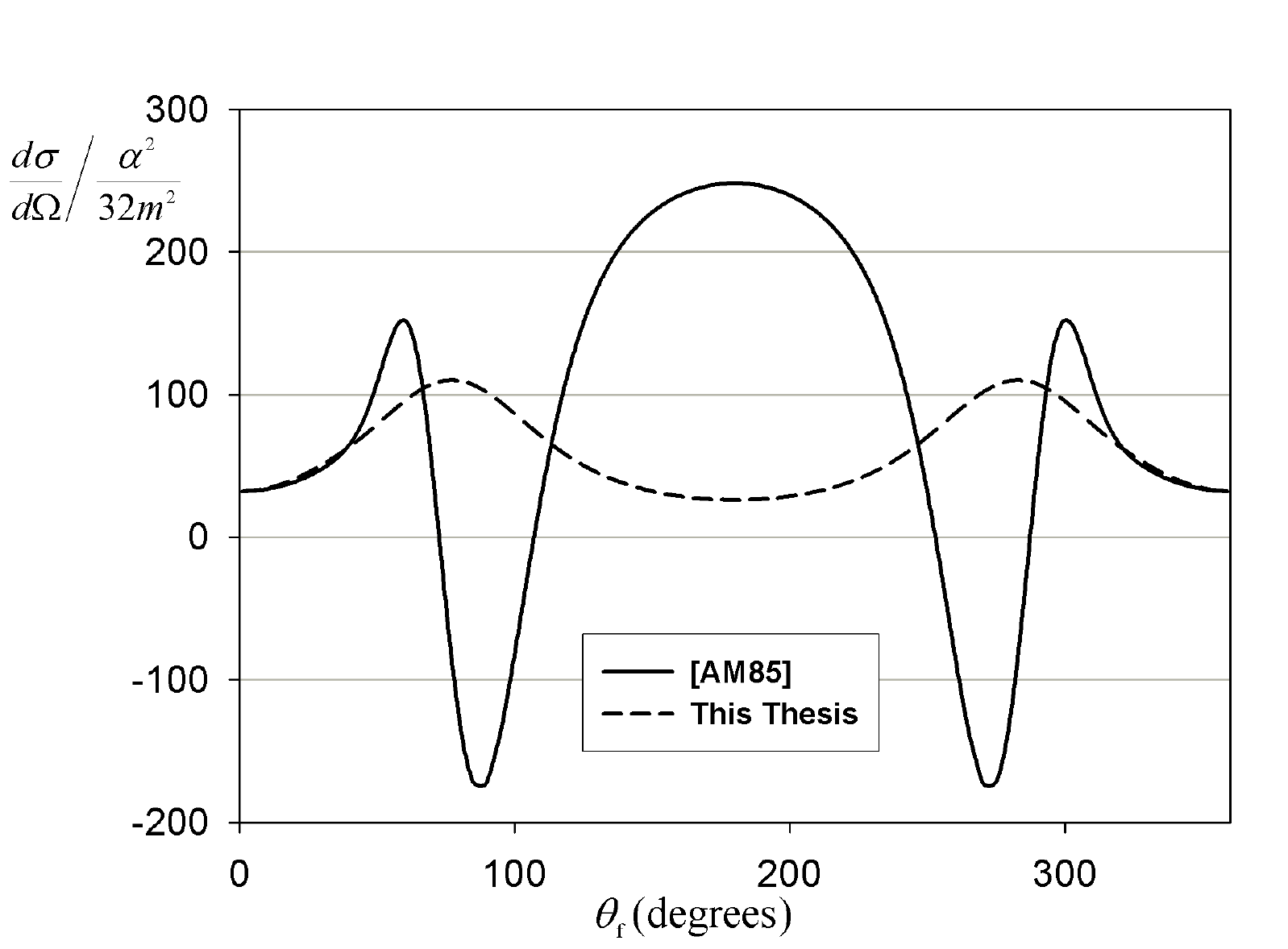}}
%\centerline{\psfig{figure=./compton/initdata/akh_me_comp/akh_me_comp_nu0.7_wi0.8_w1_theti0.pdf,height=8cm,width=10cm}}
\caption{\bf\bm SCS differential cross section comparison between 
\cite{AkhMer85} and the expressions of Chapter 3 for $\omega=0.512$ MeV, 
$\,\omega_i=0.41$ MeV, $\,\theta_i=0^{\circ}$ and $\nu^2=0.7$.} \end{figure}

Small but significant differences were found between analytic expressions of \cite{AkhMer85} and 
the analytic expressions obtained here. Numerical values obtained from the expressions of 
\cite{AkhMer85} became negative for certain parameter ranges; a non physical result. In contrast, 
numerical results from the expressions obtained in this thesis remain positive at all times (see 
figure \ref{c3lim.fig1})

The second limiting case to be studied is that of vanishing external field achieved by allowing 
the external field intensity parameter to approach zero, $\nu ^2\rightarrow 0$. In this limit,
the SCS process reduces to ordinary Compton scattering \cite{KleNis28}.
As  $\nu ^2\rightarrow 0$ all four Bessel function arguments $\bar z_i,\bar z_f,\dbr{z}_i,\dbr{z}_f$ 
reduce to zero, and the infinite summations ${\displaystyle \sum_{rr'}}$ can be performed immediately. 
The occurrence of Bessel functions with zero argument in the SCS differential cross section expressions 
ensure that no laser photons $l=0$ contribute to the process. The trace expressions reduce to

\begin{equation}
\label{c3.limit.eq2} 
\begin{array}{lll}
{\limfunc{Tr}\,}Q_1(\bar p_r,\bar p_{r\prime }) & = & 32\{-m^2(p_f\bar
p)+(p_ik_i)(p_fk_i)+2m^2(p_ik_i)+2m^4\} \\[12pt] 
\limfunc{Tr}\,Q_2(\bar p_r,\dbr{p}_{r\prime }) & = & 16m^2[(p_i\bar p)+(p_i 
\dbr{p})+(p_f\bar p)+(p_f\dbr{p}) \\  &  & +(p_ip_f)+(\bar p\dbr{p}%
)-2m^4]-32(\bar p\dbr{p})(p_ip_f) 
\end{array}
\end{equation}

\medskip\ 

Equation \ref{c3.limit.eq2} is identical with equations (14a) and (14b) of \cite{Oleinik68} who 
considered the
SCS process with linearly polarised external electromagnetic field in the
same limit of vanishing external field.

With substitution of equation \ref{c3.limit.eq2} into the SCS differential cross section (equation 
\ref{c3.symev.eq1} and \ref{c3smat.eqlast}), the Klein-Nishina result for an incoming photon of 
initial energy $\omega _i$
and final energy $\omega _f$ scattered through an angle $\theta_f$ is obtained.

\begin{equation}
\label{c3.limit.eq3}
\begin{array}{rl}
\dfrac{d\sigma}{d\Omega } & = \dfrac{r_o^2}2\left( 
\dfrac{\omega _f}{\omega _i}\right) ^2\left( \dfrac{\omega _f}{\omega _i}+
\dfrac{\omega _i}{\omega _f}-\sin {}^2\theta_f \right) \\[10pt]
\text{where} \quad\quad r_o &=\dfrac{e^2}{4\pi m }
\end{array}
\end{equation}

\bigskip\
\section{Stimulated Two Photon $e^{+}e^{-}$ pair production (STPPP) in an External Field}
\label{c3pprod}

The stimulated two photon pair production process is the production of a Volkov electron/positron pair, 
$p_{-}$ and $p_{+}$, by an initial state consisting of two photons, $k_1$ and $k_2$, with a 
number of laser field photons contributing.
The STPPP matrix element can be written down from Wick's
theorem, and consists of two channels, each of which can be represented
by the Feynman diagrams shown in figure 3.4

\begin{figure}[!b]
\label{c3ppf1}
\centerline{\includegraphics[height=4.5cm,width=11cm]{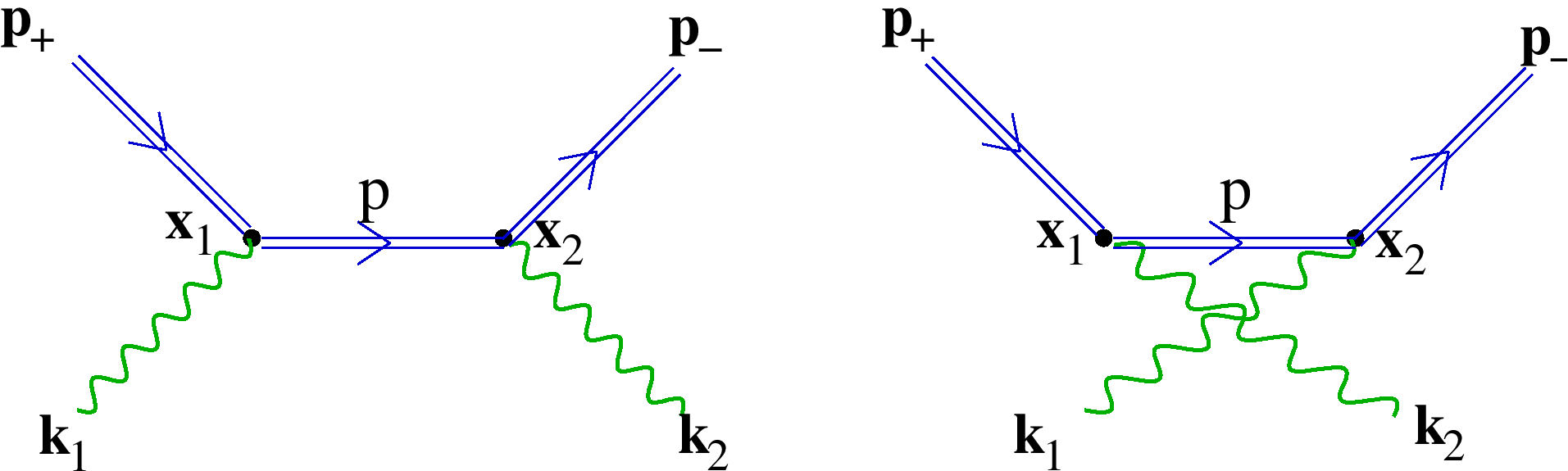}}
\caption{\bf Feynman diagrams for stimulated two photon pair production.}
\end{figure}

%\begin{figure}[!b]
%\hbox{ 
%\psfig{figure=./tex/feyndiag/stim_pprod1.eps,height=6.0cm,width=6.5cm}
%       \hspace{1cm}
%       
%\psfig{figure=./tex/feyndiag/stim_pprod2.eps,height=6.0cm,width=6.5cm}}
%\caption{\bf Feynman diagrams for two photon $e^{+}e^{-}$ pair production
% in an external field.}
%\label{c3.pp.fig1}
%\end{figure}

From the usual rules associated with Feynman diagrams, the STPPP matrix element is

\begin{equation}
\label{c3.pp.eq1} 
\begin{array}{rl}
S_{fi}^e = & -e^2\dint_{-\infty }^\infty \;d^4x_1\,d^4x_2\;\biggl\{\bar 
\psi_{V}^{-}(x_2,p_-)\,\slashed{A}_{k_1}^{+}(x_2)\,G^e(x_2,x_1)\,\slashed{A}_{k_2}^{-}(x_1)\,
\psi _{V}^{+}(x_1,p_+) \\[10pt]
& \quad\quad + \;\; \bar 
\psi_{V}^{-}(x_2,p_-)\,\slashed{A}_{k_2}^{-}(x_2)\,G^e(x_2,x_1)\,\slashed{A}_{k_1}^{+}(x_1)\,
\psi_{V}^{+}(x_1,p_+)\biggr\} 
\end{array}
\end{equation}

\medskip\ 

$\psi_{V}^{-}(x,p_-)$ are the electron Volkov solutions and $\psi_{V}^{+}(x,p_+)$ are the positron 
solutions. Substitution into equation \ref{c3.pp.eq1} of the expressions for the bound
electron propagator $G^e$, the Volkov solutions, $\psi_V$, and the Maxwell wave functions 
$A_k$, yields for the STPPP matrix element

\begin{equation}
\label{c3.pp.eq2} 
\begin{array}{rl}
S_{fi}^e & = -e^2\dint_{\nths -\infty }^\infty \;d^4x_1\,d^4x_2\;d^4q\;\;\bar 
u(p_{-})\,\bar
E_{p_{-}}(x_2) \\[10pt]
& \times \;\; \biggl\{\slashed{e}(k_1)e^{-ik_1x_2}E_p(x_2)\left[ \dfrac{\slashed{p}+m}
{p^2-m^2}\right] \bar E_p(x_1)\slashed{e}(k_2) e^{-ik_2x_1} \\[10pt]
& + \;\; \slashed{e}(k_2) e^{-ik_2x_2} E_p(x_2)\left[ \dfrac{\slashed{p}+m}{p^2-m^2}\right] 
\bar E_p(x_1)\slashed{e}(k_1) e^{-ik_1x_1} \biggr\} E_{-p_{+}}(x_1)v(p_{+}) 
\end{array}
\end{equation}

\medskip\

Crossing symmetry (section \ref{c2sub}) allows the STPPP matrix element to be obtained from the 
SCS matrix element (equation \ref{c3smat.eqlast}) with the following substitutions

\begin{equation}
\label{c3.pp.eq3} 
\begin{array}{ccccccc}
q_f\leftrightarrow q_{-} & , & q_i\leftrightarrow -q_{+} & , & 
k_i\leftrightarrow k_2 & , & k_f\leftrightarrow -k_1 \\ 
e(k_i)\leftrightarrow e(k_2) & , & e(k_f)\leftrightarrow e(k_1) & , & \bar 
u(p_f)\leftrightarrow \bar u(p_{-}) & , & u(p_i)\leftrightarrow v(p_{+}) 
\end{array}
\end{equation}

\medskip\

The calculation of the square of the STPPP matrix element proceeds in the same 
fashion as that for the SCS process. Fourier expansions into
infinite summations of Bessel functions, integrations over 4-space and
4-momentum variables $x_1,x_2,q$ and the summation over electron spins and
photon polarisations are all employed to write down the sum over initial
and final states of the square of the matrix element. Alternatively the substitutions 
in equation \ref{c3.pp.eq3} can be made in the square of the SCS matrix element. 
A factor of $-1$ is introduced due to the summation over positron bispinors.

\begin{equation}
\label{c3.pp.eq4} 
\begin{array}{rl}

\sum\limits_{if} |S_{fi}^e|^2 & = -\dfrac{e^4}{16m^2} \dsum\limits_l
\delta^4(k_1+k_2+lk-q_{-}-q_{+}) \Tr \, Q_{\text{STPPP}} \\[20pt]

\text{where} \quad \Tr \, Q_{\text{STPPP}} & = \dsum\limits_{rr'} \,\left[ \Tr \;Q_1(\bar p_r,\bar 
p_{r\prime })+ \Tr\;Q_1(\dbr{p}_r,\dbr{p}_{r\prime })+\Tr\;Q_2(\bar p_r,
\dbr{p}_{r\prime })+\Tr\;Q_2^{*}(\bar p_r,\dbr{p}_{r\prime }) \right] \\[20pt]

\text{and} \quad Q_1(\bar p_r,\bar p_{r'}) & = (\slashed{p}_{-}+m)
B(\gamma^\mu ,\bar N_r,\bar p,p_{-})\left[ \dfrac{\slashed{\bar{p}}_r+m}{\bar p_r^2-m^2}\right] 
B(\gamma^\nu ,\bar M_r,-p_{+},\bar p) \\[10pt]
& \times \;\; (-\slashed{p}_{+}+m)\widetilde B(\gamma _\nu ,\bar M_{r^{\prime }},-p_{+},\bar
p)\left[ \dfrac{\slashed{\bar{p}}_{r'}+m}{\bar p_{r'}^2-m^2}\right]^{*}
\widetilde B(\gamma _\mu ,\bar N_{r\prime },\bar p,p_{-}) \\[20pt]

Q_2(\bar p_r,\dbr{p}_{r'}) & = (\slashed{p}_{-}+m)B(\gamma^\mu ,\bar N_r,\bar p,p_{-})
\left[ \dfrac{\slashed{\bar{p}}_r+m}{\bar p_r^2-m^2}\right] 
B(\gamma^\nu ,\bar M_r,-p_{+},\bar p) \\[10pt]
& \times \;\; (-\slashed{p}_{+}+m)\widetilde{B}(\gamma_\mu,\dbr{M}_{r^{\prime }},-p_{+},\dbr{p})
\left[ \dfrac{\slashed{\dbr{p}}_{r'}+m}{\dbr{p}_{r^{\prime }}^2-m^2}\right]^{*}
\widetilde{B}(\gamma _\nu ,\dbr{N}_{r\prime },\dbr{p},p_{-}) \\[20pt]

\bar p_r & = -p_{+}+\dfrac{e^2a^2}2\left[ \dfrac 1{(kp_{+})}+\dfrac 1{(k\bar p)}\right] k+k_2+rk \\[10pt]
\dbr{p}_r & = -p_{+}+\dfrac{e^2a^2}2\left[ \dfrac 1{(kp_{+})}+\dfrac 1{(k\dbr{p})}\right] k+k_1+rk

\end{array}
\end{equation}

\medskip\ 

The applicability of a crossing symmetry makes the analytic computation of the STPPP simple. 
The computer programs that generate numerical results for the SCS differential 
cross section can be used with the appropriate symbolic substitutions. The only additional 
analytic work required is to choose a reference frame and to transform the STPPP phase space integral.

\begin{figure}[!b]

\centerline{\includegraphics[height=6cm,width=6cm]{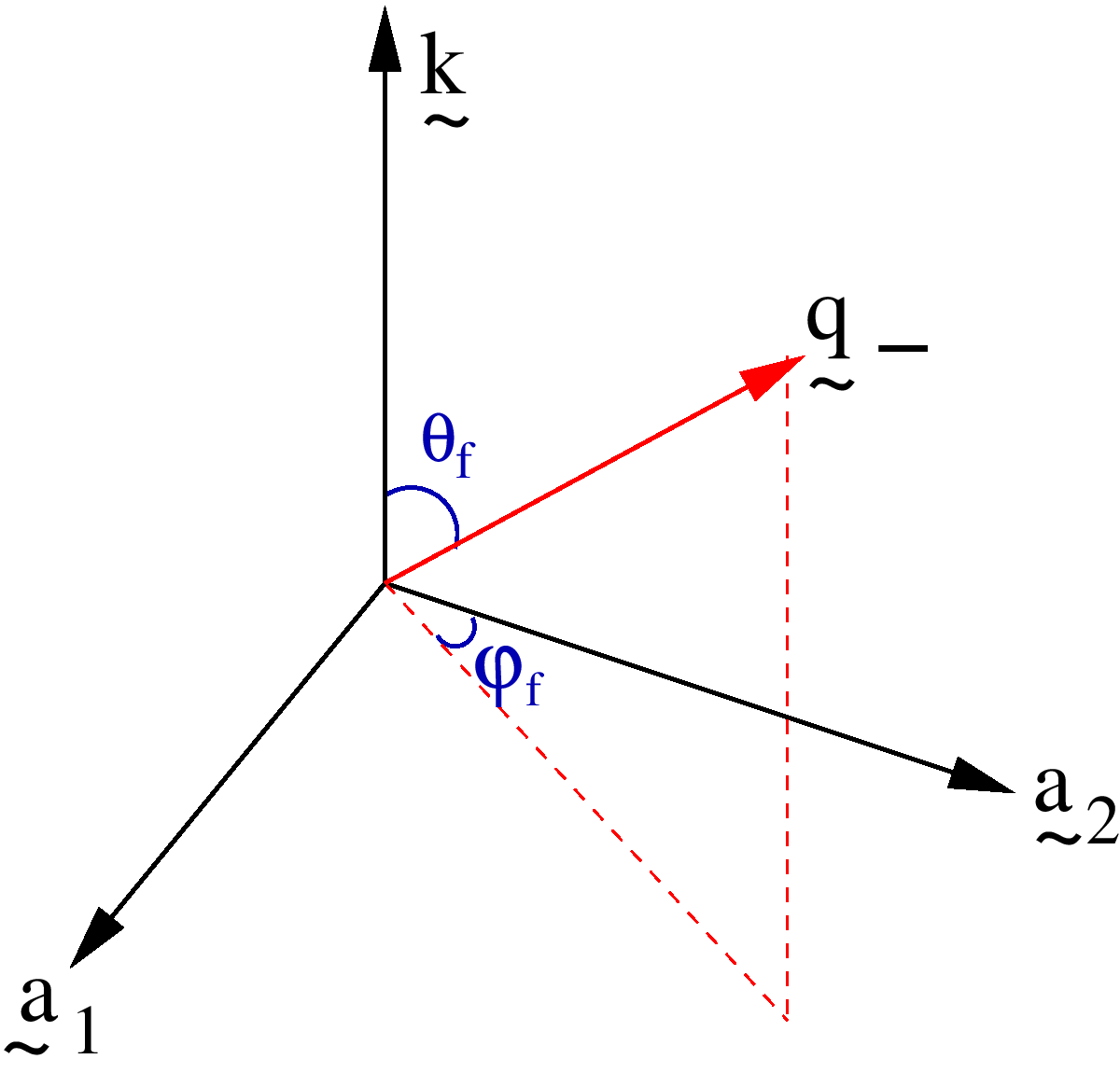}}
\caption{\bf scattering Geometry for stimulated two photon pair production}
\label{c3.pp.fig2}
\end{figure}

Choosing the same gauge and coordinate axes to specify the external field 4-vectors, as for the SCS 
process, the STPPP process geometry is represented in figure \ref{c3.pp.fig2}. The 
bound electron 3-momentum $\squ{q}_{-}$ is scattered into an element
of solid angle $d\Omega _{q_{-}}$ characterised by angles $\theta _f$ and 
$\varphi _f$.

The STPPP differential cross section calculation is greatly simplified by use of the centre of 
mass-like reference frame which includes the contribution from the external field and is expressed by

\begin{equation}
\label{c3.pp.eq5}
\begin{array}{rl}
  \sql{k}_1 + \sql{k}_2 &= \squ{q}_- + \squ{q}_+ - l \sql{k} = 0 \\
  \omega_1 + \omega_2   &= \varepsilon_{q_-} + \varepsilon_{q_+} - l \omega
\end{array}
\end{equation}

\medskip\ 

The electron and positron energies must each be at least the external field shifted 
rest mass of the fermion, $m_{*}$. With the aid of equation \ref{c3.pp.eq5}, a 
condition for the minimum number of laser photons required to create the $e^+e^-$ pair is

\begin{equation}
\label{c3.pp.eq6}
\begin{array}{c}
 l \geq - \dfrac{\omega_1}{m} \left( \dfrac{\omega_1 - m_{\ast}}{\omega_1 
- \frac{1}{2} m_{\ast}} \right)
\end{array}
\end{equation}

\medskip\ 

Other parameters in the STPPP differential cross section can be expressed in terms of 
photon energies $\omega$ and $\omega_1$, the number of laser photons $l$, 
and the angle $\theta_f$.

\begin{equation}
\label{c3.pp.eq7}
\begin{array}{rl}
\omega_2 & = \omega_1 \\
\varepsilon_{q_+} &= 2 \omega_1 - \varepsilon_{q_-} + l \omega \\ 
\varepsilon_{q_-} &=\dfrac{1}{2 (\mathcal{A}^2-\mathcal{C}^2)} (-2 \mathcal{B}\mathcal{C}
+ \sqrt{(2\mathcal{B}\mathcal{C})^2+ 4(\mathcal{A}^2-\mathcal{C}^2)(\mathcal{A}^2 
(1+\nu^2)+\mathcal{B}^2)}) \\[15pt]
\text{where} &\quad\quad \mathcal{A}=l \omega \cos \theta_f \\
                  &\quad\quad \mathcal{B}=2 \omega_1 (\omega_1 + l \omega) \\
                  &\quad\quad \mathcal{C}=2 \omega_1 + l\omega
\end{array}
\end{equation}

\medskip\ 

The integrations over final particle momenta expressed by the phase space integral can be transformed 
in a similar fashion to those of the SCS process. The Jacobian of the transformation
between free fermion momenta $d\,\squ{p}_-$ , $d\,\squ{p}_+$ to corrected 
fermion momenta $d\,\squ{q}_-$ , $d\,\squ{q}_+$ is unity (Appendix \ref{jacob}) and the phase 
space integral transforms into the element of solid angle into which the created bound electron 
3-momentum is directed

\begin{equation}
\label{c3.pp.eq8}
\begin{array}{c}
\dint \dfrac{d^3\squ{p}_-\;d^3\squ{p}_+}{\epsilon_{p_-}\epsilon_{p_+}}\delta 
^4(P_i-q_{-}-q_{+})  = \dfrac{1}{2}\dint d\Omega _{q_-}\;\dfrac{|\squ{q}_-|}
{\epsilon_{q_-}}
\end{array}
\end{equation}

\medskip\ 

The STPPP differential cross section is written with the aid of the square of the STPPP matrix 
element (equation \ref{c3.pp.eq4}), the general differential cross section expression 
(equation \ref{c2.eq13}), the results of the SCS calculation with appropriate substitutions 
(equation \ref{c3.pp.eq3}), and the phase space integral transformation just obtained. The relative 
velocity of the two initial photons $k_1$ and $k_2$ is $v_{ab}=
\frac 1{\omega _1\omega _2}\sqrt{(k_1k_2)^2}=2$ in the centre of mass frame.

\begin{equation}
\label{c3.pp.eq9}
\begin{array}{c}

\dfrac{d\sigma}{d\Omega_{q_-}}  = -\dfrac{e^4}{128(2\pi)^2}\dsum\limits_{lrr'} 
\dfrac 1{\sqrt{(k_1k_2)^2}}\;\dfrac{|\squ{q}_-|}{2\epsilon_{q_-}}\; \Tr \, Q_{\text{STPPP}}

\end{array}
\end{equation}

\medskip\ 

In the limit of vanishing external field intensity parameter $\nu^2 \rightarrow 0$, the STPPP
differential cross section reduces to one term in which no external field photons contribute, 
$l=0$. The trace expression reduces to

\begin{equation}
\label{c3.pp.eq10}
\begin{array}{rl}
\Tr \; Q_{\text{STPPP}} \rightarrow & -\dfrac{16}{(1-\beta ^2\cos^2\theta_f)^2}
\left[ 1-\beta ^4\cos^4\theta _f+2\left( \dfrac m{\omega_1}\right)^2\beta^2\sin^2\theta_f\right] \\[10pt] 
& \text{where} \quad\quad \beta ^2=1-\dfrac{m^2}{\omega _1^2}
\end{array}
\end{equation}

\medskip\ 

In the same limit of vanishing external field intensity the STPPP differential cross section 
reduces to the Breit-Wheeler result for non external field two photon 
pair production in the centre of mass frame \cite{BreWhe34,JauRoh76}.

\begin{equation}
\label{c3.pp.eq11}
\begin{array}{rl}

\dfrac{d\sigma }{d\Omega _{p_{-}}} = & \dfrac{r_0^2}4\left(
\dfrac m{\omega _1}\right)^2\dfrac \beta {(1-\beta ^2\cos^2\theta_f)^2}
\left[ 1-\beta ^4\cos^4\theta _f+2\left( \dfrac m{\omega _1}\right)^2\beta ^2\sin^2\theta _f\right] \\[10pt]
& \text{where} \quad\quad r_0 = \dfrac{e^2}{4\pi m}

\end{array}
\end{equation}

\bigskip

%% file: chap4.tex
\section{Introduction}

In this chapter we present numerical calculations of the SCS differential
cross section analytic expressions obtained in Chapter 3. The numerical
results are contained in Section 4.2 and an analysis of these results is
contained in Section 4.3.
The SCS differential cross section as expressed in equation 
\ref{c3smat.eqlast} is a triple
infinite summation over integer variables $l,r$ and $r^{\prime }$. On 
occasion many of these summation terms contribute significantly to the SCS 
differential cross section, producing potentially complex behaviour.
The task of analysing numerical results is made simpler by
dividing figures into two groups. 

The first group, contained in Section 4.2.1, represents the summation terms
of the SCS differential cross section separately (or summed in part). The
analysis for this group of results is contained in Section 4.3.1. The
terminology "$l$ contribution" simply means the $l$ summation term in SCS 
differential cross section. "Longitudinal" and "transverse" are also terms often 
used in the discussion. When not qualified these refer respectively to directions 
parallel and perpendicular to the direction of propagation of the external field.
The second group of results, presented in Section 4.2.2 (with a
corresponding analysis in Section 4.3.2) are those of the complete SCS
differential cross section in which all summations have been performed.

The general structure of the analysis will be to introduce a set of figures with 
common parameter sets, describe the main features and to provide an explanation 
of how the features arise. The explanation is physical (as opposed to 
purely mathematical) as often as possible and analogies are sometimes drawn. 
Sometimes the behaviour of plots is complicated and an appeal to the analytic 
structure of equations is made. Generally speaking, the variation of the SCS differential 
cross section with azimuthal angle $\varphi_f$ is much simpler than the variation with other 
parameters. Consequently only a couple of figures are devoted to $\varphi_f$ 
variation. Elsewhere $\varphi_f$ is set to zero.

\bigskip
\subsection{Differential Cross section $l$ Contributions}
\enlargethispage*{4cm}

Table 4.1 displays the parameter values of the SCS process
investigated in this section. The parameters $\omega $ and $\omega _i$ are,
respectively, the particle energies of the external field and of the initial
photon. The initial photon enters the scattering region at an angle of 
$\theta_i$ to the direction of the propagation of the external field, and
exits in a direction specified by angles $\theta_f$ and $\varphi_f$ (see figure 
\ref{c3phase.fig1}). The external field
intensity is represented by the parameter $\nu ^2=\frac{e^2a^2}{m^2}$, which
is dimensionless in natural units.
The vertical axes of all figures in Sections 4.2.1 and 4.2.2 show the
SCS differential cross section divided by a function of the fine structure
constant $\alpha $ and electron mass $m$, and the units are Steradian$^{-1}$.
All angles represented on horizontal axes use units of degrees.

\bigskip\

\input{./tex/tables/c_table1}

\clearpage
\begin{figure}[H]
 \centerline{\includegraphics[height=8cm,width=15cm]{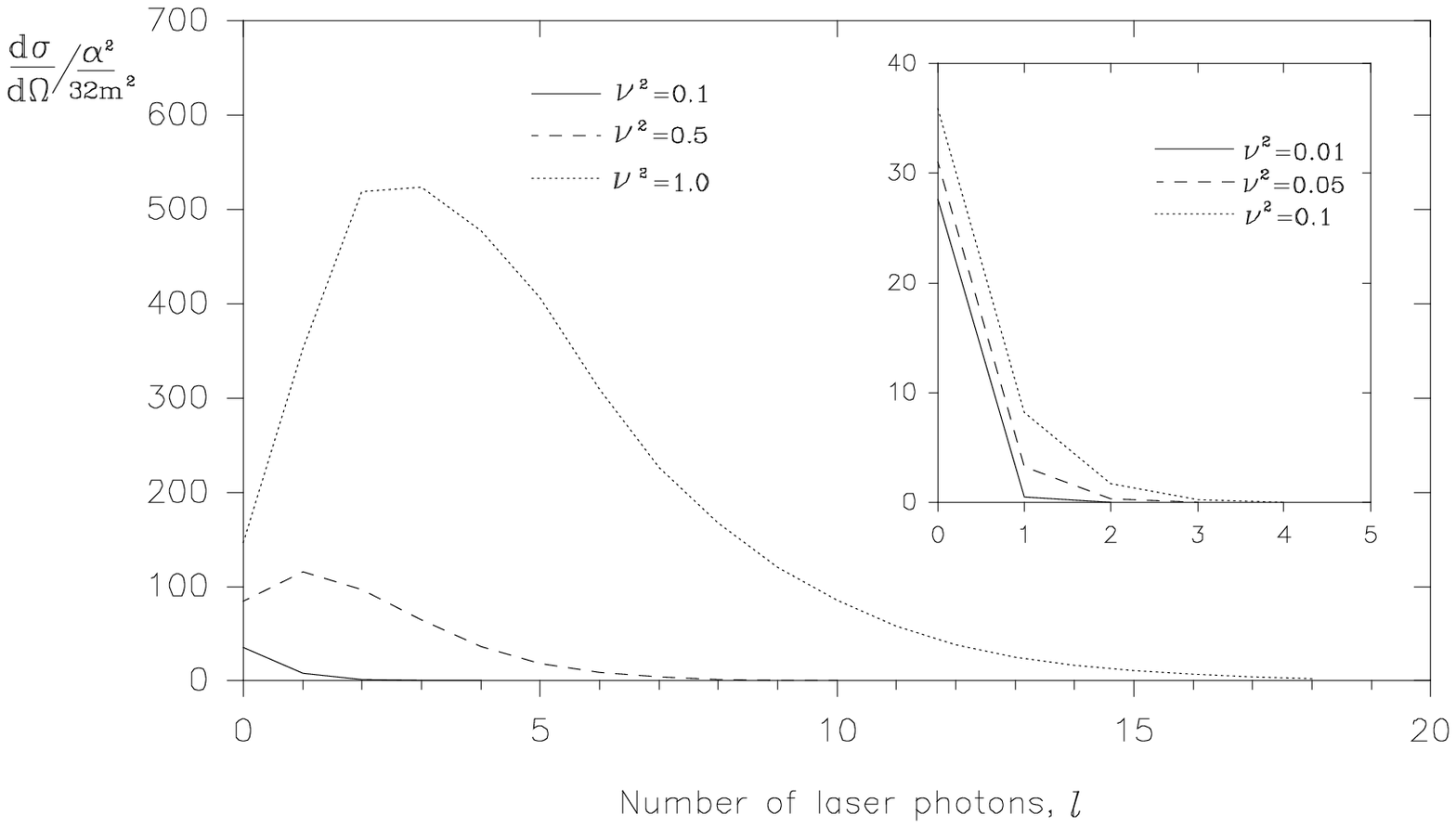}}
\caption{\bf\bm The SCS differential cross section vs $l$ external field photons for $\,\omega=0.001$ 
keV, $\omega_i=0.001$ keV, $\theta_i=10^{\circ}$, $\theta_f=45^{\circ}$, $\varphi_f=0^{\circ}$ and 
various $\nu^2$.}
\label{cga1}
\end{figure}

\begin{figure}[H]
 \centerline{\includegraphics[height=8cm,width=15cm]{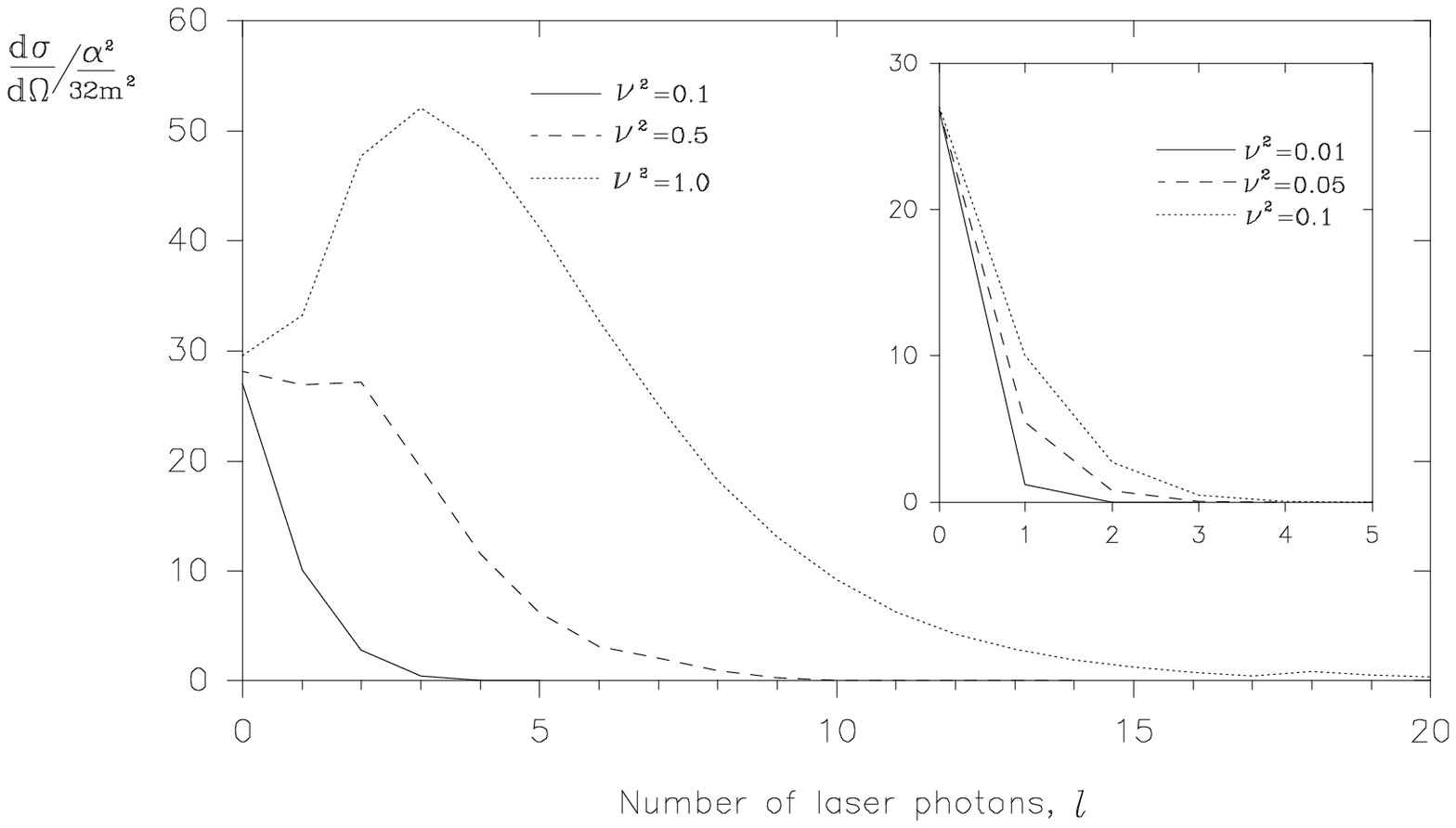}}
\caption{\bf\bm The SCS differential cross section vs $l$ external field 
photons for $\,\omega=0.512$ keV, $\omega_i=0.1$ keV, $\theta_i=80^{\circ}$,
$\theta_f=45^{\circ}$, $\varphi_f=0^{\circ}$ and various $\nu^2$.}
\label{cga2}
\end{figure}

\clearpage

\begin{figure}[H]
 \centerline{\includegraphics[height=8cm,width=15cm]{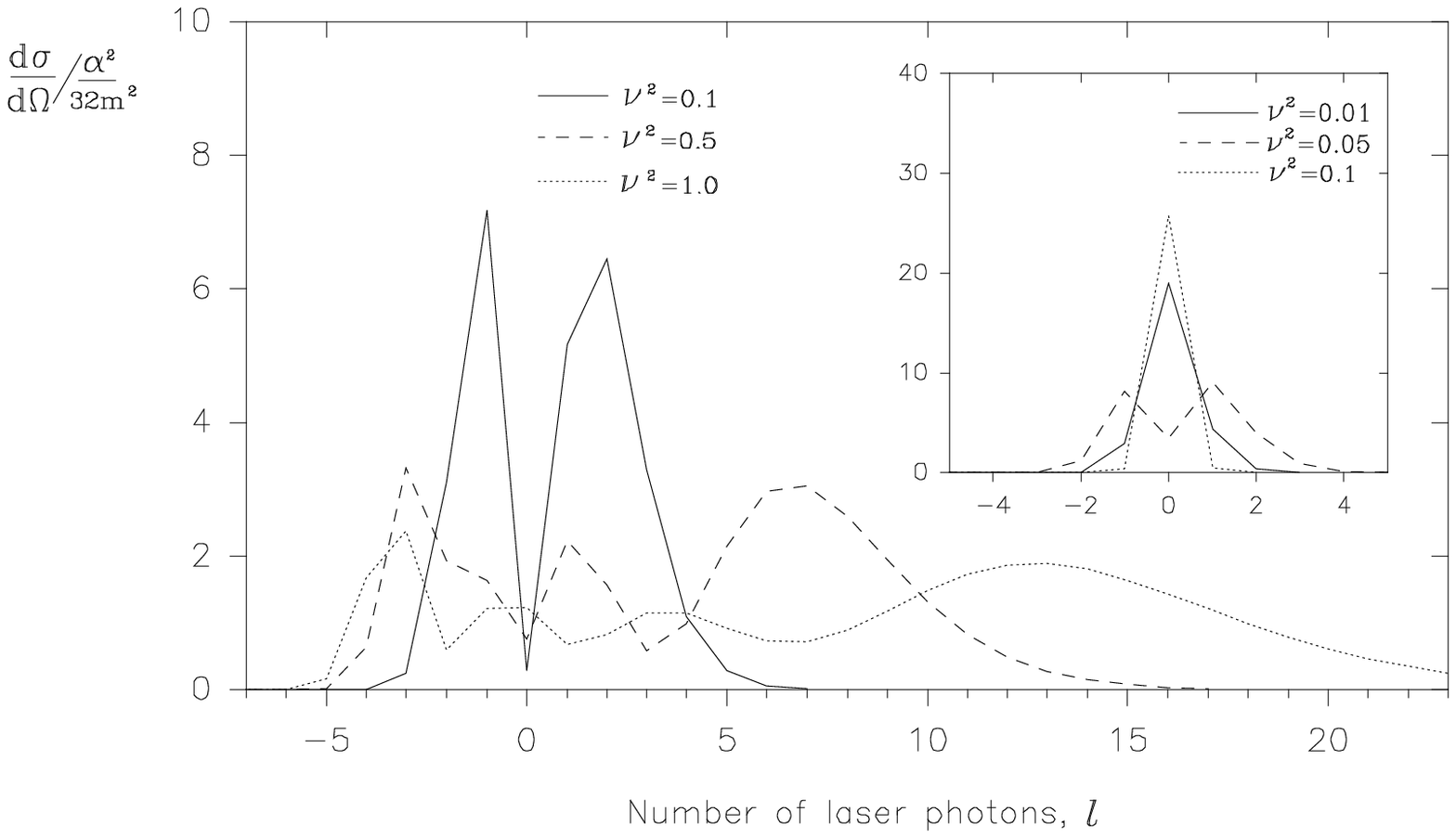}}
\caption{\bf\bm The SCS differential cross section vs $l$ external field 
photons for $\,\omega=0.512$ keV, $\omega_i=7.68$ keV, $\theta_i=10^{\circ}$,
$\theta_f=45^{\circ}$, $\varphi_f=0^{\circ}$ and various $\nu^2$.}
\label{cga3}
\end{figure}

\begin{figure}[H]
 \centerline{\includegraphics[height=8cm,width=15cm]{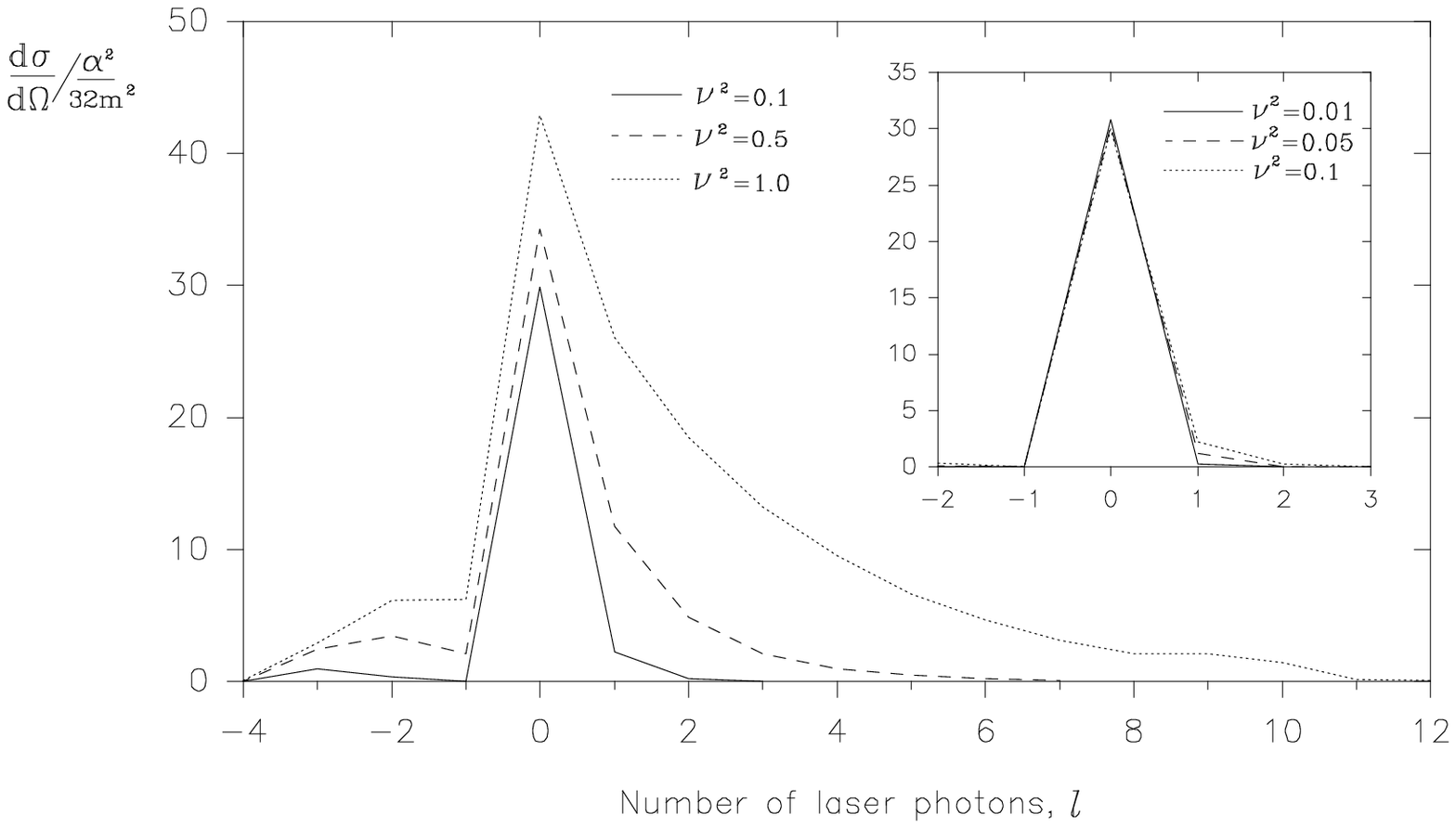}}
\caption{\bf\bm The SCS differential cross section vs $l$ external field 
photons for $\,\omega=0.512$ keV, $\omega_i=2.047$ keV, $\theta_i=30^{\circ}$,
$\theta_f=45^{\circ}$, $\varphi_f=0^{\circ}$ and various $\nu^2$.}
\label{cga4}
\end{figure}

\clearpage

\begin{figure}[t]
 \centerline{\includegraphics[height=8cm,width=10cm]{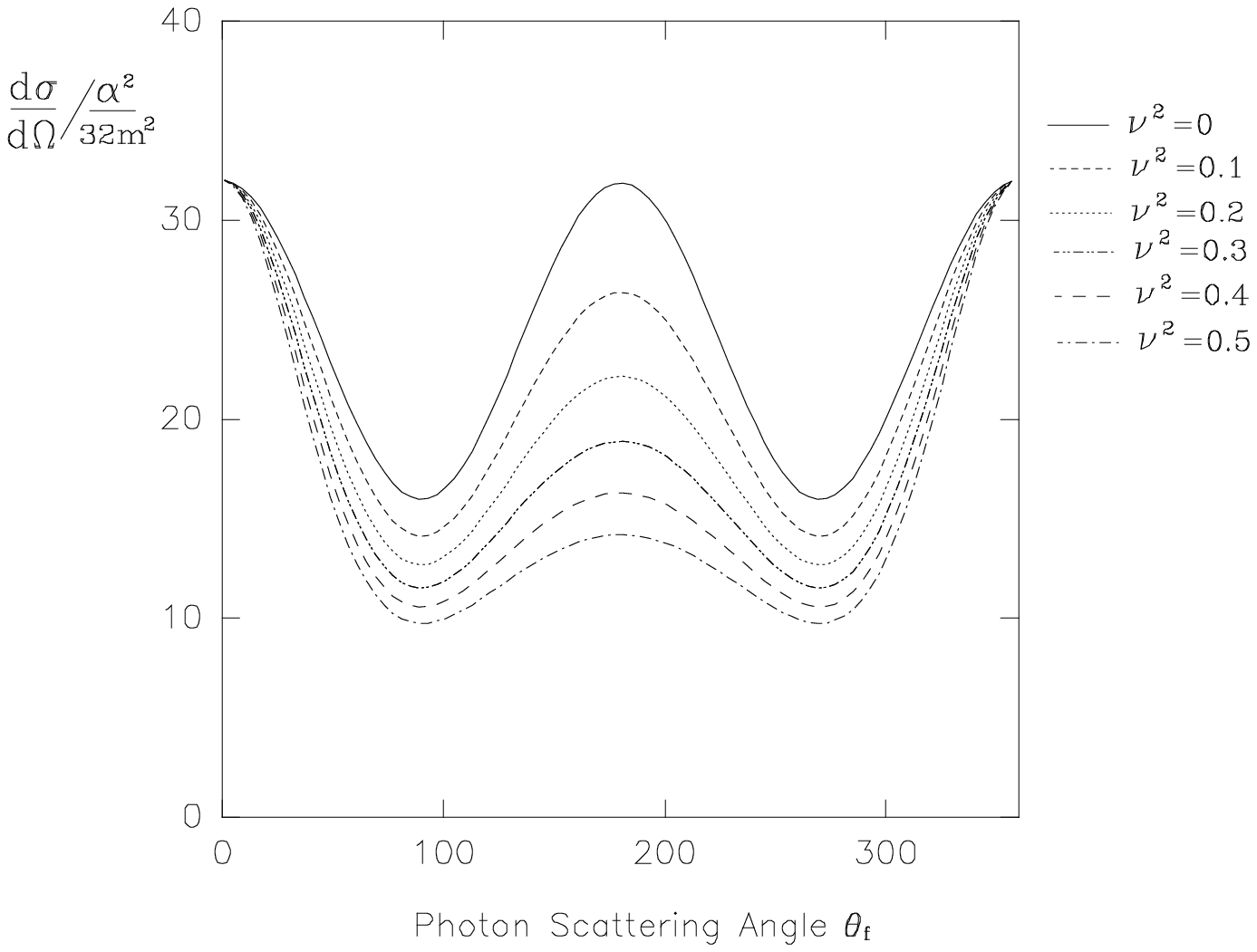}}
\caption{\bf\bm The SCS $l=0,r=0$ differential cross section vs $\theta_f$ for
$\,\omega=0.512$ keV, $\omega_i=0.409$ keV, $\theta_i=0^{\circ}$,
$\varphi_f=0^{\circ}$ and various $\nu^2$.}
\label{cgb1}
\end{figure}

\begin{figure}[t]
 \centerline{\includegraphics[height=8cm,width=10cm]{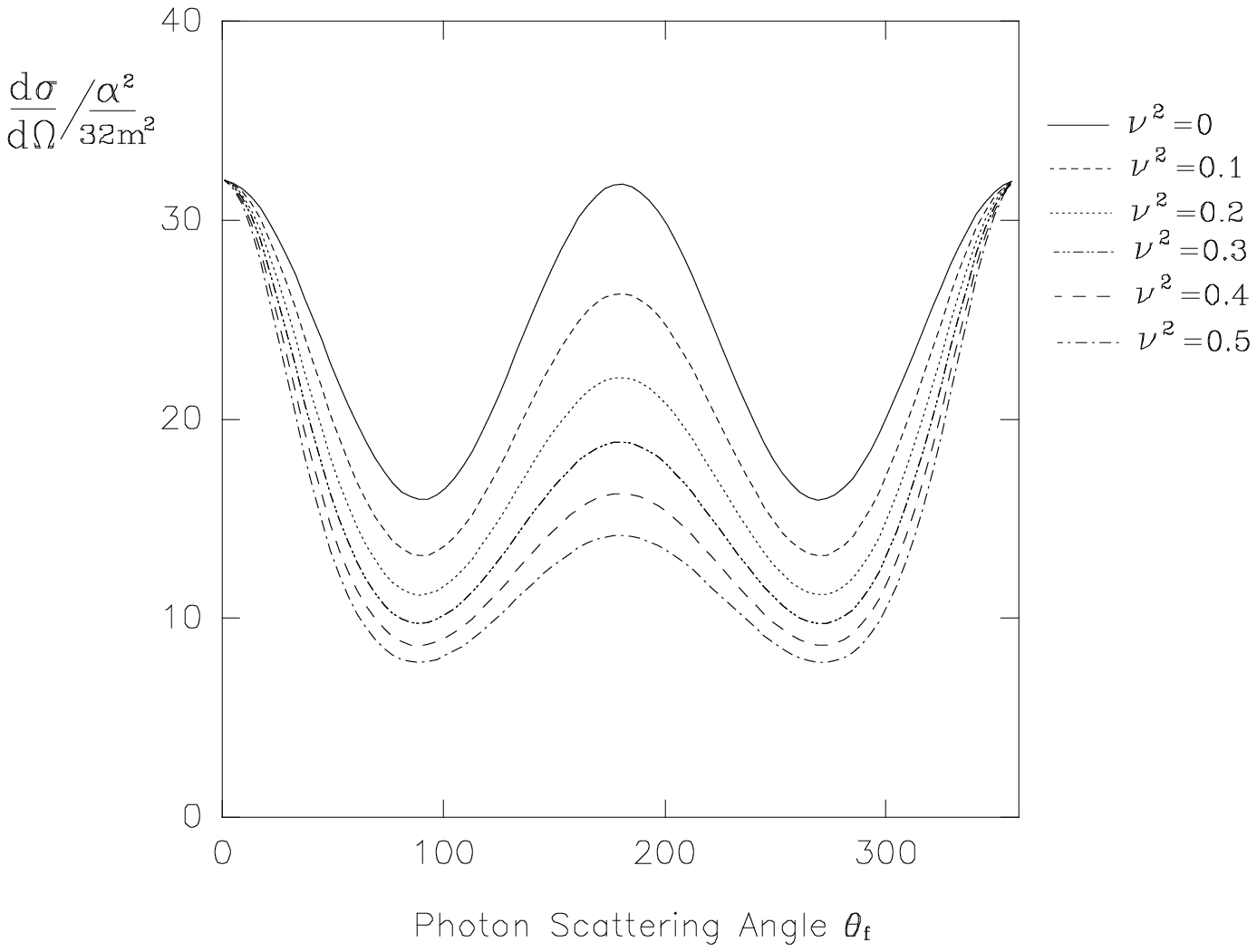}}
\caption{\bf\bm The SCS $l=0,r=0$ differential cross section vs $\theta_f$ for
$\,\omega=0.512$ keV, $\omega_i=0.768$ keV, $\theta_i=0^{\circ}$,
$\varphi_f=0^{\circ}$ and various $\nu^2$.}
\label{cgb2}
\end{figure}

\begin{figure}[t]
 \centerline{\includegraphics[height=8cm,width=10cm]{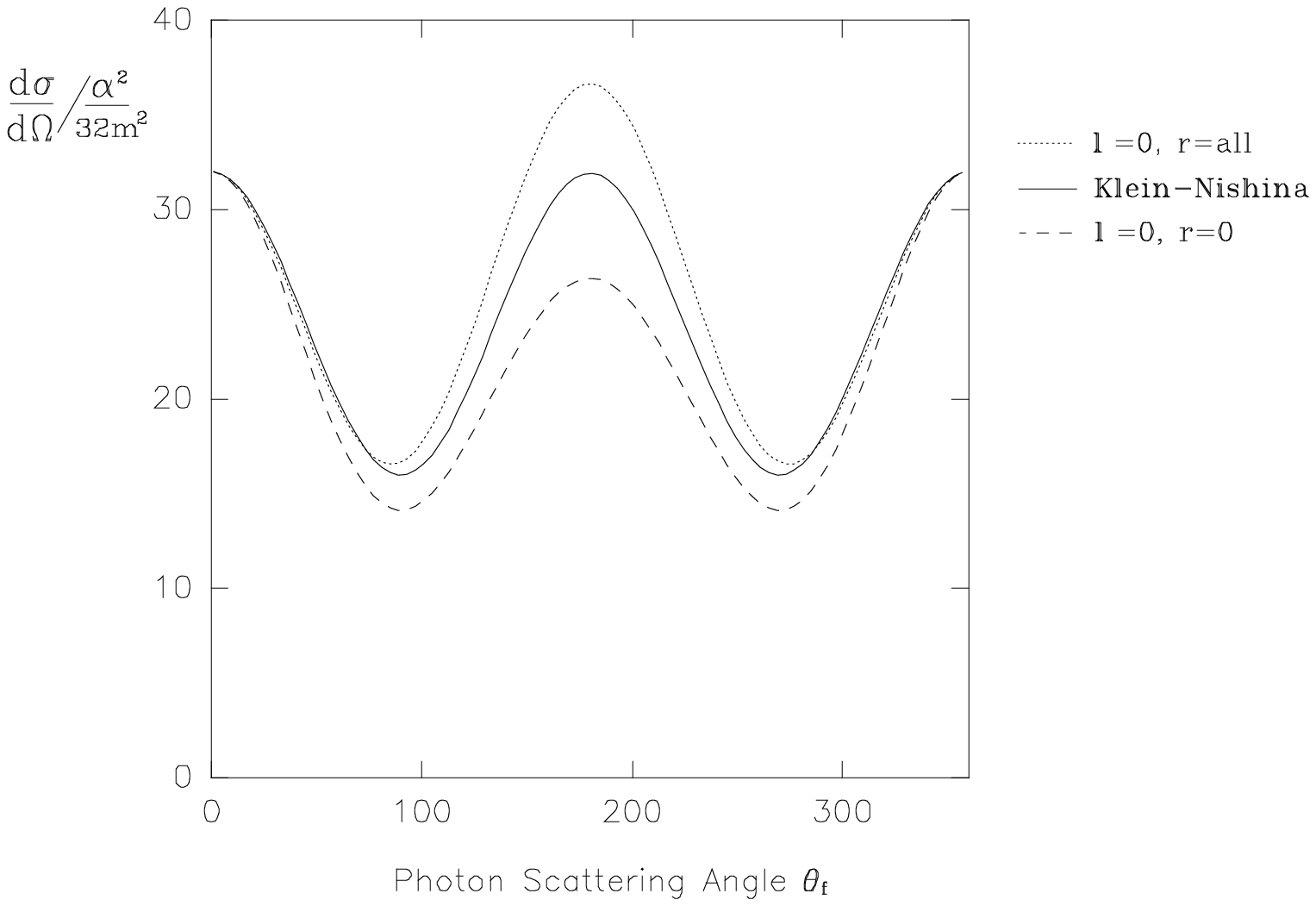}}
\caption{\bf\bm Comparison of The STPPP $l=0,r=0$, $l=0,r=\text{all}$ and Klein-Nishina
differential cross section vs $\theta_f$ for
$\,\omega=0.512$ keV, $\omega_i=0.409$ keV, $\theta_i=0^{\circ}$,
$\varphi_f=0^{\circ}$ and $\nu^2=0.1$.}
\label{cgb4a}
\end{figure}

\begin{figure}[t]
 \centerline{\includegraphics[height=8cm,width=10cm]{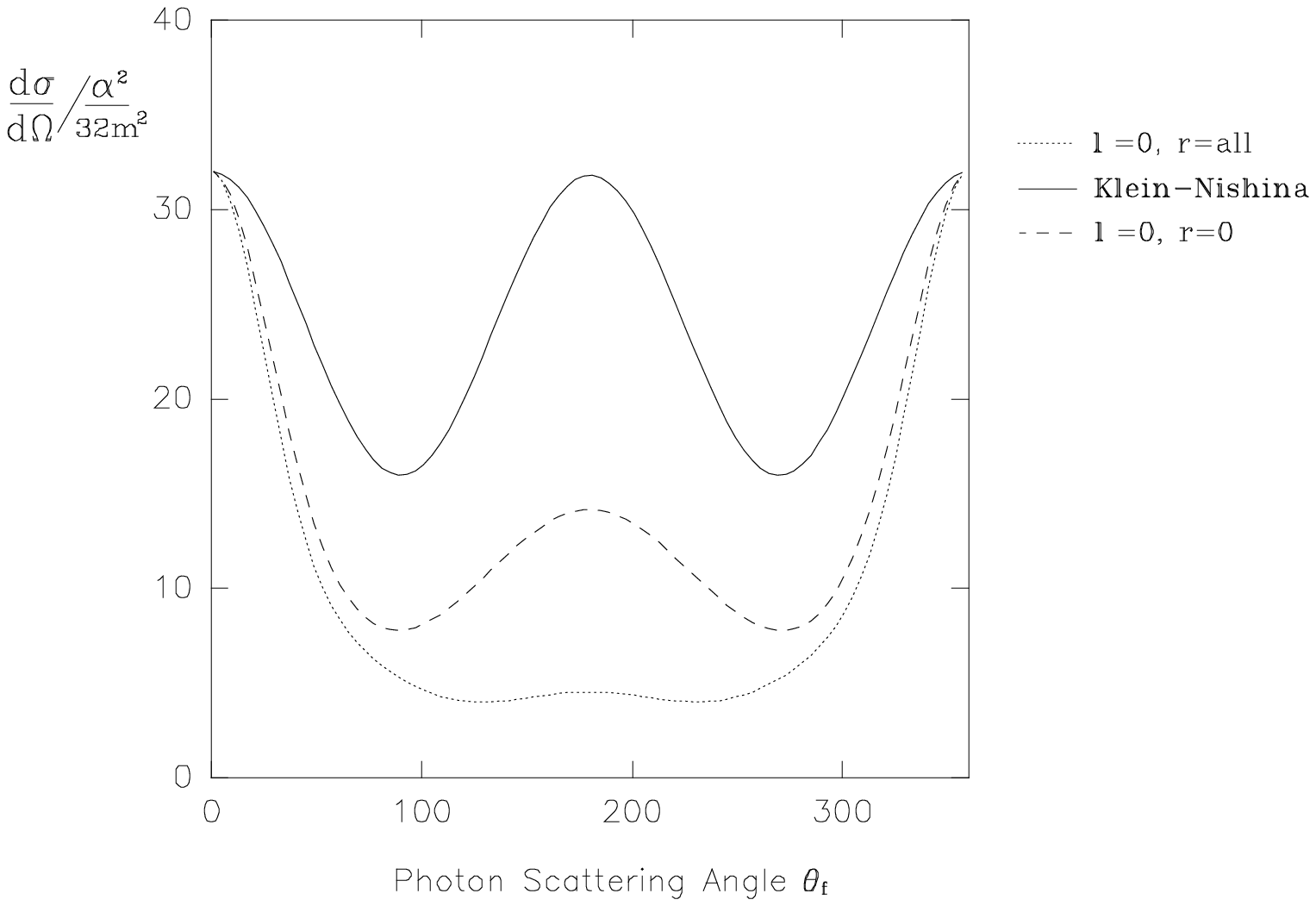}}
\caption{\bf\bm Comparison of The STPPP $l=0,r=0$, $l=0,r=\text{all}$ and Klein-Nishina
differential cross section vs $\theta_f$ for
$\,\omega=0.512$ keV, $\omega_i=0.768$ keV, $\theta_i=0^{\circ}$,
$\varphi_f=0^{\circ}$ and $\nu^2=0.1$.}
\label{cgb4b}
\end{figure}

\clearpage

\begin{figure}[t]
 \centerline{\includegraphics[height=8cm,width=15cm]{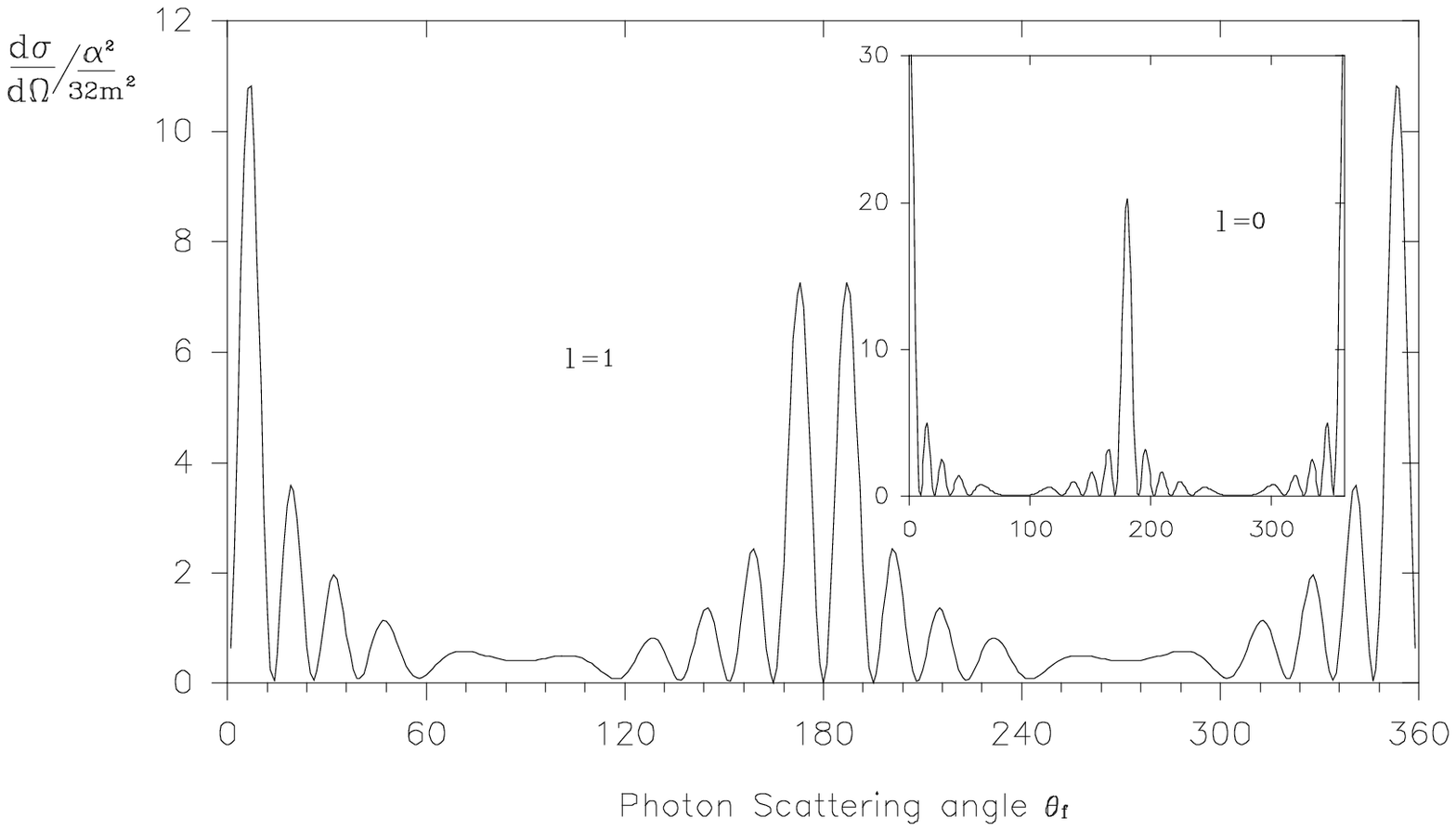}}
\caption{\bf\bm The SCS differential cross section vs $\theta_f$ for
$\,\omega=0.512$ keV, $\omega_i=2.559$ keV, $\theta_i=0^{\circ}$,
$\varphi_f=0^{\circ}$, $\nu^2=0.1$ and various $l$.}
\label{cgb5}
\end{figure}

\begin{figure}[t]
 \centerline{\includegraphics[height=8cm,width=10cm]{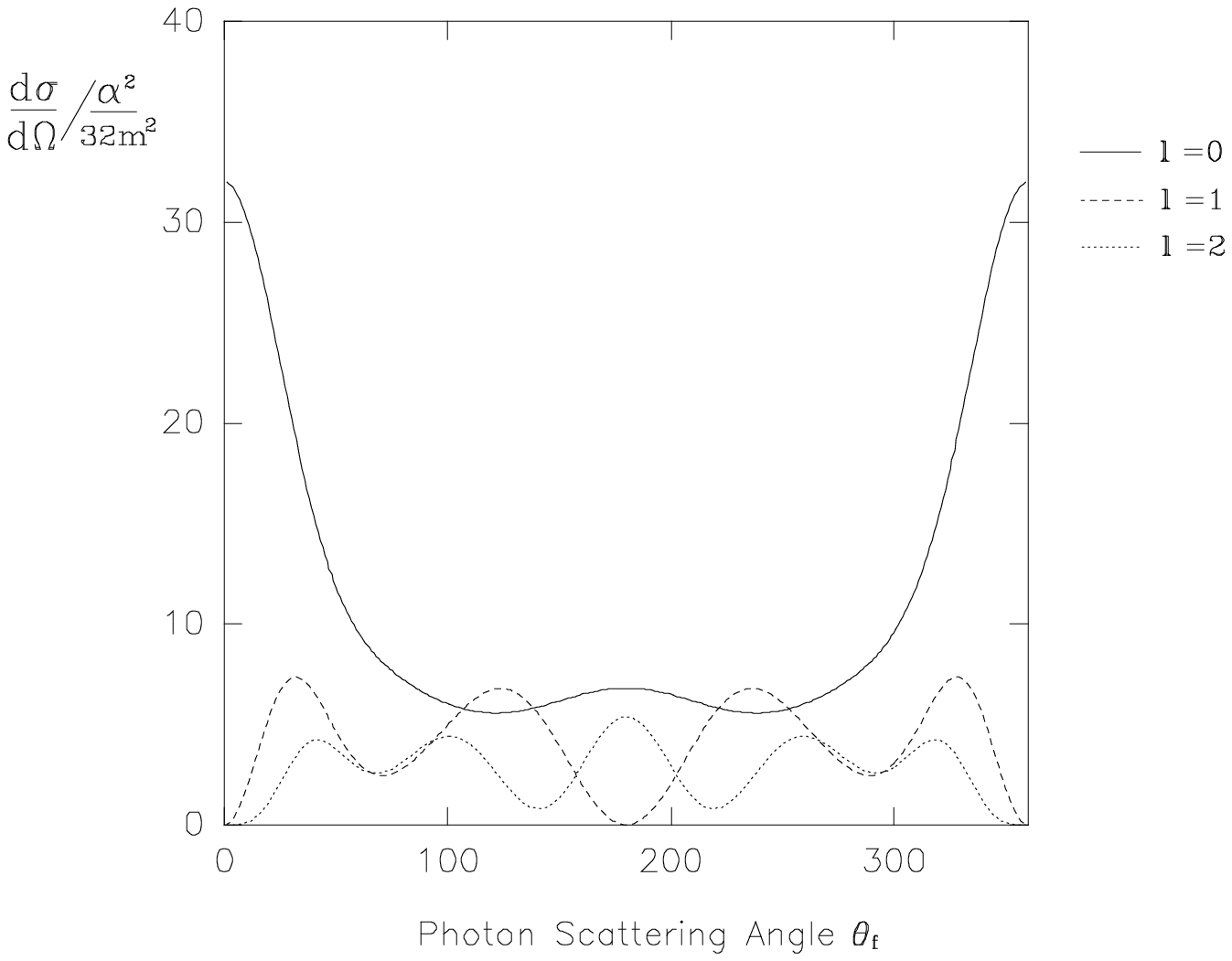}}
\caption{\bf\bm The SCS differential cross section vs $\theta_f$ for
$\,\omega=0.512$ keV, $\omega_i=0.768$ keV, $\theta_i=0^{\circ}$,
$\varphi_f=0^{\circ}$, $\nu^2=0.5$ and various $l$.}
\label{cgb6}
\end{figure}

\begin{figure}[t]
 \centerline{\includegraphics[height=8cm,width=10cm]{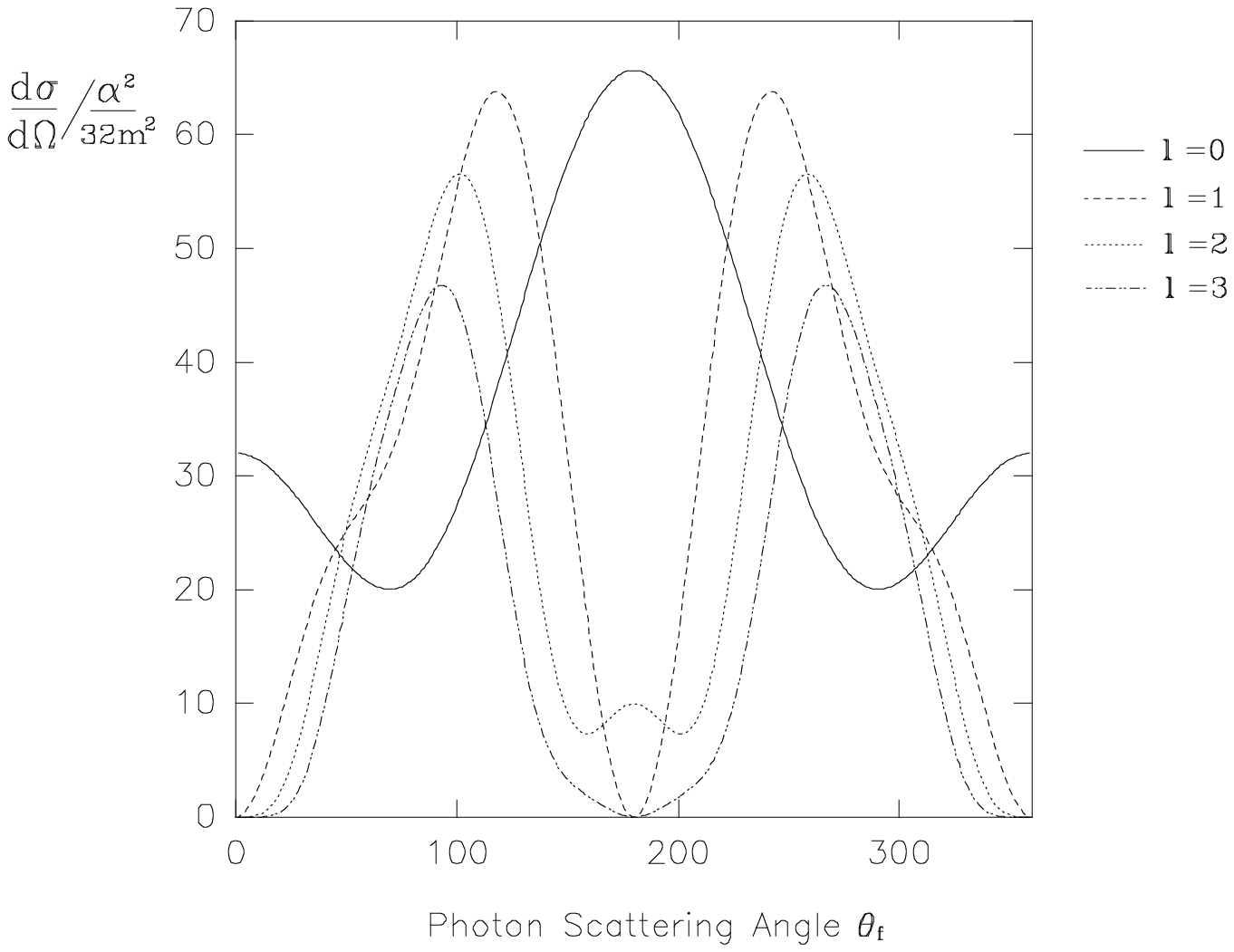}}
\caption{\bf\bm The SCS differential cross section vs $\theta_f$ for
$\,\omega=0.512$ keV, $\omega_i=0.409$ keV, $\theta_i=0^{\circ}$,
$\varphi_f=0^{\circ}$, $\nu^2=0.5$ and various $l$.}
\label{cgb7}
\end{figure}

\begin{figure}[t]
 \centerline{\includegraphics[height=8cm,width=10cm]{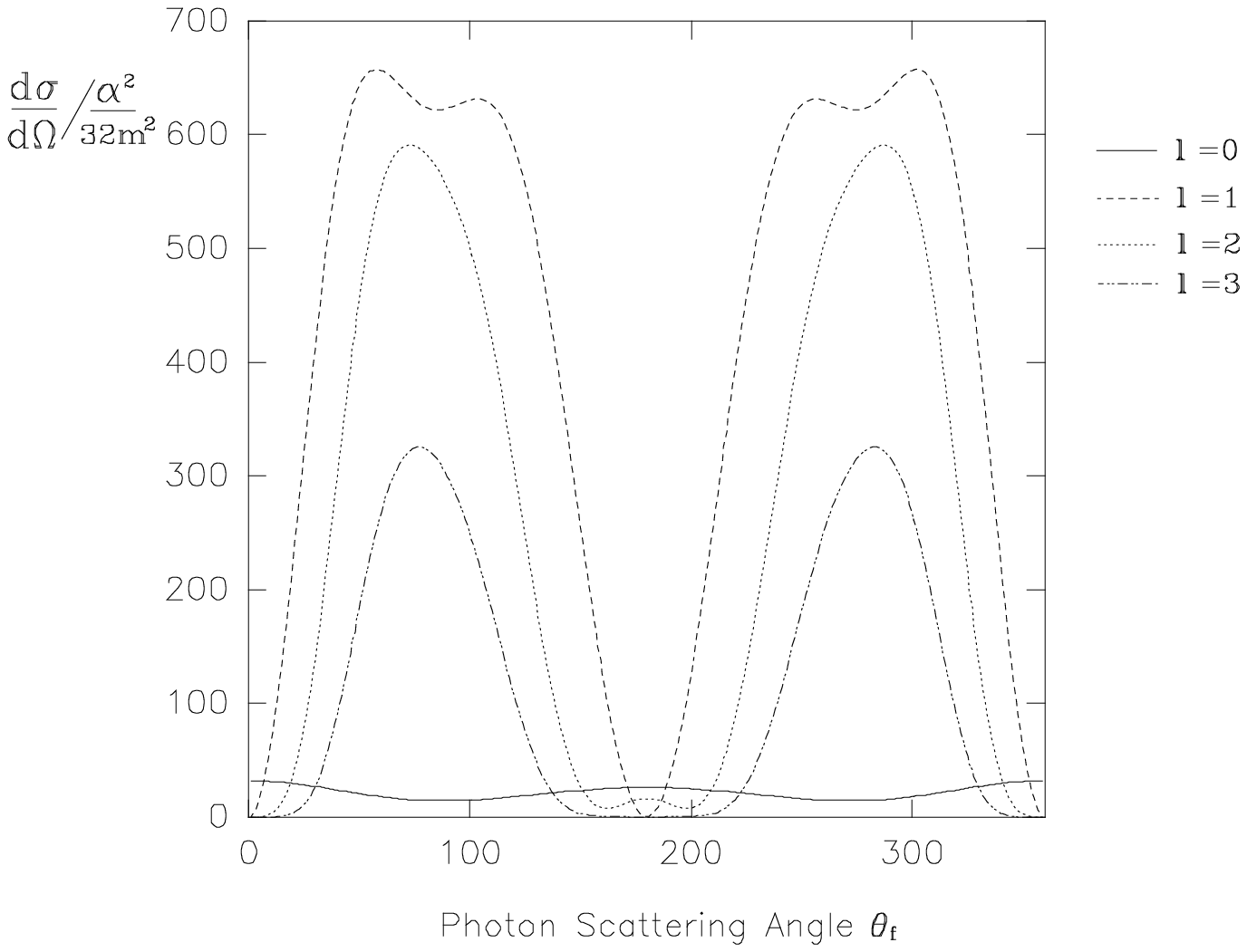}}
\caption{\bf\bm The SCS differential cross section vs $\theta_f$ for
$\,\omega=51.2$ keV, $\omega_i=5.12$ keV, $\theta_i=0^{\circ}$,
$\varphi_f=0^{\circ}$, $\nu^2=0.2$ and various $l$.}
\label{cgb8}
\end{figure}

\clearpage

\begin{figure}[t]
 \centerline{\includegraphics[height=8cm,width=10cm]{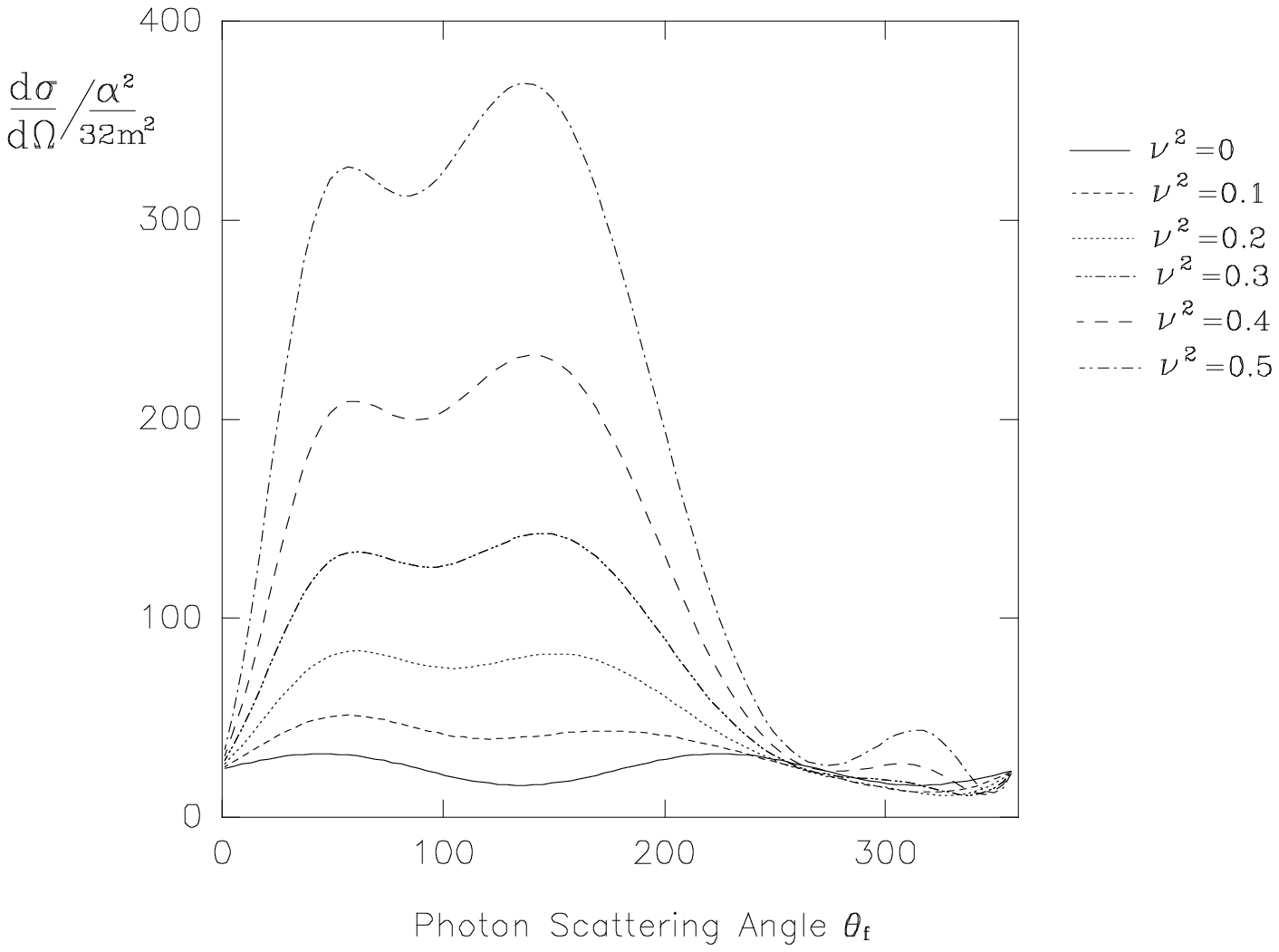}}
\caption{\bf\bm The SCS $l=0$ differential cross section vs $\theta_f$ for
$\,\omega=0.512$ keV, $\omega_i=0.409$ keV, $\theta_i=45^{\circ}$,
$\varphi_f=0^{\circ}$ and various $\nu^2$.}
\label{cgb9}
\end{figure}

\begin{figure}[t]
 \centerline{\includegraphics[height=8cm,width=10cm]{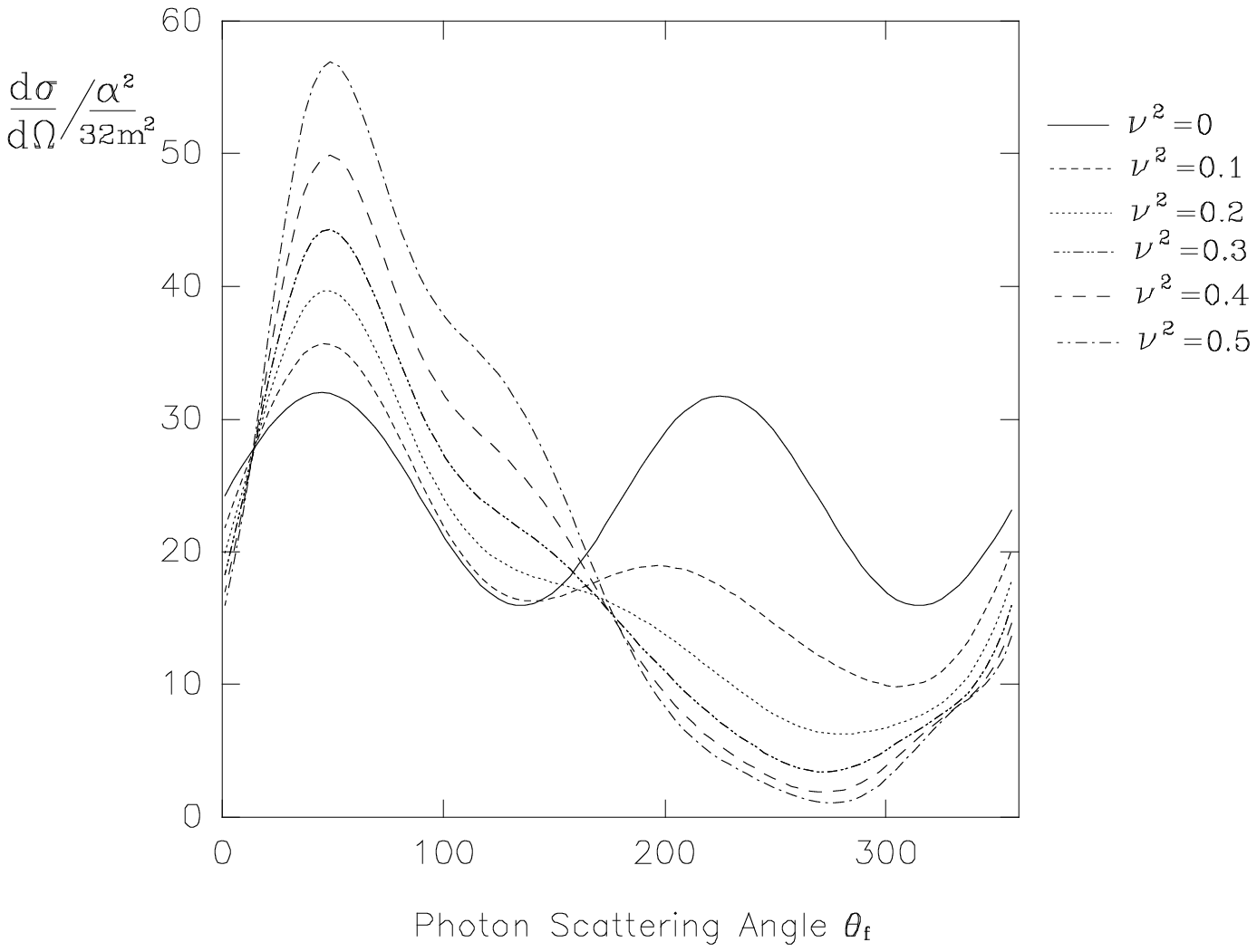}}
\caption{\bf\bm The SCS $l=0$ differential cross section vs $\theta_f$ for
$\,\omega=0.512$ keV, $\omega_i=0.768$ keV, $\theta_i=45^{\circ}$,
$\varphi_f=0^{\circ}$ and various $\nu^2$.}
\label{cgb10}
\end{figure}

\begin{figure}[t]
 \centerline{\includegraphics[height=8cm,width=10cm]{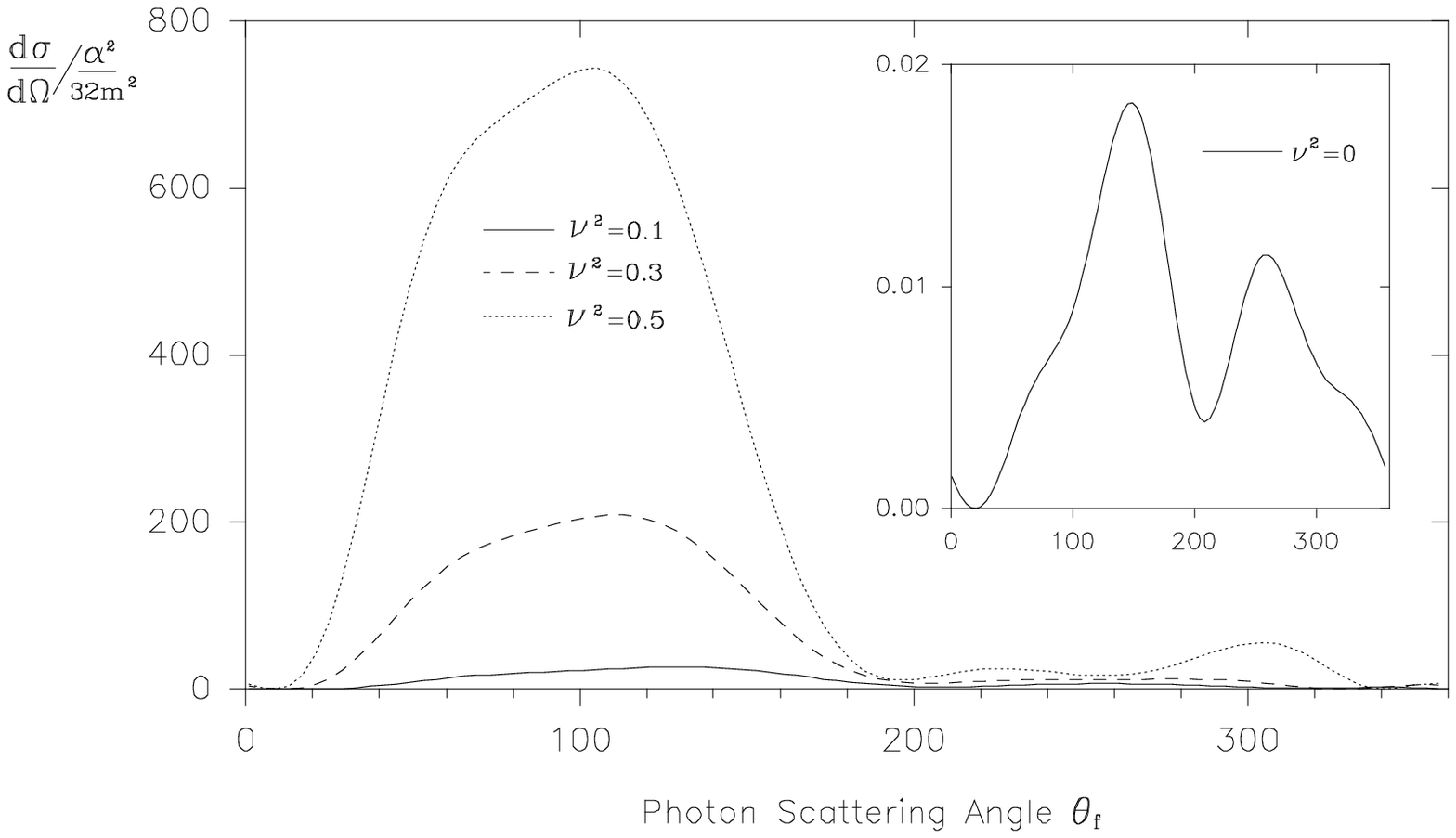}}
\caption{\bf\bm The SCS $l=1$ differential cross section vs $\theta_f$ for
$\,\omega=0.512$ keV, $\omega_i=0.409$ keV, $\theta_i=45^{\circ}$,
$\varphi_f=0^{\circ}$ and various $\nu^2$.}
\label{cgb11}
\end{figure}

\begin{figure}[t]
 \centerline{\includegraphics[height=8cm,width=10cm]{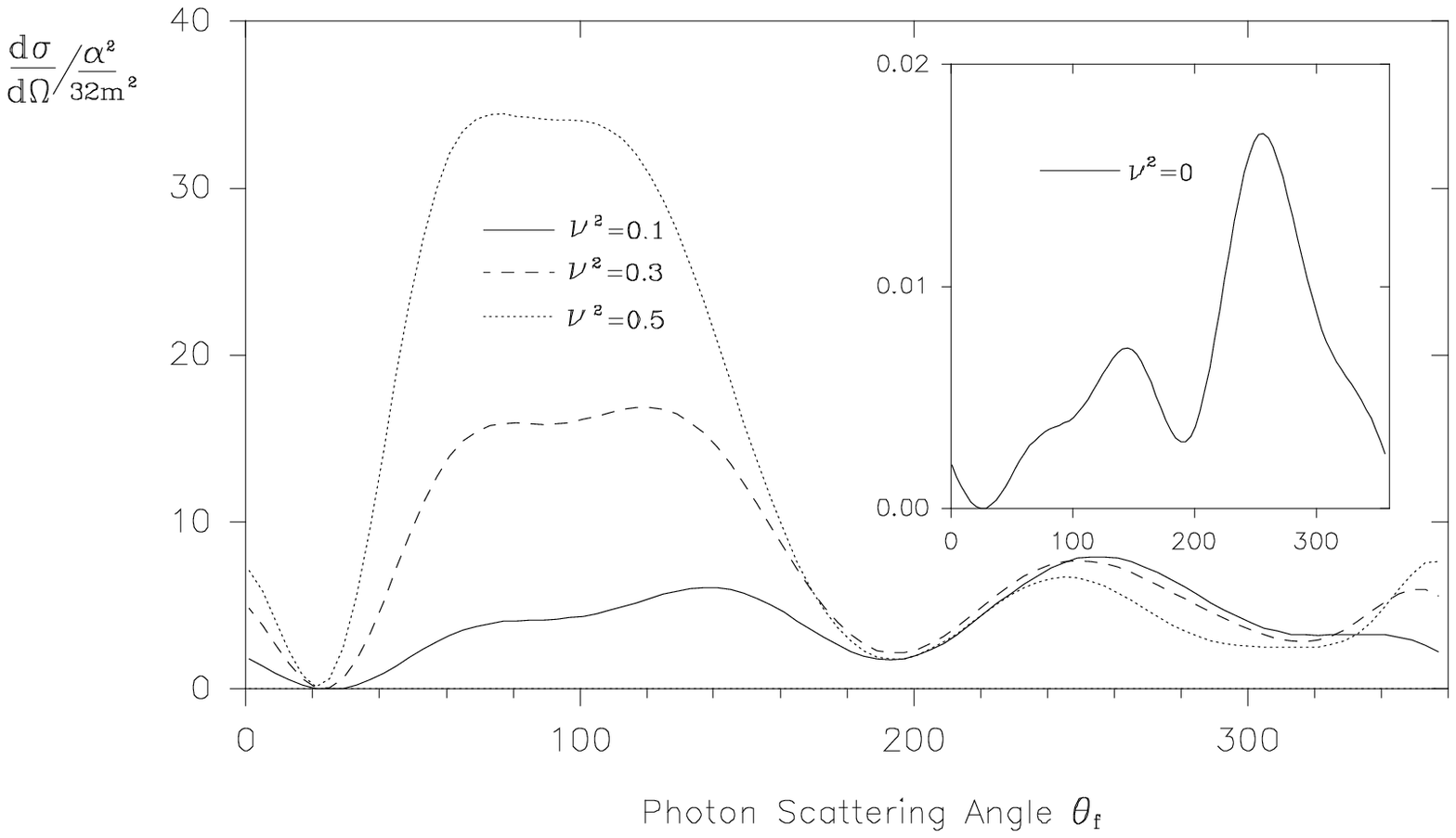}}
\caption{\bf\bm The SCS $l=0$ differential cross section vs $\theta_f$ for
$\,\omega=0.512$ keV, $\omega_i=0.768$ keV, $\theta_i=45^{\circ}$,
$\varphi_f=0^{\circ}$ and various $\nu^2$.}
\label{cgb12}
\end{figure}

\begin{figure}[t]
 \centerline{\includegraphics[height=8cm,width=10cm]{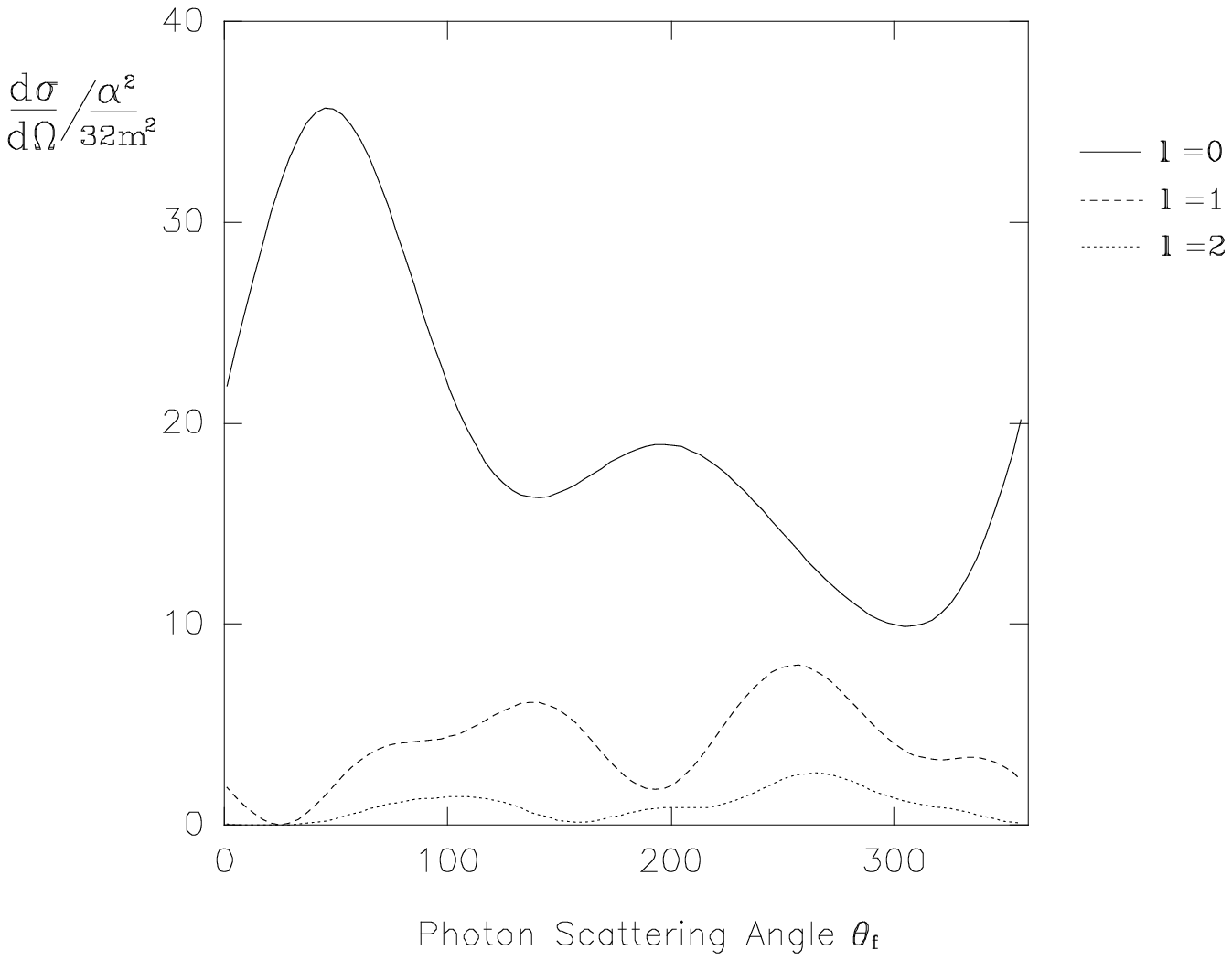}}
\caption{\bf\bm The SCS differential cross section vs $\theta_f$ for
$\,\omega=0.512$ keV, $\omega_i=0.768$ keV, $\theta_i=45^{\circ}$,
$\varphi_f=0^{\circ}$, $\nu^2=0.1$ and various $l$.}
\label{cgb13}
\end{figure}

\begin{figure}[t]
 \centerline{\includegraphics[height=8cm,width=10cm]{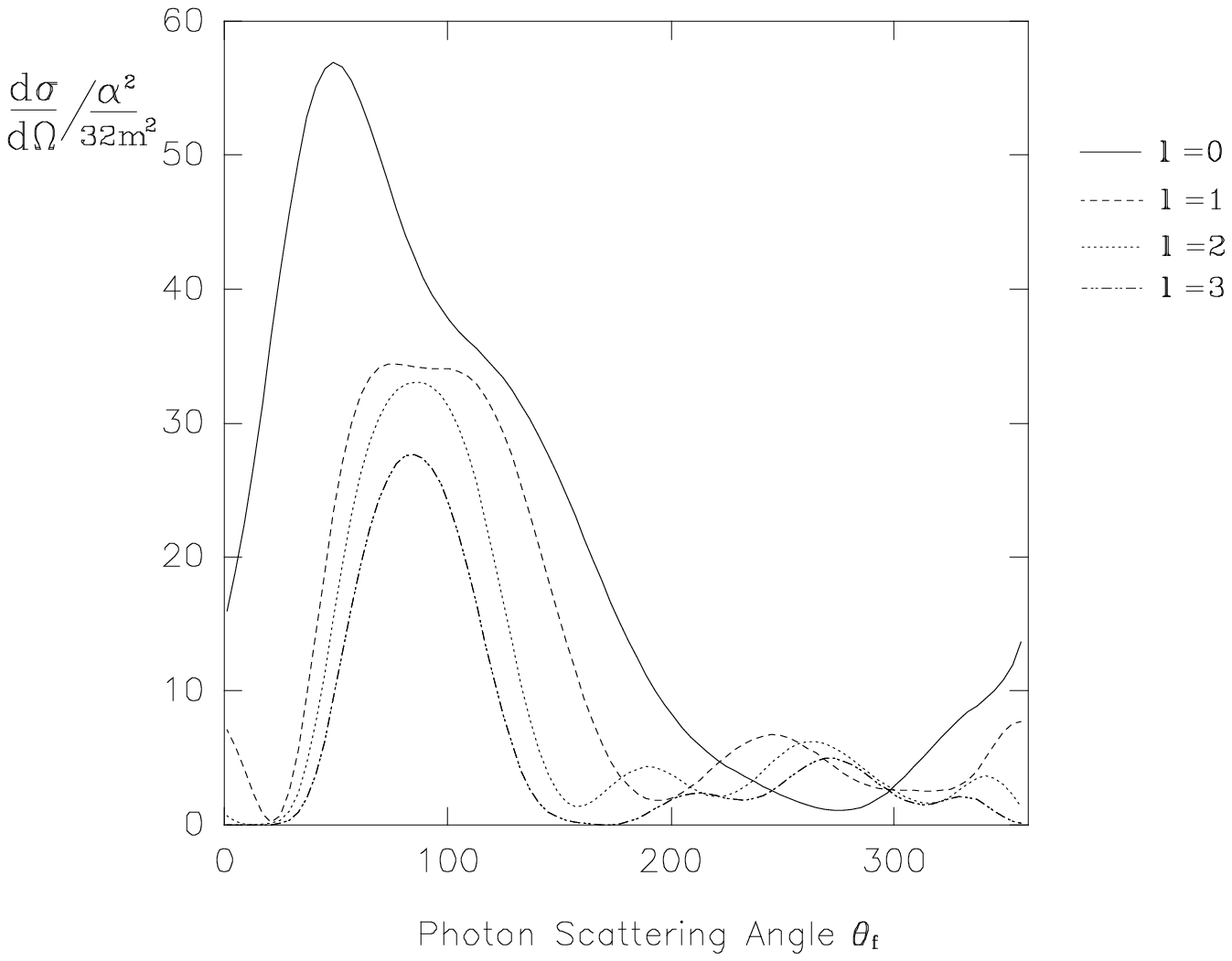}}
\caption{\bf\bm The SCS differential cross section vs $\theta_f$ for
$\,\omega=0.512$ keV, $\omega_i=0.768$ keV, $\theta_i=45^{\circ}$,
$\varphi_f=0^{\circ}$, $\nu^2=0.5$ and various $l$.}
\label{cgb14}
\end{figure}

\clearpage

\begin{figure}[t]
 \centerline{\includegraphics[height=8cm,width=10cm]{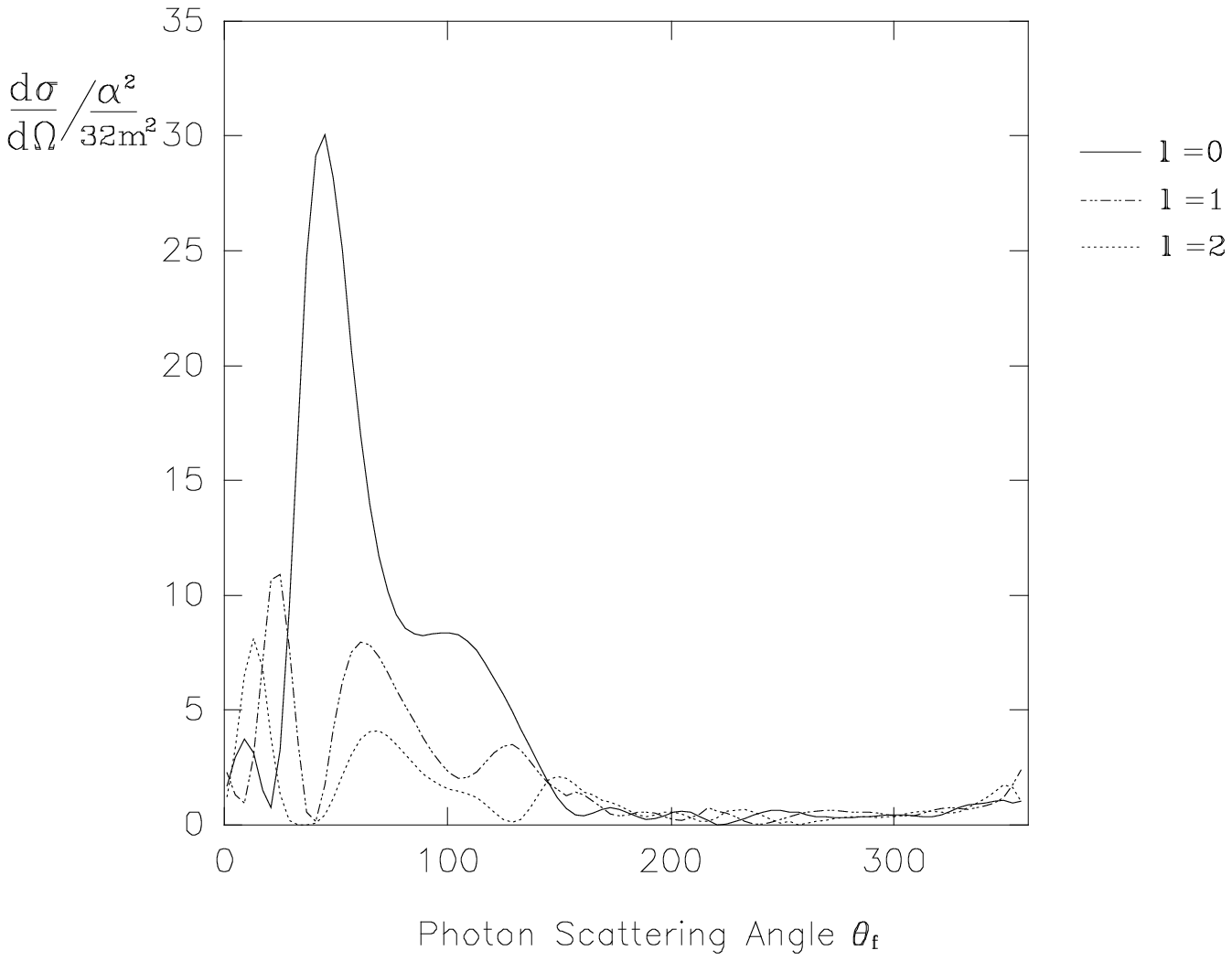}}
\caption{\bf\bm The SCS differential cross section vs $\theta_f$ for
$\,\omega=0.512$ keV, $\omega_i=5.12$ keV, $\theta_i=45^{\circ}$,
$\varphi_f=0^{\circ}$, $\nu^2=0.5$ and various $l$.}
\label{cgb15}
\end{figure}

\begin{figure}[t]
 \centerline{\includegraphics[height=8cm,width=10cm]{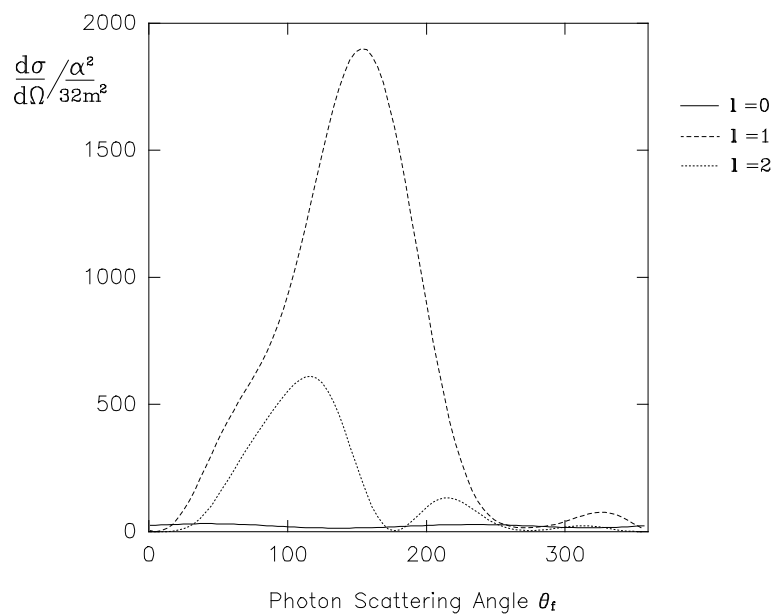}}
\caption{\bf\bm The SCS differential cross section vs $\theta_f$ for
$\,\omega=0.512$ keV, $\omega_i=0.051$ keV, $\theta_i=45^{\circ}$,
$\varphi_f=0^{\circ}$, $\nu^2=0.1$ and various $l$.}
\label{cgb16}
\end{figure}

\clearpage
\subsection{Differential Cross sections summed over all $l$}

Table 4.2 displays the parameter values of the SCS scattering process
investigated in this section. The parameters here have exactly the same
meaning as in Section 4.2.1. The integer parameter values $l$ and $r$ are
not present in this section due to the summations over these parameters
having been performed numerically.

\bigskip\bigskip\

\input{./tex/tables/c_table2}

\clearpage
\begin{figure}[H]
 \centerline{\includegraphics[height=8cm,width=10cm]{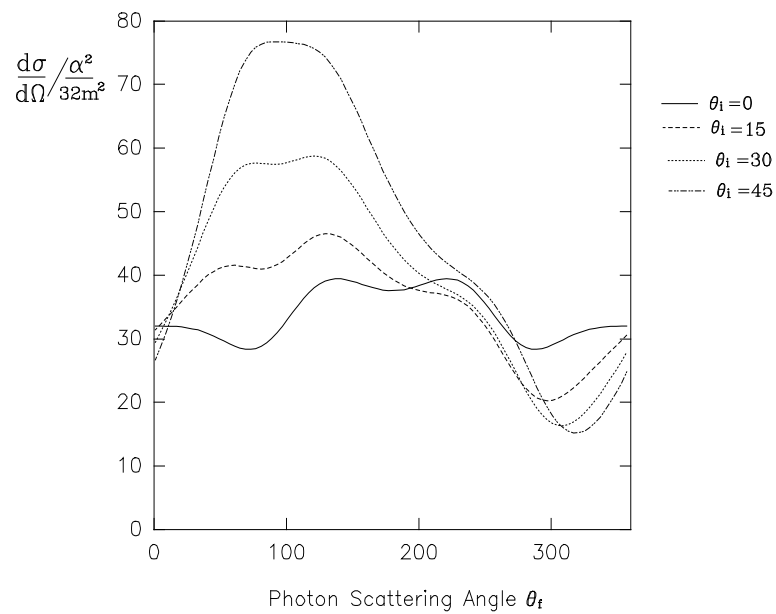}}
\caption{\bf\bm The SCS differential cross section vs $\theta_f$
for $\,\omega=0.512$ keV, $\omega_i=0.409$ keV, $\varphi_f=0^{\circ}$, $\nu^2=0.1$ and 
various $\theta_i$.}
\label{cg1}
\end{figure}

\begin{figure}[H]
 \centerline{\includegraphics[height=8cm,width=10cm]{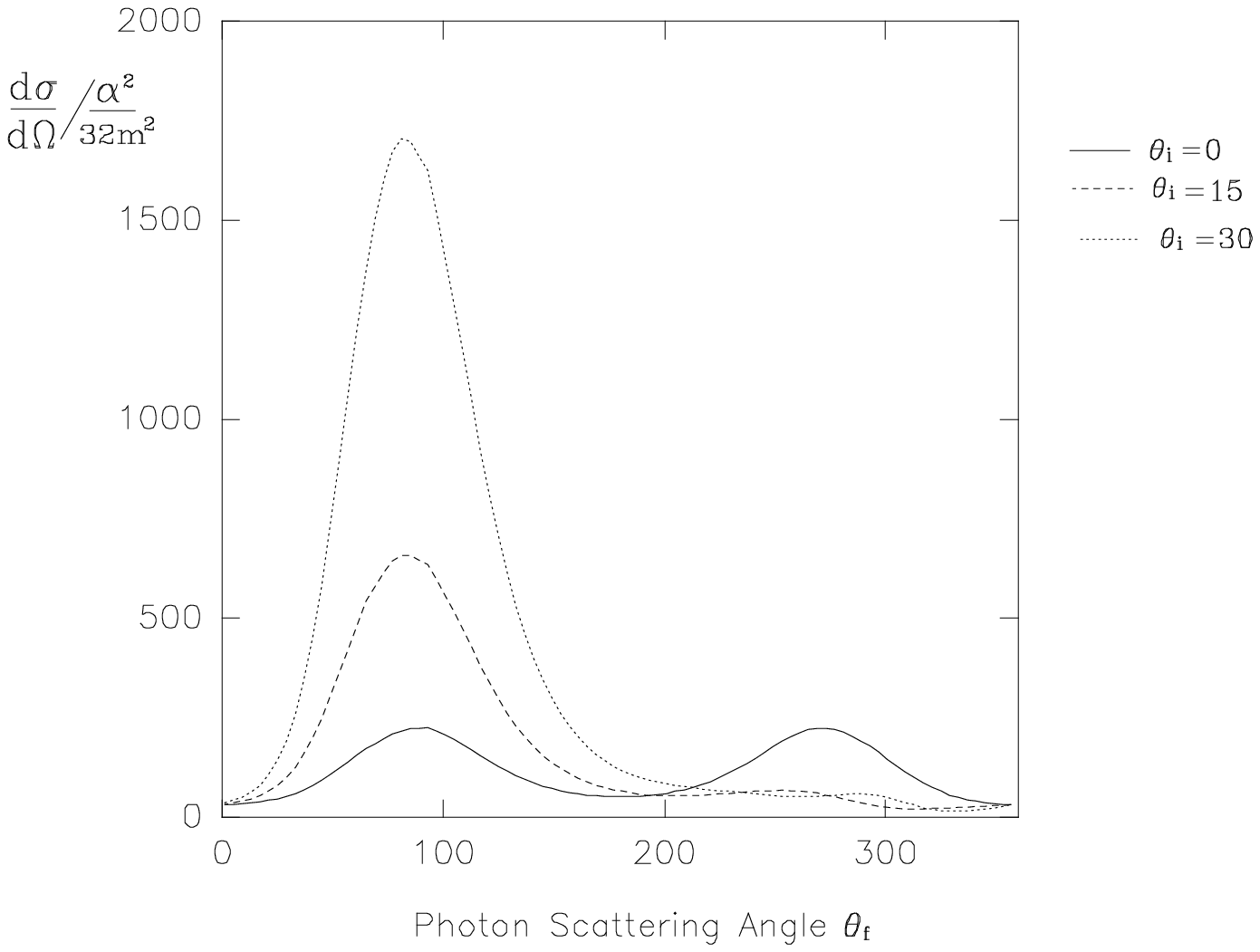}}
\caption{\bf\bm The SCS differential cross section vs $\theta_f$
for $\,\omega=0.512$ keV, $\omega_i=0.409$ keV, $\varphi_f=0^{\circ}$, $\nu^2=0.5$ and
various $\theta_i$.}
\label{cg2}
\end{figure}

\clearpage

\begin{figure}[H]
 \centerline{\includegraphics[height=8cm,width=10cm]{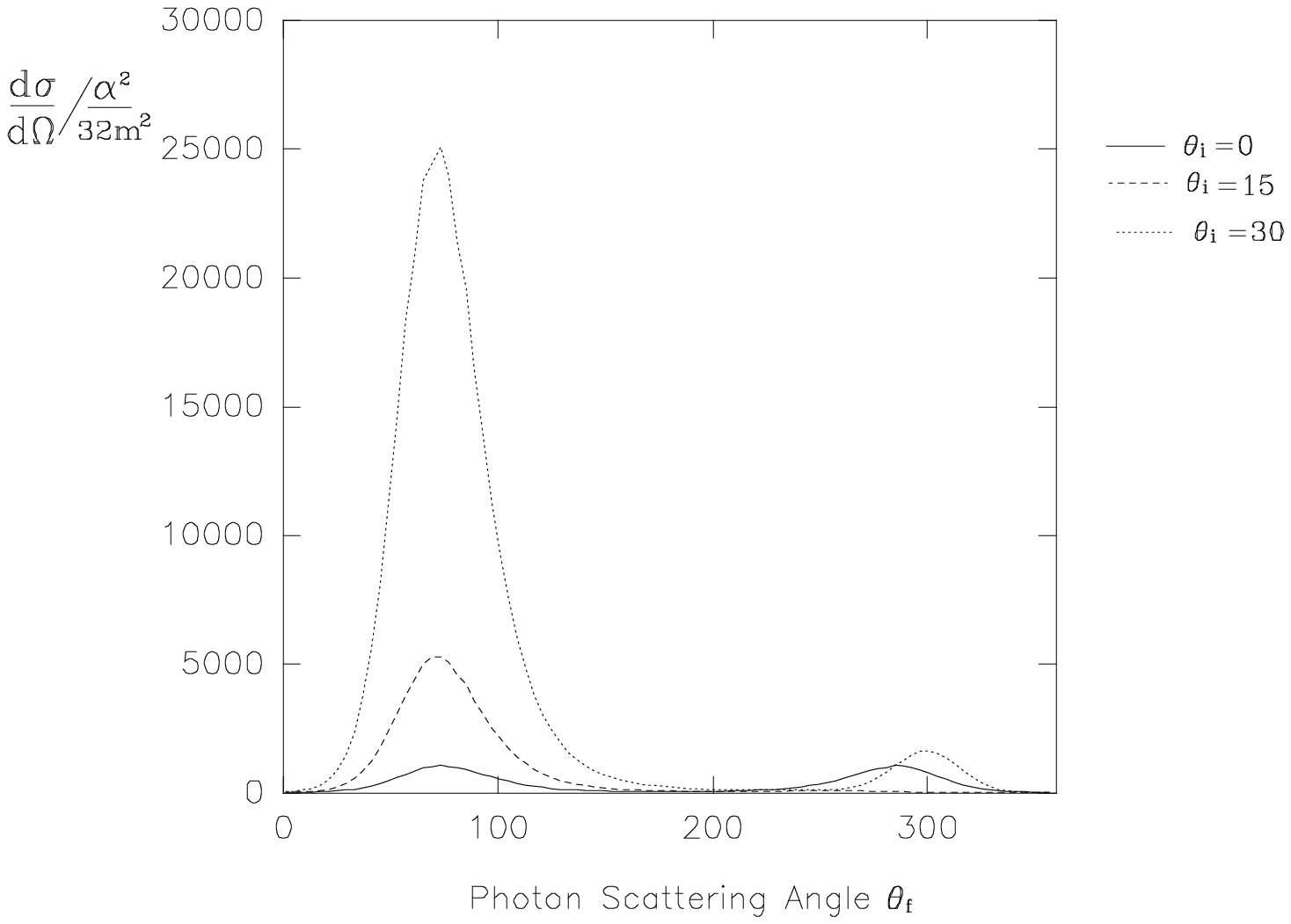}}
\caption{\bf\bm The SCS differential cross section vs $\theta_f$
for $\,\omega=0.512$ keV, $\omega_i=0.409$ keV, $\varphi_f=0^{\circ}$, $\nu^2=1$ and
various $\theta_i$.}
\label{cg3}
\end{figure}

\begin{figure}[H]
 \centerline{\includegraphics[height=8cm,width=10cm]{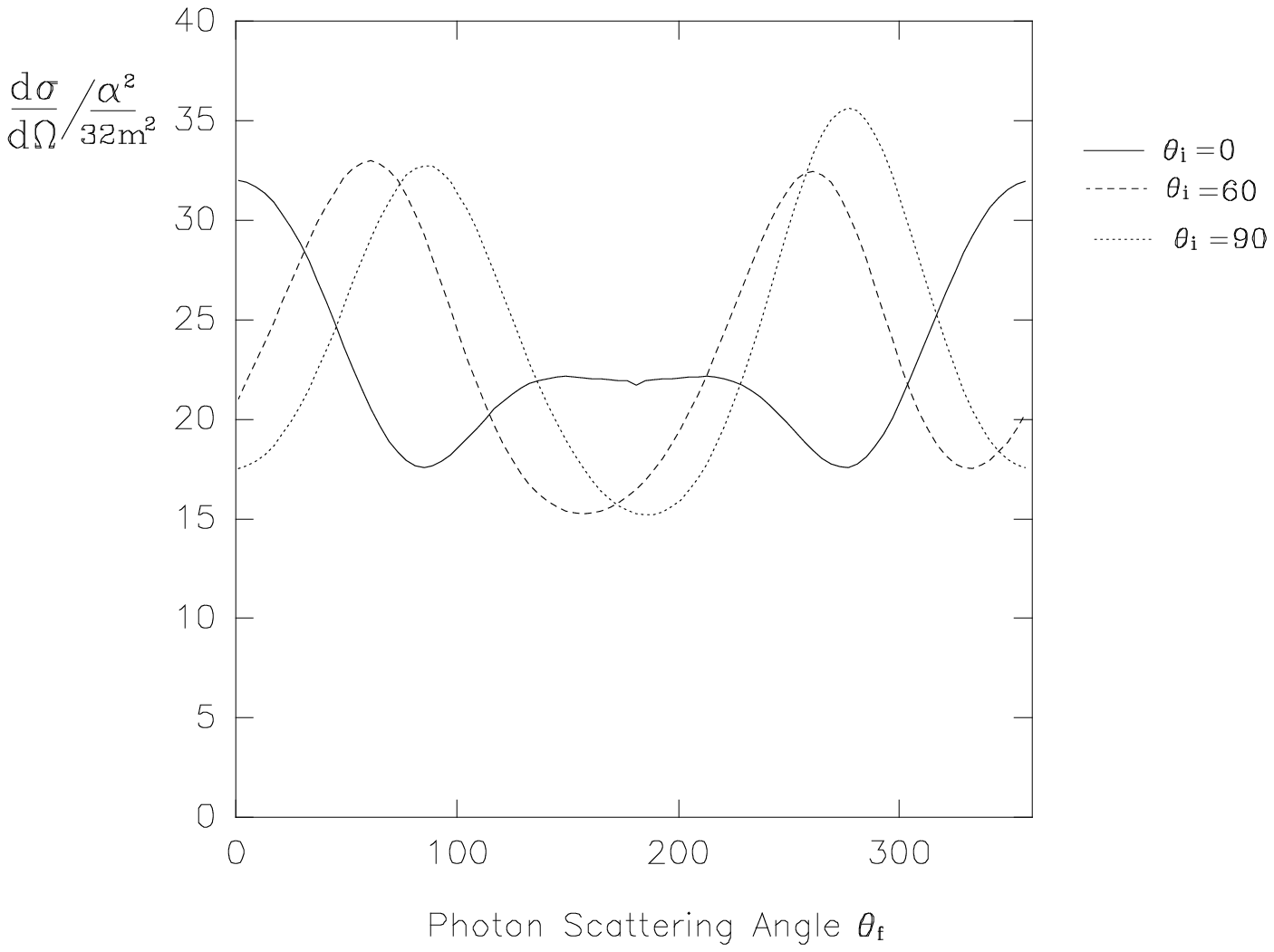}}
\caption{\bf\bm The SCS differential cross section vs $\theta_f$
for $\,\omega=0.005$ keV, $\omega_i=0.026$ keV, $\varphi_f=0^{\circ}$, $\nu^2=0.1$ and
various $\theta_i$.}
\label{cg4}
\end{figure}

\clearpage

\begin{figure}[H]
 \centerline{\includegraphics[height=8cm,width=10cm]{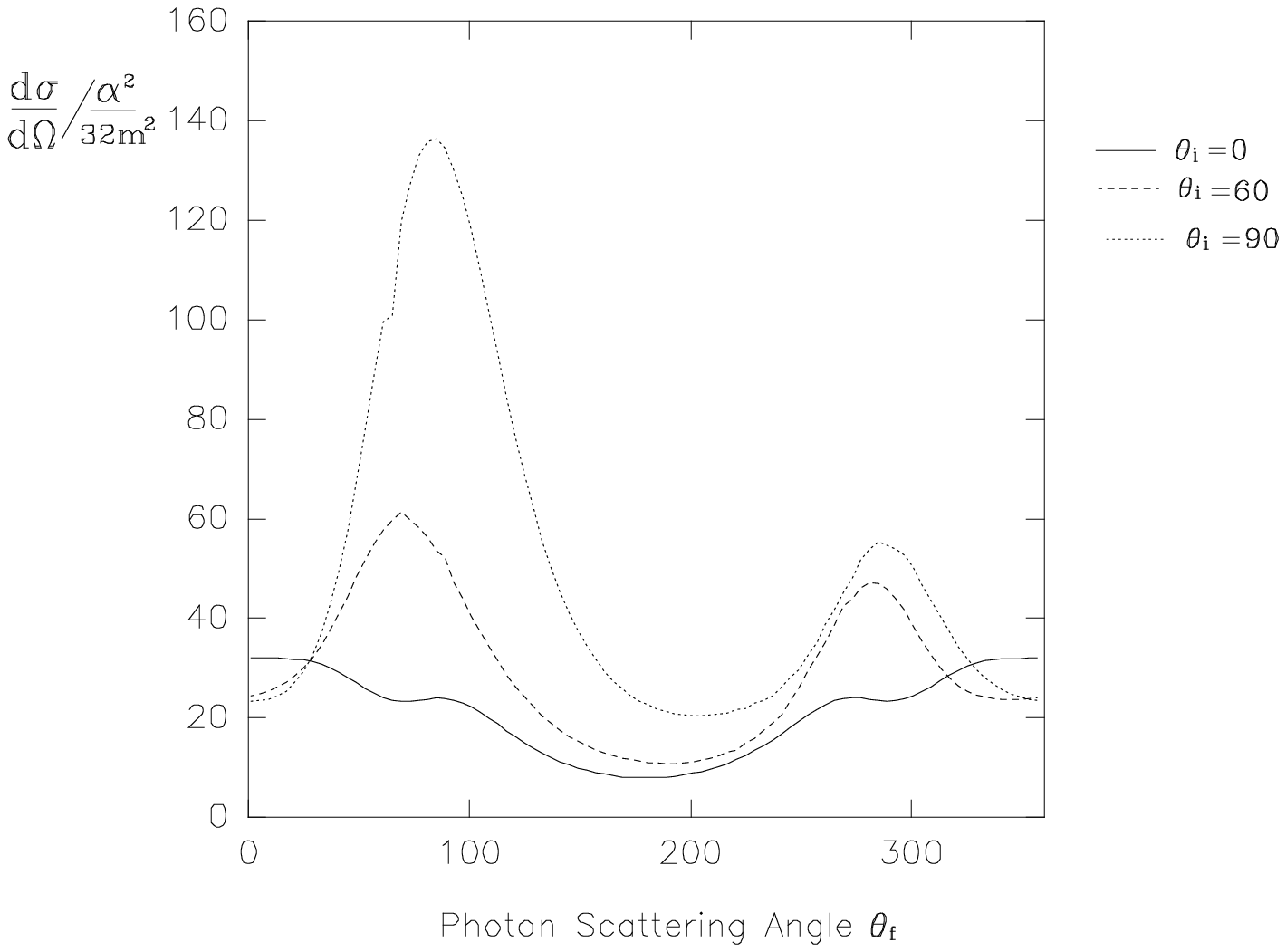}}
\caption{\bf\bm The SCS differential cross section vs $\theta_f$
for $\,\omega=0.005$ keV, $\omega_i=0.026$ keV, $\varphi_f=0^{\circ}$, $\nu^2=0.5$ and
various $\theta_i$.}
\label{cg5}
\end{figure}

\begin{figure}[H]
 \centerline{\includegraphics[height=8cm,width=10cm]{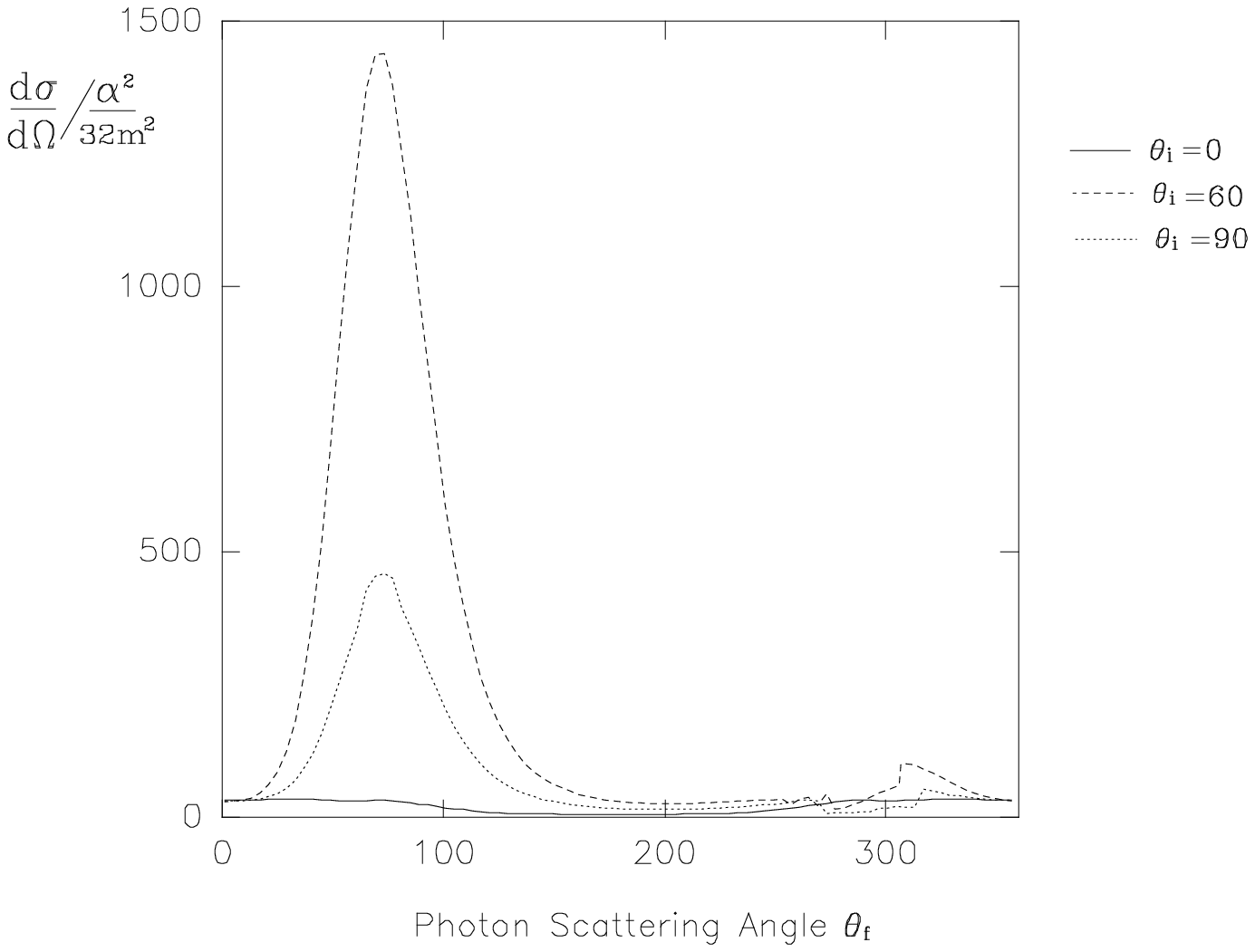}}
\caption{\bf\bm The SCS differential cross section vs $\theta_f$
for $\,\omega=0.005$ keV, $\omega_i=0.026$ keV, $\varphi_f=0^{\circ}$, $\nu^2=1$ and
various $\theta_i$.}
\label{cg6}
\end{figure}

\clearpage

\begin{figure}[H]
 \centerline{\includegraphics[height=8cm,width=10cm]{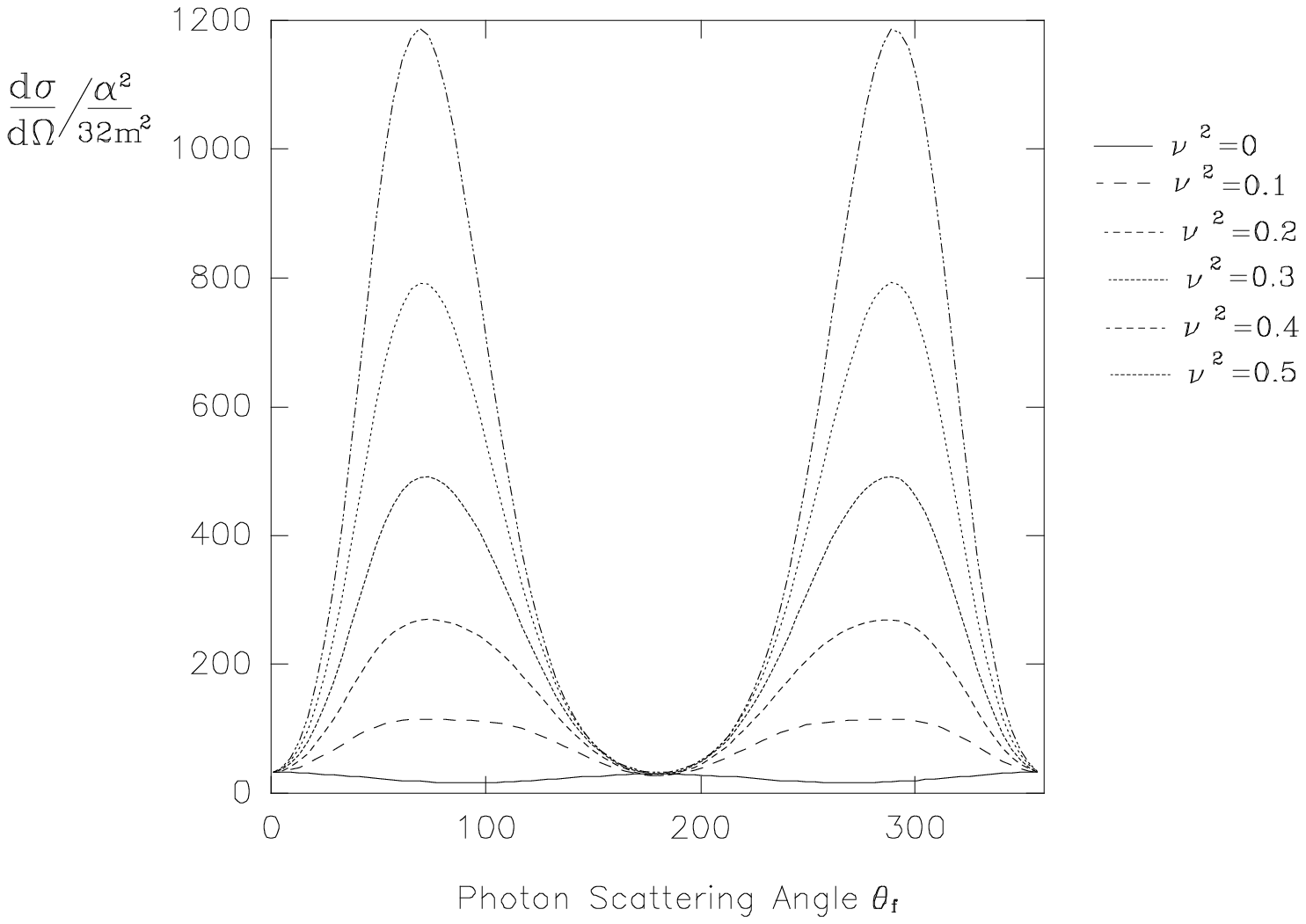}}
\caption{\bf\bm The SCS differential cross section vs $\theta_f$
for $\,\omega=51.2$ keV, $\omega_i=10.2$ keV, $\theta_i=0^{\circ}$, $\varphi_f=0^{\circ}$, 
and various $\nu^2$.}
\label{cg7}
\end{figure}

\begin{figure}[H]
 \centerline{\includegraphics[height=8cm,width=10cm]{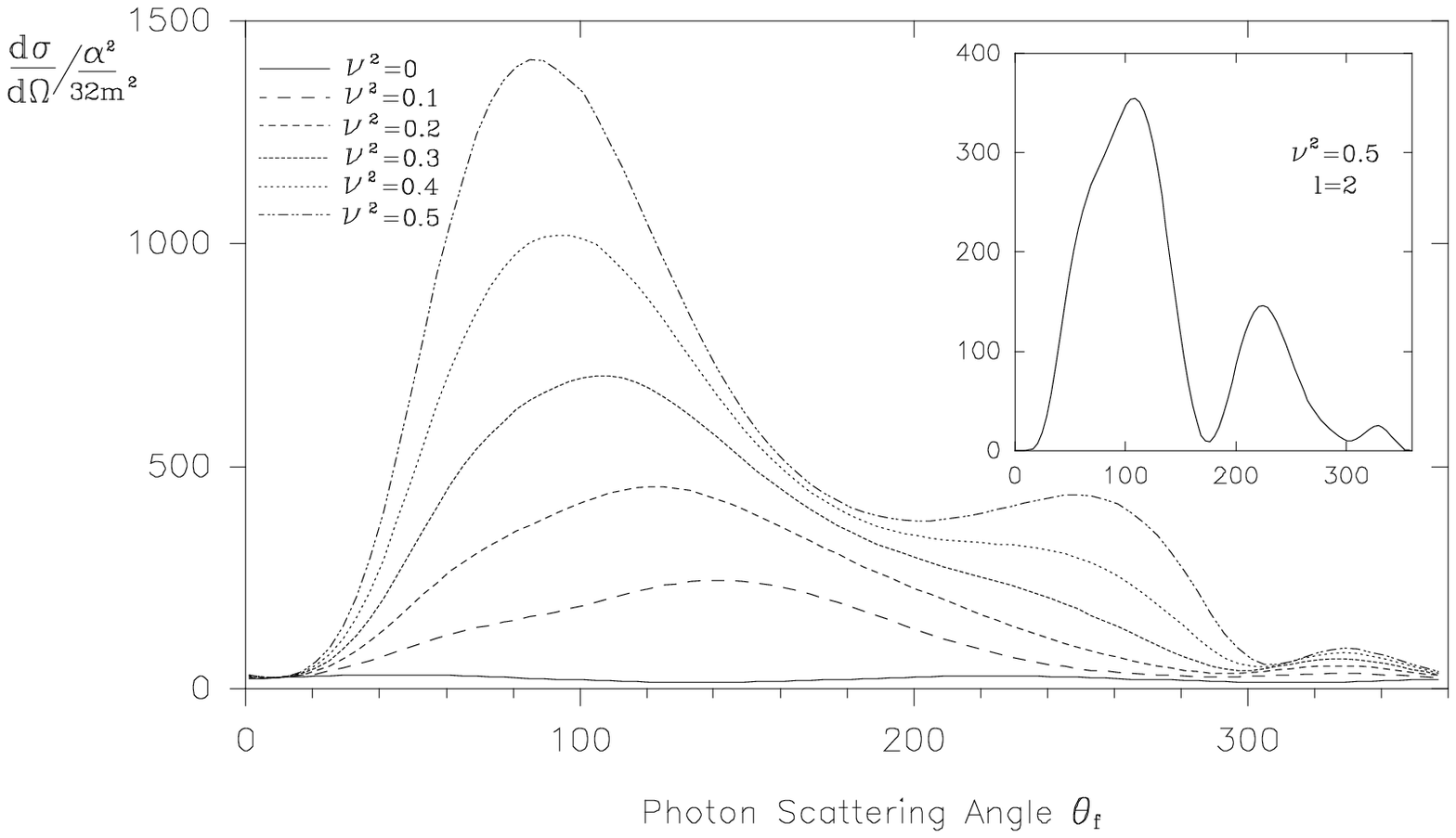}}
\caption{\bf\bm The SCS differential cross section vs $\theta_f$
for $\,\omega=51.2$ keV, $\omega_i=10.2$ keV, $\theta_i=45^{\circ}$, $\varphi_f=0^{\circ}$,
and various $\nu^2$.}
\label{cg8}
\end{figure}

\clearpage

\begin{figure}[H]
 \centerline{\includegraphics[height=8cm,width=10cm]{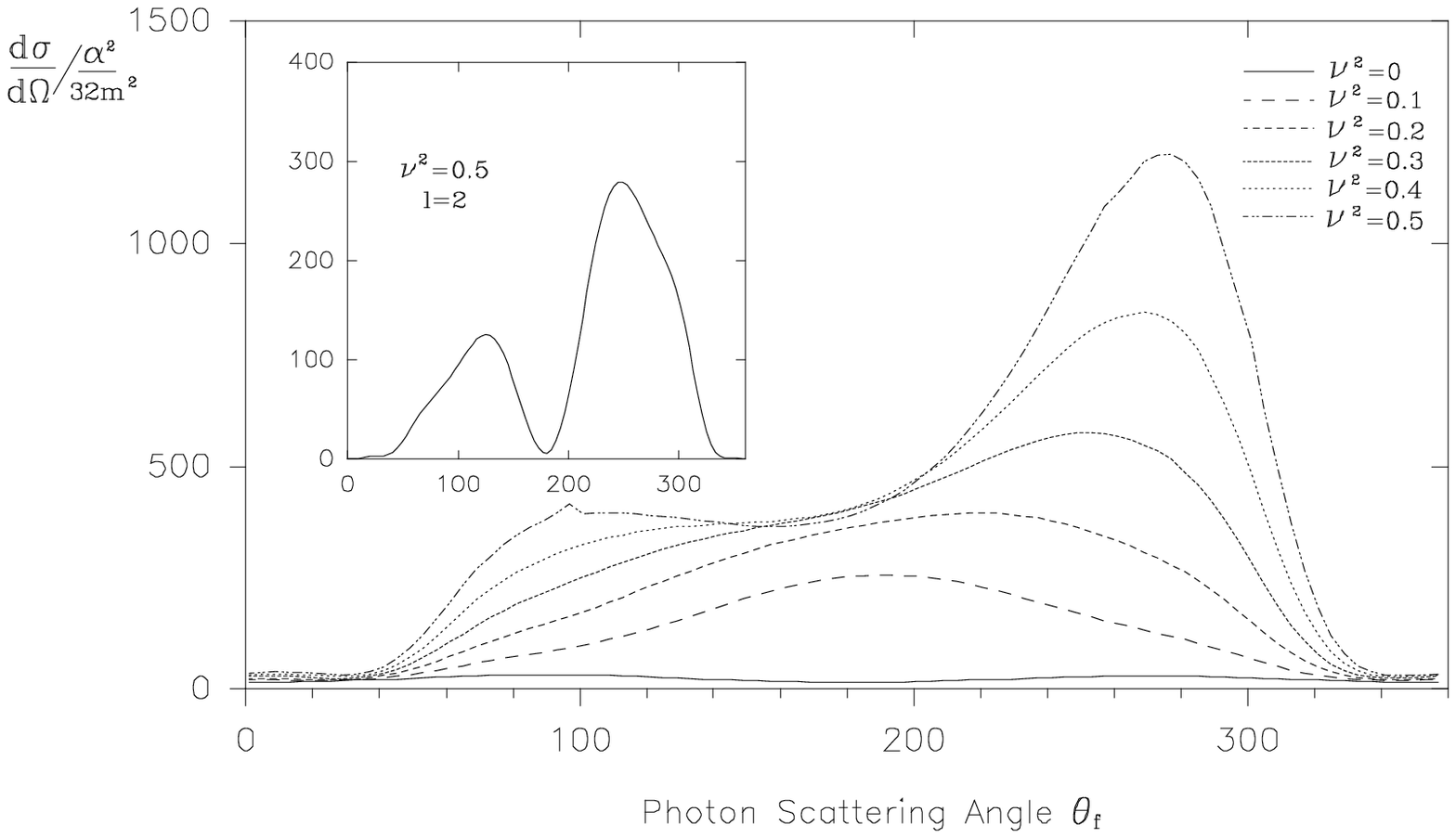}}
\caption{\bf\bm The SCS differential cross section vs $\theta_f$
for $\,\omega=51.2$ keV, $\omega_i=10.2$ keV, $\theta_i=90^{\circ}$, $\varphi_f=0^{\circ}$,
and various $\nu^2$.}
\label{cg9}
\end{figure}

\begin{figure}[H]
 \centerline{\includegraphics[height=8cm,width=10cm]{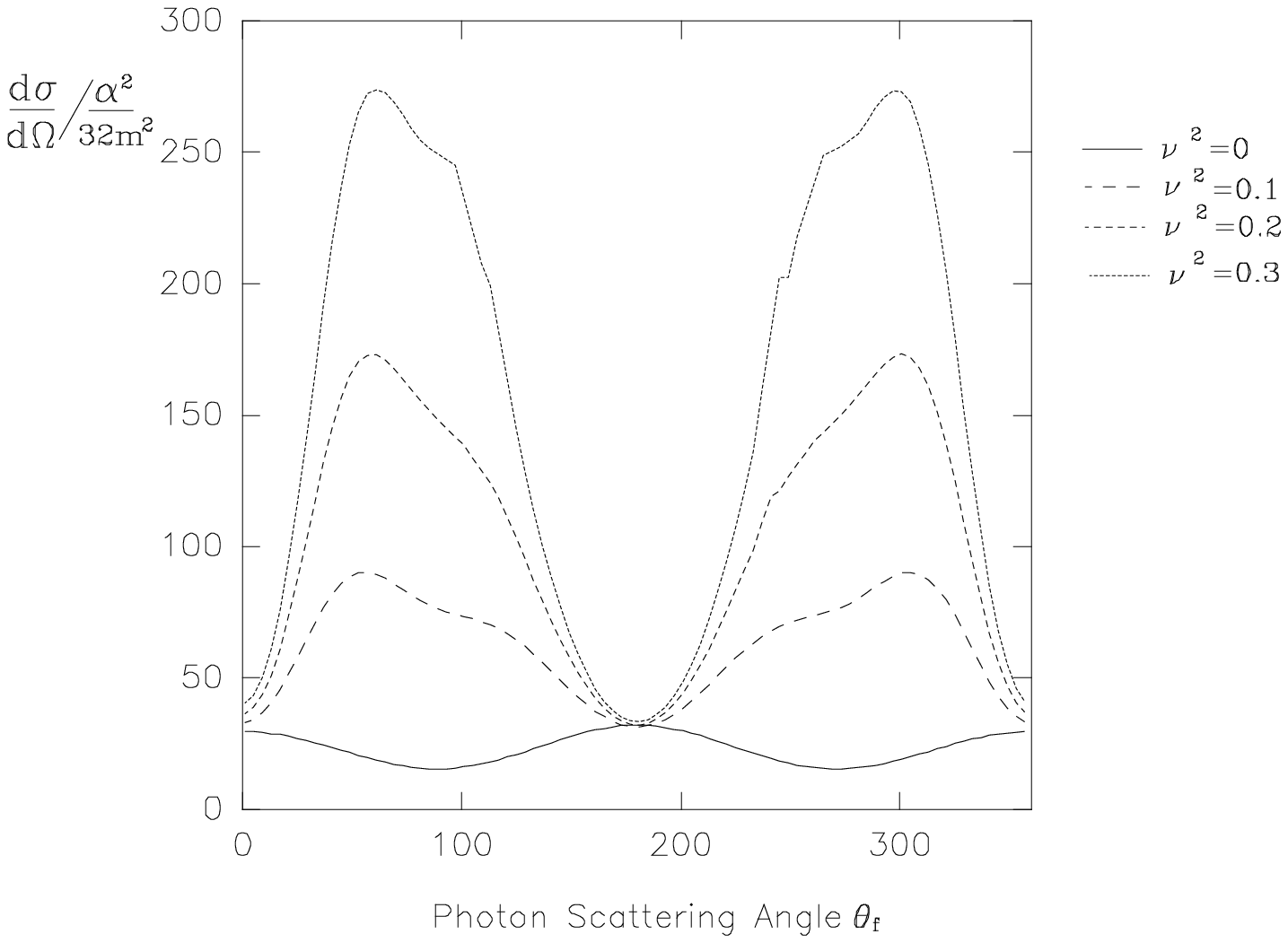}}
\caption{\bf\bm The SCS differential cross section vs $\theta_f$
for $\,\omega=51.2$ keV, $\omega_i=10.2$ keV, $\theta_i=180^{\circ}$, $\varphi_f=0^{\circ}$,
and various $\nu^2$.}
\label{cg10}
\end{figure}

\clearpage

\begin{figure}[H]
 \centerline{\includegraphics[height=8cm,width=10cm]{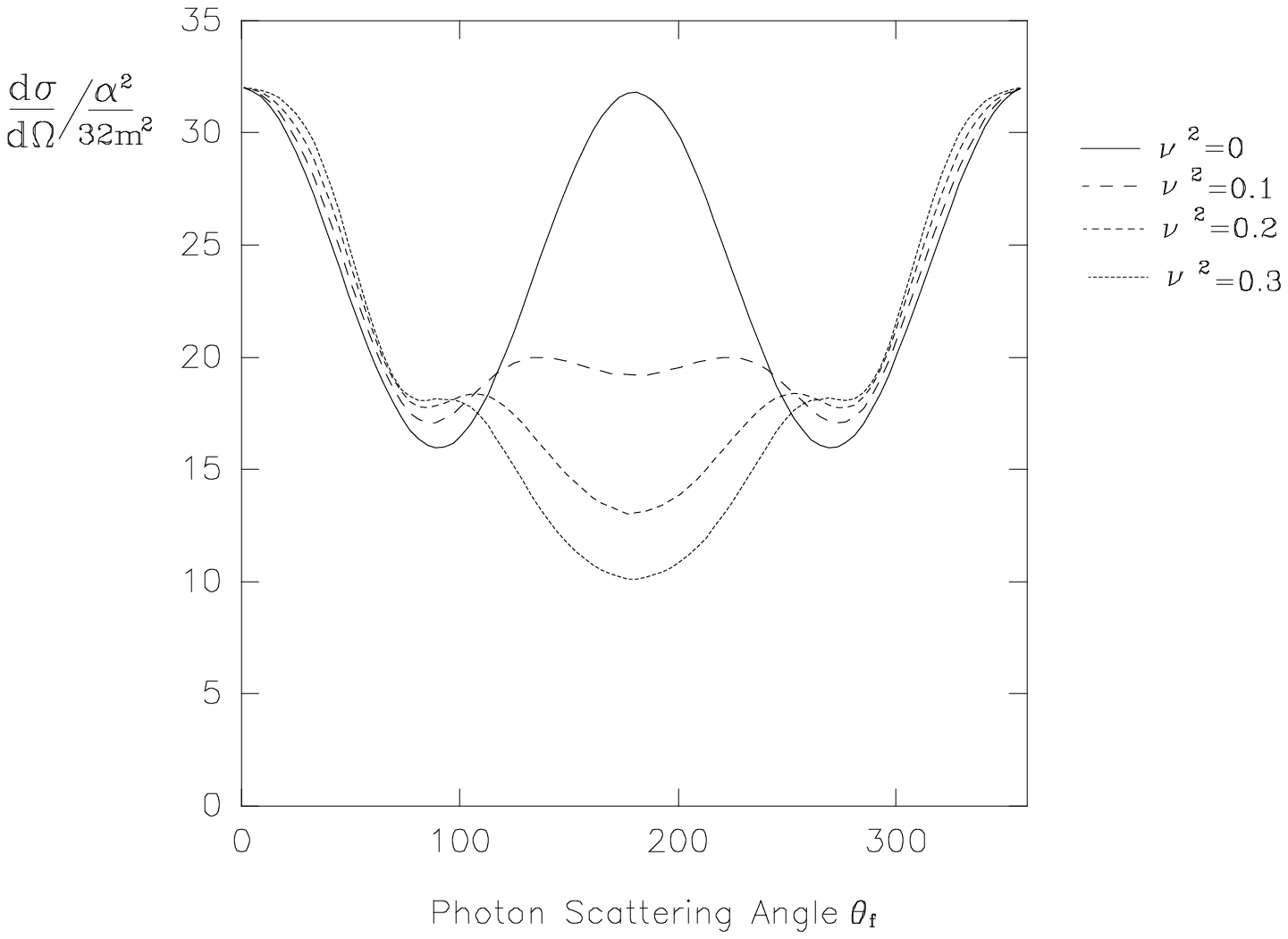}}
\caption{\bf\bm The SCS differential cross section vs $\theta_f$
for $\,\omega=0.512$ keV, $\omega_i=0.768$ keV, $\theta_i=0^{\circ}$, $\varphi_f=0^{\circ}$,
and various $\nu^2$.}
\label{cg11}
\end{figure}

\begin{figure}[H]
 \centerline{\includegraphics[height=8cm,width=10cm]{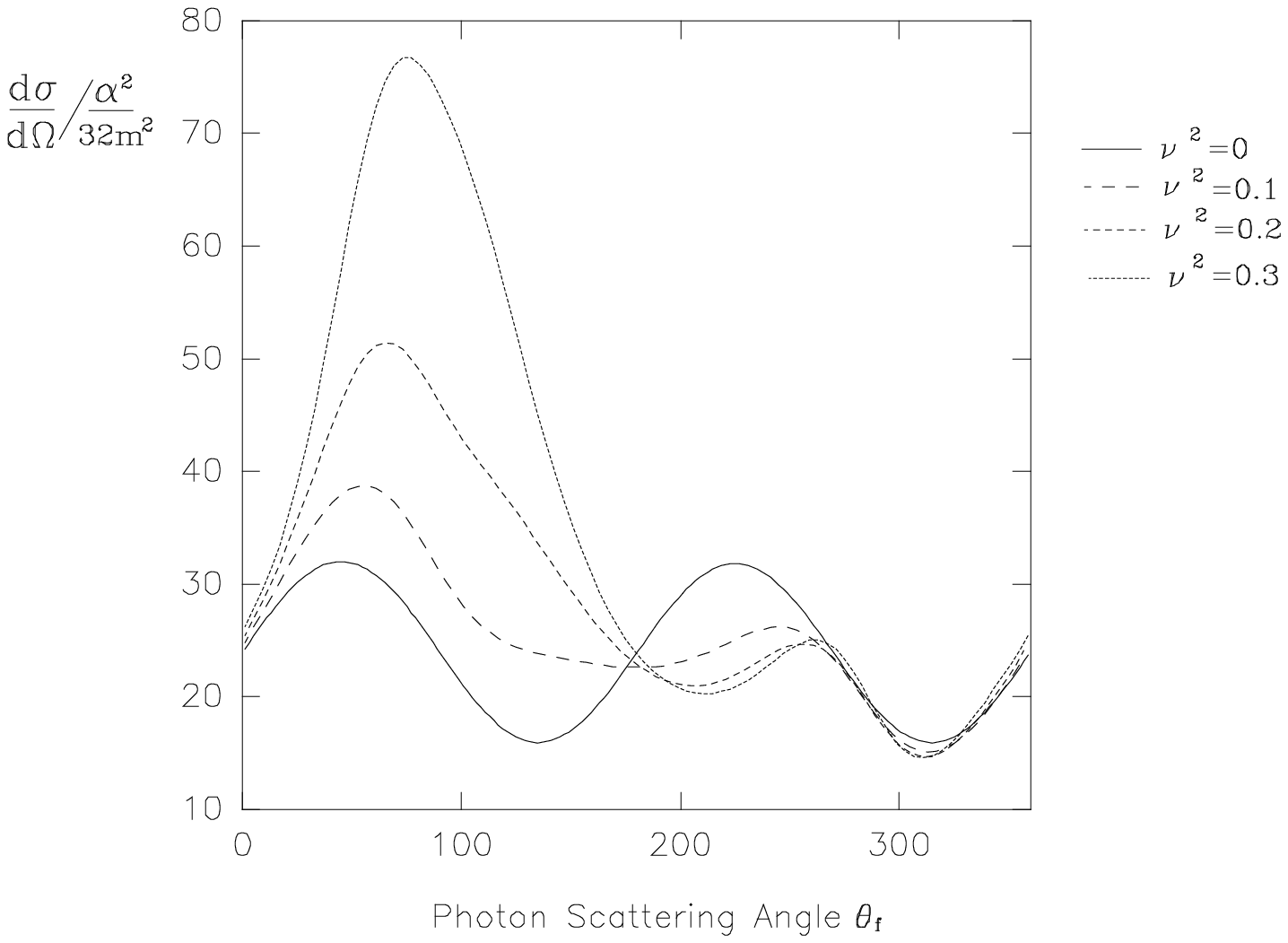}}
\caption{\bf\bm The SCS differential cross section vs $\theta_f$
for $\,\omega=0.512$ keV, $\omega_i=0.768$ keV, $\theta_i=45^{\circ}$, $\varphi_f=0^{\circ}$,
and various $\nu^2$.}
\label{cg12}
\end{figure}

\clearpage

\begin{figure}[H]
 \centerline{\includegraphics[height=8cm,width=10cm]{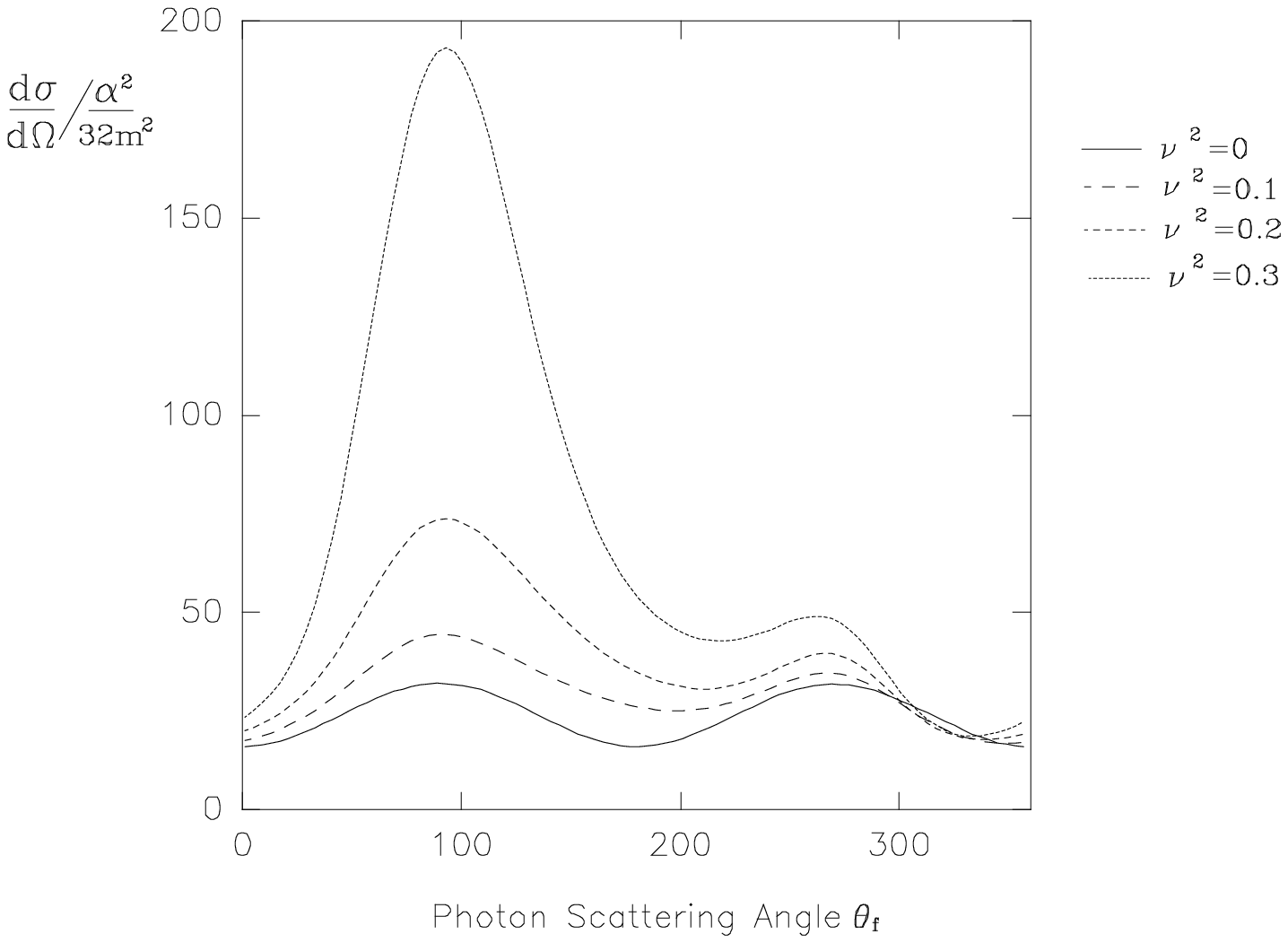}}
\caption{\bf\bm The SCS differential cross section vs $\theta_f$
for $\,\omega=0.512$ keV, $\omega_i=0.768$ keV, $\theta_i=90^{\circ}$, $\varphi_f=0^{\circ}$,
and various $\nu^2$.}
\label{cg13}
\end{figure}

\begin{figure}[H]
 \centerline{\includegraphics[height=8cm,width=10cm]{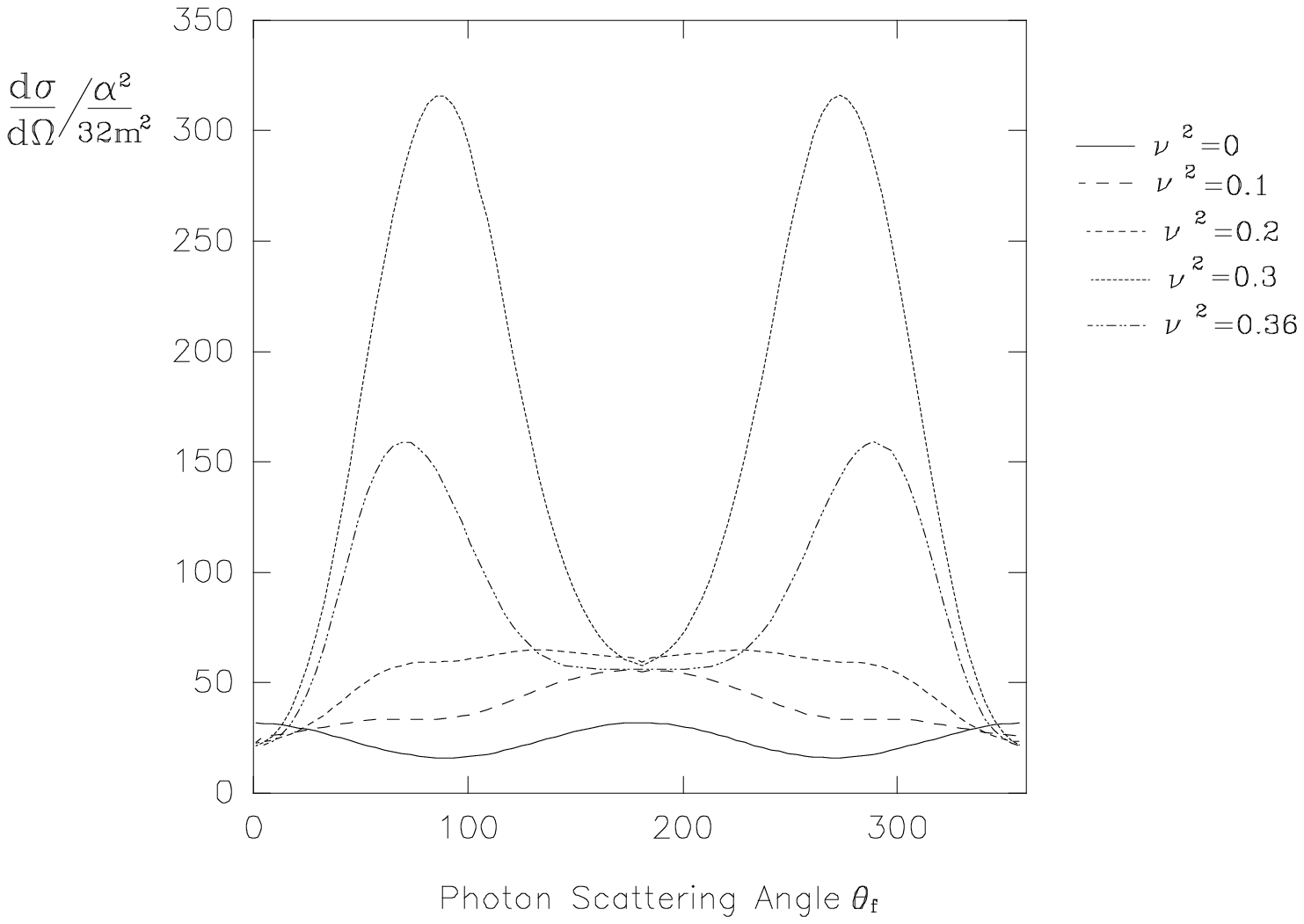}}
\caption{\bf\bm The SCS differential cross section vs $\theta_f$
for $\,\omega=0.512$ keV, $\omega_i=0.768$ keV, $\theta_i=180^{\circ}$, $\varphi_f=0^{\circ}$,
and various $\nu^2$.}
\label{cg14}
\end{figure}

\clearpage

\begin{figure}[H]
 \centerline{\includegraphics[height=8cm,width=10cm]{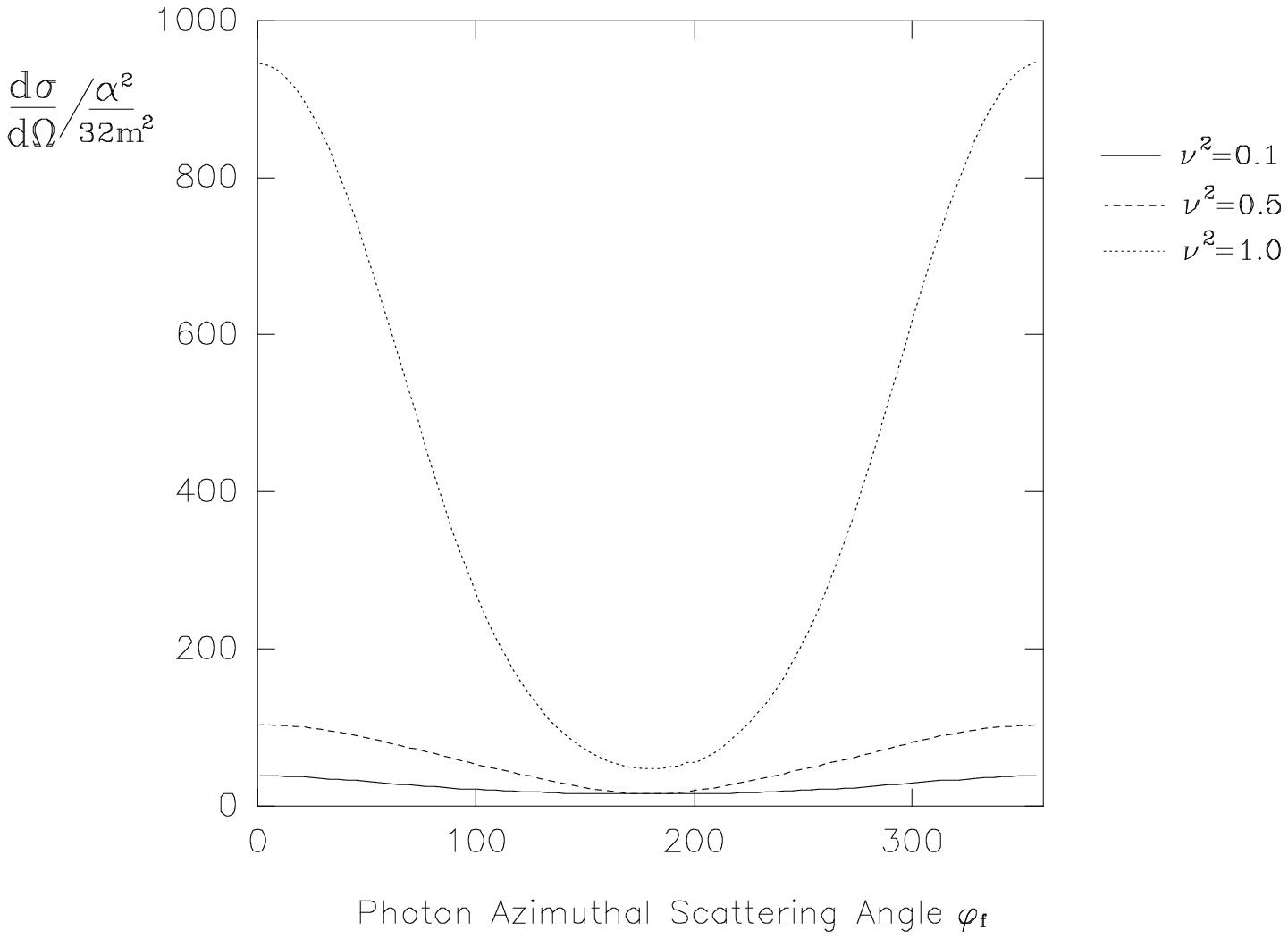}}
\caption{\bf\bm The SCS differential cross section vs $\varphi_f$
for $\,\omega=0.512$ keV, $\omega_i=0.768$ keV, $\theta_i=45^{\circ}$, $\theta_f=45^{\circ}$,
and various $\nu^2$.}
\label{cg15}
\end{figure}

\begin{figure}[H]
 \centerline{\includegraphics[height=8cm,width=10cm]{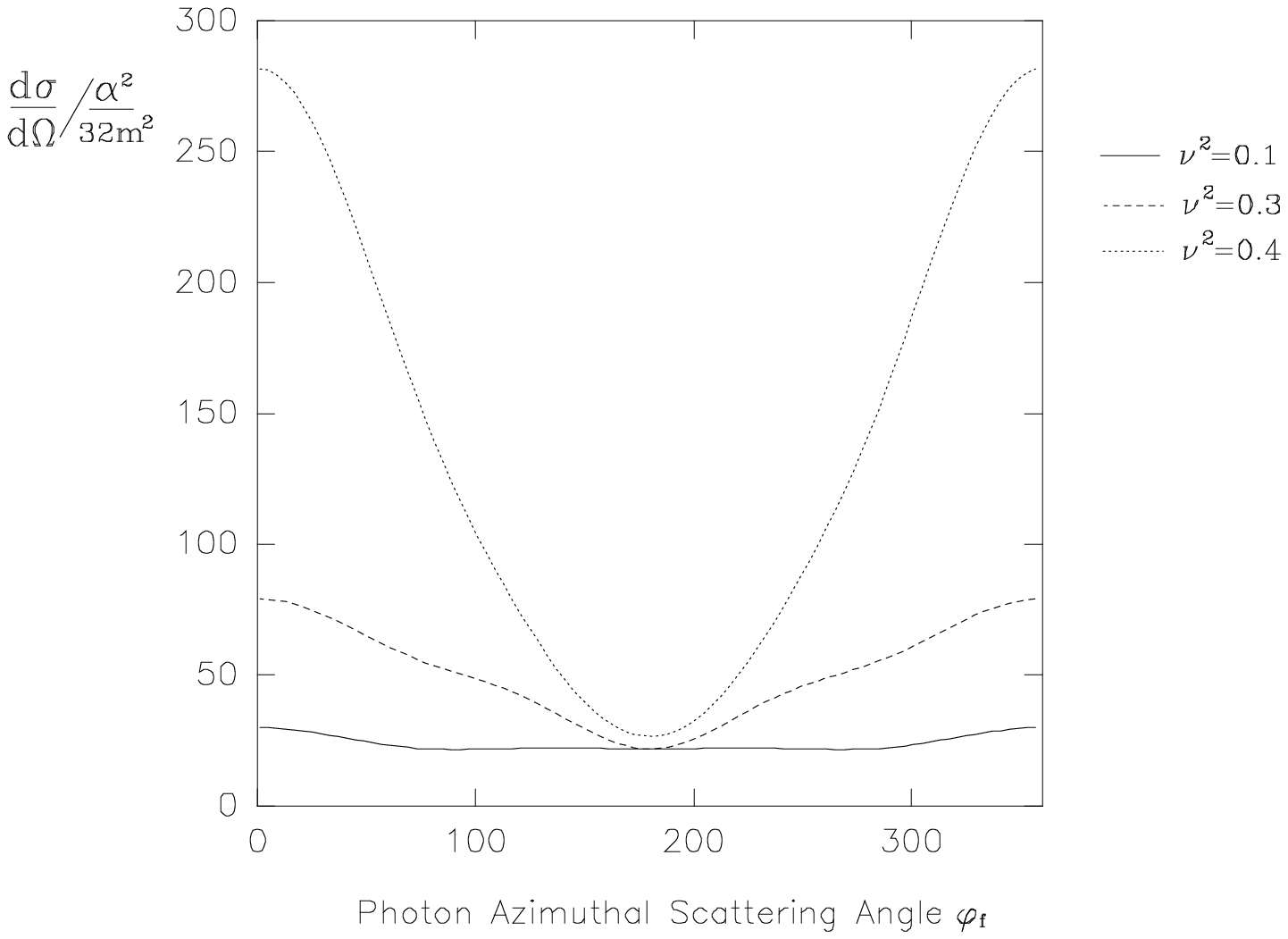}}
\caption{\bf\bm The SCS differential cross section vs $\varphi_f$
for $\,\omega=0.512$ keV, $\omega_i=0.768$ keV, $\theta_i=90^{\circ}$, $\theta_f=45^{\circ}$,
and various $\nu^2$.}
\label{cg16}
\end{figure}

\clearpage

\begin{figure}[H]
 \centerline{\includegraphics[height=8cm,width=10cm]{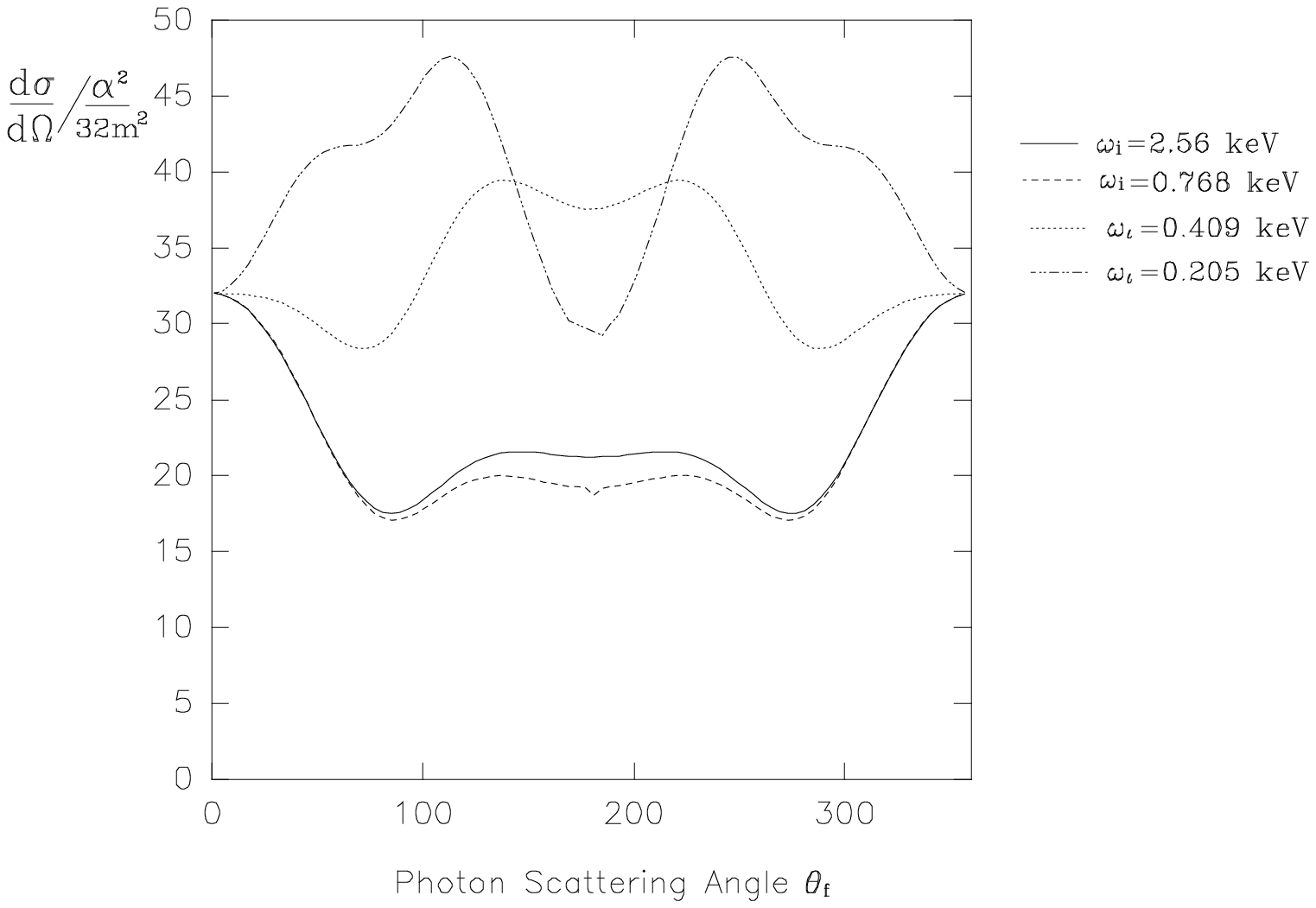}}
\caption{\bf\bm The SCS differential cross section vs $\theta_f$
for $\,\omega=0.512$ keV, $\theta_i=0^{\circ}$, $\varphi_f=0^{\circ}$, $\nu^2=0.1$ 
and various $\omega_i$.}
\label{cg17}
\end{figure}

\begin{figure}[H]
 \centerline{\includegraphics[height=8cm,width=10cm]{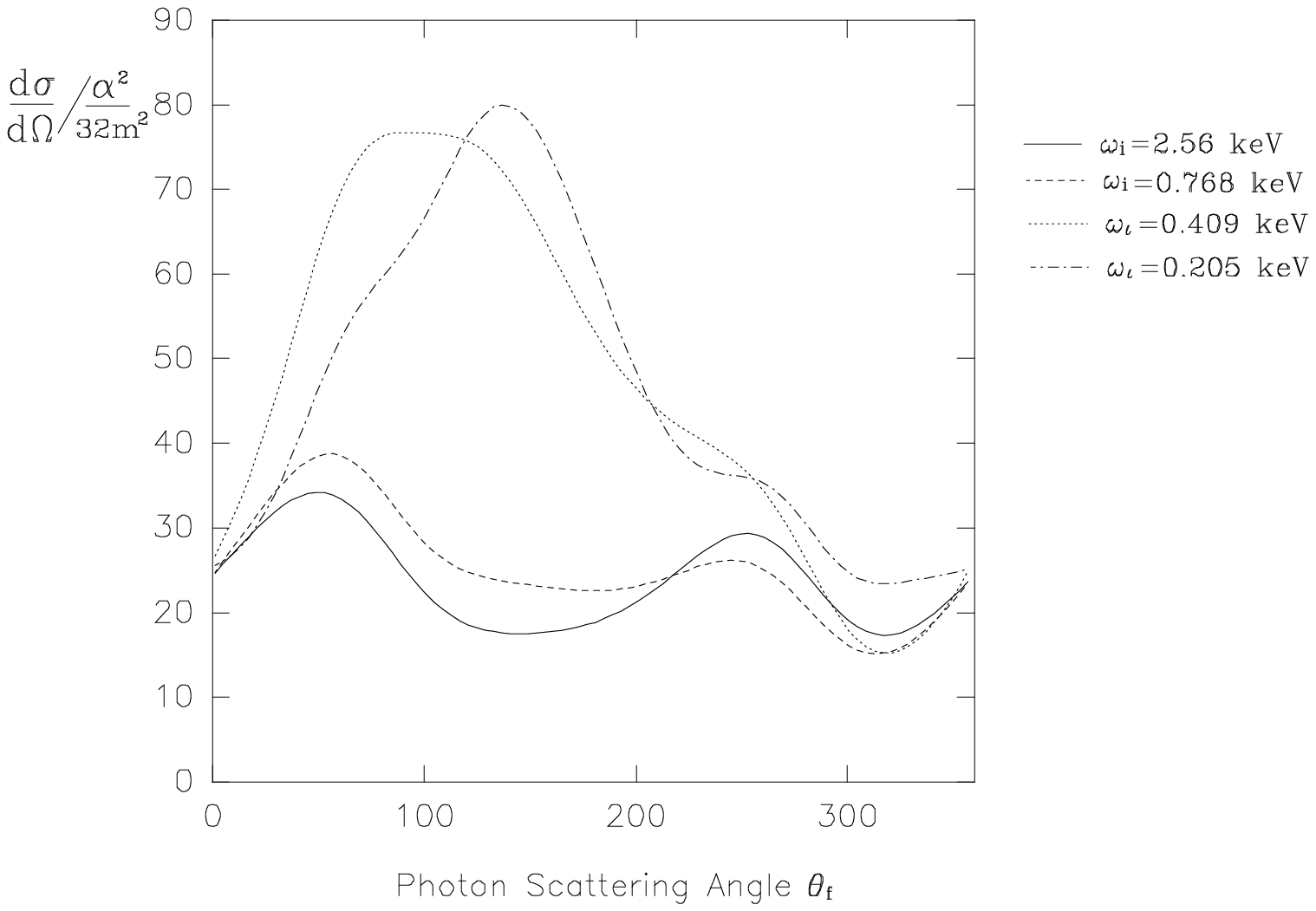}}
\caption{\bf\bm The SCS differential cross section vs $\theta_f$
for $\,\omega=0.512$ keV, $\theta_i=45^{\circ}$, $\varphi_f=0^{\circ}$, $\nu^2=0.1$
and various $\omega_i$.}
\label{cg18}
\end{figure}

\clearpage

\begin{figure}[H]
 \centerline{\includegraphics[height=8cm,width=10cm]{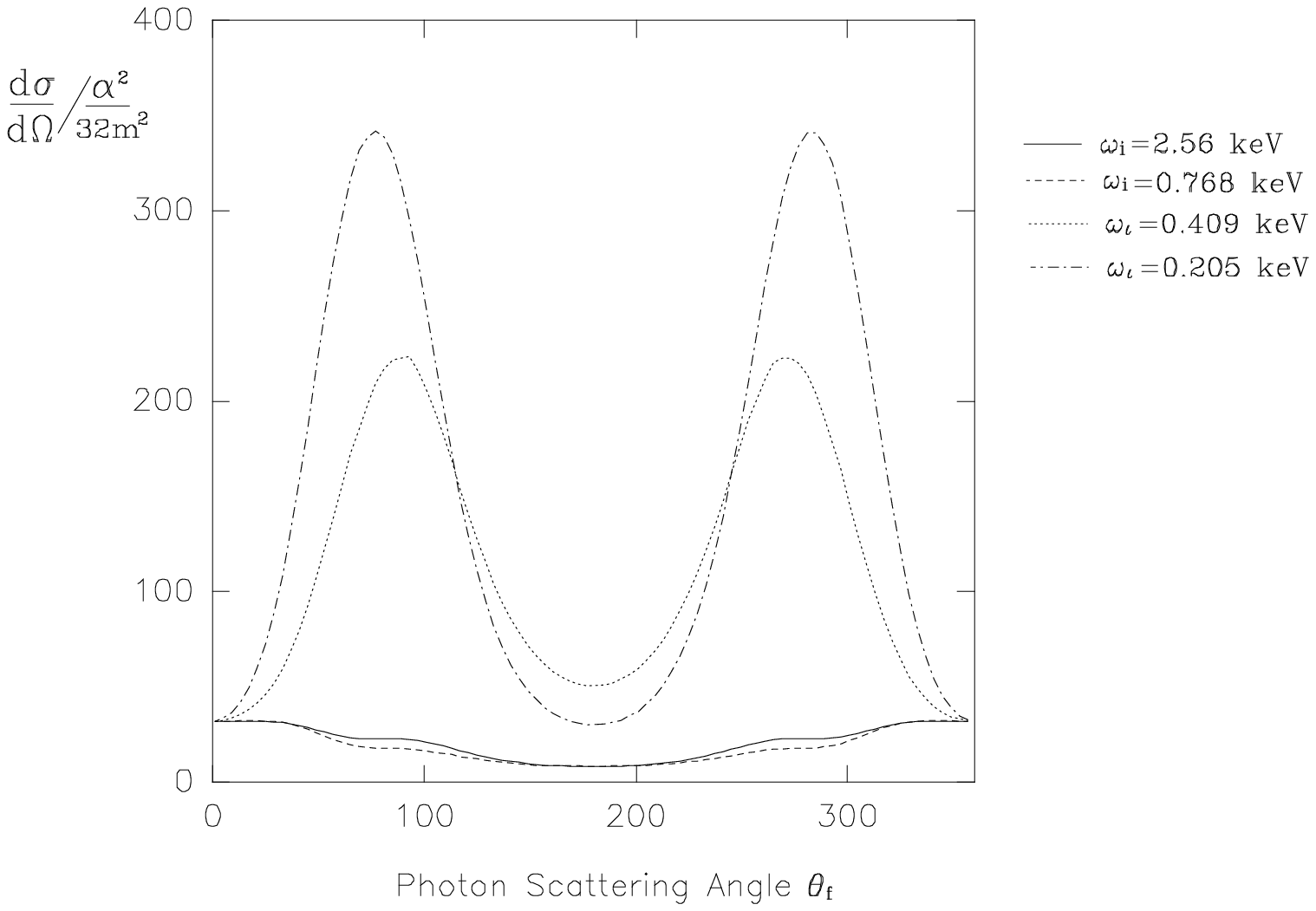}}
\caption{\bf\bm The SCS differential cross section vs $\theta_f$
for $\,\omega=0.512$ keV, $\theta_i=0^{\circ}$, $\varphi_f=0^{\circ}$, $\nu^2=0.5$
and various $\omega_i$.}
\label{cg19}
\end{figure}

\begin{figure}[H]
 \centerline{\includegraphics[height=8cm,width=10cm]{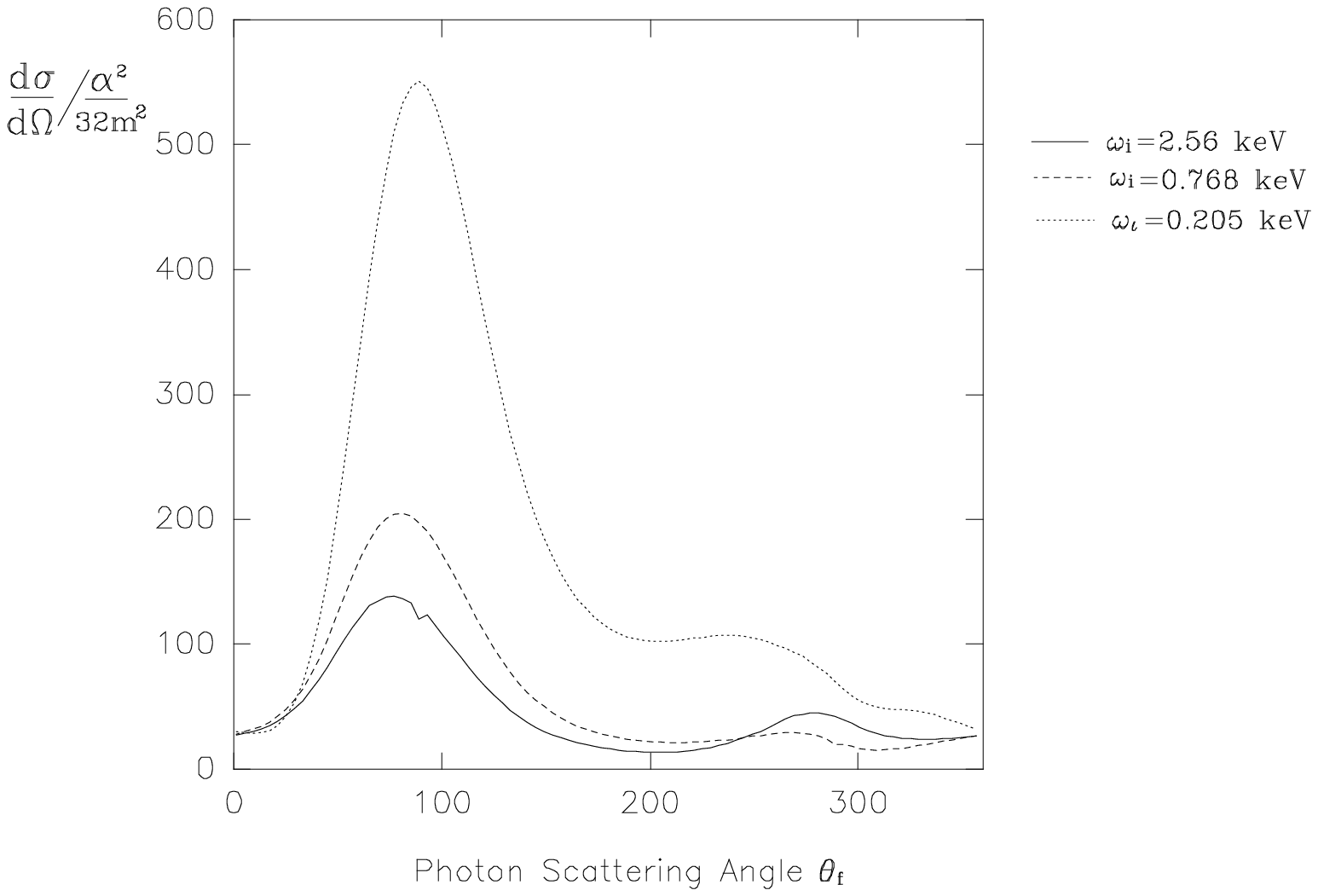}}
\caption{\bf\bm The SCS differential cross section vs $\theta_f$
for $\,\omega=0.512$ keV, $\theta_i=45^{\circ}$, $\varphi_f=0^{\circ}$, $\nu^2=0.5$
and various $\omega_i$.}
\label{cg20}
\end{figure}

\clearpage

\begin{figure}[H]
 \centerline{\includegraphics[height=8cm,width=10cm]{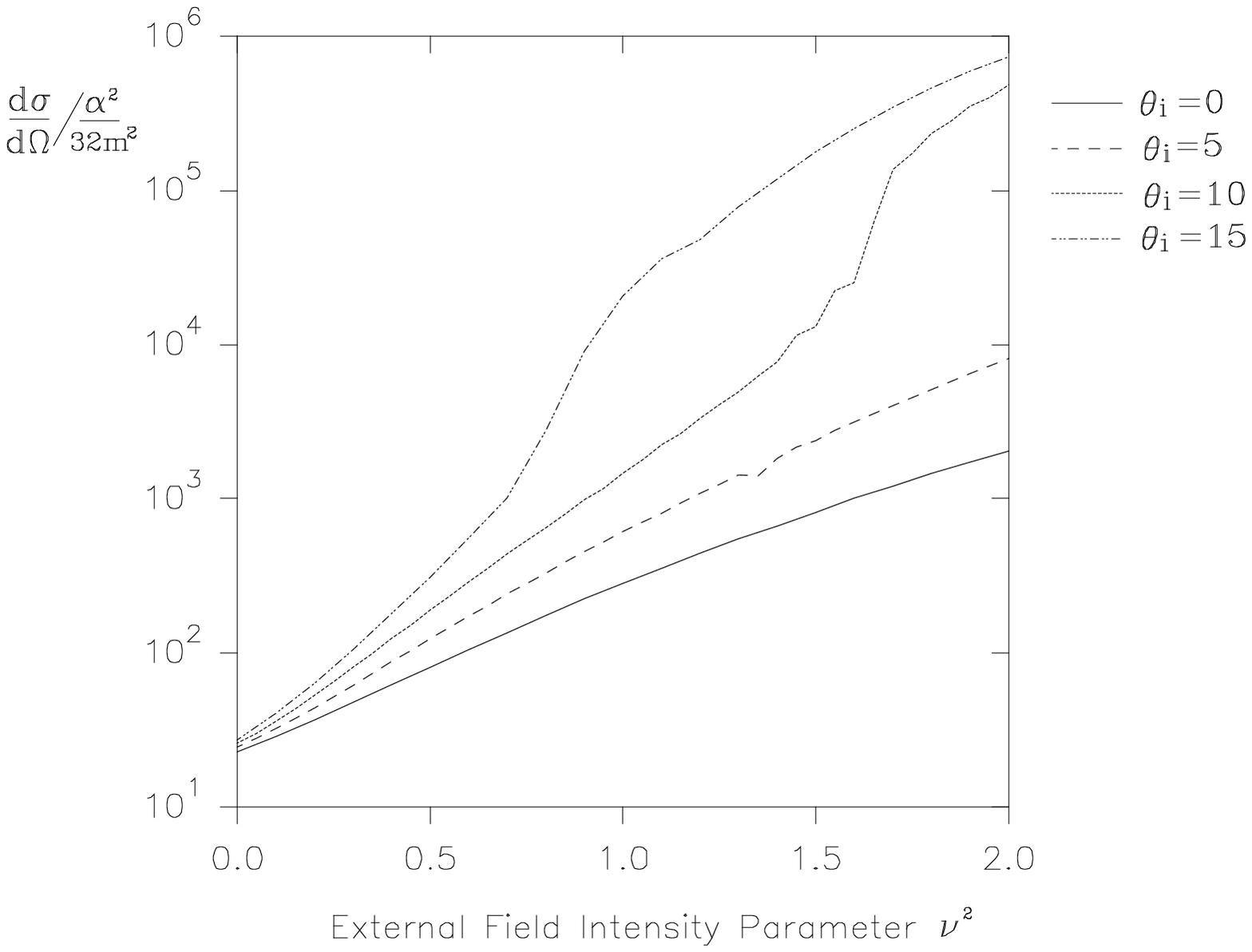}}
\caption{\bf\bm The SCS differential cross section vs $\nu^2$
for $\,\omega=51.2$ keV, $\omega_i=40.9$ keV, $\theta_f=45^{\circ}$, $\varphi_f=0^{\circ}$ and 
various $\theta_i$.}
\label{cg21}
\end{figure}

\begin{figure}[H]
 \centerline{\includegraphics[height=8cm,width=10cm]{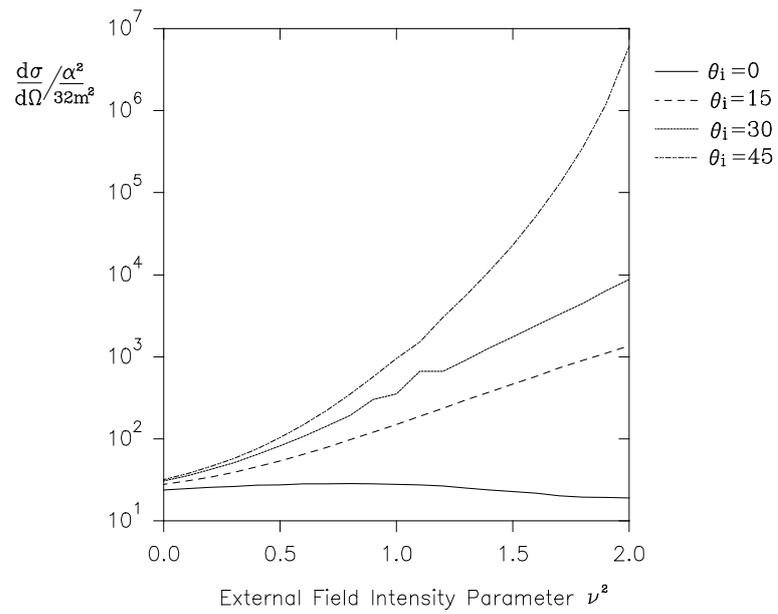}}
\caption{\bf\bm The SCS differential cross section vs $\nu^2$
for $\,\omega=0.512$ keV, $\omega_i=0.768$ keV, $\theta_f=45^{\circ}$, $\varphi_f=0^{\circ}$ and
various $\theta_i$.}
\label{cg22}
\end{figure}

\clearpage
\subsection{Differential Cross Section $l$ Contributions}

Figures \ref{cga1} - \ref{cga4} show the relative contributions of the
SCS differential cross section according to the number, $l$ of external
field quanta that take part in the scattering process. The differential
cross section contributions are calculated for various particle energies,
scattering angles and external field intensities $\nu ^2$. The insets of
figures \ref{cga1} - \ref{cga4} represent the same variation of parameters
as their parent plots, except that the parameter $\nu^2$ approaches
zero.

We note first of all that the graphs are not ''smooth''. This is so because
of the discrete nature of the horizontal data points and the relatively 
rapid convergence of the SCS differential cross section
with respect to the infinite summation over $l$ for the parameter values
considered ($\frac{d\sigma }{d\Omega }\rightarrow 0\quad $for$\quad l\leq 25$).
Figures \ref{cga1} and \ref{cga2} concern particle parameters for which the
initial photon is less energetic than laser photons ($\omega _i<\omega $).
The initial state of the scattering process consisting of
an electron at rest and the initial photon $k_i$, does not contain enough
energy to allow the emission of a quantum of energy, $\omega $ , to the
external field. Consequently the only $l$ contributions that are possible are those in
which the external field gives up quanta to the initial state of the
electron, hence $l\geq 0$.
Figures \ref{cga3} and \ref{cga4} on the other hand, concern particle
parameters for which the initial photon energy exceeds that of the laser
photons ($\omega _i>\omega $). Here the SCS initial state can give up energy
to the laser field and differential cross section contributions with $l<0$
are energetically permitted.

A general feature of all four figures is the "broadening out" of the
graphs as $\nu ^2$ increases. A greater number of $l$ contributions become significant to the 
overall SCS differential cross section. This behaviour makes sense physically since there is 
a direct relationship between $\nu ^2$ and the photon number density. With increasing number 
of external field photons in the fixed region of space in which the SCS process takes place, 
the probability that a greater number of these photons will combine with the SCS
electron also increases.

The insets of figures \ref{cga1} - \ref{cga4} show the opposite effect in
that the graphs display a sharpening peak at $l=0$ with decreasing external
field intensity. As $\nu ^2$ (and the external field photon number density)
decreases to zero, the only significant contribution to the differential
cross section is the one in which $l=0$. The relative value of the $l$
contributions is dependent in part on the scattering angles and particle energies. 
The region $-4\leq l\leq 4$ in figure \ref{cga3} displays a more complicated 
oscillatory behaviour and a more thorough examination of the variation of the SCS differential
cross section with scattering angles $\theta _i$ and $\theta _f$ and photon
energies $\omega $ and $\omega _i$, which follows in the rest of this chapter, is required. 

The behaviour of the SCS differential cross section is simplified for the special 
kinematic case where the incoming photon is directed parallel to the
direction of propagation of the external field $\theta _i=0^{\circ }$ and this is investigated 
in figures \ref{cgb1} - \ref{cgb8}. For these kinematics the SCS process gains an additional 
symmetry with respect to the azimuthal angle $\phi _f$ and the form of the differential
cross section is considerably simplified.
So Figures \ref{cgb1} - \ref{cgb8} show the $l=0,1,2$ contributions for
the initial scattering angle $\theta _i=0^{\circ }$ with various values of $\nu ^2$
and particle energies $\omega $ and $\omega _i$. Each $l$ contribution
consists of an infinite number of additional $r$ contributions corresponding
to the number of ways in which $l$ external field photons can be included in
the scattering process.
\footnote{Actually there are two infinite summations over $r$ and $r'$, however for 
$\theta_{i}=0$ these reduce to just one.} 
For example if the net number of external field photons
contributing to the process is $l=0$, then the $r=2$ contribution
corresponds to two external field photons being absorbed by the initial electron and two 
external field photons being emitted by the final electron.

The SCS differential cross section for $\theta _i=0^{\circ }$ was considered in section 3.5 
with the important result that the argument of some Bessel functions go to zero. 
Inspection of the order of the Bessel function co-products of the trace calculations (Appendix 
\ref{trace}) require then that the only non zero contributions to the differential 
cross section are those with $r=0$ ($l$ external field photons contribute to the process), 
$r=1$ ($l+1$ external field photons are absorbed and then one emitted), and $r=-1$ (an 
external field photon is given up to the field and then $l+1$ reabsorbed).
This result is due to the structure of the Volkov solution in a circularly polarised 
electromagnetic field. With the initial electron 3-momentum zero the only effect the external
field has on initial states is an increase in electron rest
mass. However, in the final states, the full range of electron 3-momenta are permitted 
and the Volkov solution contains contributions corresponding to $r=-1,0,1$ from $l-1$, $l$ or 
$l+1$ external field photons.

Figures \ref{cgb1} and \ref{cgb2} show the variation of the $l=0,r=0$
contribution with the final scattering angle $\theta _f$ and initial scattering
angle $\theta_i=0^{\circ }$. Each figure contains several plots
corresponding to a variation in intensity parameter $\nu ^2$. The central
feature is a double peak structure with peaks at $\theta _f=0^{\circ
},180^{\circ }$. At $\nu^2=0$ the peaks are of equal height. As $\nu
^2$ increases, the peak at $\theta _f=180^{\circ }$ broadens and decreases
in height, with the $\theta _f=0^{\circ }$ peak narrowing and remaining the
same height. Neglecting $\nu^2$, the $l=0,r=0$ contribution of the SCS differential 
cross section is identical to the Klein-Nishina differential cross section and some 
comments can be made based on its form

\begin{equation}
\label{c4eq1}
\frac{d\sigma }{d\Omega }=\frac{r_o^2}2\left( \frac{\omega _f}{%
\omega _i}\right) ^2\left( \frac{\omega _f}{\omega _i}+\frac{\omega _i}{%
\omega _f}-\sin {}^2\theta _f\right) \quad ;\quad \;\omega _f=\frac{\omega
_i }{1+\omega _i(1-\cos \theta _f)} 
\end{equation}

\medskip\ 

When $\omega _f\simeq \omega _i$, the energy of the initial photon is much less than the rest 
mass of the electron and the angular dependence of
the differential cross section is approximately proportional to the term 
$2-\sin {}^2\theta _f$. This is the source of the peaks at $\theta
_f=0^{\circ },180^{\circ }$. As the energy of the initial electron becomes significant the 
angular dependence of the $\frac{\omega_f}{\omega_i}$ term, which is a minimum at 
$\theta_f=180^{\circ}$ also becomes significant. The competing angular dependences of terms 
within the differential cross section results in a reduced peak at $\theta_f=180^{\circ}$. 
The same differential cross section peaks can be determined from physical arguments by 
considering the electromagnetic field associated with the initial SCS photon.

It is well known that an electron, undergoing simple harmonic motion radiates energy at right 
angles to its direction of propagation \cite{Jackson75}. The electromagnetic wave associated 
with the initial photon causes the SCS electron to vibrate transversely. Since the differential 
cross section was summed over all photon polarisations the predominant radiation will be 
parallel and anti-parallel to the direction of propagation of the initial photon, 
i.e. $\theta _f=0^{\circ},180^{\circ }$. If the energy of the initial photon 
increases from a low value, a significant amount of forward momentum is
transferred to the electron and radiation in the forward direction is more likely. 
Less final states are available for the scattering and the $\theta_f=180^{\circ}$ differential 
cross section decreases.

When the external field is significant $\nu^2>0$ extra momentum is contributed to the 
electron. The net momentum transferred is the difference
between that of the final state electron $p_f$ and that of the initial state electron $p_i$. 

\begin{equation}
\label{c4eq2} 
\begin{array}{rl}
\dfrac{\nu ^2}{2(kp_f)}k_\mu -\dfrac{\nu ^2}{2(kp_i)}k_\mu &=\dfrac{\nu ^2}2 
\dfrac{\omega _f(1-\cos \theta _f)}{\omega (1-\omega _f(1-\cos \,\theta _f))}
\\[10pt] \text{where} \quad \quad \quad \omega _f &=\dfrac{\omega _i+l\omega }
{1+( \frac{\nu ^2}2+l\omega +\omega _i)(1-\cos \theta _f)} 
\end{array}
\end{equation}

\medskip\ 

The net momentum contribution increases with increasing $\nu^2$ as long as 
$\theta _f\neq 0^{\circ }$. Consequently at $\theta _f=0^{\circ }$
there is no variation of the SCS differential cross section.
The maximum increase in net momentum transferred to the electron is at 
$\theta_f=180^{\circ }$. The situation here is
equivalent to the Klein-Nishina process in which a not insignificant amount
of momentum is transferred to the final electron. Here the momentum comes,
not from an energetic initial photon but from the external field. As before, extra momentum 
transferred to the electron results in a broadened and flattened differential cross section 
peak at $\theta _f=180^{\circ }$, as evidenced from figures \ref{cgb1} and \ref{cgb2}.

Figures \ref{cgb4a} and \ref{cgb4b} reveal the effect the external field has
on the $l=0$ contribution for the initial scattering geometry $\theta
_i=0^{\circ }$. These figures compare the Klein-Nishina differential
cross section with the SCS $l=0,r=0$ contribution and the SCS $l=0$
contribution summed over all $r$ contribution ($r=-1,0,1$).

The effect is dependent on the particle energy ratio $\frac{\omega_i}{\omega}$ and 
the $l=0,r=$all contribution shows an increase in the differential cross section 
reaching a maximum at $\theta _f=180^{\circ }$ when $\frac{\omega_f}{\omega}<1$
(figure \ref{cgb4a}), and a decrease reaching a minimum at $\theta
_f=180^{\circ }$ when $\frac{\omega_i}{\omega}>1$ (figure \ref{cgb4b}).

The behaviour of the $l=0,r=\text{all}$ contribution is clearly different from the 
$l=0,r=0$ contribution. Like the latter, the former can be explained partially by 
comparison with the Klein-Nishina differential cross section. However the fill explanation 
requires an appeal to the quasi-level structure of the electron embedded in the external field 
\cite{Zeldovich67}. The probability of scattering increases if the kinematics are such that the 
final bound electron energy reaches a quasi-energy level. The quasi-energy levels are given 
by integer numbers of external field photon energy $n\omega$. For the energies considered in 
figures \ref{cgb4a} - \ref{cgb4b} the first quasi-energy level is 0.512 keV.

In figure \ref{cgb4a} the incident photon energy at 0.409 keV ($\frac{\omega_i}{\omega}<1$) 
supplies energy to the SCS electron just short of the first quasi-energy level. Though momentum 
is supplied from the external field as described by \ref{c4eq2}, this simply accounts for the 
increased rest mass of the bound electron. scattering of the SCS photon in the backward 
($\theta_f=180^{\circ}$) direction becomes more probable because the rebounding final bound 
electron gains extra energy to carry it closer to the quasi-energy level. In figure \ref{cgb4b} 
the energy of the initial photon $\omega_i=0.768\,\text{keV}$ ($\frac{\omega_i}{\omega}>1$) 
exceeds the quasi-energy level. This time, backwards scattering of the SCS photon is suppressed 
as the gained energy carries the final bound electron away from the quasi-level. 

Figures \ref{cgb5} - \ref{cgb8} compare the angular dependence on the final
scattering angle $\theta _f$ of the $l=0,1,2$ contributions for various
particle energies and external field intensity for the initial scattering
geometry $\theta _i=0^{\circ}$. Figures \ref{cgb5} and \ref{cgb6} concern
scattering situations in which the initial photon is more energetic than the
laser photons. The central feature of figure \ref{cgb5} $(\frac{\omega_i}{\omega}=5)$ are 
oscillatory cross section contributions. Maximum peaks lie in the region around $\theta
_f=0^{\circ }$ and $\theta _f=180^{\circ }$, diminishing in height as the
final photon scatters at angles perpendicular to the direction of
propagation of the initial photon $(\theta _f=90^{\circ },270^{\circ })$.

In figure \ref{cgb6} $(\frac{\omega_i}{\omega}=1.33)$, the $l=0$
contribution has a primary peak at $\theta _f=0^{\circ }$ and a second,
broadened peak at $\theta _f=180^{\circ }$. This angular variation is
similar in form to the Klein-Nishina differential cross section where
significant momentum has been transferred to the electron in the course of
the scattering. Indeed its identical to the plot in figure \ref{cgb4b}. The $l\neq 0$ 
contributions oscillate and fall to zero at $\theta _f=0^{\circ }$. 
In both figures \ref{cgb5} and \ref{cgb6}, the $l=0$ contribution exceeds in magnitude the 
$l\neq 0$ contributions.

Figures \ref{cgb7} and \ref{cgb8} concern scattering situations in which the
laser photon energy exceeds the initial photon energy. In figure \ref{cgb7}, 
$\frac{\omega_i}{\omega}=0.8$ and for figure \ref{cgb8} $\frac{\omega_i}{\omega}=0.1$. 
Both figures show $l=0$ contributions with peaks at $\theta _f=0^{\circ },180^{\circ }$ (Its 
difficult to see in figure \ref{cgb8} but its there). The $l\neq 0$ contributions show the
development of peaks at angles approximately perpendicular to the direction
of the initial photon (and the direction of propagation of the external
field), $\theta _f=90^{\circ },270^{\circ }$. These peaks show evidence of
structure, i.e. secondary and tertiary peaks. In figure \ref{cgb8} the $%
l\neq 0$ contributions clearly exceed the $l=0$ contribution%
\footnote{the peak of the $l=1$ contribution is 20.6 times the magnitude of the $l=0$
contribution peaks}, whereas, in figure \ref{cgb7}, they are of the same
order of magnitude.

The peaks associated with the $l=0$ contribution have already been discussed. The precise form 
of the peaks associated with the $l\neq 0$ contributions is the result of many terms in the 
differential cross section equations and cannot be discussed in detail. However general
comments can be made about the origin of these peaks, approximately where we can expect
to find them, and the relative magnitude of the $l=0$ and $l\neq 0$
contributions.

The origins of the peaks associated with the $l\neq 0$ contributions can be
determined by considering the momentum of an electron at space-time $x_\mu $ in the presence of
the external electromagnetic field $A_\mu (x)$ (see equation 2.29).

\medskip\

\begin{equation}
\label{c4eq4}
\Pi _\mu \left( x\right) =p_\mu -eA_\mu ^e(x)+k_\mu \left( e 
\frac{(Ap)}{(kp)}+\frac{e^2A^2}{2(kp)}\right) 
\end{equation}

\medskip\ 

The origin of the term $e\frac{(Ap)}{(kp)}k_\mu $ can be explained with the
aid of classical electromagnetic theory. When the electron crosses the plane
of vibration of the magnetic field components of the external field the electron 
momentum gains additional, oscillatory components along the direction of propagation of the 
external field ($\sql{k}$).
When the electron momentum is directed along $\sql{k}$, there is no
intersection with the plane of the magnetic field components and the
longitudinal momentum component is zero. The oscillatory nature of the
longitudinal momentum component allows it to be written in a Fourier series
of discrete external field contributions corresponding to $l=1,2,3,\ldots $.
Each contribution is interpreted as the addition of a discrete number of
laser field photons, and is proportional to squares of Bessel functions of
order $l$ and argument $z$.

The amplitude of the longitudinal momentum contributions along the direction
of propagation of the external field is given by the numerical value of the
Bessel function arguments described by

\medskip\

\begin{equation}
\label{c4eq4b}
z=\sqrt{e^2\left( \frac{(a_1p)^2}{(kp)^2}+\frac{(a_2p)^2}{(kp)^2}\right) }
\end{equation}

\medskip\ 

The variation of $z$ with the parameters of the scattering process largely
determines the behaviour of the $l\neq 0$ contributions to the SCS
differential cross section. For the direct channel (first Feynman diagram in figure 
\ref{c3.comp.fig1}) and exchange channel (second Feynman diagram in figure
\ref{c3.comp.fig1}) intermediate electrons, $z$ can be written in terms of particle energies, 
scattering angles and external field intensity

\begin{equation}
\label{c4eq5}
\begin{array}{rll}
\bar z & = \nu \dfrac{\omega _i}\omega \left| \sin \,\theta _i\right| \quad\quad\quad\quad\quad
&\text{direct channel} \\[10pt] 
\dbr{z} & = \nu \dfrac{\omega _f\left| \sin \,\theta _f\right| }{\omega
(1-\omega _f(1-\cos \,\theta _f))} &\text{exchange channel} \\[10pt]
\dfrac{\omega _f}\omega & \simeq \left( \frac{\omega _i}\omega +l\right)
\dfrac 1{1+\frac{\nu ^2}2(1-\cos \,\theta _f)} &
\end{array}
\end{equation}

\medskip\ 

The magnitude of longitudinal contributions to the momentum is
proportional to the quantities $\frac{\omega_i}\omega ,\left( \frac{\omega
_i}\omega +l\right)$. When $\,\frac{\omega _i}\omega>1$, the maximum value of $\dbr{z}$
is relatively large as long as $\nu $ is not too small
\footnote{the maximum value of $\dbr{z}$ is 1.8 for the $l=1$ plot of figure 4.9, 
and 1.44 for the $l=1$ plot of figure 4.10} ($\bar z=0$ when $\theta_i=0^{\circ }$) 
and many longitudinal contributions have a significant
impact on the differential cross section. With variation in $\theta _f$,
the associated variation in $\dbr{z}$ is also relatively large and each
longitudinal contribution to the differential cross section oscillates many times (see
figure \ref{cgb5}). This behaviour is a consequence of the mathematical
behaviour of the associated Bessel functions.

The oscillatory nature of the differential cross section can be justified physically by an 
appeal to the quasi-energy level structure. The contribution of incident photon energy such 
that $\frac{\omega_i}{\omega}=5$ in figure \ref{cgb5} allows the SCS electron to potentially 
traverse several energy levels corresponding to the differential cross section peaks. The 
general Klein-Nishina trend of maximum differential cross section in the region of $\theta 
_f=0^{\circ },180^{\circ }$ is still evident.

Conversely in figures \ref{cgb7} and \ref{cgb8} particle energies are such that 
$\frac{\omega _i}\omega<1$, the maximum variation in and value of $\dbr{z}$ is
relatively small.
\footnote{the maximum value of $\dbr{z}$ is 1.0 for the $l=1$ plot of figure 4.11, 
and 1.44 for the $l=0.44$ plot of figure 4.12}
Longitudinal momentum contributions and oscillatory behaviour of the differential cross section 
is also small. The maximum value of the differential cross section occurs at transverse 
angles when $\dbr{z}$ is largest ($\theta _f\sim 90^{\circ },270^{\circ }$). This trend is 
evident in the $l=1,2,3$ plots of figures \ref{cgb7} and \ref{cgb8}. The $l=0$ contribution 
shows Klein-Nishina type longitudinal peaks.

Comparison of figures \ref{cgb7} and \ref{cgb8} indicate that the differential 
cross section $l\neq 0$ contributions dominate the $l=0$ contribution when 
$\frac{\omega_i}{\omega}\ll 1$. Physical justification for this trend can be made by 
considering electron motion due to the fields of both the SCS photon and the external field 
photons.

As already noted, the field of the photon $k_i$ introduces a transverse
vibration to the electron which in turn radiates in the $\theta _f=0^{\circ },180^{\circ }$
directions (after summing over polarisations) with radiated power proportional to $\omega 
_i^2$. The circularly polarised wave of the external field photon introduces a circular motion 
to the electron which in turn radiates in the $\theta _f=90^{\circ },270^{\circ }$ direction
(provided the electron has no significant forward momentum) with radiated
power proportional to $\nu^2\omega^2$ \cite{Jackson75}. With the factor 
$\frac{\omega _i^2}{\nu ^2\omega ^2}<1$ ($\frac{\omega _i^2}{\nu ^2\omega ^2}=0.05$ in figure 
\ref{cgb8}) we expect the radiation produced by the electron subjected to these two fields to 
be mainly the result of the influence of the external field.

Figures \ref{cgb9} - \ref{cgb16} show differential cross section variation with various 
parameters for an initial photon angle $\theta_i=45^{\circ}$. Other $\theta_i$ values are 
considered in section 4.3.2. Figures \ref{cgb9} and \ref{cgb10} show $(\theta_f)$ variation of 
the $l=0$ contribution for various external field intensities and incident photon
energies. The main feature of both figures is a double peak differential cross section 
structure in the first half of the $\theta_f$ parameter range.
Figure \ref{cgb9} with $\frac{\omega_i}{\omega}=0.8$ also shows a slight
increase in the differential cross section with increase in the external field
intensity in the second half of the $\theta_f$ range. However in the same range figure 
\ref{cgb10} with $\frac{\omega_i}{\omega}=1.5$ shows a diminution
of the differential cross section.
At $\nu^2=0$ both figures show two peaks of the same height at 
$\theta_f=45^{\circ },225^{\circ }$. These are simply Klein-Nishina type peaks in the 
forward and backward direction of the initial photon.

An explanation of the features in figures \ref{cgb9} and \ref{cgb10} is made with the 
expression for bound electron momentum (equation \ref{c4eq4}). The electron gains two momentum 
components in the direction of external field propagation, one static and one oscillatory. For 
the direct channel of the scattering, with $\phi_0$ representing a scalar product of particle 
4-momenta, they are

\begin{equation}
\label{c4eq5c}
\begin{array}{cl}
\dfrac{\nu ^2}2\dfrac m\omega \,k_\mu \quad\quad\quad\quad\quad\quad &\text{static term} \\[10pt]  
\nu \dfrac{\omega _i}\omega \left| \sin \,\theta _i\right| \sin (kx+\phi_0)\,k_\mu   
&\text{oscillatory term} 
\end{array}
\end{equation}

\medskip\ 

As the magnitude of the external field intensity $\nu ^2$ increases, both of these momentum 
components also increase. The static term favours angles $\theta _f$ such that the final 
electron momentum is more collinear with the propagation direction of the
external field. The final photon should carry off the momentum of the initial photon so the 
differential cross section is maximised in the range $0^{\circ }<\theta_f<180^{\circ }$ and 
minimised in the range $180^{\circ}<\theta _f<360^{\circ }$. This behaviour has been 
noted before for the M\"oller scattering process in the presence of an external field 
\cite{Bos79a,Bos79b}. The oscillatory term results in radiation perpendicular to its 
longitudinal direction of oscillation producing peaks in the cross section of equal height at 
$\theta _f=90^{\circ},270^{\circ }$. The momentum requirements of the static term ensure that 
the $\theta_f=90^{\circ}$ peak is enhanced and the $\theta_f=270^{\circ}$ peak reduced. 
Additionally the oscillatory momentum contribution is at least a factor $\frac{\omega _i}{2m\nu }$ 
smaller than the static term, and for the particle energies considered $\frac{\omega _i}m\ll 1$ 
and $\nu ^2$ is not too small. The end result is that the differential cross section peaks in 
the $0^{\circ}<\theta_f < 180^{\circ}$ range dominate.

The dissimilar behaviour of figures \ref{cgb9} and \ref{cgb10} in the range 
$180^{\circ }<\theta_f<360^{\circ}$ is due in part to the quasi-energy level structure. Though 
complicated by the values of $\nu^2$ and $\theta_i$, the determining factor is still the ration 
$\frac{\omega_i}{\omega}$ which is less than one in figure \ref{cgb9} and greater than one in 
\ref{cgb10}

Figures \ref{cgb11} and \ref{cgb12} show the variation of the $l=1$
contribution with increasing external field intensity for the initial
angle $\theta _i=45^{\circ }$. The main feature is a general
increase in the cross section with an increase in $\nu ^2$ for scattering
angles $0^{\circ}<\theta_f<180^{\circ}$.
This increase has the same explanation as the one given for the $l=0$
contribution (figures \ref{cgb10} and \ref{cgb9}). Increases in $\nu^2$ shift the 
electron momentum more nearly parallel to the direction of propagation of the external field. 
Final photon scattering directions which favour this shift ($0^{\circ }<\theta
_f<180^{\circ }$) become more probable, and the cross section increases
accordingly.

Figures \ref{cgb13} - \ref{cgb16} compare the $l=0$ contributions and the $%
l\neq 0$ contributions for various values of the external field intensity
and particle energies. The relative magnitude is most distinct
in figures \ref{cgb15} and \ref{cgb16}. In figure \ref{cgb15}, with relative
particle energies $\frac{\omega i}\omega =10$, the $l=0$ contribution
clearly dominates, whereas in figure \ref{cgb16} ($\frac{\omega _i}\omega =0.1$) it is the 
$l\neq 0$ contributions which are dominant. As in the discussion of figures \ref{cgb7} and 
\ref{cgb8}, the relative impact of the electromagnetic field of the initial photon compared to 
the electromagnetic field of the external wave is the physical explanation.
The relevant factor $\frac{\omega _i^2}{\nu ^2\omega ^2}$ is 200 for figure \ref{cgb15} which 
leads to the $\theta_f=theta_i=45^{\circ}$ peak in the $l=0$ contribution and 0.02 for figure 
\ref{cgb16} in which the $l\neq 0$ contributions are largest. 

\bigskip\
\subsection{Differential Cross Sections Summed Over All $l$}

Figures \ref{cg1} - \ref{cg6} show the $\theta_f$ variation
of the complete SCS differential cross section for various initial
scattering angles $\theta _i$ and external field intensities $\nu ^2$. For
figures \ref{cg1} - \ref{cg3} an initial photon of energy 0.409 keV is
incident on an electron embedded in an 0.512 keV external field. For
figures \ref{cg4} - \ref{cg6} a 0.026 keV initial photon is incident on
electron embedded in an 0.005 keV external field.

The main feature of all six figures (with the exception of figure \ref{cg4}) is the development 
of a peak at $\theta _f\sim 90^{\circ }$. Since the differential cross sections here are 
combinations of $l=0$ and $l\neq 0$ contributions, much of the previous section's analysis 
applies. The $l=0$ contribution produces peaks in the $\theta_f=\theta_i,\theta_i + 
180^{\circ}$ directions, and the $l\neq 0$ contributions in transverse directions. The relative 
strength of the contributions is once again given by the factor $\frac{\omega _i^2}{\nu 
^2\omega ^2}$, which is smallest where the $\theta_i=90^{\circ}$ peak is strongest 
($\frac{\omega _i^2}{\nu ^2\omega ^2}=0.638$ and maximum peak height $\sim 25000$ in figure 
\ref{cg3}). The competing trends are complicated by transverse $\theta_i$ angles. The 
$\theta_f=90^{\circ}$ peaks increase with increasing $\theta_i$ and $\nu^2$ except for figure 
\ref{cg4} where the influence of the external field is weak. An explanation can be given by 
considering the longitudinal momentum the SCS electron gains from the field 
$\frac{\nu^2}{2(kp)}\;k_\mu$. As the longitudinal momentum of the SCS electron increases with 
$\nu^2$, the probability that it will carry off the transverse momentum supplied by a 
transverse initial photon decreases. Consequently, the photon has to carry off the transverse 
momentum of the initial state.

\input{./tex/tables/c_table4}

A point of interest is the relative magnitude of the peak maxima for figures 
\ref{cg1} - \ref{cg3} compared to those for figures \ref{cg4} - \ref{cg6} (table 
\ref{c4.table4}). For instance, the peak maxima for figure \ref{cg3} is a factor of 
17.25 larger than that of figure \ref{cg6}. The crucial factor here is the
denominator of the electron propagator which expresses the quasi-energy levels of
the electron embedded in the external field. Figure \ref{cg3} represents a
scattering where the addition of an initial photon of energy $409\,\,keV$
raises the SCS electron to an energy close to its first excited level and
the cross section $\theta _f=90^{\circ }$ is relatively large. Figure \ref
{cg6}, in contrast, represents a scattering where the initial photon raises
the SCS electron to an energy half way between its 7th and 8th excited level
and the cross section $\theta _f=90^{\circ }$ peak is relatively small.

Figures \ref{cg7} - \ref{cg14} show the variation of SCS differential cross section with final 
scattering angle $\theta _f$ for various initial scattering angles $\theta _i$, and 
particle energies $\omega $ and $\omega _i$. Each figure contains several plots with differing 
external field intensity $\nu^2$ allowing ease of comparison with the Klein-Nishina case 
($\nu^2=0$).

Figures \ref{cg7} - \ref{cg10} show the scattering of a 10.2 keV
photon incident on an electron embedded in a 51.2 keV external field for four initial photon 
angles ($\theta _i=0^{\circ },45^{\circ },90^{\circ},180^{\circ }$).
The main feature is the development of peaks at $\theta _f\sim 90^{\circ
},270^{\circ }$, which increase with increasing $\nu ^2$. In figures \ref
{cg7} and \ref{cg10} ($\theta _i=0^{\circ },180^{\circ }$) these peaks are
of equal height. In figure \ref{cg8} ($\theta _i=45^{\circ }$)
the $\theta _f\sim 90^{\circ }$ is dominant, whereas in figure \ref{cg9} ($%
\theta _i=90^{\circ }$) the $\theta _f\sim 270^{\circ }$ peak is dominant.
The factor $\frac{\omega _i^2}{\nu ^2\omega ^2}$ is less than one for all four figures 
accounting for the characteristic transverse differential cross section peaks.

Figures \ref{cg8} and \ref{cg9} for transverse $\theta_i$ show enhancement of one of the 
$\theta_f=90^{\circ},270^{\circ}$ peaks. Naively we would expect the final photon to carry off 
the transverse momentum enhancing the $\theta_f= 90^{\circ}$ peaks in both cases. Figure 
\ref{cg9} however shows the opposite. A more detailed analysis of the quasi-energy levels in 
terms of the bound electron propagator is required.

The energy levels can be discussed by considering the expressions for the denominators
of electron propagators for the direct and exchange channels of the
scattering. Starting from equation \ref{c3.smat.eq4} and expanding scalar products in 
terms of energy-momenta and scattering angles, the propagator denominator for each of the 
scattering channels is

\begin{equation}
\label{c4eq10}
\begin{array}{lll}
2(q_ik_i)\left[ l\,\dfrac \omega {\omega _i}\dfrac{1+\omega _i(1-\cos
\,\theta _i)}{1+\frac{\nu ^2}2(1-\cos \,\theta _i)}\;+\;1\right] &  & 
\text{direct channel} \\  &  &  \\ 
2(q_ik_f)\left[ l\,\dfrac \omega {\omega _f}\dfrac{1-\omega _f(1-\cos
\,\theta _f)}{1+\frac{\nu ^2}2(1-\cos \,\theta _f)}\;-\;1\right] &  & 
\text{exchange channel} 
\end{array}
\end{equation}

\medskip\ 

The energy levels of the exchange channel are dependent on the final scattering angle 
$\theta_f$ and the contribution to the values of the differential cross section varies with 
$\theta_f$ variation. For the inset of figure \ref{cg9} ($\theta_i=90^{\circ }$) 
the contribution is 17.2 at $\theta _f=90^{\circ }$ and 
$22.07$ at $\theta _f=277^{\circ }$. scattering in the $\theta _f\sim 270^{\circ }$ direction 
is preferred because the SCS electron moves closer to a quasi-energy level.
The relative heights of the $\theta _f\sim 90^{\circ }$ and $\theta _f\sim
270^{\circ }$ peaks in figures \ref{cg8} and \ref{cg9} are a result of two
competing physical processes. The scattering geometry and momentum
contributed by the external field, preferences scattering in the $\theta_f\sim 90^{\circ }$ 
direction, whereas the energy level structure of the SCS
electron preferences $\theta _f\sim 270^{\circ }$ scattering.

The same quasi-energy level analysis can be applied to figures \ref{cg11} - \ref{cg14} in which 
a 0.768 keV photon is incident on an electron embedded in an 0.512 keV external field.
Figures \ref{cg12} and \ref{cg13} show the $\theta_i=45^{\circ }$ and $90^{\circ}$ scattering 
geometries respectively, and reveal a $\theta _f\sim 90^{\circ }$ peak which increases with
increasing $\nu ^2$. Figure \ref{cg13} reveals a secondary peak at $\theta
_f=261^{\circ }$, however its maximum in the $\nu ^2=0.3$ plot is only $25\%$
that of the $\theta _f\sim 90^{\circ }$ peak.

The exchange channel propagator denominator is calculated for $l=2,\,\nu^2=0.3$ 
since this is the dominant term. It contributes a factor of 319.5 to the SCS
differential cross section at $\theta _f=90^{\circ }$, and 320.7 at $\theta _f=270^{\circ}$ 
for figure \ref{cg12} ($\theta _i=45^{\circ}$). For figure \ref{cg13} ($\theta_i=90^{\circ }$) 
the contribution is 289.4 at $\theta _f=90^{\circ }$ and 291 at $\theta _f=270^{\circ }$. 
The electron energy level relative weight is approximately even and the dominant scattering is 
$\theta_f=90^{\circ}$ in which the final photon carries the transverse momentum of the initial 
photon.

Figure \ref{cg11} shows the initial scattering geometry $\theta_i=0^{\circ }$, and its main 
feature is a decrease of the differential cross section
with increasing $\nu ^2$ in the angular region $90^{\circ }\leq \theta
_f\leq 270^{\circ }$. The differential cross section is minimum at $\theta_f=180^{\circ}$. 
Here the factor $\frac{\omega _i^2}{\nu ^2\omega ^2}$ is always greater then unity,
\footnote{$\frac{\omega_i^2}{\nu^2\omega^2}=22.5$ for $\nu^2=0.1$ and 
$\frac{\omega_i^2}{\nu^2\omega^2}=7.5$ for $\nu^2=0.3$} 
the incident photon $k_i$ dominates the process and the $\theta _f$ variation of the
differential cross section is similar to that of the Klein-Nishina process with
significant momentum transferred to the electron in the forward direction ($%
\theta _f=0^{\circ }$).

Figure \ref{cg14} shows the initial scattering geometry $\theta_i=180^{\circ }$, and the 
differential SCS cross section peaks increase as $\nu^2$ increases to =0.3, thereafter 
decreasing. The $\nu ^2=0.1$ plot shows characteristic
Klein-Nishina scattering with an enhanced $\theta _f=180^{\circ }$ peak. The final photon 
must carry the initial state momentum because bound electron momentum in that direction is 
inhibited by the propagation direction of the external field.
As $\nu ^2$ approaches 0.3, the cross section takes on a $\theta _f\sim
90^{\circ },270^{\circ }$ peak structure characteristic of the SCS electron
under the dominant influence of the external field. Examination of the
direct channel electron energy levels reveals that $\nu ^2=0.33$ provides the closest 
approach to a quasi-energy level. As $\nu ^2$ increases
beyond this point the scattering cross section begins to decrease again.

Figures \ref{cg15} and \ref{cg16} represent the variation of the SCS
differential cross section with the final azimuthal scattering angle $\phi_f$ 
for a 0.768 keV photon incident on the SCS electron embedded in a 0.512 keV external field. 
scattering is most favoured at $\phi_f=0^{\circ}$ and least favoured at $\phi _f=180^{\circ }$. 
The variation increases as the external field intensity increases. 
For these figures the factor  $\frac{\omega _i^2}{\nu^2\omega ^2}$ is greater than one and the 
scattering is favoured when the final photon carries the transverse momentum of the initial 
photon. This condition is most closely met at $\varphi_f=0^{\circ }$ and least met at 
$\varphi_f=180^{\circ}$.

Figures \ref{cg17} - \ref{cg20} reveal the effect on the SCS differential
cross section of changing the ratio of initial photon energy to external
field photon energy. Figures \ref{cg17} and \ref{cg18} show the SCS differential
cross section with an external field intensity of $\nu ^2=0.1$, and initial
scattering geometry $\theta _i=0^{\circ }$ and $45^{\circ }$ respectively.
Figures \ref{cg19} - \ref{cg20} show the SCS differential cross section with the
same parameters as figures \ref{cg17} and \ref{cg18} except for an increased
external field intensity $\nu ^2=0.5$.
The pivotal factor is again $\frac{\omega_i^2}{\nu^2\omega^2}$. When 
$\frac{\omega _i^2}{\nu ^2\omega ^2}$ is much greater
than unity (in the $\omega _i=12.8$ keV and $\omega _i=7.68$ keV plots of
figures \ref{cg17} and \ref{cg18}), the scattering is characteristic of a
Klein-Nishina-like process (i.e. primary peak at $\theta _f=\theta _i$ and a
flattened, secondary peak at $\theta _f=\theta _i+180^{\circ }$). When $
\frac{\omega _i^2}{\nu ^2\omega ^2}$ is much less than unity (in the 
$\omega_i=0.409$ keV and $\omega _i=0.205$ keV plots of figure \ref{cg19} and the 
$\omega _i=0.205\;keV$ plot of figure \ref{cg20}), the scattering is
characteristic of the SCS electron dominated by the circular polarised
external field (i.e. double peaks at $\theta _f\sim 90^{\circ },270^{\circ }$
for $\theta _i=0^{\circ }$ and a single peak at $\theta _f\sim 90^{\circ }$
for $\theta _i=45^{\circ }$).

Figures \ref{cg21} and \ref{cg22} represent the variation of the SCS
differential cross section with external field intensity $\nu ^2$ for various
initial photon angles. Figure \ref{cg21} shows a 40.9 keV photon incident on an electron 
embedded in a $51.2\;keV$ external field.
Figure \ref{cg22} shows a 0.768 keV photon incident on an electron
embedded in a $0.512\;keV$ external field. The vertical axes of these
figures have a logarithmic scale.
Except for the $\theta _i=0^{\circ }$ plot of figure \ref{cg22}, all plots
show an increase in the SCS differential cross section with an increase in $\nu ^2$. This
SCS differential cross section increase is physically justified by the increase in
external field photon number density associated with larger values of the
external field intensity parameter. The probability that more external field photons will take 
part in the process also increases.

A limiting factor in all experimental work on these external field
phenomena, is the intensity of the external field. It is possible that at
the highest external field intensities attainable at present second order IFQED 
cross sections may become large enough to be experimentally detected. 
Consideration of the SCS differential cross section for the scattering geometry $\theta 
_i=0^{\circ }$, results in analytic expressions which are considerably simplified 
\cite{AkhMer85}. However the results of this chapter indicated that the 
$\theta _i=0^{\circ }$ geometry produces small differential cross section values which are 
least likely to be experimentally detected. Indeed for some scattering geometries the SCS 
cross section diminishes with increasing external field intensity. Therefore our full
analytic expressions and numerical analysis for general kinematics will become important for 
future experimental work.

%% file: tex/tables/c_table1.tex
\begin{table}[!b]
\label{ctab2}
\begin{tabular}{|c|c|c|c|c|c|c|c|c|} \hline
$l$&$r$&$ \nu^{2}$ & $ \omega (\text{keV})$ & $ \omega_{i}(\text{keV})$ &$ \theta_{i}$ & 
$ (\theta_{f},\phi)$ & \;\; figure(s) \;\; \\ \hline\hline

$0\rightarrow 20 $&$ \text{all} $&$ 0.1,0.5,1.0 $&$ 0.001 $&$ 0.001 $&
$ 10^{\circ } $&$ (45^{\circ },0^{\circ }) $& \ref{cga1} \\ \hline

$0\rightarrow 20 $&$ \text{all} $&$ 
0.1,0.5,1.0 $&$ 0.512 $&$ 0.1 $&$ 80^{\circ } $&$ (45^{\circ },0^{\circ }) $&
 \ref{cga2} \\ \hline

$-7\rightarrow 23 $&$ \text{all} $&$ 0.1,0.5,1.0 $&$ 0.512 $&$ 7.68 $&$ 
10^{\circ } $&$ (45^{\circ },0^{\circ }) $& \ref{cga3} \\ \hline

$-4\rightarrow 12 $&$ 
\text{all} $&$ 0.1,0.5,1.0 $&$ 0.512 $&$ 2.047 $&$ 30^{\circ } $&$ (45^{\circ
},0^{\circ }) $& \ref{cga4} \\ \hline

$0 $&$ 0 $&$ 
\begin{array}{c}
0,0.1,0.2 \\
0.3,0.4,0.5
\end{array}
$&$ 0.512 $&$ 0.409,0.768 $&$ 
0^{\circ } $&$ (0^{\circ }\rightarrow 360^{\circ },0^{\circ }) $&
\ref{cgb1},\ref{cgb2} \\ \hline 

$0 $&$ 0,\text{all} $&$ 0.1$&$ 0.512 $&$ 0.409,0.768 $&$ 0^{\circ } $&$ 
(0^{\circ }\rightarrow 360^{\circ},0^{\circ }) $&
\ref{cgb4a},\ref{cgb4b} \\ \hline

$0,1 $&$ \text{all} $&$ 0.1 $&$ 0.512 $&$ 25.6 $&$ 0^{\circ } $&$ 
(0^{\circ }\rightarrow 360^{\circ },0^{\circ }) $& \ref{cgb5} \\ \hline

$0,1,2 $&$ \text{all} $&$ 0.5 $&$ 0.512 $&$ 0.768 $&$ 0^{\circ } $&$ 
(0^{\circ }\rightarrow 360^{\circ },0^{\circ }) $& \ref{cgb6} \\ \hline

$0,1,2 $&$ \text{all} $&$ 0.5 $&$ 0.512 $&$ 0.409 $&$ 0^{\circ } $&$ (0^{\circ
}\rightarrow 360^{\circ },0^{\circ }) $& \ref{cgb7} \\ \hline

$0,1,2 $&$ \text{all} $&$ 
0.2 $&$ 51.2 $&$ 5.12 $&$ 0^{\circ } $&$ (0^{\circ }\rightarrow 360^{\circ
},0^{\circ }) $& \ref{cgb8} \\ \hline

$0,1 $&$ \text{all} $&$ 
\begin{array}{c}
0,0.1,0.2, \\ 
0.3,0.4,0.5
\end{array}
$&$ 0.512 $&$ 0.409,0.768 $&$ 45^{\circ } $&$ (0^{\circ }\rightarrow 
360^{\circ },0^{\circ}) $&$ 
\begin{array}{c} 
\ref{cgb9},\ref{cgb10} \\ 
\ref{cgb11},\ref{cgb12} 
\end{array}$ \\ \hline

$0,1,2 $&$ \text{all} $&$ 0.1,0.5 $&$ 0.512 $&$ 0.768 $&$ 45^{\circ } $&$ 
(0^{\circ}\rightarrow 360^{\circ },0^{\circ }) $&
\ref{cgb13},\ref{cgb14} \\ \hline

$0,1,2 $&$ \text{all} $&$ 0.1,0.5 $&$ 0.512 $&$ 0.05,5.12 $&$ 45^{\circ } $&$ (0^{\circ
}\rightarrow 360^{\circ },0^{\circ }) $& \ref{cgb15},\ref{cgb16} \\ \hline
\end{tabular}
\caption{\bf\bm The parameter range for which the SCS differential cross section
$l$ and $r$ contributions are investigated.}
\end{table}

%% file: tex/tables/c_table2.tex
\begin{table}[!h]
\label{ctab1}
\begin{tabular}{|c|c|c|c|c|c|c|} \hline
$ \nu^{2}$ & $ \omega (keV)$ & $ \omega_{i}(keV)$ &$ \theta_{i}$ & 
$ (\theta_{f},\phi)$ & figure(s) \\ \hline\hline

$0.1,0.5,1.0 $&$ 0.005 $&$ 0.026 $&$ 0^{\circ },15^{\circ },30^{\circ } $&$ \left(
0^{\circ }\rightarrow 360^{\circ },0^{\circ }\right)  $&
\ref{cg1},\ref{cg2},\ref{cg3} \\ \hline

$0.1,0.5,1.0 $&$ 0.512 $&$ 0.409 $&$ 0^{\circ },60^{\circ },90^{\circ } $&$ 
\left( 0^{\circ }\rightarrow 360^{\circ },0^{\circ }\right)  $& \ref{cg4},\ref
{cg5},\ref{cg6} \\ \hline

$\begin{array}{l}
0,0.1,0.2, \\ 
0.3,0.4,0.5
\end{array}
$&$ 51.2 $&$ 10.2 $&$ 
\begin{array}{l}
0^{\circ },45^{\circ }, \\ 
90^{\circ },180^{\circ }
\end{array}
$&$ \left( 0^{\circ }\rightarrow 360^{\circ },0^{\circ }\right)  $&$
\begin{array}{l}
\ref{cg7},\ref{cg8}, \\ \ref{cg9},\ref{cg10}
\end{array} $
\\ \hline

$\begin{array}{l}
0,0.1,0.2, \\ 
0.3,0.4,0.5
\end{array}
$&$ 0.512 $&$ 0.768 $&$ 
\begin{array}{l}
0^{\circ },45^{\circ }, \\ 
90^{\circ },180^{\circ }
\end{array}
$&$ \left( 0^{\circ }\rightarrow 360^{\circ },0^{\circ }\right)  $&$
\begin{array}{l}
\ref{cg11},\ref{cg12}, \\ \ref{cg13},\ref{cg14}
\end{array} $
\\ \hline

$0.1,0.5,1.0 $&$ 0.512 $&$ 0.768 $&$ 45^{\circ } $&$ \left( 45^{\circ },0^{\circ
}\rightarrow 360^{\circ }\right)  $& \ref{cg15} \\ \hline

$0.1,0.3,0.4 $&$ 0.512 $&$ 0.768 $&$ 
90^{\circ } $&$ \left( 45^{\circ },0^{\circ }\rightarrow 360^{\circ }\right) 
$& \ref{cg16} \\ \hline

$0.1,0.5 $&$ 0.512 $&$ 
\begin{array}{l}
2.56,0.768, \\ 
0.409,0.205
\end{array}
$&$ 0^{\circ },45^{\circ } $&$ \left( 0^{\circ }\rightarrow 360^{\circ
},0^{\circ }\right)  $&$
\begin{array}{l}
\ref{cg17},\ref{cg18}, \\ \ref{cg19},\ref{cg20}
\end{array} $
\\ \hline

$0\rightarrow 2 $&$ 51.2 $&$ 0.409 $&$ 0^{\circ },5^{\circ },10^{\circ
},15^{\circ } $&$ \left(45^{\circ },0^{\circ }\right)  $& \ref{cg21} \\ \hline 

$0\rightarrow 2 $&$ 0.512 $&$ 0.768 $&$ 0^{\circ },15^{\circ },30^{\circ }
,45^{\circ } $&$ \left(45^{\circ },0^{\circ }\right)  $& \ref{cg22} \\ \hline
\end{tabular}
\caption{\bf\bm The parameter range for which the SCS differential cross section
summed over all $l$ is investigated.}

\end{table}

%% file: tex/tables/c_table4.tex
\begin{table}[!b]
\center{
\begin{tabular}{|c|c|c|c|c|c|c|} \hline
 figure & \ref{cg1} & \ref{cg2} & \ref{cg3} & \ref{cg4} & \ref{cg5} &
 \ref{cg6} \\ \hline\hline
peak maximum & 58 & 1700 & 25000 & 32.5 & 138 & 1450 \\ \hline
$ \nu^2 $ & 0.1 & 0.5 & 1 & 0.1 & 0.5 & 1 \\ \hline
\end{tabular} }
\caption{\bf\bm Peak maximums for the $\theta_i=30^\circ$ plots of figures 
\ref{cg1} - \ref{cg3}, and the $\theta_i=90^\circ$ plots of figures \ref{cg4} - \ref{cg6}.}
\label{c4.table4}
\end{table}

%% file: chap5.tex
\section{Introduction}

In this chapter we present numerical calculations of the STPPP differential
cross section obtained in section \ref{c3pprod}. The numerical
results are contained in section 5.2 and an analysis of these results is
contained in section 5.3.

We preface our remarks in this chapter by drawing a comparison with the data
and analysis presented for the SCS process.  The order of data presented 
and the phenomena highlighted is similar to that presented in Chapter 4. This is
the case because the external electromagnetic field is identical and the 
STPPP process is related to the SCS process through the Substitution Law.

Also in similarity to the SCS process, the STPPP differential cross section
is expressed as a triple infinite summation over integer variables $l$, $r$
and $r^{\prime }$. Analysis is again facilitated by dividing figures into
two groups; those representing summation terms of the STPPP differential
cross section separately or summed in part (section 5.2.1 with the
accompanying analysis in section 5.3.1), and those in which all
contributions have been summed over (section 5.2.2 with the accompanying
analysis in section 5.3.2). Sets of figures with common parameter presentations will be 
introduced and analysed in turn.

A FORTRAN program was used to perform calculational work in this chapter.
It was constructed from a similar FORTRAN program used in chapter 4 using the 
Substitution Law to make a simple reassignment of program variables.

\newpage
\subsection{Differential cross section $l$ contributions}
\enlargethispage*{4cm}

Table \ref{c52tab1} displays the parameter values of the STPPP process investigated in
this section. The parameter $\omega $ is the energy of external field
quanta. The parameters $\omega_1$ and $\omega_2$ are the particle energies of the 
initial photons and are of equal magnitude for the centre of mass reference frame in 
which the STPPP differential cross section expressions were calculated.
The initial photon $\,\sql{k}_1$ enters the interaction region at an angle of $\theta_1$ 
to the direction of the propagation of the external field. 
The direction of the created electron $\,\squ{q}_-$ is specified by the final 
polar angle $\theta_f$ and the final azimuthal scattering angle $\phi _f$. The external 
field intensity is represented by the parameter $\nu ^2=\frac{e^2a^2}{m^2}$.

The vertical axes of all figures in Sections 5.2.1 and 5.2.2 show the
STPPP differential cross section divided by a function of the fine structure
constant $\alpha $ and electron mass $m$, and the units are Steradian$^{-1}$. 
All angles represented on horizontal axes use units of degrees.

\input{./tex/tables/p_table1} 
\clearpage
\begin{figure}[H]
 \centerline{\includegraphics[height=8cm,width=15cm]{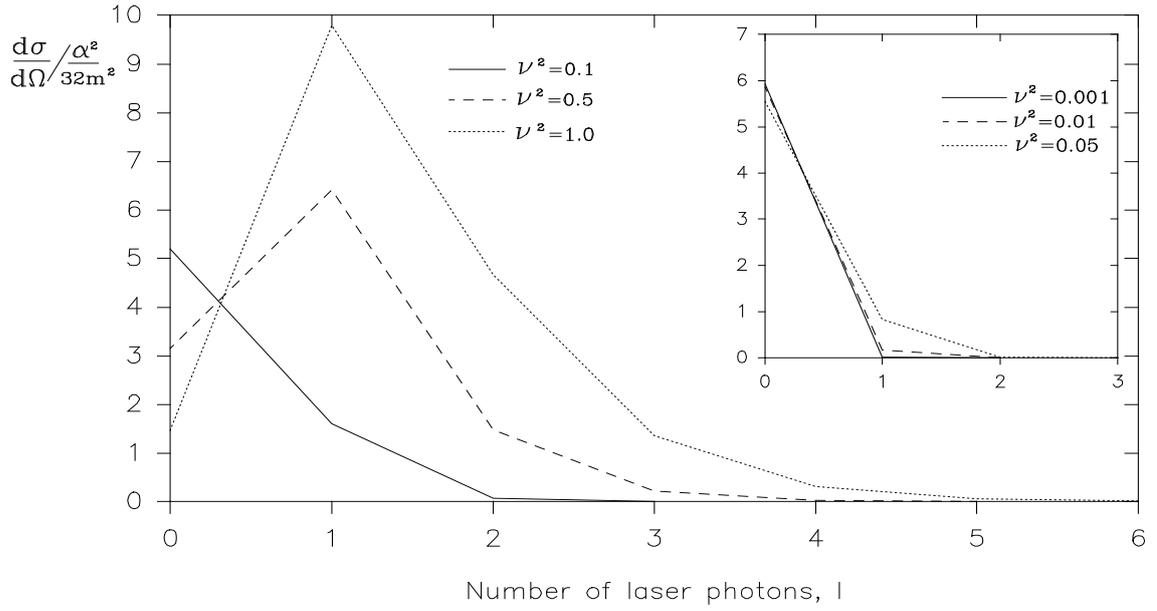}}
\caption{\bf\bm The STPPP differential cross section vs $l$ external field 
photons for $\,\omega=2.56$ MeV, $\omega_1$,$\omega_2=0.768$ MeV, $\theta_1=0^{\circ}$, 
$\theta_f=45^{\circ}$, $\varphi_f=0^{\circ}$ and various $\nu^2$.}
\label{pga1}
\end{figure}

\begin{figure}[H]
 \centerline{\includegraphics[height=8cm,width=15cm]{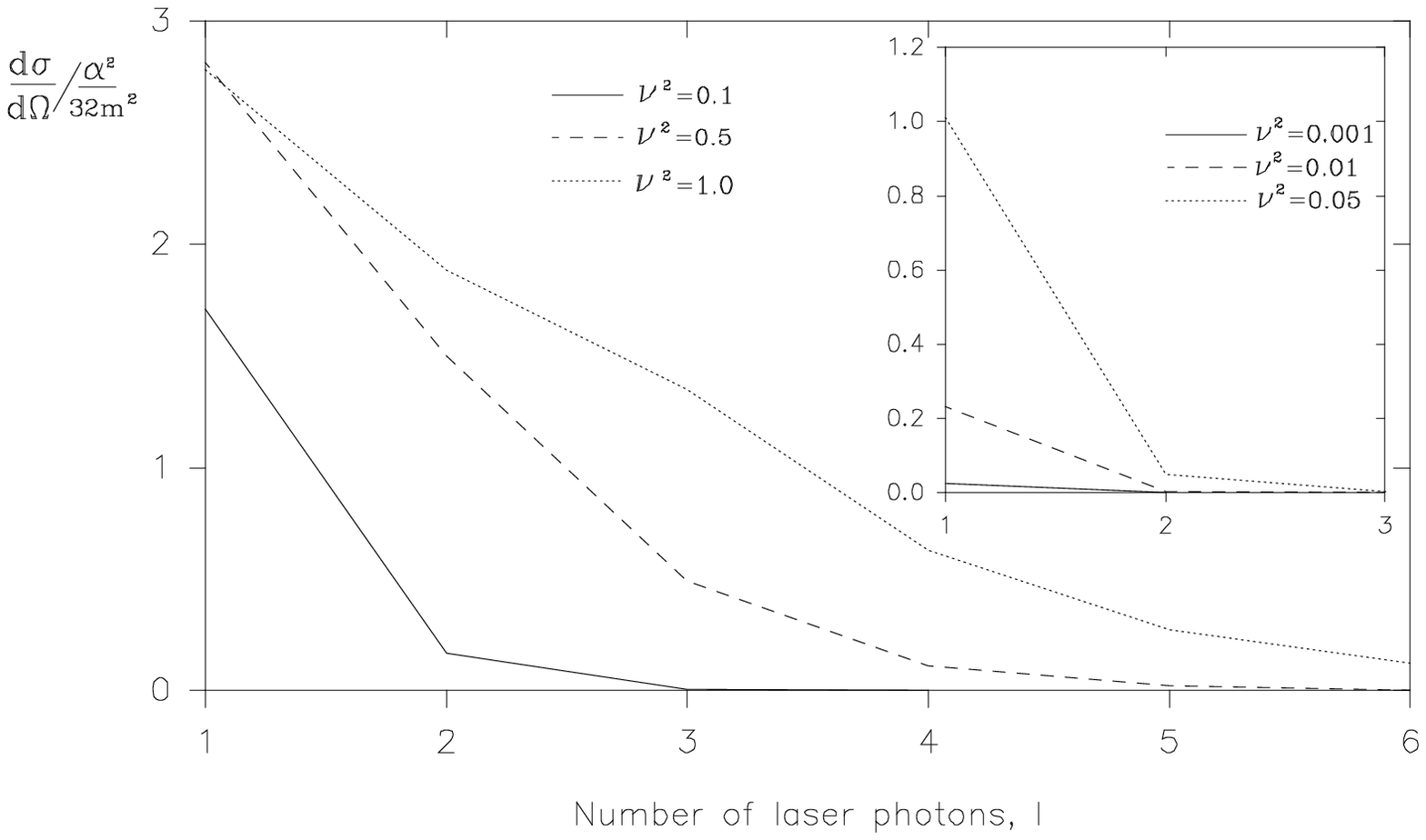}}
\caption{\bf\bm The STPPP differential cross section vs $l$ external field 
photons for $\,\omega=1.024$ MeV, $\omega_1$,$\omega_2=0.512$ MeV, $\theta_1=45^{\circ}$, 
$\theta_f=45^{\circ}$, $\varphi_f=0^{\circ}$ and various $\nu^2$.}
\label{pga2}
\end{figure}

\clearpage

\begin{figure}[H]
 \centerline{\includegraphics[height=8cm,width=15cm]{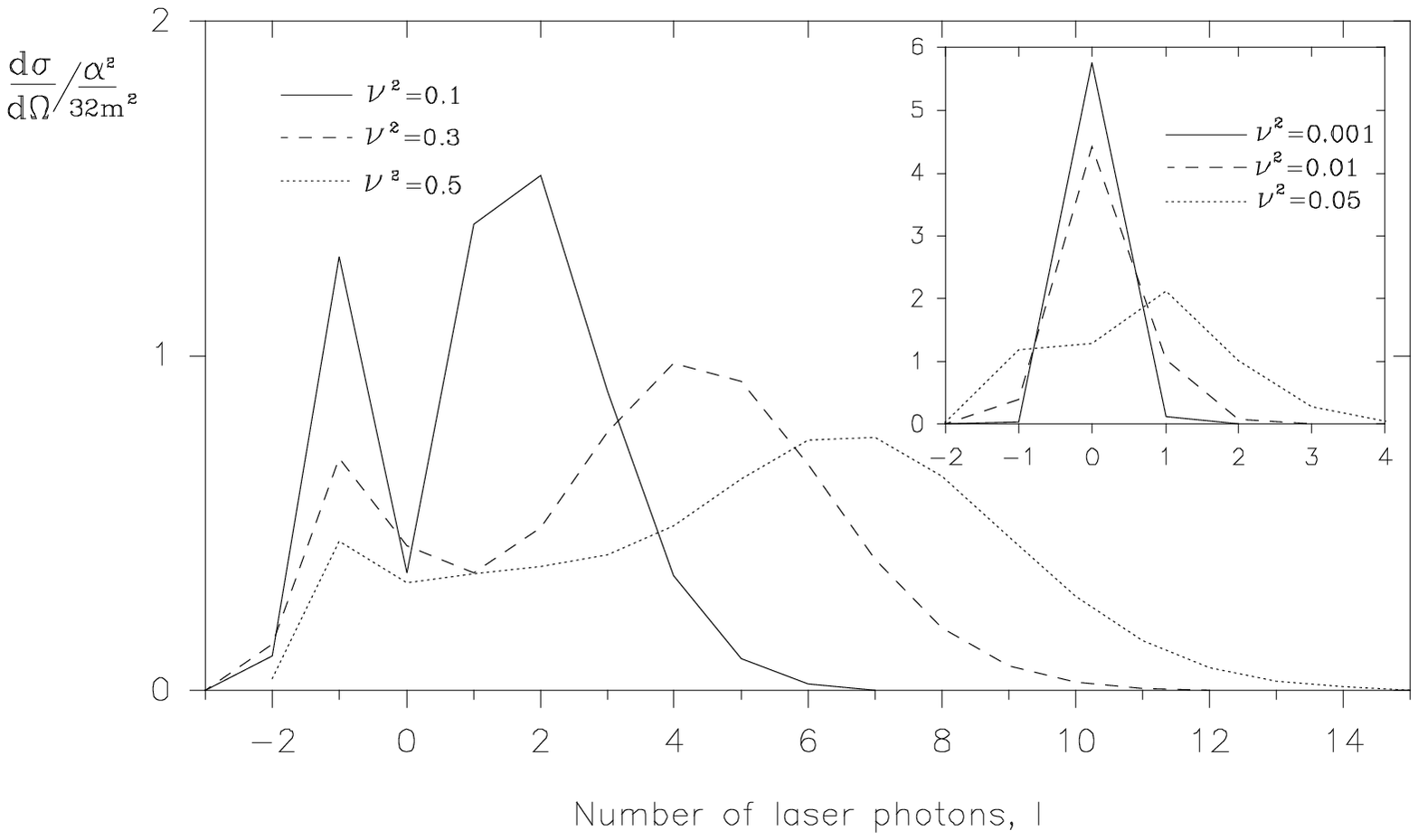}}
\caption{\bf\bm The STPPP differential cross section vs $l$ external field 
photons for $\,\omega=0.102$ MeV, $\omega_1$,$\omega_2=0.768$ MeV, $\theta_1=0^{\circ}$, 
$\theta_f=45^{\circ}$, $\varphi_f=0^{\circ}$ and various $\nu^2$.}
\label{pga3}
\end{figure}

\begin{figure}[H]
 \centerline{\includegraphics[height=8cm,width=15cm]{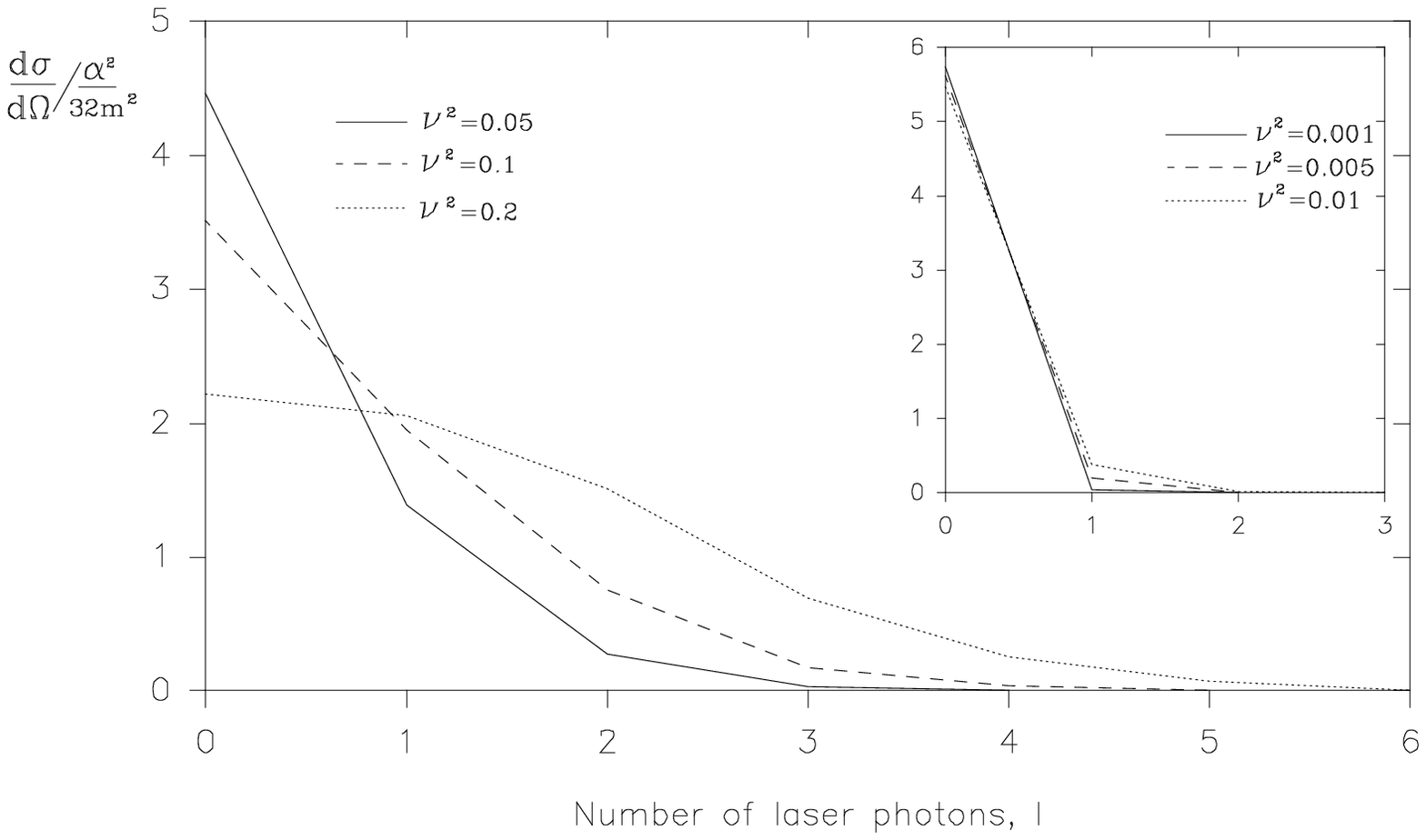}}
\caption{\bf\bm The STPPP differential cross section vs $l$ external field 
photons for $\,\omega=0.256$ MeV, $\omega_1$,$\omega_2=0.614$ MeV, $\theta_1=45^{\circ}$, 
$\theta_f=45^{\circ}$, $\varphi_f=0^{\circ}$ and various $\nu^2$.}
\label{pga4}
\end{figure}

\clearpage

\begin{figure}[t]
 \centerline{\includegraphics[height=8cm,width=10cm]{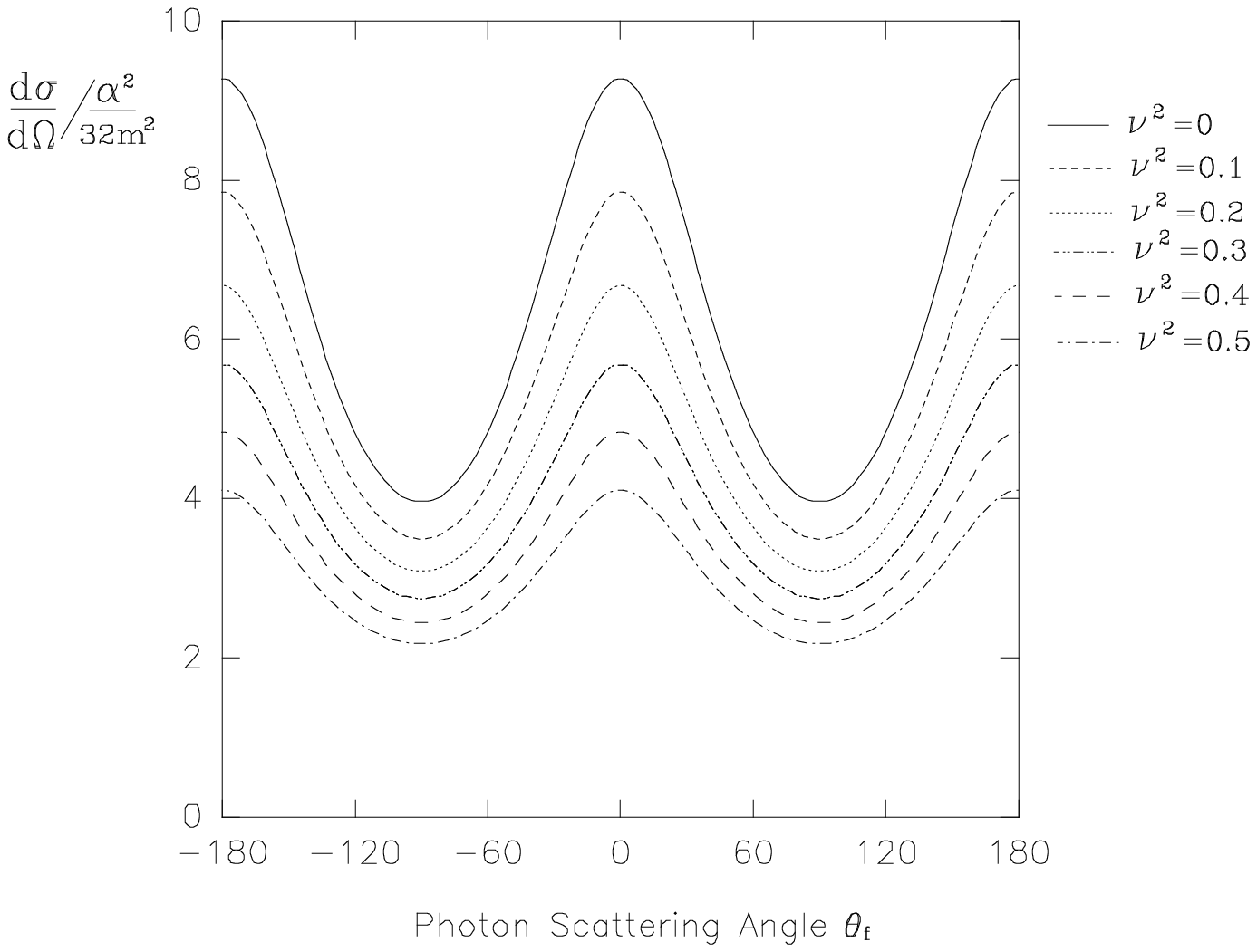}}
\caption{\bf\bm The STPPP $l=0,r=0$ differential cross section vs $\theta_f$ for 
$\,\omega=2.56$ MeV, $\omega_1$,$\omega_2=0.768$ MeV, $\theta_1=0^{\circ}$,
$\varphi_f=0^{\circ}$ and various $\nu^2$.}
\label{pgb1}
\end{figure}

\begin{figure}[t]
 \centerline{\includegraphics[height=8cm,width=10cm]{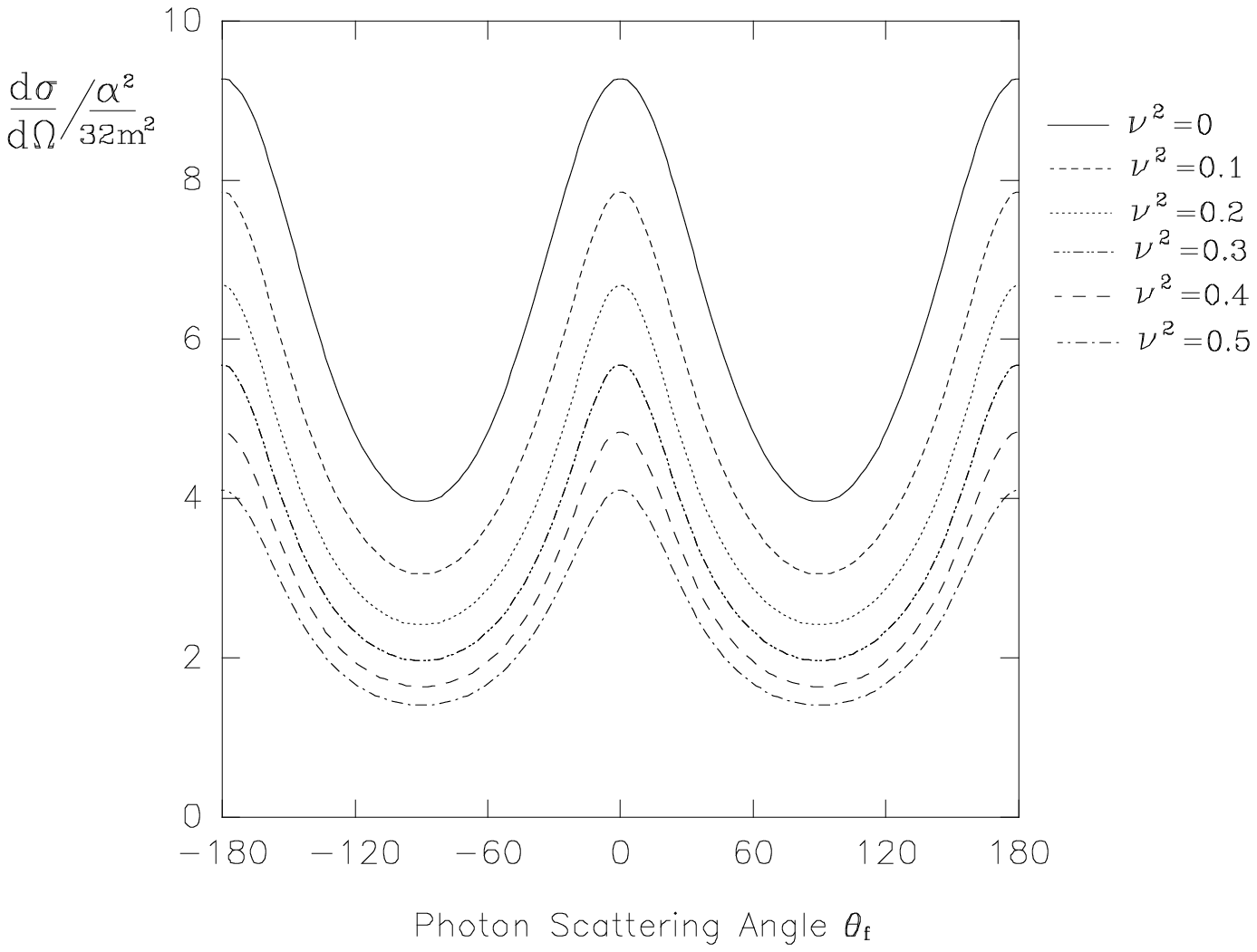}}
\caption{\bf\bm The STPPP $l=0,r=0$ differential cross section vs $\theta_f$ for
$\,\omega=0.102$ MeV, $\omega_1$,$\omega_2=0.768$ MeV, $\theta_1=0^{\circ}$,
$\varphi_f=0^{\circ}$ and various $\nu^2$.}
\label{pgb2}
\end{figure}

\begin{figure}[t]
 \centerline{\includegraphics[height=8cm,width=10cm]{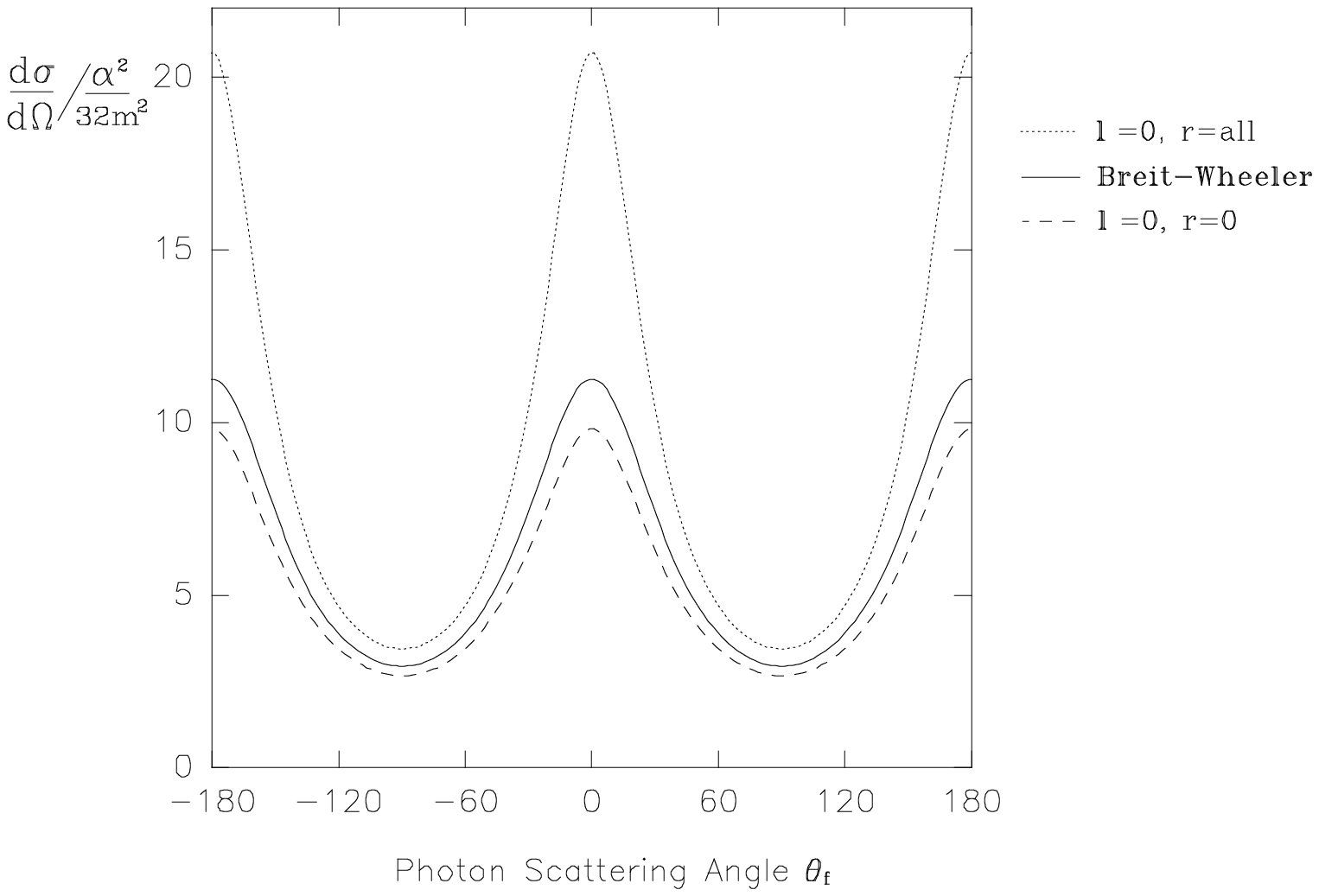}}
\caption{\bf\bm Comparison of The STPPP $l=0,r=0$, $l=0,r=\text{all}$ and Breit-Wheeler 
differential cross section vs $\theta_f$ for $\,\omega=2.56$ MeV, $\omega_1$,$\omega_2=0.768$ 
MeV, $\theta_1=0^{\circ}$, $\varphi_f=0^{\circ}$ and $\nu^2=0.1$.}
\label{pgb4a}
\end{figure}

\begin{figure}[t]
 \centerline{\includegraphics[height=8cm,width=10cm]{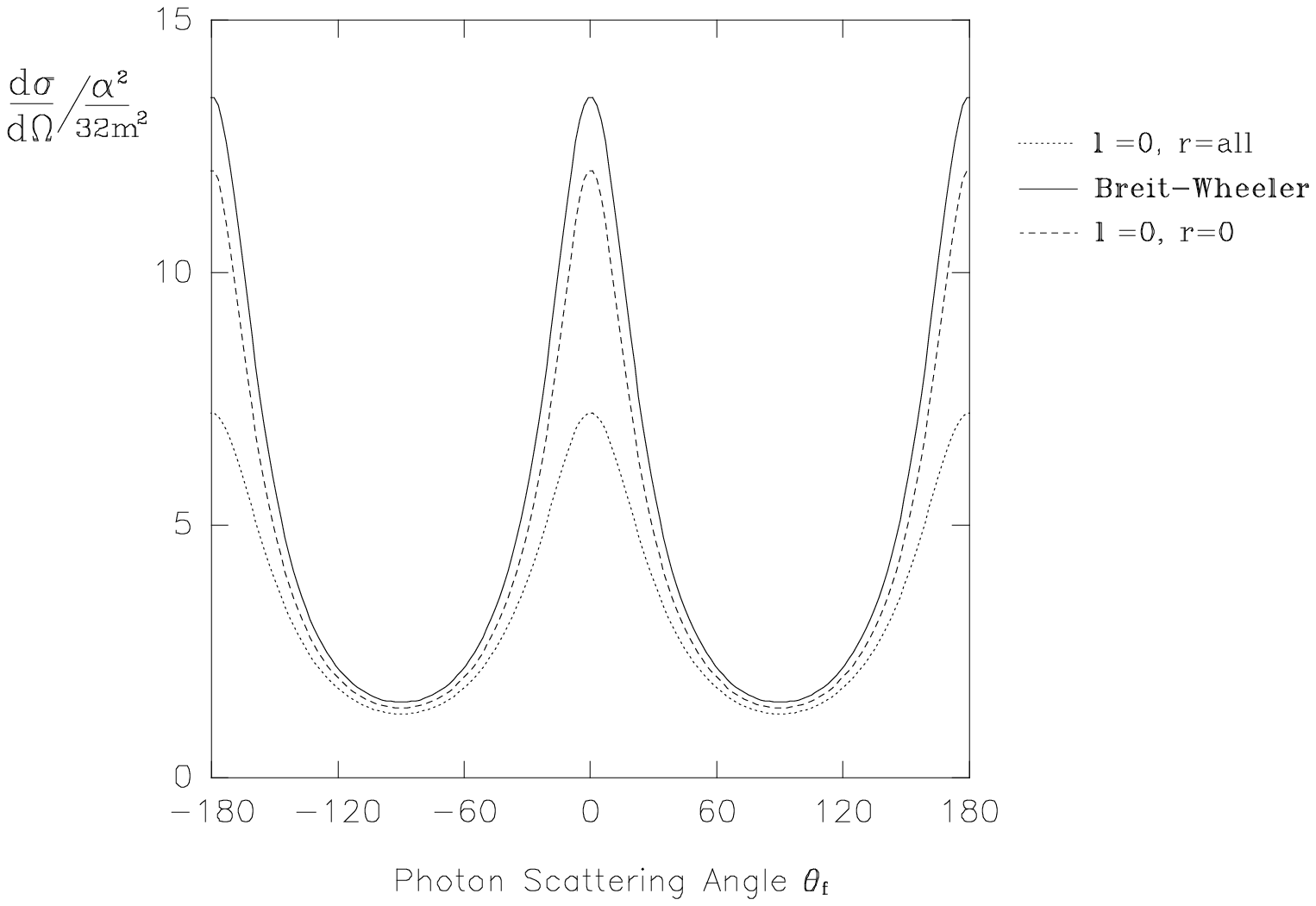}}
\caption{\bf\bm Comparison of The STPPP $l=0,r=0$, $l=0,r=\text{all}$ and Breit-Wheeler
differential cross section vs $\theta_f$ for $\,\omega=0.102$ MeV, $\omega_1$,$\omega_2=0.768$
MeV, $\theta_1=0^{\circ}$, $\varphi_f=0^{\circ}$ and $\nu^2=0.1$.}
\label{pgb4b}
\end{figure}

\clearpage

\begin{figure}[t]
 \centerline{\includegraphics[height=8cm,width=10cm]{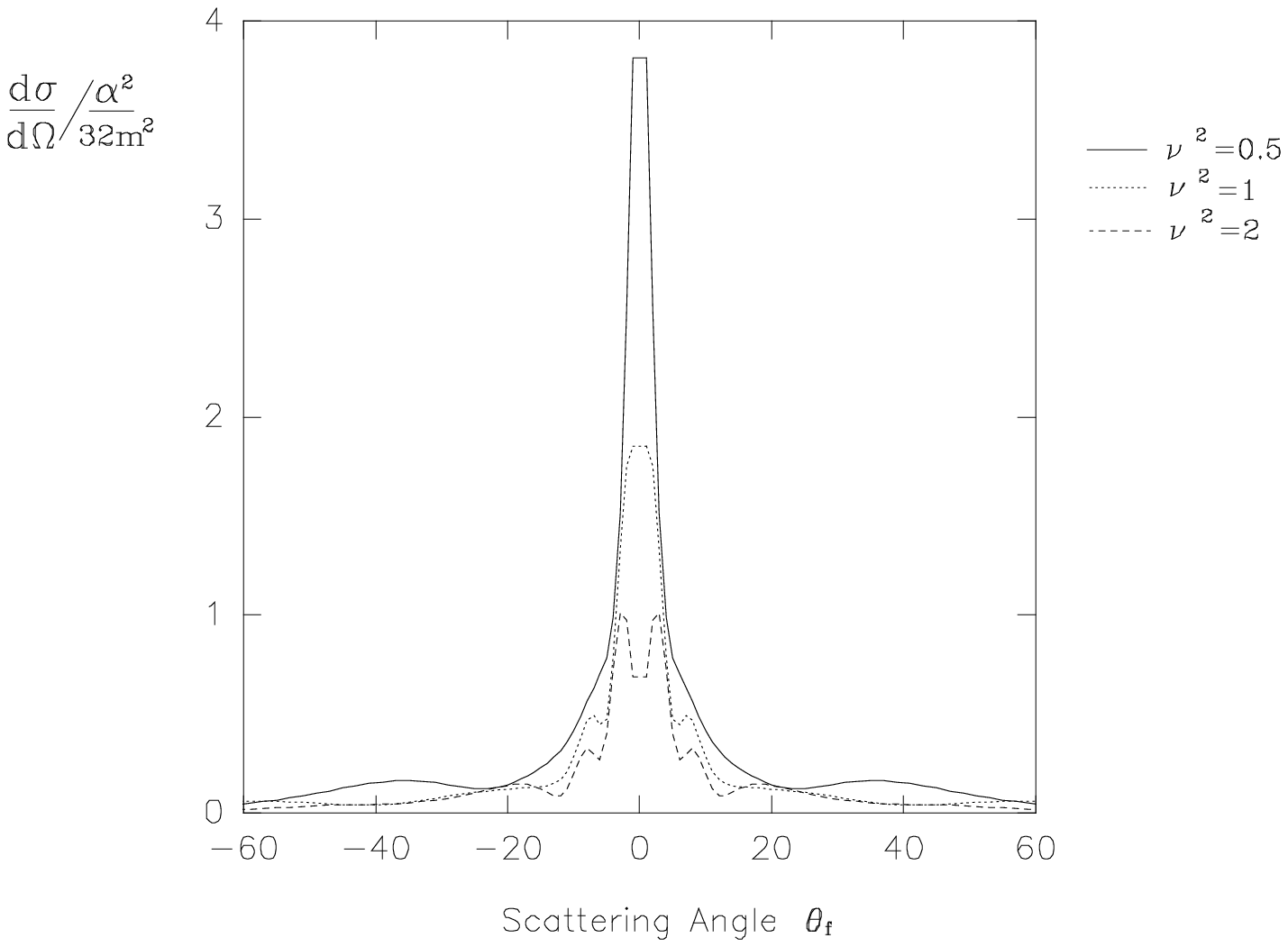}}
\caption{\bf\bm The STPPP $l=0$ differential cross section vs $\theta_f$ for
$\,\omega=0.256$ MeV, $\omega_1$,$\omega_2=2.56$ MeV, $\theta_1=0^{\circ}$,
$\varphi_f=0^{\circ}$ and various $\nu^2$.}
\label{pgb4e}
\end{figure}

\begin{figure}[t]
 \centerline{\includegraphics[height=8cm,width=10cm]{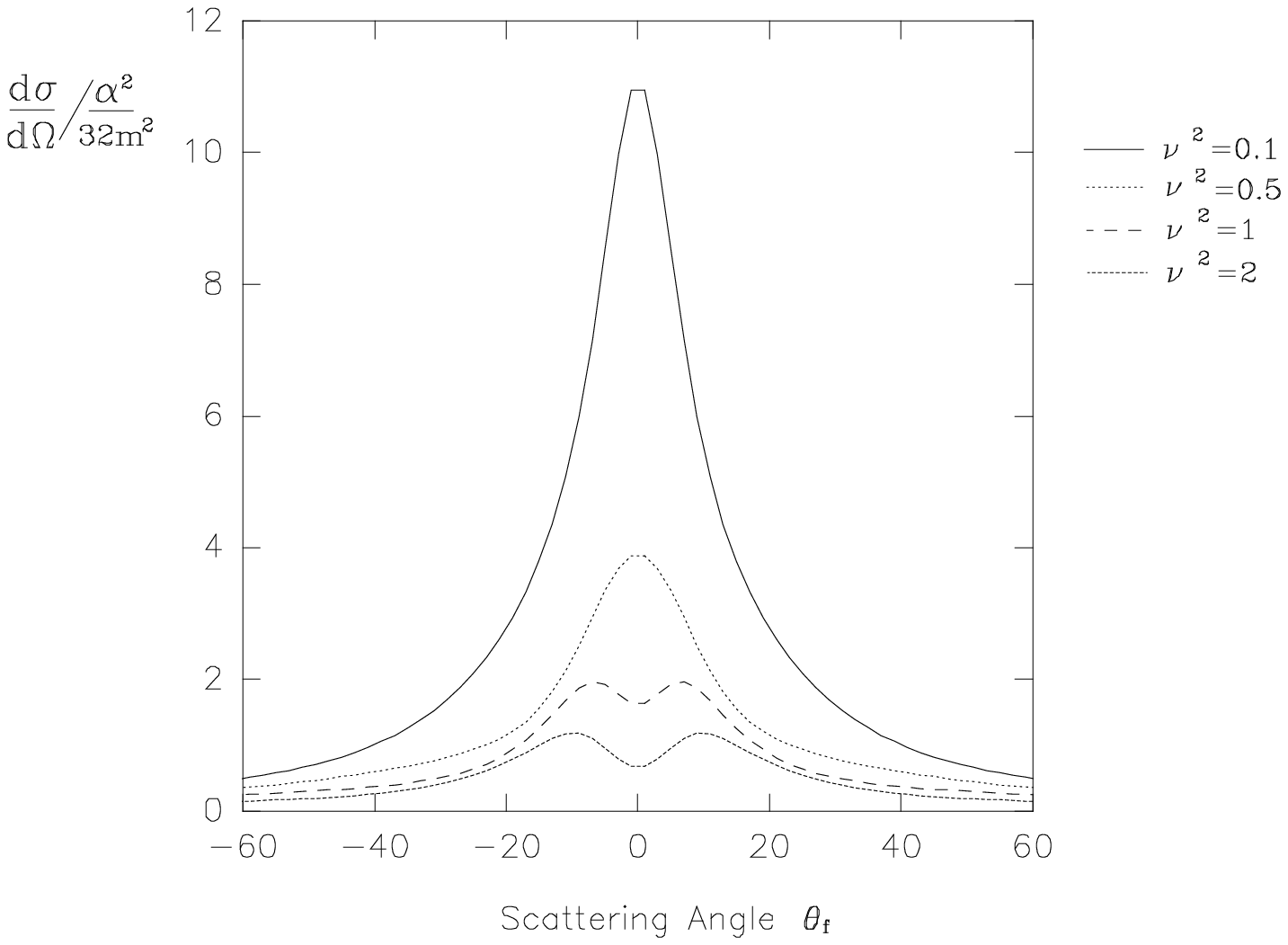}}
\caption{\bf\bm The STPPP $l=0$ differential cross section vs $\theta_f$ for
$\,\omega=1.02$ MeV, $\omega_1$,$\omega_2=2.56$ MeV, $\theta_1=0^{\circ}$,
$\varphi_f=0^{\circ}$ and various $\nu^2$.}
\label{pgb4c}
\end{figure}

\begin{figure}[t]
 \centerline{\includegraphics[height=8cm,width=10cm]{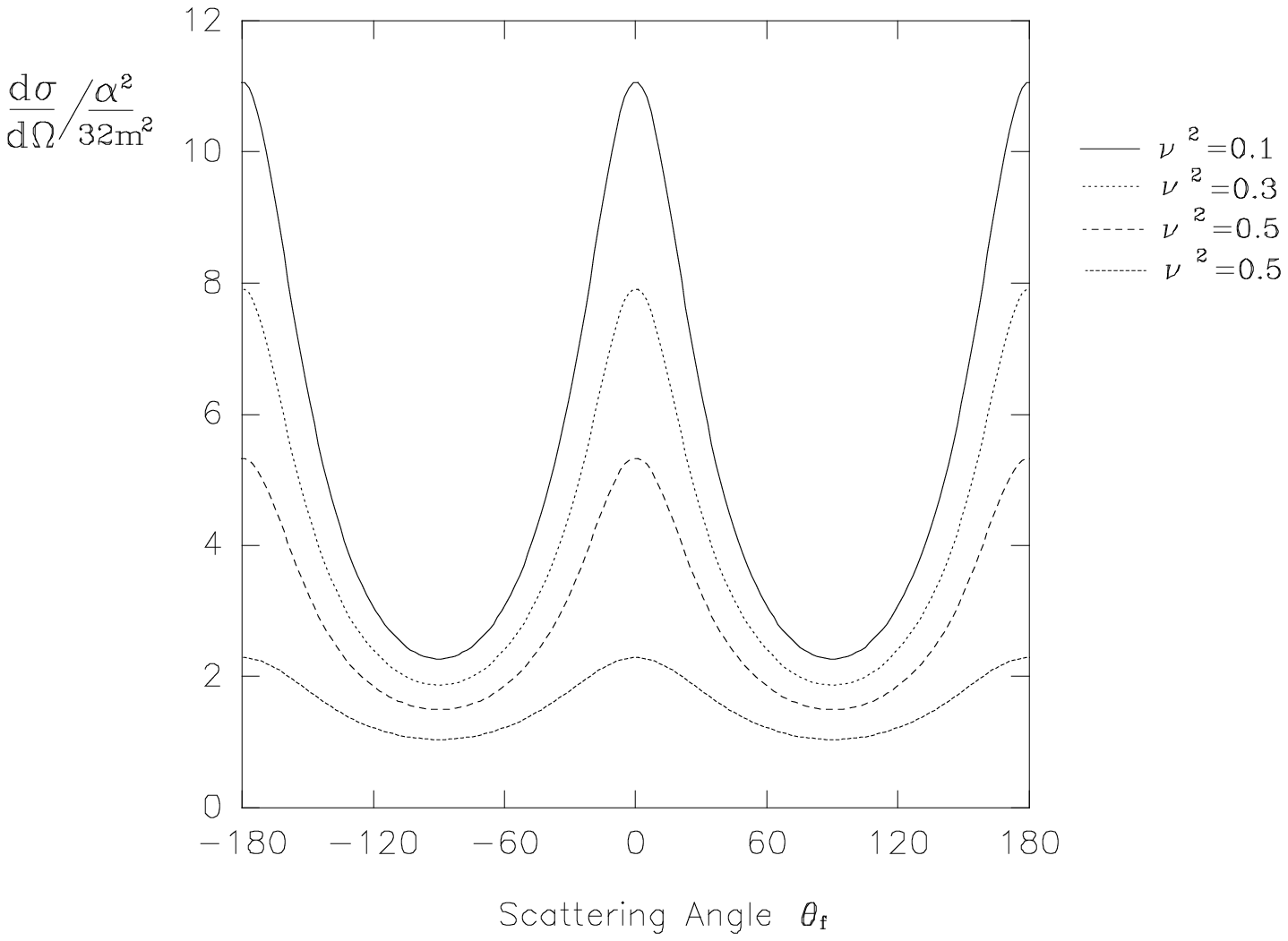}}
\caption{\bf\bm The STPPP $l=0$ differential cross section vs $\theta_f$ for
$\,\omega=1.02$ MeV, $\omega_1$,$\omega_2=0.409$ MeV, $\theta_1=0^{\circ}$,
$\varphi_f=0^{\circ}$ and various $\nu^2$.}
\label{pgb4d}
\end{figure}

\begin{figure}[t]
 \centerline{\includegraphics[height=8cm,width=10cm]{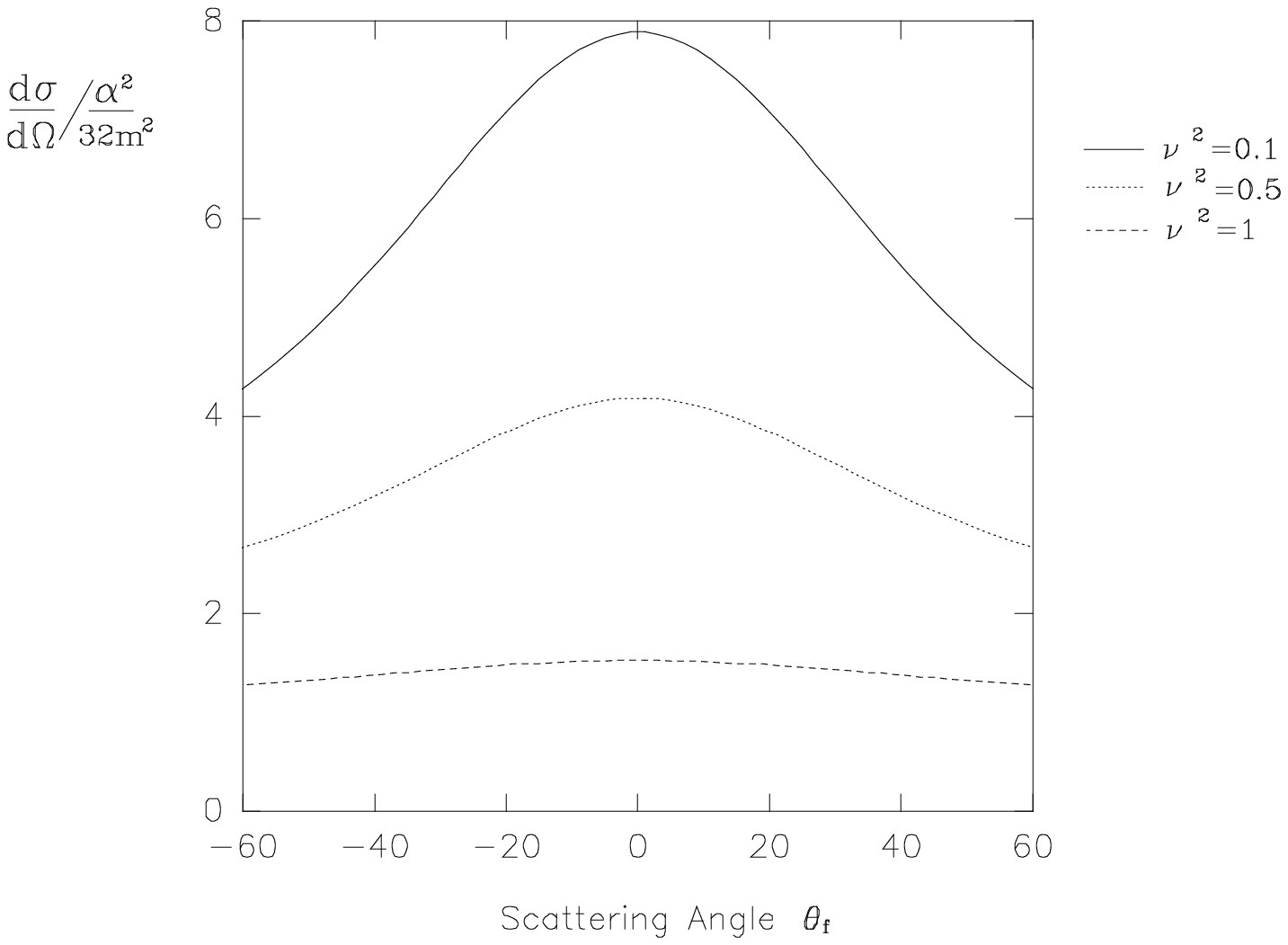}}
\caption{\bf\bm The STPPP $l=0$ differential cross section vs $\theta_f$ for
$\,\omega=5.12$ MeV, $\omega_1$,$\omega_2=0.768$ MeV, $\theta_1=0^{\circ}$,
$\varphi_f=0^{\circ}$ and various $\nu^2$.}
\label{pgb4f}
\end{figure}

\begin{figure}[t]
 \centerline{\includegraphics[height=8cm,width=15cm]{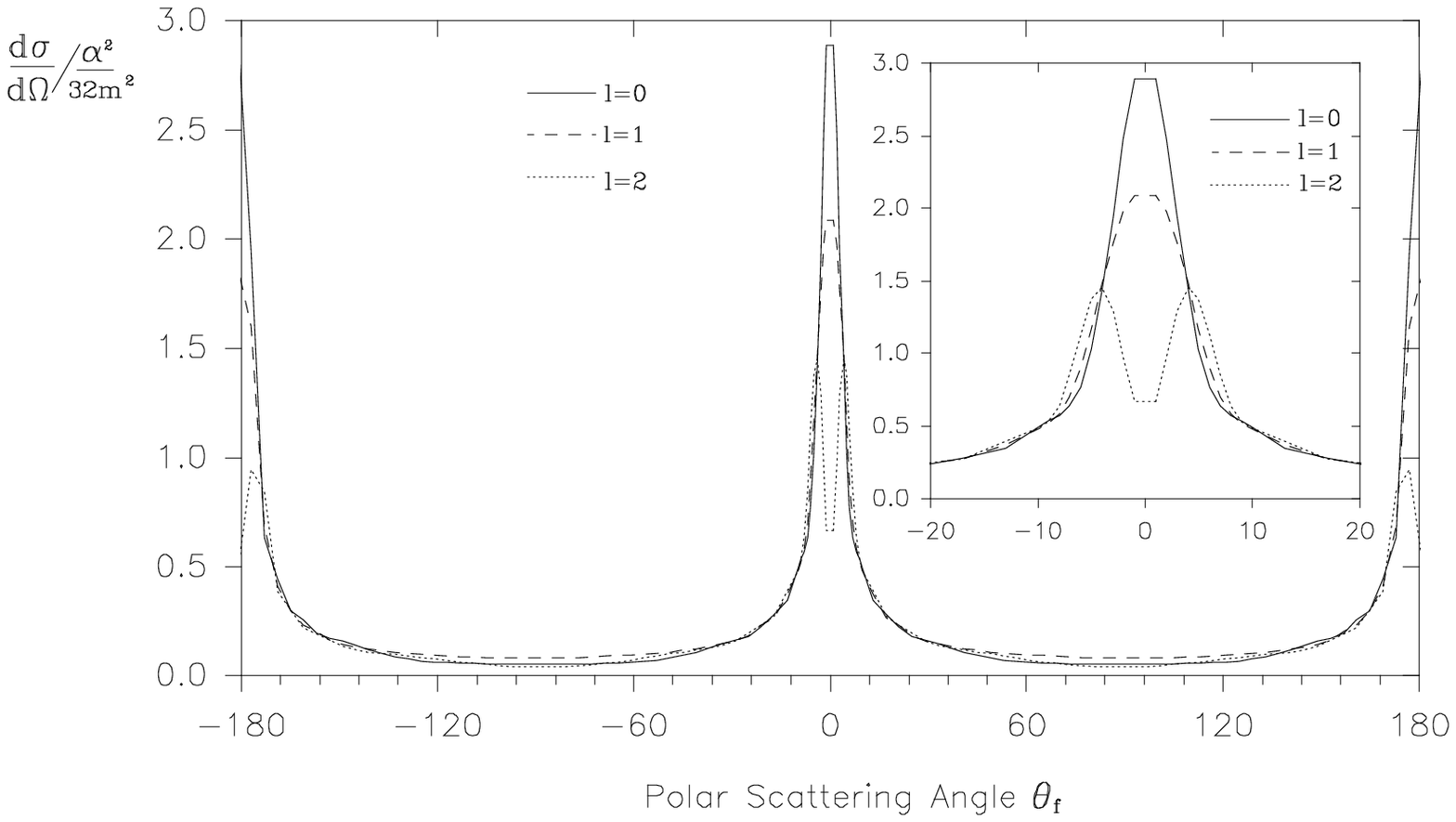}}
\caption{\bf\bm The STPPP differential cross section vs $\theta_f$ for
$\,\omega=0.05$ MeV, $\omega_1$,$\omega_2=1.024$ MeV, $\theta_1=0^{\circ}$,
$\varphi_f=0^{\circ}$ $\nu^2=0.5$ and various $l$.}
\label{pgb5}
\end{figure}

\begin{figure}[t]
 \centerline{\includegraphics[height=8cm,width=15cm]{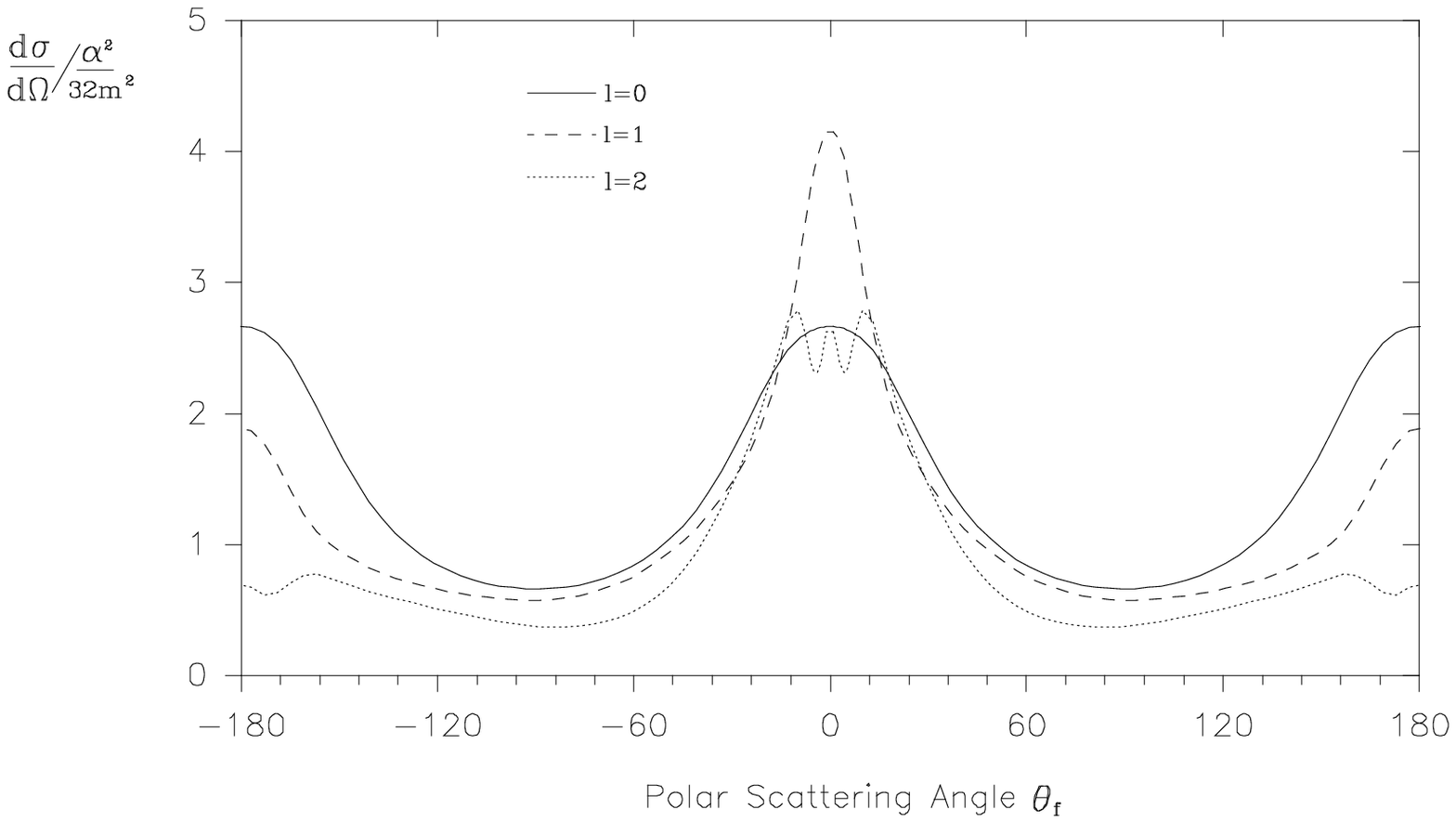}}
\caption{\bf\bm The STPPP differential cross section vs $\theta_f$ for
$\,\omega=0.41$ MeV, $\omega_1$,$\omega_2=1.024$ MeV, $\theta_1=0^{\circ}$,
$\varphi_f=0^{\circ}$ $\nu^2=0.5$ and various $l$.}
\label{pgb6}
\end{figure}

\begin{figure}[t]
 \centerline{\includegraphics[height=8cm,width=15cm]{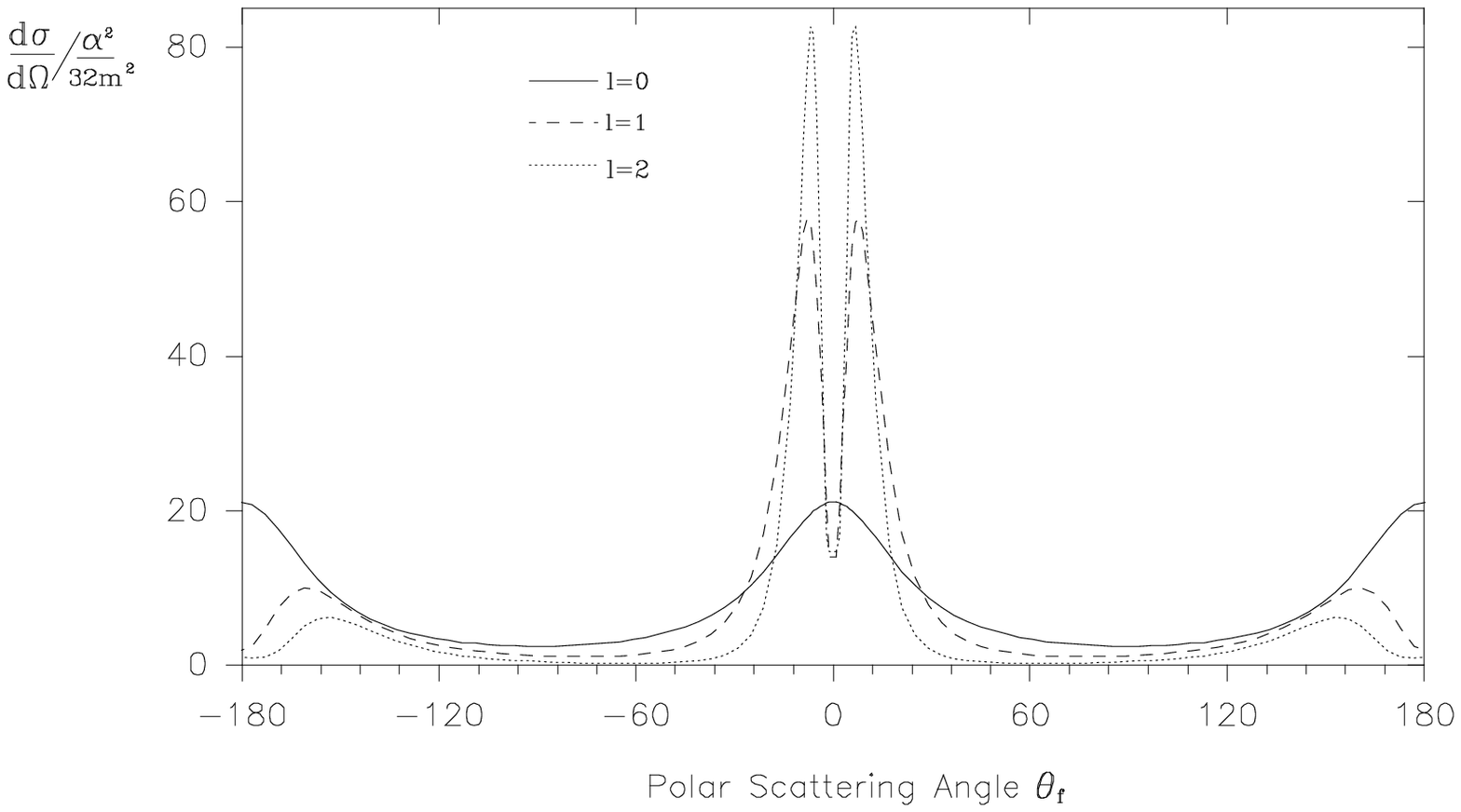}}
\caption{\bf\bm The STPPP differential cross section vs $\theta_f$ for
$\,\omega=1.28$ MeV, $\omega_1$,$\omega_2=1.024$ MeV, $\theta_1=0^{\circ}$,
$\varphi_f=0^{\circ}$ $\nu^2=0.5$ and various $l$.}
\label{pgb7}
\end{figure}

\begin{figure}[t]
 \centerline{\includegraphics[height=8cm,width=15cm]{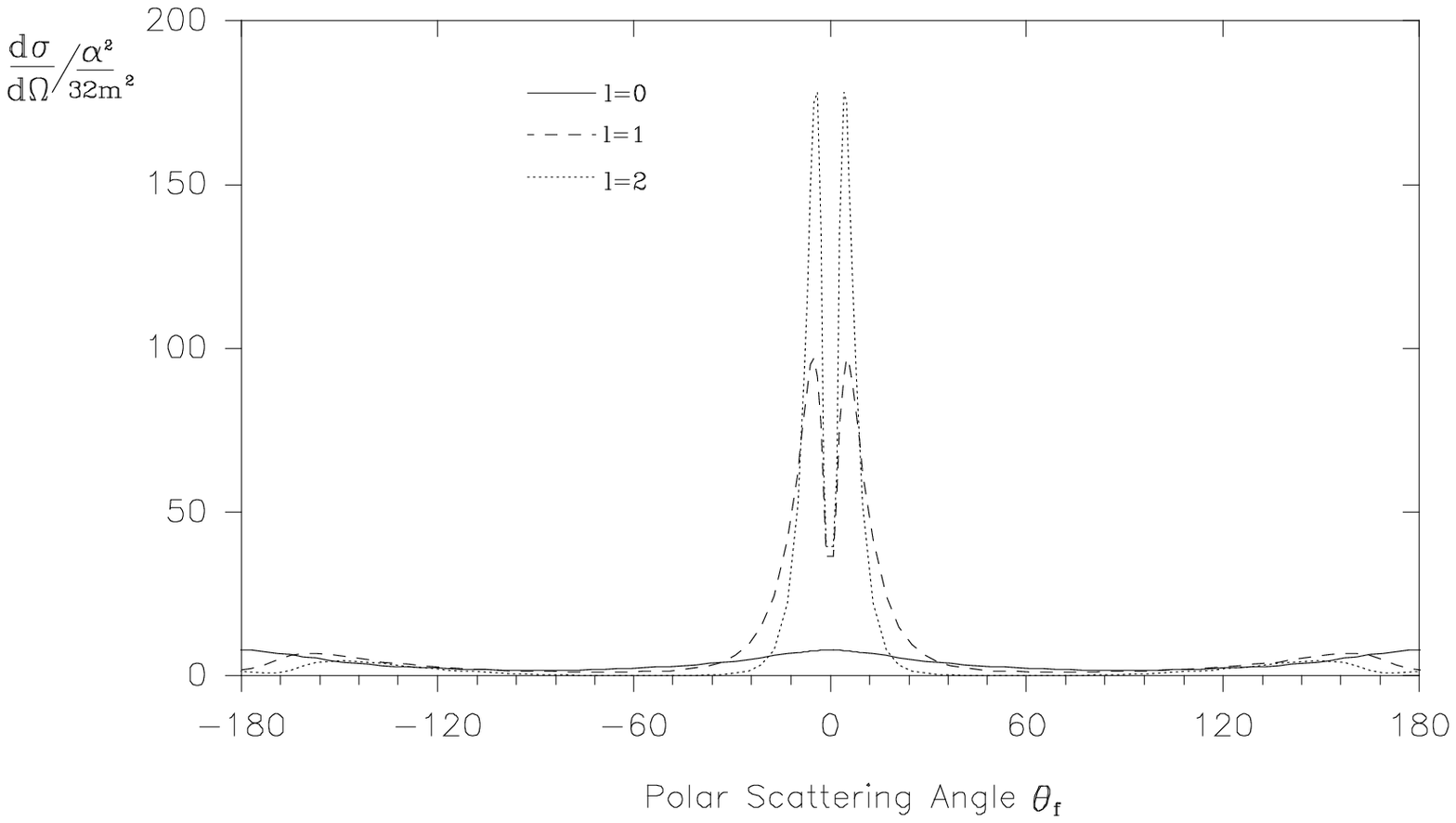}}
\caption{\bf\bm The STPPP differential cross section vs $\theta_f$ for
$\,\omega=2.56$ MeV, $\omega_1$,$\omega_2=1.024$ MeV, $\theta_1=0^{\circ}$,
$\varphi_f=0^{\circ}$ $\nu^2=0.5$ and various $l$.}
\label{pgb8}
\end{figure}

\clearpage

\begin{figure}[t]
 \centerline{\includegraphics[height=8cm,width=15cm]{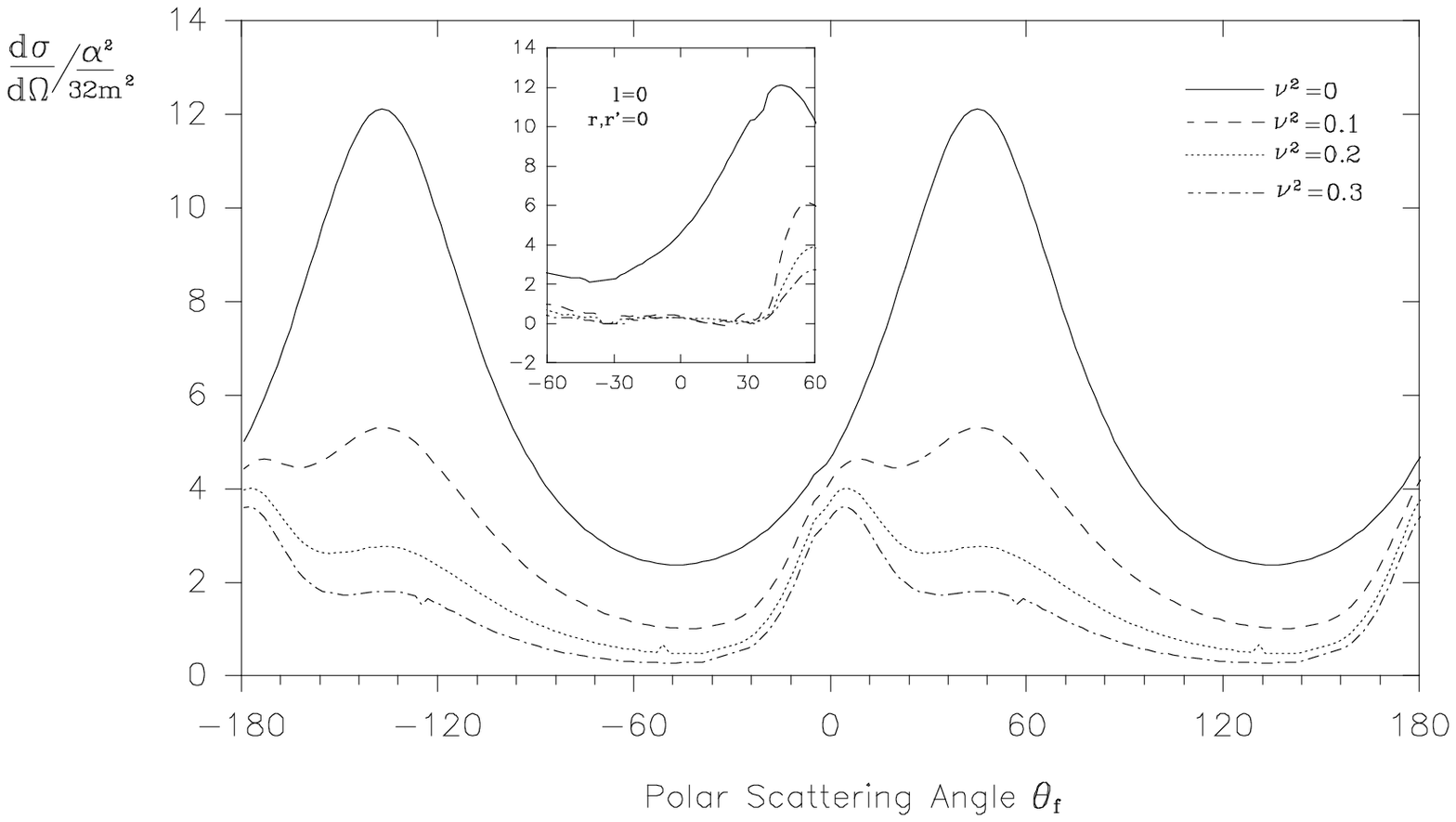}}
\caption{\bf\bm The STPPP $l=0$ differential cross section vs $\theta_f$ 
for $\,\omega=0.256$ MeV, $\omega_1$,$\omega_2=1.024$ MeV, 
$\theta_1=45^{\circ}$, $\varphi_f=0^{\circ}$ and various $\nu^2$.}
\label{pgb14}
\end{figure}

\begin{figure}[t]
 \centerline{\includegraphics[height=8cm,width=15cm]{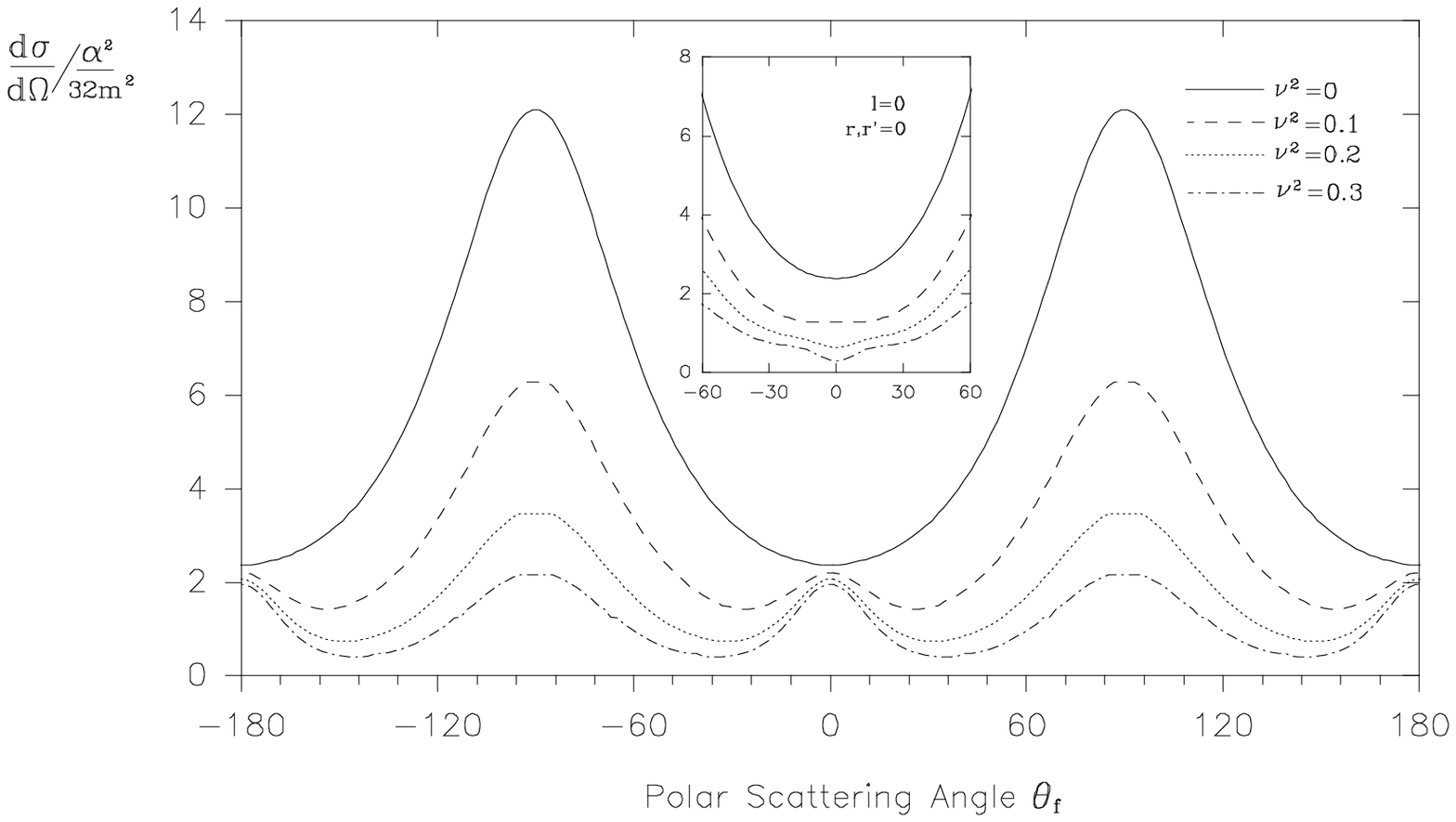}}
\caption{\bf\bm The STPPP $l=0$ differential cross section vs $\theta_f$
for $\,\omega=0.256$ MeV, $\omega_1$,$\omega_2=1.024$ MeV,
$\theta_1=90^{\circ}$, $\varphi_f=0^{\circ}$ and various $\nu^2$.}
\label{pgb13}
\end{figure}

\begin{figure}[t]
 \centerline{\includegraphics[height=8cm,width=15cm]{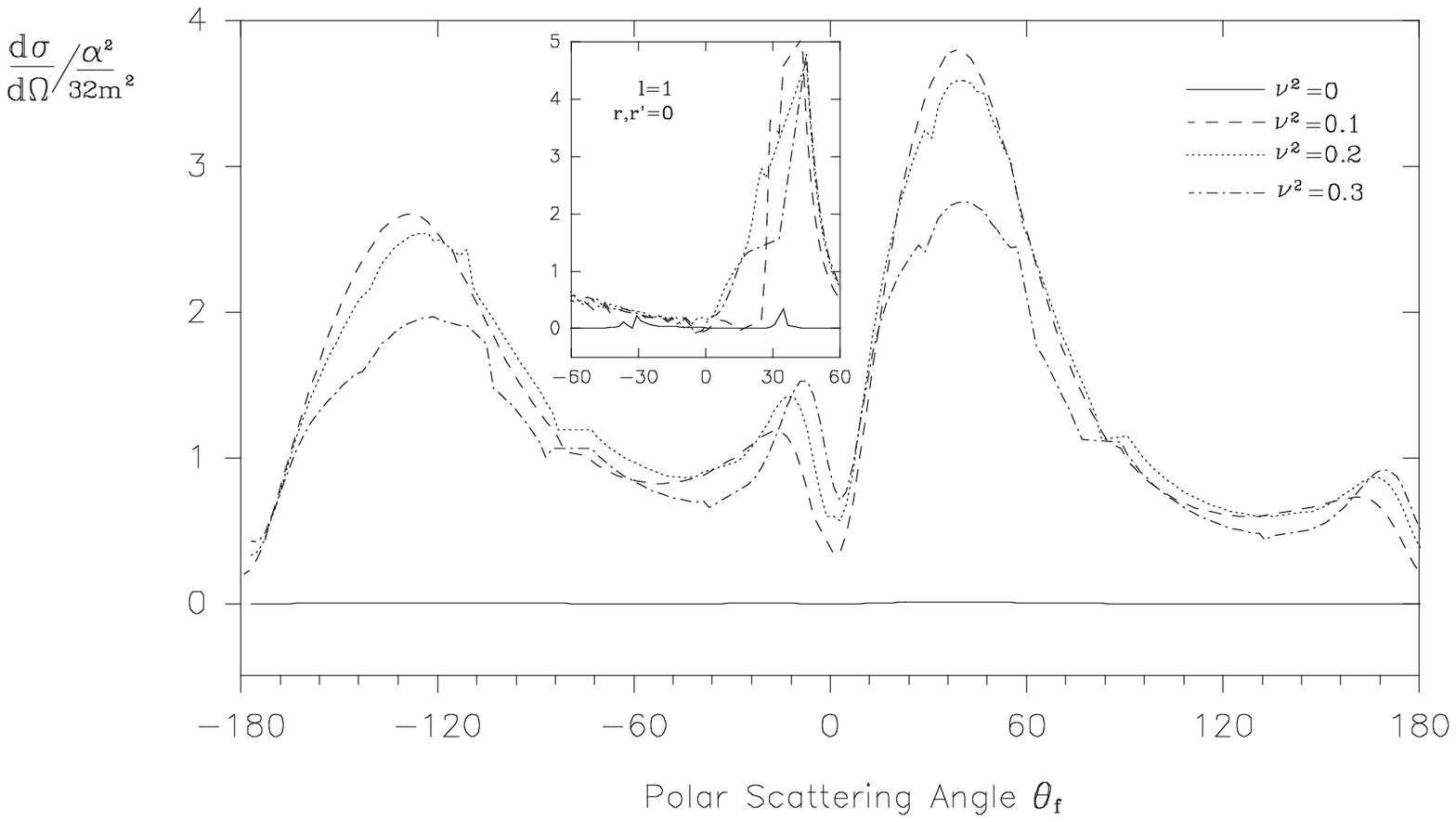}}
\caption{\bf\bm The STPPP $l=1$ differential cross section vs $\theta_f$
for $\,\omega=0.256$ MeV, $\omega_1$,$\omega_2=1.024$ MeV,
$\theta_1=45^{\circ}$, $\varphi_f=0^{\circ}$ and various $\nu^2$.}
\label{pgb15}
\end{figure}
\begin{figure}[t]
\centerline{\includegraphics[height=8cm,width=15cm]{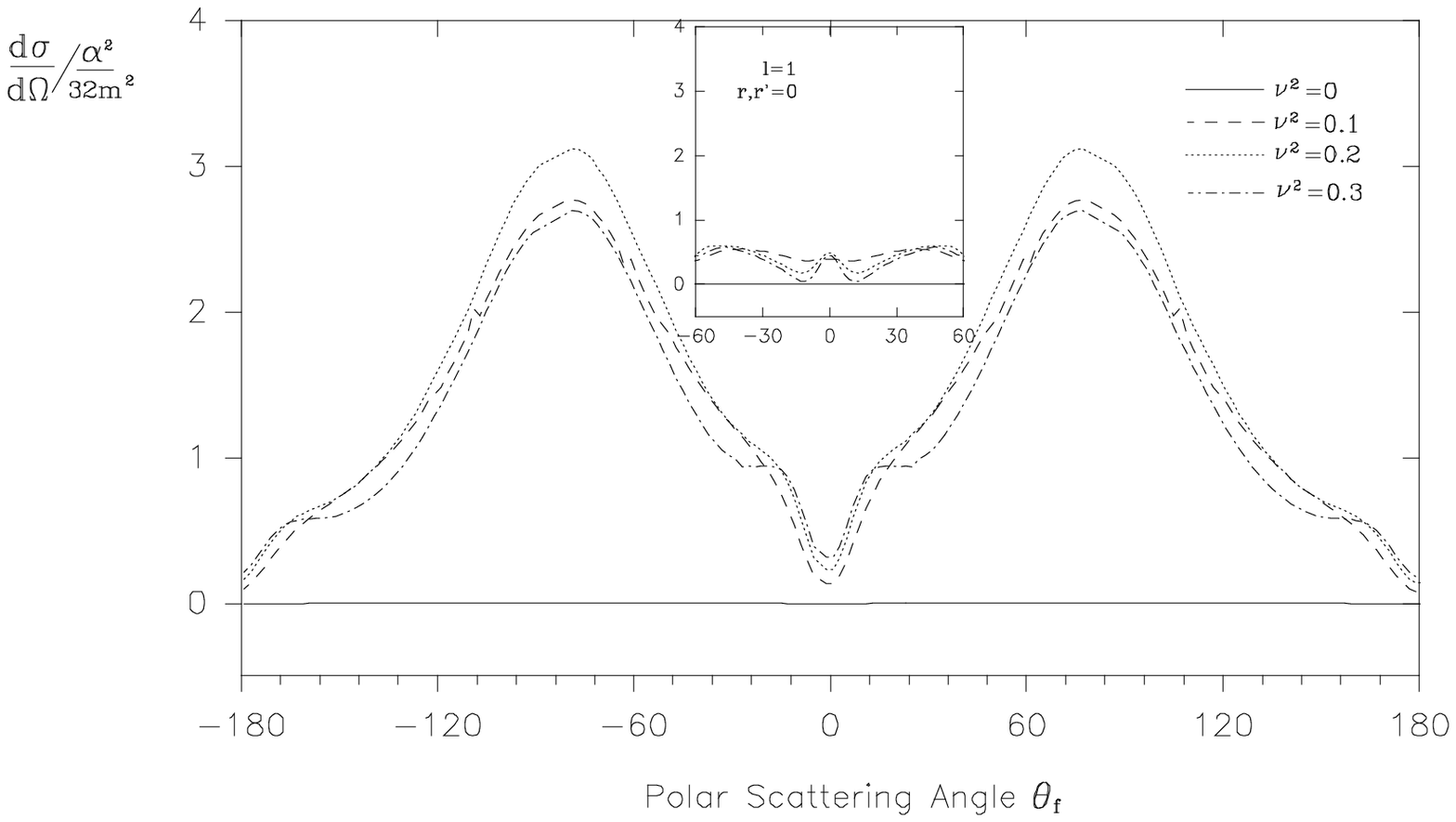}}
\caption{\bf\bm The STPPP $l=1$ differential cross section vs $\theta_f$
for $\,\omega=0.256$ MeV, $\omega_1$,$\omega_2=1.024$ MeV,
$\theta_1=90^{\circ}$, $\varphi_f=0^{\circ}$ and various $\nu^2$.}
\label{pgb16}
\end{figure}

\begin{figure}[t]
 \centerline{\includegraphics[height=8cm,width=15cm]{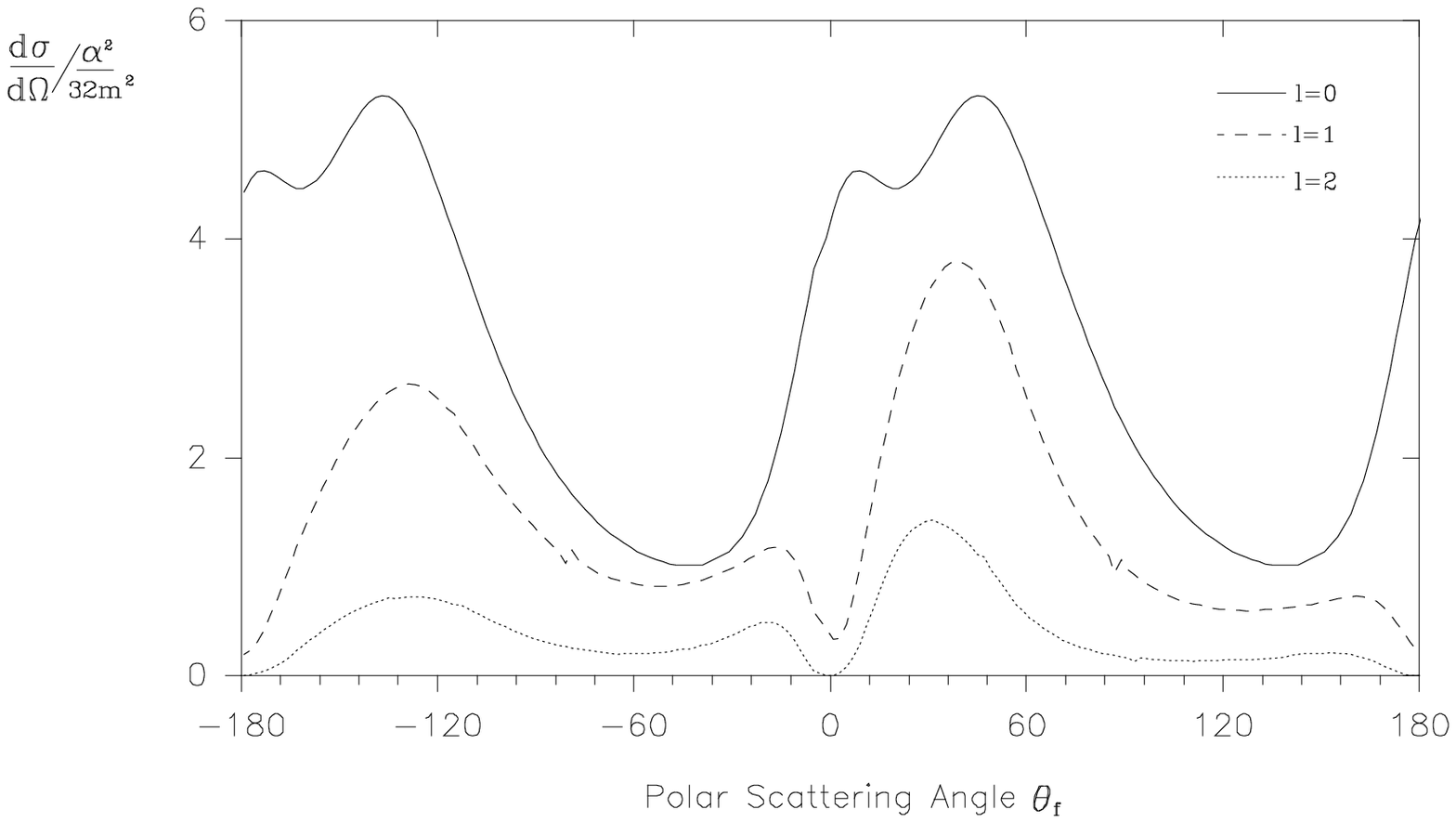}}
\caption{\bf\bm The STPPP differential cross section vs $\theta_f$
for $\,\omega=0.256$ MeV, $\omega_1$,$\omega_2=1.024$ MeV,
$\theta_1=45^{\circ}$, $\varphi_f=0^{\circ}$, $\nu^2=0.1$ and various $l$.}
\label{pgb11}
\end{figure}

\begin{figure}[t]
 \centerline{\includegraphics[height=8cm,width=15cm]{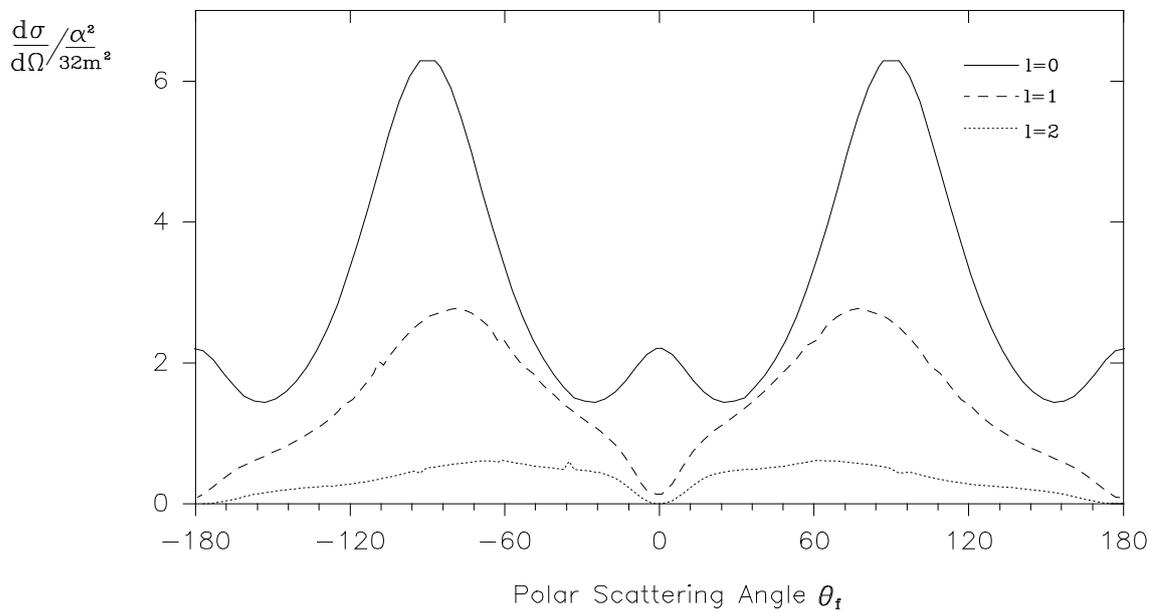}}
\caption{\bf\bm The STPPP differential cross section vs $\theta_f$
for $\,\omega=0.256$ MeV, $\omega_1$,$\omega_2=1.024$ MeV,
$\theta_1=90^{\circ}$, $\varphi_f=0^{\circ}$, $\nu^2=0.1$ and various $l$.}
\label{pgb12}
\end{figure}

\clearpage
\subsection{Differential Cross sections summed over all $l$}

Table 5.2 displays the parameter values of the STPPP scattering process
investigated in this section. The parameters here have exactly the same
meaning as in Section 5.2.1. The STPPP differential cross section has been 
summed over all values $l$ and $r$.

\bigskip\bigskip\

\input{./tex/tables/p_table2}

\clearpage

\begin{figure}[H]
 \centerline{\includegraphics[height=8cm,width=15cm]{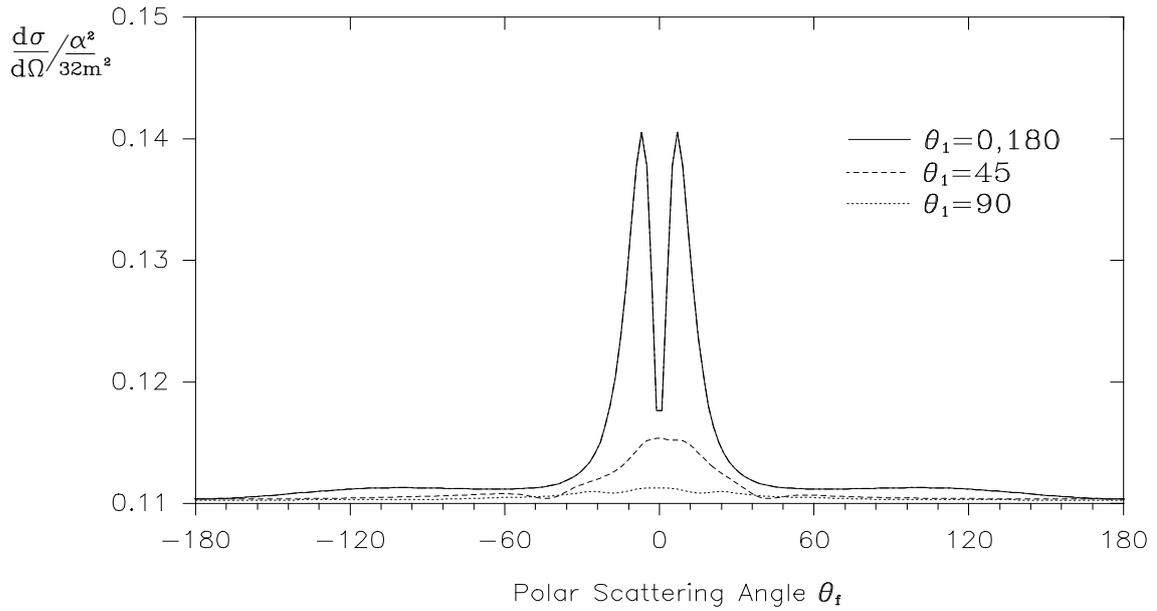}}
\caption{\bf\bm The STPPP differential cross section vs $\theta_f$
for $\,\omega=2.56$ MeV, $\omega_1$,$\omega_2=0.512$ MeV,
$\varphi_f=0^{\circ}$, $\nu^2=0.00001$ and various $\theta_1$.}
\label{pg1}
\end{figure}

\begin{figure}[H]
 \centerline{\includegraphics[height=8cm,width=15cm]{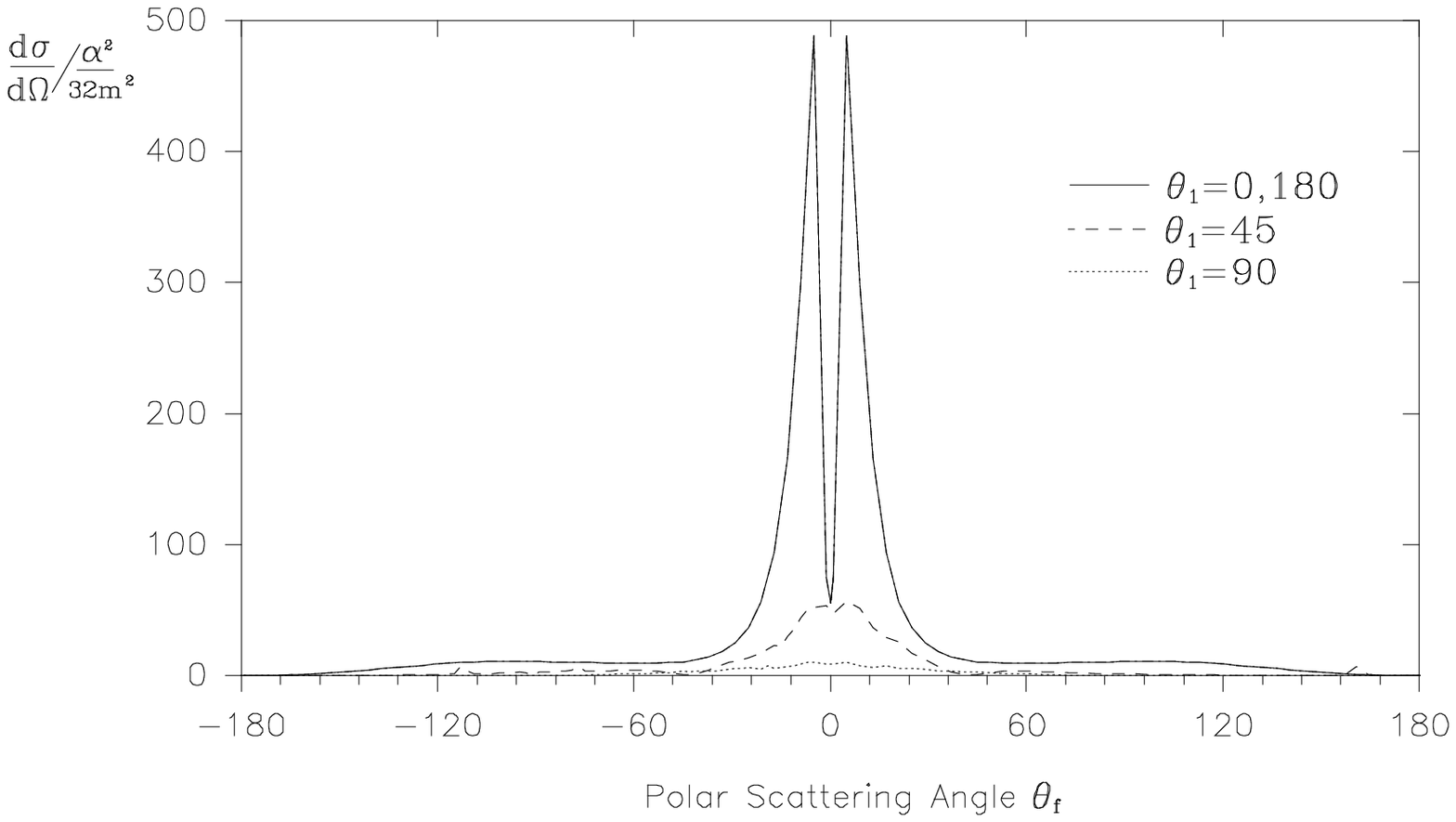}}
\caption{\bf\bm The STPPP differential cross section vs $\theta_f$
for $\,\omega=2.56$ MeV, $\omega_1$,$\omega_2=0.512$ MeV,
$\varphi_f=0^{\circ}$, $\nu^2=0.1$ and various $\theta_1$.}
\label{pg2}
\end{figure}

\clearpage

\begin{figure}[H]
 \centerline{\includegraphics[height=8cm,width=15cm]{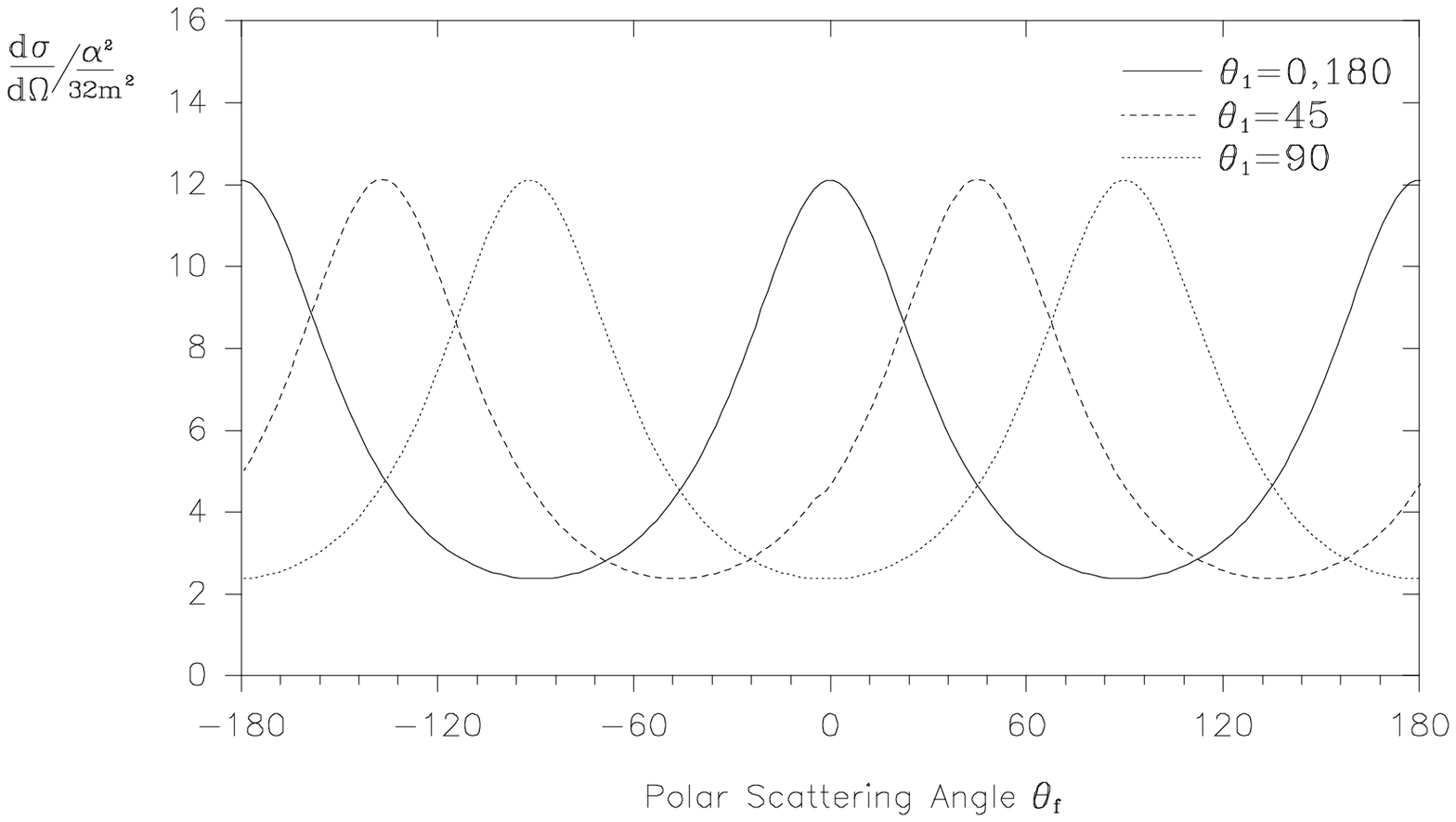}}
\caption{\bf\bm The STPPP differential cross section vs $\theta_f$
for $\,\omega=0.256$ MeV, $\omega_1$,$\omega_2=1.024$ MeV,
$\varphi_f=0^{\circ}$, $\nu^2=0.0001$ and various $\theta_1$.}
\label{pg3}
\end{figure}

\begin{figure}[H]
 \centerline{\includegraphics[height=8cm,width=15cm]{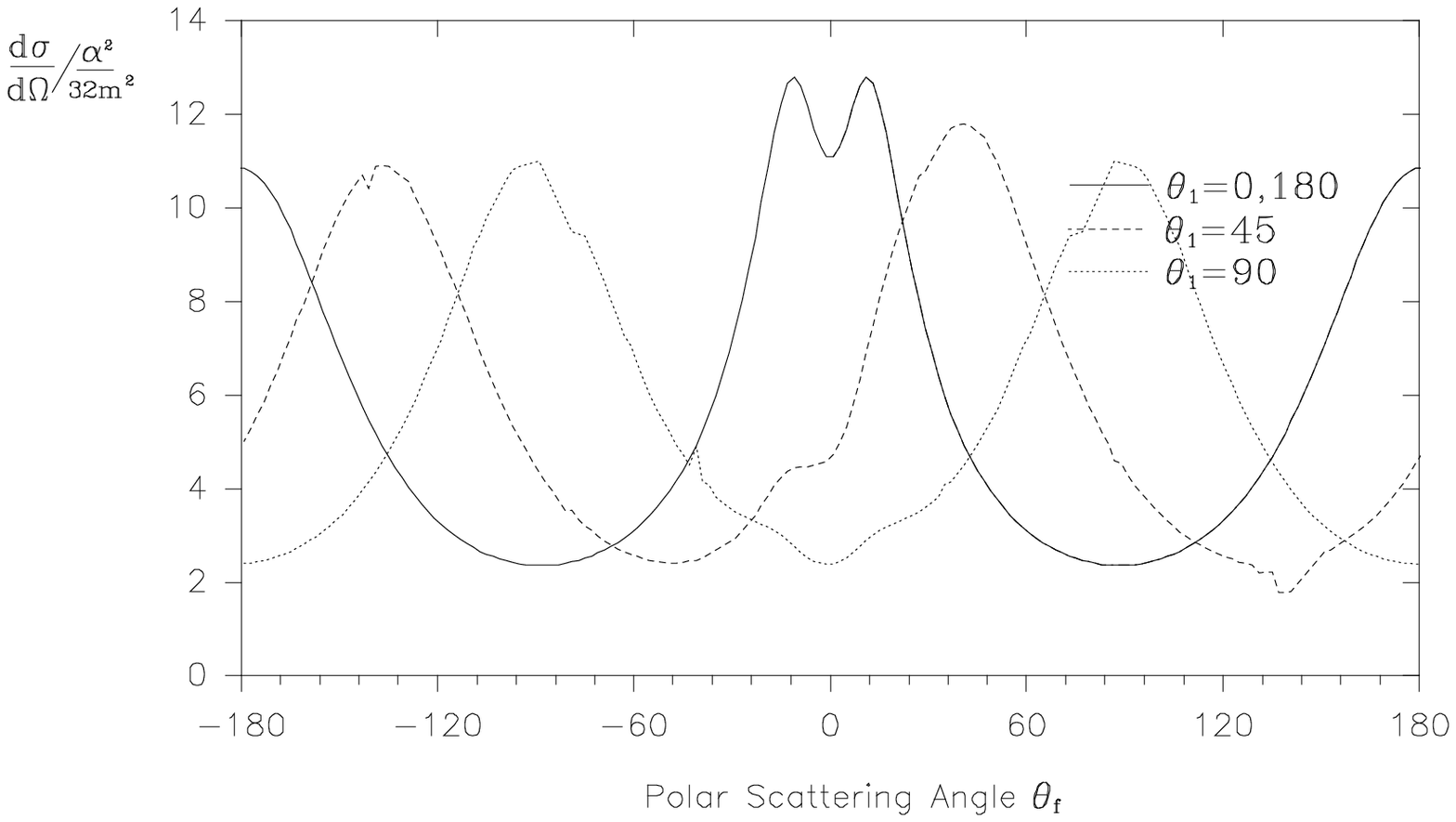}}
\caption{\bf\bm The STPPP differential cross section vs $\theta_f$
for $\,\omega=0.256$ MeV, $\omega_1$,$\omega_2=1.024$ MeV,
$\varphi_f=0^{\circ}$, $\nu^2=0.1$ and various $\theta_1$.}
\label{pg4}
\end{figure}

\clearpage

\begin{figure}[t]
 \centerline{\includegraphics[height=8cm,width=15cm]{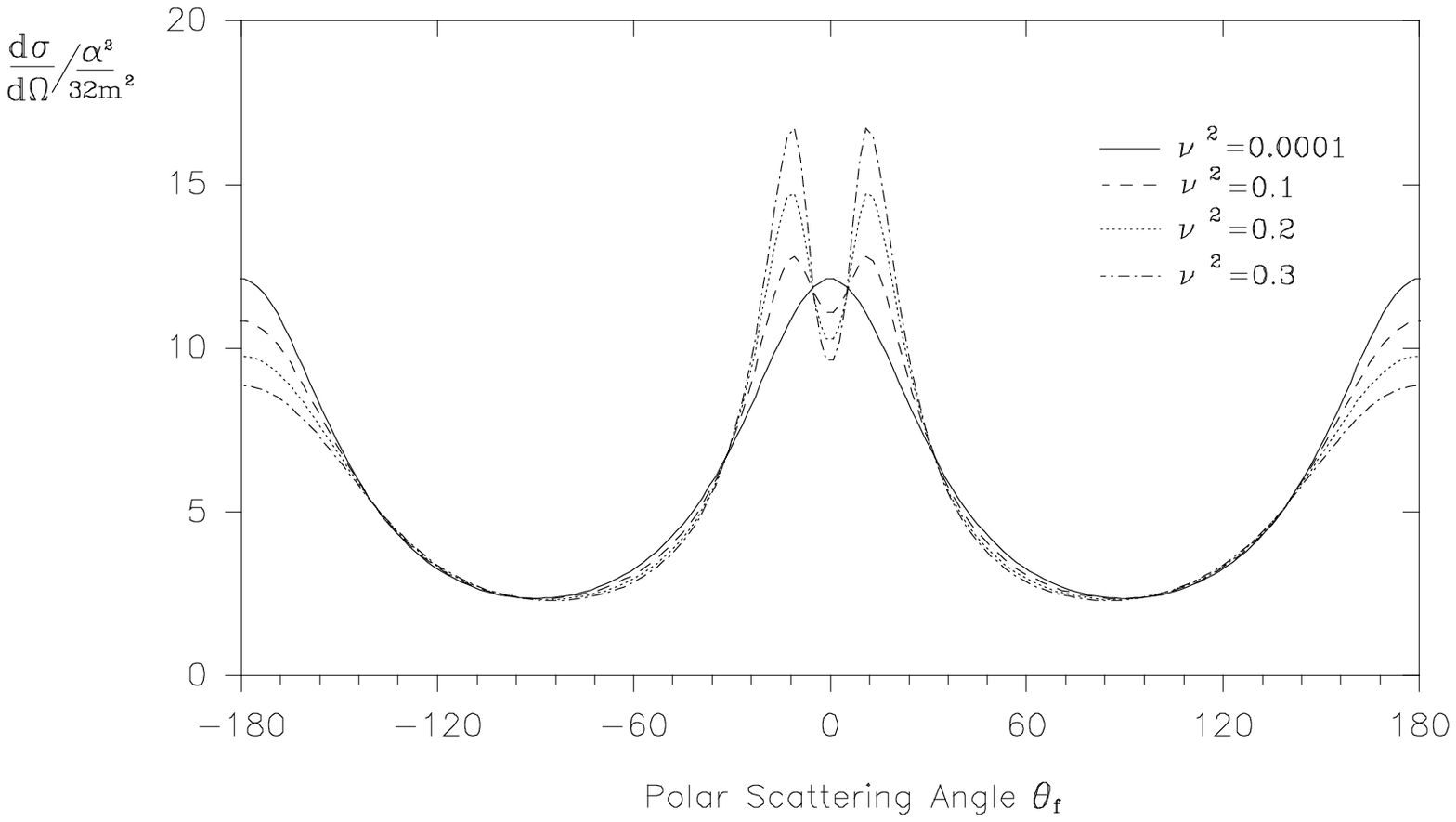}}
\caption{\bf\bm The STPPP differential cross section vs $\theta_f$
for $\,\omega=0.256$ MeV, $\omega_1$,$\omega_2=1.024$ MeV, 
$\theta_1=0^{\circ}$, $\varphi_f=0^{\circ}$ and various $\nu^2$.}
\label{pg5}
\end{figure}

\begin{figure}[t]
 \centerline{\includegraphics[height=8cm,width=15cm]{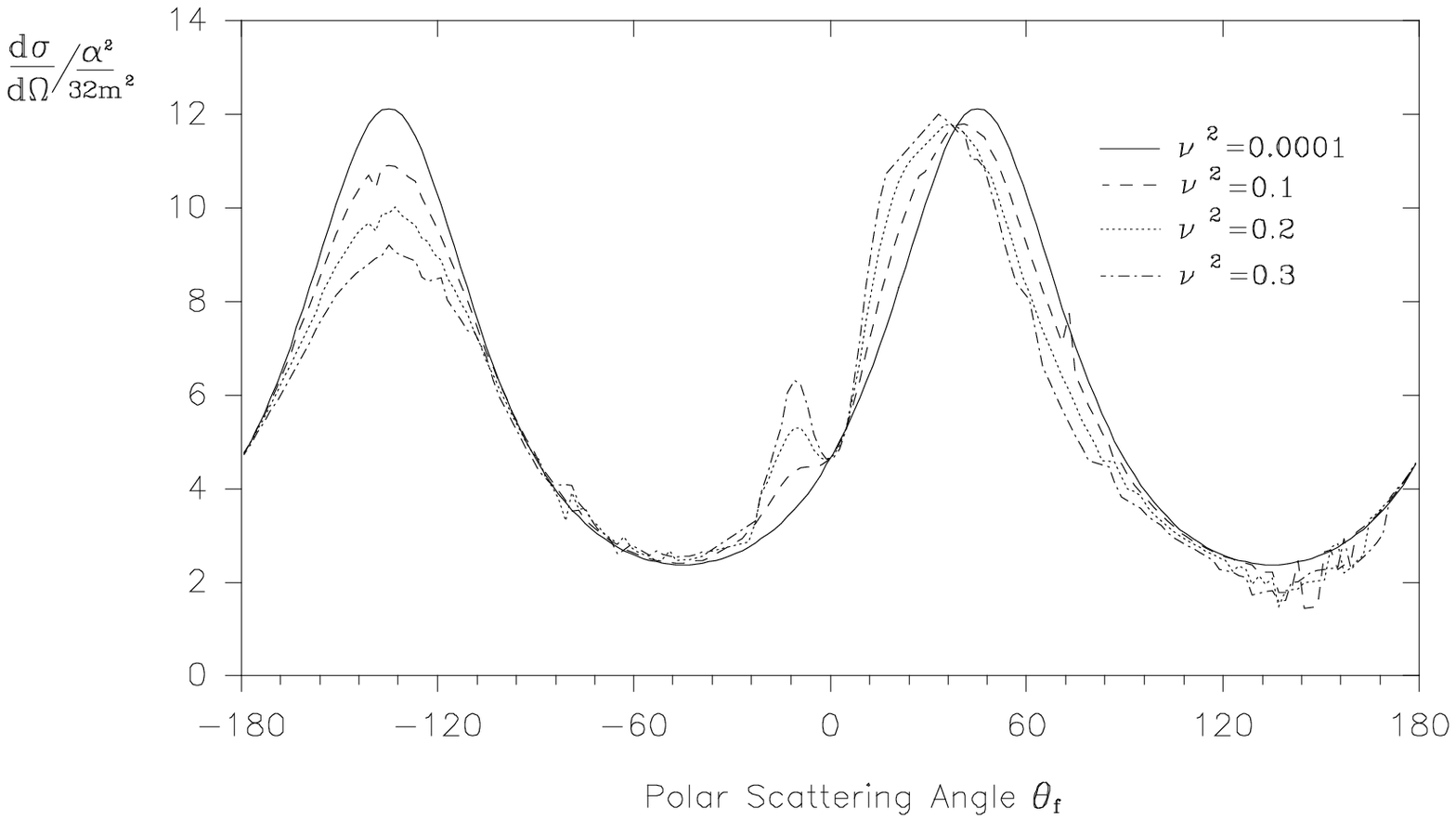}}
\caption{\bf\bm The STPPP differential cross section vs $\theta_f$
for $\,\omega=0.256$ MeV, $\omega_1$,$\omega_2=1.024$ MeV,
$\theta_1=45^{\circ}$, $\varphi_f=0^{\circ}$ and various $\nu^2$.}
\label{pg6}
\end{figure}

\clearpage

\begin{figure}[t]
 \centerline{\includegraphics[height=8cm,width=15cm]{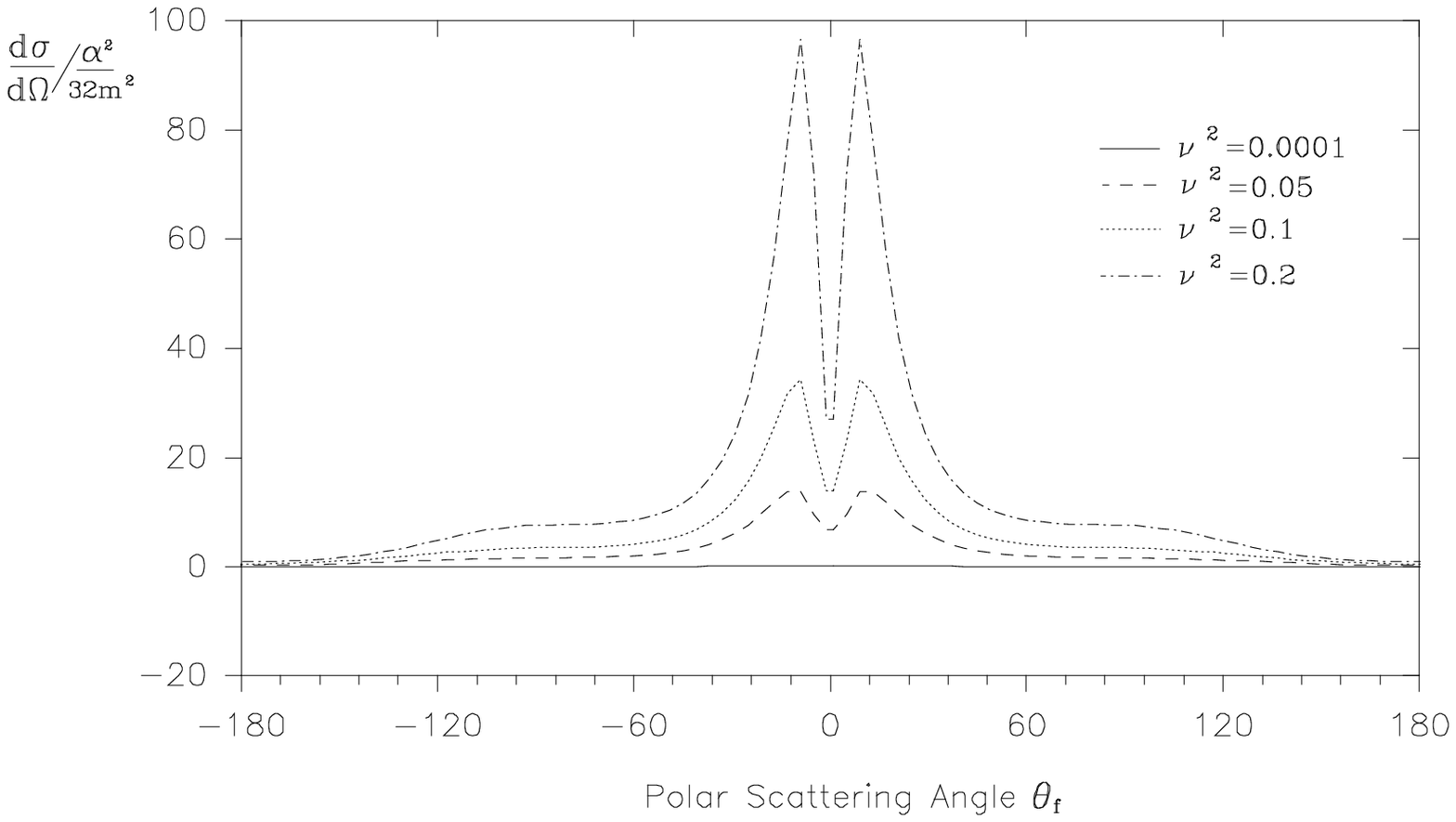}}
\caption{\bf\bm The STPPP differential cross section vs $\theta_f$
for $\,\omega=1.024$ MeV, $\omega_1$,$\omega_2=0.512$ MeV,
$\theta_1=0^{\circ}$, $\varphi_f=0^{\circ}$ and various $\nu^2$.}
\label{pg7}
\end{figure}

\begin{figure}[t]
 \centerline{\includegraphics[height=8cm,width=15cm]{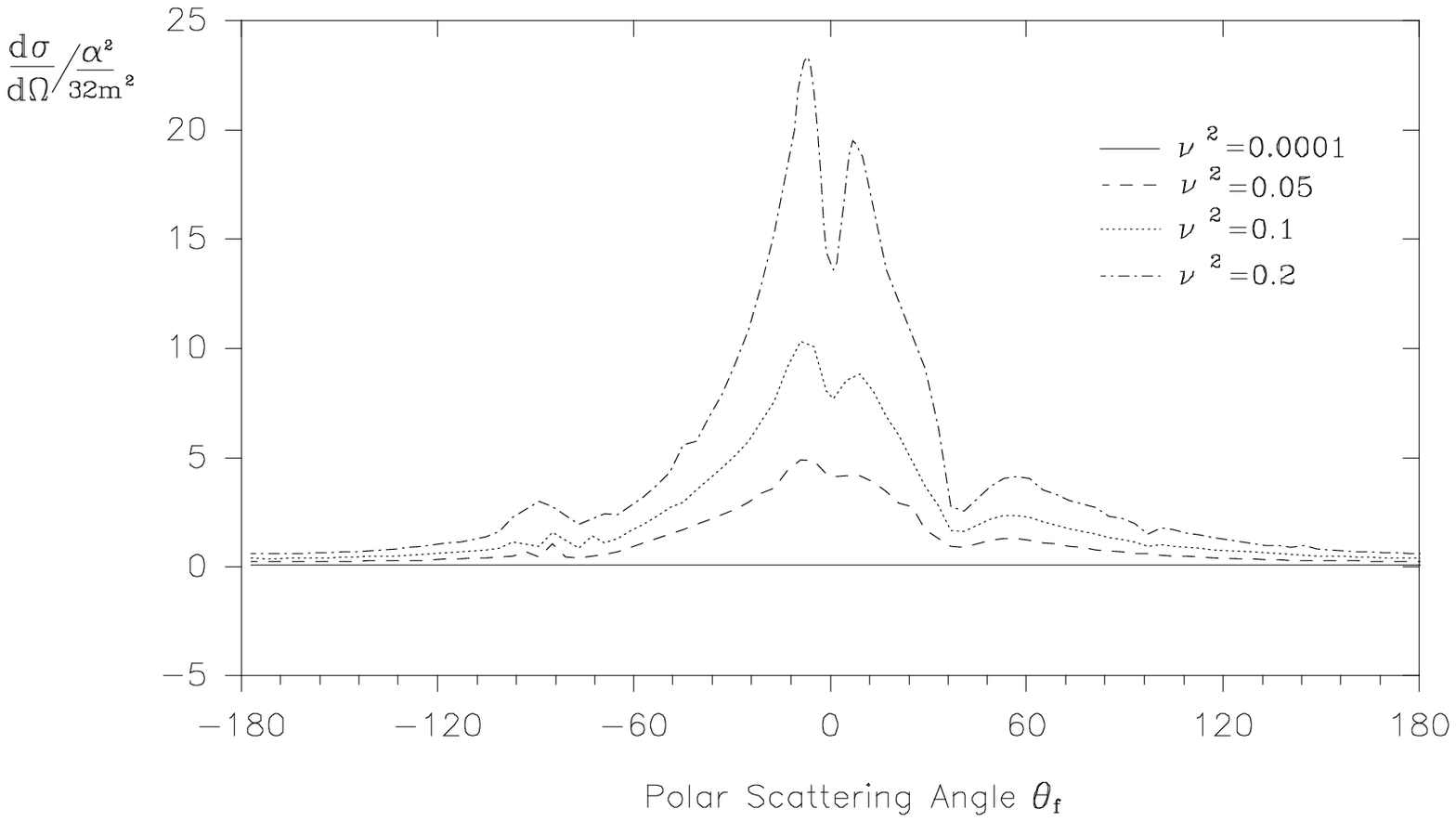}}
\caption{\bf\bm The STPPP differential cross section vs $\theta_f$
for $\,\omega=1.024$ MeV, $\omega_1$,$\omega_2=0.512$ MeV,
$\theta_1=45^{\circ}$, $\varphi_f=0^{\circ}$ and various $\nu^2$.}
\label{pg8}
\end{figure}

\clearpage

\begin{figure}[t]
 \centerline{\includegraphics[height=8cm,width=10cm]{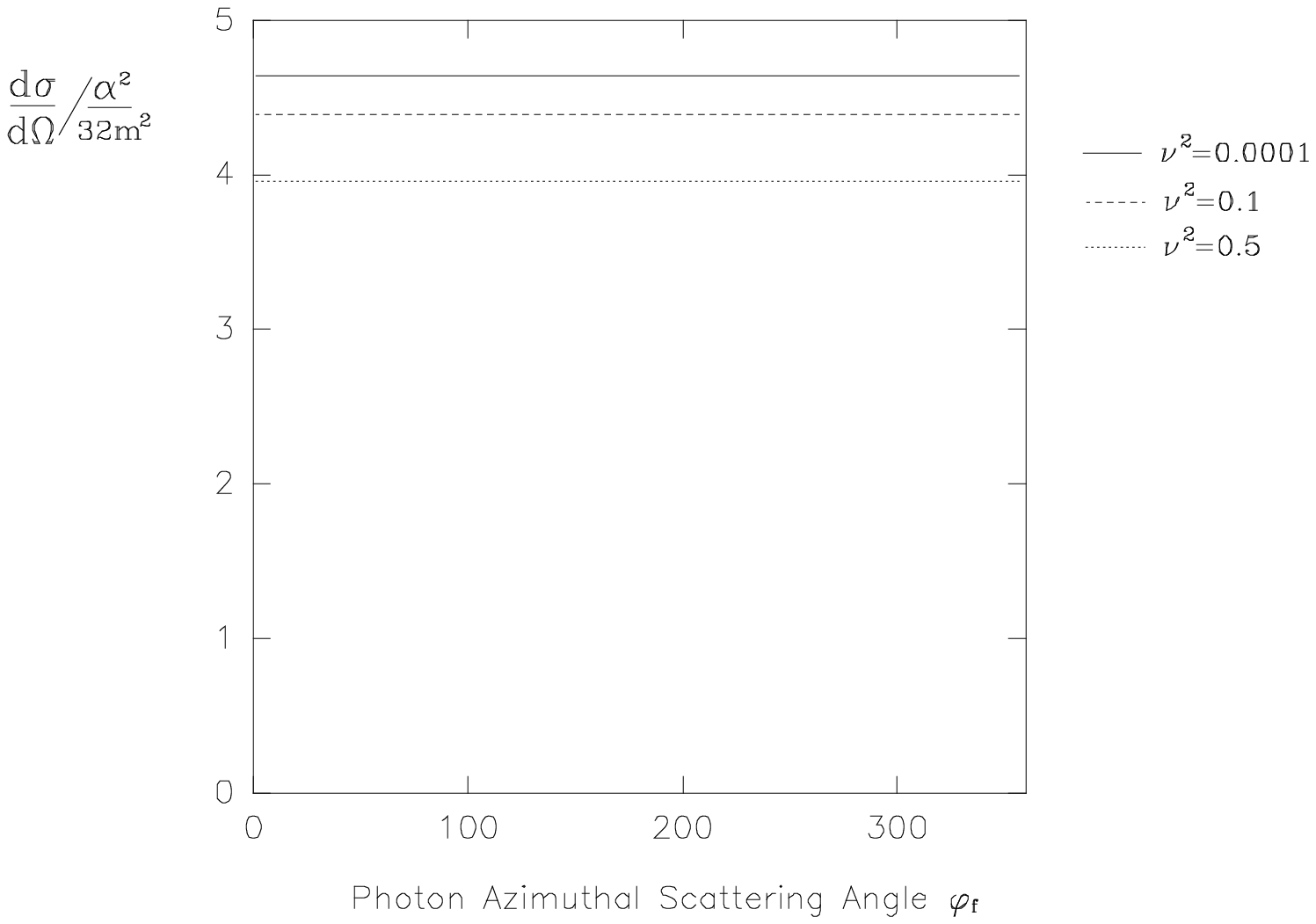}}
\caption{\bf\bm The STPPP differential cross section vs $\varphi_f$
for $\,\omega=0.256$ MeV, $\omega_1$,$\omega_2=1.024$ MeV,
$\theta_1=0^{\circ}$, $\theta_f=45^{\circ}$ and various $\nu^2$.}
\label{pg11}
\end{figure}

\begin{figure}[t]
 \centerline{\includegraphics[height=8cm,width=10cm]{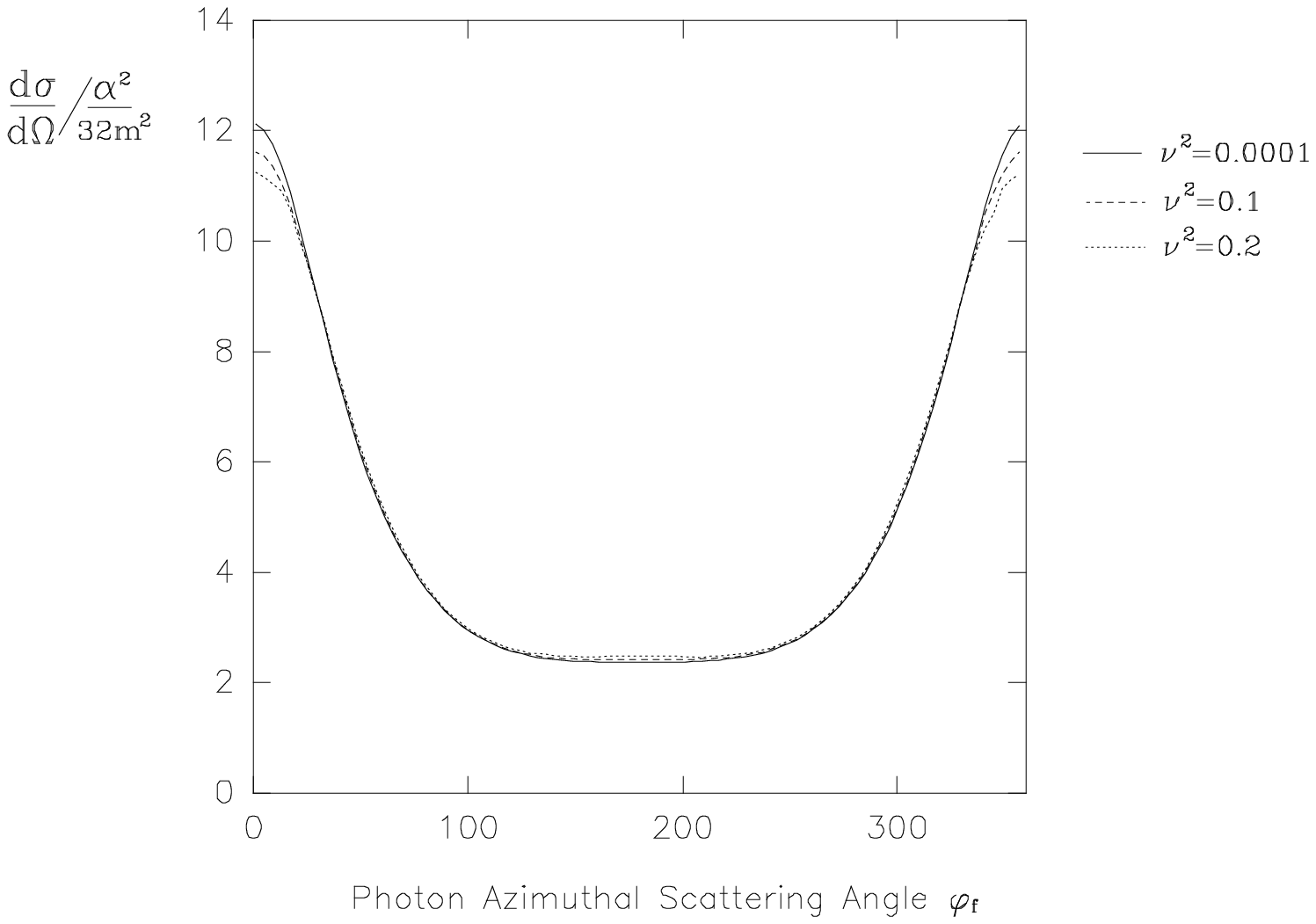}}
\caption{\bf\bm The STPPP differential cross section vs $\varphi_f$
for $\,\omega=0.256$ MeV, $\omega_1$,$\omega_2=1.024$ MeV,
$\theta_1=45^{\circ}$, $\theta_f=45^{\circ}$ and various $\nu^2$.}
\label{pg12}
\end{figure}

\begin{figure}[t]
 \centerline{\includegraphics[height=8cm,width=15cm]{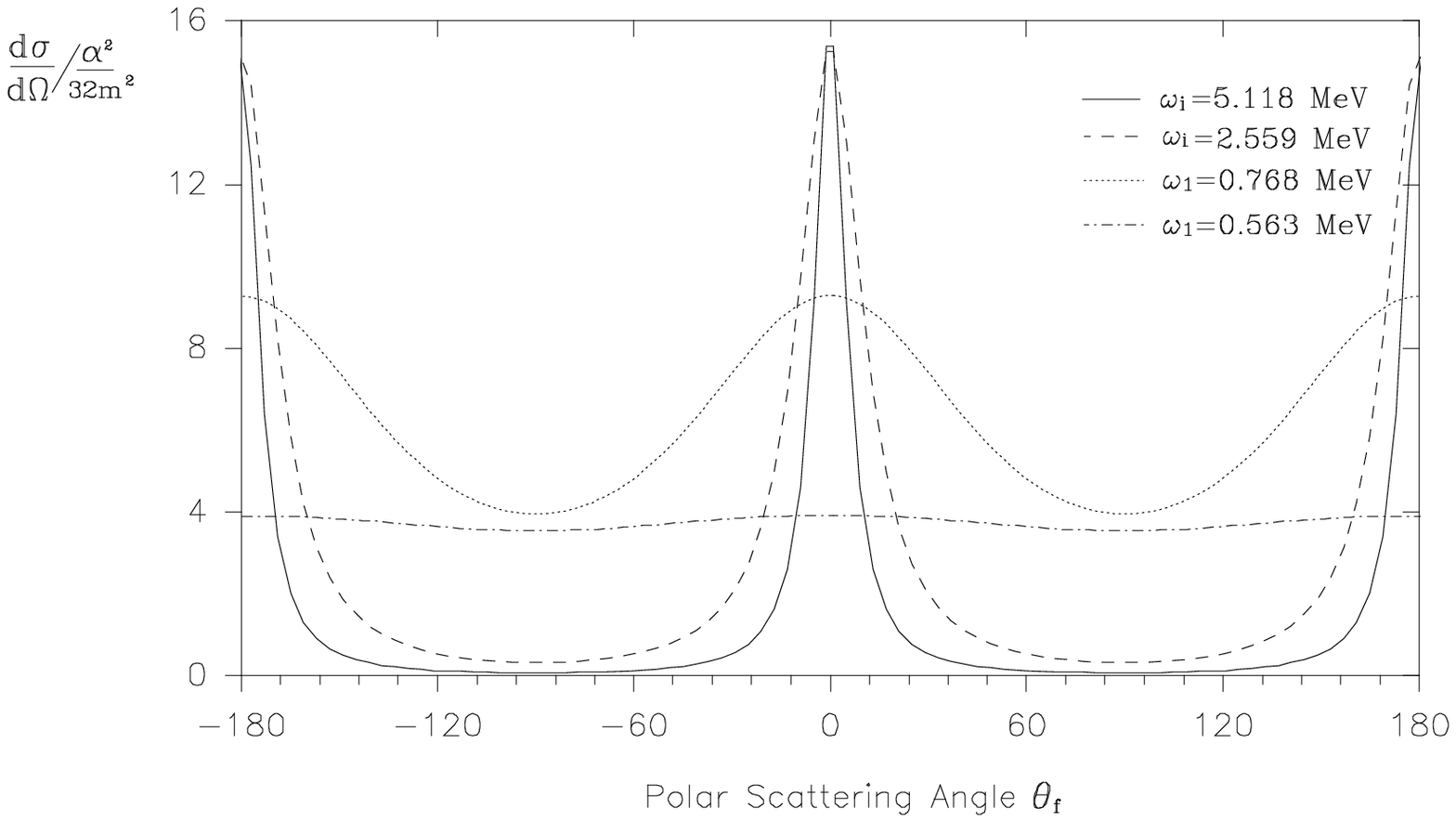}}
\caption{\bf\bm The STPPP differential cross section vs $\theta_f$
for $\,\omega=1.024$ MeV, $\theta_1=0^{\circ}$, $\varphi_f=0^{\circ}$, 
$\nu^2=0.0001$ and various $\omega_1$.}
\label{pg13}
\end{figure}

\begin{figure}[t]
 \centerline{\includegraphics[height=8cm,width=15cm]{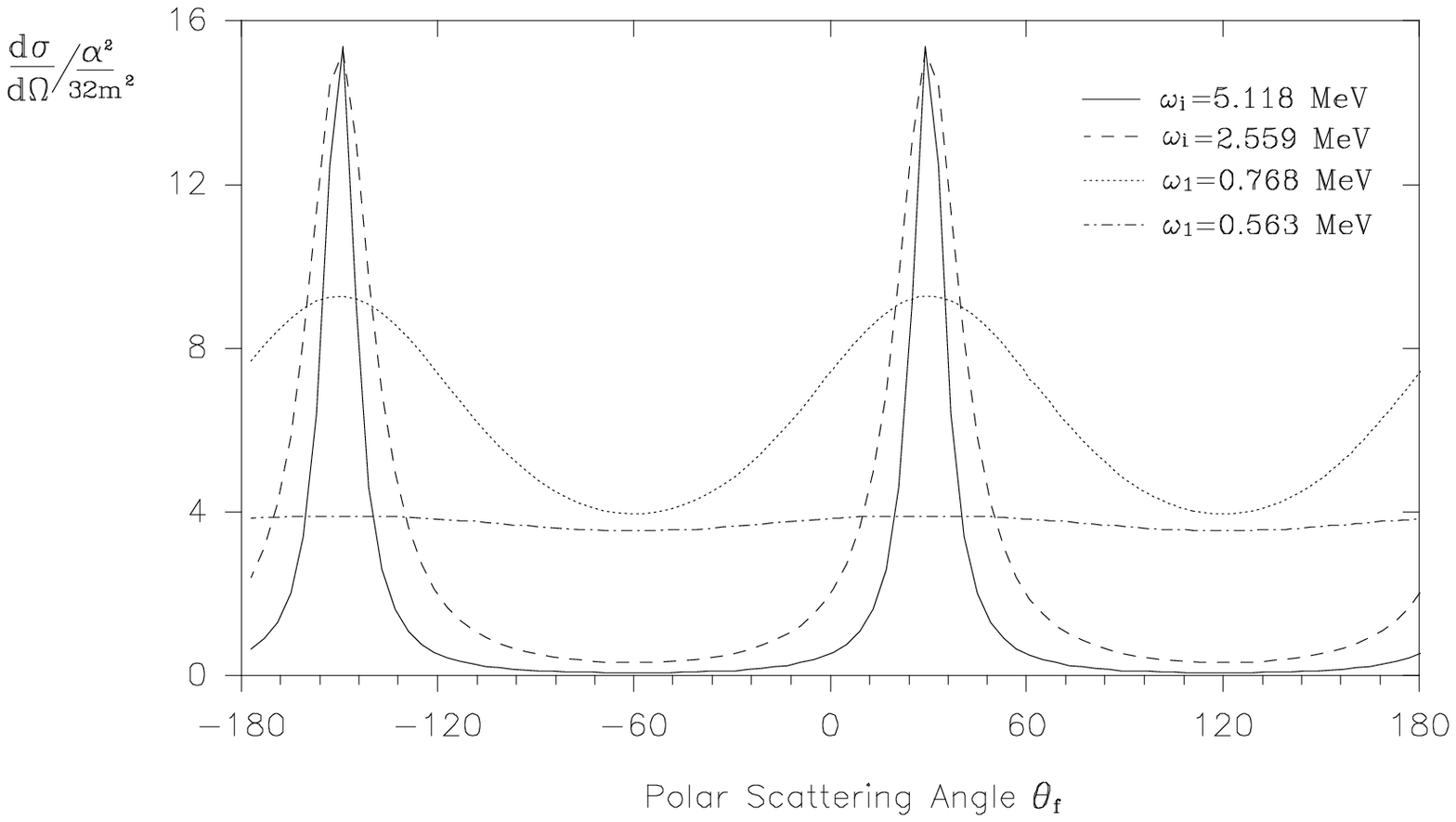}}
\caption{\bf\bm The STPPP differential cross section vs $\theta_f$
for $\,\omega=1.024$ MeV, $\theta_1=30^{\circ}$, $\varphi_f=0^{\circ}$,
$\nu^2=0.0001$ and various $\omega_1$.}
\label{pg14}
\end{figure}

\clearpage

\begin{figure}[t]
 \centerline{\includegraphics[height=8cm,width=15cm]{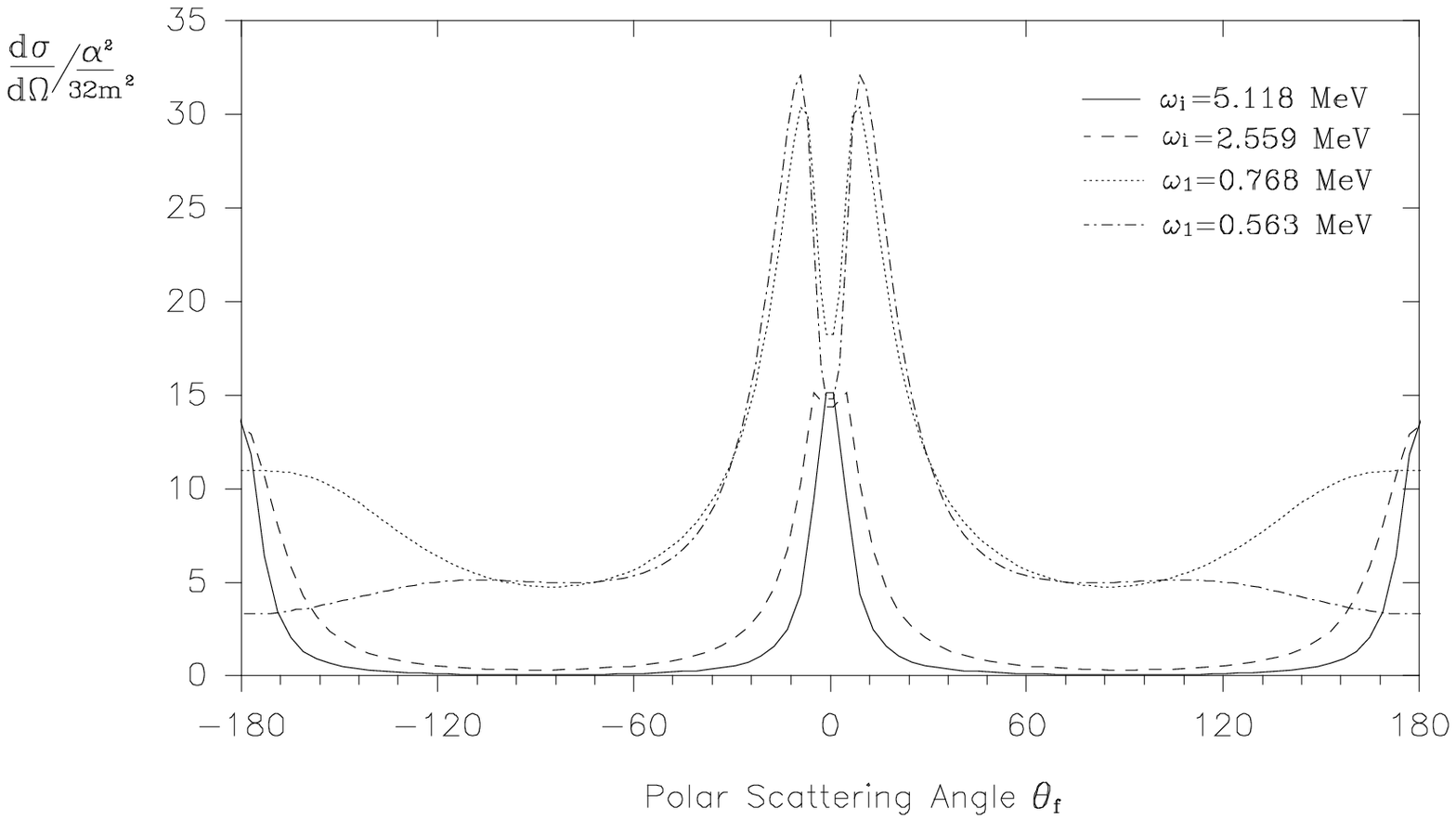}}
\caption{\bf\bm The STPPP differential cross section vs $\theta_f$
for $\,\omega=1.024$ MeV, $\theta_1=0^{\circ}$, $\varphi_f=0^{\circ}$,
$\nu^2=0.1$ and various $\omega_1$.}
\label{pg15}
\end{figure}

\begin{figure}[t]
 \centerline{\includegraphics[height=8cm,width=15cm]{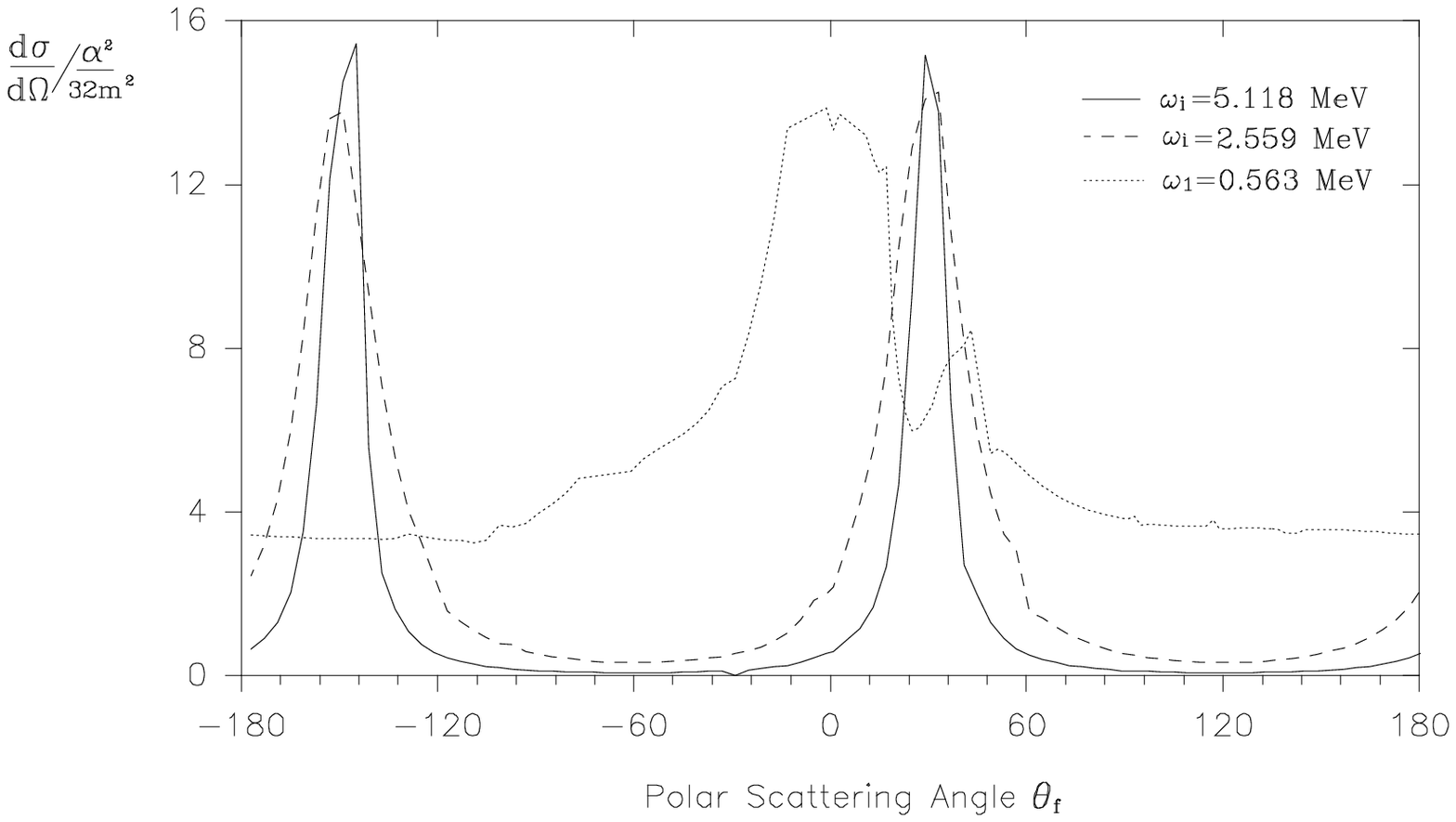}}
\caption{\bf\bm The STPPP differential cross section vs $\theta_f$
for $\,\omega=1.024$ MeV, $\theta_1=30^{\circ}$, $\varphi_f=0^{\circ}$,
$\nu^2=0.1$ and various $\omega_1$.}
\label{pg16}
\end{figure}

\begin{figure}[t]
 \centerline{\includegraphics[height=8cm,width=10cm]{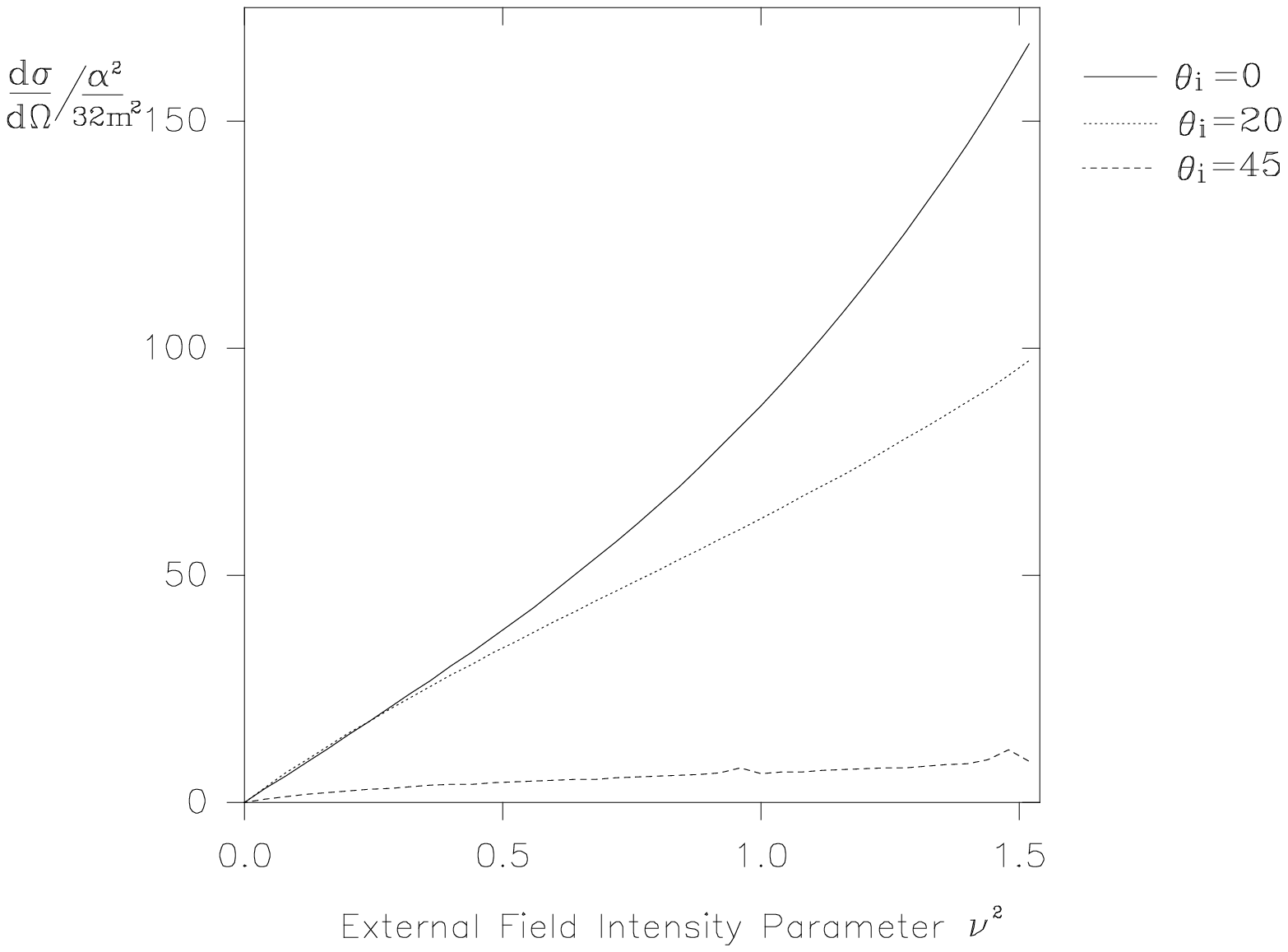}}
\caption{\bf\bm The STPPP differential cross section vs $\nu^2$
for $\,\omega=1.536$ MeV, $\omega_1$,$\omega_2=0.512$ MeV, 
$\theta_f=45^{\circ}$, $\varphi_f=0^{\circ}$ and various $\theta_1$.}
\label{pg17}
\end{figure}

\begin{figure}[t]
 \centerline{\includegraphics[height=8cm,width=10cm]{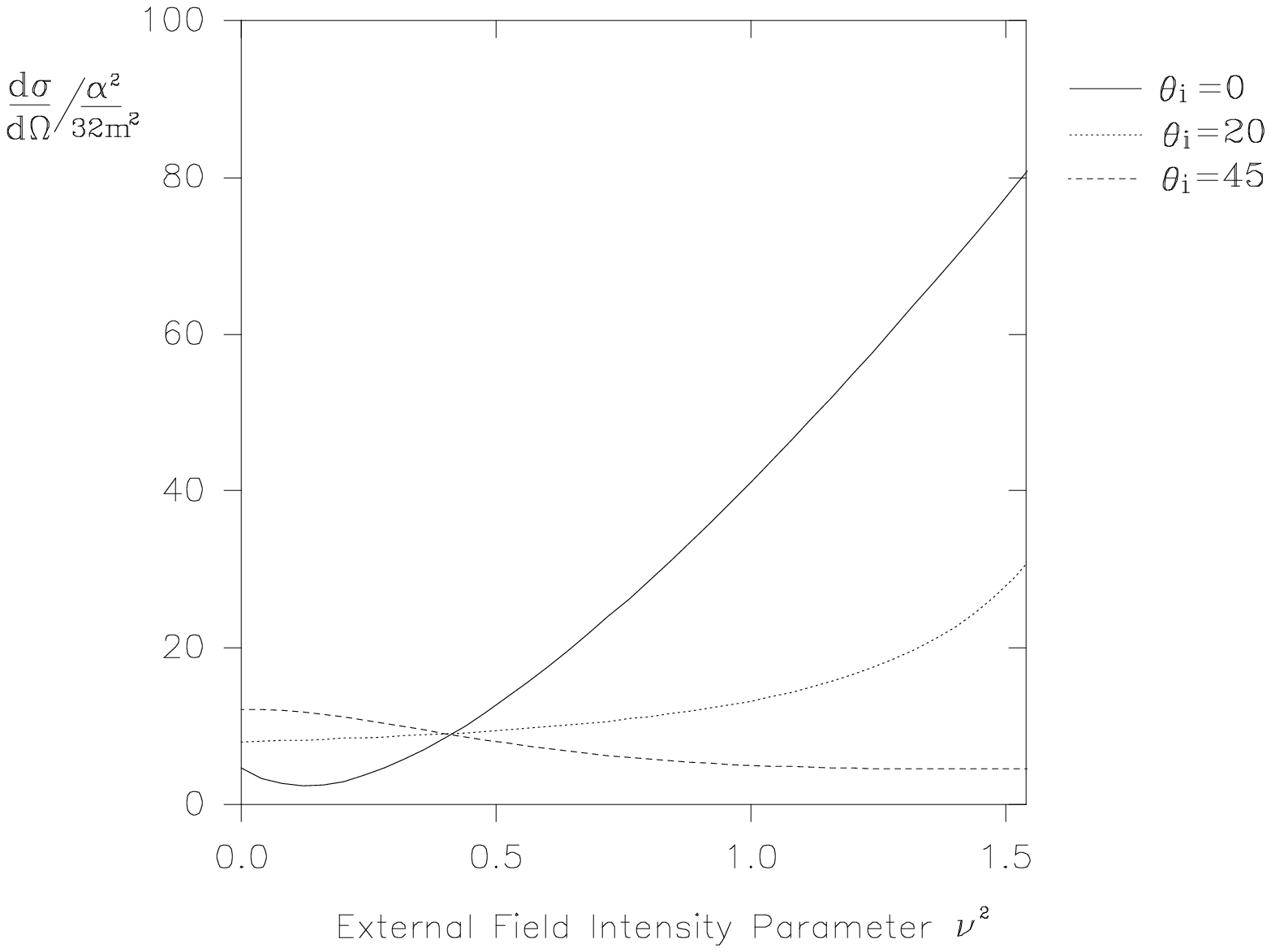}}
\caption{\bf\bm The STPPP differential cross section vs $\nu^2$
for $\,\omega=0.973$ MeV, $\omega_1$,$\omega_2=1.024$ MeV,
$\theta_f=45^{\circ}$, $\varphi_f=0^{\circ}$ and various $\theta_1$.}
\label{pg18}
\end{figure}

\begin{figure}[t]
 \centerline{\includegraphics[height=8cm,width=10cm]{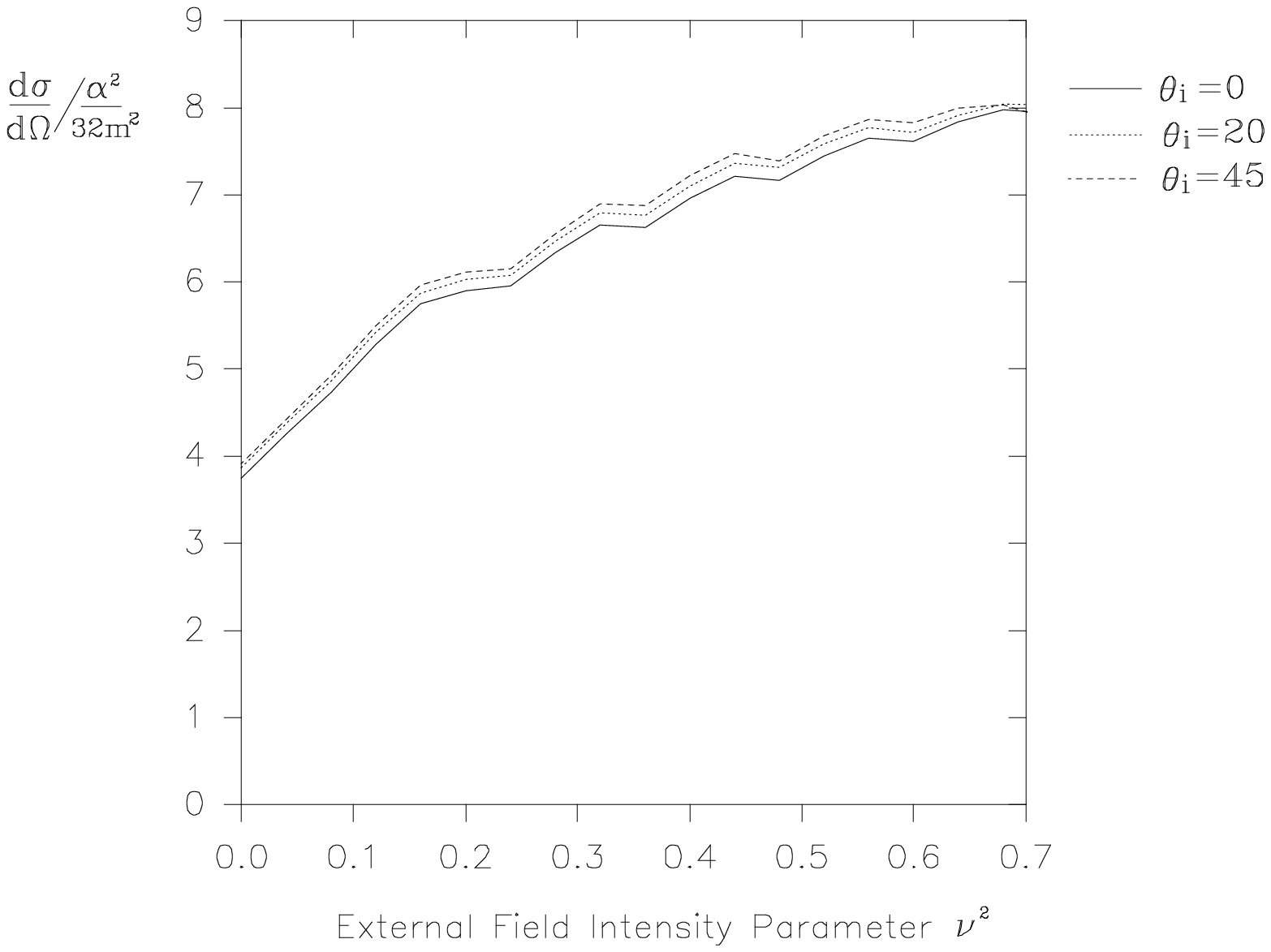}}
\caption{\bf\bm The STPPP differential cross section vs $\nu^2$
for $\,\omega=0.061$ MeV, $\omega_1$,$\omega_2=0.563$ MeV,
$\theta_f=45^{\circ}$, $\varphi_f=0^{\circ}$ and various $\theta_1$.}
\label{pg19}
\end{figure}

\begin{figure}[t]
 \centerline{\includegraphics[height=8cm,width=15cm]{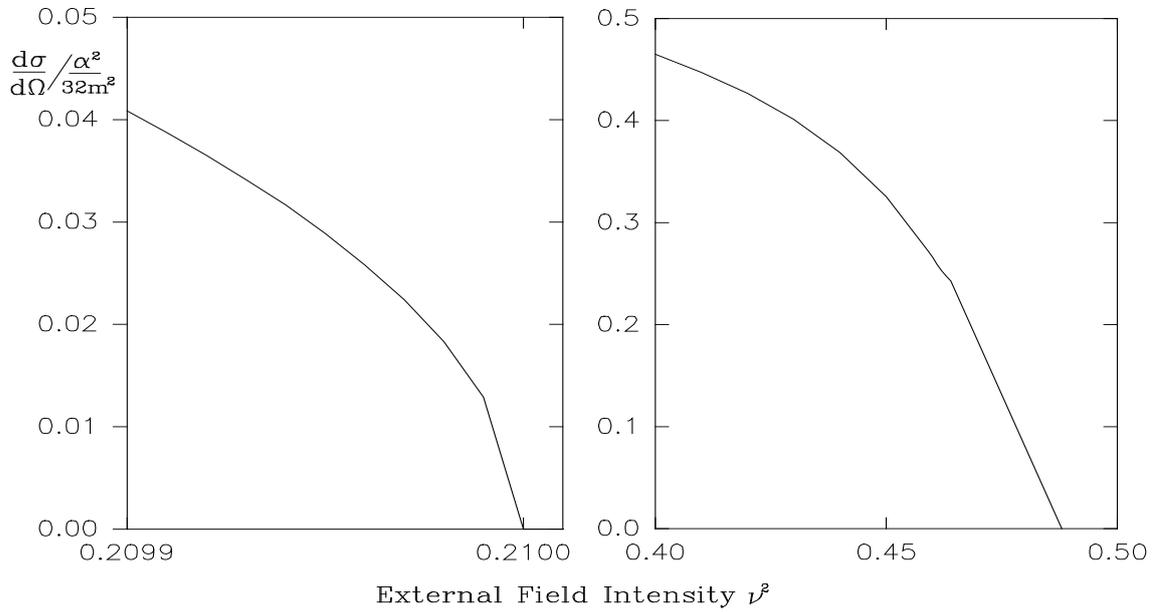}}
\caption{\bf\bm The STPPP $l=0$ and $l=2$ differential cross section vs 
$\nu^2$ for $\,\omega=0.061$ MeV, $\omega_1$,$\omega_2=0.563$ MeV,
$\theta_1=0^{\circ}$, $\theta_f=45^{\circ}$, $\varphi_f=0^{\circ}$.}
\label{pg20}
\end{figure}

\clearpage
\subsection{Differential Cross Section $l$ Contributions}

Figures \ref{pga1}-\ref{pga4} show the variation of the STPPP
differential cross section with the number $l$ of external field photons
that contribute to the process. Each figure includes several plots
corresponding to different values of the external field intensity parameter 
$\nu ^2$.

Figures \ref{pga1}-\ref{pga2} deal with particle energies such that the
ratio of initial photon energy to external field energy, $\frac{\omega _1}\omega $, 
is less than one.
\footnote{Due to the reference frame used the energies of the two incident 
photons are identical.} Under these conditions the only STPPP process permitted is one in 
which $l\geq0$. Conversely, Figures \ref{pga3}-\ref{pga4}, for which 
$\frac{\omega _1}\omega$ is greater than unity, permit STPPP processes in which 
quanta is given up to the external field as long as sufficient energy is left to create the 
$e^+e^-$ pair. The precise requirement for the number of laser
photons energetically permitted to take part in the STPPP process is given
by the inequality

\begin{equation}
\label{c5eq1}
l\geq 2\frac{\sqrt{1+\nu ^2}}{\omega /m}-2\frac{\omega _1}%
\omega 
\end{equation}

\medskip\ 

Figures \ref{pga1}-\ref{pga4} show the same general broadening with
increasing $\nu ^2$ as the equivalent figures of Chapter 4 (figures \ref{cga1}-%
\ref{cga4}). Increasing $\nu^2$ leads to increasing external field photon 
number density and it is more likely that $l$ external field quanta will contribute to 
the STPPP process.

For a fixed value of $\nu ^2$ however, figures \ref{pga1}-\ref{pga4} reveal
an optimum value for $l$ at which the differential cross section is a maximum and beyond which the
differential cross section decreases 
\footnote{$l=1$ for the $\nu^2 = 0.5,1.0$ plots and $l=0$ for the $\nu^2 =0.1$ plot 
of figure \ref{pga1}. $l=1$ for all plots of figure \ref{pga2}. $l=2$ for the 
$\nu^2 =0.1$ plot, $l=4$ for the $\nu^2 =0.3$ plot, and $l=6$ for the 
$\nu^2 =0.5$ plot of figure \ref{pga3}. $l=0$ for all plots of figure \ref{pga4}.}.

This is entirely reasonable since no matter how large the external field photon density may be, 
it is still finite and the contribution of a still larger number of external field photons is 
unlikely.

The variation of the parameter $\nu ^2$ is not the sole determining factor
of the variation in differential cross section with the number of laser
quanta $l$ taking part in the STPPP process. To proceed, the
simplest initial scattering geometry in which the initial photons are
collinear with the direction of propagation of the external field ($\theta
_1=0^{\circ }$), is considered.

Figures \ref{pgb1}-\ref{pgb8} consider the variation of the $l=0,1,2$ STPPP differential 
cross section contributions with other parameters for $\theta_1=0^{\circ}$. 
As was the case for the SCS process, the condition $\theta_1=0^{\circ }$ leads to 
the two infinite summations over $r$ and $r^{\prime}$ in the STPPP analytic expressions 
shrinking to one. The only non zero contributions to the STPPP differential cross section 
come from the $r=-1,0,1$ terms.

Figures \ref{pgb1}-\ref{pgb2} show the $l=0$, $r=0$ contribution to
the STPPP differential cross section when $\theta _1=0^{\circ }$ at different values of the 
initial particle energy-momenta, $\omega $ and $\omega _1$. 
Each figure contains several plots corresponding to different
values of the external field intensity parameter $\nu ^2$. The main feature
is a $\theta _f=-180^{\circ },0^{\circ },180^{\circ }$ peak
structure, with differential cross section minimums occurring at $%
\theta _f=-90^{\circ },90^{\circ }$. As $\nu ^2$ increases from zero, the
differential cross section generally decreases while maintaining the same
peak structure. The greatest rate of decrease of the cross section occurs
at $\theta _f=-180^{\circ },0^{\circ },180^{\circ }$. The $\nu ^2=0$ plot of
figures \ref{pgb1}-\ref{pgb2} corresponds to two photon pair production in the absence of an 
external field. The $\theta _f=-180^{\circ },0^{\circ },180^{\circ }$ peak structure can be
explained by the Breit-Wheeler equation in the centre of mass reference frame

\begin{equation}
\label{c5eq2}
\frac{d\sigma }{d\Omega _{p_{-}}}=\frac{r_0^2}4\left( \frac
m{\omega _1}\right) ^2\frac \beta {(1-\beta ^2\cos {}^2\theta _f)^2}\left[
1-\beta ^4\cos {}^4\theta _f+2\left( \frac m{\omega _1}\right) ^2\beta
^2\sin {}^2\theta _f\right] 
\end{equation}

\medskip\ 

The $e^{+}e^{-}$ pair are most likely created such that their 3-momenta are
collinear with the initial photon 3-momenta. That peaks of equal height
occur at $\theta _f=-180^{\circ },180^{\circ }$ as well as $\theta
_f=0^{\circ }$ is due to the identical energy of the initial photons in the
centre of mass frame. As the external field intensity $\nu ^2$ increases, the $e^{+}e^{-}$ pair
gain extra momentum due to their interaction with the field of the external
electromagnetic wave. At $\theta _1=0^{\circ }$ the net momentum gained is given by

\begin{equation}
\label{c5eq3}
\dfrac{\nu ^2}{\omega \omega _1}\,\frac 1{1-(1-\frac{1+\nu ^2}{%
\omega _1^2})\cos {}^2\theta _f} 
\end{equation}

\medskip\ 

Since the $l=0$ term of the STPPP differential cross section does not
involve a momentum contribution from external field quanta, then the
momentum required to produce the $e^{+}e^{-}$ pair embedded in the external
field must come from the initial photon momenta. Less energy is available to create the 
pair and the result is a decrease in the differential cross section. 
That this decrease is not constant for
all final scattering geometries is due to the angular dependence of equation 
\ref{c5eq3}. At $\theta _f=-180^{\circ },0^{\circ },180^{\circ }$ the net
momentum gain due to the external field is a maximum, and for fixed $\nu
^2,\omega _1$ and $\omega $, the decrease in the differential cross section
is also a maximum. The converse is true at $\theta _f=-90^{\circ },90^{\circ}$.

Figures \ref{pgb4a}-\ref{pgb4b} compare the $l=0$, $r=0$ and $l=0$, $r=$all
contributions to the STPPP differential cross section with the Breit-Wheeler differential 
cross section for initial scattering angle $\theta _1=0^{\circ }$.
The main feature is the $\theta _f=-180^{\circ },0^{\circ },180^{\circ }$ peak structure, 
the origin of which has been discussed previously. Comparison of figure \ref{pgb4a} to 
figure \ref{pgb4b} reveals that the magnitude of the differential cross section seems to 
depend on the ratio $\frac{\omega _1}\omega$. Figure \ref{pgb4a} with $\frac\omega{\omega_1}=3.3$ 
displays a $l=0$, $r=\text{all}$ contribution in excess of the Breit-Wheeler plot, and 
figure \ref{pgb4b} with $\frac\omega{\omega _1}=0.13$ displays a $l=0$, $r=$all plot
of lower magnitude than the Breit-Wheeler plot. As the external field is
switched on, the $e^{+}e^{-}$ pair produced in the STPPP process require a
greater initial momentum to be created, consequently the $l=0$, $r=0$ plots
of figures \ref{pgb4a}-\ref{pgb4b} lie below the Breit-Wheeler plots. 

The contribution from the $l=0$, $r=\pm 1$ plots is actually dependent on the comparative 
values of $\omega,\omega_1$ and $m_\ast$. In figure \ref{pgb4a} the initial photon does not 
have enough energy to create the pair and enable a quantum to be given up to the external field. 
So the $r=-1$ contribution is suppressed and the $r=1$ contribution alone increases the 
$\theta_f=-180^{\circ },0^{\circ },180^{\circ }$ peaks due to the availability of extra 
momentum. In figure \ref{pgb4b} there is more than enough energy for quanta to be given up. The 
$r=-1$ contribution outweighs the effect of the $r=1$ contribution and there is a diminished 
momentum available for the $e^+e^-$ pair.

Figures \ref{pgb4e}-\ref{pgb4f} display the effect on the total $l=0$
contribution of increasing the external field
intensity for differing values of the ratio of particle energy $\frac{\omega_1}\omega$. 
The main feature of all four figures is a central peak centred
at $\theta _f=0^{\circ }$ which diminishes in height as the external field
intensity increases. Peak widths vary with $\frac{\omega_1}\omega$, the narrowest
obtained for figure \ref{pgb4e} with $\frac{\omega_1}\omega =10$ increasing
in size to the largest width in figure \ref{pgb4f} with $\frac{\omega_1}\omega=0.15$. 
Another feature dependent on the value of $\frac{\omega_1}\omega$ is a secondary and 
more complicated peak structure.
\footnote{the $\nu^2=1,2$ plots of figure \ref{pgb4e} 
($\frac{\omega_1}{\omega}=10$) and figure \ref{pgb4c} 
($\frac{\omega_1}{\omega}=2.5$) display a twin peak structure.}

The behaviour here can be explained by examination of the arguments of the
Bessel functions which are nonzero when fermions involved in the scattering process 
propagate in directions other than the direction of propagation of the external field. The 
fermion momenta gain a longitudinal oscillatory components which can be interpreted as the 
contribution of external field quanta to the STPPP process. With $l=0$, $\theta 
_i=0^{\circ }$ and $F(\cos\theta_f)$ being a function that is minimised at  $\theta 
_f=-180^{\circ },0^{\circ },180^{\circ }$, the Bessel function arguments are 

\begin{equation}
\label{c5eq5} 
\begin{array}{lcl}
\bar z=\nu \dfrac{\omega _1}\omega F(\cos \,\theta _f) &  & \text{%
direct channel} \\ \dbr{z}=0 &  & \text{exchange channel} 
\end{array}
\end{equation}

\medskip\ 

The onset of secondary (and more complicated) peak structure in figures \ref
{pgb4e}-\ref{pgb4c} coincides with large $\frac{\omega _1}\omega$ (and therefore large $\bar z$). 
In the range $-60^{\circ}<\theta_f<60^{\circ}$ $\bar{z}$ varies greatly and the Bessel function 
along with the differential cross section is oscillatory. This oscillatory behaviour is enhanced by
increasing external field intensity due to the direct relationship between $%
\bar z$ and $\nu $.

Where the ratio $\frac{\omega _1}\omega $ is small ($\frac{\omega _1}\omega
=0.4$ for figure \ref{pgb4d} and $\frac{\omega _1}\omega =0.15$ for figure 
\ref{pgb4f}) the argument $\bar z$ achieves a small maximum value and the
form of the differential cross section is a single peak centred at $\theta _f=0^{\circ}$. 

Figures \ref{pgb5}-\ref{pgb8} compare the $l=0,1,2$ contributions to the
STPPP differential cross section for a field intensity of $\nu ^2=0.5$ and
varying particle energies. The main features of these figures are the development
of a dual $\theta_1\sim 0^{\circ}$ peak structure for higher values of $l$,
peak heights which become larger at $\theta _f\sim 0^{\circ }$ than at 
$\theta _f=-180^{\circ },180^{\circ }$ when $l$ increases, and
relative $l$ contributions to the differential cross section which vary with
the value of $\frac{\omega _1}\omega $.

We examine the differing $\theta _f=-180^{\circ },0^{\circ },180^{\circ }$ peak heights 
first. The $e^+e^-$ pair are more likely produced with 3-momenta collinear with initial photon 
momenta. When external field quanta contribute to the STPPP process momentum is 
available for fermion production and the differential cross section peak increases in the 
direction of propagation of the external field ($\theta_f=0^{\circ}$). The disparity between 
peak heights increases as the energy of external field quanta increase relative to the energy of 
the initial photons (i.e. $\frac{\omega_1}{\omega}$ decreases) and the disparity is larger for 
the $l=2$ contributions than for the $l=1$ contributions. Table \ref{c5.table3} displays the 
trend.

\input{./tex/tables/p_table3}

Mathematically the dual peak arises from the numerical value of $\bar{z}$ (equation \ref{c5eq5}) 
which is related to longitudinal components of fermion momentum obtained from interaction with 
the external field. $\bar{z}$ is small at $\theta _f=-180^{\circ },0^{\circ },180^{\circ }$. The 
dual peak structure at $\theta_f\sim 0^{\circ}$ is evidence of two competing trends. On one hand 
extra momentum is available for fermion production. On the other, momentum is not absorbed from 
the external field very well at $\theta _f=-180^{\circ },0^{\circ },180^{\circ }$. The two 
trends have different angular spreads resulting in a dual peak. This dual peak also appears in 
figures \ref{pgb5}-\ref{pgb8} at $\theta_f\sim-180^{\circ},180^{\circ}$ however its diminished by 
the preferential direction of propagation of the external field.

The relative contributions to the STPPP differential cross section for $l=0,1
$ or $2$ external field quanta contributing to the scattering process, can
only be determined precisely by a numerical calculation of the complete
differential cross section expressions. However a correlation can be
established between the number of final states available for the scattering
process and the relative peak heights of figures \ref{pgb5}-\ref{pgb8} in
the $\theta _f=0^{\circ }$ region (see Table \ref{c5.table5}). 

\input{./tex/tables/p_table5}

STPPP processes for angles other than $\theta _1=0^{\circ }$ are now considered. 
Differential cross section peaks are generally broad so only $\theta_1=45^{\circ},90^{\circ}$ 
are considered. In order to avoid resonances (which are studied in Chapters 6 and 7) $\omega_1$ 
was chosen to be considerably larger than $\omega$.

Figures \ref{pgb14}, \ref{pgb15} and \ref{pgb11} consider the STPPP process
at $\theta_1=45^{\circ }$. Figure \ref{pgb14} and \ref{pgb15} show the
variation of the $l=0$ and $l=1$ contributions for various $\nu^2$, and figure \ref{pgb11} 
compares the $l=0$, $l=1$ and $l=2$ contribution when $\nu ^2=0.1$. Figures \ref{pgb13}, \ref
{pgb16} and \ref{pgb12} are identical to figures \ref{pgb14}, \ref{pgb15}
and \ref{pgb11} except that $\theta _1=90^{\circ }$. The insets of each of 
figures \ref{pgb14}-\ref{pgb16} show the $l=0$, $r,r^{\prime }=0$ contributions to the 
STPPP process in the region $-60^{\circ }\leq \theta _f\leq 60^{\circ }$ for the same parameters
considered in the parent plots. For all six figures $2$ MeV initial photons combine with $0.5$ 
MeV external field quanta to produce the $e^{+}e^{-}$ pair. 

The main feature of figures 
\ref{pgb14}-\ref{pgb13} is a twin transverse peak structure ($\theta _f=-135^{\circ },45^{\circ}$ 
for figure \ref{pgb14} and $\theta _f=-90^{\circ },90^{\circ }$ for figure \ref{pgb13}) 
which diminishes as $\nu ^2$ increases. Secondary to both figures is the appearance of a
smaller longitudinal ($\theta _f=-180^{\circ },0^{\circ },180^{\circ }$) peak structure as 
$\nu ^2$ increases from zero. As $\nu ^2$ reaches $0.3$ the longitudinal peak structure 
dominates, having approximately twice the height of the transverse peaks.
The origin of the transverse peaks is due to the likelihood that the $e^{+}e^{-}$ pair will be 
created with 3-momenta collinear to that of the initial photons. As $\nu ^2$ increases so does 
the $e^{+}e^{-}$ rest mass and the initial energy required to produce the pair, and 
there is a general decrease in the differential cross section. 

The emergence of the $\theta _f=-180^{\circ },0^{\circ },180^{\circ }$ peaks
have their origin in differential cross section terms other than the $%
r,r^{\prime }=0$ contribution. This is indicated by the
insets of figures \ref{pgb14} and \ref{pgb13}. Once again these peaks are due to the net 
momentum gained by the $e^{+}e^{-}$ pair due to exchange of quanta with the external 
field (equation \ref{c5eq3}). The effect is enhanced as $\nu^2$ increases.

Figures \ref{pgb15} and \ref{pgb16} which display the $l=1$ contribution
reveal a zero differential cross section when $\nu ^2=0$. This is a trivial result
relating to the inability of the external field to contribute quanta to the
STPPP process when the field intensity is zero. Limits of computational accuracy were reached 
whilst generating the $\nu^2 \neq 0$ plots resulting in small "jittery" features. These can be 
ignored.  The main feature of these figures is a transverse peak structure with the 
$\theta _f=45^{\circ }$ peak approximately $1.4$ times the height of the $\theta _f=-135^{\circ }$ 
peak for figure \ref{pgb15}, and $\theta _f=-90^{\circ },90^{\circ }$ peaks of equal height for
figure \ref{pgb16}. The contribution of quanta from the external field ensures that scattering 
in the forward direction ($-90^{\circ }\leq \theta _f\leq 90^{\circ }$) is preferred.
A second feature of both figures is a localised decrease or trough at 
$\theta _f=-180^{\circ },0^{\circ },180^{\circ }$ regions. This is related once again to the 
ability of the fermions to absorb external field quanta at these points.

Figures \ref{pgb11} and \ref{pgb12}, which compare the $l=0$, $l=1$ and $l=2$
contributions at $\nu ^2=0.1$ show all the features discussed in the previous four 
figures. It is reasonably clear that summing the contributions together will produce dual 
longitudinal peaks and transverse peaks collinear with the initial photons. The next section 
will examine total STPPP differential cross sections for all $l$ contributions added together.

\bigskip\
\subsection{Differential Cross Sections Summed Over All $l$}

Figures \ref{pg1}-\ref{pg4} show the effect on the STPPP process of
varying the initial angle $\theta _1$. Each figure contains several plots
corresponding to $\theta _1=0^{\circ },45^{\circ },90^{\circ }$ and $%
180^{\circ }$. Figures \ref{pg1}-\ref{pg2} show a STPPP process in which $0.512$ MeV photons 
combine in a $2.56$ MeV external field for $\nu ^2=0.00001$ and $\nu ^2=0.1$ respectively.

The main feature of both figures is a peak structure centered around $\theta _f=0^{\circ }$.
For both figures the $\theta _1=0^{\circ }$ plot (identical to the $\theta
_1=180^{\circ }$ plots in the centre of mass reference frame) has a double peak at 
$\theta_f\sim 0^{\circ}$. The $\theta_1=45^{\circ }$ and $\theta _1=90^{\circ }$ plots reveal 
a single, flattened peak. The $\theta _1=0^{\circ },180^{\circ }$ plot peaks are of greatest
height, having a value of $0.14$ in figure \ref{pg1}
and approximately $500$ in figure \ref{pg2}. The $\theta _1=45^{\circ }$ and 
$\theta _1=90^{\circ }$ plots show a large diminishment in peak height, with
the $\theta _f=0^{\circ }$ peak almost vanishing for the $\theta
_1=90^{\circ }$ plots. We note that the vertical axis of figure \ref{pg1}
has a minimum bound of $0.11$.

The energy of initial photons, being precisely the rest mass of the 
$e^{+}e^{-}$ pair in the absence of the external field, do not create the $e^{+}e^{-}$ pair in 
the presence of the external field due to the higher fermion rest mass required. 
Consequently the only non zero contributions
to the STPPP differential cross section are those in which one or more
external field quanta are absorbed. In figure \ref{pg1}, which represents an
STPPP process in which the external field intensity is very small, the
probability of the external field giving up quanta to the STPPP process is
correspondingly small and differential cross section peaks don't exceed 0.14.
The STPPP process represented in figure \ref{pg2},
with higher external field intensity, has a correspondingly higher
probability. 

The $\theta _f=0^{\circ }$ location of differential cross section peaks no matter what the 
value of $\theta_1$ is explained by the momentum contribution from external field quanta which
dominate the STPPP process in figures \ref{pg1} and \ref{pg2}. 
Since the initial photons $k_1$ contribute no momentum to the fermion
pair, the most likely direction of both electron and positron 3-momenta is
parallel to the direction of propagation of the external field. The double
peak structure of the $\theta _1=0^{\circ }$ plots of both figures is
explained by a combination of two competing factors. On one hand the
probability of pair production is increased in directions parallel to the
3-momenta of external field quanta ($\theta _f=0^{\circ }$). On the other
hand, when $\theta _1=0^{\circ }$ there is a decreased probability that
external field quanta will participate in the STPPP process when the $%
e^{-}e^{+}$ pair are co-linear with the direction of propagation of the
external field ($\theta _f=0^{\circ }$). 

For an analysis of relative peak heights the Breit-Wheeler equation for 
two photon pair production in the centre of mass frame is of use (equation \ref{c5eq2})
For energetic photons the Breit-Wheeler differential cross section is large but strongly 
dependent on how collinear the produced fermions are with initial photons. The external field 
supplies plenty of energy to create the pair, but the momenta of external field quanta 
require the pair be created at $\theta_f=0^{\circ}$. Consequently the STPPP process is highly 
favoured for $\theta_1=0^{\circ}$ and disfavoured elsewhere.

Figures \ref{pg3}-\ref{pg4} show a STPPP process in which $1.024$ MeV
photons combine with $0.256$ MeV external field quanta to produce the 
$e^{+}e^{-}$ pair. For this ratio of particle energies ($\frac{\omega _1}\omega =4$), $l=0$ 
contributions to the differential cross section dominate the $l\neq 0$ contributions.

Figure \ref{pg3} shows a STPPP process with a very low intensity
external field ($\nu ^2=0.0001$) and is essentially a Breit-Wheeler process
with a differential cross section that peaks whenever the $e^{+}e^{-}$
3-momenta is collinear with the 3-momenta of initial photons. As $\theta_1$ changes, 
the differential cross section $\theta _f$ peaks shift by the same amount and remain the same 
height. This indicates the negligible impact of the
external field on the STPPP process in this case.

By comparison, figure \ref{pg4} with an external field intensity parameter of $0.1$ reveals
the effect of switching on the external field. The $\theta _1=0^{\circ }$ plot of 
figure \ref{pg4} reveals a $\theta _f=-180^{\circ },0^{\circ },180^{\circ }$ 
peak structure with the $\theta _f=-180^{\circ },180^{\circ }$ peaks diminished due to the 
direction of propagation of the external field. The $\theta _f\sim 0^{\circ }$ twin
peak is generally enhanced due to the contribution of external field quanta,
but diminished at $\theta _f=0^{\circ }$ due to the lower probability that
the external field will contribute quanta at that point. In comparison with
the behaviour of the $\theta _1=0^{\circ }$ STPPP differential cross section
in the $\theta _f=0^{\circ }$ region of figures \ref{pg1}-\ref{pg2}, the
double peak behaviour of figure \ref{pg4} is minimal. This is again due to
relative particle energies.
The other plots of figure \ref{pg4} show a similarly small
impact from the external field. The $\theta _1=45^{\circ }$ plot shows a
slight enhancement of the $\theta _f=45^{\circ }$ peak relative to the $%
\theta _f=-135^{\circ }$ peak, due to components of the external field
quanta 3-momenta contributing positively in the $\theta _f=45^{\circ }$
direction. The $\theta _1=90^{\circ }$ plot shows peaks of equal height 
shifted slightly in the forward direction ($\theta_f=-88^\circ,88^\circ$). 

Figures \ref{pg5}-\ref{pg8} show the effect of increasing external field 
intensity in more detail. Figures \ref{pg5}-\ref{pg6} show a STPPP 
process in which $1.024$ MeV photons are incident on a $0.256$ MeV external 
field. In figure \ref{pg5} the incident photons are collinear with the direction 
of propagation of the external field, in figure \ref{pg6} their 3-momenta 
intersect at $45^{\circ}$. 
Figures \ref{pg7}-\ref{pg8} have the same respective $\theta_1$ values
but with $0.512$ MeV photons incident on a $1.024$ MeV external field.

The $\nu ^2=0.0001$ plots of all four figures displays the differential
cross section $\theta _f$ variation expected of the Breit-Wheeler process.
Figures \ref{pg5}-\ref{pg6} with incident photons twice the energy of the 
$e^{+}e^{-}$ pair rest mass, display the usual peaks collinear with initial 
photons. Figures \ref{pg7}-\ref{pg8} show a STPPP process in which the 
initial photons have just sufficient energy to create the $e^{+}e^{-}$ pair, and 
the $\nu ^2=0.0001$ plot is consequently very close to zero.

The main feature of figures \ref{pg5} and \ref{pg7} ($\theta _1=0^{\circ }$) is
the development of a twin peak structure in the $\theta _f\simeq 0^{\circ }$
region. As has been discussed previously, this behaviour is attributable to
contributions to the STPPP process from external field quanta. This feature
is more marked in figure \ref{pg7}
\footnote{The $\nu^2=0.2$ plot of figure \ref{pg5} reaches a maximum height of 
$14.72$ whereas the $\nu^2=0.2$ plot of figure \ref{pg7} reaches a maximum height 
of $96.74$.} 
which has relative particle energies ($\frac{\omega _1}\omega =0.5$) 
in which external field quanta contributions
dominate. Figure \ref{pg5} shows a differential cross section that
diminishes with increasing $\nu ^2$ at $\theta _f=-180^{\circ },0^{\circ
},180^{\circ }$. This is due to a combination of increasing $e^{+}e^{-}$ rest 
mass and low probability of $l\neq 0$ contributions when 
$\frac{\omega_1}{\omega}>1$.

Figures \ref{pg6} and \ref{pg8} display a loss of azimuthal
symmetry associated with initial scattering geometries in which 
$\theta _1\neq 0^{\circ }$. The main feature
of figure \ref{pg6} is the development of a dual asymmetric 
$\theta_f\sim 0^{\circ }$ peak. The positive $\theta_f$ peak merges with the 
$\theta_f=45^{\circ}$ peak collinear with initial photons, seemingly shifting it. 
The ratio $\frac{\omega_1}{\omega}>1$ and the dual longitudinal peaks are small.

The diminishment of the $\theta _f\simeq -135^{\circ }$ peak height is again due
to the 3-momenta of external field quanta, components of which diminish the
electron 3-momenta for scattering geometries in which the electron travels
in backward directions.
\footnote{i.e. whenever $-180^\circ\leq\theta_f\leq -90^\circ$ and 
$90^\circ\leq\theta_f\leq 180^\circ$.} The ''noisy'' behaviour of the $\nu
^2=0.1,0.2$ and $0.3$ plots in the region $120^{\circ }<\theta _f<170^{\circ
}$ is not a feature of the STPPP process. It's origin lies in the accuracy
restrictions imposed on the computation of the differential cross section.

The main feature dominating figure \ref{pg8} is the development of a dual 
peak centered at $\theta _f=0^{\circ }$, which is to be expected for the energy 
regime in which external field quanta are more energetic than the initial 
photons. The twin peak is not symmetrical, with the $\theta _f=-7^{\circ }$ peak 
of greater height than the $\theta _f=7^{\circ }$ peak. The 
$\theta_f=-135^{\circ },45^{\circ }$ peaks have been shifted towards the forward
direction under the impact of the relatively more energetic external field
quanta. The $\theta _f=-135^{\circ }$ peak shifts to $\theta_f=-90^{\circ }$ 
and the $\theta _f=45^{\circ }$ peak is partially obscured by
the positive $\theta _f$ peak. A secondary feature of figure \ref{pg8} is
the development of a smaller peak at $\theta _f\simeq 55^{\circ }$.

The asymmetric $\theta _f\sim 0^{\circ }$ dual peak is explained by
competing factors. Positive $\theta_f$ pair production is enhanced because its 
more nearly collinear with initial photons. However when the electron is 
directed to negative $\theta _f$ directions ($\theta_f\sim -7^{\circ }$) and 
further away from the collinear direction, the probability
of external field quanta contributing to the process increases. This is so 
because with $\theta_1\neq 0$ the relations expressing the longitudinal fermion 
momentum contributions (equation \ref{c5eq5}) become asymmetric with respect 
to $\theta_f$. In the relative particle energy regime considered, the second factor 
outweighs the first and the $\theta _f\sim -7^{\circ }$ peaks are of greater 
height than the $\theta _f\sim 7^{\circ }$ peaks.

The $\theta _f\sim 55^{\circ }$ peak is explained by the expression for
the net momentum gained by the $e^{+}e^{-}$ pair due to the interaction with
the external field, an expression considerably more complicated than that
for the $\theta _1=0^{\circ }$ case. A numerical evaluation reveals that the net
momentum is at its minimum at $\theta _f=-53^{\circ }$ and $\theta
_f=53^{\circ }$ for the $l=1$ contribution, and at these points the
probability of pair production is consequently maximised. The effect is
small with only the $\theta _f\simeq 55^{\circ }$ peak visible due to the
positive contribution from external field quanta at that point.

Figures \ref{pg11}-\ref{pg12} show the azimuthal ($\phi _f$) variation of
the STPPP differential cross section for $1.024$ MeV photons incident on a $%
0.256$ MeV external field. For figure \ref{pg11} $\theta_1=0^{\circ}$ and 
$\theta_f=45^{\circ}$. With this geometry, the differential cross section is
unchanged as $\phi _f$ ranges from $0^{\circ }$ to $360^{\circ }$. This
is due to the circular polarisation of the external 
field which, along with $\theta_1=0^{\circ}$ initial photons remain unchanged
by rotation through azimuthal angles. That the differential cross section
decreases with increasing $\nu ^2$ is due to the increase
in the $e^{+}e^{-}$ pair rest mass.

Figure \ref{pg12} examines azimuthal variation when $\theta_1,\theta_f=45^{\circ}$ 
The STPPP process is most favourable at $\phi _f=0^{\circ }$, and least 
favourable at $\phi_f=180^{\circ }$. The explanation here is a geometrical one. 
At $\phi _f=0^{\circ }$ the $e^{+}e^{-}$ pair and the initial photons are 
collinear, and at $\phi _f=180^{\circ }$ perpendicular.

Figures \ref{pg13}-\ref{pg16} show the effect of varying the ratio of initial 
photon energy to the energy of external field quanta. Each figure contains four 
plots with incident photons of varying energy incident on a $1.024$ MeV
external field. Figures \ref{pg13}-\ref{pg14} show a STPPP process with
a very small external field intensity $\nu ^2=0.0001$ and an incident angle of 
$\theta _1=0^{\circ }$ and $\theta _1=30^{\circ }$ respectively. Figures 
\ref{pg15}-\ref{pg16} have the same scattering geometry with an external
field intensity $\nu ^2=0.1$.

The main feature of figures \ref{pg13}-\ref{pg14} is a Breit-Wheeler, 
collinear peak structure ($\theta _f=-180^{\circ },0^{\circ },180^{\circ}$
for $\theta _1=0^{\circ }$ and at $\theta _f=-150^{\circ },30^{\circ }$
for $\theta _1=30^{\circ }$). The peaks broaden out and decrease in height as 
$\omega_1$ approaches the rest mass of the STPPP fermion ($0.512$ MeV for figure 
\ref{pg13}, and $0.537$ MeV for figure \ref{pg14}). 

That the STPPP differential cross section behaves like that of the Breit-Wheeler 
process is due to the very low external field intensity which ensures that the 
external field plays little part. The peak structure is 
due to the 3-momenta of the initial photons, and the decrease in peak height with 
decreasing $\omega _1$ is a consequence of the decrease in 3-momenta available 
to the $e^{+}e^{-}$ pair from the initial photons. 

Figures \ref{pg15}-\ref{pg16} have the same parameters as the previous two 
figures but with significant external field intensity. The main feature 
of figure \ref{pg15} is $\theta_f=-180^{\circ },0^{\circ },180^{\circ }$ peaks which
take on a $\theta_f\sim 0^{\circ}$ dual peak structure as $\omega_1$ decreases.
The $\theta_f=-180^{\circ },180^{\circ }$ peaks decrease. As we have seen
previously, the probability of participation from external field quanta is
dependent on the ratio $\frac{\omega _1}\omega$, increasing the
differential cross section in the propagation direction of the external
field ($\theta _f\simeq 0^{\circ }$) and decreasing it at $\theta
_f=-180^{\circ },180^{\circ }$. The development of the dual peak structure
is once again due to the probability of participation of 
external field quanta which remains low at $\theta _f=0^{\circ }$ but increases 
in the region near $\theta _f=0^{\circ }$.

Figure \ref{pg16} displays $\theta _f=-150^{\circ },30^{\circ }$ peaks for the 
$\omega _1=5.12$ MeV and $\omega _1=2.56$ MeV plots in which the ratio 
$\frac{\omega _1}\omega $ is greater than unity and the
STPPP process is dominated by the 3-momenta of the initial photons. For the 
$\omega _1=0.563$ MeV plot in which external field quanta contributions dominate, 
a $\theta_f=0^{\circ}$ peak develops.
The small secondary peak at $\theta _f=44^{\circ }$ is due to the angular
dependence of the net momentum obtained by the $e^{-}e^{+}$ pair from the
external field, which falls to its minimum value, and therefore maximises
the differential cross section at that point. We note that the $\omega
_1=0.768$ MeV plot has not been included in figure \ref{pg16}. This is due
to the onset of resonance behaviour which is to be considered in chapter 7.

Figures \ref{pg17}-\ref{pg20} show the variation of the STPPP
differential cross section with external field intensity for various 
$\theta_1$. Figures \ref{pg17}-\ref{pg19} consider different values for $\omega$, 
$\omega_1$ and figure \ref{pg20} shows the $l=0$ and $l=2$ contributions of 
figure \ref{pg19}. For all figures $\theta_f=45^{\circ }$ and 
$\varphi_f=0^{\circ }$.

The main feature of figures \ref{pg17}-\ref{pg18} is a differential
cross section that generally increases with increasing $\nu ^2$. The
increase is greatest for $\theta _1=0^{\circ }$ and least for $\theta
_1=45^{\circ }$. At work here are two competing tendencies. An increasing $%
\nu ^2$ increases the rest mass of the $e^{-}e^{+}$ pair and diminishes the
STPPP differential cross section. However an increasing $\nu ^2$ increases
the probability that the external field quanta will contribute to the
process thereby increasing the differential cross section. For most of the
plots of figures \ref{pg17}-\ref{pg18}, the second tendency is dominant and
the differential cross section increases. For the $0\leq \nu ^2\leq 0.12$
region of the $\theta _1=0^{\circ }$ plot and the entire $\theta
_1=45^{\circ }$ plot of figure \ref{pg18}, the first tendency is dominant
and the differential cross section decreases.
That the $\theta _1=0^{\circ }$ plots of figures \ref{pg17}-\ref{pg18} show
the greatest increase in differential cross section is due to the
contribution of external field quanta which for the relative particle energy
regime considered $\frac{\omega _1}\omega \leq 1$, favour scattering
geometries in which initial photons are collinear with the propagation
direction of the external field.

Figure \ref{pg19} shows an increase in differential cross section that 
proceeds in a series of "steps". This phenomena has been observed before
in computations of the OPPP process \cite{NarNikRit65,Lyulka75} and relates to the
existence of a threshold for pair production (expressed by equation \ref
{c5eq1}). For the particle energies represented by figure \ref{pg19}, an
increase in $\nu ^2$ increases the $e^{-}e^{+}$ pair rest mass and increases
the number of external field quanta required to create the pair. For instance in 
the region $0<\nu^2 <0.21$ no external field quanta are required, 
in the region $0.21<\nu^2 <0.35$ one external field quanta is required, 
and in the region $0.35<\nu^2 <0.49$ two external field quanta are required.

Whether or not the differential cross section is discontinuous at the "steps" is 
a matter of concern. However, figure \ref{pg20}, which shows the 
$l=0$ and $l=2$ contributions at their respective "cut-off" points, reveals
that the differential cross section approaches zero as the cut-off point is
reached. This is as it should be since a STPPP process with an initial state
which contains just enough energy to create the $e^{-}e^{+}$ pair, but no
energy to provide the pair with a non zero 3-momenta, yields a zero
probability that $e^{-}e^{+}$ 3-momenta will be observed.

%% file: tex/tables/p_table1.tex
\begin{table}[!b]
\label{c52tab1}
\begin{tabular}{|c|c|c|c|c|c|c|c|c|} \hline
$l$&$r$&$ \nu^{2}$ & $ \omega (\text{MeV})$ & 
$ \omega_{1,2}(\text{MeV})$ &$ \theta_{1}$ & 
$ (\theta_{f},\phi_{f})$ & figure(s) \\ \hline\hline
$0\rightarrow 6 $&$ \text{all} $&$ 0.1,0.3,0.5 $&$ 2.56 $&$ 0.768 $&
$ 0^{\circ } $&$ (45^{\circ },0^{\circ }) $& \ref{pga1} \\ \hline
$1\rightarrow 9 $&$ \text{all} $&$ 
0.1,0.3,0.5 $&$ 1.024 $&$ 0.512 $&$ 45^{\circ } $&$ (45^{\circ },0^{\circ }) $&
 \ref{pga2} \\ \hline
$-3\rightarrow 15 $&$ \text{all} $&$ 0.1,0.3,0.5 $&$ 0.102 $&$ 0.768 $&$ 
0^{\circ } $&$ (45^{\circ },0^{\circ }) $& \ref{pga3} \\ \hline
$0\rightarrow 7 $&$ 
\text{all} $&$ 0.1,0.3,0.5 $&$ 0.256 $&$ 0.614 $&$ 45^{\circ } $&$ (45^{\circ
},0^{\circ }) $& \ref{pga4} \\ \hline
$0 $&$ 0 $&$
\begin{array}{c}
0,0.1,0.2 \\ 
0.3,0.4,0.5
\end{array} 
$&$ 2.56,0.102 $&$ 0.768 $&$ 
0^{\circ } $&$ (0^{\circ }\rightarrow 360^{\circ },0^{\circ }) $&
\ref{pgb1},\ref{pgb2}
\\ \hline 

$0 $&$ \text{all} $&$
\begin{array}{l}
0.5,1.0, \\ 
2.0
\end{array}
$&$ 0.256 $&$ 2.56 $&$ 0^{\circ } $&$ 
(0^{\circ }\rightarrow 360^{\circ },0^{\circ }) $& \ref{pgb4e} \\ \hline
$0 $&$ \text{all} $&$
\begin{array}{l}
0.1,0.5, \\ 
1.0,2.0
\end{array}
$&$ 1.024 $&$ 2.56 $&$ 0^{\circ } $&$ 
(0^{\circ }\rightarrow 360^{\circ },0^{\circ }) $& \ref{pgb4c} \\ \hline
$0 $&$ \text{all} $&$
\begin{array}{l}
0.1,0.5, \\ 
1.0,2.0
\end{array}
$&$ 1.024 $&$ 0.409 $&$ 0^{\circ } $&$ 
(0^{\circ }\rightarrow 360^{\circ },0^{\circ }) $& \ref{pgb4d} \\ \hline
$0 $&$ \text{all} $&$
\begin{array}{l}
0.1,0.5, \\ 
1.0
\end{array}
$&$ 5.12 $&$ 0.768 $&$ 0^{\circ } $&$ 
(0^{\circ }\rightarrow 360^{\circ },0^{\circ }) $& \ref{pgb4f} \\ \hline

$0,1,2 $&$ \text{all} $&$ 
0.5 $&$ \begin{array}{c} 0.05,0.77 \\ 1.28,2.56 \end{array} $&$ 1.024 
$&$ 0^{\circ } $&$ (0^{\circ }\rightarrow 360^{\circ
},0^{\circ }) $&$ \begin{array}{c} \ref{pgb5},\ref{pgb5} \\ \ref{pgb7},\ref{pgb8} 
\end{array} $\\ \hline

$0,1 $&$ \text{all} $&$
\begin{array}{c}
0,0.1, \\ 
0.2,0.3
\end{array}
$&$ 0.256 $&$ 1.024 $&$ 45^{\circ},90^{\circ } $&$ (0^{\circ }\rightarrow 360^{\circ },
0^{\circ}) $&$ \begin{array}{c}
\ref{pgb14},\ref{pgb13} \\ 
\ref{pgb15},\ref{pgb16}
\end{array} $ \\ \hline

$0,1,2 $&$ \text{all} $&$ 0.1 $&$ 0.256 $&$ 1.024 $&$ 
45^{\circ},90^{\circ } $&$ 
(0^{\circ}\rightarrow 360^{\circ },0^{\circ }) $&
\ref{pgb11},\ref{pgb12} \\ \hline
\end{tabular}
\caption{\bf The parameter range for which the STPPP differential cross section 
$l$ and $r$ contributions are investigated.} 

\end{table}

%% file: tex/tables/p_table2.tex
\begin{table}[!h]
\label{c53tab1}
\begin{tabular}{|l|l|l|l|l|l|l|} \hline
$ \nu^{2}$ & $ \omega (MeV)$ & $ \omega_{1},\omega_{2}(MeV)$ &$
\theta_{1}$ & $ (\theta_{f},\phi_{f})$ & figure(s) \\ \hline\hline
$0.00001,0.1 $&$ 2.56 $&$ 0.512 $&$ 
\begin{array}{l}
0^{\circ },45^{\circ }, \\ 
90^{\circ },180^{\circ }
\end{array}
$&$ \left( 0^{\circ }\rightarrow 360^{\circ },0^{\circ }\right)  $& \ref{pg1},
\ref{pg2} \\ \hline
$0.0001,0.1 $&$ 0.256 $&$ 1.024 $&$
\begin{array}{l}
0^{\circ },45^{\circ }, \\ 
90^{\circ },180^{\circ }
\end{array}
$&$ 
\left( 0^{\circ }\rightarrow 360^{\circ },0^{\circ }\right)  $& \ref{pg3},\ref
{pg4} \\ \hline
$0,0.1,0.2,0.3 $&$ 0.256 $&$ 1.024 $&$ 0^{\circ },45^{\circ } $&$ \left(
0^{\circ }\rightarrow 360^{\circ },0^{\circ }\right)  $& \ref{pg5},\ref{pg6}
\\ \hline
$\begin{array}{l}
0,0.05, \\ 
0.1,0.2
\end{array}
$&$ 1.024 $&$ 0.512 $&$ 0^{\circ },45^{\circ } $&$ \left( 0^{\circ }\rightarrow
360^{\circ },0^{\circ }\right)  $& \ref{pg7},\ref{pg8} \\ \hline
$0,0.1,0.5 $&$ 
0.256 $&$ 1.024 $&$ 0^{\circ },45^{\circ} $&$ 
\left( 45^{\circ },0^{\circ }\rightarrow
360^{\circ }\right)  $& \ref{pg11},\ref{pg12} \\ \hline
$0.0001 $&$ 1.024 $&$ 
\begin{array}{l}
5.12,2.56, \\ 
0.768,0.563
\end{array}
$&$ 0^{\circ },30^{\circ} $&$ \left( 0^{\circ }\rightarrow 360^{\circ },
0^{\circ }\right) $& 
\ref{pg13},\ref{pg14}
\\ \hline
$0.1 $&$ 1.024 $&$ 
\begin{array}{l}
5.12,2.56, \\ 
0.768,0.563
\end{array}
$&$ 0^{\circ },30^{\circ} $&$ \left( 0^{\circ }\rightarrow 360^{\circ },
0^{\circ }\right) $& 
\ref{pg15},\ref{pg16}
\\ \hline
$0\rightarrow 1.5 $&$ 1.536 $&$ 0.512 $&$ 0^{\circ },20^{\circ},45^{\circ } $&$
 \left( 45^{\circ},0^{\circ }\right)  $& \ref{pg17} \\ \hline
$0\rightarrow 1.5 $&$ 0.973 $&$ 1.024 $&$ 0^{\circ
},20^{\circ },45^{\circ } $&$ \left( 45^{\circ },0^{\circ }\right)  $& \ref
{pg18} \\ \hline
$0\rightarrow 1.5 $&$ 0.563 $&$ 0.061 $&$ 0^{\circ} $&$ 
\left( 45^{\circ },0^{\circ }\right)  $& \ref{pg19},\ref{pg20} \\ \hline
\end{tabular}
\caption{\bf The parameter range for which the STPPP differential cross section
summed over all $l$ is investigated.}

\end{table}

%% file: tex/tables/p_table3.tex
\begin{table}[!b]
\center{
\begin{tabular}{|c|c|c|c|} \hline
figure & $l=1:l=0$ & $l=2:l=0$ & $\frac{\omega}{\omega_1}$ \\ \hline\hline
\ref{pgb5} &$ 0.723 $&$ 0.5 $&$ 0.05 $ \\ \hline
\ref{pgb6} &$ 1.56 $&$ 1.046 $&$ 0.4 $ \\ \hline
\ref{pgb7} &$ 2.77 $&$ 3.96 $&$ 1.25 $ \\ \hline
\ref{pgb8} &$ 12.26 $&$ 22.52 $&$ 2.5 $ \\ \hline
\end{tabular} }
\caption{\bf\bm The ratio of peak heights and particle energies for
figures \ref{pgb5} - \ref{pgb8}.}
\label{c5.table3}
\end{table}

%% file: tex/tables/p_table5.tex
\begin{table}[!b]
\center{
\begin{tabular}{|c||c|c||c|c||c|c||} \hline
figure & $l=1$:$l=0$ & $l=1$:$l=0$  & $l=2$:$l=1$ &
$l=2$:$l=1$  &  $l=2$:$l=0$ & $l=2$:$l=0$  \\ 
 & density of & differential  & density of &
differential  &  density of & differential  \\
 & final states & cross section  & final states &
cross section  &  final states & cross section  \\ \hline\hline
\ref{pgb5}  &$ 1.077 $&$ 0.72 $  &$ 1.071 $&$ 0.7 $
 &$ 1.154 $&$ 0.5 $  \\ \hline
\ref{pgb6}  &$ 1.59 $&$ 1.56 $  &$ 1.35 $&$ 0.67 $
 &$ 2.14 $&$ 1.05 $  \\ \hline
\ref{pgb7}  &$ 2.74 $&$ 2.74 $  &$ 1.59 $&$ 1.43 $
 &$ 4.36 $&$ 3.91 $  \\ \hline
\ref{pgb8}  &$ 4.36 $&$ 12.27 $  &$ 1.73 $&$ 1.84 $
 &$ 7.55 $&$ 22.54 $  \\ \hline
\end{tabular} }
\caption{\bf\bm A comparison of the number of final states and the
differential cross section values for various ratios of l contributions
for the STPPP process represented in figures \ref{pgb5} - \ref{pgb8}.}
\label{c5.table5}
\end{table}

%% file: chap6.tex
\section{Introduction}

Both the SCS and STPPP differential cross sections contain resonant
infinities corresponding to the points where the denominator of the electron
propagator goes to zero. These infinities arise because the interaction of
the electrons with the cloud of virtual photons has not been taken into
account. This interaction gives rise to a ''spreading'' of electron energy
levels, i.e. they gain a finite width. The width, containing both an
imaginary and a real part, has the effect of restricting the cross section to
finite values at points of resonance. A calculation of this resonance width
requires a calculation of the electron self energy, or mass operator as it
is also known.

The electron mass operator in the presence of a
circularly polarised electromagnetic field was calculated by \cite{Baier75}. 
They use an operator technique to derive the mass operator. After
taking an operator average, they write the imaginary part as a double
integration with upper bounds at infinity.
\cite{BecMit76} perform a more complete calculation by obtaining the
corrected electron propagator in the external field via a calculation of the
electron self energy using Schwinger's proper-time electron Green's
functions, choosing the Feynman gauge for the radiation field and using
light-like coordinates. The corrected propagator so obtained contains
modifications in both numerator and denominator and, as has been recognised
by the authors, is too complex to be included in other external field
processes (see also \cite{AffKru87}).

\cite{BecMit76} approximate the corrected propagator by expanding both
numerator and denominator in powers of the fine structure constant $\alpha $. 
They obtain a mass correction in the denominator containing both imaginary
and real parts which are functions of the parameter $\rho =\frac{2(kp)}{m^2}$. 
The form of these mass corrections are the same as those obtained by \cite{Baier75b}, 
i.e. a double integration with infinite upper bounds.

\cite{BecMit76} write down a further approximation to the imaginary and
real parts of the mass correction as a simple algebraic function of the
parameter $\rho$ and the external field intensity $\nu ^2$. These further
approximations are only valid for a restricted range of parameter values 
($0.1\leq \rho \leq 0.3$ and $\nu^2 \leq 0.1$). Moreover, \cite{BecMit76}
find their expansion invalid for the real part whenever $\rho \leq \alpha (=\frac 1{137})$. 
This condition falls within the range of parameter values to be considered.

In this chapter the calculations are repeated using a different approach. The
external field electron energy shift (EFEES) is calculated directly by using
Schwinger's well know equation for an electron interacting with a quantised
Maxwell field and a classical plane wave \cite{Schwinger54}. The
EFEES is simply the external field electron self energy 
(EFESE)\footnote{Please note the difference - EFEES and EFESE are two different acronyms} 
sandwiched between Dirac bispinors. The form of the bound electron propagator to be used is 
that obtained by \cite{Ritus72} and real and imaginary parts of the EFEES are separated 
by use of dispersion relations. An EFEES expression that is an infinite
summation of terms corresponding to the number of external field photons
that contribute to the self energy process will be obtained. 

This calculation is necessary for two reasons. Firstly, the EFEES is to be inserted into the 
expressions for the SCS and STPPP cross sections considerably lengthening the computational time
required to obtain a numerical result.\footnote{The EFEES must be recalculated 
at each data point in the plots presented in Chapters 4,5} 
The expressions obtained in this chapter for the EFEES require considerably less
computational time than the expressions obtained by 
\cite{Baier75,BecMit76}.\footnote{This is due largely to the rapid convergence of the summation} 
It is possible to obtain the representation for the EFEES derived here from
that obtained by \cite{BecMit76}, by using several transformations \cite{Baier75b}. 
However these are not immediately obvious and the representation in
terms of an infinite summation is not written down in the literature.

Secondly, these calculation lend themselves to physical insights not obvious in other calculations. 
This is the case because, from the outset, the influence of the external field upon the process 
is expressed as a series of contributions corresponding to the number of external field photons 
that take part. As \cite{Zeldovich67} and others have shown, IFQED processes
are best understood by transitions between electron energy levels characterised by exchange of 
external field quanta. The work performed in this chapter take as a starting point the 
calculation of the electron energy shift in the absence of an external field (\cite{ItzZub80}  
pg.329) and the mass operator in a constant crossed electromagnetic field \cite{Ritus72}.

The full gamut of external field radiative corrections, though important in
their own right, were not considered in this work. This is purely in regards to
limitation of space. Calculation of the EFEES alone is sufficient to obtain the
important characteristics of the resonant cross section and serves as a good
first approximation to the nonlinear effect of the external field on the SCS and
STPPP cross sections.

\bigskip\
\section{The external field electron energy shift in a circularly polarised external 
field}

The general form of the energy shift $\Delta \epsilon _{p,s}$ of an
electron of momentum $p$ and spin $s$ in the presence of the external field,
is the external field electron self energy (EFESE) in the external field sandwiched between Dirac
bi-spinors (see section \ref{c2efees}).

\begin{equation}
\label{c6.eq1}\Delta \epsilon _{p,s}=\frac m{\epsilon _p}\,\bar
u_{p,s}\,\Sigma ^e(p)\,u_{p,s} 
\end{equation}

\medskip\ 

The EFESE in momentum space is a
Volkov transform of the EFESE in position space \cite{BecMit76}. This
representation can be obtained directly from the orthogonality and
completeness properties of the Volkov function $E_p(x)$ .

\begin{equation}
\label{c6.eq2} 
\begin{array}{c}
\Sigma ^e(p)=\dfrac 1{VT(2\pi )^4}\dint d^4x_1d^4x_2\,\bar E_p(x_2)\,\Sigma
^e(x_2,x_1)\,E_p(x_1) \\ 
\Sigma ^e(x_2,x_1)=ie^2\gamma^\mu \,G^e(x_2,x_1)\gamma _\nu \,D_{\mu \nu
}(x_2-x_1) 
\end{array}
\end{equation}

\medskip\ 

\begin{figure}[!b]
\centerline{\includegraphics[height=4.5cm,width=2cm]{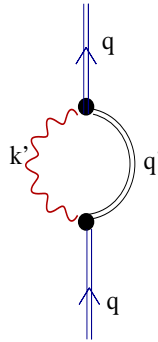}}
\caption{\bf\bm Feynman diagram for the electron self-energy in an external field.}
\label{c6fig1}
\end{figure}

\medskip\ 

The precise form of this representation can be written with the aid of the
EFESE Feynman diagram (Figure \ref{c6fig1}) and the expressions for the
fermion propagator in the external field $G^e(x_2,x_1)$ and the free photon
propagator $D_{\mu \nu }(x_2-x_1)$.

\begin{equation}
\label{c6.eq3} 
\begin{array}{rl}
\Sigma ^e(p) & = \dfrac{i4e^2}{VT(2\pi )^4}\dint d^4x_1\,d^4x_2\,d^4k'\,d^4q'\; \\[10pt]
& \times \;\; \bar E_p(x_2)\gamma^\mu \,E_{p'}(x_2)\left( \slashed{q}'+
\dfrac{e^2a^2}{2(kq^{\prime })}\slashed{k}+m\right) \bar E_{p'}(x_1)\gamma _\mu \,E_p(x_1)\,\dfrac
1{q^{\prime 2}-m_{*}^2}\,\dfrac 1{k^{\prime 2}}\,e^{ik^{\prime }(x_2-x_1)} 
\end{array}
\end{equation}

\medskip\ 

A representation is extracted from equation \ref{c6.eq3} in terms of a
discrete number of external field contributions in direct analogy with the
procedure in Chapters 4 and 5 for the SCS and STPPP processes. The exponential functions are 
grouped together, oscillatory functions expanded in an infinite
summation of Bessel functions and the orthogonality properties of the
4-vectors $a_1,a_2,k$ are used to write

\begin{equation}
\label{c6eq4} 
\begin{array}{rl}
\Sigma^e(p) & = \dfrac{i4e^2}{VT(2\pi )^4}\dsum\limits_{bc}\dint
d^4x_1\,d^4x_2\,d^4k'\,d^4q'\,A(\gamma^\mu ,C,q',q)
\left( \slashed{q}'+\frac{e^2a^2}{2(kq')}\slashed{k}+m\right) \\[10pt] 
& \times \;\; A(\gamma _\mu ,B,q',q)\dfrac 1{q^{\prime 2}-m_{*}^2}\,\dfrac
1{k^{\prime 2}}\;e^{i(q-q^{\prime }+k^{\prime }+ck)x_2}\,e^{i(q^{\prime}-q-bk)x_1} \\[20pt]

\text{where} \quad & A(\gamma _\mu ,C,q^{\prime },q) = \gamma_\mu C_1-
\dfrac e{2(kq')}\left( \gamma _\mu\slashed{a}_1\slashed{k}C_2+\gamma _\mu 
\slashed{a}_2\slashed{k}C_3\right) \\[10pt]
 + & \dfrac e{2(kq)}\left( \slashed{a}_1\slashed{k}\gamma _\mu \,C_2+\slashed{a}_2
\slashed{k}\gamma _\mu \,C_3\right) -\dfrac{e^2a^2}{4(kq^{\prime })(kq)}\slashed{k}
\gamma _\mu \slashed{k}C_1 \\[15pt]
& B_1 = J_b(z)e^{ib\phi _0} \\[10pt]
& B_2 = \dfrac 1{2z}\left[ J_{b-1}(z)\,(\alpha _1-i\alpha_2)+
J_{b+1}(z)\,(\alpha _1+i\alpha _2)\right] e^{ib\phi _0} \\[10pt]
& B_3 = \dfrac 1{2z}\left[ J_{b-1}(z)\,(\alpha _2+i\alpha_1)+
J_{b+1}(z)\,(\alpha _2-i\alpha _1)\right] e^{ib\phi _0} \\[10pt]
& C_1 = J_c(z)e^{-ic\phi _0} \\[10pt]
& C_2 = \dfrac 1{2z}\left[ J_{c-1}(z)\,(\alpha _1+i\alpha
_2)+J_{c+1}(z)\,(\alpha _1-i\alpha _2)\right] e^{-ic\phi _0} \\[10pt]
& C_3 = \dfrac 1{2z}\left[ J_{c-1}(z)\,(\alpha _2-i\alpha
_1)+J_{c+1}(z)\,(\alpha _2+i\alpha _1)\right] e^{-ic\phi _0} \\[12pt]

\text{and} \quad & z = \sqrt{\alpha _1^2+\alpha _2^2} \quad;\quad 
\alpha_1=e\left[\dfrac{(a_1 p)}{(kp)}-\dfrac{(a_1p')}{(kp')}\right] \quad;\quad
\alpha_2=e\left[\dfrac{(a_2 p)}{(kp)}-\dfrac{(a_2p')}{(kp')}\right] \\[10pt]
& \quad\quad \cos \phi _0 = \dfrac{\alpha_1}z \quad;\quad \sin \phi _0 = \dfrac{\alpha _2}z

\end{array}
\end{equation}

\medskip 

Integrations over $x_1,x_2$ and $q'$ yield a single delta function $\delta^4(bk-ck)$ and the 
EFESE is only non zero if $b=c$. The number of infinite summations reduces to one, and the 
following holds

\begin{equation}
\label{c6.eq4b} 
\begin{array}{c}
C_1=B_1^{*} \quad;\quad C_2=B_2^{*} \quad;\quad C_3=B_3^{*} 
\end{array}
\end{equation}

\medskip\ 

Proceeding with the calculation at this stage presents a problem in that the
EFESE contains the term $\frac 1{VT}$ which goes to zero as all space-time is included. 
This can be resolved by setting $b=c$ at the outset. The integrations over $x_1$ and $x_2$ are 
then transformed into integrations over $x_1-x_2$ and $x_1+x_2$, with the first yielding a delta
function expressing conservation of 4-momentum for the process, and the other yielding a space-time 
volume $VT$.

\begin{equation}
\label{c6.eq5} 
\begin{array}{rl}
\Sigma ^e(p) & = \dfrac{i4e^2}{(2\pi )^4}{\displaystyle \dsum\limits_b}\dint d^4k^{\prime }\,d^4q'
\,\delta^4(q^{\prime }-q-k^{\prime }-bk)\, \\[10pt]
& \times \;\; A(\gamma^\mu,B^{*},q',q)\left( 
q'+\frac{e^2a^2}{2(kq')}k+m\right) A(\gamma _\mu ,B,q^{\prime },q)
\,\dfrac 1{q^{\prime 2}-m_{*}^2}\,\dfrac 1{k^{\prime 2}} 
\end{array}
\end{equation}

The EFESE is a summation\ of contributions corresponding to the number of
external field photons $b$ that contribute to the process. There is no lower limit on $b$ as was 
the case for the summations involved
with the SCS and STPPP processes.\footnote{A negative value for $b$ corresponds to energy being given up to the field.}
This is so because the initial and final states of the EFESE electron are virtual, 
permitting energy states to reach negative values.

An average over electron spins in equation \ref{c6.eq1} leads to the appearance of a  trace 
expression. The spin averaged EFEES is 

\begin{equation}
\label{c6.eq6} 
\begin{array}{rl}
\Delta \epsilon _p & = \dfrac 12{\displaystyle \dsum\limits_{s=1}^2}\Delta \epsilon _{p,s} \\[10pt]  
& = \dfrac{i4e^2}{(2\pi )^4}\dfrac m{\epsilon _p}{\displaystyle \dsum\limits_b}\dint 
d^4k'\,d^4q'\,\delta ^4(q^{\prime }-q-k^{\prime }-bk)\, \\[10pt]
& \times \;\; \dfrac 1{4m} \Tr \left\{ (\slashed{p}+m)\,Q(B,q,q')\right\} 
\,\dfrac 1{q^{\prime 2}-m_{*}^2}\,\dfrac 1{k^{\prime 2}} \\[20pt]
\text{where} \quad & Q(B,q,q')=A(\gamma^\mu ,B^{*},q',q)\left(q'+\frac{e^2a^2}{2(kq')}k+m\right) 
A(\gamma _\mu ,B,q',q)
\end{array}
\end{equation}

In equation \ref{c6.eq6} the integrations over $k^{\prime }$ and $q'$ are performed, and 
real and imaginary parts are extracted using a dispersion relation method outlined in Appendix 
\ref{app.kallen}. Introducing new 4-vectors $K_\mu,D_\mu$ and defining $\theta$ to be the angle 
between $\sql{K}$ and $\sql{k}$, the real and imaginary parts of the EFEES are (see equation 
\ref{app6.eq7})

\begin{equation}
\label{c6.eq7} 
\begin{array}{rl}
\Re \Delta \epsilon _p(D^2) & = \dfrac 1\pi P\dint_{\nths -\infty }^\infty 
\dfrac{\Im \Delta \epsilon _p(\sigma ^2)}{\sigma ^2-D^2}d\sigma ^2 \\[15pt]  

\Im \Delta \epsilon _p(D_\mu ) & = \dfrac{e^2m^2}{16\pi \epsilon _p}
{\displaystyle \sum\limits_b}\dint_{\nths -1}^1 d\cos \theta \,\left( 1-\dfrac{m_{*}^2}{D^2}\right) 
\\[10pt]
& \times \;\; \biggl\{ -2\,J_b^2(z)+\nu ^2\left( 1+\frac 12 
\frac{[(kD)-(kK)]^2}{(kD)(kK)}\right)  \\[10pt]
& \times \;\; J_{b-1}^2(z)+J_{b+1}^2(z)-2J_b^2(z) \biggr\} \Theta \left(D^2-m_{*}^2\right) \\[20pt]
\text{where} \quad z & = \sqrt{\frac{e^2a^2}{(kK)}(D^2+m_{*}^2)
\left[ \frac 1{(kD)}-\frac1{(kK)} \right] } \\[10pt]
D_\mu & \equiv q_\mu + b\, k_\mu \\
K_\mu & \equiv q'_\mu + \frac{1}{2}\, k_\mu 

\end{array}
\end{equation}

\medskip

The form of the real part of the EFEES requires the imaginary part to be written in terms of a 
length $D^2$. However the imaginary part of equation \ref{c6.eq7} contains vector dependences 
$(kD)$. The EFEES is to be inserted in propagator denominators and becomes important at resonance 
$q^2=m_{*}^2$ and can be neglected elsewhere. The resonance condition allows the replacement

\begin{equation}
\dfrac 1{(kD)}\rightarrow \dfrac{2b}{D^2-m_{*}^2} 
\end{equation}

\medskip\

The imaginary part of the EFEES becomes dependent only on $D^2$. A shift of integration variable 
can be made and the EFEES is

\begin{equation}
\label{c6.eq8}
\begin{array}{rl}
\Re \Delta \epsilon _p(D^2) & = \dfrac 1\pi P\dint_{\nths (m_{*}+\epsilon)^2}^\infty 
\dfrac{\Im \Delta \epsilon _p(\sigma ^2)}{\sigma ^2-D^2}d\sigma ^2 \\[20pt]  

\Im \Delta \epsilon _p(D^2) & = \dfrac{e^2m^2}{16\pi \epsilon _p}
{\displaystyle \dsum\limits_b}\dint_{\nths 0}^{u(D^2)}\dfrac{du}{(1+u)^2} \\[10pt]
& \times \;\; \left\{-4\,J_b^2(z)+\nu ^2\left( 2+ \dfrac{u^2}{1+u}\right) 
\left( J_{b-1}^2(z)+J_{b+1}^2(z)-2J_b^2(z)\right)
\right\} \\[10pt]
& \times \;\; \Theta \left( D^2-m_{*}^2\right) \\[15pt]
\text{where} \quad u(D^2) & = \dfrac{D^2}{m_{*}^2}-1 \\[10pt]
z(D^2) & = \sqrt{\nu ^2(1+\nu ^2)\frac{m^4}{(kp)^2}\;u\left(u(D^2)-u\right) }
\end{array}
\end{equation}

\medskip\ 

The lower bound, $(m_{*}+\epsilon)^2$, on the integration in the real part of the EFEES is due 
to the step function $\Theta (D^2-(m_*+\epsilon)^2)$ in the imaginary part. Self energy 
calculations in QED are prone to divergences. Indeed it is known that the electron self energy in 
the absence of the external field is divergent. Taking the limit of vanishing external field
intensity, $\nu^2\rightarrow 0$, and a contribution of zero external field
photons, $b=0$, a logarithmic divergence appears

\begin{equation}
\label{c6.eq9}
\begin{array}{rcl}
\Im \Delta \epsilon_p(D^2,\nu^2 \rightarrow 0,b=0) & = & 
\dfrac{e^2m^2}{4\pi\varepsilon_p} \left( \dfrac{m^2}{D^2}-1 \right) \\[10pt]
\Re \Delta \epsilon_p(\nu ^2\rightarrow 0,b=0) & = &
-\dfrac{e^2m^2}{4\pi^2\varepsilon_p} \dint^{\infty}_{\nths m^2} \dfrac{\text{d}\sigma^2}{\sigma^2}

\end{array} 
\end{equation}

\medskip\ 

A regularisation procedure is required. In \cite{BecMit76}'s study of the electron 
self energy in a circularly polarised external field, divergences were found to exist only in 
unregularised expressions that contained no dependence on the external field. 
\cite{Ritus72} found a similar result for the case of a constant crossed external 
electromagnetic field. Indeed, numerical evaluation of equation \ref{c6.eq8} reveals that 
divergences are only present in the field free parts. The regularisation procedure for the 
EFEES consequently reduces to the regularisation procedure for the same process in the absence of 
the external field.

Renormalisation allows the divergences of the EFEES to be absorbed into the physical 
mass of the electron. A discussion of renormalisation with the external field present 
is left to section \ref{c7renorm}. Assuming here that the EFEES has already been properly 
renormalised, the regularised EFEES, $\Delta \epsilon _p^R(D^2)$ is simply the difference
between the unregularised EFEES and the $b=0\,$ term of the unregularised
EFEES at the mass shell, $p^2=m^2$

\begin{equation}
\label{c6.eq9a}
\Delta \epsilon _p^R(D^2)=\Delta \epsilon _p(D^2)-\Delta\epsilon _p(D^2,p^2=m^2,b=0) 
\end{equation}

\medskip\ 

As it stands, the divergent part $\Delta\epsilon_p(\sigma ^2,\nu ^2\rightarrow0,b=0)$ is also undefined since the 
argument of the Heaviside step function goes to zero at that point. A small photon mass (cf. equation
\ref{appsum.eq5}) is inserted to avoid the pole so that

\begin{equation}
\begin{array}{c}
\Theta (D^2-m_*^2) \rightarrow \Theta (D^2-(m_*+\epsilon)^2) \\
q^2=m_*^2 \, , \, b=0 \Rightarrow D^2=m_*^2
\end{array}
\end{equation}

\medskip\

The argument of the step function ensures that the lower bound of the summation is $b=1$. Transforming the 
integration variable in the real part to $y\equiv\sqrt{1-\dfrac{m_*^2}{\sigma^2}}$, the regularised 
EFEES near the resonance ($q^2 \sim m_*^2$) is

\begin{equation}
\label{c6.eq11}
\begin{array}{rl}
\Re \Delta \epsilon _p^R(D^2) & = \dfrac{e^2m^2}{16\pi ^2\epsilon _p}
{\displaystyle \dsum\limits_{b=1}^\infty} \dint_{\nths 0}^1\dfrac{dw}w\dfrac{q^2-m_{*}^2}
{m_{*}^2w^2+2b(kp)(1-w^2)}\dint_{\nths 0}^{u(w^2)}\dfrac{du}{(1+u)^2} \\[10pt] 
& \times \;\;  \left\{ -4\,J_b^2(z)+\nu ^2\left( 2+ 
\dfrac{u^2}{1+u}\right) \left( J_{b-1}^2(z)+J_{b+1}^2(z)-2J_b^2(z)\right)
\right\} \\[20pt] 

\Im \Delta \epsilon _p^R(D^2) & = \dfrac{e^2m^2}{16\pi \epsilon _p}
{\displaystyle \dsum\limits_{b=1}^\infty} \dint_{\nths 0}^{u(D^2)}\dfrac{du}{(1+u)^2} \\[10pt]
& \times \;\; \biggl\{ -4\,J_b^2(z)+\nu ^2\left( 2+ 
\dfrac{u^2}{1+u}\right) \\[10pt]
& \times \;\; J_{b-1}^2(z)+J_{b+1}^2(z)-2J_b^2(z) \biggr\} \Theta \left(
D^2-(m_{*}+\epsilon )^2\right) 
\end{array}
\end{equation}

\medskip\ 

Confirmation of the expression obtained for the regularised imaginary part
can be found by use of the optical theorem which connects the regularised
imaginary part with the probability for emission of a photon from an
electron embedded in the external electromagnetic field \cite{Ritus72}.
Equation \ref{c6.eq11} indeed agrees with the HICS transition probability in a 
circularly polarised field $W^e_{\text{brem}}$ \cite{NarNikRit65}, and

\begin{equation}
\label{c6.eq12}W_{brem}^e=\frac 12\Im \Delta \epsilon _p^R(D^2) 
\end{equation}

\medskip\ 

A numerical investigation and analysis will be made in section \ref{c6fig} and \ref{c6anal}. 
Numerical comparison will also be made with the expressions obtained by \cite{Baier75,BecMit76} 
which are only valid in certain limits. The real part of the EFEES will be neglected since it 
reduces to zero at the mass shell. The imaginary part on the other hand, plays a vital role in 
determining the SCS and STPPP resonant differential cross sections which will investigated in 
Chapter 7.

\newpage 
\section{EFEES Plots}
\label{c6fig}

\begin{figure}[H]
 \centerline{\includegraphics[height=8cm,width=10cm]{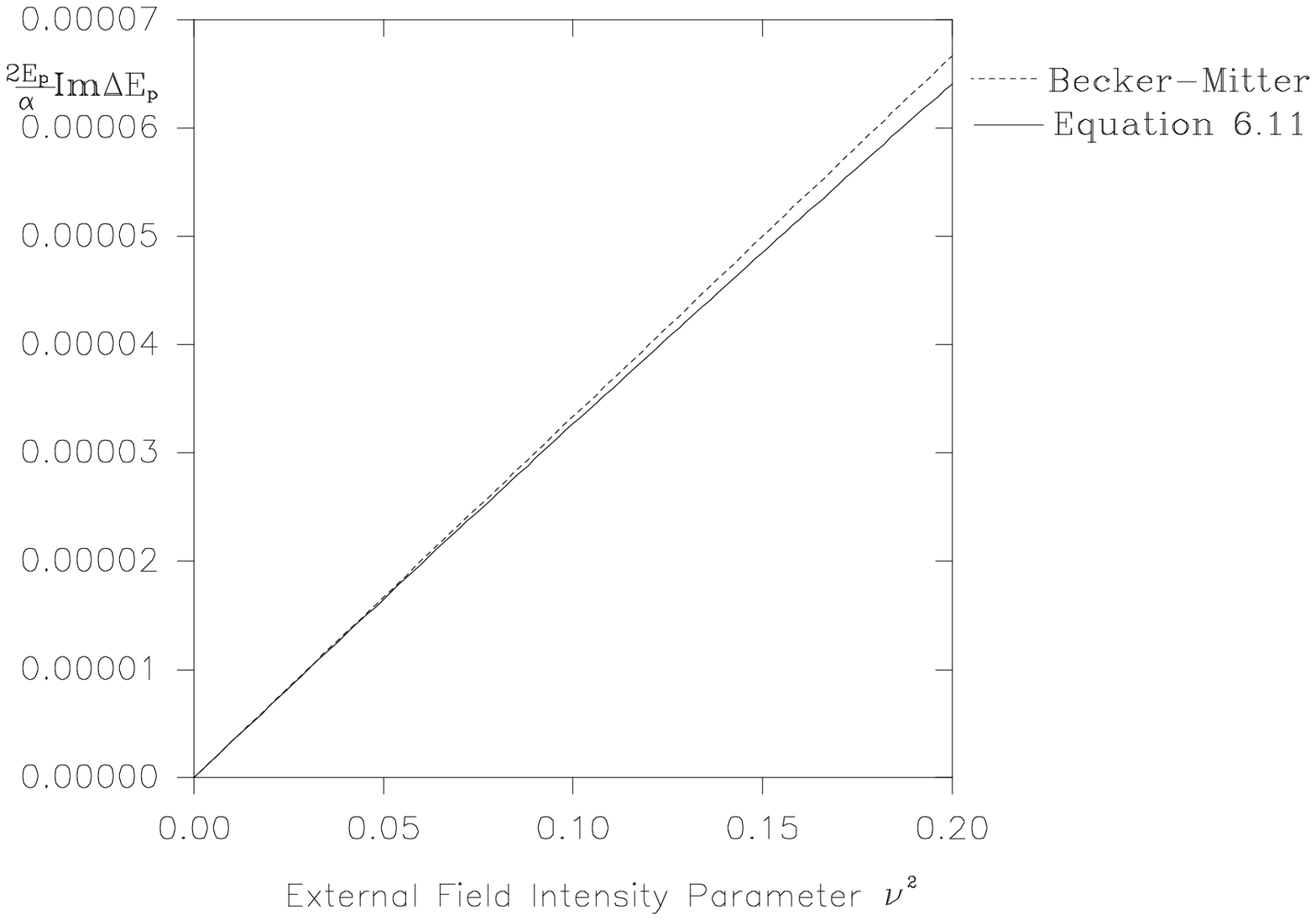}}
\caption{\bf Comparison of the EFEES imaginary part of Chapter 6 and that of \cite{BecMit76} vs 
$\nu^2$ for $\rho=0.001$.}
\label{c6g7}
\end{figure}

\begin{figure}[H]
 \centerline{\includegraphics[height=8cm,width=10cm]{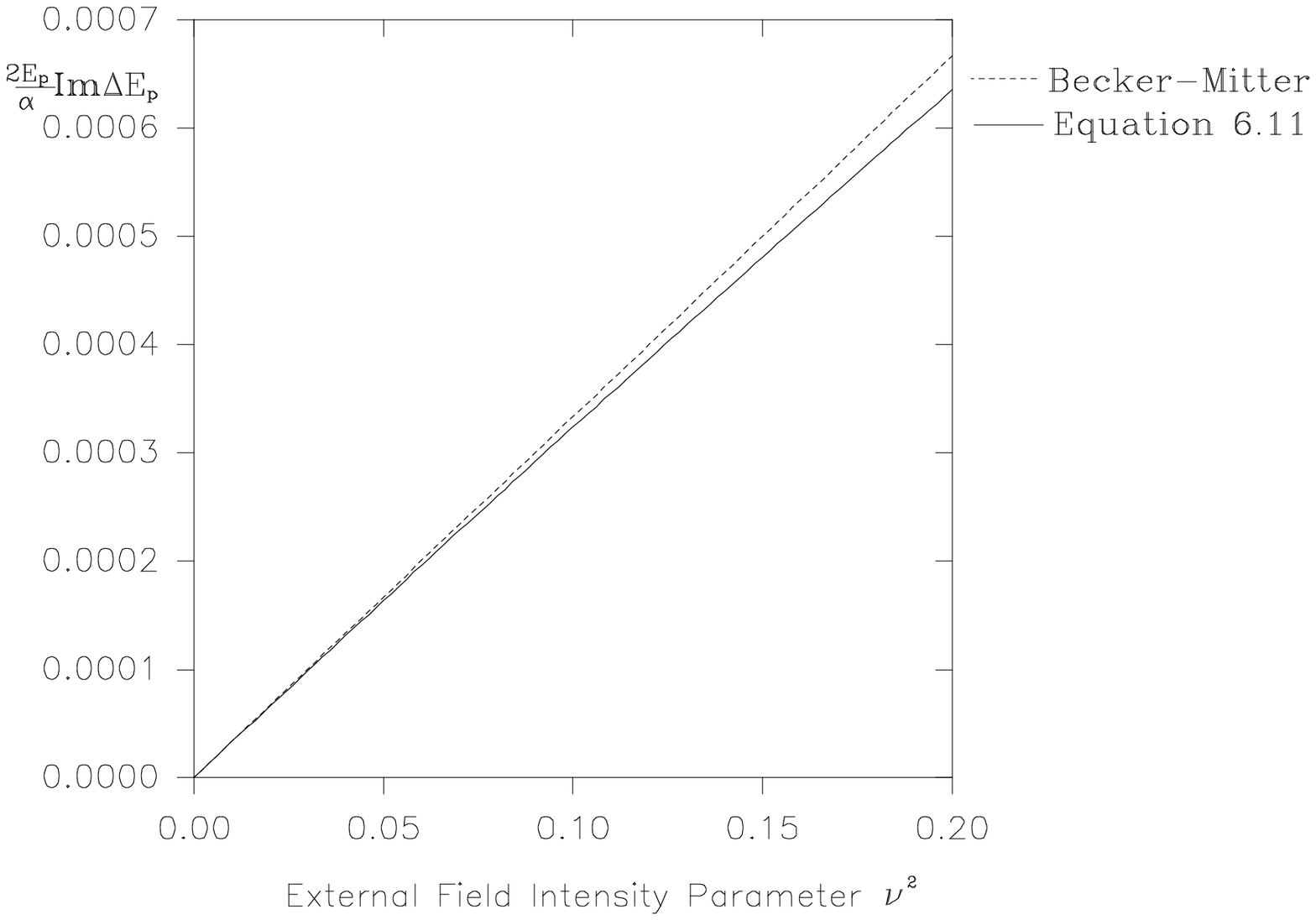}}
\caption{\bf Comparison of the EFEES imaginary part of Chapter 6 and that of \cite{BecMit76} vs
$\nu^2$ for $\rho=0.01$.}
\label{c6g8}
\end{figure}

\begin{figure}[H]
 \centerline{\includegraphics[height=8cm,width=10cm]{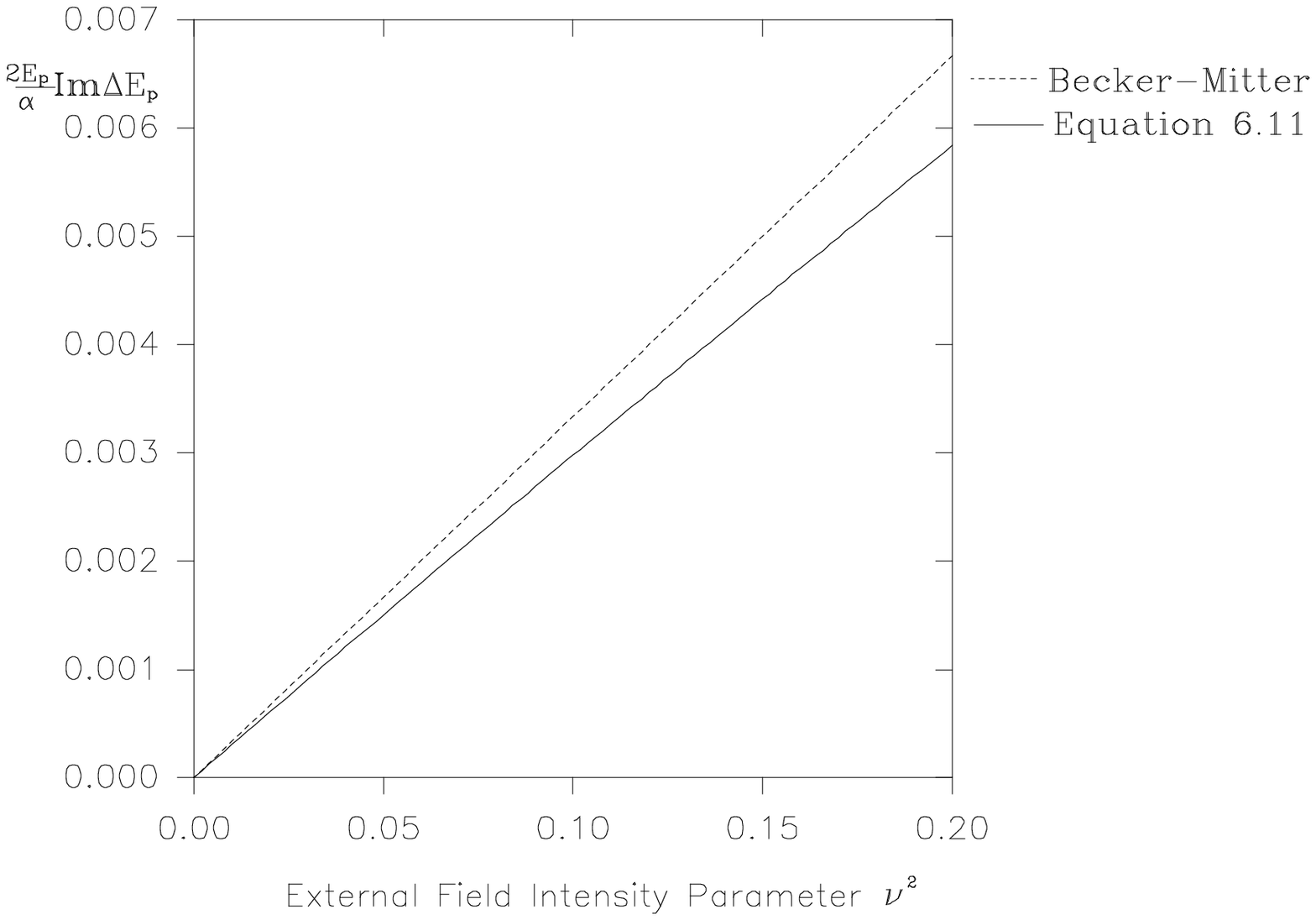}}
\caption{\bf Comparison of the EFEES imaginary part of Chapter 6 and that of \cite{BecMit76} vs
$\nu^2$ for $\rho=0.1$.}
\label{c6g9}
\end{figure}

\begin{figure}[H]
 \centerline{\includegraphics[height=8cm,width=10cm]{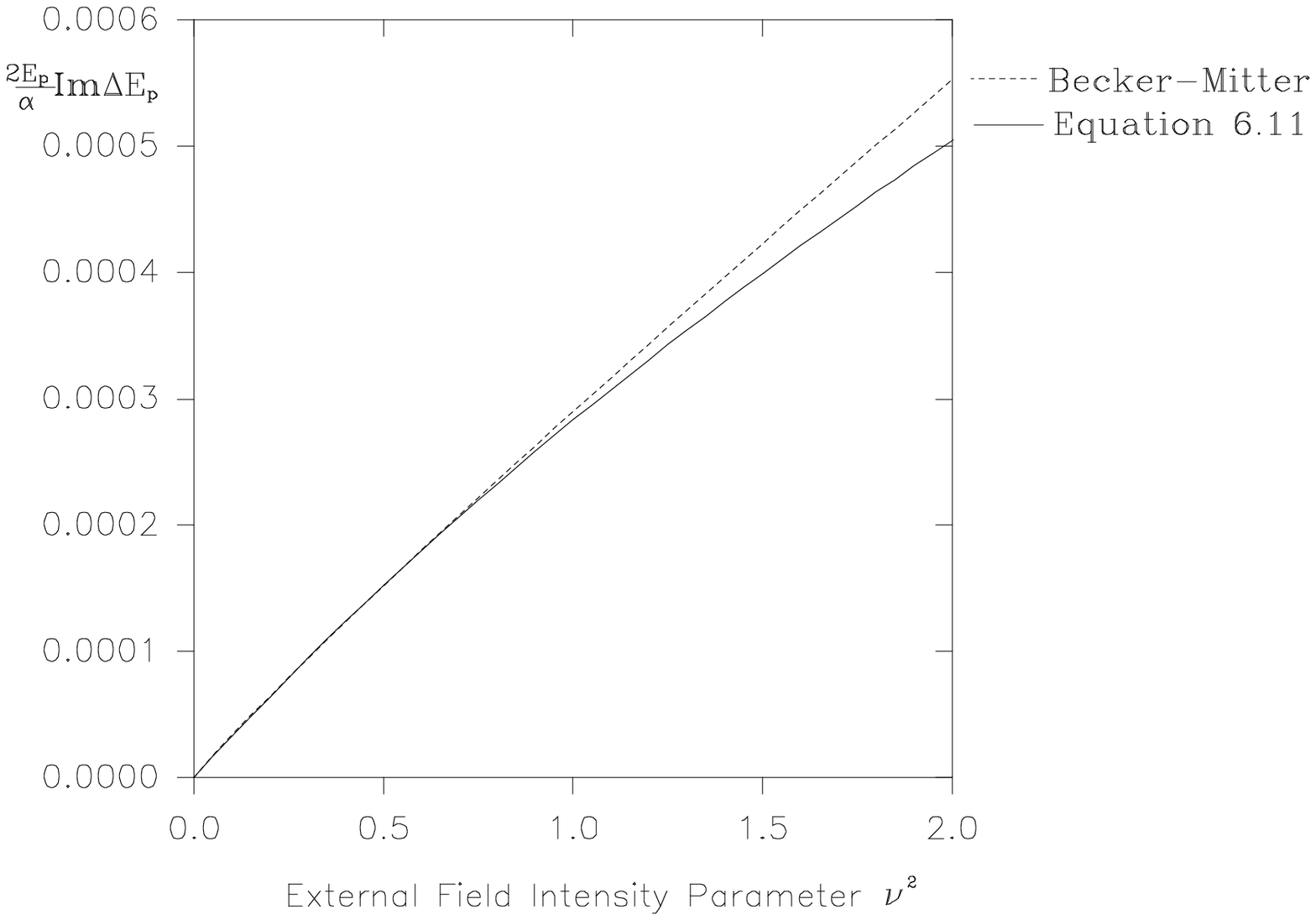}}
\caption{\bf Comparison of the EFEES imaginary part of Chapter 6 and that of \cite{BecMit76} vs
$\nu^2$ for $\rho=0.001$ and $\nu^2<2$.}
\label{c6g10}
\end{figure}

\begin{figure}[H]
 \centerline{\includegraphics[height=8cm,width=10cm]{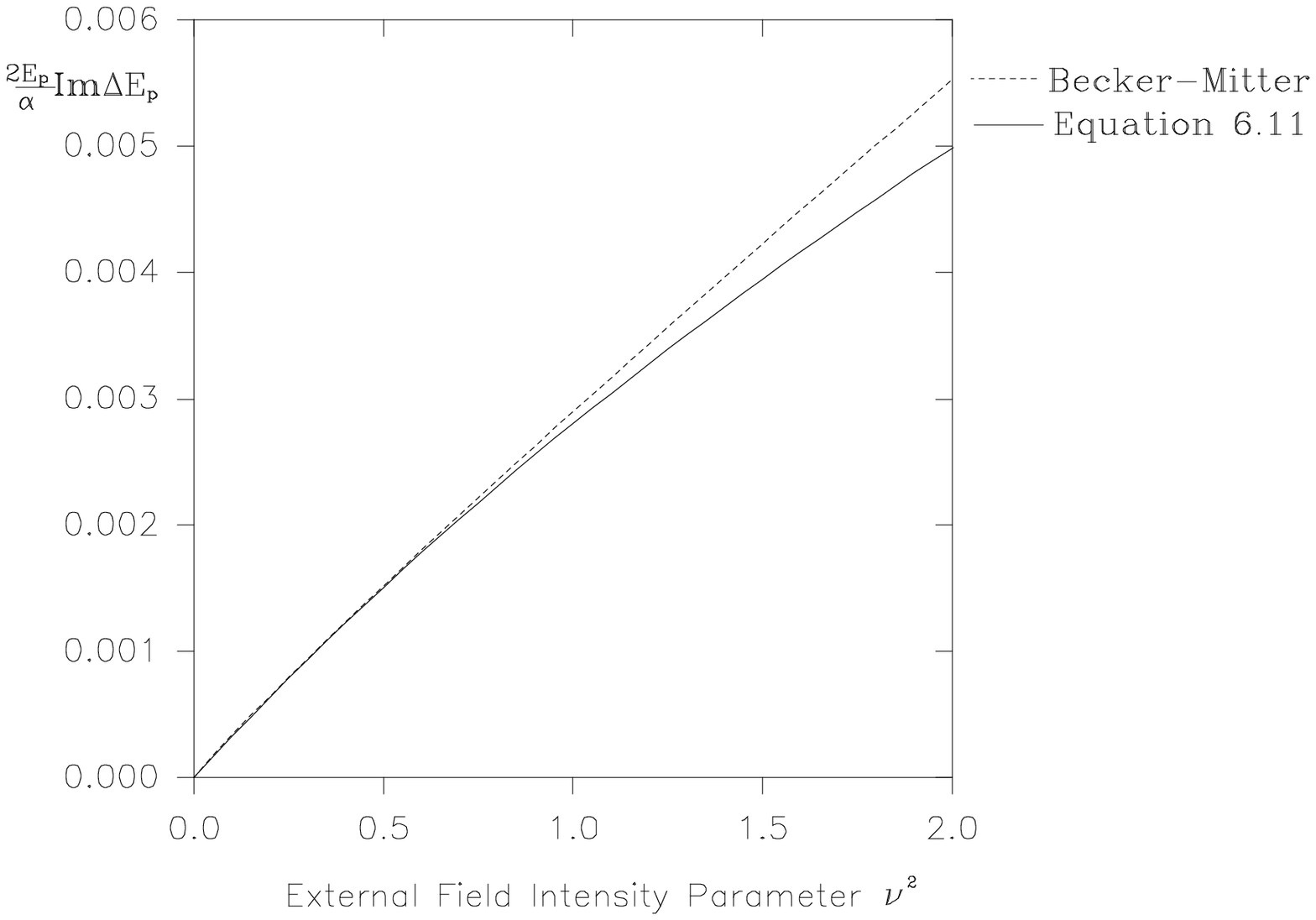}}
\caption{\bf Comparison of the EFEES imaginary part of Chapter 6 and that of \cite{BecMit76} vs
$\nu^2$ for $\rho=0.01$ and $\nu^2<2$.}
\label{c6g11}
\end{figure}

\begin{figure}[H]
 \centerline{\includegraphics[height=8cm,width=10cm]{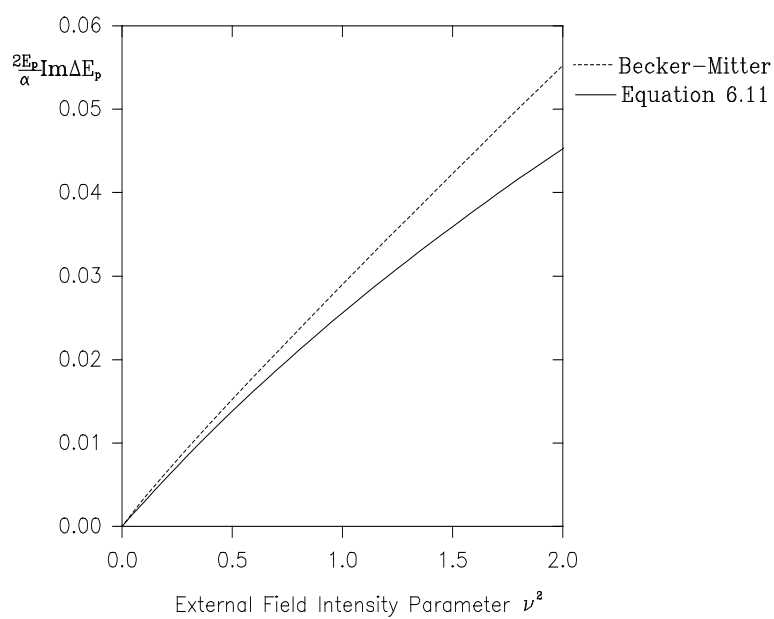}}
\caption{\bf Comparison of the EFEES imaginary part of Chapter 6 and that of \cite{BecMit76} vs
$\nu^2$ for $\rho=0.1$ and $\nu^2<2$.}
\label{c6g12}
\end{figure}

The regularised imaginary part of the EFEES is investigated in this section for the 
parameter range set out in Table \ref{c6tab1} expressed by equation \ref{c6.eq11}.
The integer variable $b$ corresponds to the number of external field photons
that contribute to the process. $\rho =\frac{2(kp)}{m^2}$ is a scalar
product function of the external field and electron 4-momenta. The external
field intensity is represented by the parameter $\nu ^2=\frac{e^2a^2}{m^2}$.
Both $\rho$ and $\nu^2$ are dimensionless in natural units. Figures \ref{c6g7} - \ref{c6g12} 
represent a comparison between approximation formulae obtained by \cite{BecMit76} and the expression 
contained in equation \ref{c6.eq11}.

The vertical axes of all figures in this sections show the regularised imaginary part of the EFEES 
multiplied by a factor $\frac{2\epsilon_p}{\alpha}$ where $\alpha$ is the fine structure constant 
and $\epsilon_p$ the electron energy.
Generally, the summation over $b$ converges rapidly and the typical computational time required 
to generate a single plot containing 50 data points was of the order of 10
seconds. However plots representing parameter values at the high end of $\rho$ and $\nu^2$ 
required approximately 10 minutes of computational time.

\medskip\ 

\input{./tex/tables/c6_table}

\clearpage

\begin{figure}[H]
 \centerline{\includegraphics[height=8cm,width=10cm]{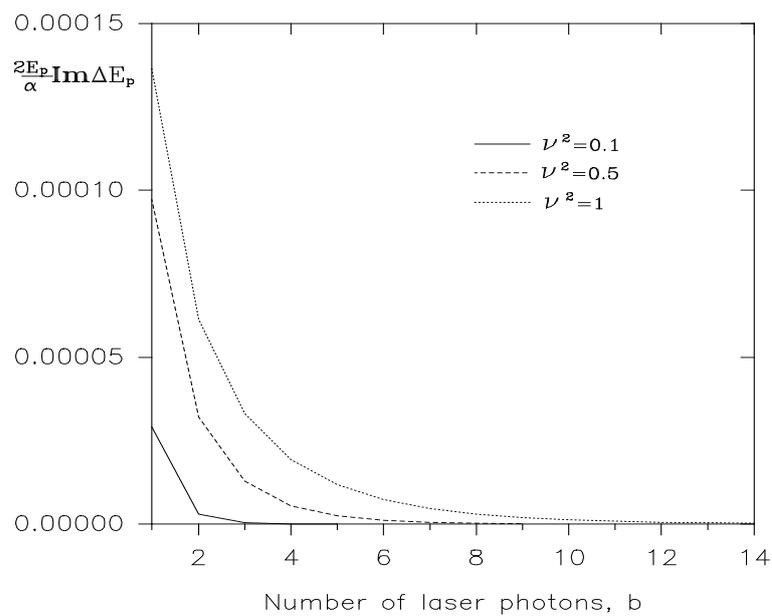}}
\caption{\bf The EFEES imaginary part vs $b$ external field photons for $\rho=0.001$ and various 
$\nu^2$.}
\label{c6g1}
\end{figure}

\begin{figure}[H]
 \centerline{\includegraphics[height=8cm,width=10cm]{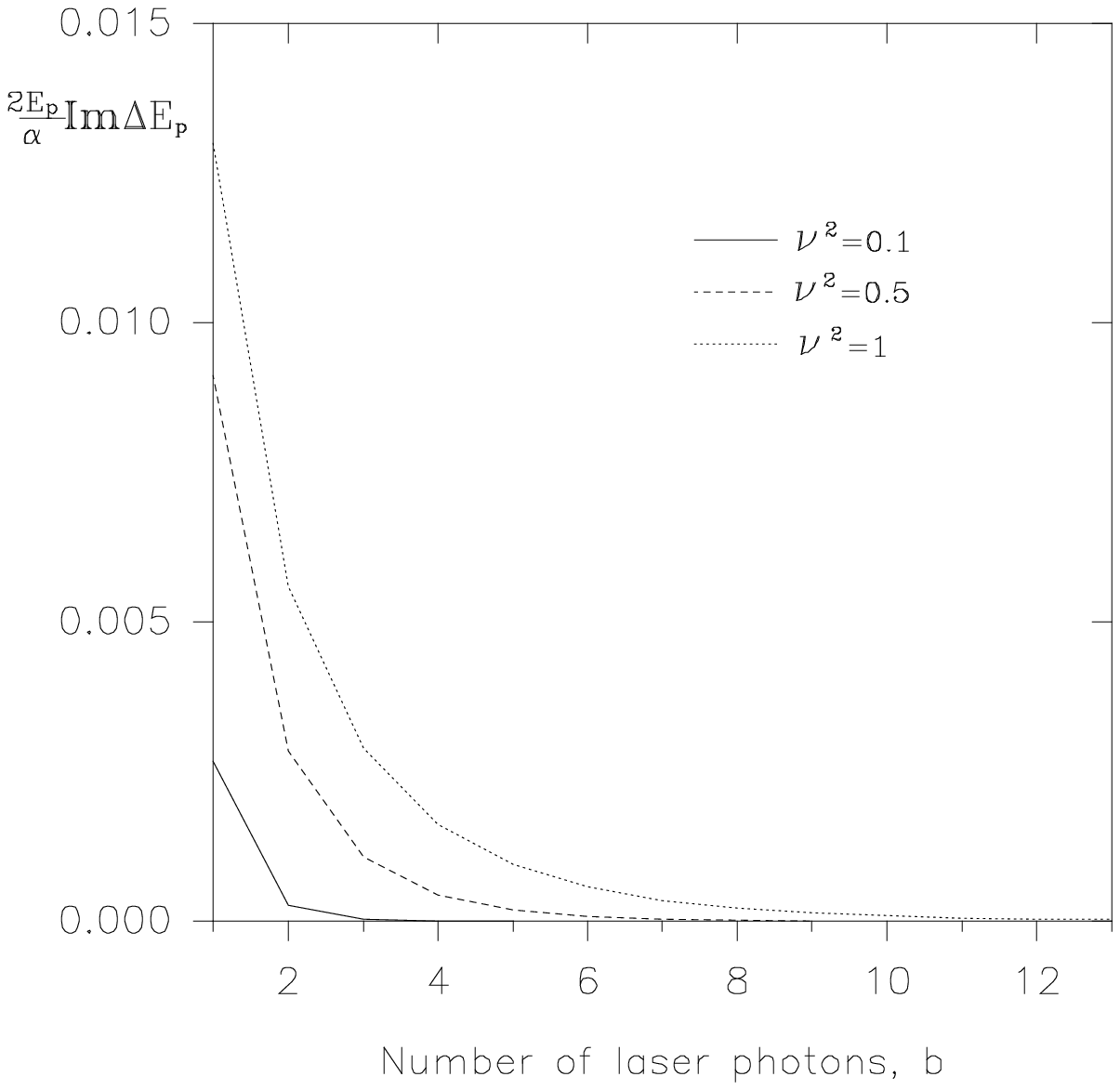}}
\caption{\bf The EFEES imaginary part vs $b$ external field photons for $\rho=0.1$ and various 
$\nu^2$.}
\label{c6g2}
\end{figure}

\clearpage

\begin{figure}[H]
 \centerline{\includegraphics[height=8cm,width=10cm]{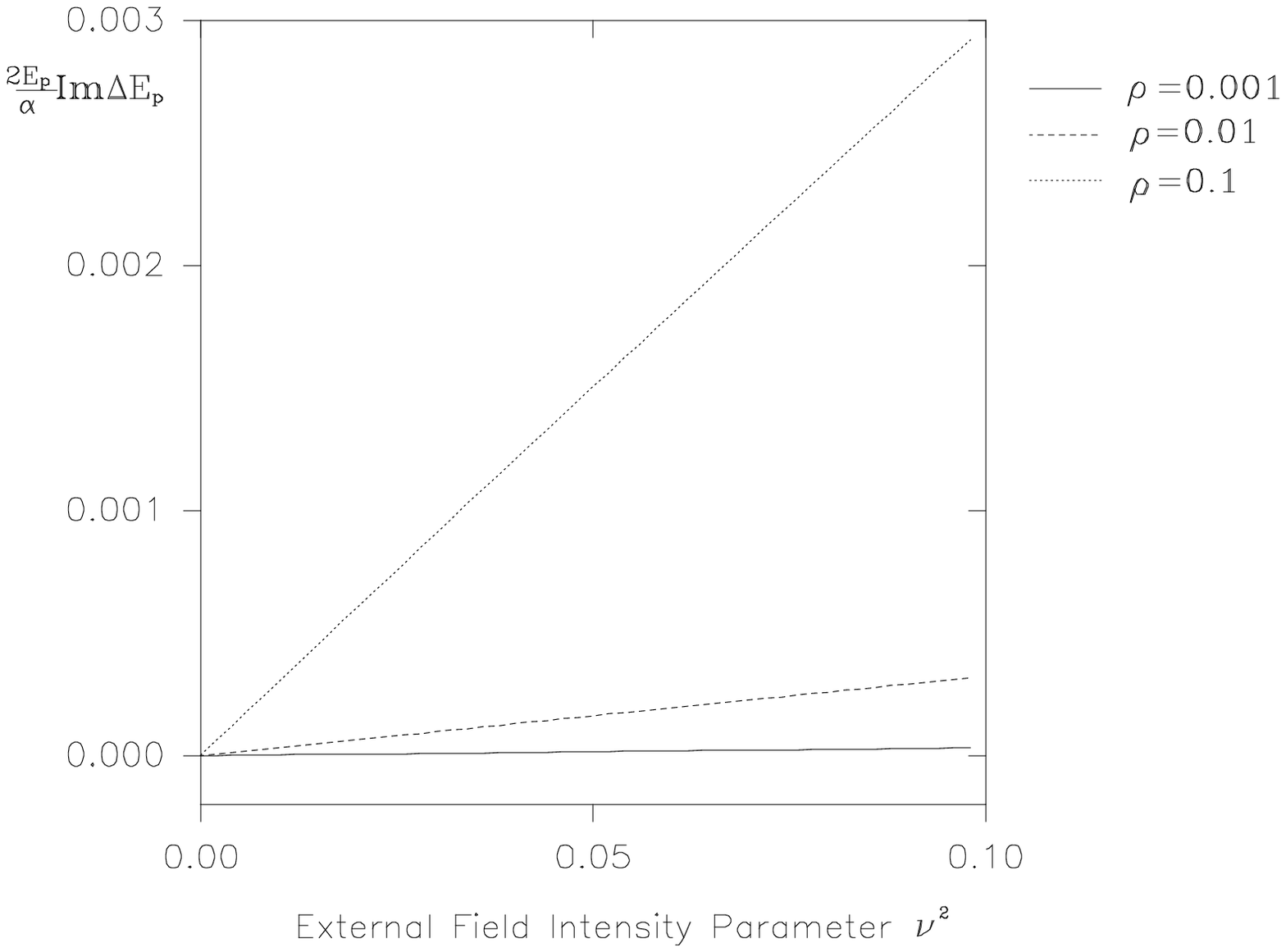}}
\caption{\bf The EFEES imaginary part vs $\nu^2<0.1$ for various $\rho$.}
\label{c6g3}
\end{figure}

\begin{figure}[H]
 \centerline{\includegraphics[height=8cm,width=10cm]{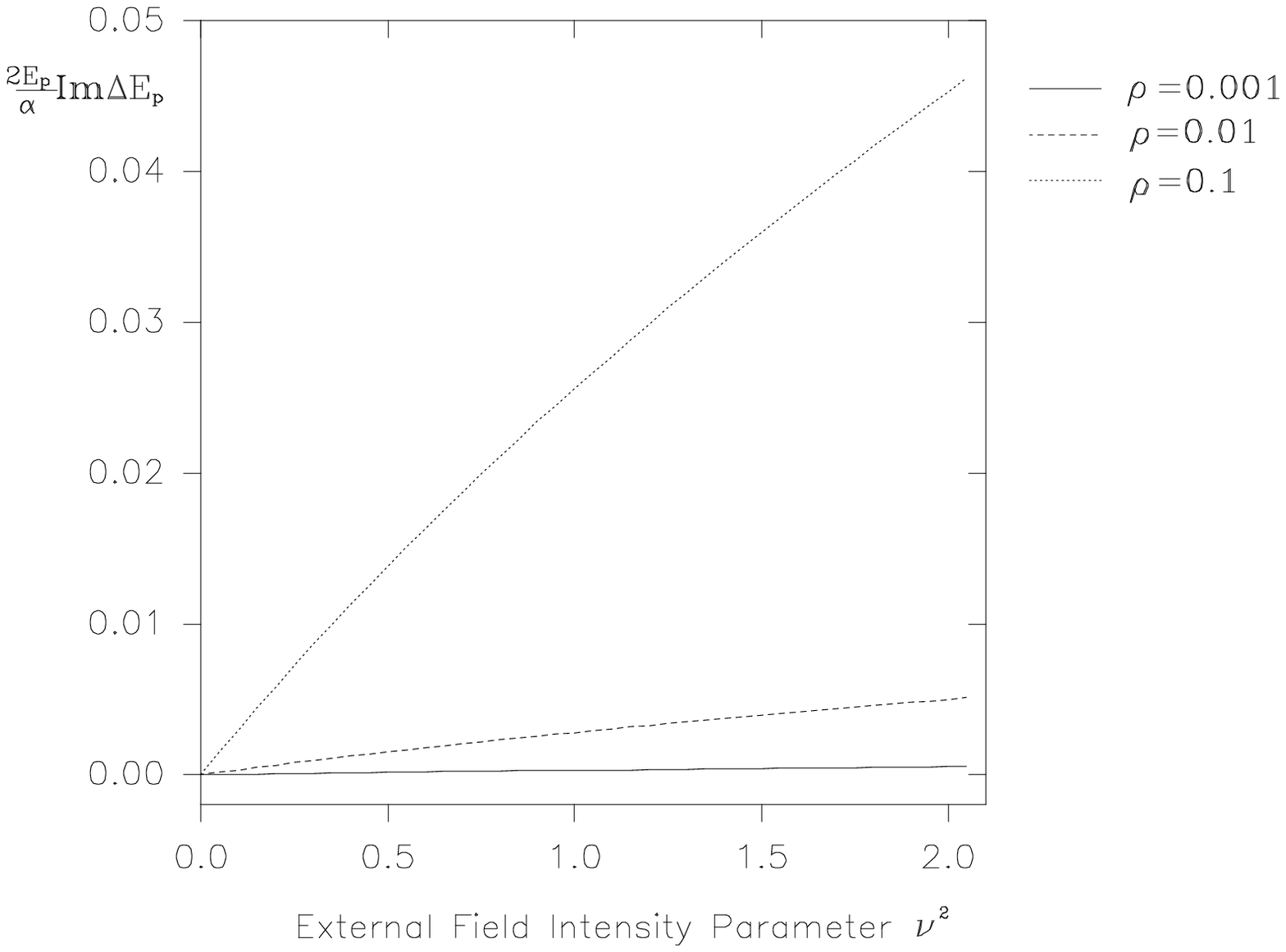}}
\caption{\bf The EFEES imaginary part vs $\nu^2<2$ for various $\rho$.}
\label{c6g4}
\end{figure}

\clearpage

\begin{figure}[H]
 \centerline{\includegraphics[height=8cm,width=10cm]{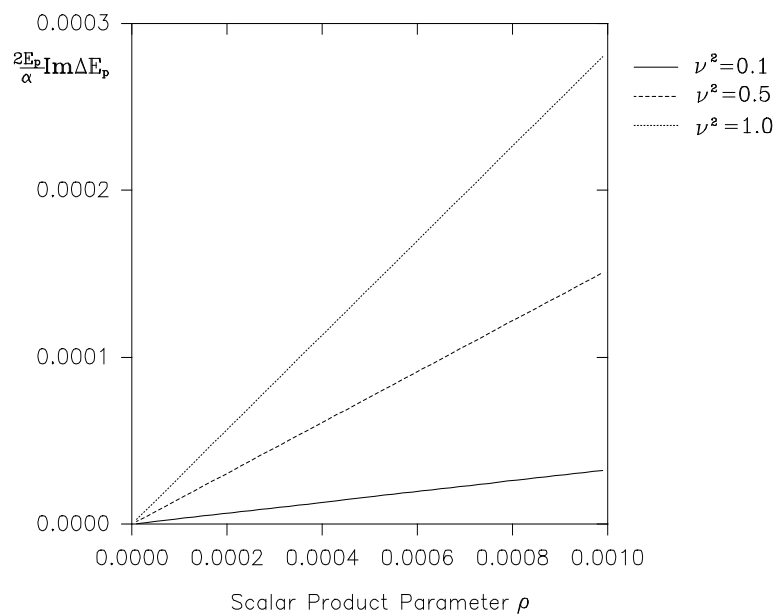}}
\caption{\bf The EFEES imaginary part vs $\rho<0.001$ for various $\nu^2$.}
\label{c6g5}
\end{figure}

\begin{figure}[H]
 \centerline{\includegraphics[height=8cm,width=10cm]{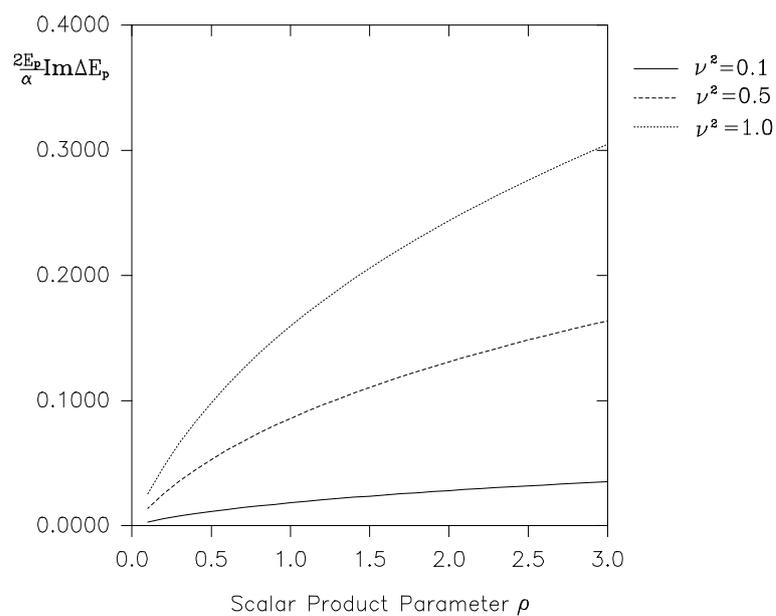}}
\caption{\bf The EFEES imaginary part vs $\rho<3$ for various $\nu^2$.}
\label{c6g6}
\end{figure}

\clearpage

\clearpage

\clearpage

\clearpage

\newpage\ 
\section{EFEES Analysis}
\label{c6anal}

The variation of the imaginary part of the EFEES displayed in the 
figures of section 6.3 is analysed here. The main 
purpose in calculating the imaginary part of the EFEES is to include it in the propagator 
denominators of the SCS and STPPP differential cross sections. Therefore only brief remarks 
are made about the physical origin of the variation in magnitude of the EFEES imaginary part
and the relevance of the numerical range of the parameters considered.

Figures \ref{c6g1} - \ref{c6g2} show the variation of the EFEES imaginary part with
the number of external field photons $b$ that contribute to the process. The
figures display several plots corresponding to the differing values of the
scalar product parameter $\rho $ and the external field intensity $\nu ^2$.
The horizontal scale of both figures has $b=1$ as its lowest value. This
point corresponds to the minimum value of contributing external field
photons required for the regularised EFEES to be non zero.

The main feature of both figures is a seemingly inverse relationship
between the EFEES imaginary part and the parameter $b$. Maximum values are
obtained at $b=1$, decreasing towards zero as $b$ increases. 
The parameter $\rho $ has little bearing on the EFEES imaginary part variation
with $\nu ^2$, though it increases its magnitude in figure \ref{c6g2} ($\rho =0.1$)
by a factor of approximately 100 compared with its magnitude in figure \ref{c6g1} 
($\rho =0.001$).

The discussion makes use of the optical theorem (equation \ref{c6.eq12}) which allows 
interpretation of the behaviour of the EFEES imaginary part to be given
in terms of the probability of emission of radiation from the electron
embedded in the external field. The discussion is facilitated by drawing a 
comparison with the probability of radiation by an electron in the field of a nucleus,
which bears an inverse relationship to the recoil momentum of the external
field (\cite{Nachtmann90} p.153, \cite{Heitler54} p.246). It is assumed here that a similar 
relationship exists when the external field is electromagnetic.

The probability of emission of an electron in an external field has an inverse relationship to 
the recoil momentum (\cite{ManSha84} p.167). The recoil momentum is the energy of the emitted 
photon $\omega_f$ which has a dependence on $\nu^2,b,\rho$ and the angle $\theta$ its 3-momentum makes with 
the direction of propagation of the external field. This relation can be obtained from equation 
\ref{c3.symev.eq4} by setting $\sql{k}_i,\omega_i$ to zero and substituting $b$ for $l$

\begin{equation}
\label{c6.eq13}
\omega_f=\frac{\rho \;b}{1+(b\omega +\frac{\nu^2}2)(1-\cos\,\theta )} 
\end{equation}

\medskip\ 

At least one external field photon contributes and the process can be considered a Compton 
scattering. The contributed photon is most likely scattered collinear to its initial state 
($\theta \sim 0^{\circ}$) and the probability it does so increases as $\nu^2$ (see Chapter 4). 
At $\theta\sim 0^{\circ}$ the recoil momentum is almost directly proportional to $b$ and 
therefore the probability of emission (which is equivalent to the vertical axes of figures 
\ref{c6g1} and \ref{c6g2}) should be inversely proportional to $b$.

Figures \ref{c6g3} and \ref{c6g4} show the variation of the EFEES imaginary part with
external field intensity for different values of the scalar product parameter $\rho$. The 
relationship between the EFEES imaginary part and $\nu^2$ appears to be linear. However its not 
quite so for the $\rho=0.1$ plot of figure \ref{c6g4}. The behaviour can be justified again by 
considering the recoil momentum (equation \ref{c6.eq13}). All else being constant, an increasing 
$\nu^2$ decreases $\omega_f$ and therefore increases the probability of emission. An 
understanding of the relationship with $\nu^2$ is also gained by considering an electron 
accelerated by the circularly polarised external field. The power radiated (and probability of 
emission) is directly proportional to the square of the electron acceleration \cite{Jackson75}.

The variation of the EFEES imaginary part with the scalar product parameter $\rho$ is 
considered in figures \ref{c6g5} and \ref{c6g6} which shows a seemingly linear relationship when 
$\rho$ is small (figure \ref{c6g5}) though non linear when $\rho$ is large (figure \ref{c6g6}). 
A linear relationship is suggested for the recoil momentum which is directly proportional to 
$\rho$. However a closer examination of the analytic expressions for the imaginary part of the 
EFEES reveal terms which are not linear. The two tendencies combine to produce the numerical 
variation shown.

Figures \ref{c6g7} - \ref{c6g12} contain a comparison between numerical values for the
EFEES imaginary part and the two approximation formulae given by \cite{BecMit75,BecMit76} in 
their treatment of the same process. These figures are divided in two groups.

Figures \ref{c6g7} - \ref{c6g9} consider the variation of the EFEES imaginary part for 
$0\leq \nu ^2\leq 0.2$ and contain a comparison with the second of \cite{BecMit76} approximation 
formulae which are quoted as valid for $\nu ^2\leq 0.1$ and $\rho \leq 0.1$. For this range of
parameters, in figures \ref{c6g7} and \ref{c6g8}, good agreement is obtained with 
\cite{BecMit76}. Variation is obtained as expected where \cite{BecMit76} is no longer valid at 
$\nu^2>0.1$ and $\rho=0.1$ (figure \ref{c6g9}).

Figures \ref{c6g10} - \ref{c6g12}, which contain a comparison with the first of \cite{BecMit76}'s 
approximation formulae valid for $0.1\leq \nu ^2\leq 1.0$ and $\rho\leq 0.1$, 
consider the variation of the EFEES imaginary part for $0\leq \nu ^2\leq 2.0$. 
Again, good agreement is obtained whenever the parameters considered fall within the range of 
validity for \cite{BecMit76}'s expressions. The variation obtained with \cite{BecMit76} 
shows the necessity of using the expressions for the EFEES imaginary part (equation \ref{c6.eq11}) in further
calculations of SCS and STPPP resonances. At the $\nu ^2=1.0$ point in
figure \ref{c6g12}, the variation obtained represents a $13.4\%$ error in the
heights of resonant peaks which have an inverse square relationship to the EFEES imaginary part. 

A potential problem emerges as $\nu^2\rightarrow 0$. Here the imaginary part of the EFEES falls 
to zero resulting in potentially resonant cross sections again becoming infinite. However 
resonances do not occur in the SCS and STPPP differential cross sections when $\nu^2\sim 0$. A 
detailed examination follows in Chapter 7.

%% file: tex/tables/c6_table.tex
%\LaTeXparent{c:/thesis/tex/thesis.tex}

\begin{table}[h]
\center{
\begin{tabular}{|c|c|c|c|} \hline
$\pmb b$&$\pmb \nu^{\pmb 2}$ & $\pmb \rho $ & \bf figure(s) \\ \hline\hline
$1\rightarrow 20 $&$ 0.1,0.5,1.0 $&$ 0.001 $& \ref{c6g1} \\ \hline
$1\rightarrow 20 $&$ 0.1,0.5,1.0 $&$ 0.1 $& \ref{c6g2} \\ \hline
$ all $&$ 0\rightarrow 0.1 $&$ 0.001,0.01,0.1 $& \ref{c6g3} \\ \hline
$ all $&$ 0\rightarrow 0.2 $&$ 0.001,0.01,0.1 $& \ref{c6g7},\ref{c6g8},
\ref{c6g9} \\ \hline
$ all $&$ 0\rightarrow 2.0 $&$ 0.001,0.01,0.1 $& \ref{c6g4},\ref{c6g10},
\ref{c6g11},\ref{c6g12} \\ \hline
$ all $&$ 0.1,0.5,1.0 $&$ 0\rightarrow 0.1 $& \ref{c6g5} \\ \hline
$ all $&$ 0.1,0.5,1.0 $&$ 0.1\rightarrow 3.0 $& \ref{c6g6} \\ \hline
\end{tabular} }
\caption{\bf\bm The parameter range for which the regularised imaginary part of the EFEES
is investigated.}
\label{c6tab1}
\end{table}

%% file: chap7.tex
\section{Introduction}
\enlargethispage*{1cm}

As will be seen in this chapter, the virtual particle exchanged in both the SCS
and STPPP processes reaches the mass shell. Uncorrected, the SCS and STPPP cross
sections would contain resonant infinities at those points. The results of 
chapter 6 for the electron self
energy will be used to render these resonant infinities finite.
The one loop electron self energy is only one of many radiative corrections, including the vertex
correction, the correction of particle wave functions and the propagator
corrections, that chould be calculated and included in the SCS and STPPP cross
sections. However there are constraints imposed by the
difficulties of performing these corrections in the presence of the external
electromagnetic field. Also, computational restrictions require radiative corrections of 
immediate concern to be prioritised. So the focus in this chapter is on the insertion of the 
imaginary part of the EFEES into the bound fermion propagator present in both the SCS and STPPP
processes.

In section \ref{c7renorm} the regularisation and renormalisation procedures
required to deal with the divergences present in the electron self energy are revisited.
The regularisation and renormalisation procedure in the presence of the external field reduces
to the equivalent one in the absence of the external field \cite{BecMit76}. The insertion of the 
EFEES into the denominator of the bound fermion propagator yields expressions identical to those 
of \cite{BecMit76}'s fully corrected propagator to order one in the fine structure constant.

The possible experimental detection of QED processes in external electromagnetic fields is of 
great interest. In the last decade experimental apparatus has been
developed which allows observation of some first order IFQED processes. At resonance however, the 
detection of the second order IFQED processes may become more likely. With possible experimental 
detection in mind, sections \ref{c7reson} - \ref{c7exp} are devoted to the calculation of the 
SCS and STPPP differential cross sections at points of resonance.
In section \ref{c7reson} analytic expressions for the widths, locations, spacings and heights 
of the resonant differential cross sections are obtained. In section \ref{c7fig} numerical 
results are presented for SCS and STPPP resonances which are analysed in section \ref{c7anal}. 
Finally in section \ref{c7exp} the likely experimental constraints in the detection
of these resonances are discussed.
\section{Renormalisation in the external field}
\label{c7renorm}

\begin{figure}[!b]
\centerline{\includegraphics[height=4.5cm,width=8cm]{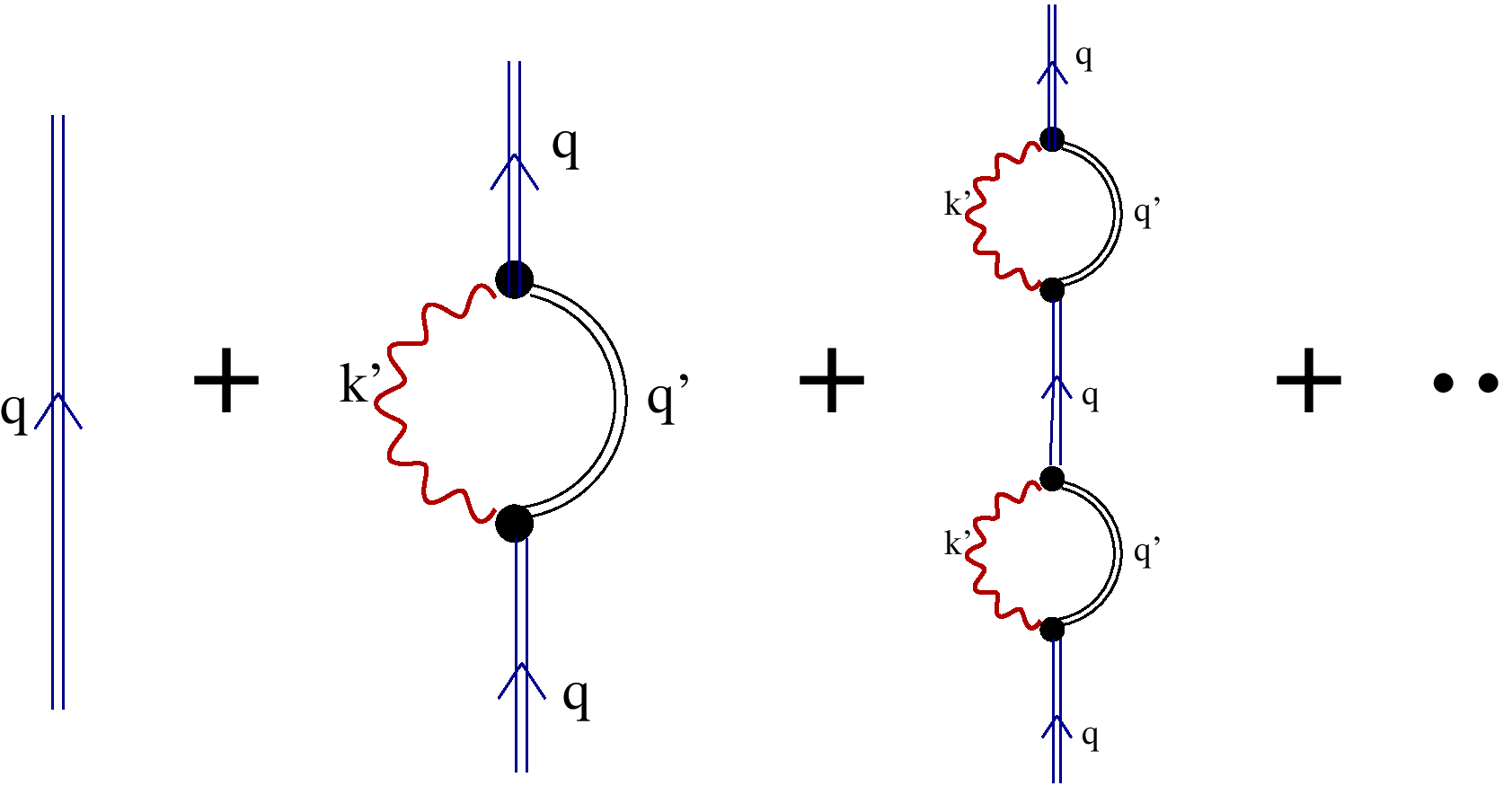}}
\caption{\bf Feynman diagrams for the external field electron
self energy series.}
\label{c7.fig1}
\end{figure}

The exact, radiatively corrected fermion propagator embedded in the external field can be
obtained from an infinite series in direct analogy to the procedure used to
obtain the exact fermion propagator in the absence of the external field.
The Dyson equation can be used to write the exact external field propagator $G_{SE}^e(x_2,x_1)$ 
as a function of the external field electron self energy $\Sigma ^e(x_2,x_1)$ and the external 
field fermion propagator in the absence of radiative corrections $G^e(x_2,x_1)$. The expression 
is contained in equation \ref{c7.eq1} and can be written down directly from the series of
Feynman diagrams represented in figure \ref{c7.fig1}.

\begin{equation}
\label{c7.eq1} 
\begin{array}{rl}
G_{SE}^e(x_2,x_1) & = G^e(x_2,x_1) \\  
& + \dint d^4x_a \, d^4x_b G^e(x_2,x_a)\Sigma ^e(x_a,x_b)G^e(x_b,x_1) \\  
& + \dint d^4x_a \, d^4x_b \, d^4x_c \, d^4x_d \, G^e(x_2,x_a)\Sigma 
^e(x_a,x_b)G^e(x_b,x_c)\Sigma^e(x_c,x_d)G^e(x_d,x_1) \\  
& + \ldots 
\end{array}
\end{equation}

\medskip\ 

The integrations over space-time points can be absorbed by using the expression for the momentum 
space EFESE, $\Sigma^e(p)$ and the series becomes

\begin{equation}
\label{c7.eq2} 
\begin{array}{rcl}
G_{SE}^e(x_2,x_1) & = & \dint d^4p\;E_p(x_2)\dfrac 1{\slashed{p}-m}\bar E_p(x_1) \\[10pt]
& + & \dint d^4p\;E_p(x_2)\dfrac 1{\slashed{p}-m}\Sigma^e(p) \dfrac 1{\slashed{p}-m} \bar E_p(x_1) \\  
& + & \ldots 
\end{array}
\end{equation}

\medskip\ 

The sum in equation \ref{c7.eq2} can be reduced and the momentum
space EFESE, $\Sigma ^e(p)$ appears in the denominator of the propagator

\begin{equation}
\label{c7.eq3}
G_{SE}^e(x_2,x_1)=\int d^4p\;E_p(x_2)\frac 1{\slashed{p}%
-m-\Sigma ^e(p)}\bar E_p(x_1) 
\end{equation}

\medskip\ 

The EFESE $\Sigma ^e(p)$ contains divergences and a
renormalisation procedure is required. The exact bound bound fermion propagator 
$G_{SE}^e(x_2,x_1)$ is the exact fermion propagator in the absence of the
external field sandwiched between Volkov $E_{p}(x)$ functions.
A renormalisation procedure equivalent to that of the field free case is suggested. Renormalisation of 
the fermion mass and charge is obtained by an expansion of the field free electron self energy 
as a function of the slash vector $\slashed{p}$. However with the external field
present, the EFESE has vector dependencies on
external field parameters $\slashed{a}_1,\slashed{a}_2,\slashed{k}$ as well as a
dependence on the electron momentum vector $\slashed{p}$.

\cite{BecMit76} used the axial symmetry of a circularly polarised external field and gauge
invariance to find that the divergences present in the EFESE have no
dependence on external field parameters. This result was made use of in Chapter 6 and is so used again 
here. The EFESE $\Sigma ^e(p)$ is separated into a non external field part $\Sigma ^0(p)$ 
and an external field dependent part $\Sigma ^L(p,a_1,a_2,k)$. 

\begin{equation}
\label{c7.eq4}\Sigma ^e(p)=\Sigma ^0(p)+\Sigma ^L(p,a_1,a_2,k) 
\end{equation}

\medskip 

The divergences present in $\Sigma ^0(p)$ are dealt with by the standard
regularisation and renormalisation procedure in the absence of the external
field (section \ref{c2regn}). The regularised EFESE is

\begin{equation}
\label{c7.eq4a}\Sigma _R^e(p)=\Sigma ^0(p)-\Sigma ^0(p,p^2=m^2)+\Sigma
^L(p,a_1,a_2,k)
\end{equation}

\medskip\ \ 

In Chapter 6, the external field electron energy shift (EFEES) was calculated. The EFEES can be 
inserted into the electron energy in the propagator denominator. However before doing so, it is 
necessary to show that the procedure is equivalent to the insertion of the EFESE as in equation 
\ref{c7.eq3}. In previous work the general structure of the EFESE was obtained and expansions to order 
$\alpha$ used \cite{BecMit76}.

The general structure of the EFESE can be constructed from products of the
slash vectors $\slashed{p},\slashed{a}_1,\slashed{a}_2,\slashed{k}$. Taking linear
combinations of these with the aim of achieving orthogonality, the EFESE can
be reduced to a dependence on products of three slash vectors

\begin{equation}
\label{c7.eq5} 
\begin{array}{c}
\slashed{p} \quad;\quad \slashed{F}_1=\slashed{a}_1-\dfrac{(a_1p)}{(kp)}\slashed{k} 
\quad;\quad \slashed{F}_2=\slashed{a}_2-\dfrac{(a_2p)}{(kp)}\slashed{k} 
\end{array}
\end{equation}

\medskip\ 

Each of the slash vectors defined in equation \ref{c7.eq5} are orthogonal in
the sense that a scalar product of any two of these is zero. In general the EFESE can
be written

\begin{equation}
\label{c7.eq6} 
\begin{array}{c}
\Sigma ^e(p)=A_1 
\slashed{p}+A_2\slashed{F}_1+A_3\slashed{F}_2+A_4\slashed{p}\slashed{F}_1
+A_5\slashed{p}\slashed{F}_2 \\ +A_6\slashed{F}_1\slashed{F}_2+A_7\slashed{p}\slashed{F}_1\slashed{F}%
_2+A_8 
\end{array}
\end{equation}

\medskip\ 

Furry's theorem (\cite{JauRoh76}, pg.160) requires that $\Sigma^e(p) $ be an even function of the 
4-vectors $F_1$ and $F_2$. Because of orthogonality then

\begin{equation}
\label{c7.eq7}A_2,A_3,A_4,A_5=0 
\end{equation}

\medskip\ 

Since $F_1,F_2$ can only appear in the remaining coefficients in a
scalar product with itself (i.e. as $F_1^2=F_2^2=a^2$) then the coefficients 
$A_1,A_6,A_7,A_8$ must be independent of vectors $a_1$ and $a_2$. Equation 
\ref{c7.eq6} can be simplified further by performing an operation in which
the 4-vectors $a_1$ and $a_2$ are exchanged. Examination of the complete expression of the EFESE 
(see equation \ref{c6eq4}) reveals that the effect of swapping $a_1$ and $a_2$ is to obtain the 
complex conjugate of $\Sigma ^e(p)$. The remaining $A$ coefficients of equation \ref{c7.eq6} can then 
be written

\begin{equation}
\label{c7.eq8} 
\begin{array}{ccc}
A_6 & = & -A_6^{*} \\ 
A_7 & = & -A_7^{*} \\ 
A_1 & = & A_1^{*} \\ 
A_8 & = & A_8^{*} 
\end{array}
\end{equation}

\medskip\ 

With the restrictions on the coefficients $A_i$ contained in equations \ref
{c7.eq7} and \ref{c7.eq8}, the general expression for the EFESE contained in
equation \ref{c7.eq6} can be rewritten in terms of coefficients $Y_i$ which
are functions of real quantities and are each proportional to the fine
structure constant $\alpha $.

\begin{equation}
\label{c7.eq9}\Sigma ^e(p)=Y_1\slashed{p}+iY_2\slashed{F}_1\slashed{F}_2+iY_3%
\slashed{p}\slashed{F}_1\slashed{F}_2+Y_4 
\end{equation}

\medskip\ 

Insertion of the EFESE into the exact electron propagator (equation \ref
{c7.eq3}) is achieved by separating $\Sigma ^e(p)$ into a vector part and a
scalar part, and operating twice on the numerator and denominator by the
conjugate of the denominator. The result for the propagator denominator, to
first order in the fine structure constant is

\begin{equation}
\label{c7.eq10}(p^2-m^2)^2+2(p^2-m^2)(2p^2Y_1+2mY_4) 
\end{equation}

\medskip\ 

Alternatively, the EFEES can be inserted into the electron energy contained
in the denominator of the external field propagator, $p^2-m^2=\epsilon
_p^2-\left(|\squ{p}| ^2+m^2\right) $. After squaring and
expanding to first order in the fine structure constant, the
denominator becomes

\begin{equation}
\label{c7.eq11}(p^2-m^2)^2+2(p^2-m^2)2\epsilon _p\Delta \epsilon _p 
\end{equation}

\medskip\ 

The equivalence of equations \ref{c7.eq10} and \ref{c7.eq11} is obtained by
use of the expression relating the EFESE and the EFEES (see equations 6.1
and 6.6)

\begin{equation}
\label{c7.eq12} 
\begin{array}{lll}
2\epsilon _p\Delta \epsilon _p & = & m\dsum\limits_{s=1}^2\,\bar
u_{p,s}\,\Sigma ^e(p)\,u_{p,s} \\  
& = & \frac 12 
\limfunc{Tr}\left\{ (\slashed{p}+m)\Sigma ^e(p)\right\} \\  & = & 
2p^2Y_1+2mY_4 
\end{array}
\end{equation}

\medskip\ 

The regularised EFEES $\Delta \epsilon _p^R$ (which is equivalent to the
regularised EFESE), is inserted into the denominator of the external field
propagator to produce, finally, an expression for the exact external field
electron propagator (to first order in the fine structure constant)

\begin{equation}
\label{c7.eq13}G_{SE}^e(x_2,x_1)=\int d^4p\;E_p(x_2)\frac{\slashed{p}+m}{%
p^2-m^2+2\epsilon _p\Delta \epsilon _p^R}\bar E_p(x_1) 
\end{equation}

\bigskip\
\section{Resonance Conditions}
\label{c7reson}

The replacement of the bare fermion propagator in the SCS and STPPP
differential cross sections with the corrected fermion propagator obtained in Section \ref{c7renorm}
(equation \ref{c7.eq13}) has the effect of rendering the resonant
infinities present in the uncorrected cross sections finite.
The corrected propagator denominator, from equation \ref{c7.eq13}, is

\begin{equation}
\label{c7.eq14}p^2-m^2+2\epsilon _p\Re \Delta \epsilon _p^R+i2\epsilon _p\Im
\Delta \epsilon _p^R 
\end{equation}

\medskip\ 

For the purposes of experimental detection of the SCS and STPPP differential
cross sections, the characteristics of the resonant peaks, particularly the
location, separation, height and width of the peaks are of interest. For
numerical evaluation of these characteristics the precise value of the
imaginary part of the regularised EFEES is of central importance. The real
part provides only a small correction to the peak location and for the sake
of simplicity can be neglected.

Analytic expressions developed for resonance locations and resonance widths
proved useful computationally. Resonance heights
were determined numerically by running the computer programs that generated the results of Chapters 4
and 5 at the resonant parameter region indicated by analytic expressions. 
The resonance peaks locations, obtained from the mass shell condition $p^2=m^2$, 
are determined by the solution of two equations corresponding to
the direct channel and the exchange channel of the scattering.
For the SCS process, using the same notation as Chapter 3, these are

\begin{equation}
\label{c7.eq15} 
\begin{array}{rrl}
2(q_ik_i)+2r\left[ (kp_i)+(kk_i)\right] & = & 0\quad \quad 
\text{direct channel} \\ -2(q_ik_f)+2r\left[ (kp_i)-(kk_f)\right] & = & 
0\quad \quad \text{exchange channel} 
\end{array}
\end{equation}

\medskip\ 

For a particular set of particle energy-momenta, external field intensity
and summation variables $r$ and $l$, equation \ref{c7.eq15} yields
expressions for the resonant scattering angles $\theta _i^{res}$ and $\theta
_f^{res}$

\begin{equation}
\label{c7.eq16}
\begin{array}{rl}
\cos \,\theta _i^{res} & = 1+ \dfrac{1+r\frac \omega {\omega _i}}
{\frac 12\nu^2+r\frac{\omega}{m} } \\[12pt] 
\cos\,\theta _f^{res} & = \dfrac 1{2(b^2+c^2)}\left( 2ab\pm \sqrt{4c^2(b^2+c^2)-4a^2c^2}\right) 
\\[20pt] 
\text{where\quad \quad }a & = r\frac{\omega}{m} + \frac{\nu ^2}2+1-\dfrac{rm\omega }{(q_ik_i)+l(k\bar p)}
(1+\frac{\nu ^2}2+l\frac{\omega}{m} +\frac{\omega_i}{m}) \\[10pt]
 b & = r\frac{\omega}{m} + \frac{\nu ^2}2-\dfrac{rm\omega }{(q_ik_i)+l(k\bar p)}
( \frac{\nu ^2}2+l\frac{\omega}{m} + \frac{\omega_i}{m}) \\[10pt]
 c & = -\,\dfrac{r\frac{\omega}{m} \frac{\omega_i}{m}\sin
\theta _i\cos \phi _f}{[1+\frac{\nu ^2}2(1-\cos \theta _i)]} 
\end{array}
\end{equation}

\medskip\ 

For the STPPP process, the two equations to be solved for resonant scattering angles are

\begin{equation}
\label{c7.eq17} 
\begin{array}{rrl}
(r-l)\dfrac \omega {\omega _1}\,\dfrac{\epsilon _{q_{-}}-\left| q_{-}\right|
\cos \theta _f-\omega _1(1-\cos \theta _1)}{\epsilon _{q_{-}}-\left|
q_{-}\right| (\cos \theta _1\cos \theta _f-\sin \theta _1\sin \theta _f\cos
\phi _f)}=1 &  & \quad \quad 
\text{direct channel} \\[12pt]
 (r-l)\dfrac \omega {\omega _1}\,\dfrac{\epsilon
_{q_{-}}-\left| q_{-}\right| \cos \theta _f-\omega _1(1+\cos \theta _1)}{%
\epsilon _{q_{-}}+\left| q_{-}\right| (\cos \theta _1\cos \theta _f-\sin
\theta _1\sin \theta _f\cos \phi _f)}=1 &  & \quad \quad \text{exchange
channel} 
\end{array}
\end{equation}

\medskip\ 

After substituting the expressions for the electron energy-momenta (see
equation \ref{c3.pp.eq7}), equation \ref{c7.eq17} yields a polynomial to 4th order in 
$\cos \,\theta _i^{res}$ and $\cos \,\theta _f^{res}$. The solution is most
easily achieved numerically.

The angular spacing between resonances ($\delta \,\theta _i$ and $\delta
\,\theta _f$) is obtained by solving equations \ref{c7.eq16} and \ref
{c7.eq17} for different values of the summation variables $r$ and $l$. For
instance, assuming that the SCS direct channel resonance condition is
satisfied for $r=n$, the location of the closest resonant peak is determined
by the solution of equation \ref{c7.eq16} for $r=n+1$ (or $r=n-1$). The
resonance spacing is obtained by a solution of the equation set

\begin{equation}
\label{c7.eq18} 
\begin{array}{rl}
\cos (\theta _i^{res}+\delta \,\theta _i) & = 
1+\dfrac{1+(n+1)\frac \omega {\omega _i}}{\tfrac 12\nu ^2+(n+1)\frac{\omega}{m} } \\[12pt] 
\cos \,\theta _i^{res} & = 1+\dfrac{1+n\frac \omega {\omega _i}}{\tfrac
12\nu ^2+n\frac{\omega}{m} } 
\end{array}
\end{equation}

\medskip\ 

The combination of values obtained for resonance spacing and resonance width
determine whether or not resonance peaks can be resolved using currently
available experimental apparatus, and whether or not resonances overlap. By
definition, the angular resonance width $\Delta \theta $ is obtained by the
condition that the differential cross section at the points $\theta =\theta
^{res}\pm \Delta \theta $ be half its resonant value at $\theta =\theta
^{res}$.

Generally, the differential cross section for either the SCS or STPPP
process is of the form

\begin{equation}
\label{c7.eq19}\frac{d\sigma }{d\Omega }\sim \left| \sum_{r=-\infty }^\infty
\left( \frac A{\bar p_r^2-m^2+i2\epsilon _p\Im \Delta \epsilon _p^R(\bar
\rho )}+\frac B{{\dbr{p}}_r^2-m^2+i2\epsilon _p\Im \Delta \epsilon _p^R({%
\dbr{\rho}})}\right) \right| ^2 
\end{equation}

\medskip\ 

where the notation $\bar p_r$ and $\bar \rho $ refer to the direct channel
scattering, and ${\dbr{p}}_r$ and ${\dbr{\rho}}$ the exchange channel
scattering. Assuming the resonance width is small, the quantities $A$ and $B$
can be considered constant over the range of angles $\theta ^{res}-\Delta
\theta <\theta <\theta ^{res}+\Delta \theta $. The resonance width is then
determined by a solution of

\begin{equation}
\label{c7.eq20} 
\begin{array}{rcll}
\left| \bar p_r^2-m^2\right| & = & \left| 2\epsilon _p\Im \Delta \epsilon
_p^R(\bar \rho )\right| & \quad \quad 
\text{direct channel} \\ \left| {\dbr{p}}_r^2-m^2\right| & = & \left|
2\epsilon _p\Im \Delta \epsilon _p^R({\dbr{\rho}})\right| & \quad \quad 
\text{exchange channel} 
\end{array}
\end{equation}

\medskip\ 

For the SCS process, assuming that a resonance occurs for parameters $\omega
,\omega _i,\nu ^2$ and $r$, the $\theta _i$ resonance width is approximately
given by

\begin{equation}
\label{c7.eq21}\cos \,\Delta \theta _i\approx \left| 1-\dfrac{2\epsilon
_p\Im \Delta \epsilon _p^R(\bar \rho )}{\omega _i(\frac 12\nu ^2+r\frac{\omega}{m}
)+\omega _i+r\omega }\right| 
\end{equation}

\medskip\ 

The SCS $\theta _f$ resonance width and the equivalent expressions for the
STPPP process are considerably more complicated and are not written down
here. However these are easily determined computationally. From the analysis
of Chapter 6 the numerical value of $\tfrac{2\epsilon _p\Im \Delta \epsilon
_p^R(\bar \rho )}{\omega _i(\frac 12\nu ^2+r\frac{\omega}{m} )+\omega _i+r\omega }$ is
much less than unity, assuming that $\frac{\omega _i}m$ and $\frac \omega m$
are not too small. Therefore the numerical value of $\Delta \theta _i$ is
also very small. Whether the resonances
are too narrow to be experimentally detected requires the precise results of
section 7.4.

It is worthwhile comparing the second order SCS and STPPP resonant differential 
cross sections with the differential cross sections of the
equivalent first order process. For the SCS process the equivalent process
is HICS, and for the STPPP process it is the OPPP process. 
With an external field consisting of a circularly polarised electromagnetic wave, the 
transition probability for the HICS and OPPP processes is due to
\cite{NarNikRit65}. To convert to a differential cross section
the flux density of incoming particles is required. This is the reciprocal of the three
dimensional volume $V$ for one incident photon (OPPP) or one incident
electron at rest (HICS).

For the HICS process, an incident electron combines with $s$ external field
quanta of energy $\omega $ to produce a scattered photon of energy $\omega
_f $. The differential cross section is

\begin{equation}
\label{c7.eq22} 
\begin{array}{rl}
\dfrac{d\sigma }{d\Omega _{k_f}} & = \dfrac{\alpha ^2}{2m^2}\dsum\limits_{s=1}^\infty 
\dfrac 1s\left( \dfrac{\omega _f}\omega \right)^2 \biggl[ -\dfrac 4{\nu ^2}J_s^2(z)+  
\left(2u-1\right) \left( J_{s-1}^2(z)+J_{s+1}^2(z)-2J_s^2(z)\right) \biggr]  \\[20pt]  
\text{where} \quad  z & = \nu^2\dfrac m\omega\sqrt{(1+\nu ^2)u(u_s-u)} \\[10pt]
u & =\dfrac{(kk_f)}{(kp_f)} \quad;\quad u_s =\dfrac{2s\frac{\omega}{m} }{1+\nu ^2}
\end{array}
\end{equation}

\medskip\ 

For the OPPP process, an incident photon of energy $\omega _1$ combines with 
$s$ external field quanta $\omega $ to produce an electron-positron pair of
4-momenta $q_{-}$ and $q_{+}$. The differential cross section is

\begin{equation}
\label{c7.eq23}
\begin{array}{rl}
\dfrac{d\sigma }{d\Omega _{q_{-}}} & = \dfrac{2\alpha ^2}{m^2}
\dsum\limits_{s>s_0}\left( \dfrac{m^2}{\omega \omega _1}\right) \dfrac{|q_{-}| }{\omega _1+s\omega }
\biggl[ \dfrac 2{\nu^2}J_s^2(z)+( 2u-1) 
\left(J_{s-1}^2(z)+J_{s+1}^2(z)-2J_s^2(z)\right) \biggr] \\[20pt]  
\text{where} \quad z & =4\nu^2 \sqrt{1+\frac 1{\nu ^2}}\dfrac{m^2}{(kk_1)}\sqrt{u(u_s-1)} \\[10pt]
  u &= \dfrac{(kk_1)^2}{4(kq_{-})(kq_{+})} \quad;\quad u_s=\dfrac{s(kk_1)}{2m_{*}^2} 
\quad;\quad s_0=\dfrac{2m_{*}^2}{(kk_1)} 
\end{array}
\end{equation}

\newpage
\section{Resonance Figures}
\label{c7fig}

The figures contained within section \ref{c7fig} are divided into two main groups. 
The SCS resonances are investigated in section \ref{c7figscssub} and the STPPP process 
in section \ref{c7figstpppsub}. Tables \ref{c7tab1} and \ref{c7tab2} present the 
range of parameters to be investigated. These parameters were defined in chapters 4 and 5.
The majority of the figures presented in section \ref{c7fig} show the angular location and 
differential cross section height of resonance peaks. The remaining figures show the 
variation of resonance height and width with the external field intensity parameter $\nu ^2$.
Vertical axes representing differential cross sections unless otherwise labelled are 
understood to be the differential cross section 
$\left.\frac{d\sigma }{d\Omega }\right/ \negthinspace \negthinspace \frac{\alpha ^2}{32m^2}$ in 
units of $\text{steradian}^{-1}$. Resonant differential cross sections are generally large and vary 
greatly in height. Consequently a logarithmic scale is used.
Resonances arise in specific summation terms, characterised by the $l$ and $%
r,r^{\prime }$ parameters, of the SCS and STPPP differential cross section
expressions. Some figures show specific summation terms. Most however are summed over all permissible 
values. At the end of each subsection a full cross section is presented. This
was obtained by numerically integrating over final angles using the DIVONNE
implementation in the numerical integration library CUBA \cite{Hahn05}. The
usefulness of this implementation is that assistance is accepted from initial
specification of peak locations. This proved particularly useful for the STPPP
cross-section which contained only narrow peaks.

\enlargethispage*{2cm}

\bigskip\

\subsection{SCS Resonances}
\label{c7figscssub}

\input{./tex/tables/c7_tab1}
\clearpage
\input{./tex/c7cfig}

\subsection{STPPP Resonances}
\label{c7figstpppsub}

\bigskip\

\enlargethispage*{2cm}

\input{./tex/tables/c7_tab2} 
\clearpage  

\input{./tex/c7ppfig}
\section{Analysis of Resonance Plots}
\label{c7anal}

The differential cross section resonances of both the SCS and STPPP processes are discussed in this 
section. Analysis is in terms of the four quantities considered in section \ref{c7reson}, namely 
resonance height, width, spacing and location. Generally the analysis follows the figure order - SCS 
resonances first, and STPPP following. However comparison is made between groups of 
SCS resonance figures and groups of STPPP resonance figures. The figures showing variation with 
respect to $\nu^2$ for both SCS and STPPP processes are discussed at the end of
the section. The variation of the full cross section with $\nu^2$ for both
processes will also be discussed.

The $\theta_f$ distribution of resonance peaks of the SCS (figures \ref{resc1} - \ref{resc8} 
and STPPP (figures \ref{resp1} - \ref{resp8}) processes are discussed first. 
The logarithmic scale on all vertical axes of
figures \ref{resc1} - \ref{resc8} and figures \ref{resp1} - \ref{resp8},
facilitates comparison of the SCS and STPPP resonances with the
corresponding first order external field processes and the corresponding
second order process in the absence of the external field. The SCS resonance
differential cross section is compared with the differential cross sections
of the HICS process (equation \ref{c7.eq22}) and the Klein-Nishina process
(equation \ref{c3.limit.eq3}) in the reference frame in which the initial electron
is stationary. The resonant STPPP differential cross section is compared
with the differential cross section of the OPPP process in the reference
frame in which the sum of initial photon 3-momentum and the 3-momenta sum
total of contributing external field quanta is zero (equation \ref{c7.eq23}), 
and the differential cross section of the Breit-Wheeler process (equation 
\ref{c3.pp.eq11}).

Each of figures  \ref{resc1} - \ref{resc8} display several resonances of differing heights. Figures 
 \ref{resc1} and \ref{resc2} show an external field intensity parameter of $\nu ^2=0.1$ at relative 
particle energies such that the ratio $\frac{\omega _i}\omega $ is above and below
unity.\footnote{$\frac{\omega}{\omega_1}=2$ for figure \ref{resc1} and $\frac{\omega}
{\omega_1}=0.83$ for figure \ref{resc2}.} The maximum resonances of figure \ref{resc2}
reach a value of approximately $10^7$ compared to approximately $10^6$ for
figure \ref{resc1}. This result appears consistent with the analysis of chapter 4
which found, generally, that the SCS differential cross section is enhanced
for a relative particle energy regime in which $\frac{\omega _i}\omega <1$.
On the other hand, figures \ref{resc3} - \ref{resc4} with an external field intensity 
$\nu ^2=1$, reveal a reverse behaviour. The maximum resonance of figure \ref{resc3}
with $\frac{\omega _i}\omega =2$ is larger in value than the maximum
resonance of figure \ref{resc4} with $\frac{\omega _i}\omega =0.83$. This 
is due to the appearance of the maximum resonance in differing
summation terms.
The maximum resonance of figure \ref{resc3} occurs in the $l=-1$ contribution which is more 
probable than the $l=3$ contribution in which the maximum resonance of figure \ref{resc4} appears.
Generally, the resonance heights of figures \ref{resc3} and \ref{resc4} are lower than those of 
figures \ref{resc1} and \ref{resc2}. This is due to the increase of the EFEES imaginary part with 
increased $\nu^2$ (see figures \ref{c6g3} and \ref{c6g4}).

Figures \ref{resc5} - \ref{resc8} show the SCS resonances with variation in $\theta_i$.
The maximum resonances of all four figures occur at $\theta _i=61.726^{\circ }$ indicating
that the resonance occurs in the direct channel of the scattering, which has a dependence only on initial
state quantities. The $\theta _i=61.726^{\circ },298.271^{\circ }$ resonance peaks
dominate in each of figures \ref{resc5} - \ref{resc8} since they appear in every $l$ 
contribution of the SCS differential cross section.

Figures \ref{resp1} - \ref{resp8} display the resonances of the STPPP process with a
similar grouping of parameters as figures \ref{resc1} - \ref{resc8} for the SCS
process. The heights of maximum STPPP resonance peaks fall in the range 
$10^5\leq \left.\frac{d\sigma }{d\Omega }\right/ \negthinspace 
\negthinspace \frac{\alpha ^2}{32m^2}\,\leq 10^6$. That the resonance peaks generally differ greatly 
in height, is due to the summation terms of the STPPP differential cross section in which 
they appear and whether they appear in the direct or exchange channel of the process.

The point of greatest interest in the consideration of resonance peak heights of both
the SCS and STPPP processes is that they exceed by several orders of
magnitude the differential cross sections of the equivalent first order
processes. Naively it would be expected that the first order process cross sections which are 
proportional to the fine structure constant $\alpha$ would dominate the second order processes which 
are proportional to $\alpha^2$. Indeed most experimental attempts to detect QED
processes in the external electromagnetic field of similar form to that
considered in this thesis have dealt with the first order processes. 
However the second order processes at resonance
provide a much more likely candidate for experimental detection. For the
range of parameters considered in section \ref{c7fig}, the maximum resonant
differential cross sections expressed as a percentage increase over the equivalent first order
processes are

\begin{equation}
\label{c7.ana.eq1} 
\begin{array}{ccl}
1.19\times 10^8\;\% &  & \text{for the }\theta _i=61.726^{\circ }\text{ SCS
resonance of figure }7.6 \\ 4.49\times 10^6\;\% &  & \text{for the }\theta
_1=239.95^{\circ }\text{ STPPP resonance of figure }7.19 
\end{array}
\end{equation}

\medskip\ 

For the purposes of experimental detection the number of resonances that
occur and their angular location, is also of interest. The number of resonances that occur for a 
particular set of parameters and all possible scattering geometries ($0^{\circ }\leq \theta
_f\leq 360^{\circ }$ and $0^{\circ }\leq \theta _i\leq 360^{\circ }$) is generally larger for the 
STPPP resonances. An examination of the energy level structure of the electron
embedded in the external field reveals why this should be the case. 
For an electron of 4-momentum $q$ embedded in an external field of intensity $\nu ^2$ and 4-momentum 
$k$, the $n$th energy level is characterised by

\begin{equation}
\label{c7.ana.eq2}(q+nk)^2=m^2(1+\nu ^2) 
\end{equation}

\medskip\ 

Using the notation of chapters 4 and 5, the energy levels of the
intermediate SCS and STPPP electron with $l$ external field quanta
contributing can be written

\begin{equation}
\label{c7.ana.eq3}
\begin{array}{cclcl}
n-l & = & \dfrac{(q_ik_i)}{(kp_i)+(kk_i)} &  & \text{SCS direct channel} \\[10pt] 
n-l & = & \dfrac{-(q_ik_f)}{(kp_i)-(kk_f)} &  & \text{SCS exchange channel}\\[10pt] 
n-l & = & \dfrac{(q_{-}k_1)}{(kp_{-})-(kk_1)} &  & \text{STPPP direct channel} \\[10pt]
n-l & = & \dfrac{(q_{-}k_2)}{(kp_{-})-(kk_2)} &  & \text{STPPP exchange channel} 
\end{array}
\end{equation}

\medskip\ 

The number of resonances available for any given set of parameters is
related to the number of energy levels the intermediate electron can
traverse, which is dependent on the numerical value of the right hand side
of equation \ref{c7.ana.eq3}. These are, in full

\begin{equation}
\label{c7.ana.eq4}
\begin{array}{rclcl}
\dfrac{(q_ik_i)}{(kp_i)+(kk_i)} & = & \dfrac{\omega _i}\omega \dfrac{1+\frac{%
\nu ^2}2(1-\cos \theta _i)}{1+\frac{\omega_i}{m}(1-\cos \theta _i)} &  & \text{SCS
direct channel} \\[10pt]
\dfrac{-(q_ik_f)}{(kp_i)-(kk_f)} & = & -\dfrac{\omega _f}\omega 
\dfrac{1+\frac{\nu ^2}2(1-\cos \theta _f)}{1-\frac{\omega_f}{m}(1-\cos \theta
_f)} &  & \text{SCS exchange channel} \\[10pt] 
\dfrac{(q_{-}k_1)}{(kp_{-})-(kk_1)} & = & \dfrac{\omega _1}\omega \dfrac{\epsilon _{q_{-}}-\left| 
q_{-}\right|(\cos \theta _1\cos \theta _f-\sin \theta _1\sin \theta _f\cos \phi _f)}
{\epsilon _{q_{-}}-\left| q_{-}\right| \cos \theta _f-\omega _1(1-\cos \theta
_1)} &  & \text{STPPP direct channel} \\[10pt] 
\dfrac{(q_{-}k_2)}{(kp_{-})-(kk_2)} & = & \dfrac{\omega _1}\omega \dfrac{\epsilon _{q_{-}}+\left| 
q_{-}\right|(\cos \theta _1\cos \theta _f-\sin \theta _1\sin \theta _f\cos \phi _f)}
{\epsilon _{q_{-}}-\left| q_{-}\right| \cos \theta _f-\omega _1(1+\cos \theta_1)} &  & 
\text{STPPP exchange channel} 
\end{array}
\end{equation}

\medskip\ 

The energy levels correspond to points where the numerical value of the
expressions of equation \ref{c7.ana.eq4} are integer values. Using figure \ref{resc5} 
for the SCS process and figure \ref{resp5} for the STPPP process as examples

\begin{equation}
\label{c7.ana.eq5} 
\begin{array}{ccccccl}
0.833 & \leq & \dfrac{(q_ik_i)}{(kp_i)+(kk_i)} & \leq & 1.389 &  & \text{SCS
direct channel} \\[10pt]
-6.389 & \leq & \dfrac{-(q_ik_f)}{(kp_i)-(kk_f)} & \leq & 
-0.833 &  & \text{SCS exchange channel} \\[10pt] 
-47 & \leq & \dfrac{(q_{-}k_1)}{(kp_{-})-(kk_1)} & \leq & 132 &  & 
\text{STPPP direct channel} \\[10pt] 
-47 & \leq & \dfrac{(q_{-}k_2)}{(kp_{-})-(kk_2)} & \leq & 132 &  & 
\text{STPPP exchange channel} 
\end{array}
\end{equation}

\medskip\ 

The STPPP intermediate electron traverses a greater number of energy
levels and can therefore reach more points of resonance.

The expressions of equation \ref{c7.ana.eq4} suggest a greater number of
resonances when the ratio $\frac{\omega_i}{\omega},\frac{\omega_1}\omega$ is large.
The validity of that suggestion can be established by comparing pairs of figures with similar 
parameter sets (table \ref{c7tab3})

\input{./tex/tables/c7_tab3}

Figure \ref{resc5}, with $\theta _f=0^{\circ }$, shows a special case for the
SCS process. For this final scattering geometry the direct channel resonance
terms reduce to a dependence purely on the ratio $\frac{\omega _i}\omega $.
Since this ratio is never unity, no direct channel resonances exist. Of note
also for the SCS process, is that no resonances exist when $\theta _i=0^{\circ
} $, except when the ratio $\frac{\omega _i}\omega$ is integral. 
This is confirmed by the work of \cite{AkhMer85}.

Figures \ref{resc9} - \ref{resc12} (SCS process) and figures \ref{resp9} - \ref{resp12}
(STPPP process) show the variation of resonance height and widths with external field 
intensity $\nu^2$. The trend for figure \ref{resc9} which shows $l=3,r=6$ and $l=4,r=7$ SCS resonant 
peak heights is difficult to determine due to the
existence of sharp peaks at $\nu ^2=0.2$ and $\nu ^2=1.3$. These points
correspond to instances in which direct channel resonances coincide with the
exchange channel resonance considered.

Figure \ref{resc10} shows increasing $l=4\;r,r'=5$ peak height up to $\nu ^2=1.5$ 
and remaining approximately constant thereafter. The $l=1$ peak of the same figure, 
apart from the coincidence of an exchange
channel resonance peak at $\nu ^2=0.55$, remains approximately constant
throughout the range of $\nu ^2$ values considered. There are two competing tendencies at work in 
figures \ref{resc9} and \ref{resc10}. The SCS differential cross section increases with increasing 
$\nu^2$ and resonance heights diminish with increasing resonance width.

The SCS resonance widths of figures \ref{resc11} and \ref{resc12} also increase with increasing 
$\nu^2$. Exchange channel resonance widths (figure \ref{resc11}) reach a value of
approximately $0.5^{\circ }$ at $\nu ^2=1$, and direct channel resonance
widths reach approximately $2.5^{\circ }$ at $\nu ^2=1$. Resonances of this width should 
easily be resolved experimentally.

The $\nu^2$ variation of STPPP resonance peak heights represented in
figures \ref{resp9} and \ref{resp10} reveal an increase up until an optimum $\nu^2$
value, thereafter decreasing. Maximum resonance peak heights are obtained at 
$\nu^2=0.4$ for the $l=1\;r,r'=3$ resonance of figure \ref{resp9}, $\nu^2\sim0.25$ 
for the $l=1\;r,r'=-1$ resonance of figure \ref{resp10}, and $\nu ^2\sim0.3$ for the $l=1\;r,r'=2$ 
resonance of figure \ref{resp10}. The $l=1\;r,r'=-1$ peak height of figure \ref{resp9} has 
no maximum, but begins to flatten off at upper limits. The angular widths of the same resonances,
considered in figures \ref{resp11} and \ref{resp12} show an almost linear increase 
with $\nu^2$. At $\nu^2=0.5$ the $l=1\;r,r'=3$ and $l=1\;r,r'=-1$ resonance widths of figure 
\ref{resp11} are approximately $0.02^\circ$ and $0.05^\circ$ respectively, 
while the $l=1\;r,r^{\prime }=-1$ and $l=1\;r,r'=2$ resonance widths of figure \ref{resp12} are both 
about $0.02^\circ$ at the same $\nu^2$. Generally STPPP resonances are much narrower than the SCS 
resonances.

The variation of the full cross section of the SCS process with external field intensity $\nu^2$ is 
considered in figure \ref{rescfull}. The initial photon has an energy of 51.2 keV and the external
field photon energy is 25.6 keV. These are the same parameters as those considered for the $\nu^2$
variation of figure \ref{resc9} and the same trend of a large peak at $\nu^2=0.2$ is obtained. The
full SCS cross section exceeds the Klein-Nishina cross section by 5 to 6 orders of magnitude at that
point and the extent of the increase is generally due to the fact that the SCS differential cross
section resonances are broad. A full STPPP cross section
variation with $\nu^2$ is shown in figure \ref{respfull}. This plot displays a much more moderate
increase in cross section with increasing $\nu^2$ which is due to the generally narrower resonances
of the STPPP differential cross section. The STPPP cross section increases more rapidly at low
values of $\nu^2$ with the increase levelling off as $\nu^2$ increases. This has the same explanation
as the $\nu^2$ variation of figures \ref{resp9}-\ref{resp10}. The increase of the STPPP cross section
over the Breit-Wheeler process is slightly more than an order of magnitude by $\nu^2=0.5$. 
\bigskip\
\section{Experimental Considerations}
\label{c7exp}

In this section all results obtained thus far for the SCS and
STPPP differential cross sections are discussed in light of
experimental work due to \cite{Bamber99} and \cite{EngRin83}. A numerical comparison 
between parameters used in this work with those that have been used experimentally is 
pertinent.

The initial photon energies employed in calculation of SCS differential
cross sections ranges from UV ($\omega _i=26$ eV) to x-rays ($\omega _i=10.2$
keV), whereas those employed in STPPP differential cross section
calculation, the requirement being that they must be at least the electron
rest mass, range from low energy gamma rays ($\omega _1=0.512$ MeV) to high
energy gamma rays ($\omega _1=5.12$ MeV). Since the photon beam containing the initial photons does 
not need to be ultra intense, there is no experimental difficulty in producing a full range of initial 
photon energies. These can be produced in the IR \cite{Sauteret91}, UV \cite{Luk89,Bamber99} or 
x-ray \cite{Murnane91} directly from a laser source, or higher
energy photons are obtained by the scattering of lower energy photons from
high energy electron beams. Using this method \cite{Federici80} have obtained $%
80$ MeV photons from a $1.5$ GeV electron beam and more recently \cite{Bamber99}
obtained $36.5$ GeV photons from an incident UV ($\lambda =350$ nm) laser
pulse and a $50$ GeV electron beam.

Due to computational time restrictions the external field energies considered were never smaller than 
initial photon energies divided by a factor of 10. The number
of $l$ contribution terms required to obtain data points is roughly
proportional to twice the ratio $\frac{\omega _i}\omega,\frac{\omega _1}\omega$. 
The external field energies examined range from the visible ($\omega =5$ eV) to x-rays ($\omega =51.2$
keV) for the SCS process, and from hard x-rays ($\omega =0.256$ MeV) to
gamma rays ($\omega =5.12$ MeV) for the STPPP process. However, ultra high intensity lasers 
operate at much lower photon energies. The TableTop Terawatt (T$^3$) generation of lasers produce $3.55$
eV photons at an intensity of $4\times 10^{17}$ W cm$^{-2}$ \cite{McDonald91}
or 5 eV photons at an intensity of $2\times 10^{19}$ W cm$^{-2}$.
Higher photon energy lasers are available but at lower intensity. 50-1200 keV photons can be produced 
with an intensity of $6.1\times10^{15}-1.5\times 10^{17}$ W cm$^{-2}$ \cite{Sprangle92}.

The T$^3$ lasers which produce the greatest energy flux are considered the
most interesting for investigating QED processes in the external field. For
these lasers the external field intensity parameter $\nu ^2$ approaches
unity, at which point perturbative QED is of limited validity \cite{Bamber99}.
Further, the cross sections of the first order external field QED processes
are optimal when the parameter $\frac 14(\frac \omega m)^2\nu ^2\sim 1$
\cite{Becker91}.\footnote{Based on the electron embedded in the external field being in its 
ground state.} 
The second order processes investigated here, however, return
large differential cross section values at points of resonance. The figures
of section 7.4 reveal resonance differential cross sections several orders
of magnitude in excess of those of first order processes for external field
intensities as low as $\nu ^2=0.1$. Available laser intensity is more than adequate for experimental 
detection of the SCS and STPPP resonances.

Another parameter of experimental significance is the momentum of
incoming particles. Experiments use relativistic electron beams or electron plasmas \cite{Becker91}. 
The SCS calculations in this thesis were performed in the laboratory frame and there is little difficulty 
in repeating calculations with a non zero initial electron momentum.
The STPPP differential cross section calculations, performed in the centre of mass frame of
incoming photons, are immediately amenable to experimental study due to the
availability of counter propagating photon beams.

In determining the observability of the second order IFQED processes values for the full cross section 
and the number density of incident particles are required. The cross section calculation can 
proceed in one of two ways. Assuming that the resonant differential cross section dominates, we
can calculate the contribution to the total cross section from each resonant
peak can be calculated, or alternatively the probability of detection of a particular
resonant peak by a real particle detector of finite aperture in a fixed
position can be determined. The second alternative is considered here.

\cite{EngRin83} use an experimental setup such that their particle
detector aperture subtends $0.024$ of the total solid angle, which
translates to a $5.66^{\circ }$ angular resolution for a circular detector
aperture. In \cite{McDonald91} an angular resolution of $4\times 10^{-5\,\circ }$ is obtained. Assuming 
angular resolution is such that the top $50\%$ of one resonant peak falls
within the aperture of a detector in a fixed position, the cross section
measured is obtained by an integration of the differential cross section
across the range of angles subtended by the detector's aperture.%
\footnote{$\Delta\Gamma$ is the resonance width and $\Delta\cos\Gamma$ is 
difference in cosine angles across the resonance width.}

\begin{equation}
\label{c7.exp.eq1}
\sigma _{\text{detector}}=\dint_{-\frac 12\Delta \Gamma}^{\frac 12\Delta \Gamma }\;
d\phi _f\,\int_{-\Delta \cos \Gamma }^{\Delta
\cos \Gamma }\;d\cos \theta _f\;\left. \frac{d\sigma }{d\Omega }\right| _{\text{resonance}} 
\end{equation}

\medskip\ 

Taking, as a first approximation, a linear relationship between the
differential cross section near resonance and the cosine of the polar angle 
$\theta _f$, and no variation of the differential cross section with the
azimuthal angle $\phi _f$ near 
resonance,\footnote{This is not a bad assumption since, generally, 
the SCS and STPPP differential cross section vary much more slowly with 
$\phi_f$ than with $\theta_f$.}
the cross section measured by the detector is the resonant differential
cross section multiplied by the factor

\begin{equation}
\label{c7.exp.eq2}\frac 34\Delta \Gamma \,\Delta \cos \Gamma 
\end{equation}

\medskip\ 

Choosing the largest resonance peaks for the SCS and STPPP processes and including the conversion
factor from natural units to SI units, the cross section measured by the detector is

\begin{equation}
\label{c7.exp.eq3} 
\begin{array}{cclcc}
\sigma _{\text{detector}} & = & 1.528\times 10^{-21}\;\text{cm}^2 &  & \text{%
SCS process} \\ \sigma _{\text{detector}} & = & 5.95\times 10^{-25}\,\text{cm%
}^2 &  & \text{STPPP process} 
\end{array}
\end{equation}

\medskip\ 

To convert these into the number of particles detected per second 
the number density of incoming particles is required. Real photon sources are pulsed and only one 
detection per pulse is possible \cite{McDonald91}. The number density is calculated from the flux
density of the photon source $I$, pulse length $\Delta t$ and photon energy $\omega $

\begin{equation}
\label{c7.exp.eq4}\text{number density}=\frac{I\,\Delta t}\omega 
\end{equation}

\medskip\ 

For the maximum resonances of the SCS and STPPP processes considered 
($\omega _i=51.2$ keV and $\omega _1=0.768$ MeV respectively) an incident photon source in the hard 
x-ray/low energy gamma ray range is required. Such a source with a photon number density 
of $2.4\times 10^{14}$ per pulse is available at a pulse repetition rate of 1 Hz \cite{McDonald91}.

The number of events per second (which is a product of the cross section and
particle number density) is increased by introducing the second incident
particle (electron for the SCS process and photon for the STPPP process) via
a particle beam. For the purposes of the SCS process, a $50$ GeV pulsed
electron beam at $10^9-10^{10}$ electrons per pulse is available. The consequent 
increase in the event rate cannot be included here since a recalculation
of the SCS differential cross section allowing for initial electron
3-momentum, is required. For the STPPP process two counter propagating photon beams of
similar energy is easily achieved experimentally. For two incident
particle beams a factor of $10^{-2}$ is assumed as a loss due to the spatial
coincidence of the beams and the detector efficiency \cite{McDonald91}.

The event rates obtained for the SCS and STPPP maximum resonances are

\begin{equation}
\label{c7.exp.eq5} 
\begin{array}{rcc}
3.7\times 10^{-7}\;\text{s}^{-1} &  & \text{SCS process} \\ 342.72\;\text{s}%
^{-1} &  & \text{STPPP process} 
\end{array}
\end{equation}

\medskip\ 

As it stands, the event rate obtained for the maximum resonant peak of the
SCS process is too low to be experimentally observable. For the first order
HICS process, however, the allowance for an initial electron 3-momentum can
lead to an enhancement of the differential cross section \cite{Derlet95,Ginzburg83b}. 
A factor of $10^7-10^8$ for the electron number density plus spatial coincidence of particle beams can be 
included, leading to an observable SCS event rate. The STPPP event rate is easily observable and in fact 
larger than the OPPP event rate \cite{McDonald91}.

In order to achieve maximum photon flux densities, laser beams are focused
resulting in a non plane wave electromagnetic field. To determine the
relevance of our results we need to estimate the extent to which SCS and
STPPP differential cross section calculations performed in a plane wave
external field and a non plane wave focused field differ.

No work in a focused external field has been performed for the second order
processes considered in this work, however the first
order HICS and OPPP processes in which one external field quanta
participates has been considered. The plane wave external field constitutes a valid
approximation to the focused external field whenever the energy of the quanta
associated with the field is several orders of magnitude less than that of other incident
particles. When particle energies are equivalent, differential transition
rates obtained with the focused external field are $40\%$ smaller than those
obtained with a plane wave external field \cite{Derlet95}.

An estimate of the effect on second order processes can be obtained by
examining the focused field equations and the extent to which they differ
from a plane wave. If a focussed laser beam can be considered Gaussian, then in
the paraxial approximation (which neglects longitudinal components of the 
electric and magnetic fields), the equation governing the radius of curvature of
the wavefront $R(z)$ is simple. The relationship is in terms of the beam waist
width $w_0$, laser wavelength $\lambda$ and the longitudinal distance from the
beam waist, $z$. This relationship reveals that the beam is a plane wave
($R=\infty$) at the focal plane ($z=0$) (\cite{Alda02} p.999-1013).

\begin{equation}
R(z)=z\left[ 1+\left(\dfrac{\pi w_{0}^{2}}{z\lambda}\right)^2\right]
\end{equation}

\medskip\

However the QED calculations consider Volkov wave functions far from the focal
plane. The paraxial approximation again provides relatively simple expressions
however now the wave fronts are curved. Corrections to the paraxial
approximation have been made \cite{Lax75,Davis79} and the most general
expression of the electromagnetic field at the focus of a lens were provided by 
\cite{BoiWol64}. In the general expression the electric and magnetic field strengths at a point $P$ near the
focus of an ideal lens are given by

\begin{equation}
\label{c7.exp.eq6} 
\begin{array}{c}
E(P,t)=Re\left[ e(P)\,e^{-i\omega t}\right] \\ 
H(P,t)=Re\left[ h(P)\,e^{-i\omega t}\right] 
\end{array}
\end{equation}

$$
\begin{array}{cclcl}
\text{where} &  & e_x(P)=-i\alpha \,\left( I_0+I_2\,\cos 2\phi \right) &  & 
h_x(P)=-i\alpha \,I_2\,\sin 2\phi \\  
&  & e_y(P)=-i\alpha \,I_2\,\sin 2\phi &  & h_y(P)=-i\alpha \,\left(
I_0-I_2\,\cos 2\phi \right) \\  
&  & e_z(P)=-2\alpha \,I_1\,\cos \phi &  & h_z(P)=-2\alpha \,I_1\,\sin \phi 
\end{array}
$$

\medskip\ 

Here $\alpha =k\,f\,E_0$ , $E_0$ is the amplitude of the electric field
strength in the incident plane wave, $f$ is the focal length of the lens, 
$k=\frac \omega c$. With $J_\mu$ a Bessel function of order $\mu $, 
$\theta _0=\arctan \,\frac d{2f}$ , $d$ the lens diameter, and $(\rho ,z,\phi )$ are the coordinates
of a point in the cylindrical coordinate system with centre at the focus and
polar axis coinciding with the optical axis of the lens, the functions $I_\mu $ are described by 

\begin{equation}
\label{c7.exp.eq7}I_\mu (k\rho ,kz,\theta _0)=4\int_0^{\theta _0}d\theta
\,\cos {}^{1/2}\theta \,\sin {}^{\mu +1}\left( \tfrac \theta 2\right) \cos
{}^{3-\mu }\left( \tfrac \theta 2\right) J_\mu (k\rho \,\sin \theta
)\;e^{i\,kz\,\cos \theta } 
\end{equation}

\medskip\ 

With the use of a $\frac fd=3$ lens, a focused laser beam
of spot size $2.8\lambda \;(\lambda =1054$ nm$)$ for which the parameter $k\rho$ 
varies from $0$ to $17.59$ can be obtained \cite{McDonald91}. The extent to which the focused field can be
approximated by a plane wave is the extent to which the function $I_0$ is
sinusoidal and the functions $I_1$ and $I_2$ are negligible. The plane wave
approximation for the focused field is best for low $k\rho $ (i.e. near the
centre of the focused laser beam), and for the upper limit of $k\rho$ considered by \cite{McDonald91} the 
approximation remains reasonably valid.

The SCS and STPPP differential cross section results in this thesis were obtained with laser quanta of 
$0.061-1.024$ MeV. The beam spot sizes available for
these laser sources result in the range of parameters $0\leq k\rho \leq
1.52\times 10^8$ for which a plane wave approximation cannot be expected to
remain valid. What is required is a more tightly focused high energy
laser beam or a recalculation of the SCS and STPPP differential cross sections at lower laser quanta 
energies.

\bigskip\

%\begin{figure}[h]
%\centerline{\psfig{figure=./tex/graphs/focus3.eps,height=8cm,width=10cm}}
%\caption{The functions I for $k\rho =0$.} 
%\label{c7.exp.fig1}
%\end{figure}

%\begin{figure}[h]
%\centerline{\psfig{figure=./tex/graphs/focus4.eps,height=8cm,width=10cm}}
%\caption{The functions I for $k\rho =10$.} 
%\label{c7.exp.fig2}
%\end{figure}

%\begin{figure}[h]
%\centerline{\psfig{figure=./tex/graphs/focus5.eps,height=8cm,width=10cm}}
%\caption{The functions I for $k\rho =20$.} 
%\label{c7.exp.fig3}
%\end{figure}

%% file: tex/tables/c7_tab1.tex
\begin{table}[h]
\label{c7tab1}
\center{
\begin{tabular}{|c|c|c|c|c|c|c|c|c|} \hline
$\bm l$&$\bm r,r'$&$\bm \nu^{\bm 2}$ & $\bm \omega (keV)$ & $\bm \omega_{i} (keV)$ &$ 
\bm \theta_{i}$ & 
$ \bm (\theta_{f},\phi_{f})$ & \bf figure(s) \\ \hline\hline
$ \text{all} $&$ \text{all} $&$ 0.1 $&$ 25.6,61.4 $&$ 51.2 $&
$ 90^{\circ } $&$ (0^{\circ}\rightarrow 360^{\circ },0^{\circ }) $& 
\ref{resc1},\ref{resc2} \\ \hline
$ \text{all} $&$ \text{all} $&$ 1 $&$ 25.6,61.4 $&$ 51.2 $&
$ 90^{\circ } $&$ (0^{\circ}\rightarrow 360^{\circ },0^{\circ }) $& 
\ref{resc3},\ref{resc4} \\ \hline

$ \text{all} $&$ \text{all} $&$ 1 $&$ 61.4 $&$ 51.2 $&
$ 0^{\circ}\rightarrow 360^{\circ } $&$ (0^{\circ },0^{\circ }) $& 
\ref{resc5} \\ \hline
$ \text{all} $&$ \text{all} $&$ 1 $&$ 61.4 $&$ 51.2 $&
$ 0^{\circ}\rightarrow 360^{\circ } $&$ (45^{\circ },0^{\circ }) $& 
\ref{resc6} \\ \hline
$ \text{all} $&$ \text{all} $&$ 1 $&$ 61.4 $&$ 51.2 $&
$ 0^{\circ}\rightarrow 360^{\circ } $&$ (90^{\circ },0^{\circ }) $& 
\ref{resc7} \\ \hline
$ \text{all} $&$ \text{all} $&$ 1 $&$ 61.4 $&$ 51.2 $&
$ 0^{\circ}\rightarrow 360^{\circ } $&$ (180^{\circ },0^{\circ }) $& 
\ref{resc8} \\ \hline
$ 3,4 $&$ 6,7 $&$  $&$ 61.4 $&$ 51.2 $&
$ 90^{\circ} $&$  $& \ref{resc9},\ref{resc11} \\ \hline
$ 3,4 $&$ 6,7 $&$  $&$ 61.4 $&$ 51.2 $&
$  $&$ 45^{\circ} $& \ref{resc10},\ref{resc12} \\ \hline
$ \text{all} $&$ \text{all} $&$ 0.05\rightarrow 0.6 $&$ 25.6 $&$ 51.2 $&
$  $&$ 90^{\circ} $& \ref{rescfull} \\ \hline
\end{tabular} }
\caption{\bf\bm The parameter range for which the SCS cross section resonances are 
investigated.} 
\end{table}

%% file: tex/c7cfig.tex
\begin{figure}[H]
 \centerline{\includegraphics[height=7.5cm,width=15cm]{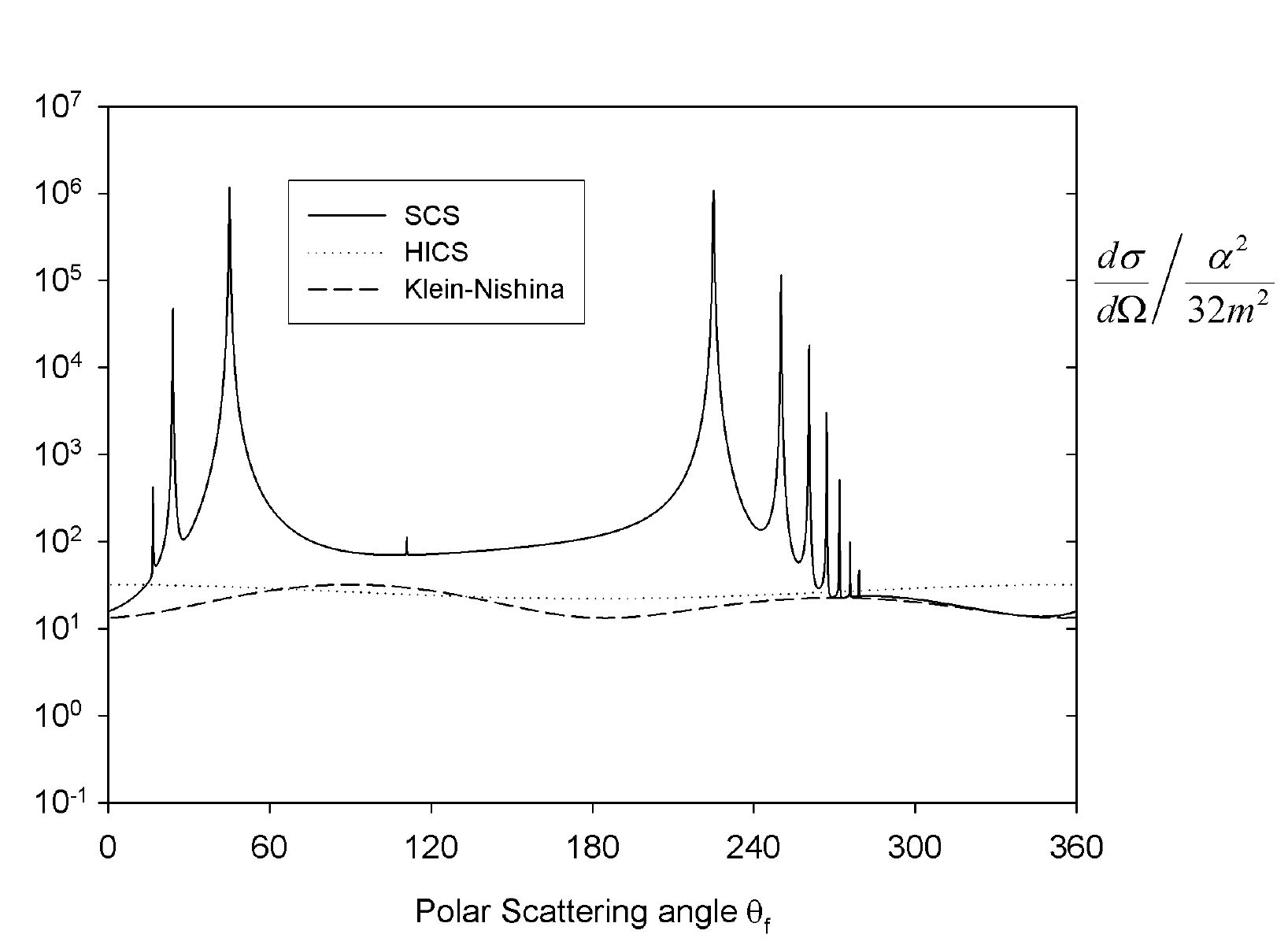}}
% \centerline{\psfig{figure=./compton/reson_thesis_final/nu0.1_w_0.05_wi_0.1_ti90.pdf,height=7.5cm,width=15cm}}
\caption{\bf\bm Comparison of the Klein-Nishina, HICS and SCS differential cross section 
resonances vs $\theta_f$ for $\omega=25.6$ keV, $\omega_i=51.2$ keV, $\theta_i=90^{\circ}$, 
$\varphi_f=0^{\circ}$ and $\nu^2=0.1$.}
\label{resc1}
\end{figure}

\begin{figure}[H]
 \centerline{\includegraphics[height=7.5cm,width=15cm]{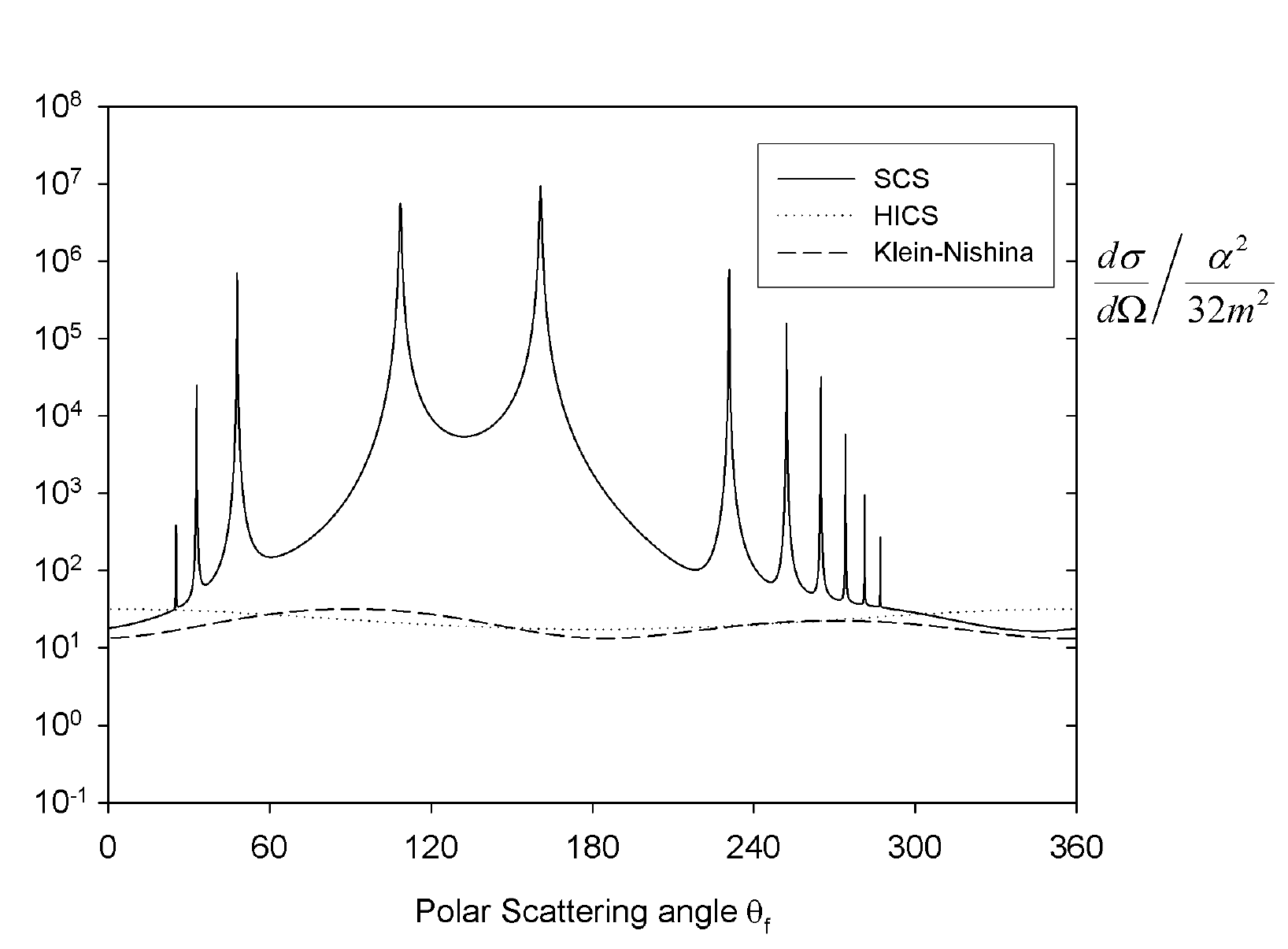}}
% \centerline{\psfig{figure=./compton/reson_thesis_final/nu0.1_w_0.12_wi_0.1_ti90.pdf,height=7.5cm,width=15cm}}
\caption{\bf\bm Comparison of the Klein-Nishina, HICS and SCS differential cross section 
resonances vs $\theta_f$ for $\omega=61.4$ keV, $\omega_i=51.2$ keV, $\theta_i=90^{\circ}$, 
$\varphi_f=0^{\circ}$ and $\nu^2=0.1$.}
\label{resc2}
\end{figure}

\clearpage

\begin{figure}[H]
 \centerline{\includegraphics[height=8cm,width=15cm]{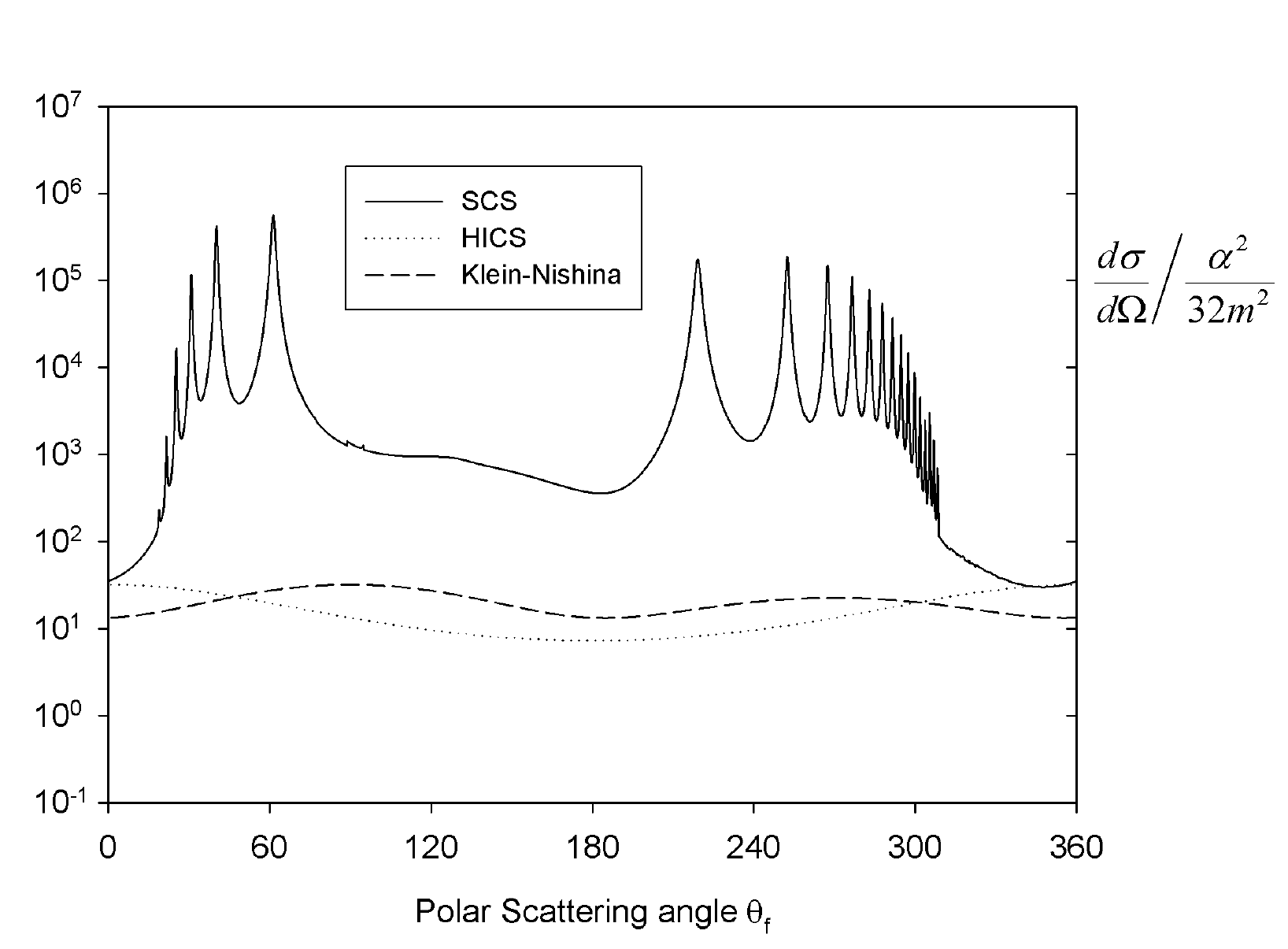}}
% \centerline{\psfig{figure=./compton/reson_thesis_final/nu1_w_0.05_wi_0.1_ti90.pdf,height=8cm,width=15cm}}
\caption{\bf\bm Comparison of the Klein-Nishina, HICS and SCS differential cross section
resonances vs $\theta_f$ for $\omega=25.6$ keV, $\omega_i=51.2$ keV, $\theta_i=90^{\circ}$,
$\varphi_f=0^{\circ}$ and $\nu^2=1$.}
\label{resc3}
\end{figure}

\begin{figure}[H]
 \centerline{\includegraphics[height=8cm,width=15cm]{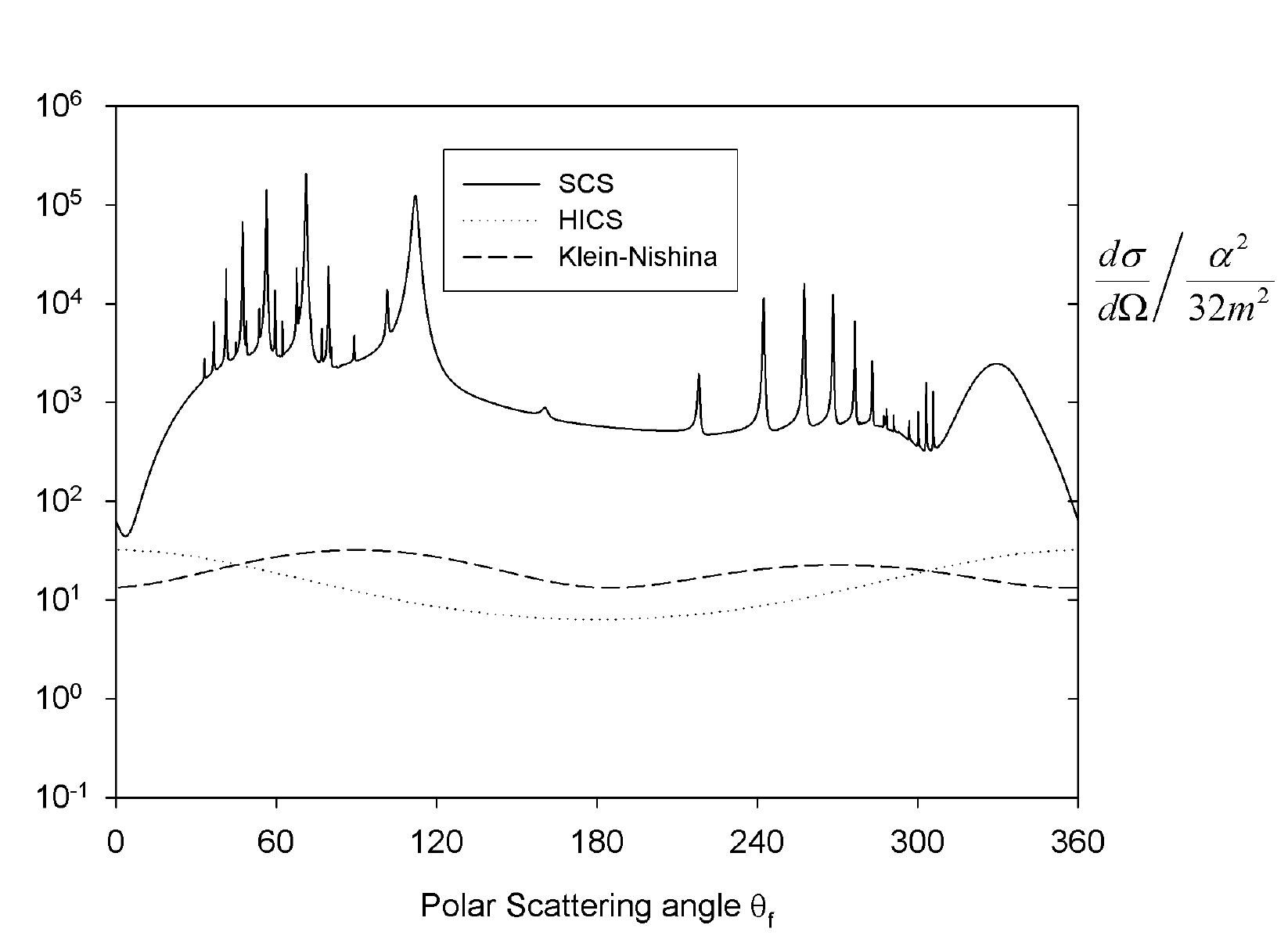}}
% \centerline{\psfig{figure=./compton/reson_thesis_final/nu1_w_0.12_wi_0.1_ti90.pdf,height=8cm,width=15cm}}
\caption{\bf\bm Comparison of the Klein-Nishina, HICS and SCS differential cross section
resonances vs $\theta_f$ for $\omega=61.4$ keV, $\omega_i=51.2$ keV, $\theta_i=90^{\circ}$,
$\varphi_f=0^{\circ}$ and $\nu^2=1$.}
\label{resc4}
\end{figure}

\clearpage

\begin{figure}[H]
 \centerline{\includegraphics[height=8cm,width=15cm]{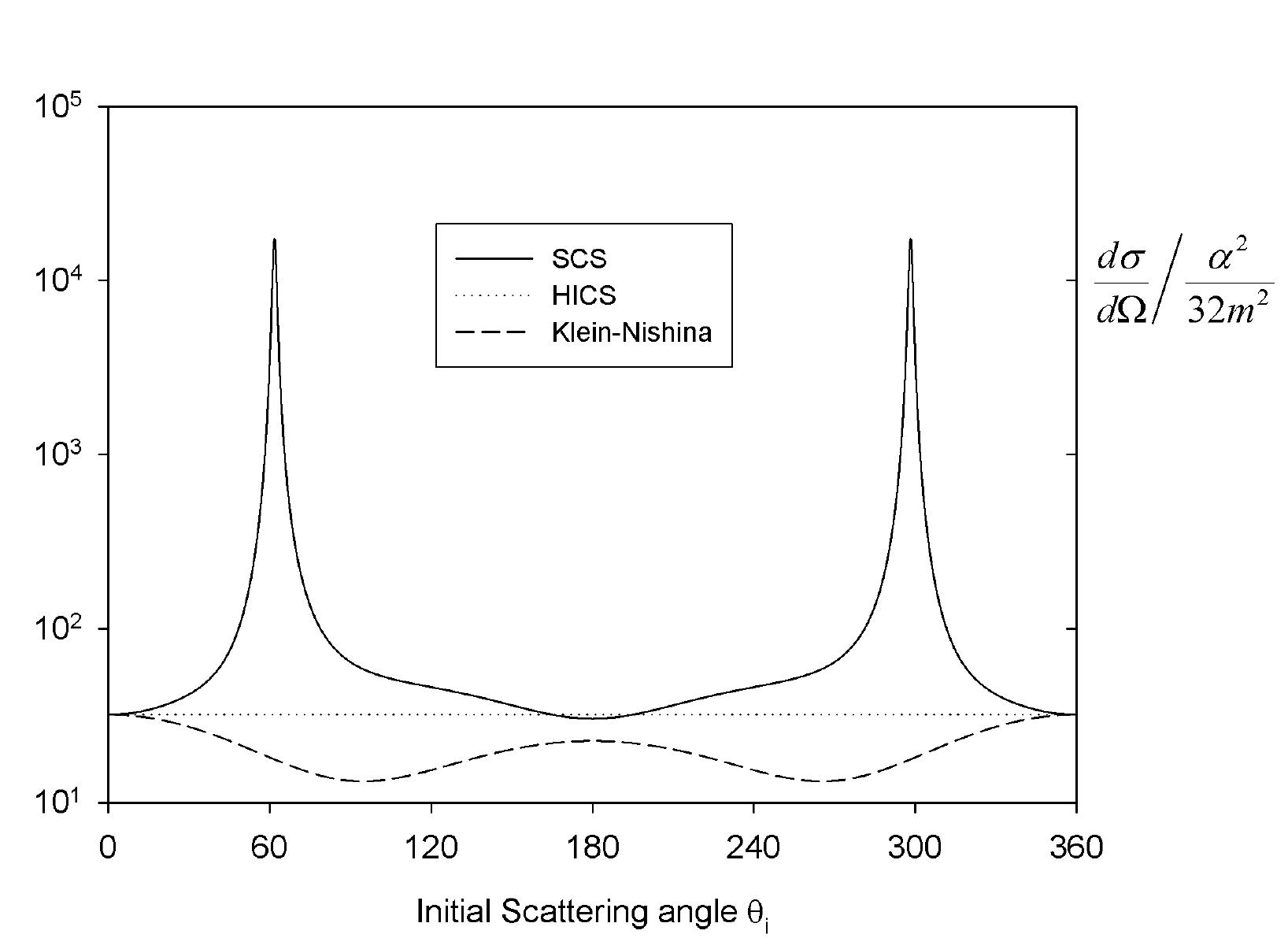}}
% \centerline{\psfig{figure=./compton/reson_thesis_final/vsti_nu1_w0.12_w1_0.1_tf0.pdf,height=8cm,width=15cm}}
\caption{\bf\bm Comparison of the Klein-Nishina, HICS and SCS differential cross section
resonances vs $\theta_f$ for $\omega=25.6$ keV, $\omega_i=51.2$ keV, $\theta_i=0^{\circ}$,
$\varphi_f=0^{\circ}$ and $\nu^2=1$.}
\label{resc5}
\end{figure}

\begin{figure}[H]
 \centerline{\includegraphics[height=8cm,width=15cm]{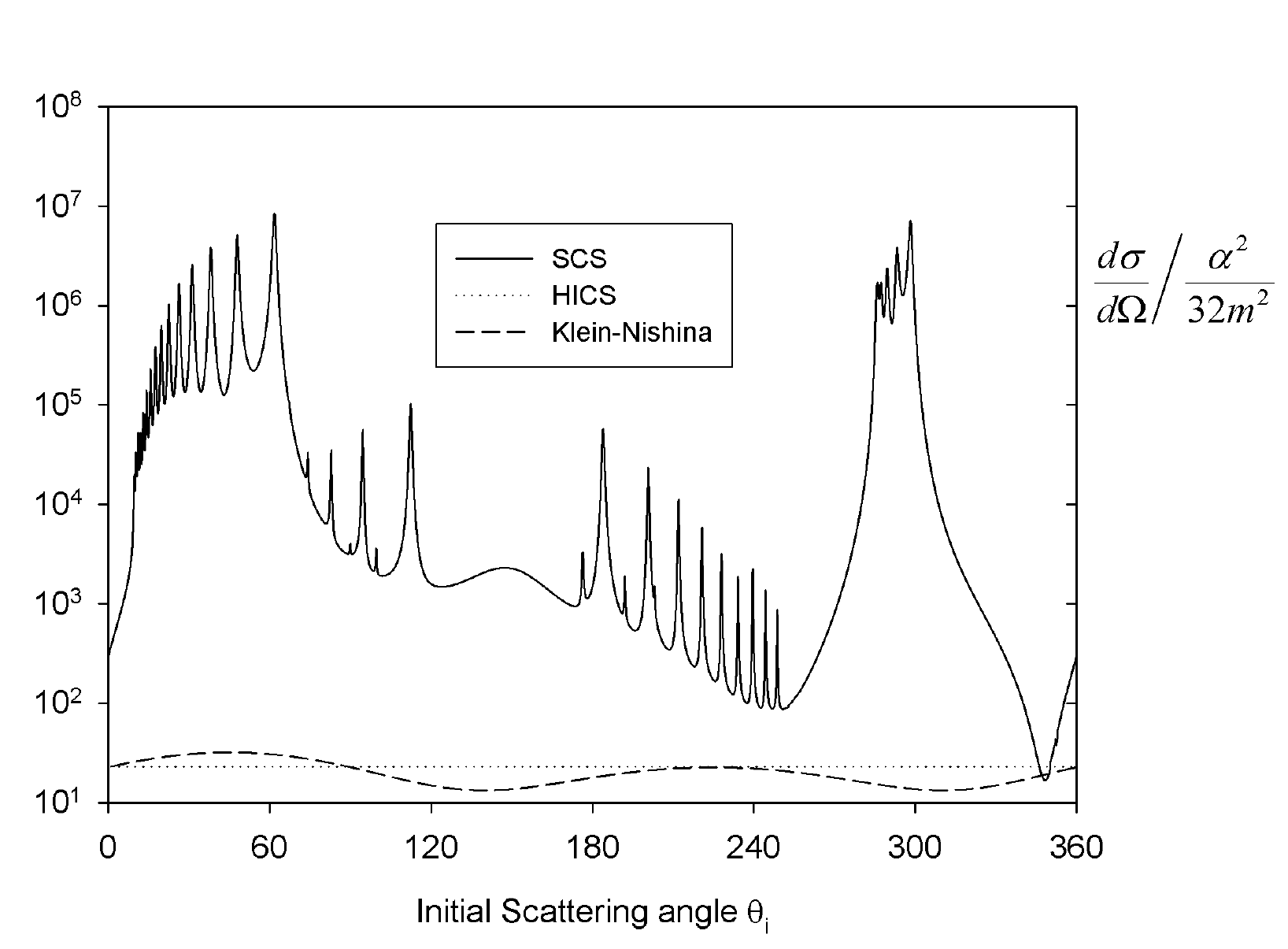}}
% \centerline{\psfig{figure=./compton/reson_thesis_final/vsti_nu1_w0.12_w1_0.1_tf45.pdf,height=8cm,width=15cm}}
\caption{\bf\bm Comparison of the Klein-Nishina, HICS and SCS differential cross section
resonances vs $\theta_f$ for $\omega=25.6$ keV, $\omega_i=51.2$ keV, $\theta_i=45^{\circ}$,
$\varphi_f=0^{\circ}$ and $\nu^2=1$.}
\label{resc6}
\end{figure}

\clearpage

\begin{figure}[H]
 \centerline{\includegraphics[height=8cm,width=15cm]{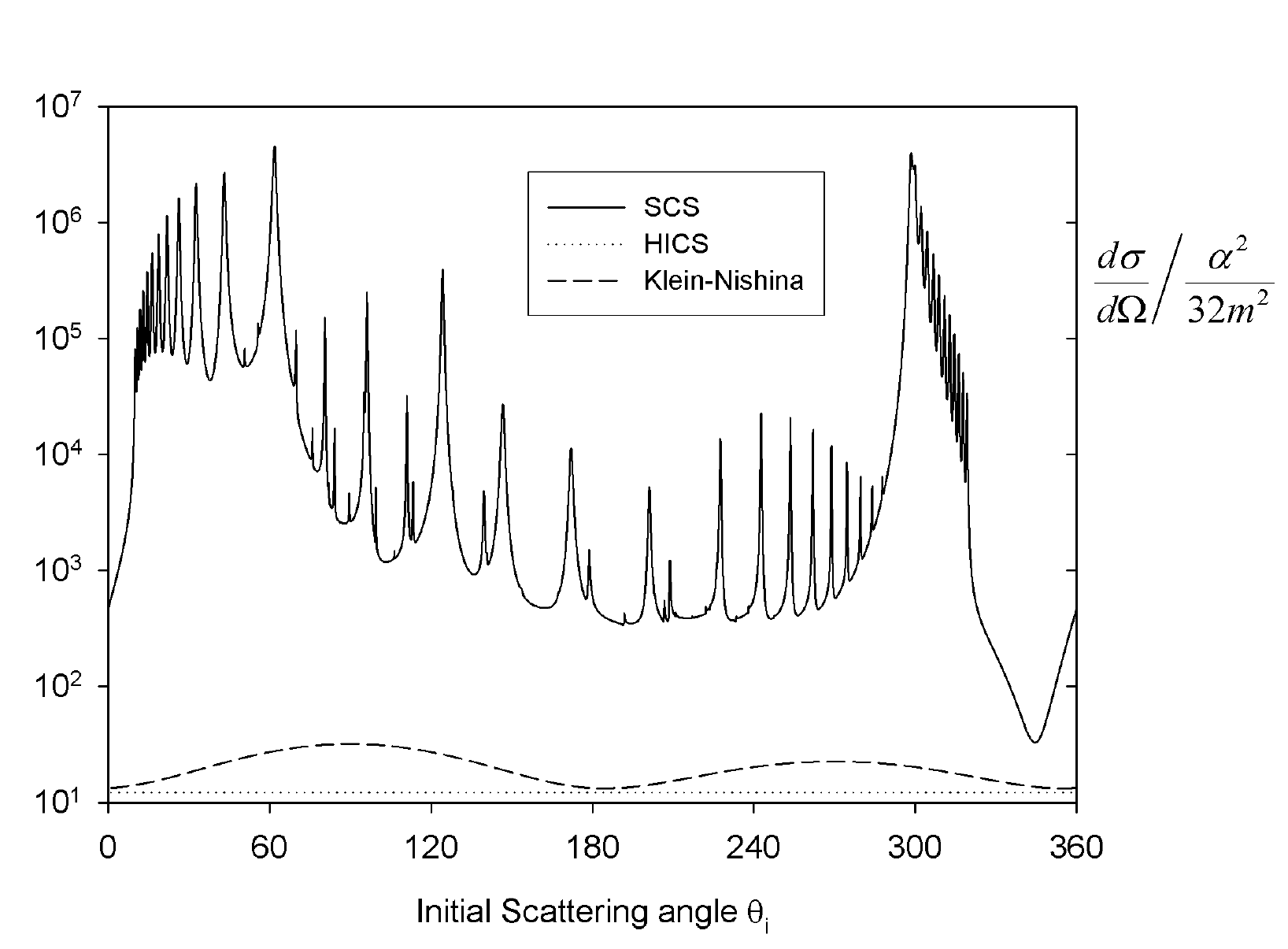}}
\caption{\bf\bm Comparison of the Klein-Nishina, HICS and SCS differential cross section
resonances vs $\theta_f$ for $\omega=25.6$ keV, $\omega_i=51.2$ keV, $\theta_i=90^{\circ}$,
$\varphi_f=0^{\circ}$ and $\nu^2=1$.}
\label{resc7}
\end{figure}

\begin{figure}[H]
 \centerline{\includegraphics[height=8cm,width=15cm]{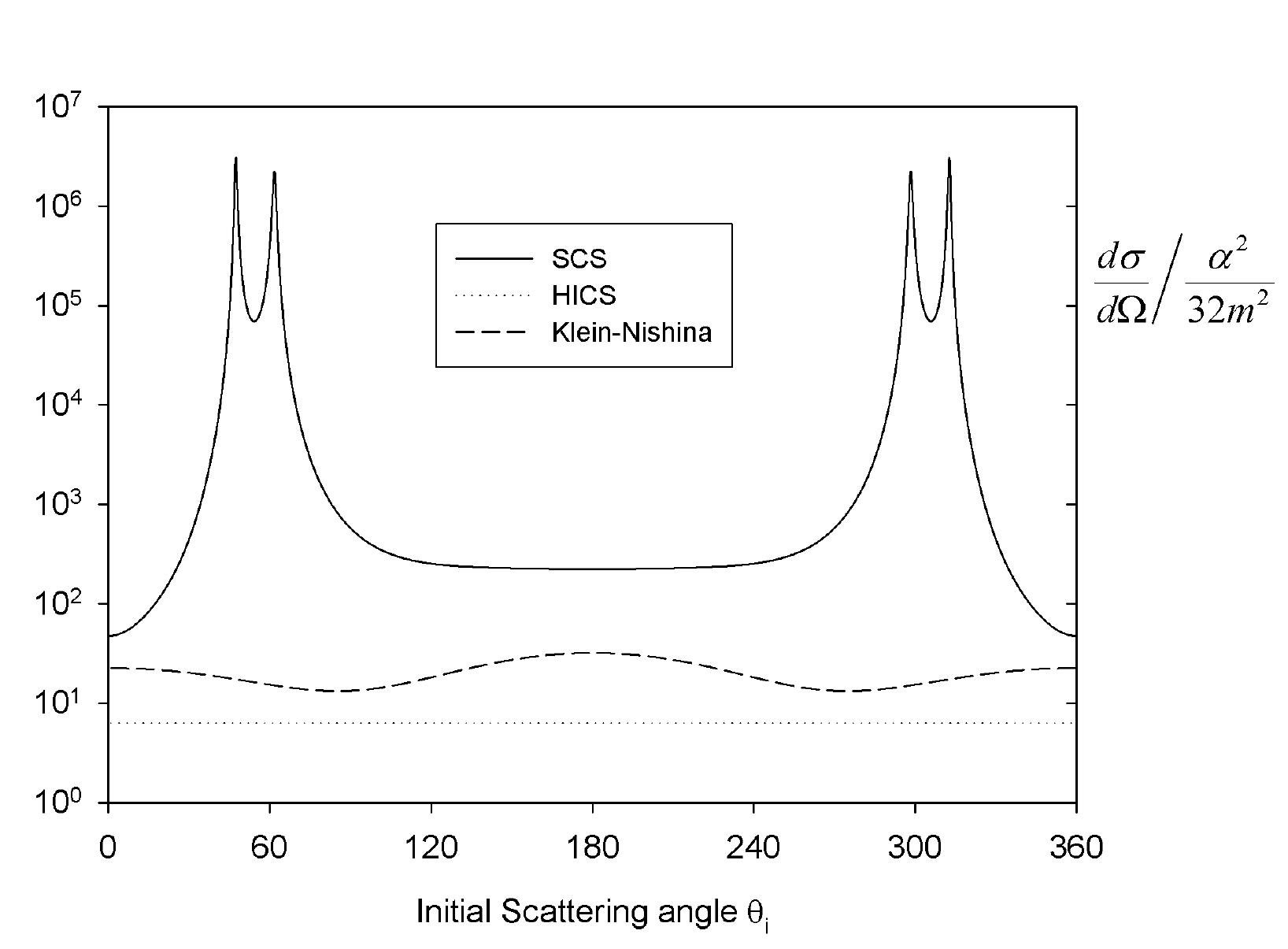}}
\caption{\bf\bm Comparison of the Klein-Nishina, HICS and SCS differential cross section
resonances vs $\theta_f$ for $\omega=25.6$ keV, $\omega_i=51.2$ keV, $\theta_i=180^{\circ}$,
$\varphi_f=0^{\circ}$ and $\nu^2=1$.} 
\label{resc8}
\end{figure}

\clearpage

\begin{figure}[H]
 \centerline{\includegraphics[height=8cm,width=15cm]{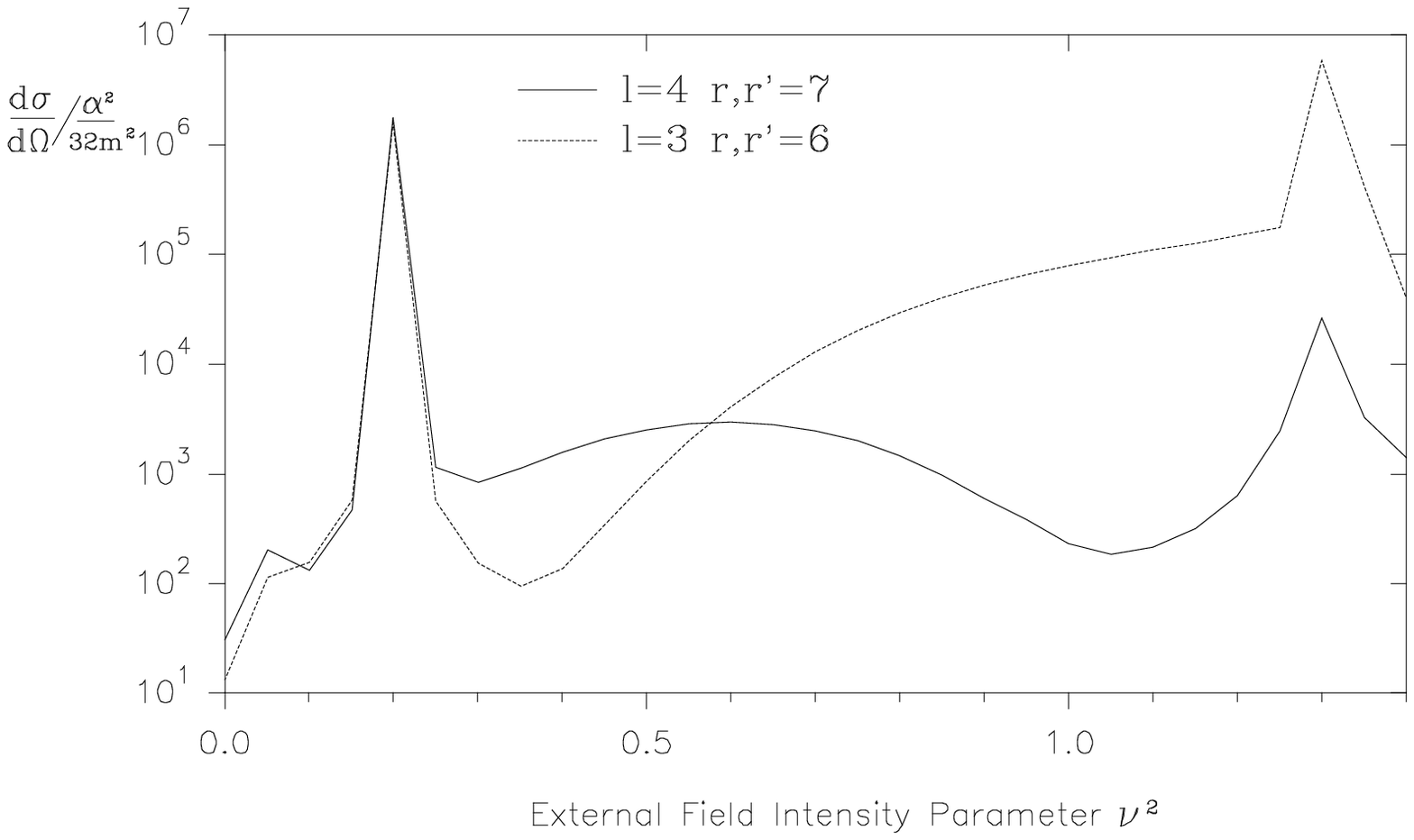}}
\caption{\bf\bm SCS differential cross section resonances vs $\nu^2$ for $\omega=25.6$ keV, 
$\omega_i=51.2$ keV, $\theta_i=90^{\circ}$,
$\varphi_f=0^{\circ}$ and various $l,r,r'$ terms.}
\label{resc9}
\end{figure}

\begin{figure}[H]
 \centerline{\includegraphics[height=8cm,width=15cm]{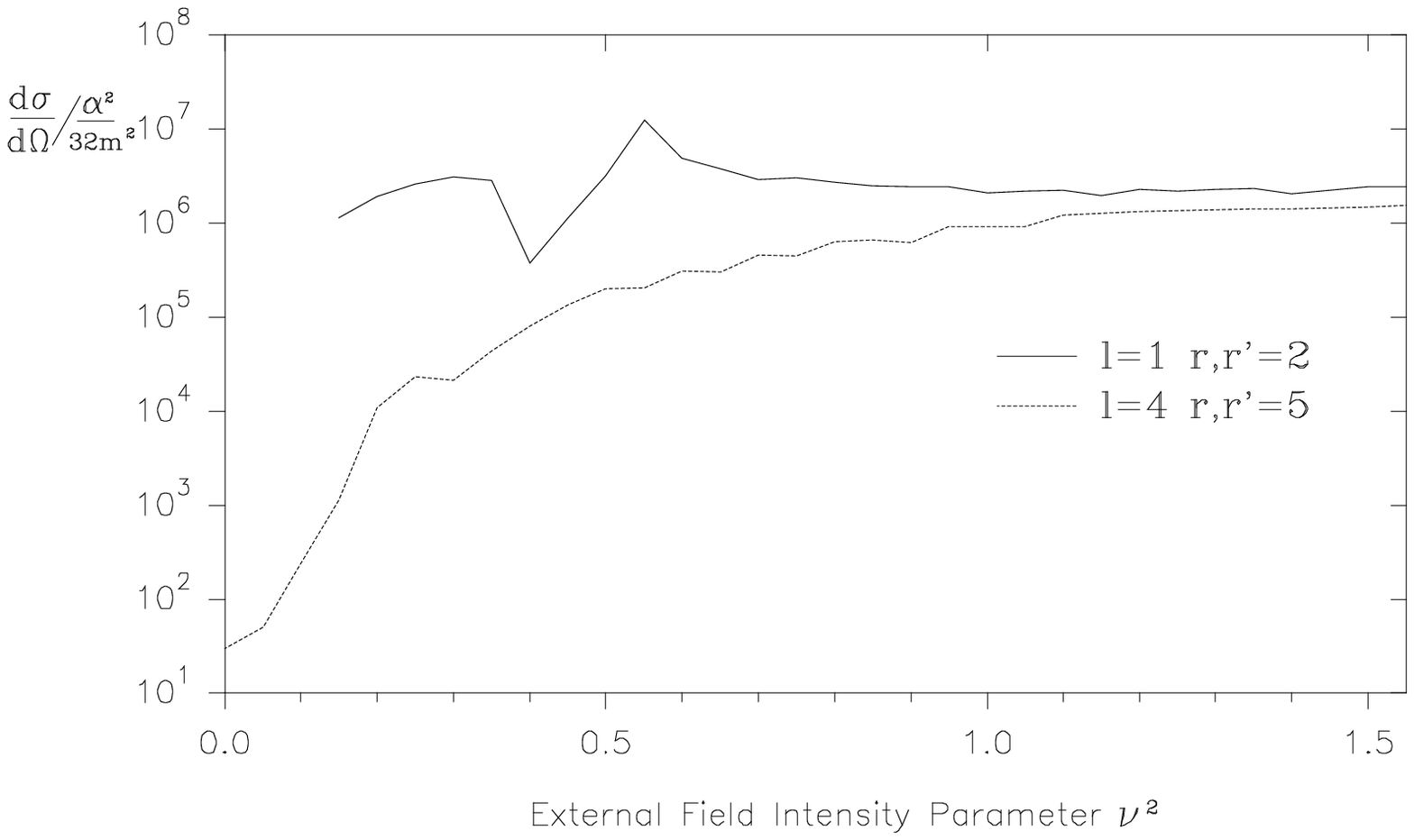}}
\caption{\bf\bm SCS differential cross section resonances vs $\nu^2$ for $\omega=61.4$ keV,
$\omega_i=51.2$ keV, $\theta_f=45^{\circ}$, $\varphi_f=0^{\circ}$ and various $l,r,r'$ terms.}
\label{resc10}
\end{figure}

\clearpage

\begin{figure}[H]
 \centerline{\includegraphics[height=8cm,width=15cm]{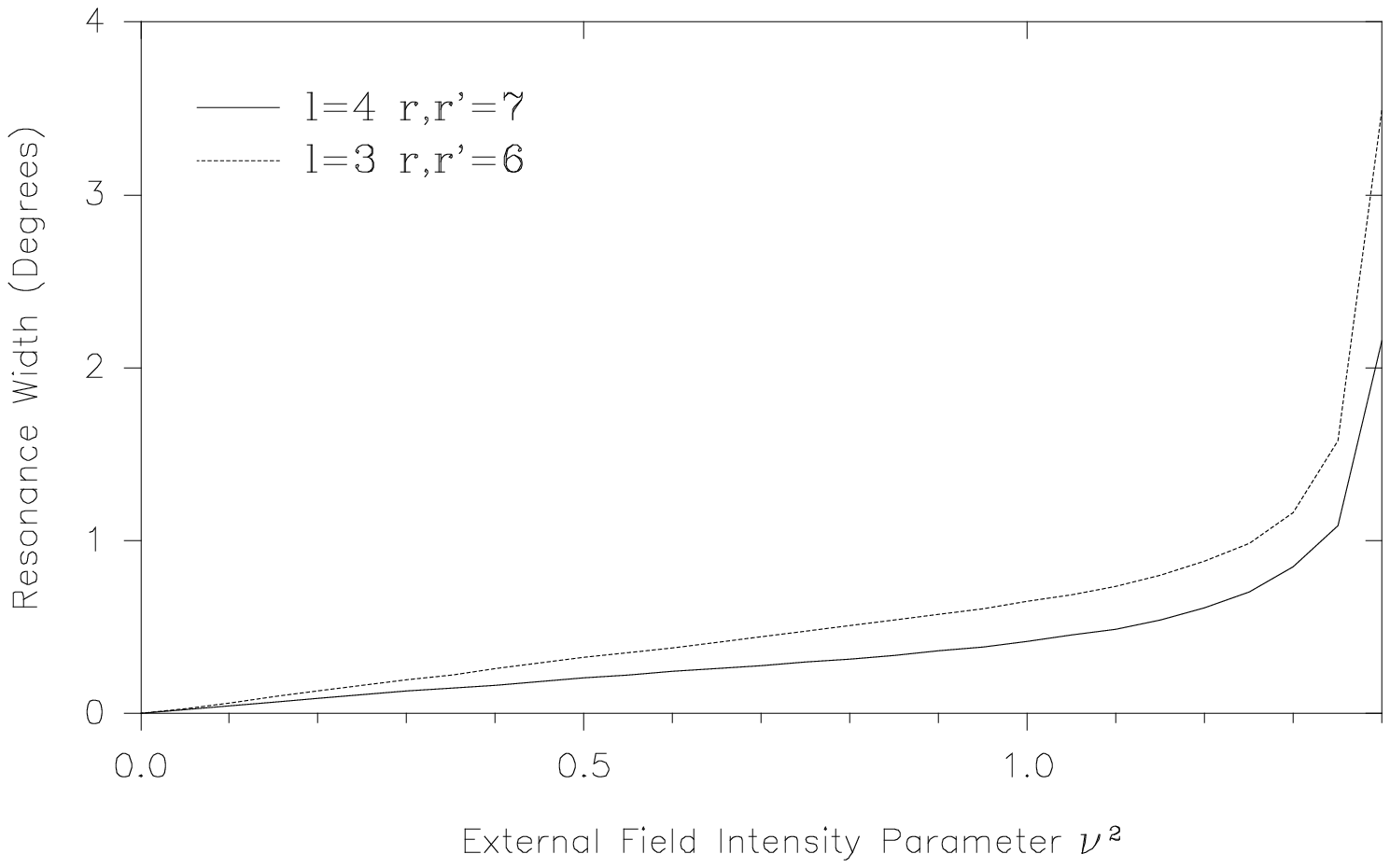}}
\caption{\bf\bm SCS differential cross section resonances vs $\nu^2$ for $\omega=25.6$ keV,
$\omega_i=51.2$ keV, $\theta_f=45^{\circ}$, $\varphi_f=0^{\circ}$ and various $l,r,r'$ terms.}
\label{resc11}
\end{figure}

\begin{figure}[H]
 \centerline{\includegraphics[height=8cm,width=15cm]{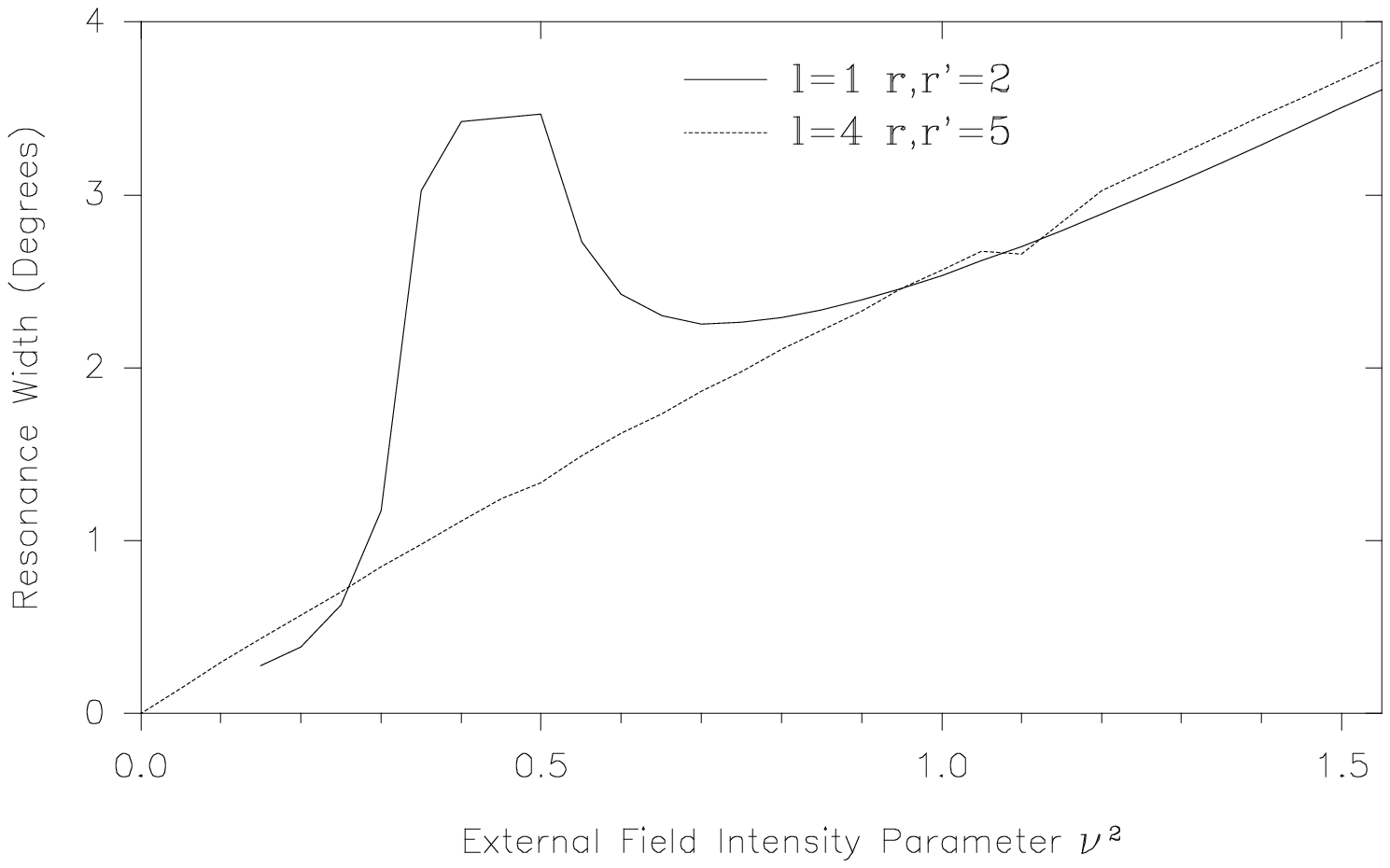}}
\caption{\bf\bm SCS differential cross section resonances vs $\nu^2$ for $\omega=61.4$ keV,
$\omega_i=51.2$ keV, $\theta_i=90^{\circ}$, $\varphi_f=0^{\circ}$ and various $l,r,r'$ terms.}
\label{resc12}
\end{figure}

\clearpage

\begin{figure}[H]
 \centerline{\includegraphics[height=8cm,width=15cm]{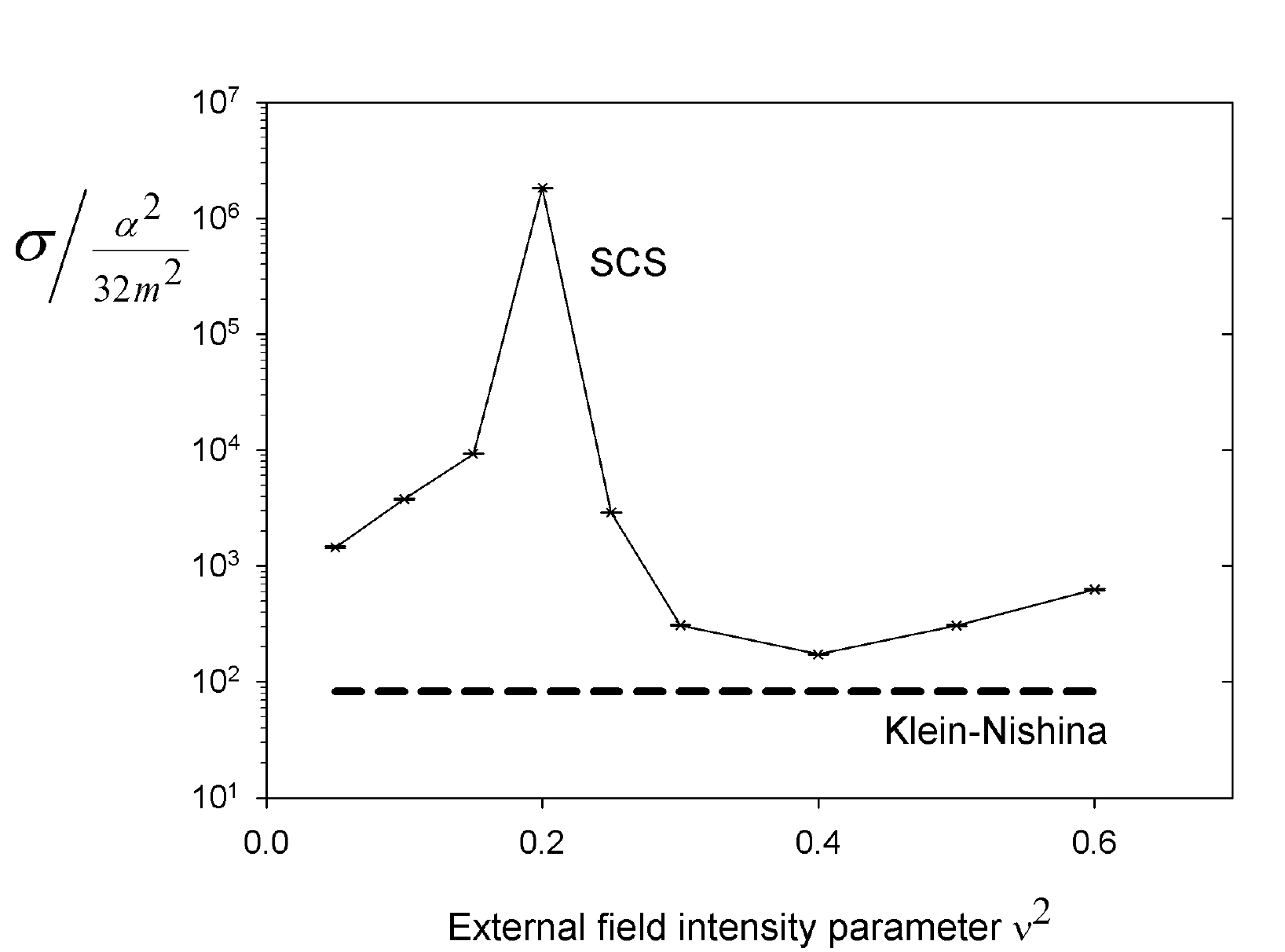}}
\caption{\bf\bm Comparison of the full Klein-Nishina and SCS cross sections vs $\nu^2$ for 
$\omega=1.28$ MeV,$\omega_1,omega_2=0.92$ MeV, $\theta_f,=45^{\circ}$ and $\varphi_f=0^{\circ}$.}
\label{rescfull}
\end{figure}

\clearpage

%% file: tex/tables/c7_tab2.tex
\begin{table}[h]
\label{c7tab2}
\center{
\begin{tabular}{|c|c|c|c|c|c|c|c|c|} \hline
$\bm l$&$\bm r,r'$&$\bm \nu^{\bm 2}$ & $\bm  \omega (\text{MeV})$ & $\bm
\omega_{1,2},(\text{MeV}) $ &$ 
\bm \theta_{1}$ & 
$\bm (\theta_{f},\phi_{f})$ &\bf figure(s) \\ \hline
$ \text{all} $&$ \text{all} $&$ 0.1 $&$ 0.409,1.28 $&$ 0.768 $&
$ 45^{\circ } $&$ (0^{\circ}\rightarrow 360^{\circ },0^{\circ }) $& 
\ref{resp1},\ref{resp2} \\ \hline
$ \text{all} $&$ \text{all} $&$ 0.5 $&$ 0.409,1.28 $&$ 0.768 $&
$ 45^{\circ } $&$ (0^{\circ}\rightarrow 360^{\circ },0^{\circ }) $& 
\ref{resp3},\ref{resp4} \\ \hline
$ \text{all} $&$ \text{all} $&$ 0.5 $&$ 1.024 $&$ 0.768 $&
$ 0^{\circ}\rightarrow 360^{\circ } $&$ (0^{\circ },0^{\circ }) $& 
\ref{resp5} \\ \hline
$ \text{all} $&$ \text{all} $&$ 0.5 $&$ 1.024 $&$ 0.768 $&
$ 0^{\circ}\rightarrow 360^{\circ } $&$ (45^{\circ },0^{\circ }) $& 
\ref{resp6} \\ \hline
$ \text{all} $&$ \text{all} $&$ 0.5 $&$ 1.024 $&$ 0.768 $&
$ 0^{\circ}\rightarrow 360^{\circ } $&$ (90^{\circ },0^{\circ }) $& 
\ref{resp7} \\ \hline
$ \text{all} $&$ \text{all} $&$ 0.5 $&$ 1.024 $&$ 0.768 $&
$ 0^{\circ}\rightarrow 360^{\circ } $&$ (180^{\circ },0^{\circ }) $& 
\ref{resp8} \\ \hline
$ 2 $&$ 3,-1 $&$  $&$ 0.768 $&$ 0.409 $&
$ 45^{\circ} $&$  $& \ref{resp9},\ref{resp11} \\ \hline
$ 1 $&$ -1,2 $&$  $&$ 1.024 $&$ 0.768 $&
$  $&$ 45^{\circ} $& \ref{resp10},\ref{resp12} \\ \hline
$ \text{all} $&$ \text{all} $&$ 0.05\rightarrow 1.0 $&$ 1.28 $&$ 0.92 $&
$ 45^{\circ} $&$  $& \ref{respfull} \\ \hline
\end{tabular} }
\caption{\bf\bm The parameter range for which the STPPP cross section resonances are
investigated.}
\end{table}

%% file: tex/c7ppfig.tex
\begin{figure}[H]
 \centerline{\includegraphics[height=7.5cm,width=15cm]{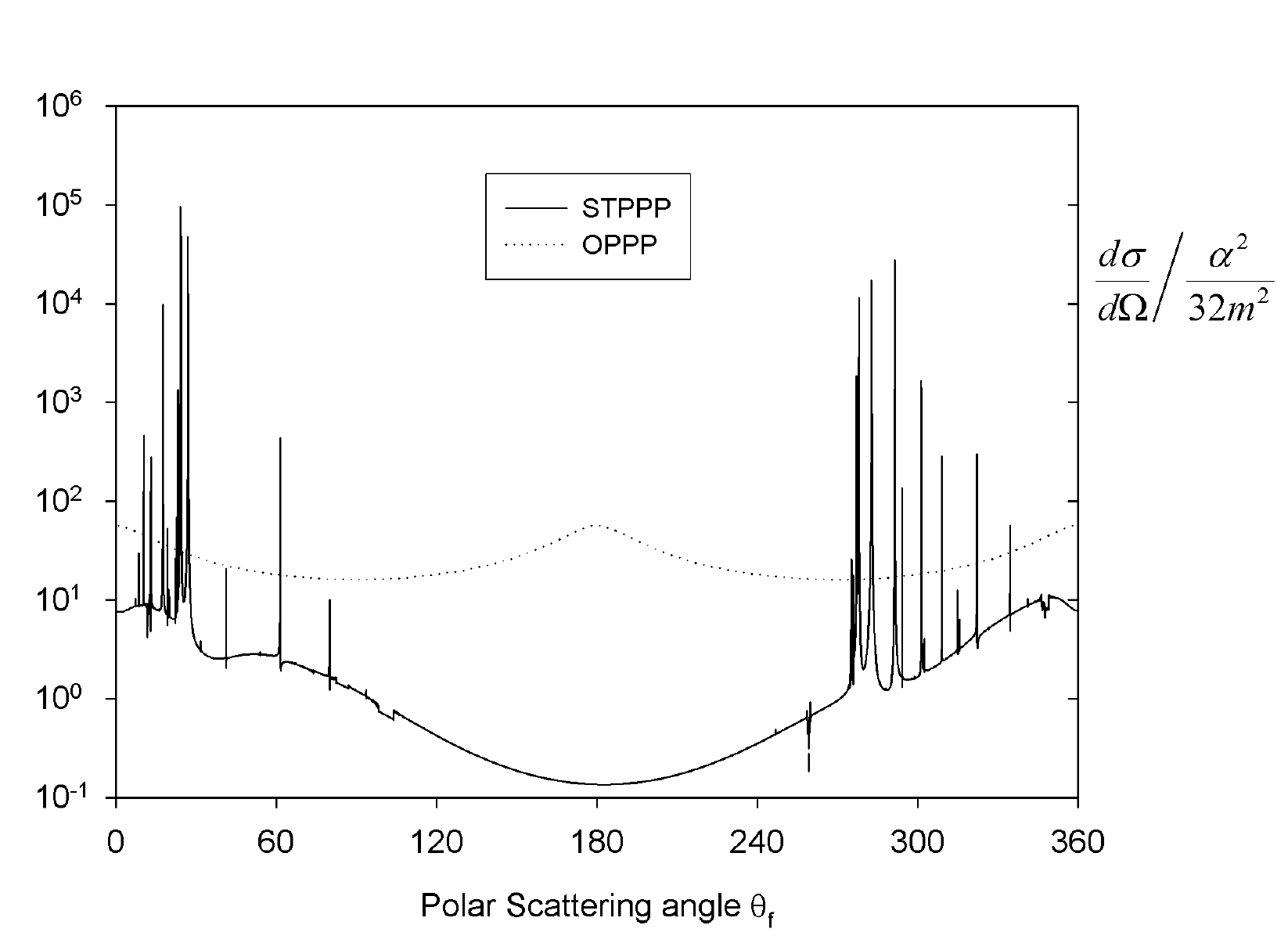}}
\caption{\bf\bm Comparison of the OPPP and STPPP differential cross section
resonances vs $\theta_f$ for $\omega=0.409$ MeV, $\omega_1,\omega_2=0.768$ MeV, $\theta_1=45^{\circ}$,
$\varphi_f=0^{\circ}$ and $\nu^2=0.1$.}
\label{resp1}
\end{figure}

\begin{figure}[H]
 \centerline{\includegraphics[height=7.5cm,width=15cm]{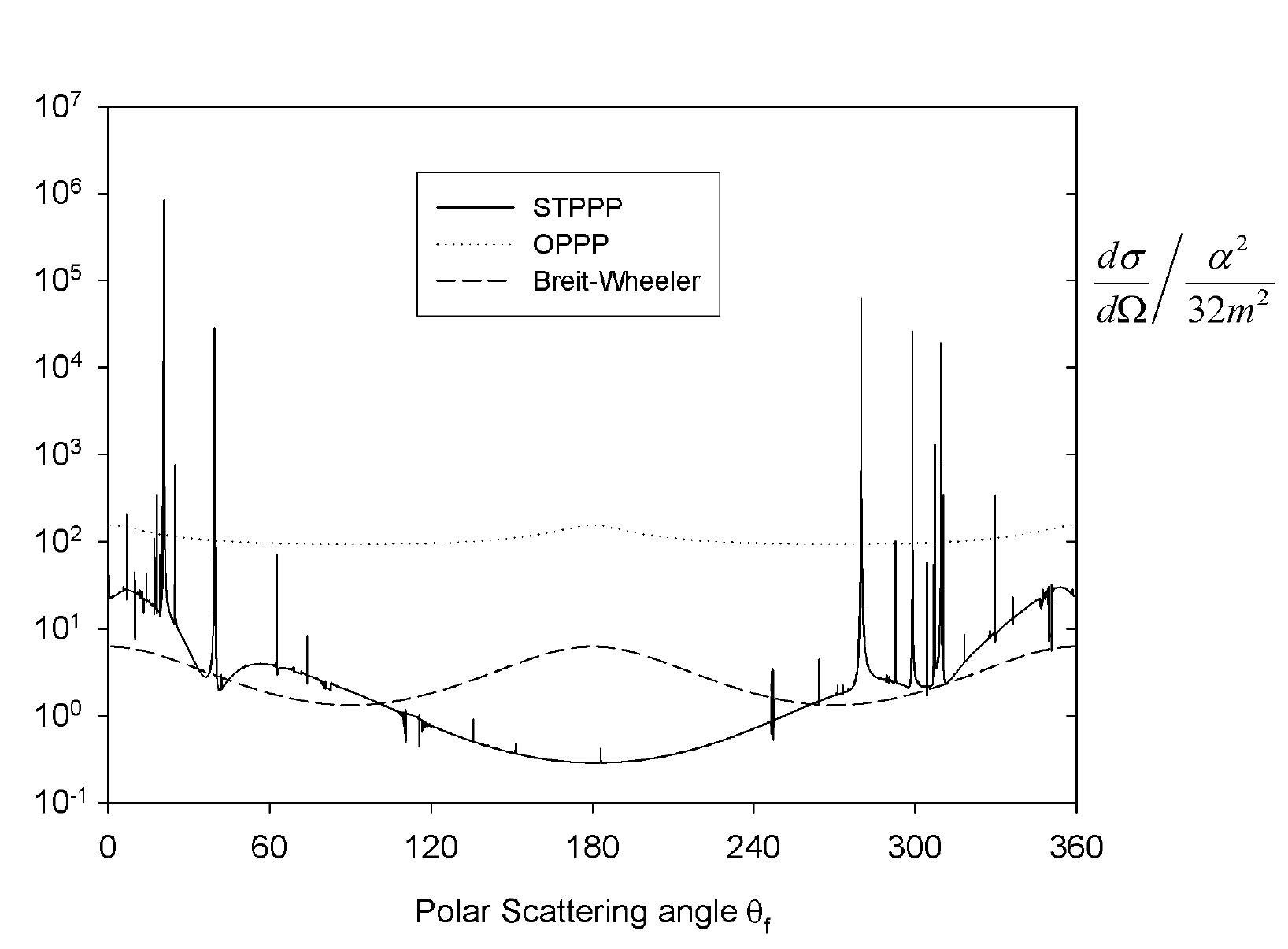}}
\caption{\bf\bm Comparison of the Breit-Wheeler, OPPP and STPPP differential cross section
resonances vs $\theta_f$ for $\omega=0.409$ MeV, $\omega_1,\omega_2=1.28$ MeV, $\theta_1,=45^{\circ}$,
$\varphi_f=0^{\circ}$ and $\nu^2=0.1$.}
\label{resp2}
\end{figure}

\clearpage

\begin{figure}[H]
 \centerline{\includegraphics[height=8cm,width=15cm]{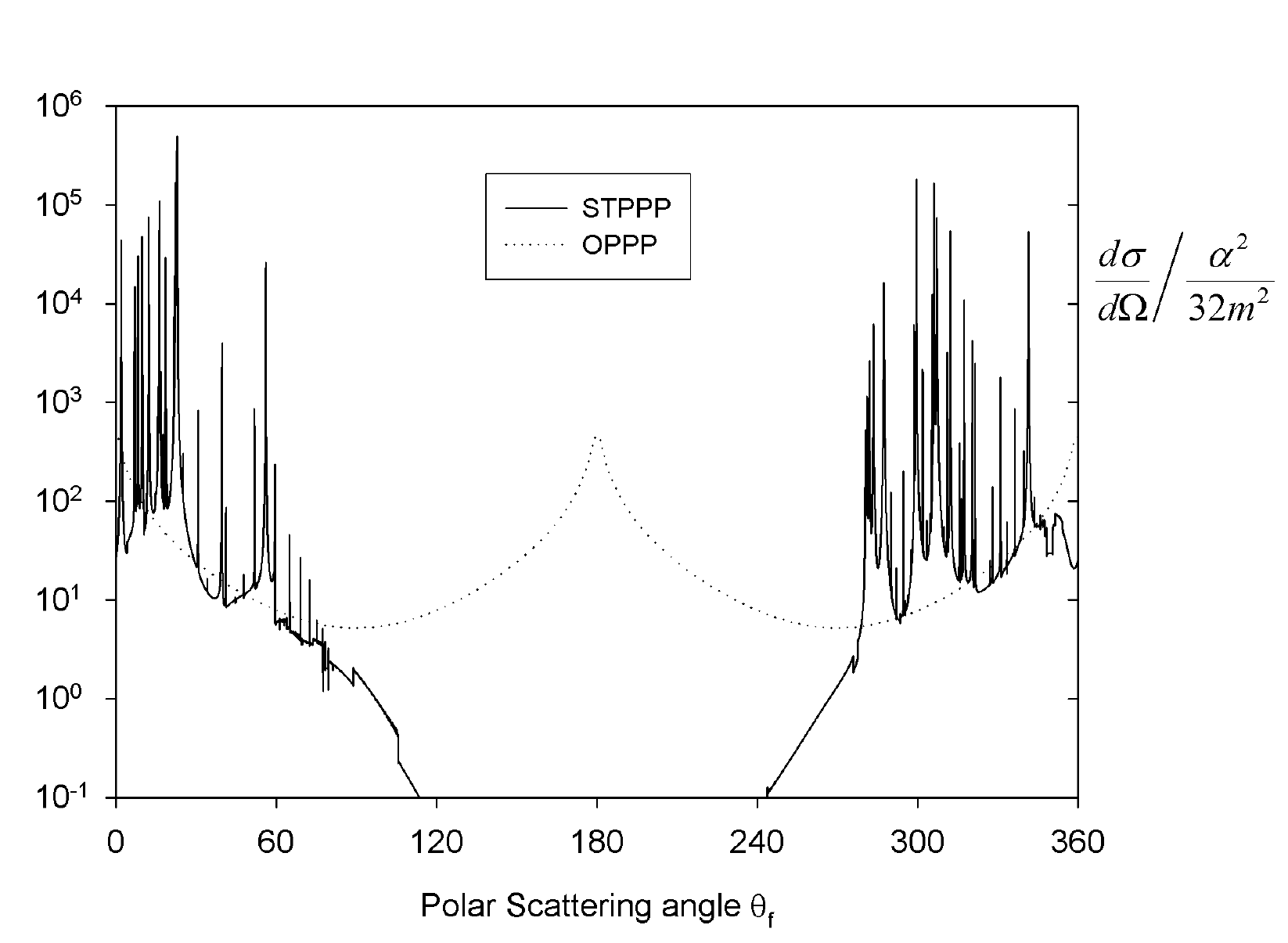}}
\caption{\bf\bm Comparison of the OPPP and STPPP differential cross section
resonances vs $\theta_f$ for $\omega=0.409$ MeV, $\omega_1,\omega_2=0.768$ MeV, $\theta_1,=45^{\circ}$,
$\varphi_f=0^{\circ}$ and $\nu^2=0.5$.}
\label{resp3}
\end{figure}

\begin{figure}[H]
 \centerline{\includegraphics[height=8cm,width=15cm]{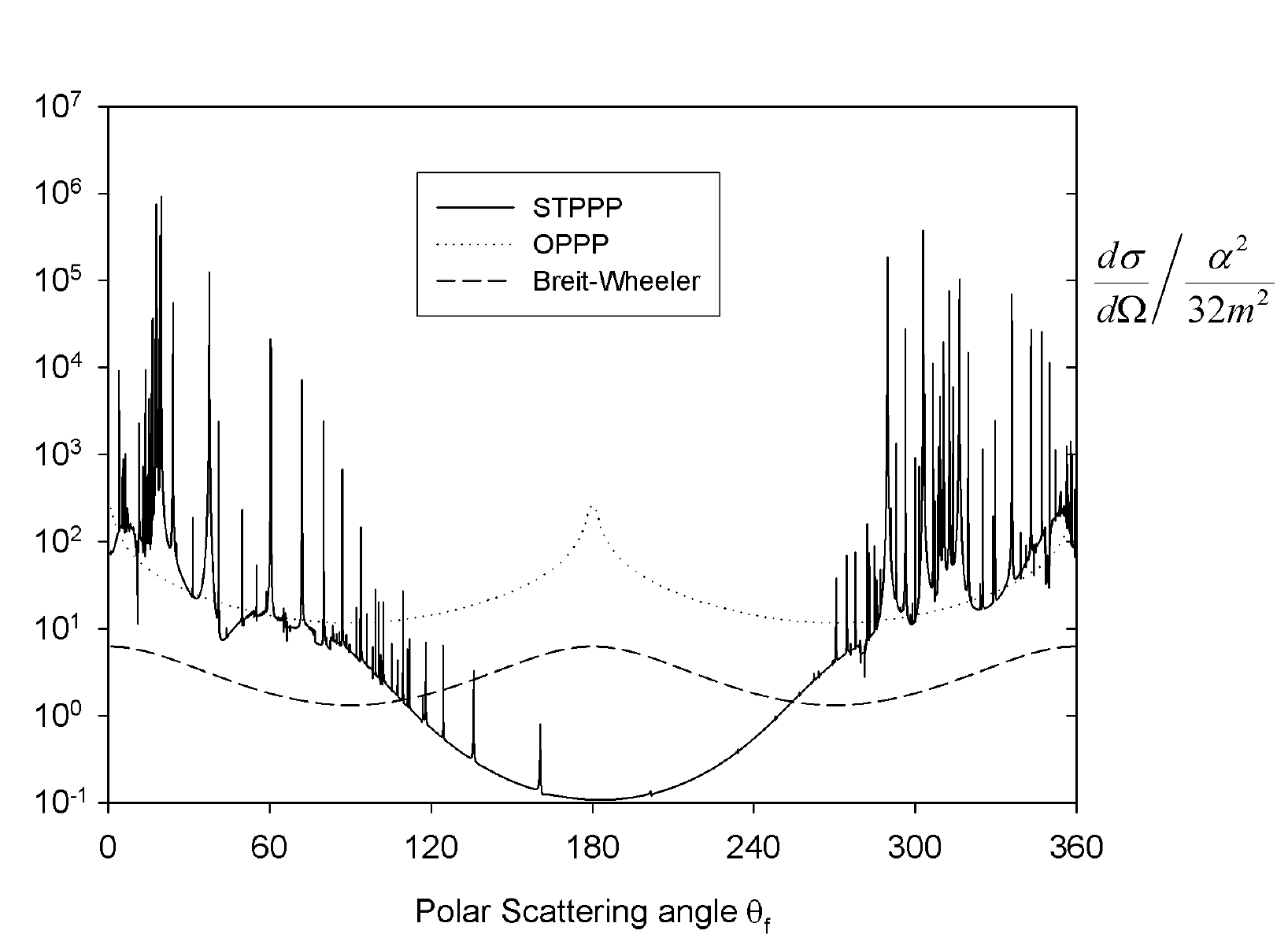}}
\caption{\bf\bm Comparison of the Breit-Wheeler, OPPP and STPPP differential cross section
resonances vs $\theta_f$ for $\omega=0.409$ MeV, $\omega_1,\omega_2=1.28$ MeV, $\theta_1,=45^{\circ}$,
$\varphi_f=0^{\circ}$ and $\nu^2=0.5$.}
\label{resp4}
\end{figure}

\clearpage

\begin{figure}[H]
 \centerline{\includegraphics[height=8cm,width=15cm]{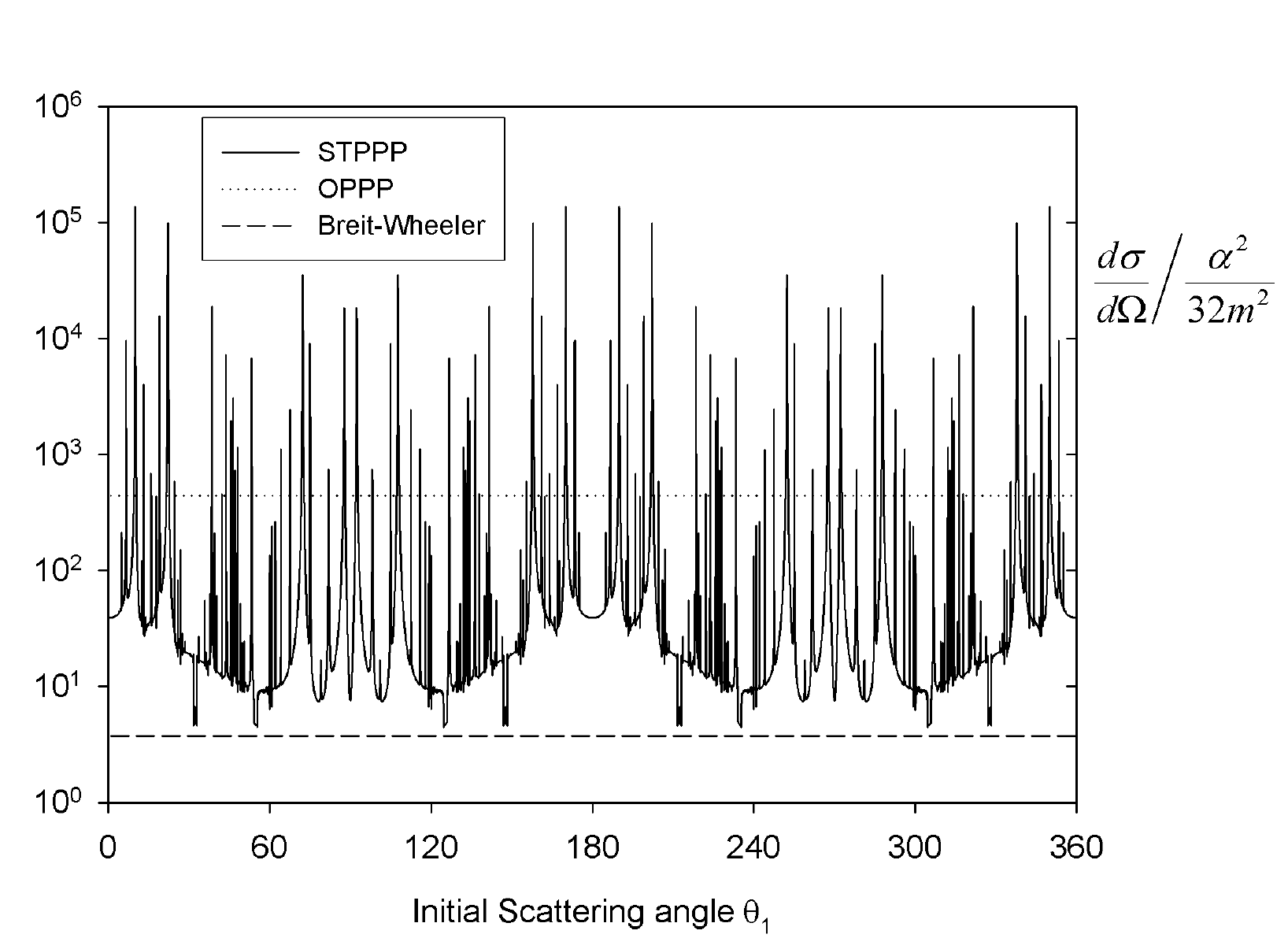}}
\caption{\bf\bm Comparison of the Breit-Wheeler, OPPP and STPPP differential cross section
resonances vs $\theta_1$ for $\omega=1.024$ MeV, $\omega_1,\omega_2=0.768$ MeV, $\theta_f,=0^{\circ}$,
$\varphi_f=0^{\circ}$ and $\nu^2=0.5$.}
\label{resp5}
\end{figure}

\begin{figure}[H]
 \centerline{\includegraphics[height=8cm,width=15cm]{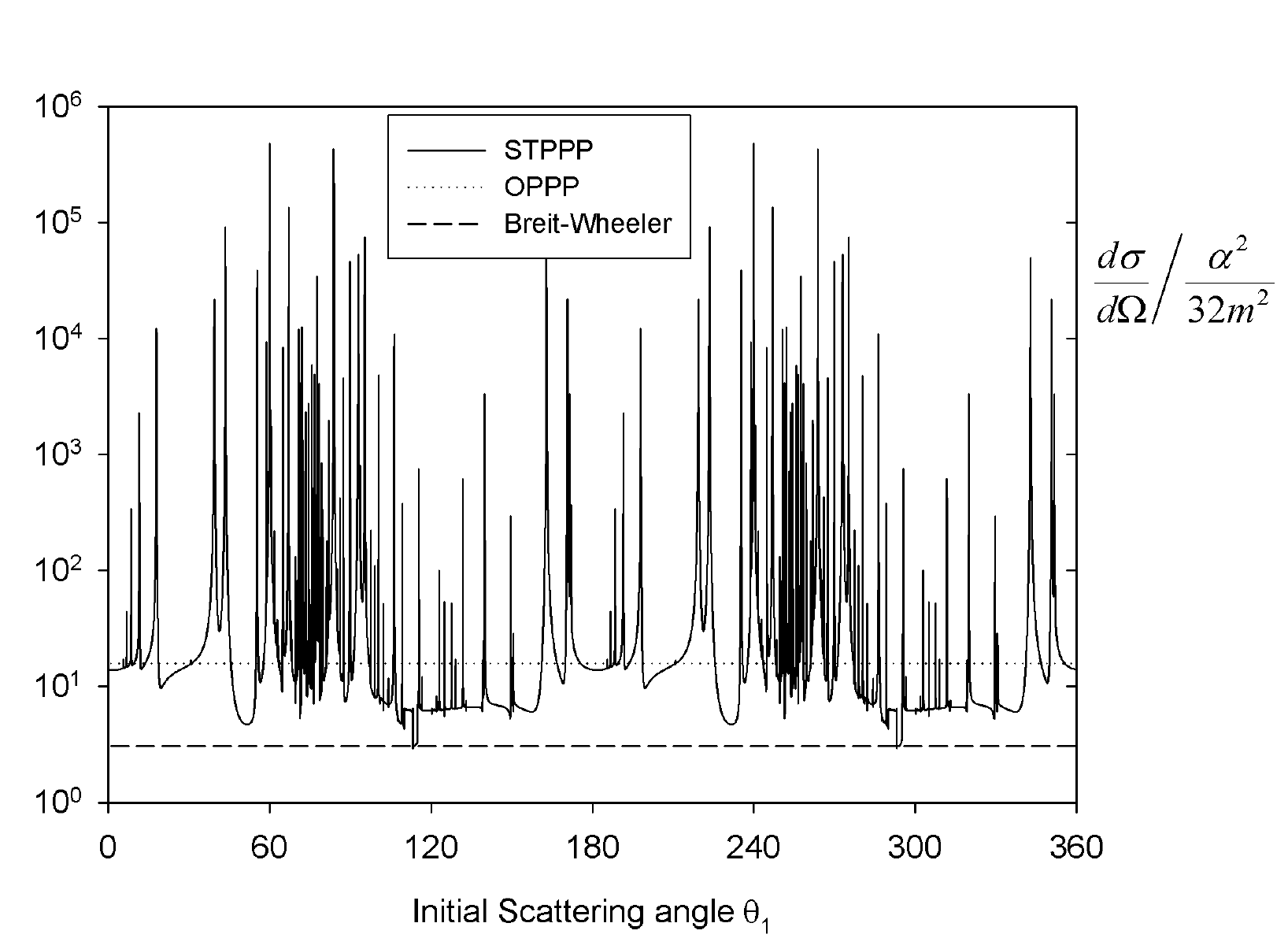}}
\caption{\bf\bm Comparison of the Breit-Wheeler, OPPP and STPPP differential cross section
resonances vs $\theta_1$ for $\omega=1.024$ MeV, $\omega_1,\omega_2=0.768$ MeV, $\theta_f,=45^{\circ}$,
$\varphi_f=0^{\circ}$ and $\nu^2=0.5$.}
\label{resp6}
\end{figure}

\clearpage

\begin{figure}[H]
 \centerline{\includegraphics[height=8cm,width=15cm]{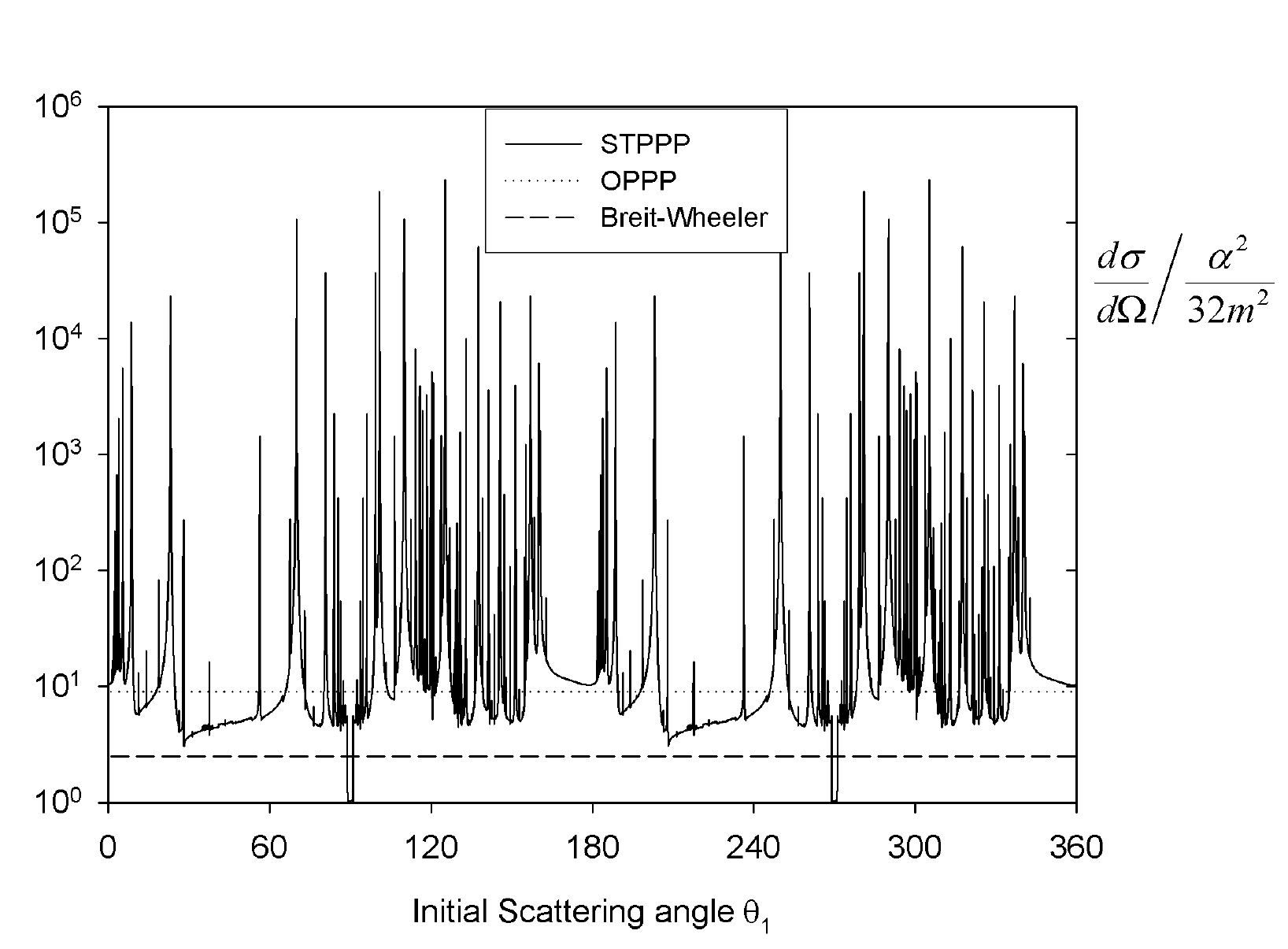}}
\caption{\bf\bm Comparison of the Breit-Wheeler, OPPP and STPPP differential cross section
resonances vs $\theta_1$ for $\omega=1.024$ MeV, $\omega_1,\omega_2=0.768$ MeV, $\theta_f,=90^{\circ}$,
$\varphi_f=0^{\circ}$ and $\nu^2=0.5$.}
\label{resp7}
\end{figure}

\begin{figure}[H]
 \centerline{\includegraphics[height=8cm,width=15cm]{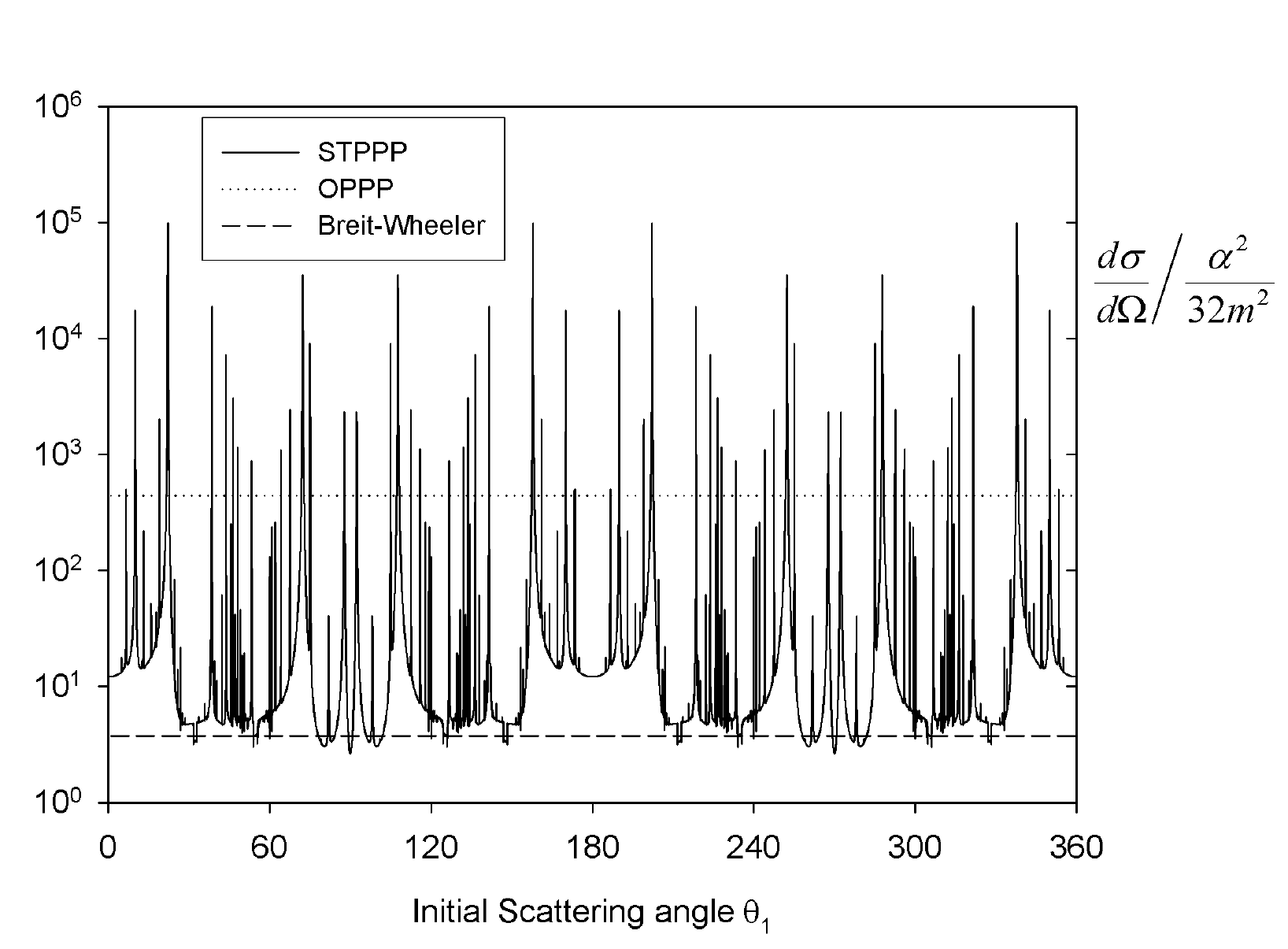}}
\caption{\bf\bm Comparison of the Breit-Wheeler, OPPP and STPPP differential cross section
resonances vs $\theta_1$ for $\omega=1.024$ MeV, $\omega_1,\omega_2=0.768$ MeV, 
$\theta_f,=180^{\circ}$,$\varphi_f=0^{\circ}$ and $\nu^2=0.5$.}
\label{resp8}
\end{figure}

\clearpage

\begin{figure}[H]
 \centerline{\includegraphics[height=8cm,width=15cm]{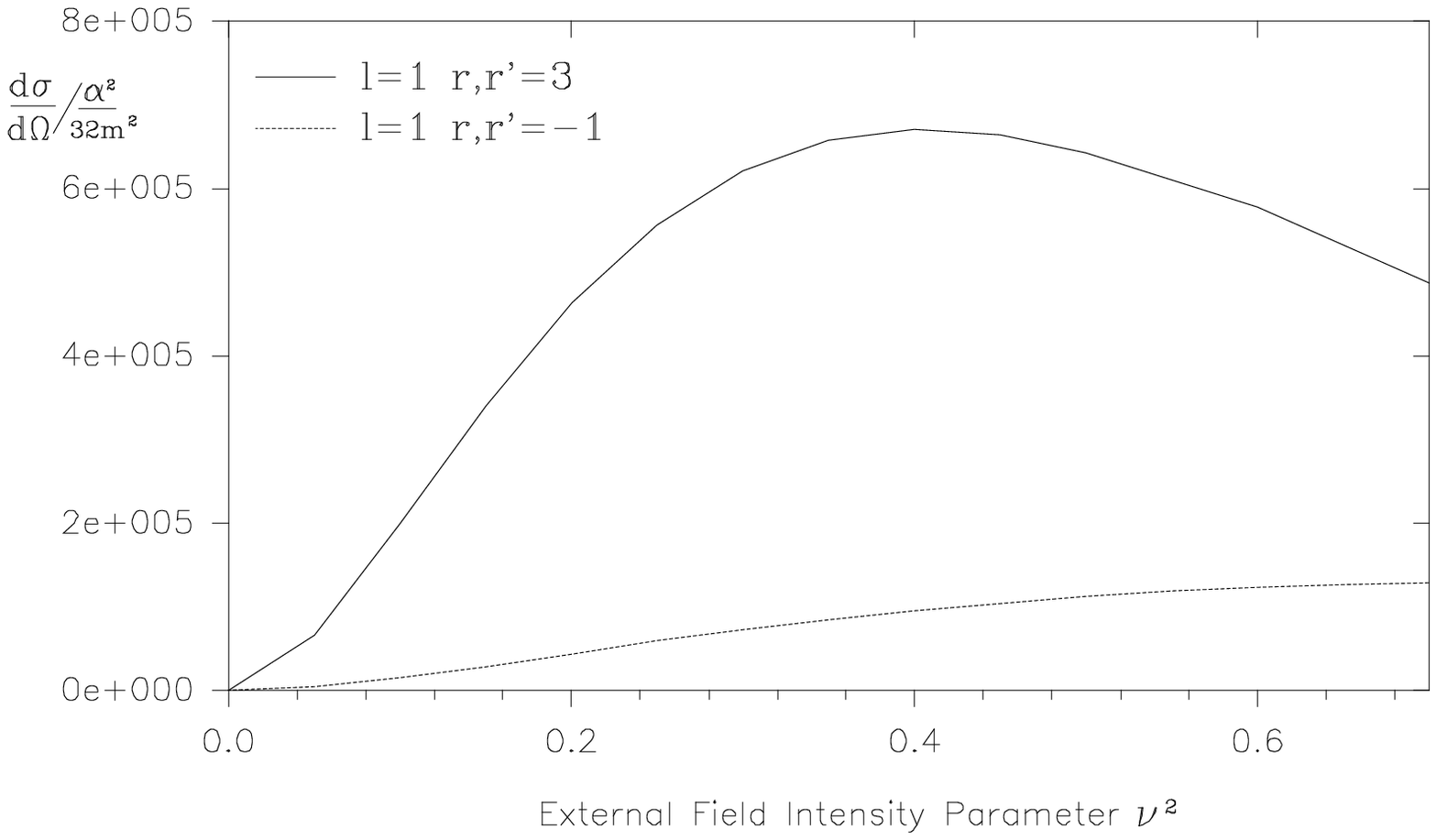}}
\caption{\bf\bm The STPPP differential cross section resonances vs $\nu^2$ for $\omega=0.768$ MeV, 
$\omega_1,\omega_2=0.409$ MeV, $\theta_1,=45^{\circ}$,
$\varphi_f=0^{\circ}$ and various $l,r,r'$ terms.}
\label{resp9}
\end{figure}

\begin{figure}[H]
 \centerline{\includegraphics[height=8cm,width=15cm]{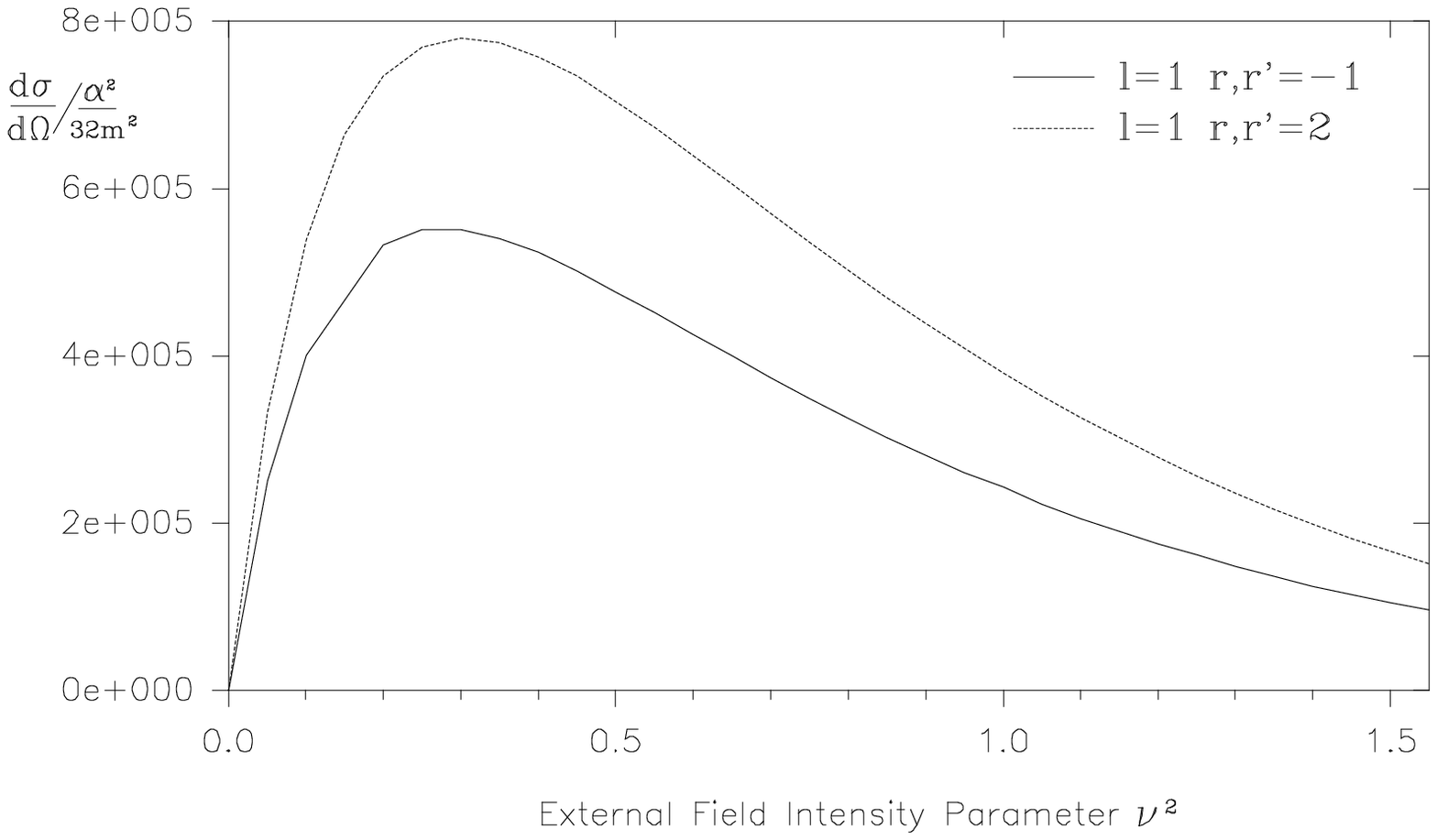}}
\caption{\bf\bm The STPPP differential cross section resonances vs $\nu^2$ for $\omega=1.024$ MeV,
$\omega_1,\omega_2=0.768$ MeV, $\theta_f,=45^{\circ}$,
$\varphi_f=0^{\circ}$ and various $l,r,r'$ terms.}
\label{resp10}
\end{figure}

\clearpage

\begin{figure}[H]
 \centerline{\includegraphics[height=8cm,width=15cm]{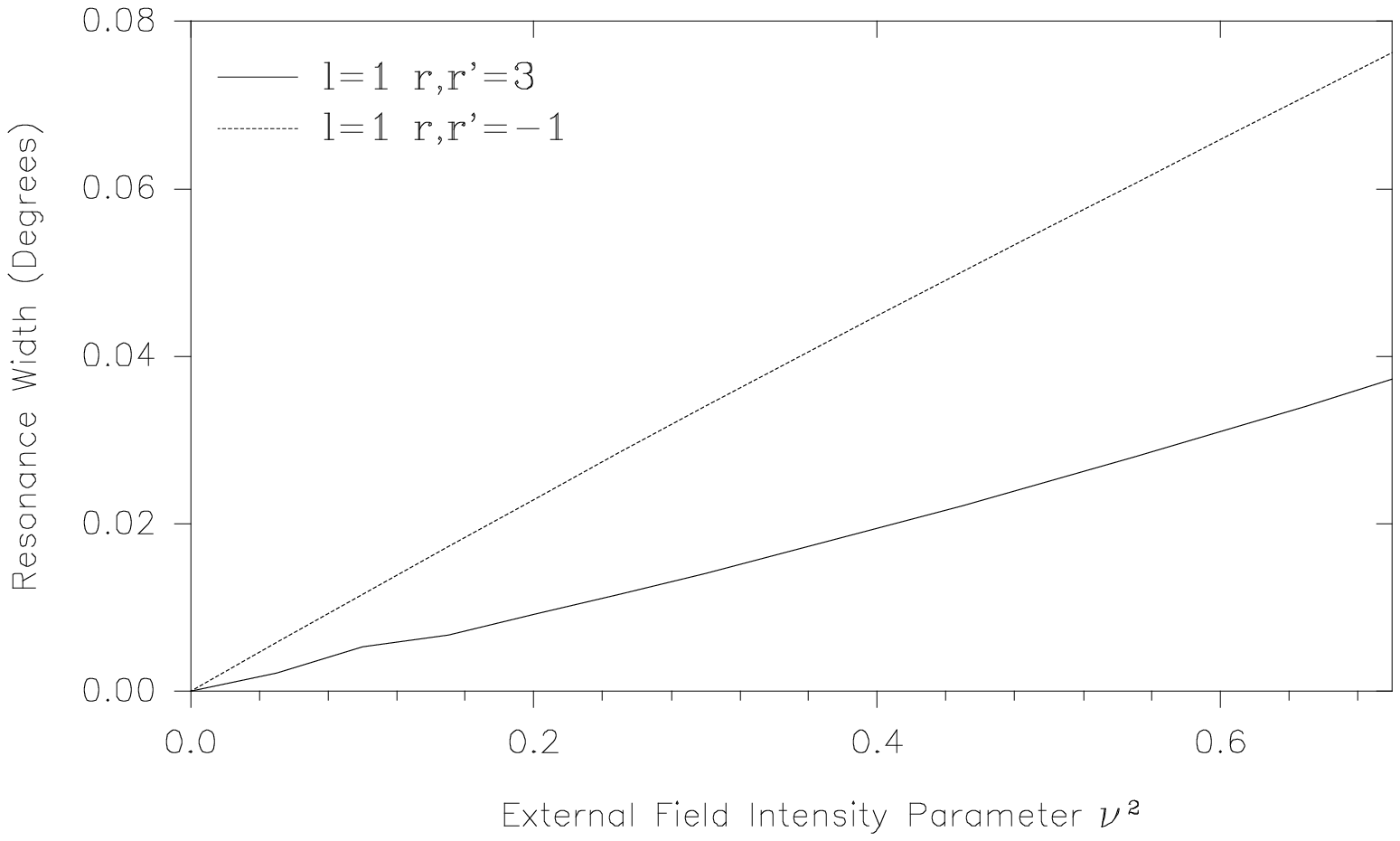}}
\caption{\bf\bm The STPPP differential cross section resonances vs $\nu^2$ for $\omega=0.768$ MeV,
$\omega_1,\omega_2=0.409$ MeV, $\theta_1,=45^{\circ}$,
$\varphi_f=0^{\circ}$ and various $l,r,r'$ terms.}
\label{resp11}
\end{figure}

\begin{figure}[H]
 \centerline{\includegraphics[height=8cm,width=15cm]{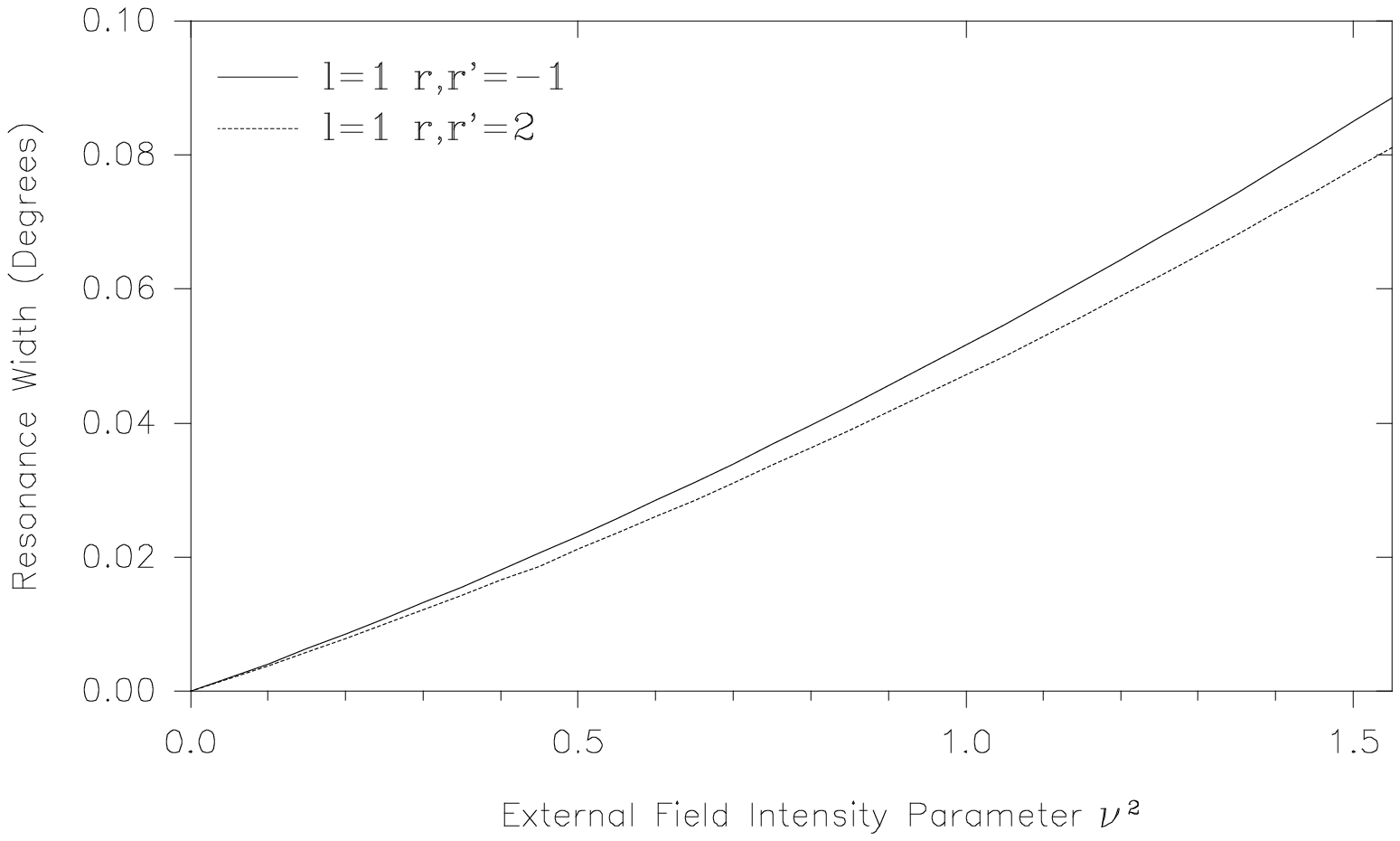}}
\caption{\bf\bm The STPPP differential cross section resonances vs $\nu^2$ for $\omega=1.024$ MeV,
$\omega_1,\omega_2=0.768$ MeV, $\theta_f,=45^{\circ}$,
$\varphi_f=0^{\circ}$ and various $l,r,r'$ terms.}
\label{resp12}
\end{figure}

\clearpage

\begin{figure}[H]
 \centerline{\includegraphics[height=8cm,width=15cm]{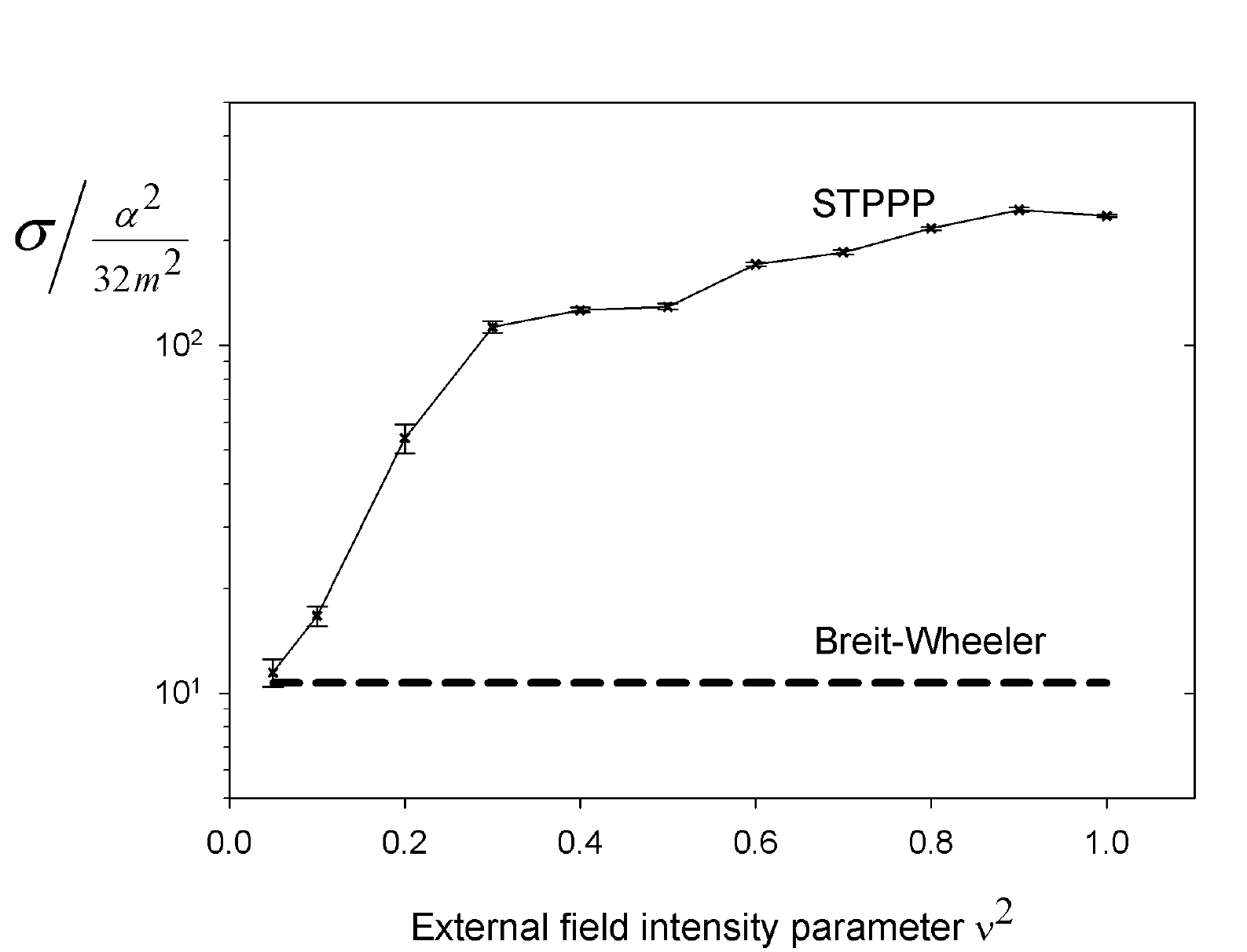}}
\caption{\bf\bm Comparison of the full Breit-Wheeler and STPPP cross sections vs $\nu^2$ for 
$\omega=1.28$ MeV, $\omega_1,\omega_2=0.92$ MeV, $\theta_f,=45^{\circ}$ and $\varphi_f=0^{\circ}$.}
\label{respfull}
\end{figure}

\clearpage

%% file: tex/tables/c7_tab3.tex
\begin{table}[!b]
\label{c7tab3}
\center{
\begin{tabular}{|c|c|c|} \hline
Figure & $ \frac{\omega_i}{\omega},\frac{\omega_1}{\omega}$ & number of resonances \\ 
\hline\hline
\ref{resc3} & 2 & 8 \\ \hline
\ref{resc4} & 0.83 & 4 \\ \hline\hline
\ref{resp1} & 1.67 & 169 \\ \hline
\ref{resp2} & 0.53 & 58 \\ \hline\hline
\ref{resp3} & 1.67 & 186 \\ \hline
\ref{resp4} & 0.53 & 86 \\ \hline
\end{tabular} }
\caption{\bf\bm Variation of the number of SCS and STPPP resonances with 
$ \frac{\omega_i}{\omega},\frac{\omega_1}{\omega}$.}
\end{table}

%% file: chap8.tex
\section{Introduction}

The design of future linear colliders includes intense energetic electron and
positron bunch collisions in order to maximise centre of mass energy and
collider luminosity. Associated with these intense bunches are strong electromagnetic fields
which affect the physics processes at and near the interaction point. Of
particular concern is unwanted pair production which causes background
radiation in detectors and other collider components. Background studies of
this pair production take as a starting point beamsstrahlung from the fermion
bunches undergoing the pinch effect. The processes studied hitherto include
the two-vertex Breit-Wheeler \cite{BreWhe34}, Bethe-Heitler \cite{betHei34} and 
Landau-Lifshitz \cite{LanLif34} processes. What has not been studied (and no 
rationale has been given for so neglecting) is the Breit-Wheeler process in the presence of 
the electromagnetic fields of the bunches in which a pair is produced from two real 
photons and a contribution from the bunch field. This process - stimulated two photon 
pair production (STPPP) in a beam field - is studied in this chapter.

The electron and positron beams at linear colliders are ultra relativistic
and the electromagnetic fields produced by ultra relativistic beams approximate well 
to a constant, crossed plane wave. The Volkov 
solution can therefore be used with an external field 4-potential of suitable form. This 
work is done in sections \ref{c8field} and \ref{c8volk}.

Much of the analytic work performed in Chapter 3 for the STPPP process is reapplied to a 
real case of beam parameters proposed in the R\&D for future linear colliders, including the International Linear Collider (ILC). 
Since differential cross section resonances were found for the STPPP in a circularly 
polarised field, their potential appearance for the STPPP process in a beam field is of 
particular concern. The resonances potentially increase the differential cross section by 
orders of magnitude. In order to quantify the resonances the electron self energy in
the constant crossed field (EFESE) is calculated. This electron self energy calculation has been 
performed before \cite{Ritus72}. The calculation here is consistent with that of Ritus, 
and a new analytic expression is found when the external field is non azimuthally symmetric 
(section \ref{c8ese}).

In section \ref{c8stppp} the expression for the STPPP process in the constant crossed field is
developed, the EFESE is included as a radiative correction to the propagator and 
numerical results are obtained. The cross section of the STPPP process is compared to the
ordinary (non external field) Breit-Wheeler cross section and an estimate is provided,
assuming azimuthal symmetry, of the overall cross section.

\bigskip\
\section{Electromagnetic Field of a Relativistic Charged Beam}
\label{c8field}

\begin{figure}[!b] 
\centerline{\includegraphics[height=4.5cm,width=8cm]{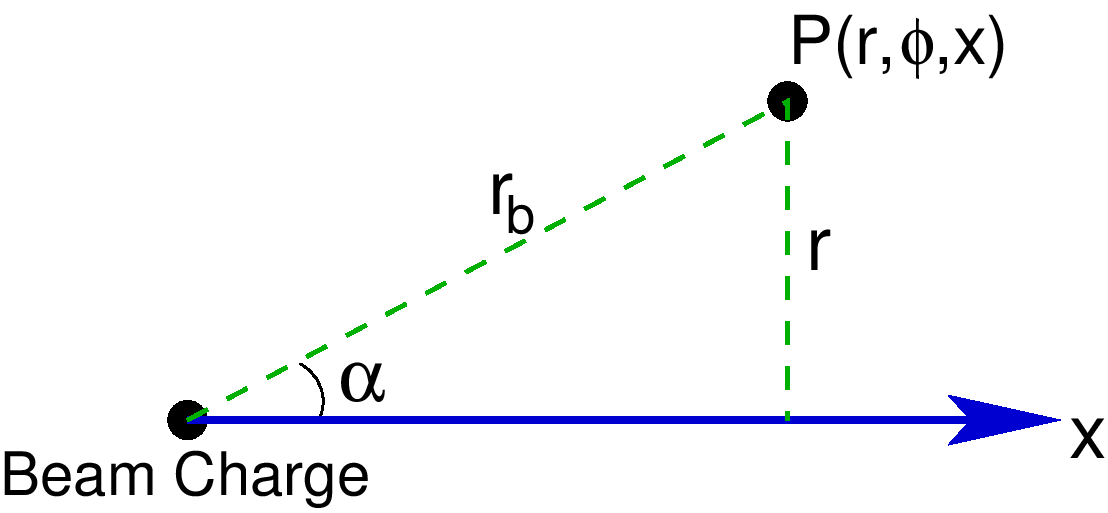}}
\caption{\bf Parameters associated with the beam charge field vectors}
\label{c8.fig1}
\end{figure}

Generally speaking, the charges of a typical charged bunch at the ILC move
uniformly and ultra relativistically along a x-axis (see figure \ref{c8.fig1}). The
field potential at a point, P, is a special case of the Lienard-Weichert
potentials (\cite{LanLif75} section 63) and the associated field vectors can be
written in cylindrical coordinates as

\begin{equation}
\label{c8.field.eq1}
\begin{array}{rl}
  \sql{E} (z) &= \dfrac{e}{4 \pi \varepsilon_0 \gamma^2} \dfrac{[\sin \alpha, 0,
  \cos \alpha]}{r_b^2 (1 - \beta^2 \sin^2 \alpha)^{3 / 2}} \\[10pt]
  \sql{B} (z) &= \dfrac{e \beta}{4 \pi \varepsilon_0 c \gamma^2} \dfrac{[0, \sin
  \alpha, 0]}{r_b^2 (1 - \beta^2 \sin^2 \alpha)^{3 / 2}}
\end{array}
\end{equation}

\medskip\

\medskip\

In the relativistic limit, $\sql{E}$ and $\sql{B}$ are predominantly equal in magnitude,
mutually orthogonal and transverse to the direction of motion of the beam
charge. That is, they take on the characteristics of a constant crossed plane
wave field described by 4-potential $A_{\mu}$ and with field vector magnitudes

\begin{equation}
\label{c8.field.eq2}
\begin{array}{rl}
  c|\sql{E}| = |\sql{B}| &= \dfrac{e \beta}{4 \pi \varepsilon_0 c \gamma^2 r^2_b} 
  \dfrac{1}{(1 - \beta^2)^{3 / 2}} \\[10pt]
  A_{\mu} &= a_{1\mu} (kx)
\end{array}
\end{equation}

\medskip\

Using Maxwell's field equations it becomes clear that the 3-potential
$\sql{a}_1$ lies in the direction of $\sql{E}$ (i.e. radially to the z-axis) and
the magnitude of the 3-potential can be related to $|\sql{B}|$. The magnitudes are related via the external field photon energy $\omega$

\begin{equation}
\label{c8.field.eq3}
\begin{array}{rl}
  |\sql{a}_{1} | \,\omega &= | \sql{B} | 
\end{array}
\end{equation}

\medskip\

The non linear first order processes considered in collider background studies
are described by the dimensionless beamsstrahlung parameter $\Upsilon$. This
parameter is written alternatively as a ratio of beam magnetic field to
Schwinger critical field, $B_c$ or the magnitude of the 4-vector product of
field tensor and particle 4-momentum. $\Upsilon$ is fixed with bunch 
dimensions ($\sigma_x, \sigma_y, \sigma_z)$, bunch charge $N$ and the relativistic factor of the bunch $\gamma$ \cite{YokChe91}

\begin{equation}
\label{c8.field.eq4}
\begin{array}{rl}
  \Upsilon = \dfrac{| \sql{B} |}{B_c} &= \dfrac{e\hbar}{m^3 c^5} \sqrt{|
  (F_{\mu \nu} p^{\nu}_{^{}})^2 |} \sim \dfrac{5}{6} \dfrac{Nr^2_e
  \gamma}{\alpha \sigma_z (\sigma_x + \sigma_y)} \quad ;\quad  \alpha \sim\dfrac{1}{137}\\[10pt]
  \text{where} \quad\quad B_c & = \dfrac{m^2}{e}=4.4 \text{ GTesla} \quad;\quad r_e = \dfrac{e^2}{4
  \pi \varepsilon_0 m_{} c^2}\approx 2.8\times 10^{-15} \text{ m}
\end{array}
\end{equation}

\medskip\

The energy spectrum of the beamsstrahlung process in the intense beam magnetic
field, given by the Sokolov-Ternov equation \cite{SokTer64} has a dependence 
on the external field completely described by a function of $\Upsilon$. The same can be achieved for the STPPP process by reexamining the constant crossed 4-potential. The form of the external field 4-momentum $k$, allows the energy to be extracted so that a unit, light-like 4-vector, $n$ remains

\begin{equation}
A_\mu=a_{1\mu}(kx)\equiv\omega a_{1\mu}(nx) \quad ; \quad k_\mu=(\omega,0,0,\omega),\; n_\mu=(1,0,0,1)
\end{equation}

\medskip\

For the constant crossed field then, the dimensionless parameter $\nu$ can be reinterpreted by including the external field photon energy with the external field 4-potential $|\sql{a}_{1\mu}|$. In the same way, the external field photon 4-momentum can be reinterpreted to be the unit 4-vector $n$. Thus $\nu$ can be made equivalent to the beamstrahlung parameter and the external field photon energy is replaced by unity. For the rest of this chapter these redefinitions will be in place

\begin{equation}
\nu\rightarrow \dfrac{e\omega|\sql{a}_{1\mu}|}{m^2}\equiv\dfrac{|\sql{B}|}{B_c}\equiv\Upsilon \quad ; \quad \dfrac{\omega}{m}\rightarrow\frac{1}{m}
\end{equation}

\medskip\

With the aid of equation \ref{c8.field.eq4} the external field 
dimensionless parameter $\nu$ can be written in terms of bunch parameters. For a typical set of bunch
parameters being considered at the ILC $(\sigma_x = 0.335 \mu\text{m}, \; \sigma_y = 2.7\text{nm}, 
\;\sigma_z = 225 \mu\text{m}, \;N = 2 \times 10^{10}$, \;bunch particle 
energy= 500 GeV), and in the centre of mass frame of the colliding bunches

\begin{equation}
\begin{array}{rl}
  \nu^2 \left( \equiv \Upsilon^2 \right) &= \left[ \dfrac{5}{6} \dfrac{Nr^2_e \gamma}{\alpha \sigma_z
  (\sigma_x + \sigma_y)} \right ]^2= 0.054
\end{array}
\end{equation}

\medskip\

A value of $\nu^2 \sim 1$ is commonly recognised as the limit at 
which first order non linear effects become important \cite{Bamber99}. The results of Chapter 7 
show that cross section resonances can occur at values as low as $\nu^2=0.1$. Future linear colliders are planned with beam parameters in which $\nu^2$ can reach, or even exceed, unity.

\bigskip\
\section{Volkov functions in a constant crossed electromagnetic field}
\label{c8volk}

The Volkov solution for the Dirac equation in a general external plane wave field was 
provided in section \ref{c2volk}. Its solution with a constant crossed electromagnetic field 
described by the 4-potential $A_\mu$ (equation \ref{c8.field.eq2}) is

\begin{equation}
\label{c8.volk.eq1}
\begin{array}{rl}
  \Psi^{V}_{p}(x) &= E_p(x)\; u_p \\[10pt]
  E_p (x) &= \left[ 1 + \dfrac{e}{2 (kp)} \slashed{k} \slashed{a}_1 (kx)
  \right] \\[10pt]
  &\times \;\; \exp \left( - iqx + i \dfrac{e^2 a^2}{2 (kp)}(kx) -
  i \dfrac{e (a_1p)}{2 (kp)} (kx)^2 - i \dfrac{e^2 a^2}{6 (kp)} (kx)^3 \right)\\[20pt]
  & \text{where} \quad\quad q = p + \dfrac{e^2 a^2}{2 (kp)} k 
\end{array}
\end{equation}

\medskip\

As was done in Chapter 3 for the circularly polarised field, a Fourier transform 
of the exponential function in equation \ref{c8.volk.eq1} is obtained. Since the 
argument of this exponential function is non periodic it will be a 
continuous Fourier transform

\begin{equation}
\label{c8.volk.eq2}
\begin{array}{rl}
  &\exp \left( iQ\,(kx) - iQ_{\varphi}\,(kx)^2 -
  iQ \dfrac{(kx)^3}{3} \right) \rightarrow \dfrac{1}{2 \pi}
  \dint^{\infty}_{\nths -\infty} F_0 (r)\exp (-ir (kx)) dr \\[20pt]
  & \text{where} \quad\quad F_0 (r) = \dint^{\infty}_{\nths -\infty} \exp \left( i (r + Q) t
  - iQ_\varphi t^2 - iQ \dfrac{t^3}{3} \right) dt \\[10pt]
  & \quad\quad\quad\quad\quad\quad   Q\equiv\dfrac{e^2 a^2}{2 (kp)} \quad ; \quad
  Q_\varphi \equiv\dfrac{e (a_1p)}{2 (kp)}  
\end{array}
\end{equation}

\medskip\

The Fourier transform variable $(kx)$ appears also as a cofactor of the slash vectors 
appearing in $E_p (x)$. After inserting the Volkov wave function into a trace calculation there 
are, potentially, several Fourier transforms involving powers of the integration variable 
which expressions need to be found for. These can be collectively labelled

\begin{equation}
F_{n, r}= \int^{\infty}_{- \infty} t^n \exp \left[ i (r + Q) t -
   iQ_\varphi t^2 - iQ \dfrac{t^3}{3} \right] dt
\end{equation}

\medskip\

\begin{figure}[!b] 
\centerline{\includegraphics[height=6cm,width=6cm]{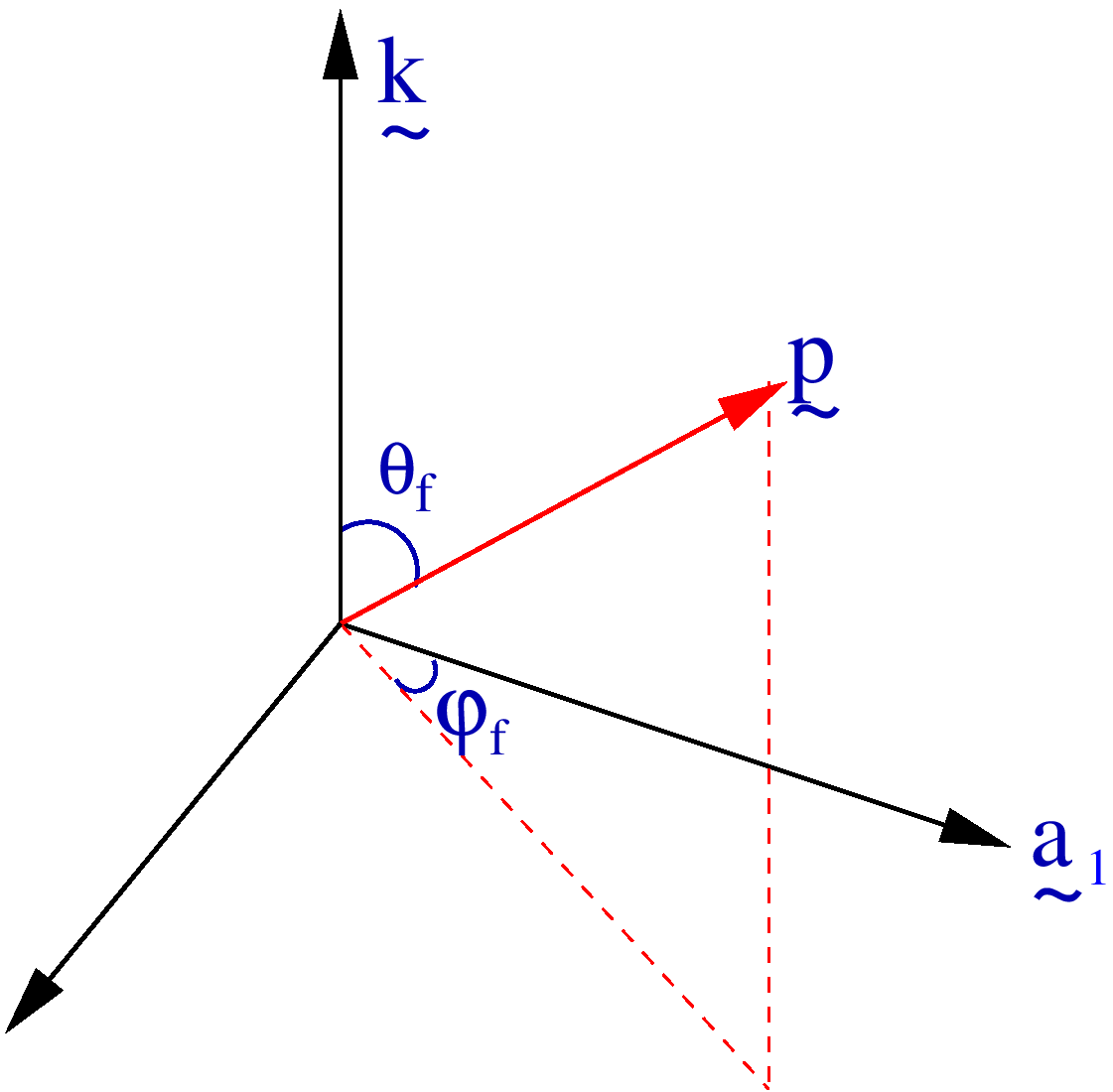}}
\caption{\bf Parameters associated with the beam charge field vectors}
\label{c8.fig2}
\end{figure}

$\displaystyle Q_\varphi$, as indicated in its subscript, is dependent on the azimuthal 
angle $\varphi_f$ and its explicit form is to be considered. From 
equation \ref{c8.volk.eq2} $\displaystyle Q_\varphi$ is vectorially dependent on 
$\sql{a}_{1}$, $\sql{k}$, and $\sql{p}$. Figure \ref{c8.fig2} represents these vectors and 
defines the azimuthal angle $\varphi_f$ around the 
external field 3-momentum $\sql{k}$, and the angle $\theta_f$ between the fermion 
3-momenta $\squ{p}$ and $\sql{k}$. 

Using the same gauge in which the 4th component of $a_{1\mu}$ is zero, the scalar 
products $(a_1p)$ can be written

\begin{equation}
\label{c8.volk.eq3}
(a_1p) = - \sql{a}_1 . \squ{p} = - a_1| \squ{p} | \sin \theta_f \cos \varphi_f
\end{equation}

\medskip\

There will be two sets of Fourier transforms - azimuthally symmetric and non azimuthally 
symmetric - depending on whether $(a_1p)$ and therefore $\displaystyle Q_\varphi$ can be neglected.
It is clear from equation \ref{c8.volk.eq3} that $(a_1p)$ can be neglected if $\sql{k}$ and 
$\squ{p}$ are collinear. As will be seen, the resultant Fourier transforms are Airy functions 
which appear in both the Ritus expression for the electron self energy in the crossed field 
\cite{Ritus72} and the Sokolov-Ternov equation (as equivalent Bessel $K$ functions of fractional order). It is pertinent to 
ask whether an assumption of azimuthal symmetry is justified.

These Fourier transforms will be inserted in both the STPPP calculation and the EFESE calculations. 
The STPPP process for ILC-like collisions will occur with initial photons which are highly energetic, 
nearly collinear with the external field, and directed oppositely to each other. It is likely from momentum 
considerations that the resultant fermions will also be collinear and azimuthal symmetry is obtained. 
However this consideration doesn't encompass all possible collisions. Even if the final state 
fermions have significant transverse momentum, azimuthal symmetry is still retained if the collision 
takes place on axis and if the beam profile is round. Neither of these is necessarily the case at the 
ILC.

In the EFESE calculation, the symmetry to be considered is between intermediate momentum states and 
the external field. Integrations have to be carried out over all these states and again azimuthal 
symmetry can only be assumed if the external field is azimuthally symmetric. If so, then the 
reference frame can be rotated around the field axis until $\sql{a_1}$ and $\squ{p}$ are 
perpendicular and $\displaystyle Q_\varphi$ goes to zero.

Both sets of Fourier transforms are written below. In both cases the integration 
over $\varphi_f$ is performed resulting in a factor of $2\pi$ for azimuthal symmetry and 
Bessel functions for non azimuthal symmetry. First the azimuthally symmetric transforms

\begin{equation}
\label{c8.volk.eq4}
\begin{array}{rl}
  F_{0, r} &= \dint^{\infty}_{\nths -\infty} \exp_{} \left[ i (r + Q) t -
  iQ \dfrac{t^3}{3} \right] = 2 \pi Q^{- \frac{1}{3}} \text{Ai}(z)\\[10pt]
  F_{1, r} &= \dint^{\infty}_{\nths -\infty} t\;\exp_{} \left[ i (r + Q) t -
  iQ \dfrac{t^3}{3} \right] = 2 \pi i Q^{- \frac{2}{3}} \text{Ai}'(z)\\[10pt]
  F_{2, r} &= \dint^{\infty}_{\nths -\infty} t^2 \;\exp_{} \left[ i (r + Q) t -
  iQ \dfrac{t^3}{3} \right] = - 2 \pi \dfrac{z}{Q} \text{Ai}(z) \\[20pt]
  &\text{where} \quad\quad Q = \dfrac{e^2 a^2}{2 (kp)} \quad;\quad z=-(r+Q) Q^{-\frac{1}{3}} 
\end{array}
\end{equation}

\medskip\

and the non azimuthally symmetric

\begin{equation}
\label{c8.volk.eq5}
\begin{array}{rl}
  F^{(\varphi)}_{0, r} &= 2 \dint^{\infty}_{\nths 0} J_0 (P t^2) \cos \left[ (r
  + Q) t - Q \dfrac{t^3}{3} \right] dt \equiv \text{Ai}_0\text{J}_0(P,Q) \\[10pt]
  F^{(\varphi)}_{1, r} &= 2 i \dint^{\infty}_{\nths 0} t \;J_0 (P t^2) \sin
  \left[ (r + Q) t - Q \dfrac{t^3}{3} \right] dt \equiv \text{Ai}_1\text{J}_0(P,Q)\\[10pt]
  F^{(\varphi)}_{2, r} &= 2 \dint^{\infty}_{\nths 0} t^2 \;J_0 (P t^2) \cos
  \left[ (r + Q) t - Q \dfrac{t^3}{3} \right] dt \equiv \text{Ai}_2\text{J}_0(P,Q)\\[10pt]
  &\\& \text{where} \quad\quad P = \dfrac{e a| \squ{p} |}{2 (kp)} \sin \theta_f 
\end{array}
\end{equation}

\medskip\

The numerical differences between the two sets of Fourier transforms will be considered in 
section \ref{c8aznonaz}. Finally, for both the EFESE and STPPP calculations, the 
Volkov wave functions appear in combinations like

\begin{equation}
\label{c8.volk.eq6}
\begin{array}{rl}
  \bar{E}_p (x) \gamma_{\mu} E_{p'} (x) &= \dfrac{1}{2 \pi} \dint^{\infty}_{\nths 
-\infty} H_r (p, p') \;\exp \left[ i (q - q' - rk) x \right] \\[15pt]
  \text{where} \quad H_r (p, p') &= \gamma_{\mu} F^{(\varphi)}_{0, r}(p, p') + \left( 
\dfrac{e}{2 (kp)}
  \slashed{a}_1 \slashed{k} \gamma_{\mu} - \dfrac{e}{2 (kp')} \gamma_{\mu}
  \slashed{a}_1 \slashed{k} \right) F^{(\varphi)}_{1, r} (p, p')\\[10pt]
  & + \;\; \dfrac{e^2 a^2}{4 (kp)(kp')} \slashed{k} \gamma_{\mu} \slashed{k} 
  F^{(\varphi)}_{2, r}(p, p')
\end{array}
\end{equation}

\medskip\

and the functions $\displaystyle F^{(\varphi)}_{n, r}(p, p')$ refer to either the azimuthally 
symmetric or non azimuthally symmetric Fourier transforms which contain

\begin{equation}
\label{c8.volk.eq7}
\begin{array}{rl}
  Q(p,p') &= \dfrac{e^2 a^2}{2} \left[\dfrac{1}{(kp')}-\dfrac{1}{(kp)}\right] \\[10pt]
  Q_{\varphi}(p,p') &= \dfrac{e}{2} \left[\dfrac{(a_1p')}{(kp')}-\dfrac{(a_1p)}{(kp)}\right]
\end{array}
\end{equation}

\bigskip\
\section{Numerical comparison of Fourier transforms $\displaystyle F_{n,r}$ and 
$\displaystyle F^{(\varphi)}_{n,r}$ }
\label{c8aznonaz}

In this section the Airy functions associated with azimuthal symmetry and the AiJ 
functions that emerged for non azimuthal symmetry, are computed numerically. 
The AiJ transforms do not reduce to known functions and will have to be computed numerically

\begin{figure}[!t]
\label{c8az.fig1}
\centerline{\includegraphics[height=9cm,width=9cm]{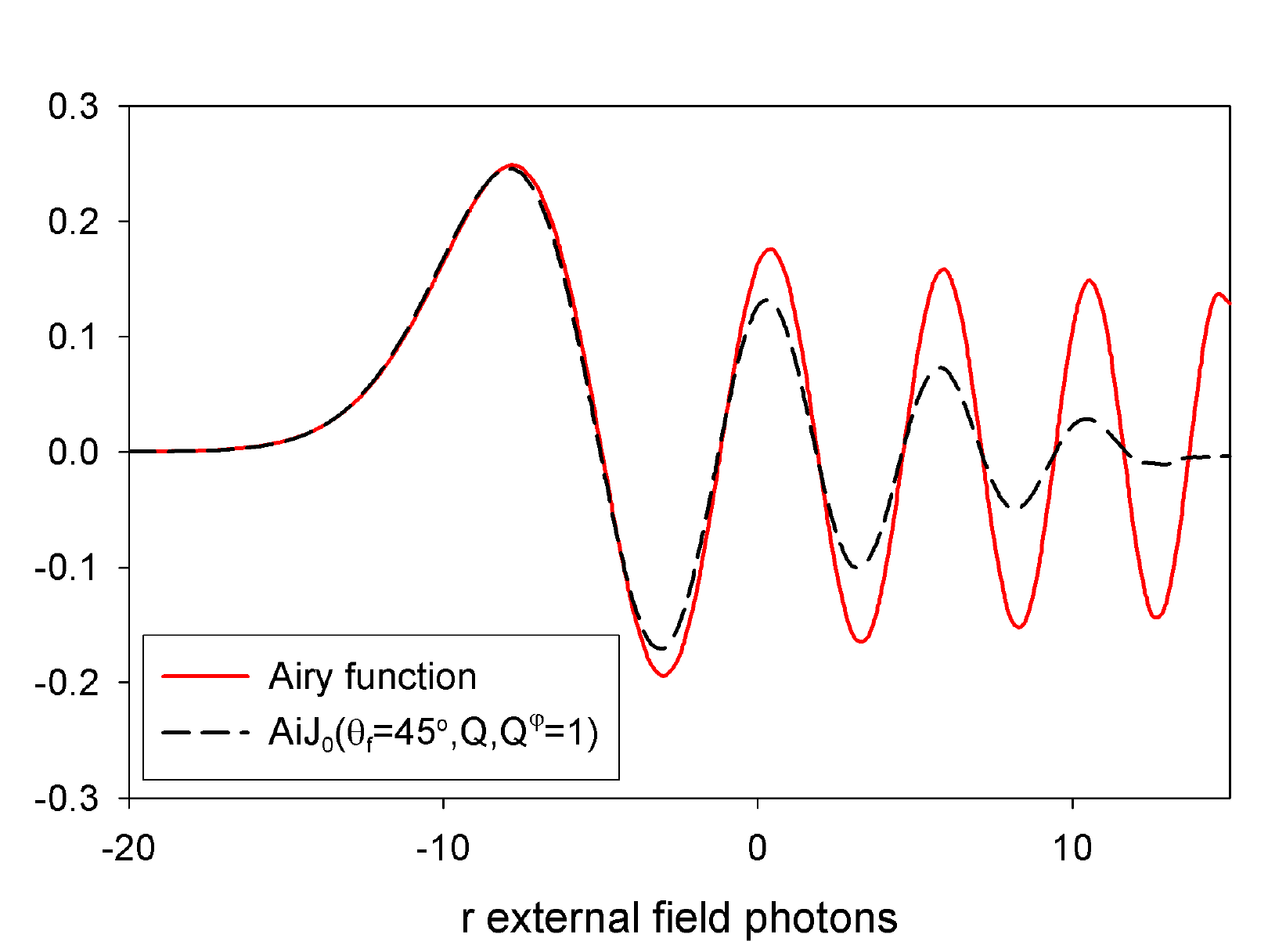}}
\caption{\bf Comparison of azimuthally symmetric and non azimuthally symmetric Fourier 
Transforms.}
\end{figure}

Figure \ref{c8az.fig1} shows a numerical comparison between $F_{0, r} (p)$ and
$F^{(\varphi)}_{0, r} (p)$. As $r$ decreases below zero, 
the argument of both functions is positive and the two functions converge and 
fall to zero. As $r$ increases above zero and the argument becomes negative, the Airy 
function remains large and oscillatory, but the AiJ function is damped. This 
will impact ultimately on process cross sections considered in this chapter.

\begin{figure}[!b]
\label{c8az.fig2}
\centerline{\includegraphics[height=9cm,width=9cm]{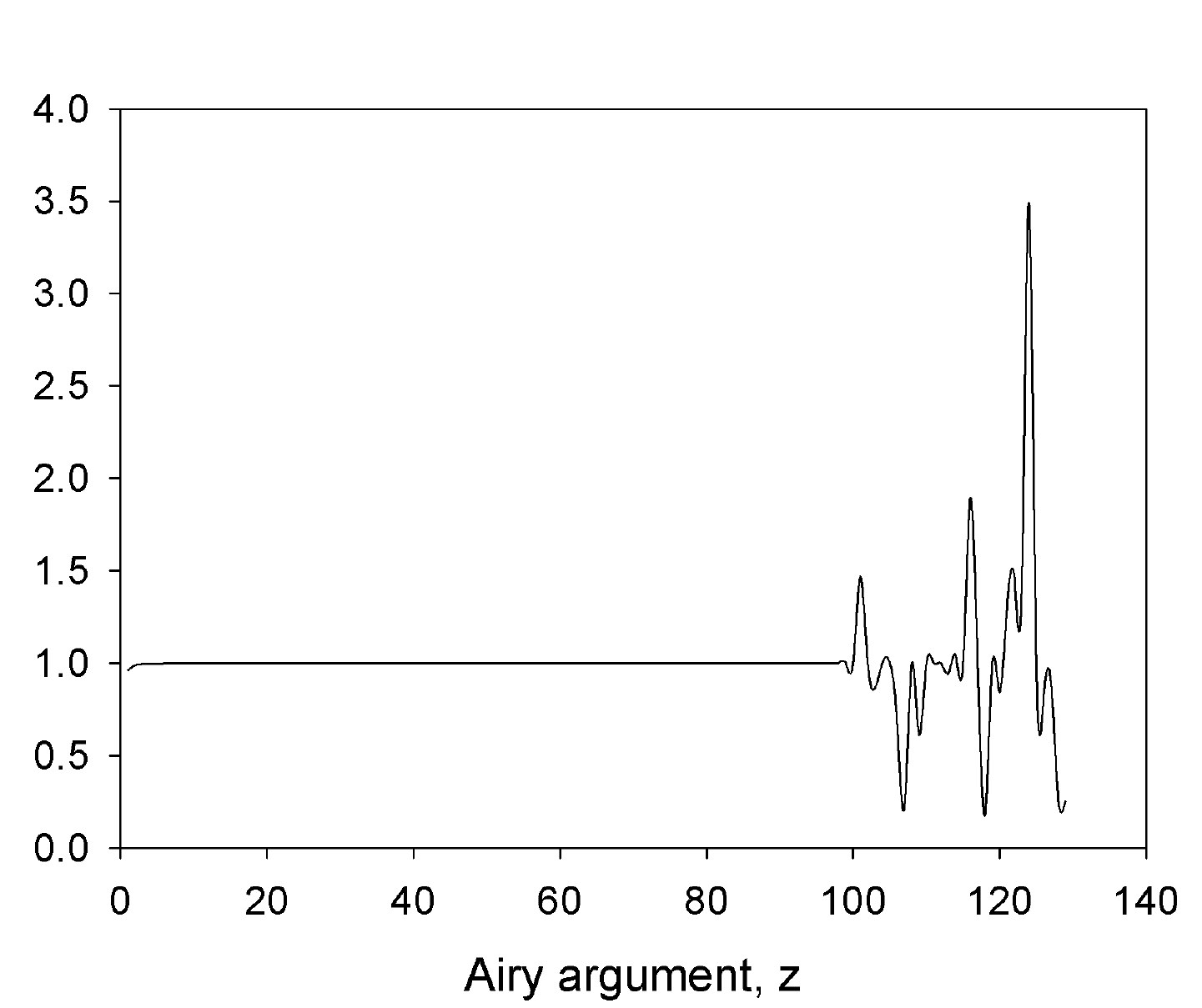}}
\caption{\bf Ratio of an Airy function to its numerically equivalent integration using 
Mathematica 5.0.}
\end{figure}

The computer program Mathematica v5.0 was used to perform the numerical integration of 
the AiJ function and its accuracy is of concern. Figure \ref{c8az.fig2} shows the 
ratio of Airy function to its equivalent numerical integration as calculated by 
Mathematica. The result should be 1 if the numerical integration performs well, 
which it does until the argument reaches $z\approx 100$. As will be seen, numerical 
analysis of the resonance cross section involves arguments of at least $10^4$, so this 
method of calculation will not suffice

An alternative is to find an asymptotic expansion for the AiJ function. Such was 
done successfully for the generalised Bessel functions that appear for 
linearly polarised \cite{LeuStr81} and elliptically polarised \cite{Panek02} external 
fields. Time constraints did not permit this to be done in this thesis, so the use of 
AiJ functions in numerical calculations of the EFEES and STPPP processes is 
curtailed for the present.

\clearpage
\section{Electron self energy in a constant crossed electromagnetic field}
\label{c8ese}

As was the case for the SCS and STPPP processes in the circularly polarised
field, the STPPP process in the constant crossed field is expected to contain
resonances. The electron self energy in the constant crossed field (EFESE) 
has been calculated before by \cite{Ritus72} and
contains Airy functions and their derivatives. \cite{Ritus72} makes use of the optical
theorem to show that the mass operator gives the previously calculated
expression for the probability of emission of a photon by an electron in a
crossed field \cite{NikRit67}.

The methods developed in Chapter 6 and Appendix 
\ref{app.kallen} can be used to obtain an expression for the imaginary part of the EFESE using 
the transforms to Airy functions (equation \ref{c8.volk.eq4}). This result should be 
consistent with that obtained by \cite{NikRit67}. The assumption of azimuthal
symmetry will then be dropped and the transforms of equation \ref{c8.volk.eq5} will be used to 
obtain a new result for the imaginary part of the EFESE in the constant crossed 
electromagnetic field. 

The general form of the EFESE was written down in equation \ref{c6.eq3}. Using the expression for products of 
Volkov functions in the presence of the constant crossed field (equation \ref{c8.volk.eq7}) 
the EFESE is

\begin{equation}
\label{c8.ese.eq1}
\begin{array}{ll}
  \Sigma^e (p) &= \dfrac{i 4 e^2}{VT(2 \pi)^4} \dint d^4 x_1
  d^4 x_2 d^4 k' d^4 q'  \dfrac{1}{(2 \pi)^4} \dint_{\nths -\infty}^{\infty} dr \; ds \\[10pt]
  & \times \;\; H_r^{} (p, p') \left( \slashed{q}' + \dfrac{e^2 a^2}{2 (kp')} \not{k} + m \right) H_s
  (p, p') \dfrac{1}{q'^2 - m^2_{\ast \text{}}}  \dfrac{1}{k'^2} \\[10pt]
  &\times \;\; exp \left[ i (q - q' + k' - r k) x_2 \right]\; 
    exp \left[ i (q' - q - k' - sk) x_1 \right]
\end{array}
\end{equation}

\medskip\

The integrations over $x_1, x_2$ and $s$ produce a delta
function whose argument expresses the conservation of 4-momenta and the condition $s = - r$.
Noting the symmetry of the functions $F_{n, r}$ and $H_r$, the momentum space EFESE becomes

\begin{equation}
\begin{array}{rcl}
  \Sigma_{}^e (p) & = &\dfrac{i 4 e^2}{VT(2 \pi)^4} \dint d^4 k'
  d^4 q'  \dfrac{1}{(2 \pi)^2} \dint^{\infty}_{\nths-\infty} d r 
  H_r (p, p') ( \slashed{q}' + \frac{e^2 a^2}{2 (kp')} \slashed{k} + m) \\[10pt]
  &\times & \bar{H}_r (p, p')  \dfrac{1}{q'^2 - m^2_{\ast \text{}}}  \dfrac{1}{k'^2}
  \delta (q - q' + k' - r k) \\[20pt]
  &\text{since}& F_{0, - r}^{\ast} = F_{0, r} \quad;\quad  F_{1, - r}^{\ast} = -F_{1, r} \quad;\quad
  F_{2, - r}^{\ast} = F_{2, r} \\[10pt]
  & & \bar{H}_r (p, p') = H_{r}^{\ast} (p', p)
\end{array}
\end{equation}

\medskip\

The identities of Appendix \ref{app.kallen} allow the expression for the
EFESE to be written in the form of a dispersion relation, and the average over 
electron yields a trace of 4-vectors. The imaginary part of the external field 
electron energy shift (EFEES) in the constant crossed field becomes (cf. equation 
\ref{c6.eq7})

\begin{equation}
\label{c8.ese.eq2}
\begin{array}{rl}
  \Im \Delta \varepsilon_p (D^2) &= - \dfrac{e^2 m^2}{16 \pi \varepsilon_p}
  \dint^{\infty}_{\nths -\infty} d r \dint^1_{\nths -1} d \cos \theta_{}  \left(
  1 - \dfrac{m^2_{\ast}}{D^2} \right) \\[10pt]
  & \biggl\{ \left( 2 - 2 \nu^2 + 2 r[(kD) - (kK)] + \nu^2 \left( \dfrac{(kD)}{(kK)} +
  \dfrac{(kK)}{(kD)} \right) \right) F^2_{0, r} \\[10pt]
  & + \nu^2 \left(\dfrac{(kD)}{ (kK)} + \dfrac{(kK)}{(kD) } \right)
  F^2_{1, r} + 2 \nu^2 F_{0, r} F_{2, r} \biggr\} \Theta (D^2 - (m_{\ast} +\epsilon)^2) \\[20pt]
  & \quad \quad \text{where}\quad\quad  K \equiv q' + \frac{1}{2} D \quad 
  \text{and} \quad D \equiv q - r k
\end{array}
\end{equation}

\medskip\

Making a transformation of variable to $u = \frac{(kD)-(kK)}{(kK)}$ and with 
$\theta$ defined as the angle between $\sql{K}$ and $\sql{k}$, we have

\begin{equation}
\left( 1 - \frac{m^2_{\ast}}{D^2} \right) d\cos \theta = \frac{2
d u}{(1 + u)^2}
\end{equation}

\medskip\

The regularisation procedure is the same as for the circularly polarised
external field. The Heaviside Step function ensures that the field free
unregularised mass shift goes to zero (see equation 6.12). The condition 
$D^2= - 2 r \rho > 0 ; \rho = (kp)$ ensues, and its convenient to
make the shift $r \rightarrow - r$. The imaginary part of the regularised
EFEES is

\begin{equation}
\label{c8.ese.eq3}
\begin{array}{rl}
  \Im \Delta \varepsilon^R_p (\rho) & = \dfrac{e^2 m^2}{16 \pi \varepsilon_p}
  \dint^{\infty}_{\nths 0+\epsilon} dr \dint^{u_r}_{\nths 0+\epsilon} \dfrac{du}{(1 + u)^2}
  Q^{- \frac{2}{3}} \\[10pt]
  & \times \;\; \left[ - 4 \text{Ai}(z)^2 - 2 \nu^2 \left( 1 -
  \dfrac{r}{Q} \right) \left( 2 + \dfrac{u^2}{1 + u} \right) \left( \text{Ai}
  (z)^2 + \dfrac{1}{z} \text{Ai}' (z)^2 \right) \right] \\[20pt]
  & \text{where} \quad\quad z = Q^{- \frac{1}{3}} (r - Q) \quad ; \quad
  Q = \dfrac{\nu^2 u}{2 \rho} \\[10pt]
  & \quad\quad\quad\quad u_r = \dfrac{2 r \rho}{1 + \nu^2} \quad ; \quad \rho = (kp)
\end{array}
\end{equation}

\medskip\

Via the optical theorem, equation \ref{c8.ese.eq3} should also be the transition 
probability for the emission of a photon by an electron in the crossed field. 
Indeed, comparison with equation 45 of \cite{NikRit67}, after a suitable change of 
integration variables, shows this to be the case.

Equation \ref{c8.ese.eq3} contains a pole as either of the integration variables reach zero. 
Mathematically the pole can be traced back 
to the Fourier transform, $F_{n, r}$. In the limit of vanishing external field 
$F_{n, r}$ reduces to a Dirac delta function which expresses the constraint
that $r = 0$ external field photons should contribute to the process. The pole
in equation \ref{c8.ese.eq3} can be avoided with consideration of the Heaviside step
function argument which requires the lower bound of the $r$ integration be
greater than zero by an infinitesimal amount $\epsilon$. The integrations could
proceed by taking the Cauchy principal value. Computationally however, it is convenient to shift integration
variables in order to absorb the poles. These shifts (equation \ref{c8.ese.shifts}) are valid because the
lower bounds of $u$ and $r$ are never quite zero.

\begin{equation}
\label{c8.ese.shifts}
u \rightarrow \dfrac{1}{r}\dfrac{\nu^2}{2\rho} u \quad\quad\quad  r \rightarrow Q^{-4/3} r^{2/3}
\end{equation}

\medskip\

Finally, the imaginary part of the EFEES is

\begin{equation}
\label{c8.ese.final}
\begin{array}{rl}
  \Im \Delta \varepsilon^R_p (\rho) & = \dfrac{e^2 m^2}{16 \pi \varepsilon_p}
  \dint^{\infty}_{\nths 0+\epsilon} dr \dint^{u_r}_{\nths 0+\epsilon} 
  \; \dfrac{3\,du}{2(\sigma + u^3r^{3/2})^2}\\[10pt]
  \times & \left[ - 4 \sigma r u^2 \text{Ai}(z)^2 + 2 \nu^2 (1 - u) 
  \left( 2\sigma + \dfrac{r^3u^6}{\sigma + u^3 r^{3/2}} \right) \left( ru\text{Ai}(z)^2 + 
  \dfrac{1}{1-u} \text{Ai}' (z)^2 \right) \right] \\[20pt]
  & \text{where} \quad z = ru(1-u) \quad ; \quad
   u_r = \dfrac{\nu^2}{1 + \nu^2} \quad ; \quad \sigma=\dfrac{\nu^2}{2\rho} \quad ; \quad \rho = (kp)
\end{array}
\end{equation}

\medskip\

\begin{figure}[!b]
\centerline{\includegraphics[height=10cm,width=14cm]{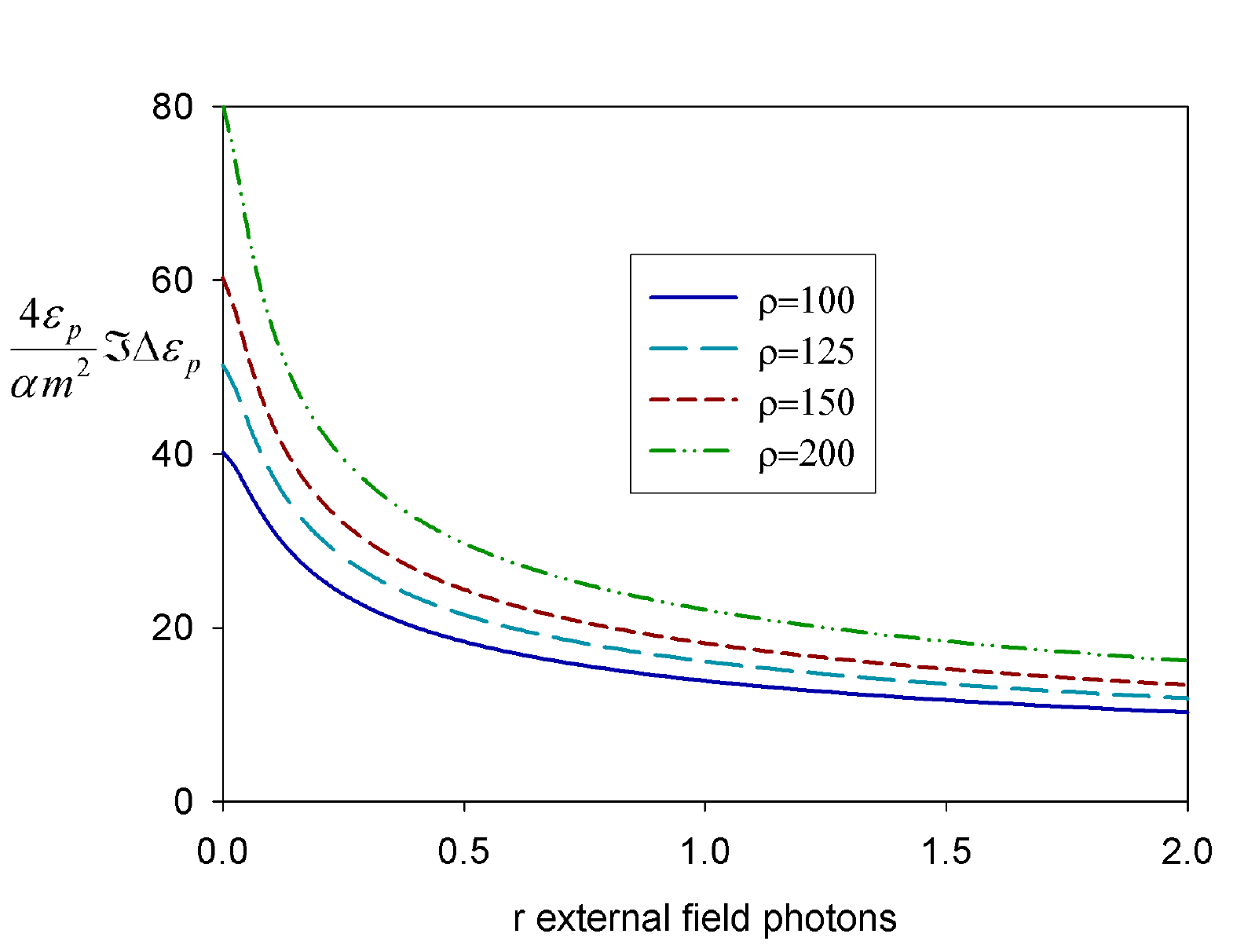}}
\caption{\bf Variation of the regularised, imaginary part of the EFEES in a crossed field with 
contribution of $r$ external field photons and different $\rho$ values}
\label{c8.ese.fig1}
\end{figure}

\begin{figure}[!t]
\centerline{\includegraphics[height=10cm,width=14cm]{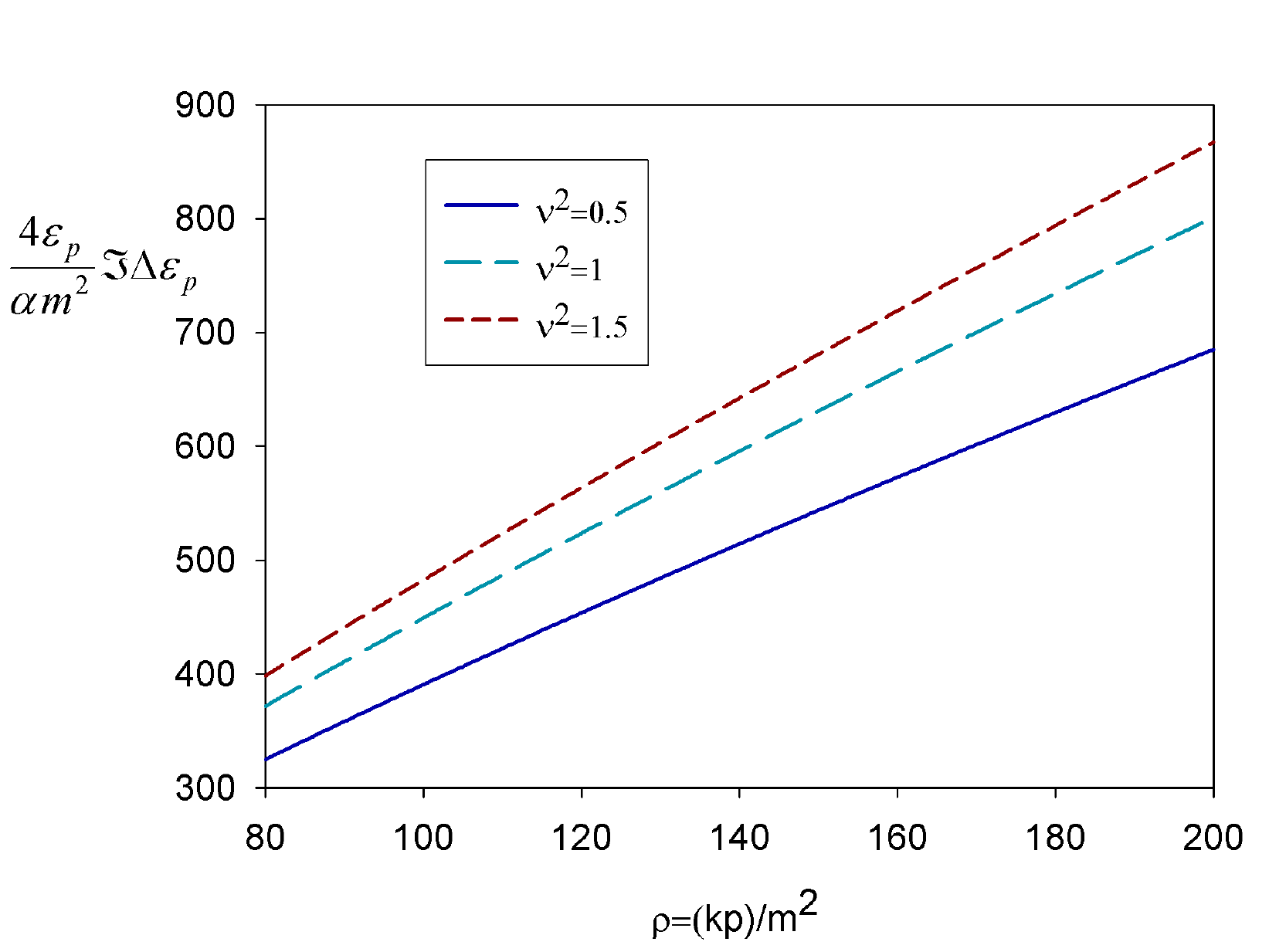}}
\caption{\bf Electron Energy shift in a crossed field.}
\label{c8.ese.fig2}
\end{figure}

Some results which show the dependence of $\Im \Delta \varepsilon^R_p (\rho)$
on the variables $r, \rho$ and $\nu^2$ are shown in figures \ref{c8.ese.fig1} 
and \ref{c8.ese.fig2}. These figures show similar trends as the equivalent ones studied in Chapter 6, 
for the imaginary part of the EFEES in the circularly polarised field. 
In particular figure \ref{c8.ese.fig1} should be compared with figures \ref{c6g1} and \ref{c6g2}, 
and figure \ref{c8.ese.fig2} should be compared with figures \ref{c6g6} and \ref{c6g7}. Much 
of the same analysis of Chapter 6 can be applied and only a few extra remarks are required.

The variation of $\Im \Delta \varepsilon^R_p (\rho)$ with the number of
external field photons in figure \ref{c8.ese.fig1} shows a peak at $r = 0$. 
For the circularly polarised case, the lower x ordinate was discrete and ended 
on $b = 1$ external field photons since the $b= 0$ point was subtracted out by the regularisation procedure. In the crossed
field case considered here the x-ordinate is continuous and the $r=0$ point was removed by 
the shift in integration variables.

The variation of $\Im \Delta \varepsilon^R_p (\rho)$ with $\rho$ in figure \ref{c8.ese.fig2} 
shows the same trend as the circularly polarised case, i.e. increasing probability of photon 
emission{\footnote{Equating the probability of photon emission with $\Im \Delta 
\varepsilon^R_p (\rho)$}} when an electron
travelling in the external field is more energetic or the external field is
more intense. The overall values of $\Im \Delta \varepsilon^R_p (\rho)$ are
greater than those of Chapter 6 because the values of $\rho$ considered are
greater. The numerical values for $\Im \Delta \varepsilon^R_p (\rho)$ for appropriate
values of $\rho$ will be inserted into the propagator denominators of the STPPP
process in the crossed field (section \ref{c8stppp}) to properly calculate the contribution 
of resonances to the STPPP cross section.

A non azimuthally symmetric version of $\Im \Delta \varepsilon^R_p (\rho)$ containing 
AiJ functions can be developed. The required expression is obtained by 
substituting the Fourier transforms of equation \ref{c8.volk.eq5} into equation \ref{c8.ese.eq2}. 
The same shift of integration variable $\cos \theta_{} \rightarrow u$ can be made 
and the argument of the Bessel function becomes

\begin{equation}
P = \sqrt{ \frac{\nu^2 u (u_r - u)}{4 \omega^2 (1 + u_{r})}}
\end{equation}

\medskip\

The non azimuthally symmetric version of $\Im \Delta \varepsilon^R_p (\rho)$ contains products
of the functions $F_{n, r}^{(\varphi)} (p, p')$ which can be simplified. Writing the 
product explicitly

\begin{equation}
\begin{array}{rl}
   & F_{n, r}^{(\varphi)} (p, p') F_{m, - r}^{(\varphi)} (p, p') = \dint dt \,dt' \; t^n \,t'^m \\[10pt]
   & \times \;\; \exp \left[ i (r + Q) (t - t') - iP (t^2 - t'^2)
   \cos \varphi_f - i \dfrac{Q}{3} (t^3 - t'^3) \right]
\end{array}
\end{equation}

\medskip\

The integration over $\varphi_f$ will result in a Bessel function of order zero
and an argument which is a difference $P t^2 - P t'^2$. The following identity is of use 
(\cite{Watson22} pg.30)

\begin{equation}
J_0 (a - b) = \sum_{m = -\infty}^{\infty} J_m (a) J_m (b)
\end{equation}

\medskip

With AiJ functions defined in equation \ref{c8.volk.eq5} , the 
product of $F_{n,r}^{(\varphi)}$ transforms can be written

\begin{equation}
\label{c8.ese.eq4}
\begin{array}{rl}
\dint d \varphi_f F_{m, r}^{(\varphi)} (p, p') F_{n, - r}^{(\varphi)}
   (p, p') &= \dsum\limits_{m = -\infty}^{\infty} \text{Ai}_n\text{J}_m (P, Q)^2 
\end{array}
\end{equation}

\medskip

The non azimuthally symmetric version of the regularised, imaginary part of the EFEES is 
obtained after substitution of equation \ref{c8.ese.eq4} into equation \ref{c8.ese.eq2} and 
use of the dispersion relations method outlined in Appendix \ref{app.kallen}  

\begin{equation}
\label{c8.ese.eq5}
\begin{array}{rl}
  \Im \Delta \varepsilon^R_p (\rho) &= \dfrac{e^2 m^2}{16 \pi
  \varepsilon_p} \dsum\limits_{m = -\infty}^{\infty}\dint^{\infty}_{\nths 0 + \epsilon} dr 
\dint^{u_r}_{\nths 0 +\epsilon} \dfrac{d u}{(1 + u)^2} \\[10pt]
  &\times \;\; \biggl[ \left(  - 4 - 4
  \nu^2 \left. \left( 1 - \dfrac{r}{Q} \right) \left( \dfrac{u^2}{1 + u} \right)
  \right) \text{Ai}_0\text{J}_m (P, Q)^2 - \right. \\[10pt]
  & 2 \nu^2 \left( 2 + \dfrac{u^2}{1 + u} \right) \text{Ai}_1\text{J}_m (P, Q)^2 
  - 4 \nu^2 \text{Ai}_2\text{J}_m (P, Q)^2 \biggr]
\end{array}
\end{equation}

\medskip\

Equations \ref{c8.ese.eq3} and \ref{c8.ese.eq5} can be identified with the probability 
of emission, $W_p$ by an electron of 4-momentum $p_{\mu}$ in the constant crossed field. To
obtain the energy spectrum of emitted photons, $W_p$ needs to be 
integrated with respect to photon energy $w'$. There will be two such expressions 
for the energy spectrum of emitted photons; one for azimuthally symmetric emission 
with respect to the external field and one for the non azimuthal symmetry. Only 
expressions for the azimuthally symmetric spectra will be written down here, though there 
is nothing especially difficult in obtaining the non azimuthally symmetric expression

Considering the integration variable $u$ as a function of $K$ and recalling that $K$ was 
used to represent the 4-momentum $q'_{\mu}$, the conservation of 4-momenta
$q_{\mu} - q'_{\mu} - k'_{\mu} - r k_{\mu} = 0$ can be used to shift the 
integration variable d$u$ to d$A$

\begin{equation}
\label{c8.ese.eq6}
\begin{array}{c}
  u = \dfrac{(kk')}{(kq')} \equiv \dfrac{A}{\rho - A} 
  \quad\quad\text{where} \quad\quad A = (kk') = \omega \omega' (1 - \cos \theta) \\[10pt]
  \dfrac{du}{(1 +u)^2} \rightarrow \dfrac{dA}{\rho}
\end{array}
\end{equation}

\medskip\

Substituting equation \ref{c8.ese.eq6} into equation \ref{c8.ese.eq3},
taking the derivative with respect to A and integrating over variable $r$ using known 
integrals of products of Airy functions \cite{Albright77}, the azimuthally symmetric 
emitted photon energy spectrum is

\begin{equation}
\begin{array}{rl}
  \dfrac{dW_p^{az}}{dA} &= - \dfrac{e^2 m^2}{4 \pi
  \rho \varepsilon_p} [Q^{\frac{1}{3}} \text{Ai} (- Q^{\frac{2}{3}}) + Q^{-
  \frac{1}{3}} \text{Ai}' (- Q^{\frac{2}{3}}) \\[10pt]
  &+ \;\;\dfrac{\nu^2}{2 Q} ( \dfrac{\rho - A}{A} + \dfrac{A}{\rho - A})
  \text{Ai} (- Q^{\frac{2}{3}}) 
  \text{Ai}' (-Q^{\frac{2}{3}})] \\
  &\\ &\text{where} \quad\quad Q = \dfrac{\nu^2}{2 \rho}  \dfrac{A}{\rho - A}
\end{array}
\end{equation}

\bigskip\
\section{STPPP in a constant crossed electromagnetic field}
\label{c8stppp}

The calculation of the imaginary part of the EFEES provides both a
correction to the bound electron propagator and the energy spectrum of
beamsstrahlung photons. Both of these are to be inserted into the STPPP cross section.
The Feynman diagrams and matrix element of the STPPP process in a general external field
were provided in section 3.7. The matrix element is

\begin{equation}
\begin{array}{rl}
  S_{f i}^e &= - \dfrac{e^2}{2} \dint^{\infty}_{\nths -\infty} d^4 x_1 d^4 x_2 d^4 q 
  \; e^{\mu} (k_1) \bar{e}^{\mu} (k_1) e^{\nu} (k_2) \bar{e}^{\nu}
  (k_2) \dfrac{1}{p^2 - m^2}  \bar{u}(p_-) (\mathcal{A}+\mathcal{B}) v(p_+) \\[20pt]
  &\text{where} \quad \mathcal{A}= \bar{E}_{p_{-}} (x_2) \gamma_{\mu} E_p (x_2) (
  \slashed{p} + m) \bar{E}_{p} (x_1) \gamma_{\nu} E_{-p_{+}} (x_1) e^{-
  ik_1 x_2 - ik_2 x_1} \\
  & \text{and} \quad\quad \mathcal{B}= \bar{E}_{p_{-}} (x_2) \gamma_{\nu} E_p (x_2) 
  ( \slashed{p}+ m) \bar{E}_{p} (x_1) \gamma_{\mu} E_{-p_{+}} (x_1) e^{-ik_2 x_2
  - ik_1 x_1} 
\end{array}
\end{equation}

\medskip\

Using the expression for products of Volkov $E_p$ functions (equation
\ref{c8.volk.eq6}), making the transform to functions $F_{n, r}$ and 
performing integrations over $x_1, x_2$ and $q$, the matrix element becomes (omitting 
polarisation vectors $e (k)$ and bispinors $u(p)$) is

\begin{equation}
\label{c8.stppp.eq1}
\begin{array}{rl}
  S_{f i}^e &= - \dfrac{e^2}{2} \dfrac{1}{(2 \pi)^2} \dint^{\infty}_{\nths -\infty}
  dr ds \biggl\{ H^{\mu}_r (p_-, \bar{p}) \dfrac{\slashed{\bar{p}} +
  m}{\bar{p}^2 - m^2} H^{\nu}_s ( \bar{p}, - p_+) \\[10pt]
  & + \;\; H^{\nu}_r (p_-, \dbr{p}) \dfrac{\slashed{\dbr{p}} + m}{\dbr{p}^2 - 
  m^2} H^{\mu}_s (\dbr{p}, - p_+) \biggr\} \delta (q_- + q_+ - k_1 - k_2 - (r + s) k)\\[20pt]
  & \text{where} \quad\quad \bar{q} = q_- - k_1 - r k \quad;\quad \dbr{q} = q_- - k_2 - r k
\end{array}
\end{equation}

\medskip\

The ILC will collide ultra relativistic charge bunches almost head on and produce 
beamsstrahlung (which serves as the initial states of the STPPP process) in the forward direction. So the vectors, $\sql{k}_1$, $\sql{k}_2$ and
$\sql{k}$ are considered (anti)collinear, and a centre of mass frame 
$\sql{k}_1+\sql{k}_2 + r\sql{k}=0$ is used. These considerations mean that one of the 
4-vector products, $(kk_1)$ or $(kk_2)$, be zero. Choosing $(kk_2)=0$, the functions 
$Q$ and $Q_\varphi$ reduce to

\begin{equation}
\begin{array}{rcl}
  Q (p_-, \dbr{p}) &=& Q ( \bar{p}, - p_+) = Q_\varphi (p_-,\dbr{p}) = 
  Q_\varphi (\bar{p}, - p_+) = 0 \\[10pt]
  Q (p_-, \bar{p}) &=& Q (\dbr{p}, - p_+) = \dfrac{e^2 a^2 (kk_1)}{2(kp_-)(kp_+)} \\[10pt]
  Q_\varphi (p_-, \bar{p}) &=& Q_\varphi (\dbr{p}, - p_+) = 
               \dfrac{e(a_1p_-)}{2}\dfrac{(kk_1)}{(kp_-) (kp_+)}
\end{array}
\end{equation}

\medskip\

Further consequences of the special kinematics and the above relations are that
the Fourier transforms $F^{\varphi}_{n,s}$ involving integration variable
$s$, contain a Dirac delta function with reduced argument, $\delta(sk)$ requiring 
$s=0$ for the second term of equation \label{c8.stppp.eq1}. For the second term of the
equation the condition is $r=0$.

The matrix element of equation \ref{c8.stppp.eq1} is squared and the product 
of Dirac delta functions ensures that there remains only one integration over 
variable $r$. The sum over polarisations and spins introduces a trace and the 
square of the matrix element becomes

\begin{equation}
\begin{array}{rl}
  \dsum\limits_{i f} |S_{f i}^e |^2 &= \dfrac{e^2}{16 m^2} \dfrac{1}{(2 \pi)^4}
  \dfrac{1}{\bar{p}^2 - m^2} \dfrac{1}{\dbr{p}^2 - m^2} \dint^{\infty}_{\nths -\infty} dr \\[10pt]
  & \times \;\; \text{Tr} \{ G_{\mu \nu \nu \mu} ( \bar{p}, \bar{p}) + G_{\mu \nu
  \nu \mu} (\dbr{p}, \dbr{p}) + G_{\mu \nu \mu \nu} ( 
\bar{p},\dbr{p}) + G_{\mu \nu \mu \nu} (\dbr{p}, \bar{p}) \} \\[20pt]
  \text{where} \quad & \quad G_{\mu \nu \nu \mu} ( \bar{p},\bar{p}) = ( \slashed{p}_- + m) H^{\mu}_r
  (p_-, \bar{p}) ( \slashed{\bar{p}} + m) H^{\nu}_0 ( \bar{p}, - p_+) \\[10pt]
  & \quad\quad\quad\quad\quad\quad\quad \times \;\;(\slashed{p}_+ + m) H^{\nu}_0 (- p_+, 
\bar{p}) (\slashed{\bar{p}} + m) H^{\mu}_{- r}( \bar{p}, p_-) 
\end{array}
\end{equation}

\medskip\

The trace calculation is performed once again with the Feyncalc Mathematica package. 
Its analytic form is 
similar to the one already written down in Appendix \ref{trace} with suitable 
substitutions of Fourier transform functions $F^{(\varphi)}_{n,s}$ for the 
Bessel functions associated with the circularly polarised field.
The STPPP differential cross section can be obtained after transforming the phase 
space integral and defining scattering angles $\theta_f$ and $\varphi_f$ by which the 
direction of 3-momentum $\vec{q}_-$ is specified.

To establish the midpoint of the range of initial 
photon energies that need to be considered, the expression for average 
beamsstrahlung energy which describes the ratio between beamsstrahlung energy 
$\omega$ and the electron energy $E$ which emits the beamsstrahlung, can be 
used \cite{YokChe91}

\begin{equation}
   \dfrac{\omega_1}{E} \sim \dfrac{4 \sqrt{3}}{15} \Upsilon \dfrac{\sqrt{1 +
   \Upsilon^{2 / 3}}}{(1 + (1.5 \Upsilon)^{2 / 3})^2} = 0.0145
\end{equation}

\medskip\

For the 250 GeV electrons planned for the default International Linear Collider 
parameter set, the average beamsstrahlung energy is $\dfrac{\langle\omega_1\rangle}{m} = 7094$.
Numerical evaluation of the STPPP differential cross section requires careful 
consideration of the integration over the variable $r$. The Dirac delta function 
describing the conservation of 4-momenta indicates that $r$ external field photons 
contribute to the process. Since $r$ is continuous it is probably more correct to 
say that the external field contributes energy $r\omega$ to the STPPP process.
The integration over $r$ cannot be performed analytically since the integrand is a 
complicated function. It is convenient to perform the calculation in the centre of mass 
reference frame defined by

\begin{equation}
\begin{array}{rl}
  \sql{k}_1 + \sql{k}_2 + r\sql{k} &= \squ{q}_- + \squ{q}_+  = 0 \\
  \omega_1 + \omega_2 + r\omega  &= \varepsilon_{q_-} + \varepsilon_{q_+} 
\end{array}
\end{equation}

\medskip\

In this frame, with $\omega_1$ and $r\omega$ a given, the other energies and momenta are found
with the aid of the relations

\begin{equation}
\label{c8stppp.frame}
\begin{array}{rcl}
\omega_2 &=& \left| \omega_1 - r\omega \right| \\[10pt]
\varepsilon_{q-} = \varepsilon_{q+} &=& \frac{1}{2}(\omega_1 + \omega_2 + r\omega)
\end{array}
\end{equation}

\medskip\
 
\begin{figure}[!b] 
\centerline{\includegraphics[height=10cm,width=12cm]{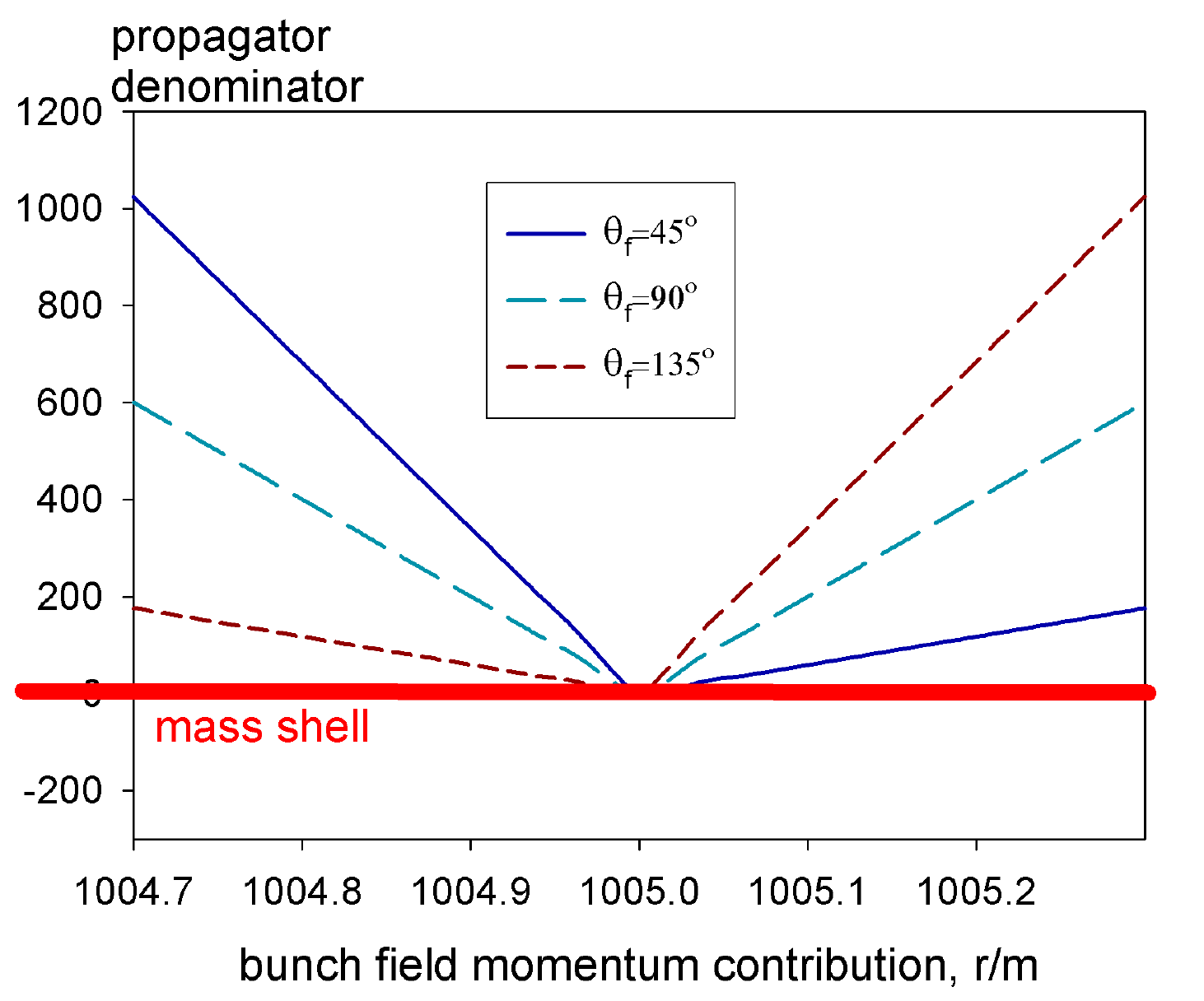}}
\caption{\bf The r value at which the peak 1 resonance occurs for
$\frac{\omega_1}{m}=1000$ and different values of scattering angle $\theta_f$.}
\label{c8.stppp.fig1}
\end{figure}

\begin{figure}[!b]
\centerline{\includegraphics[height=10cm,width=12cm]{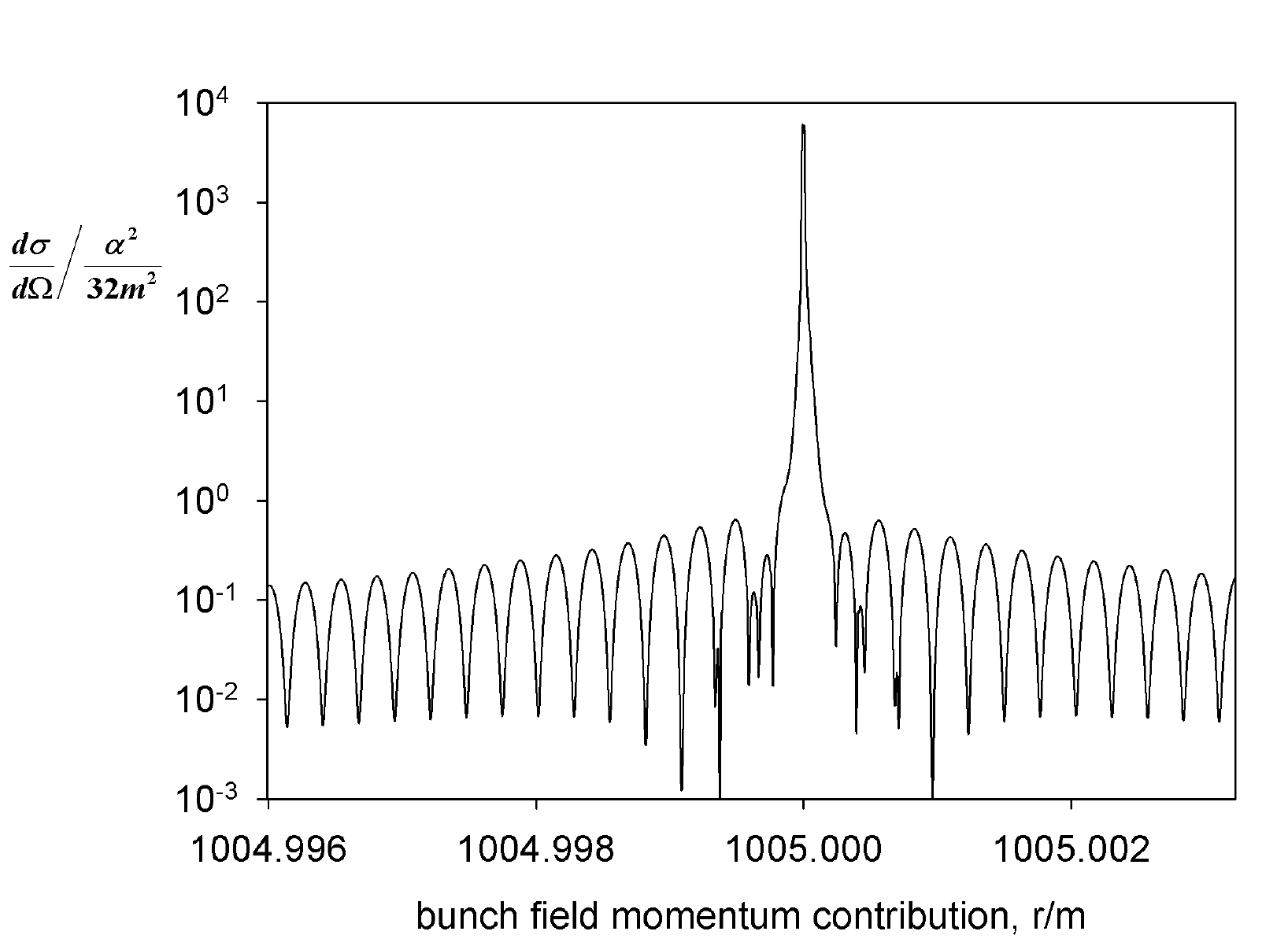}}
\caption{\bf A typical STPPP differential cross section resonance peak.}
\label{c8.stppp.fig2}
\end{figure}

Since the cross section contains an integration over $r$, there are an infinite number of reference frames.
However the frames travel at constant speed relative to each other and can be connected via
Lorentz transformations. The initial particle energies in the frame given by equation
\ref{c8stppp.frame} (the primed frame) are related to quantities in the centre of mass frame with
$r=0$ via the relativistic quantities $\gamma,\beta$ and the relations

\begin{equation}
\begin{array}{rcl}
\omega' &=& \gamma (1- \beta\,\text{sign}(r))\omega \\
\omega_1' &=& \gamma (1- \beta\,\text{sign}(r))\omega_1 \\
\omega_2' &=& \left| \omega_1' - r\omega' \right| \\
\varepsilon_{q-}' = \varepsilon_{q+}' &=& \frac{1}{2}(\omega_1' + \omega_2' + r\omega')
\end{array}
\end{equation}

\medskip\

The parameter $r$ appears in the denominators of STPPP propagators. 
Resonances in the cross section appear if these denominators 
(uncorrected by radiative corrections) reach zero. The regularised imaginary part 
of the EFEES is inserted to render resonances finite. There are two propagator 
denominators, one for each of the STPPP Feynman diagrams

\begin{equation}
\begin{array}{c}
  4 [(q_- k_1) + r (k \bar{p})]^2 + 4 (\varepsilon_{\bar{p}} \Im \Delta
  \varepsilon_{\bar{p}}^R)^2 \\[10pt]
  4 [(q_- k_2) + r (k \dbr{p})]^2 + 4 (\varepsilon_{\bar{p}} 
\Im \Delta  \varepsilon_{\bar{p}}^R)^2 
\end{array}
\end{equation}

\medskip\

and resonance occurs when

\begin{equation}
\begin{array}{c}
  r=-\dfrac{(q_-k_1)}{(k\bar{p})} \quad\quad\quad\text{peak 1 resonance}\\[10pt]
  r=-\dfrac{(q_-k_2)}{(k\dbr{p})} \quad\quad\quad\text{peak 2 resonance}
\end{array}
\end{equation}

\medskip\

The reference frame, kinematics and numerical regime considered, set the peak 1 resonance occurring
at large positive values of $r$ and the peak 2 resonance at large negative values.
At large absolute values of $r$ the cross section numerators are dominated by terms like
$r^2 Ai(r)^2$ which is very small for large negative values of $r$ and large for large 
positive values of $r$. The peak 2 resonances are therefore suppressed and the peak 1 resonances 
need to be considered further.

The variation of the peak 1 resonance point (the $r$ value) with the scattering angle $\theta_f$ is shown in figure 
\ref{c8.stppp.fig1}. The cross section values at resonance are 
determined by a combination of the highly oscillatory nature of Airy functions at 
large negative values of their argument, and the value of the imaginary part of the EFEES. 
A typical resonance is shown in figure \ref{c8.stppp.fig2}

\begin{figure}[!h] 
\centerline{\includegraphics[height=8cm,width=12cm]{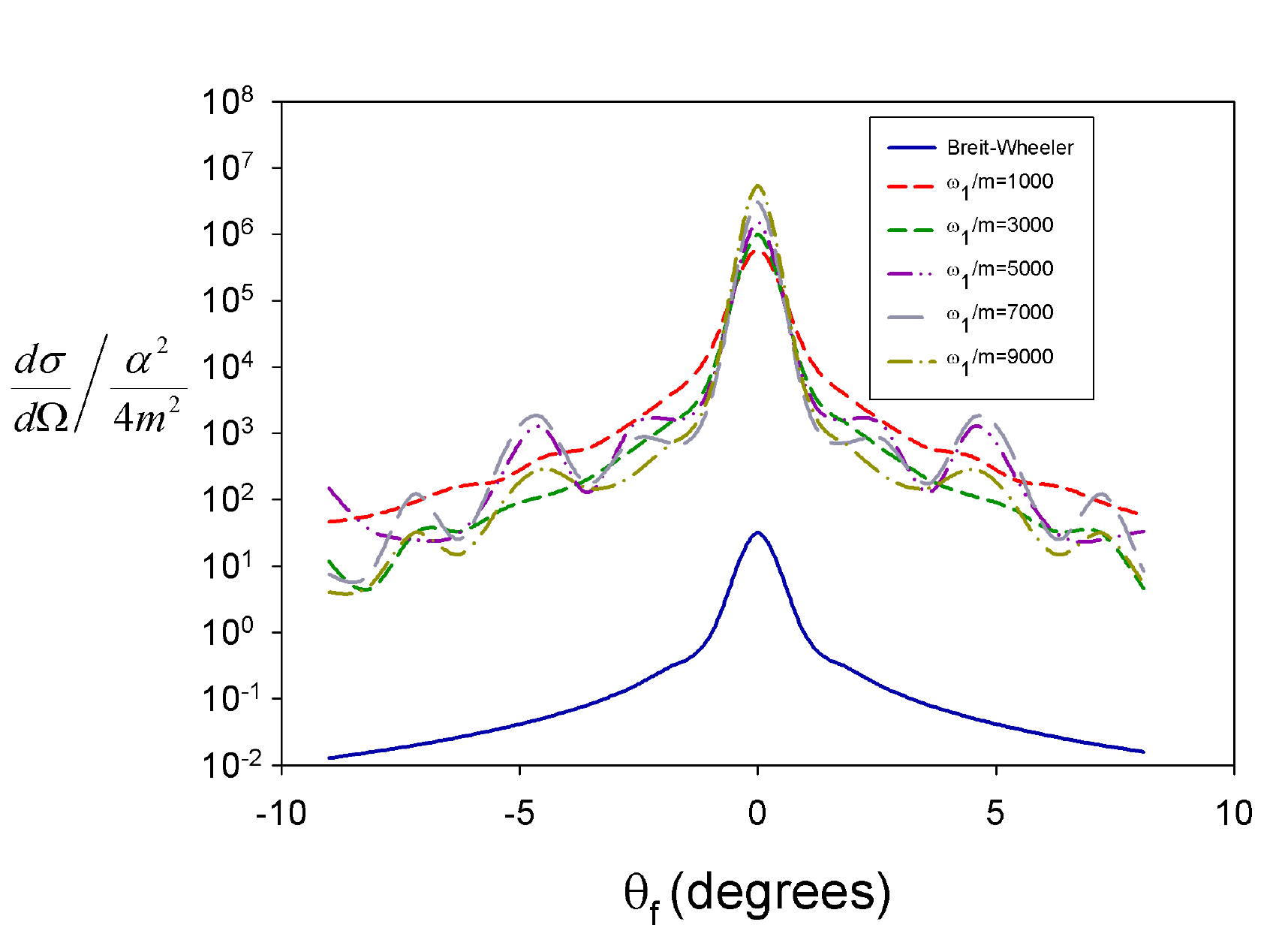}}
\caption{\bf The STPPP differential cross section for various 
values of the initial photon energy and electron scattering angle.}
\label{c8.stppp.fig3}
\end{figure}

\begin{figure}[!h]
\centerline{\includegraphics[height=8cm,width=12cm]{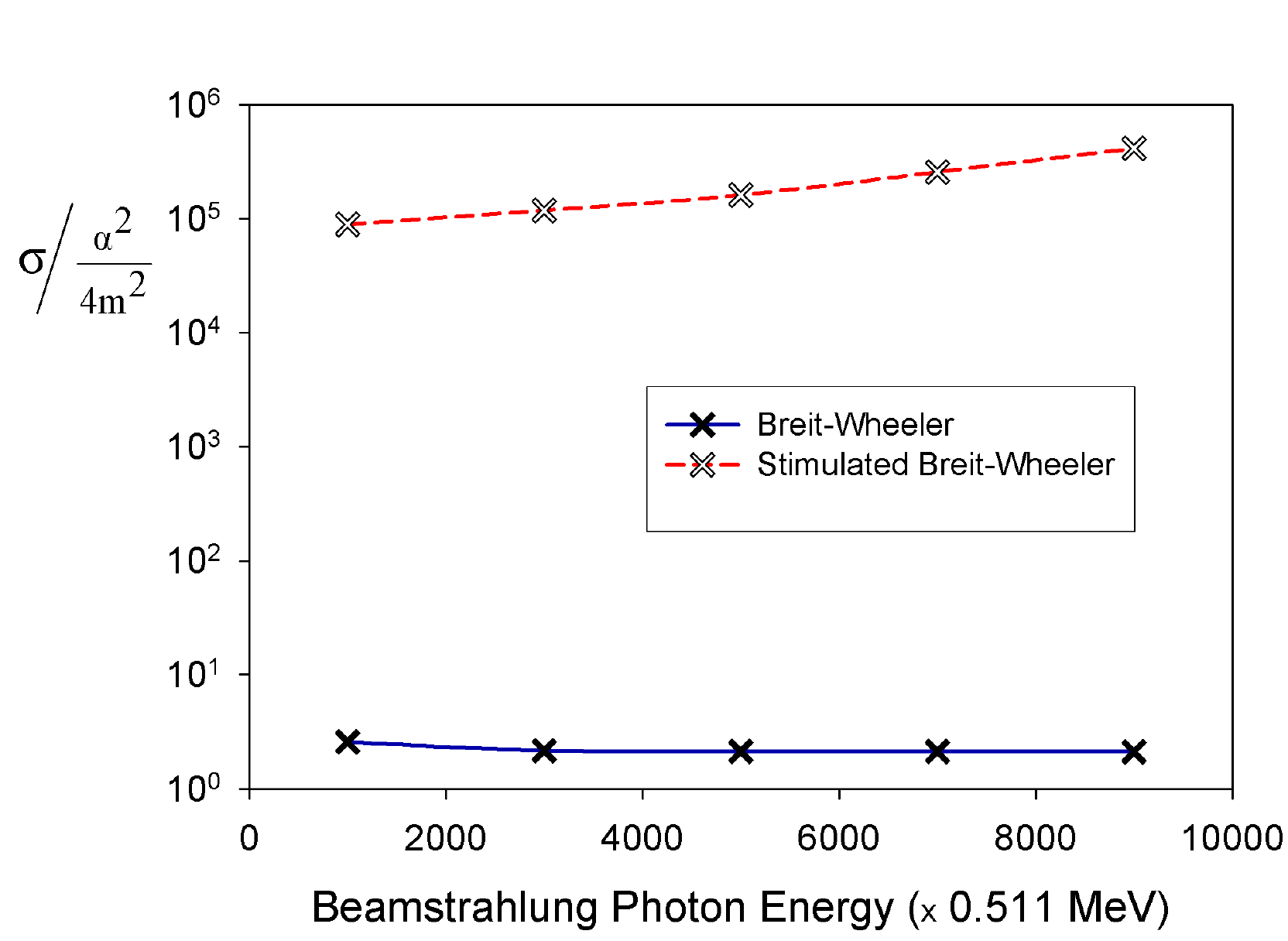}}
\caption{\bf The STPPP full cross section for various values of the initial photon energy.}
\label{c8.stppp.fig4}
\end{figure}

\clearpage

The STPPP differential cross section is to be integrated numerically over the resonance to calculate
the contribution to the overall cross section. The contribution from 
the peak 1 resonance is plotted in figure \ref{c8.stppp.fig3} 
for different scattering angles $\theta_f$ and different initial photon
energies.

The STPPP differential cross section is peaked at the same position as that of the Breit-Wheeler
process. However, as expected, it exceeds the ordinary Breit-Wheeler differential cross section by
several orders of magnitude. Integration over angles $\theta_f$ and $\varphi_f$ is straightforward
and the variation of the full cross section with beamsstrahlung energy $\omega_1$ is plotted in
figure \ref{c8.stppp.fig4}. Whereas the Breit-Wheeler cross section declines with increasing
$\omega_1$, the STPPP cross section increases. For the range of beamsstrahlung photon energies
expected at the ILC, the STPPP cross section exceeds the Breit-Wheeler process by 5 to 6 orders of
magnitude.

Overall, the Breit-Wheeler process contributes minutely to the expected number
of pairs at a typical ILC bunch collision (most come from Bethe-Heitler and
Landau-Lifshitz processes). However inclusion of the STPPP process should lead to a
significant increase in the expected number of pairs. The precise number is not quoted here 
because the numerical calculation was performed with the assumption of azimuthal symmetry and a 
separate study, taking into account real ILC bunch collisions, is required. 
It is believed that ILC processes that produce pairs from the STPPP process will not in fact be 
able to lay claim to azimuthal symmetry. It is known that the AiJ functions differ 
substantially from the Airy functions at large $r$. Therefore a  method of numerically 
calculating the AiJ functions becomes important for future work on this process.

\bigskip\
\section{Conclusion}

The differential cross sections for the external field electron energy shift 
(EFEES) and stimulated two photon pair production (STPPP) processes were analytically 
calculated for the case of a constant crossed external field. The 
EFEES calculation had been performed before \cite{Ritus72}, but the analytic results 
obtained here would agree only if azimuthal symmetry around the 
field was assumed. The Airy functions appearing in the EFEES had to be 
replaced with new AiJ functions.
The EFEES in the constant crossed electromagnetic field was investigated numerically and found 
to diminish as the external field intensity parameters $\nu^2$ 
and scalar product parameter $\rho=(kp)$ diminished. 

The STPPP in the constant crossed electromagnetic field was considered analytically and was 
found to depend on products of Airy functions (azimuthal symmetry) or products of AiJ 
functions (non azimuthal) symmetry. Two conditions for resonance were found corresponding to 
minimum values of the propagator denominator for each of the process Feynman diagrams.

Numerical calculations of the STPPP differential cross section, assuming azimuthal symmetry, 
were performed using parameter values expected at future linear colliders. A significant 
increase in the cross section of at 5 to 6 orders of magnitude was found compared 
to the equivalent, non external field, Breit-Wheeler process. As a result future linear collider bunch collisions are expected to produce 
more background pairs than were previously taken into account.

The calculations here can be improved on by further work. A satisfactory method of 
numerically calculating the AiJ functions at large values of their arguments 
should be developed and the numerical effect on the STPPP and EFEES 
cross sections calculated. The external field doesn't remain constant through a real bunch collision. The bunch is 
disrupted and the effect on the overall process cross sections should be gauged.

%% file: chap9.tex
The continued desire to test the theory of quantum electrodynamics by
theoretical and experimental investigation of the consequences of the theory
is the root motivation for the present work. With the presence of an
external electromagnetic field the basic processes of quantum
electrodynamics are altered and a new range of phenomena are introduced.

An investigation of the literature reveals that all the first order QED
processes in an external electromagnetic field have been considered.
Analytic expressions are well known and numerical consequences have been
investigated. The transition probabilities of these processes are optimised
when the external field intensity parameter is approximately unity, a
condition that became available to experimental testing within the last decade.

In contrast, the second order processes remain only partially investigated
for two main reasons. Firstly the presence of the external field renders the
analytic expressions much more complex for the second order processes than
for first order. Secondly it has been naively assumed that the transition
probabilities of these processes will be an order of the fine structure
constant smaller than those of the first order processes, thus rendering
them less amenable to experimental detection.

However some studies show that the possibility of second order IFQED resonant processes, 
due to the electron gaining a level structure when it is embedded in an electromagnetic plane wave. 
Approximate calculations reveal that differential cross sections can increase by several
orders of magnitude at resonance, suggesting experimental investigation. 
What is needed is a precise theoretical prediction of the parameters associated with the phenomena.

In this thesis, to that end, a detailed and unique examination of some of the
second order IFQED processes, and the experimental conditions under which
resonance can be achieved, was carried out. Both the stimulated Compton
scattering (SCS) and stimulated two photon pair production (STPPP) processes were 
investigated without recourse to special kinematics or non relativistic restrictions. 
The external field that was considered was a plane wave electromagnetic field 
for which the Dirac equation can be solved exactly. Two particular plane waves were considered. A 
circularly polarised field was used in both SCS and STPPP processes. A constant crossed electromagnetic 
field, being a reasonable representation of the field of an ultra relativistic charge bunch was used in 
applying the STPPP calculation to the interaction region of a $e^+e^-$ collider. 
The advantages of using a circularly polarised 
field was two-fold, in that high powered lasers which produce intense electromagnetic fields in 
experimental investigations, readily achieve this state of polarisation.
Also, the solutions of the Dirac equation for a circularly polarised electromagnetic field contain the 
well known Bessel functions. Solutions of the Dirac equation for a constant crossed electromagnetic field 
contain Airy-type functions.

The external field was chosen to have a Lorentz gauge in which the three
vectors defining the external field form a Cartesian coordinate system and
in so doing provided a simplification of cross section expressions. In order
to write down the expressions for the scattering cross section, Feynman's
formulation of S-matrix theory was used in the Bound Interaction Picture
which has proved to be the most convenient way of performing external field
problems.

The scattering matrix element for the SCS process was written down with the
aid of Feynman diagrams, and the Volkov electron wave function and the Ritus
representation for the external field electron propagator were used.
Integrations over space-time were achieved by writing the Volkov wave
functions as infinite summations of Bessel and exponential functions. The
infinite summations contain integer multiples of the external field
4-momentum which were interpreted, in the usual way, as the contribution to
the process of quanta from the external field.

The square of the matrix element and a summation over electron spin and
photon polarisation was performed to produce the usual trace expression of
Dirac gamma matrices and the 4-vectors which describe the scattering. The
scale of the trace sum required\footnote{An algebraic expansion 
of the trace sum results in over 100,000 terms.}
was reduced, in the first instance by use of the reference frame in which
the initial electron is at rest. A further simplification was achieved by
expanding the trace sum into 22 independent trace sums which served as a
base from which all other trace sums were calculated via a mapping of
scattering parameters.

Calculation of the SCS cross section required an integration over
final particle momenta which was transformed to an integration over the solid
angle into which the SCS photon was scattered. This integration could not be
performed analytically due to the complexity of the integrand. Further, the
cross section contained infinite summations over products of Bessel
functions, exponential functions and algebraic factors for which no analytic
solution is known. However an original solution of a similar infinite summation was provided 
which may provide the basis for a solution of the infinite summations of interest.
These infinite summations  are associated with all
calculations of second (and probably higher) order IFQED processes for the
case of a circularly polarised electromagnetic field. In this work these
infinite summations were performed numerically. The number of
terms required for numerical convergence was roughly proportional to twice the ratio
of initial photon energy to external field energy.

As a test of the correctness of the obtained differential cross section expressions, limiting
cases of the scattering parameters were investigated. For the cases of
vanishing external field and vanishing initial photon, the expressions obtained in this thesis agreed
with those of the literature. However for the special kinematic case of
parallel initial photon and laser field the differential cross section obtained here  differed slightly 
from that obtained by \cite{AkhMer85}. Numerical investigation showed that whereas the results of this 
thesis always return positive values for the differential cross section, the expressions obtained by 
\cite{AkhMer85} sometimes return incorrect negative values.

The stimulated two photon pair production (STPPP) process was also considered. 
The expression for the STPPP matrix element square was written down
directly with the aid of the SCS expressions and a crossing symmetry which
linked the two sets of Feynman diagrams. This crossing symmetry could be used even with 
the presence of the external field. In obtaining the STPPP differential
cross section a suitable reference frame was chosen and the phase space integral integration variables were 
transformed into an integration over the element of solid angle into which the electron passes after its 
creation. 

Numerical evaluation of the SCS and STPPP differential cross sections was
carried out for a range of process parameters. For the SCS process this parameter range included 
incident photon energy ($0.001\leq \omega _i\leq 25.6\;keV$), external
field quanta energy ($0.001\leq \omega \leq 51.2\;keV$), incident photon
angle ($0^{\circ }\leq \theta _i\leq 360^{\circ }$), final photon
scattering angles ($0^{\circ }\leq \theta _f,\phi _f\leq 360^{\circ }$) and
intensity of the external field ($0\leq \nu ^2\leq 2$). The STPPP
parameters range included incident photon energy ($0.061\leq \omega
_1,\omega _2\leq 5.12\;MeV$), external field quanta energy ($0.051\leq
\omega \leq 5.12\;MeV$), incident photon angle ($0^{\circ }\leq \theta
_i\leq 180^{\circ }$), final electron scattering angles ($-180^{\circ
}\leq \theta _f,\phi _f\leq 180^{\circ }$) and intensity of the external
field ($0\leq \nu ^2\leq 2$).

Analysis of the numerical results obtained for both processes was simplified
by considering separate summation terms in the differential cross section
expressions. These summation terms corresponded to the number, $l$ of
external field quanta which contributed to the process. Each summation term
of the differential cross section itself contained an infinite summation of terms
(denoted by $r$). The differential cross section was investigated with
respect to the simplest of these summation terms for the case of collinear
initial photons and external field propagation direction (i.e. the $r=-1,0,1$
terms of the $l=0$ external field quanta contribution for $\theta
_i=0^{\circ }$).

Analysis of the numerical data presented for both the SCS and STPPP processes
was made in terms of the effect of the external field.
A general relationship was found between the external field intensity and
the number of external field quanta terms which contribute significantly to
the differential cross section. The joint increase in both is related to the
tendency of external field quanta to participate in the interaction as the
external field photon number density increases.

The differential cross section of the SCS process for the $l=0,\theta _i=0$
case was compared with the Klein-Nishina differential cross section. The greatest difference between the 
two occurred at $\theta_f=180^{\circ }$. A maximum enhancement of the SCS differential cross section took 
place when the external field energy was in excess of the initial photon energy, and a
maximum diminishment at the same final scattering angle when the external
field energy was smaller than initial photon energy. The variations from the
Klein-Nishina differential cross section were explained in terms of the
momentum transferred from the external field to the SCS electron.

The angular variation of the SCS $l$ contributions was examined. The observed 
oscillatory behaviour of the differential cross section was explained by 
oscillatory momentum components gained by the electron due to its interaction with the external 
field. These momentum components were described as a Fourier
series of discrete harmonic contributions from the external field.
The oscillatory behaviour of the differential cross section was closely allied 
to the magnitude of the Bessel function arguments present in the differential cross section expressions. 
The larger the arguments, the more oscillatory the behaviour of the SCS differential cross section as
the polar scattering angle varied from $0^{\circ }$ to $360^{\circ }$.

Further explanation of differential cross section variation lay in the
comparison of the relative strengths of the $l$ contributions. When the
ratio $\frac{\omega _i}\omega $ was less than unity, the $l=0$ contributions
were exceeded by other contributions. The converse was true when 
$\frac{\omega _i}{\omega}$ was greater than unity. The analysis of the SCS
differential cross section variation was confirmed by comparison with theory
of radiation by an accelerated classical electron

The variation of the SCS differential cross section $l$ contributions were
also investigated for a scattering geometry in which the initial photon
momentum makes an angle of $45^{\circ }$ to the external field propagation
direction. For this geometry the electron embedded in the external field
receives a static momentum component along the direction of propagation of
the external field, an oscillatory component along the direction of
propagation of the electromagnetic field, as well as a momentum component
from the initial photon. Analysis of the SCS differential cross section was
given in terms of these competing influences and the behaviour predicted was
close to that obtained.

The investigation of the variation of the SCS differential cross section
summed over all $l$ contributions enabled the overall effect of the external
field on the process to be determined. In large part, analysis of the
behaviour for this case was in terms of the relative strengths of the $l$
contributions considered separately.

The main feature of the SCS differential cross section as the initial
scattering angle varied from $0^{\circ }$ to $90^{\circ }$ was the
development of peaks in the $\theta _f=90^{\circ }$ direction. These peaks
have their origin in the transverse component of momentum received from the
initial photon and the longitudinal component received from the external
field. The greater the longitudinal component the greater the probability
that the final states of the process with $\theta _f=90^{\circ }$ carry the
transverse momentum. The argument was supported by an examination of trends
in the relevant components.

The effect on the SCS process of increasing external field intensity was a generally enhanced 
cross section. External field photon number density increases, as does the probability that $l$ 
external field photons will participate in the process. Increase in external field 
intensity also has the effect of slightly shifting differential cross section peaks.

The location of differential cross section peaks was also determined by the
ratio of the initial photon energy with the product of external field
intensity and energy. With this ratio comparable to or less than unity,
differential cross section peaks indicate the preferred direction of
radiation from the SCS electron was transverse to the external field
propagation direction.

With variation in the initial scattering geometry, the relative peak heights
of the $\theta _f=90^{\circ }$ and $\theta _f=270^{\circ }$ peaks vary. This
variation was explained in terms of two competing processes in which
momentum contributions preference scattering in one direction whereas the
energy level structure of the electron in the external field preference
scattering in another.

The predicted non azimuthal symmetry of the SCS process whenever $\theta _i
\neq 0$ was confirmed by the numerical evaluations. Comparison of
various scattering geometries indicated that the SCS differential
cross section increased at the greatest rates with increase in the external
field intensity when the initial photon propagation direction is not
parallel with the external field propagation direction.

Numerical presentation and analysis of the STPPP differential cross section followed a similar order
to that of the SCS process. The STPPP differential cross section for
the $r=0,l=0,\theta_1=0^{\circ}$ case was investigated as the external field
intensity was increased. The zero external field case corresponded to
Breit-Wheeler scattering with the dominant feature being the characteristic 
$\theta _f=-180^{\circ },0^{\circ },180^{\circ }$ peaks. Increasing external field 
intensity led to a general decrease in the differential cross section. This was 
explained by the increased energy required to produce the fermion pair.

The complete STPPP differential cross section $l=0$ contribution was either enhanced or 
diminished with respect to the Breit-Wheeler differential cross section. This was  dependent on whether 
the ratio of initial photon energy to external field energy was greater than or less than unity.
STPPP differential cross section peaks varied with increasing
external field intensity. Peaks were broadened in
relation to the ratio of initial photon energy to external field energy. An evident 
secondary peak structure was associated with the produced electron gaining oscillatory momentum 
components parallel to the external field propagation direction.

The STPPP differential cross section $l$ contribution peaks varied depending on the value of $l$.
The $\theta _f=0^{\circ }$ peaks increased as $l$ increased and the $\theta _f=-$180$^{\circ
},180^{\circ }$ peaks decreased. The explanation was given in terms of the
momentum contribution from the external field which increased the
probability that the fermion pair momentum was parallel rather than
anti-parallel to the external field propagation direction.

The STPPP differential cross section was investigated for initial scattering geometries
other than $\theta _1=0^{\circ }$. With increasing external field
intensity differential cross section peaks appeared both in directions
parallel to initial photon momenta and parallel to the external field
propagation direction. The relative strengths of the peaks was in
part determined by the ratio of initial photon energy to external field
energy, with the largest peaks occurring in directions collinear to the most
energetic particles.

STPPP differential cross section peaks displayed a tendency to resolve into
double peaks whenever the external field intensity increased. 
The explanation was in terms of two competing trends, the increased probability that the 
fermion pair would be created with momentum parallel to the external field propagation 
direction, and the decreased probability that the external field quanta would
participate in the process.
Variation of the STPPP differential cross section with the final azimuthal
angle revealed no change when $\theta _1=0^{\circ }$, a result consistent
with the azimuthal symmetry of the process for that scattering geometry.

Further investigation of the variation of the STPPP differential
cross section with external field intensity revealed a general increase of
both, with the largest rates of differential cross section increase recorded
for scattering geometry in which $\theta _1=0^{\circ }$. In some cases
however the differential cross section diminished initially before
increasing with increasing external field intensity. The explanation was in
terms of two competing tendencies, one which increased the energy required
to produce the fermion pair and the other which increased the likelihood
that external field quanta will participate in the process.
In some circumstances a "stepped" increase in the STPPP differential
cross section with increasing external field intensity was observed. This
phenomenon was related to the existence of a energy threshold for the pair
production process to occur. As $\nu^2$ increased so did the rest mass of the bound fermion 
requiring more external field quanta to participate as a necessary condition for the fermion
creation. Close examination revealed that the ''steps'' in the differential
cross section were not discontinuous.

Both the SCS and STPPP differential cross sections contained resonant
infinities which corresponded to points at which the energy of the
intermediate electron was exactly equivalent to an energy level of the
fermion-external field system. The presence of resonant infinities indicated
that the interaction between the intermediate electron and the virtual
electron-positron field had not been taken into account. Such an interaction results in a 
shift of electron energies and the shift is affected by the presence of the external field.
This external field electron energy shift (EFEES) calculation was carried out using
the Volkov function representation for the electron propagator and an average over 
electron spin. Dispersion relations were used to separate out
real and imaginary parts. The final result consisting of a rapidly
convergent infinite summation and a single integration of a function of
products of Bessel functions was confirmed via the optical theorem and the
known results for first order IFQED processes. The real part of the EFEES reduced to zero at 
resonance points. However, only the imaginary part of the EFEES was necessary for the 
removal of the resonant infinities. The imaginary part of the EFEES was investigated 
numerically with respect to variation with the external field intensity and variation of a 
parameter $\rho $ consisting of a scalar product of the external field and electron 4-momenta.

Investigation of the summation terms of the imaginary part of the EFEES
revealed that convergence was achieved within the first ten terms for the
range of parameter values of interest. Analysis of the numerical variation 
was facilitated by considering the expression for the
recoil momentum of the electron scattered by its own field and the external
field. A comparison was drawn with the theory of radiation by an electron in
the field of a nucleus in which the probability of radiation bears an
inverse relationship to the recoil momentum.
The EFEES imaginary part increased linearly with respect to $\rho$ and $\nu^2$ 
except when $\rho$ and $\nu^2$ were large. This behaviour was explainable in terms of the 
analytic form of the recoil momentum.

A comparison was made between the numerical results achieved here and the approximation
formulae obtained by \cite{BecMit76}. Good agreement was found within the validity of the 
approximation, but as much as a $13.4\,\%$ error for other required parameter values.
The full expressions for the EFEES imaginary part obtained here were necessary for 
further calculation of the SCS and STPPP resonant differential cross sections.

SCS and STPPP resonant cross sections were rendered finite by the insertion
of EFEES imaginary part into the external field electron propagator
denominator. However the presence of divergences in the EFEES expressions 
required an appropriate regularisation and renormalisation procedure
which also took into account the external field. Regularisation and renormalisation was 
applied to the self energy (EFESE) and was shown to be the same as for the non external field 
case. The EFESE was inserted into the denominator of the bound electron propagator and was 
shown to be equivalent to the insertion of the energy shift (EFEES) to order $\alpha$.

\enlargethispage*{2cm}
Analytic expressions for the location, spacing and widths of the SCS and STPPP resonant 
differential cross sections were developed. Resonance heights
were obtained by recalculating the SCS and STPPP differential cross sections
at the parameter values at which the resonance occurred.
The resonant differential cross sections were investigated for various sets of parameters. 
These were $\nu ^2=0.1,1$, $\omega =25.6,61.4$ keV and $\omega _i=51.2$ keV for the SCS 
process, and $\nu ^2=0.1,0.5$, $\omega =0.768,1.024$ MeV and $\omega _1,\omega
_2=0.409,0.768,1.28$ MeV for the STPPP process. Some differential cross sections exceeded
the non resonance differential cross sections, the differential
cross sections of the equivalent first order processes and the differential
cross sections of the equivalent non external field processes by several orders of magnitude.

After integration over scattering angles, with particular care paid to the points of resonance, full 
SCS and STPPP cross sections were obtained. A general increase of the cross section with external
field intensity was established. However, the SCS full cross-section
sharply peaked at particular values of $\nu^2$ (0.2 for the particular parameter set considered) 
and the STPPP cross section enhancement was generally smaller. Nevertheless the SCS cross section
exceeded the Klein Nishina cross section by 5 to 6 orders of magnitude at peak, and the STPPP cross
section exceeded the Breit-Wheeler cross section by one order of magnitude by $\nu^2=0.5$.

This result was recognised as being of importance. IFQED experimental efforts to date have 
been directed towards the first order IFQED processes. It can be naively
assumed that the first order processes have larger cross sections and are
hence more detectable because of the lower order of the fine structure
constant contained in their expressions. However the extent of the SCS and STPPP 
resonant cross sections indicated that this is not the case.
The appearance of large resonant cross sections at
relatively low external field intensities is experimentally beneficial as moderately 
intense lasers should be sufficient to test theoretical predictions of the second order 
IFQED calculations.

In agreement with \cite{AkhMer85} no resonances were found at 
$\theta_i=0^{\circ }$ for the SCS process. This result vindicated the 
original decision to calculate the SCS and STPPP
differential cross sections without the benefit of special kinematics.

Generally, more resonances occur for the STPPP process than for
the SCS process. This was due to the greater range of fermion energies traversed (and 
hence fermion quasi-energy levels traversed) for the investigated STPPP parameter range.
Resonance widths were $\sim 4^{\circ }$ for the SCS differential cross section and 
$\sim 0.1^{\circ }$ for the STPPP differential cross section as $\nu ^2$ approached 
its upper limit.

Analytic work for the STPPP process was applied to collisions between $e^+e^-$ 
bunches at proposed future linear colliders. These relativistic bunches 
produce an electromagnetic field that is essentially constant and transverse to the 
bunch motion. The Volkov solution could be used once its form was found for the case of 
constant crossed external field. Fourier transforms were introduced and the 
Volkov solution contained Airy functions only if azimuthal asymmetry 
was assumed. More generally, the transforms were integrations over products of Bessel 
functions and cos functions of Airy-type arguments. These transforms were labelled 
AiJ.

Investigation of cross section resonances for the STPPP process in charge bunch collisions 
required calculation of the EFEES in a constant crossed electromagnetic field. This 
calculation contained poles at the lower limit of certain integrations.
These were avoided by introducing an infinitesimal photon mass and transforming integration
variables. As for the STPPP process in a constant crossed 
electromagnetic field, the EFEES in the same field contained products of Airy 
(or AiJ) functions. The Airy functions diverged for large negative values 
of its argument and provided difficulty in carrying out numerical calculations. 
The AiJ functions, however, are damped for both large negative and large 
positive values of their argument. In future work - once 
numerical methods are developed for computing the AiJ functions - calculation 
of the constant crossed EFEES and STPPP processes over all parameter ranges should be 
possible.

\enlargethispage*{2cm}
The constant crossed EFEES was inserted into bound propagator denominators and 
the STPPP differential and full cross sections were calculated. Using numerical values of 
parameters suggested by the expected operation of the future linear colliders, the 
contribution to the STPPP differential cross section was at 5 to 6 orders of magnitude 
greater than the Breit-Wheeler process (i.e. the STPPP process in the absence of the field). 
The expected end result is that there will be significantly more background pairs per 
bunch crossing at the ILC. More detailed investigation of the parameter range awaits future work.

Experimental conditions required for the detection of these
resonances were considered. Operating parameters of the present generation of high-powered 
lasers should be sufficient to detect the theoretically predicted SCS and STPPP resonant 
cross sections. The potential cross sections measured by a stationary particle detector which 
subtends the top half of the largest SCS and STPPP resonance peaks, were calculated. 
The predicted event rates should be easily observable.

%% file: appendicies.tex
\chapter{The Jacobian of the Transformation $d^4p\rightarrow d^4q$}
\label{jacob}

The cross section calculations in this thesis involve transformation of
integration variable between the free electron 4-momentum $p_\mu $ and the
electron 4-momentum shifted by the presence of the external field $q_\mu
\;(\,=p_\mu -\frac{\nu ^2}{2\omega (q_0-q_3)}\,k_\mu \;)$. The Jacobian of
this transformation is written in equation \ref{ap1.eq1}, where the
subscripts $1,2,3$ refer to the time, x, y and z axes respectively

\begin{equation}
\label{ap1.eq1}d^4p=J\,d^4q\quad ;\quad J=\left| 
\begin{array}{cccc}
\frac{\partial p_0}{\partial q_0} & \frac{\partial p_0}{\partial q_1} & 
\frac{\partial p_0}{\partial q_2} & \frac{\partial p_0}{\partial q_3} \\ 
\frac{\partial p_1}{\partial q_0} & \frac{\partial p_1}{\partial q_1} & 
\frac{\partial p_1}{\partial q_2} & \frac{\partial p_1}{\partial q_3} \\ 
\frac{\partial p_2}{\partial q_0} & \frac{\partial p_2}{\partial q_1} & 
\frac{\partial p_2}{\partial q_2} & \frac{\partial p_2}{\partial q_3} \\ 
\frac{\partial p_3}{\partial q_0} & \frac{\partial p_3}{\partial q_1} & 
\frac{\partial p_3}{\partial q_2} & \frac{\partial p_3}{\partial q_3} 
\end{array}
\right| 
\end{equation}

\medskip

In finding the Jacobian of this transformation we must\ make use of a
coordinate system in which the external field photons propagate along the
z-axes, $k_\mu ^{}=(\omega ,0,0,\omega )$ , with the result that the free
electron momentum can be written

\begin{equation}
\label{ap1.eq2}p_\mu =q_\mu +\frac{e^2a^2}{2\omega (q_0-q_3)}\,k_\mu 
\end{equation}

\medskip\ The Jacobian $J$ then reduces to

\begin{equation}
\label{ap1.eq3}J=\left| 
\begin{array}{cccc}
\frac{\partial p_0}{\partial q_0} & 0 & 0 & \frac{\partial p_0}{\partial q_3}
\\ \frac{\partial p_1}{\partial q_0} & 1 & 0 & \frac{\partial p_1}{\partial
q_3} \\ \frac{\partial p_2}{\partial q_0} & 0 & 1 & \frac{\partial p_2}{%
\partial q_3} \\ \frac{\partial p_3}{\partial q_0} & 0 & 0 & \frac{\partial
p_3}{\partial q_3}
\end{array}
\right| =\frac{\partial p_0}{\partial q_0}\frac{\partial p_3}{\partial q_3}-
\frac{\partial p_0}{\partial q_3}\frac{\partial p_3}{\partial q_0}=1
\end{equation}
\chapter{The Full Expressions and Trace results for $\limfunc{Tr}\,Q_1$ and $
\limfunc{Tr}\,Q_2$}
\label{trace}

In this appendix the complete trace expressions for $\limfunc{Tr}\,Q_1$ 
and $\limfunc{Tr}\,Q_2$ are presented. The trace expressions can be algebraically expanded and
subdivided into 17 independent trace coefficients for $\limfunc{Tr}\,Q_1$
and 15 independent trace coefficients for $\limfunc{Tr}\,Q_2$ . These
independent trace coefficients can be designated in terms of their
co-products, the functions $X_{ij}^{r,r'}$. 

\begin{equation}
\begin{array}{rcl}
\bar X_{ij}^r & \equiv & \dfrac{\bar M_{ir}\bar N_{jr}}{(\bar p_r^2-m^2)} \\[10pt] 
C1(\bar p,p_f,\gamma _\mu ) & \equiv & \gamma _\mu - 
\dfrac{e^2a^2}{4(k\bar p)(kp_f)}\slashed{k}\gamma _\mu \slashed{k} \\[10pt] 
C2(\bar p,p_f,\gamma _\mu ) & \equiv & -\dfrac e{2(k\bar p)}\gamma _\mu 
\slashed{a}_1\slashed{k}+\frac e{2(kp_f)}\slashed{a}_1\slashed{k}\gamma _\mu \\[10pt] 
C3(\bar p,p_f,\gamma _\mu ) & \equiv & -\dfrac e{2(k\bar p)}\gamma _\mu 
\slashed{a}_2\slashed{k}+\frac e{2(kp_f)}\slashed{a}_2\slashed{k}\gamma _\mu 
\end{array}
\end{equation}

\medskip\ 

\section{The results for $\limfunc{Tr}\,Q_1$}

$$
\begin{array}{rl}
\limfunc{Tr}\,Q_1(\bar p_r,\bar p_{r\prime }) & = \limfunc{Tr}(\slashed{p}_f+m)
\left[ C1(\bar p,p_f,\gamma _\mu )\bar N_{1r}+C2(\bar p,p_f,\gamma _\mu
)\bar N_{2r}+C3(\bar p,p_f,\gamma _\mu )\bar N_{3r}\right]  \\[10pt]
& \times \;\;  \left[ \dfrac{\slashed{\bar{p}}_r+m}{\bar p_r^2-m^2}\right] 
\left[ C1(\bar p,p_i,\gamma _\nu )\bar M_{1r}+C2(p_i,\bar p,\gamma _\nu )\bar
M_{2r}+C3(p_i,\bar p,\gamma _\nu )\bar M_{3r}\right]  \\[10pt]
& \times \;\; (\slashed{p}_i+m)\left[ C1(\bar p,p_i,\gamma _\nu )\bar M_{1r^{\prime
}}^{*}+C2(\bar p,p_i,\gamma _\nu )\bar M_{2r^{\prime }}^{*}+C3(\bar
p,p_i,\gamma _\nu )\bar M_{3r^{\prime }}^{*}\right]  \\[10pt]
& \times \;\; \left[ \dfrac{\slashed{\bar{p}}_{r\prime }+m}{\bar p_{r^{\prime 
}}^2-m^2}\right] ^{*}\left[C1(\bar p,p_f,\gamma _\mu )\bar N_{1r^{\prime }}^{*}
+C2(p_f,\bar p,\gamma_\mu )\bar N_{2r^{\prime }}^{*}+C3(p_f,\bar p,\gamma _\mu )\bar
N_{3r^{\prime }}^{*}\right] 
\end{array}
$$

\medskip\ 

\newpage
\underline{The coefficient of $\bar X_{11}^r[\bar X_{11}^{r^{\prime }}]^{*}$}

$$
\begin{array}{l}
16\left\{ 4m^2\left[ (\bar p_r\bar p_{r^{\prime }})+m^2\right] +(p_f\bar
p_{r^{\prime }})(p_i\bar p_r)+(p_f\bar p_r)(p_i\bar p_{r^{\prime
}})+2e^4a^4-8e^2a^2m^2\right\} \\[10pt] 
-16(2m^2-e^2a^2)\left[ (p_f\bar p_r)+(p_i\bar p_r)+(p_f\bar p_{r^{\prime
}})+(p_i\bar p_{r^{\prime }})\right] \\[10pt] 
-16\left[ (\bar p_r\bar p_{r^{\prime }})-m^2\right] \left\{
(p_ip_f)+e^2a^2\left[ {\dfrac{(kp_i)}{(k\bar p)}}+{\dfrac{(kp_f)}{(k\bar p)}}%
\right] \right\} 
\end{array}
$$

\medskip\

\underline{The coefficient of $\bar X_{11}^r[\bar X_{12}^{r^{\prime }}]^{*}$}

$$
\begin{array}{l}
8\left[ 
{\dfrac 1{(kp_f)}}-{\dfrac 1{(k\bar p)}}\right] \left[ (\bar p_r\bar
p_{r^{\prime }})-m^2\right] (\bar \alpha _{1i}-\bar \alpha _{{1f}})
(kp_i)(kp_f) \\[10pt]
 -8\left[ 4m^2-2e^2a^2-(p_i\bar p_r)-(p_i\bar p_{r^{\prime
}})\right] \bar \alpha _{1f} \\[10pt]
+16(r-r^{\prime })(kp_i)\left[ (k\bar p)+(kp_f)\right] \bar \alpha _{1i} 
\end{array}
$$

\medskip\ 

\underline{The coefficient of $\bar X_{11}^r[\bar X_{22}^{r^{\prime }}]^{*}$}

$$
\begin{array}{l}
-16(k\bar p)\left[ (kp_i)+(kp_f)\right] \bar \alpha _{1i}\bar \alpha _{1f}
\\[10pt]
+8e^2a^2\left\{ [\bar p^2+m^2]\left[ 
{\dfrac{(kp_i)}{(k\bar p)}}+{\dfrac{(kp_f)}{(k\bar p)}}\right] +(p_ip_f)\left[
1+{\dfrac{(kp_i)}{(k\bar p)}}\right] -(p_i\bar p_r)\left[ {\dfrac{(kp_f)}{%
(k\bar p)}}+{\dfrac{(kp_f)}{(kp_i)}}\right] \right. \\[10pt]
 \left. (p_i\bar p_{r^{\prime }})\left[ 1+{\frac{(k\bar p)}{(kp_i)}}\right] -(p_f\bar
p_r)\left[ {\frac{(kp_i)}{(k\bar p)}}+{\frac{(kp_i)}{(kp_f)}}\right]
-(p_f\bar p_{r^{\prime }})\left[ 1+{\frac{(k\bar p)}{(kp_f)}}\right]
\right\} 
\end{array}
$$

\medskip\ 

\underline{The coefficient of $\bar X_{12}^r[\bar X_{12}^{r^{\prime }}]^{*}$}

$$
\begin{array}{l}
8e^2a^2(kp_f)\left[ {\dfrac 1{(kp_f)^2}}+{\dfrac 1{(k\bar p)^2}}\right]
\left\{ \left[ 4m^2-(p_i\bar p_r)-(p_i\bar p_{r^{\prime }})-2e^2a^2\right]
(k\bar p)+\left[ (\bar p_r\bar p_{r^{\prime }})-m^2\right] (kp_i)\right\} 
\end{array}
$$

\medskip\ 

\underline{The coefficient of $\bar X_{12}^r[\bar X_{21}^{r^{\prime }}]^{*}$}
$$
\begin{array}{l}
16\left[ (kp_i)(kp_f)+(k\bar p)^2\right] \bar \alpha _{1i}\bar \alpha _{1f}
\\ 
-8e^2a\left\{ ^2[\bar p^2-m^2]\left[ 1+ 
{\dfrac{(kp_i)(kp_f)}{(k\bar p)^2}}\right] -2m^2\left[ {\dfrac{(kp_i)}{(k\bar
p)}}+{\dfrac{(kp_f)}{(k\bar p)}}\right] -(p_ip_f)\left[ 1+{\dfrac{(k\bar p)}{%
(kp_i)}}\right] \left[ 1+{\dfrac{(k\bar p)}{(kp_f)}}\right] \right. \\[10pt] \left.
+(p_i\bar p_{r^{\prime }})\left[ {\dfrac{(kp_f)}{(kp_i)}}+{\dfrac{(kp_f) }{%
(k\bar p)}}\right] +(p_i\bar p_r)\left[ 1+{\dfrac{(k\bar p)}{(kp_i)}}\right]
+(p_f\bar p_r)\left[ {\dfrac{(kp_i)}{(k\bar p)}}+{\dfrac{(kp_i)}{(kp_f)}}%
\right] +(p_f\bar p_{r^{\prime }})\left[ 1+{\dfrac{(k\bar p)}{(kp_f)}}\right]
\right\} 
\end{array}
$$

\medskip\ 
\newpage
\underline{The coefficient of $\bar X_{11}^r[\bar X_{23}^{r^{\prime }}]^{*}$}

$$
\begin{array}{l}
8\left[ (k\bar p)+(kp_f)\right] \left[ (k\bar p)+(kp_i)\right] (\bar \alpha
_{1i}\bar \alpha _{2f}-\bar \alpha _{2i}\bar \alpha _{{1f}}) 
\end{array}
$$

\medskip\ 

\underline{The coefficient of $\bar X_{12}^r[\bar X_{22}^{r^{\prime }}]^{*}$}

$$
\begin{array}{l}
8e^2a^2(kp_f)(k\bar p)\left[ \dfrac 1{(kp_f)^2}+{\dfrac 1{(k\bar p)^2}}\right]
\left[ (kp_i)-(k\bar p)\right] \bar \alpha _{1i} 
\end{array}
$$

\medskip\ 

\underline{The coefficient of $\bar X_{12}^r[\bar X_{23}^{r^{\prime }}]^{*}$}

$$
\begin{array}{l}
8e^2a^2(kp_f)(k\bar p)\left[ \dfrac 1{(kp_f)^2}-\dfrac 1{(k\bar p)^2}\right]
\left[ (kp_i)+(k\bar p)\right] \bar \alpha _{1i} 
\end{array}
$$

\medskip\

\section{The results for $\limfunc{Tr}\,Q_2$}

$$
\begin{array}{rl}
\limfunc{Tr}\,Q_2(\bar p_r,\dbr{p}_{r\prime }) & = \limfunc{Tr}(\slashed{p}%
_f+m)\left[ C1(\bar p,p_f,\gamma _\mu )\bar N_{1r}+C2(\bar p,p_f,\gamma _\mu
)\bar N_{2r}+C3(\bar p,p_f,\gamma _\mu )\bar N_{3r}\right]  \\[10pt]
& \times \;\; \left[ 
\dfrac{\slashed{\bar{p}}_r+m}{\bar p_r^2-m^2}\right] \left[ C1(\bar
p,p_i,\gamma _\nu )\bar M_{1r}+C2(p_i,\bar p,\gamma _\nu )\bar
M_{2r}+C3(p_i,\bar p,\gamma _\nu )\bar M_{3r}\right]  \\[10pt]
& \times \;\; (\slashed{p}_i+m)\left[ C1(\dbr{p},p_i,\gamma _\mu )\dbr{M}_{1r^{\prime
}}^{*}+C2(\dbr{p},p_i,\gamma _\mu )\dbr{M}_{2r^{\prime }}^{*}+C3(\dbr{p}%
,p_i,\gamma _\mu )\dbr{M}_{3r^{\prime }}^{*}\right]  \\[10pt]
& \times \;\; \left[ \dfrac{\slashed{\bar{p}}_{r\prime }+m}{\bar p_{r^{\prime 
}}^2-m^2}\right] ^{*}\left[
C1(\dbr{p},p_f,\gamma _\nu )\dbr{N}_{1r^{\prime }}^{*}+C2(p_f,\dbr{p},\gamma
_\nu )\dbr{N}_{2r^{\prime }}^{*}+C3(p_f,\dbr{p},\gamma _\nu )\dbr{N}%
_{3r^{\prime }}^{*}\right] 
\end{array}
$$

\medskip\ 

\underline{The coefficient of $\bar X_{11}^r[\dbr{X}_{11}^{r^{\prime }}]^{*}$%
}

$$
\begin{array}{l}
16m^{2} \left[ (p_{i}\bpr)+(p_{i}\dpr)+(p_{f}\bpr) +(p_{f}\dpr)+(p_{i}p_{f})+(\bpr\dpr)-2(\bpr\dpr)(p_{i}p_{f}) -2m^{2} \right]
\\ -8e^2a^2\left\{ (k_ip_i)\left[ 1-
\dfrac{(kp_f)}{(kp_i)}\right] \left[ 1-\dfrac{(k\dbr{p})}{(k\bar p)}\right]
-(k_fp_i)\left[ 1-\dfrac{(kp_f)}{(kp_i)}\right] \left[ 1-\dfrac{(k\bar p)}{(k%
\dbr{p})}\right] \right.  \\[10pt] \left. +(k_fp_f)\left[ 1-
\dfrac{(kp_i)}{(kp_f)}\right] \left[ 1-\dfrac{(k\dbr{p})}{(k\bar p)}\right]
-(k_ip_f)\left[ 1-\dfrac{(kp_i)}{(kp_f)}\right] \left[ 1-\dfrac{(k\bar p)}{(k%
\dbr{p})}\right] \right\}  \\[10pt]
 -8m^2e^2a^2\left[ (kp_i)+(kp_f)\right]^2\left[ \dfrac 1{(k\bar p)(k
\dbr{p})}+\dfrac 1{(kp_i)(kp_f)}\right]  \\[10pt] 
+32e^4a^4\left[ \dfrac{(kp_i)}{%
(kp_f)}+\dfrac{(kp_f)}{(kp_i)}+\dfrac{(k\bar p)}{(k\dbr{p})}+\dfrac{(k\dbr{p})}{%
(k\bar p)}\right] 
\end{array}
$$

\medskip\ 
\newpage
\underline{The coefficient of $\bar X_{11}^r[\dbr{X}_{12}^{r^{\prime }}]^{*}$%
}

$$
\begin{array}{l}
8m^2\left[ \dfrac 1{(k
\dbr{p})}+\dfrac 1{(kp_f)}\right] \left[ (kp_f)(k\bar p)\alpha _{\balph{1f}%
}-(kp_i)(k\dbr{p})\alpha _{\dalph{1i}}\right]  \\[10pt]
 -8m^2\left[ \dfrac 1{(k
\dbr{p})}-\dfrac 1{(kp_f)}\right] \left[ (kp_f)(k\dbr{p})\alpha _{\dalph{1f}%
}+(kp_i)(k\bar p)\alpha _{\balph{1i}}\right]  \\[10pt] 
-8(\alpha _{1i}-\alpha_{1f})\left\{ 
\dfrac{e^2a^2}{(k\dbr{p})}\biggl[
(kp_i)(k\dbr{p})+(kp_f)(k\bar p)\biggr] -m^2\left[ \dfrac 1{(k\dbr{p})}+\dfrac
1{(kp_f)}\right] (kp_i)(kp_f)+2(\bpr\dpr)(kp_i)\right\}  \\[10pt]
 -8(\alpha _{\bar
1}-\alpha _{\dbr{1}})\left\{ \dfrac{e^2a^2}{(kp_f)}\biggl[ (kp_i)(k\dbr{p}%
)+(kp_f)(k\bar p)\biggr] -m^2\left[ \dfrac 1{(k\dbr{p})}+\dfrac
1{(kp_f)}\right] (k\bar p)(k\dbr{p})+2(p_ip_f)(k\bar p)\right\} 
\end{array}
$$

\medskip\ 

\underline{The coefficient of $\bar X_{11}^r[\dbr{X}_{22}^{r^{\prime }}]^{*}$%
}

$$
\begin{array}{l}
16(k\bar p)\left[ (kp_f)-(kp_i)\right] \left[ (\alpha _{1f}-\alpha
_{1i})(\alpha _{\bar 1}-\alpha _{
\dbr{1}})\right] +16e^2a^2\left[ (p_ip_f)\dfrac{(k\bar p)}{(k\dbr{p})}-(\bpr%
\dpr)\right] \\[10pt]
 8e^2a^2m^2\left\{ 
\dfrac{(k\bar p)}{(k\dbr{p})}\left[ 1-\dfrac{(kp_f)}{(kp_i)}\right] -\dfrac{%
(k\bar p)}{(kp_f)}\left[ \dfrac{(k\dbr{p})}{(kp_i)}+\dfrac{(kp_i)}{(k\dbr{p})}%
\right] +\dfrac{(k\dbr{p})}{(kp_i)}+\dfrac{(k\dbr{p})}{(kp_f)}\right\} \\[10pt] 
+8e^2a^2\left\{ \dfrac{(p_f\bpr)}{(kp_f)}-\dfrac{(p_i\bpr)}{(kp_i)}-\dfrac{%
(k\bar p)}{(k\dbr{p})}\left[ \dfrac{(\dpr p_f)}{(kp_f)}-\dfrac{(\dpr p_i)}{%
(kp_i)}\right] \right\} \left[ (kp_f)-(kp_i)\right] 
\end{array}
$$

\medskip\ 

\underline{The coefficient of $\bar X_{12}^r[\dbr{X}_{21}^{r^{\prime }}]^{*}$%
}

$$
\begin{array}{l}
8e^2a^2m^2\left\{ 4+\left[ 
\dfrac{(k\dbr{p})}{(kp_i)}+\dfrac{(kp_i)}{(k\dbr{p})}\right] \left[ \dfrac{%
(k\bar p)}{(kp_f)}+\dfrac{(kp_f)}{(k\bar p)}\right] -\dfrac{(k\bar p)}{(kp_i)}%
- \dfrac{(kp_i)}{(k\bar p)}-\dfrac{(k\dbr{p})}{(kp_f)}-\dfrac{(kp_f)}{(k\dbr{p})%
}\right\} \\[10pt]
 -8e^2a^2\left[ (p_f \dpr)\dfrac{(kp_i)}{(kp_f)}-(p_i\dpr)\right] \left[ 
\dfrac{(kp_f)}{(kp_i)}+ \dfrac{(k\bar p)}{(k\dbr{p})}\right] \\[10pt]
 -8e^2a^2\left[ (p_i\bpr)\dfrac{(kp_f) 
}{(kp_i)}-(p_f\bpr)\right] \left[ \dfrac{(kp_i)}{(kp_f)}+\dfrac{(k\dbr{p})}{%
(k\bar p)}\right] -8e^4a^4\left[ {\dfrac{(k\dbr{p})(kp_f)}{(k\bar p)(kp_i)}}+{%
\ \dfrac{(k\bar p)(kp_i)}{(k\dbr{p})(kp_f)}}+2\right] 
\end{array}
$$

\medskip\ 

\underline{The coefficient of $\bar X_{11}^r[\dbr{X}_{23}^{r\pr}]^{*}$}

$$
\begin{array}{l}
8(k\bar p)\left[ (kp_f)-(kp_i)\right] \left[ (\alpha _{2f}-\alpha
_{2i})(\alpha _{\bar 1}-\alpha _{\dbr{1}})+(\alpha _{1f}-\alpha
_{1i})(\alpha _{\bar 2}-\alpha _{\dbr{2}})\right] 
\end{array}
$$

\medskip\ 

\underline{The coefficient of $\bar X_{12}^r[\dbr{X}_{31}^{r^{\prime }}]^{*}$%
}

$$
\begin{array}{l}
-8\left[ (kp_f)(k\dbr{p})-(kp_i)(k\bar p)\right] \left[ (\alpha _{2f}-\alpha
_{2i})(\alpha _{\bar 1}-\alpha _{\dbr{1}})-(\alpha _{1f}-\alpha
_{1i})(\alpha _{\bar 2}-\alpha _{\dbr{2}})\right] 
\end{array}
$$

\medskip\ 

\underline{The coefficient of $\bar X_{12}^r[\dbr{X}_{22}^{r\pr}]^{*}$}

$$
\begin{array}{l}
8e^2a^2\left[ (kp_f)(k\dbr{p})+(kp_i)(k\bar p)\right] \left[ \dfrac
1{(kp_f)}(\alpha _{\bar 1}-\alpha _{\dbr{1}})+\dfrac 1{(k\dbr{p})}(\alpha
_{1f}-\alpha _{1i})\right] 
\end{array}
$$

\medskip\ 

\underline{The coefficient of $\bar X_{12}^r[\dbr{X}_{32}^{r\pr}]^{*}$}

$$
\begin{array}{l}
8e^2a^2\left[ (kp_i)(k\bar p)-(kp_f)(k\dbr{p})\right] \left[ \dfrac
1{(kp_f)}(\alpha _{\bar 2}-\alpha _{\dbr{2}})-\dfrac 1{(k\dbr{p})}(\alpha
_{2f}-\alpha _{2i})\right] 
\end{array}
$$

\medskip\ 

\underline{The coefficient of $\bar X_{22}^r[\dbr{X}_{22}^{r\pr}]^{*}$}

$$
\begin{array}{l}
8e^4a^4\left[ \dfrac{(kp_i)}{(kp_f)}+\dfrac{(kp_f)}{(kp_i)}+\dfrac{(k\bar p)}{(k%
\dbr{p})}+\dfrac{(k\dbr{p})}{(k\bar p)}\right] 
\end{array}
$$

\medskip\ 

\underline{The coefficient of $\bar X_{23}^r[\dbr{X}_{23}^{r^{\prime }}]^{*}$%
}

$$
\begin{array}{l}
8e^4a^4\left[ \dfrac{(kp_i)}{(kp_f)}+\dfrac{(kp_f)}{(kp_i)}-\dfrac{(k\bar p)}{(k%
\dbr{p})}-\dfrac{(k\dbr{p})}{(k\bar p)}\right] 
\end{array}
$$
\chapter{The explicit form of certain functions of $M_j$ and $M_k$}
\label{appMM}

\begin{table}[!h]
\label{ap3.tab1}
\center{
\begin{tabular}{|c|c|c|c|} \hline
 $\alpha_{1x}M_{2} + \alpha_{2x}M_{3}$ & $\alpha_{1x}N_{2} + \alpha_{2x}N_{3}$
& $M_{2}N_{2} + M_{3}N_{3}$ & $\bar{M}_{2}\dbr{M}_{2}^{\ast} + \bar{M}_{3}
\dbr{M}_{3}^{\ast}$ \\ \hline
 $\alpha_{1x}M_{2} - \alpha_{2x}M_{3}$ & $\alpha_{1x}N_{2} - \alpha_{2x}N_{3}$
& $M_{2}N_{2} - M_{3}N_{3}$ & $\bar{N}_{2}\dbr{N}_{2}^{\ast} + \bar{N}_{3}
\dbr{N}_{3}^{\ast}$ \\ \hline
 $\alpha_{1x}M_{3} + \alpha_{2x}M_{2}$ & $\alpha_{1x}N_{3} + \alpha_{2x}N_{3}$
& $M_{2}N_{3} - M_{3}N_{2}$ & $\bar{N}_{2}\dbr{M}_{2}^{\ast} + \bar{N}_{3}
\dbr{M}_{3}^{\ast}$ \\ \hline
 $\alpha_{1x}M_{3} - \alpha_{2x}M_{2}$ & $\alpha_{1x}N_{3} - \alpha_{2x}N_{2}$
& $M_{2}N_{3} + M_{3}N_{2}$ & $\bar{N}_{2}\dbr{M}_{3}^{\ast} + \bar{N}_{3}
\dbr{M}_{2}^{\ast}$ \\ \hline
\end{tabular} }
\caption{\bf General algebraic functions involving $M_{i}$ and $N_{j}$}
\end{table} 

\medskip\

The trace calculation associated with the SCS and STPPP cross sections can be
divided into a series of terms involving trace coefficients (see Appendix \ref{trace})
and as co-products, simple algebraic functions of $M_j$ and $N_k$ defined in 
equation \ref{c3.smat.eq3}. In this appendix the most general of these algebraic functions, 
tabulated in table \ref{ap3.tab1}, are written down explicitly. The subscripts $x$ refer to 
either $i$ for initial state quantities or $f$ for final state quantities.

Evaluation of these algebraic functions of $M_j$ and $N_k$ , are based on
the basic properties of Bessel functions and involve exponential functions
of six angles defined by

\begin{equation}
\begin{array}{c}
\bar{\phi} = \bar{\phi}_{0i} - \bar{\phi}_{0f} \quad ; \quad \dbr{\phi} = \dbr{\phi}_{0i} - 
\dbr{\phi}_{0f} \quad;\quad \psi_{i} = \bar{\phi}_{0i} - \dbr{\phi}_{0i} \\
\psi_{f} = \bar{\phi}_{0f} - \dbr{\phi}_{0f} \quad;\quad \bar{\chi} = \bar{\phi}_{0i} - 
\dbr{\phi}_{0f} \quad;\quad \dbr{\chi} = \dbr{\phi}_ {0i} - \bar{\phi}_{0f} 
\end{array}
\end{equation}

\medskip\

$\alpha_{1x},\alpha_{2x},\phi_{0x} \text{ and } z_x$ are functions of scalar products of 
external field 4-vectors and obey the relations of equation \ref{c2.eq18} and \ref{c2.eq19}. 
Using the addition formulae for the Bessel functions \cite{Watson22}, $M_j$ and $N_k$ obey 
the relations of equation \ref{c3.symev.eq2} and

\newpage
\bigskip\

$$
\begin{array}{rcl}
\alpha_{1x}M_{2} + \alpha_{2x}M_{3} &=&  {\displaystyle \frac{z_{x}}{2}}
\left[J_{r-1}(z_{i})e^{i\phi_{0x}-i\phi_{0i}} + J_{r+1}(z_{i})e^{-i\phi_{0x}+i
\phi_{0i}}\right]e^{ir\phi_{0i}} \\[10pt]

\alpha_{1x}N_{2} + \alpha_{2x}N_{3}  &=& {\displaystyle \frac{z_{x}}{2}}
\left[J_{s-1}(z_{f})e^{-i\phi_{0x}+i\phi_{0f}} + J_{s+1}(z_{f})e^{i\phi_{0x}-i
\phi_{0f}}\right]e^{-is\phi_{0f}} \\[10pt]

\alpha_{1x}M_{2} - \alpha_{2x}M_{3}  &=& {\displaystyle \frac{z_{x}}{2}}
\left[J_{r-1}(z_{i})e^{-i\phi_{0x}-i\phi_{0i}} + J_{r+1}(z_{i})e^{i\phi_{0x} +
i\phi_{0i}}\right]e^{ir\phi_{0i}} \\[10pt]

\alpha_{1x}N_{2} - \alpha_{2x}N_{3}  &=&  {\displaystyle \frac{z_{x}}{2}}
\left[J_{s-1}(z_{f})e^{i\phi_{0x}+i\phi_{0f}} + J_{s+1}(z_{f})e^{-i\phi_{0x} -
i\phi_{0f}}\right]e^{-is\phi_{0f}} \\[10pt]

\alpha_{1x}M_{3} + \alpha_{2x}M_{2}  &=&  {\displaystyle i\frac{z_{x}}{2}}
\left[J_{r-1}(z_{i})e^{-i\phi_{0x}-i\phi_{0i}} - J_{r+1}(z_{i})e^{i\phi_{0x} + 
i\phi_{0i}}\right]e^{ir\phi_{0i}} \\[10pt]

\alpha_{1x}N_{3} + \alpha_{2x}N_{2}  &=&  {\displaystyle -i\frac{z_{x}}{2}}
\left[J_{s-1}(z_{f})e^{i\phi_{0x}+i\phi_{0f}} - J_{s+1}(z_{f})e^{-i\phi_{0x} -
i\phi_{0f}}\right]e^{-is\phi_{0f}} \\[10pt]

\alpha_{1x}M_{3} - \alpha_{2x}M_{2}  &=&  {\displaystyle i\frac{z_{x}}{2}}
\left[J_{r-1}(z_{i})e^{i\phi_{0x}-i\phi_{0i}} - J_{r+1}(z_{i})e^{-i\phi_{0x} +
i\phi_{0i}}\right]e^{ir\phi_{0i}} \\[10pt]

\alpha_{1x}N_{3} - \alpha_{2x}N_{2}  &=&  {\displaystyle -i\frac{z_{x}}{2}}
\left[J_{s-1}(z_{f})e^{-i\phi_{0x}+i\phi_{0f}} - J_{s+1}(z_{f})e^{i\phi_{0x} -
i\phi_{0f}}\right]e^{-is\phi_{0f}} \\[10pt]

M_{2}N_{2} + M_{3}N_{3}  &=&  \frac{1}{2}\left[J_{r-1}(z_{i})J_{s-1}(z_{f})
e^{-i\phi} + J_{r+1}(z_{i})J_{s+1}(z_{f})e^{i\phi} \right] e^{ir\phi_{0i}-is
\phi_{0f}} \\[10pt]

M_{2}N_{2} - M_{3}N_{3}  &=&  \frac{1}{2}\left[J_{r-1}(z_{i})J_{s+1}(z_{f})
e^{-i\phi_{0i}-i\phi_{0f}} + J_{r+1}(z_{i})J_{s-1}(z_{f})e^{i\phi_{0i}+i\phi
_{0f}} \right] e^{ir\phi_{0i}-is\phi_{0f}} \\[10pt]

M_{2}N_{3} + M_{3}N_{2}  &=&  i\frac{1}{2}\left[J_{r-1}(z_{i})J_{s+1}(z_{f})
e^{-i\phi_{0i}-i\phi_{0f}} - J_{r+1}(z_{i})J_{s-1}(z_{f})e^{i\phi_{0i}+i\phi
_{0f}} \right] e^{ir\phi_{0i}-is\phi_{0f}} \\[10pt]

M_{2}N_{3} - M_{3}N_{2}  &=&  -i\frac{1}{2}\left[J_{r-1}(z_{i})J_{s-1}(z_{f})
e^{-i\phi} - J_{r+1}(z_{i})J_{s+1}(z_{f})e^{i\phi} \right] e^{ir\phi_{0i}-
is\phi_{0f}} \\[10pt]

M_{2}M_{3'}^{\ast} - M_{3}M_{2'}^{\ast}  &=&  -i\frac{1}{2} \left[ J_{r-1}
(z_{i})J_{r'-1}(z_{i}) - J_{r+1}(z_{i})J_{r'+1}(z_{i}) \right] e^{ir\phi_{0i}-
is\phi_{0f}} \\[10pt]

N_{2}N_{3'}^{\ast} - N_{3}N_{2'}^{\ast}  &=&  i\frac{1}{2} \left[ J_{s-1}
(z_{f})J_{s'-1}(z_{f}) - J_{s+1}(z_{f})J_{s'+1}(z_{f}) \right] e^{ir\phi_{0i}-
is\phi_{0f}} \\[10pt]

M_{2'}^{\ast}N_{2} - M_{3'}^{\ast}N_{3}  &=&  \frac{1}{2} \left[ J_{r'-1}
(z_{i})J_{s+1}(z_{f}) e^{i\phi} - J_{r'+1}(z_{i})J_{s-1}(z_{f}) e^{-i\phi}
\right] e^{ir\phi_{0i}-is\phi_{0f}} \\[10pt]

\bar{M}_{2}\dbr{M}_{2}^{\ast} + \bar{M}_{3}\dbr{M}_{3}^{\ast}  &=&  \frac{1}{2}
\left[ J_{r-1}(z_{\bar{i}})J_{r'-1}(z_{\dbr{i}})e^{-i\psi_{i}} + J_{r+1}
(z_{\bar{i}})J_{r'+1}(z_{\dbr{i}})e^{i\psi_{i}} \right] e^{ir_{\bar{\phi}_
{0i}} - ir'_{\dbr{\phi}_{0i}}} \\[10pt]

\bar{N}_{2}\dbr{N}_{2}^{\ast} + \bar{N}_{3}\dbr{N}_{3}^{\ast}  &=&  \frac{1}{2}
\left[ J_{s-1}(z_{\bar{f}})J_{s'-1}(z_{\dbr{f}})e^{i\psi_{f}} + J_{s+1}
(z_{\bar{f}})J_{r'+1}(z_{\dbr{f}})e^{-i\psi_{f}} \right] e^{ir_{\bar{\phi}_
{0i}} - ir'_{\dbr{\phi}_{0i}}} \\[10pt]

\bar{N}_{2}\dbr{M}_{2'}^{\ast} - \bar{N}_{3}\dbr{M}_{3'}^{\ast}  &=&  
\frac{1}{2} \left[ J_{r'-1}(\dbr{z}_{i})J_{s+1}(\bar{z}_{f}) e^{i\dbr{\chi}} 
- J_{r'+1}(\dbr{z}_{i})J_{s-1}(\bar{z}_{f}) e^{-i\dbr{\chi}} \right] 
e^{ir'\dbr{\phi}_{0i}-is\bar{\phi}_{0f}} \\[10pt]

\bar{N}_{2}\dbr{M}_{3}^{\ast} - \bar{N}_{3}\dbr{M}_{2}^{\ast}  &=& -i\frac{1}{2}
\left[ J_{r\pr -1}(z_{\dbr{i}})J_{s+1}(z_{\bar{f}})e^{i\dbr{\chi}} - J_{r'+1}
(z_{\dbr{i}})J_{s-1}(z_{\bar{f}})e^{-i\dbr{\chi}} \right] e^{-ir\pr\phi_{\dalph
{0f}} - is_{\balph{0f}}} 
\end{array}
$$
\chapter{Solution to $\sum\limits_{n=-\infty}^{\infty}
\frac{1}{n+a} \left(\frac{z_1}{z_2}\right)^n J_n(z_1)J_{n-l}(z_2)$}
\label{sumsol}

Second order IFQED processes in a circularly polarised electromagnetic field involve 
infinite summations of products of Bessel functions, exponential functions and algebraic 
functions of the form

\begin{equation}
\label{appsum.eq1}
\begin{array}{c}
\dsum\limits_{n=-\infty}^{\infty} \dfrac{1}{n+a}
\left(\dfrac{z_1}{z_2}\right)^n J_n(z_1) J_{n-l}(z_2)
\end{array}
\end{equation}

\medskip\

Numerical evaluation of the second order IFQED cross sections would be substantially 
simplified if an analytic solution of equation \ref{appsum.eq1} could be obtained. 
No such solution exists in the literature. Presented here is the analytic solution of a 
related summation which is, as far as can be established, an original result.

The starting point is Ramanujan's Integral (\cite{Watson22}, pg.449). This integral is in 
the form of a Fourier transform. So taking in the inverse Fourier transform

\begin{equation}
\label{appsum.eq2}
\begin{array}{rl}

\dfrac{1}{2\pi} \dint^{\pi}_{\nths -\pi} \left[2\cos\dfrac{\phi}{2}\right]^{-l} 
\,\text{e}^{il\phi/2}\,w^l\,J_{-l}(w)\,\text{e}^{-in\phi}\;\text{d}\phi & = 
\dfrac{J_{-n}(z_1)}{z_{1}^{-n}} \, \dfrac{J_{n-l}(z_2)}{z_{2}^{n-l}} \\[20pt]
\text{where} \quad\quad w =[(z_{1}^{2}\text{e}^{i\phi/2}+z_{2}^{2}\text{e}^{-i\phi/2})2\cos 
\phi/2]^{1/2} &

\end{array}
\end{equation}

\medskip\

Next, a summation involving an exponential divided by a polynomial is required 
(\cite{Bromwich26}, pg.370)

\begin{equation}
\label{appsum.eq3}
\begin{array}{c}

\dsum\limits_{n=-\infty}^{\infty} \dfrac{(-1)^n\text{e}^{-in\phi}}{n+a} = \pi \csc \pi a \, 
\text{e}^{ia\phi} \quad;\quad -\pi<\phi<\pi 

\end{array}
\end{equation}

\medskip\

Both sides of equation \ref{appsum.eq3} are multiplied by the integrand of the left hand 
side of equation \ref{appsum.eq2} and the integration over $\phi$ is performed on the right 
hand side

\begin{equation}
\label{appsum.eq4}
\begin{array}{c}

\dint^{\pi}_{\nths -\pi} \,
\dsum\limits_{n=-\infty}^{\infty} \left[2\cos\dfrac{\phi}{2}\right]^{-l} 
\,\text{e}^{il\phi/2}\,w^l\,J_{-l}(w)\,\text{e}^{-in\phi}\;\text{d}\phi = 
\pi \csc \pi a \dfrac{J_{a}(z_1)}{z_{1}^{a}} \, \dfrac{J_{-a-l}(z_2)}{z_{2}^{-a-l}}

\end{array}
\end{equation}

\medskip\

The integrand on the left hand side of equation \ref{appsum.eq4} is uniformly convergent on 
the interval $-\pi<\phi<\pi$ as long as $n+a\neq0$. The summation 
and the summation terms not dependent on $\phi$ can be taken outside the integral. The 
integration can then be performed using equation \ref{appsum.eq2} and a common factor 
$(z_{2})^{l}$ can be removed from both sides. A condition for this identity is that $l<1$.

\begin{equation}
\label{appsum.eq5}
\begin{array}{c}

\boxed{
\dsum\limits_{n=-\infty}^{\infty} \dfrac{1}{n+a} \left(\dfrac{z_1}{z_2}\right)^n 
J_n(z_1)J_{n-l}(z_2) = (-1)^l \pi \csc \pi a \left(\dfrac{z_2}{z_1}\right)^a 
J_{a}(z_1) J_{-a-l}(z_2)
}

\end{array}
\end{equation}

%-----------------------------------------------------------------------------
\chapter{Dispersion Relation Method used in self energy Calculations}
%-----------------------------------------------------------------------------
\label{app.kallen}

The radiative corrections made to the QED processes considered in this thesis, 
involve calculation of the electron energy shift. The initial analytic expression 
contains an integration over 4 momenta of a product of electron and photon 
propagators

\begin{equation}
\label{app6.eq1}
\begin{array}{c}
\Delta \epsilon _p =\dfrac{ie^2}{(2\pi)^4} \underset{b}{\dsum}\dint 
d^4k'\,d^4q'\,\delta^4(q'-q-k'-bk)
 \text{Tr}\left\{ \mathcal{F}(q',k')\right\} \,
 \dfrac{1}{q'^2-m_{*}^2}\,\dfrac 1{k^{\prime 2}}
\end{array}
\end{equation}

\medskip\

We require expressions for both the real and imaginary parts of the electron energy Shift.
The method will be to transform equation \ref{app6.eq1} into the form of a 
dispersion relation (\cite{Kallen72} pg.112; \cite{Muirhead65} pg.410).

The integration over $k'$ can be carried out immediately. The presence of the Dirac 
delta function results in $k'$ being replaced by other 4-momenta. A shift in integration 
variable $q'$ is made so that

\begin{equation}
\label{app6.eq2}
\begin{array}{c}
 q' \rightarrow q'+\frac{1}{2}D \\
 k' \rightarrow q'-\frac{1}{2}D \\
    \text{where} \quad\quad D=q+bk
\end{array}
\end{equation}

\medskip\

Two new integrations over symbols $K(\equiv q'+\frac{1}{2}D)$ 
and $L(\equiv q'-\frac{1}{2}D)$ are introduced by use of identities

\begin{equation}
\label{app6.eq3}
\begin{array}{rl}

\frac 1{(q'+\frac{1}{2}D)^2-m_*^2+i0} &=\dint d^4K\delta (K^2-m^2)\delta^3(\squ{q'}
 +\frac{1}{2}\sql{D}-\sql{K}) \left[ \frac{\Theta (K_0)}{q'_0 + \frac{1}{2} D_0-K_0+i0}-
 \frac{\Theta (-K_0)}{q'_0 + \frac{1}{2} D_0-K_0-i0} \right]  \\[10pt]

\frac 1{(q'-\frac{1}{2}D)^2+i0} &=\dint d^4L\delta (L^2)\delta^3(\squ{q'}
 -\frac{1}{2}\sql{D}-\sql{L}) \left[ \frac{\Theta (L_0)}{q'_0 - \frac{1}{2} D_0-L_0+i0}-
 \frac{\Theta (-L_0)}{q'_0 - \frac{1}{2} D_0-L_0-i0} \right]  

\end{array}
\end{equation}

\medskip\ 

Making the shifts in equation \ref{app6.eq2} and using the identities in equation 
\ref{app6.eq3}, the electron energy shift becomes 

\begin{equation}
\label{app6.eq4}
\begin{array}{rl}
\Delta \epsilon _p(D^2) & = \dfrac{e^2}{(2\pi)^4} \underset{b}{\sum} \dint 
d^4q'\;d^4K\;d^4L\; \text{Tr}\left\{ 
\mathcal{F}(K,L)\right\} \delta (K^2-m^2)\delta (L^2) \\[10pt]  
 & \times \;\; \delta ^3(\squ{q'}+\frac{1}{2}\sql{D}-\sql{K})\delta ^3(\squ{q'}-\frac{1}{2}
   \sql{D}-\sql{L})\left[ \frac{\Theta (K_0)}{q'_0+\frac{1}{2}D_0-K_0+i0}-\frac{\Theta 
   (-K_0)}{q'_0+\frac 12D_0-K_0-i0}\right]  \\[10pt]  
 & \times \;\; \left[ \frac{\Theta (L_0)}{q'_0-\frac{1}{2}D_0-L_0+i0}-
    \frac{\Theta (-L_0)}{q'_0-\frac{1}{2}D_0-L_0-i0}\right] 
\end{array}
\end{equation}

\medskip\ 

A contour integration over $q'_0\;(d^4q'=d\squ{q'}dq'_0)$ gives zero
for terms containing $\Theta (K_0)\Theta (L_0)$ and $\Theta (-K_0)\Theta
(-L_0)$, and a $2\pi \sum \text{residues}$ form for the other terms. The integration over $\squ{q'}$ proceeds
easily with the aid of 3-delta functions and the result is

\begin{equation}
\label{app6.eq5}
\begin{array}{rl}
  \Delta \epsilon _p(D^2) &=\dfrac{2\pi ie^2}{(2\pi)^4} \underset{b}{\dsum} \dint 
d^4K\;d^4L\; 
  \text{Tr}\left\{\mathcal{F}(K,L)\right\} \delta (K^2-m^2)\delta (L^2) \\[10pt]

 & .\delta ^3(\sql{D}-\sql{K}+\sql{L})\left[ \frac{\Theta (K_0)(\Theta (-L_0)}
 {D_0-K_0+L_0+i0}-\frac{\Theta (-K_0)(\Theta (L_0)}{D_0-K_0+L_0-i0}\right] 
\end{array}
\end{equation}

\medskip\ 

The electron energy shift can be written in the form of a dispersion
relation by introducing

\begin{equation}
\label{app6.eq6}
\begin{array}{rl}
 \rho (\sigma ^2) &=\dfrac{2\pi ie^2}{(2\pi)^4} \underset{b}{\dsum} \dint d^4q'd^4Kd^4L 
\text{Tr}\left\{ \mathcal{F}(K,L)\right\} \delta (K^2-m^2)\delta (L^2) \\[10pt]

 &\times \;\; \delta ^4(\sigma -K+L)[\Theta (K_0)\Theta (-L_0)+\Theta (-K_0)\Theta (L_0)] \\[10pt]
 &\text{and} \quad\quad \Delta \epsilon _p(D^2) =\int d^4\sigma \rho (\sigma )
 \delta^3(\sql{D}-\sql{\sigma})\left[ \frac{\Theta (\sigma _{0)}}{D_0-\sigma _0+i0}-
 \frac{\Theta (-\sigma _0)}{D_0-\sigma _0-i0}\right]  \\[10pt] 
 &\text{   } \quad\quad \Delta \epsilon _p(D^2) \equiv \int_0^\infty d\sigma ^2\;
 \dfrac{\rho (\sigma ^2)}{D^2-\sigma ^2+i0}
\end{array}
\end{equation}

\medskip\ 

Comparing this last expression with the basic form of a dispersion relation
(\cite{Muirhead65} pg.410) the real and imaginary parts of the electron energy shift can be 
written 

\begin{equation}
\label{app6.eq7}
\begin{array}{rl}
\Im \Delta \epsilon _p(D^2) &= -2\pi \rho (D^2) \\ 
\Re \Delta \epsilon _p(D^2) &= \dfrac{1}{\pi} P\dint_{-\infty }^\infty 
\dfrac{\Im \Delta \epsilon _p(\sigma^2)}{\sigma ^2-D^2}d\sigma ^2
\end{array}
\end{equation}

\medskip\ 

and performing the integration over $L$ 

\begin{equation}
\label{app6.eq8}
\begin{array}{rl}
 \rho (D^2) &= \dfrac{2\pi ie^2}{(2\pi)^4} \dint d^4K \;
 \text{Tr}\left\{\mathcal{F}(K,K-D)\right\} \delta (K^2-m^2)
 \delta(D^2-2(KD)+m_*^2) \\  
 & .[\Theta (K_0)\Theta (-K_0+D_0)+\Theta (-K_0)\Theta(K_0-D_0)]
\end{array}
\end{equation}

\medskip\ 

The electron energy shift is an algebraic function of scalar products 
of 4-vectors and is therefore invariant with respect to proper 4-rotations of the 
coordinate frame. The delta functions and step functions in the expression for $\rho
(D^2)$ require that the integral goes to zero for space-like $D_\mu $, and
since proper 4-rotations leave time-vectors as time-vectors, $\rho (D^2)$
can be calculated in the reference frame in which $\sql{k}=0$ (\cite{Kallen72}, pg.112). 
The evaluation of $\rho (D^2)$ proceeds by performing the integration over K in 
4-dimensional spherical polar coordinates

\begin{equation}
\label{app6.eq9}
\begin{array}{rl}
\text{d}^4K & = -|K|^3\;\text{d}|K|\;\sinh \zeta \;\text{d}\cosh \zeta 
\;\text{d}\cos \theta \;\text{d}\phi  \\ 
|\sql{K}| & = |K|\sinh \zeta  \\ 
K_0 & = |K|\cosh \zeta 
\end{array}
\end{equation}

\medskip\ 

and rotating the coordinate system until $D^2=D_0^2$. The integrations over $|K|$ and 
$\cosh$ are straightforward and the trace calculation will result in a function 
$\mathcal{G}$ of scalar products and scattering angles. In any reference frame, the 
imaginary part of the electron energy shift turns out to be

\begin{equation}
\begin{array}{c}
\Im \Delta \epsilon _p(D^2) =-\dfrac{e^2}{16\pi} 
\underset{b}{\dsum}\dint\;\text{d}\cos\theta\;\text{d}\phi \;
\left(1-\dfrac{m_*^2}{D^2}\right)\; \mathcal{G}(D^2,(kq),\theta,\phi) \; 
\Theta (D^2-m_*^2) 
\end{array}
\end{equation}

\medskip\

The argument of the Heaviside step function proves crucial in the consideration of divergences. The 
inclusion of an infinitesimally small photon mass $\epsilon$ into the photon propagator denominator 
at the outset results in

\begin{equation}
\label{app6.eqlast}
\Theta (D^2-m_*^2) \rightarrow \Theta (D^2-(m_*+\epsilon)^2)
\end{equation}